\documentclass[twoside, 10pt, DIV12, a4paper, smallheadings, bibtotocnumbered, titlepage]{scrbook} %
\usepackage[german,american]{babel} %
\usepackage{ae}%
\usepackage[square,colon,authoryear,sort&compress]{natbib}
\usepackage[format=plain,font=small,labelfont=bf]{caption}%
\usepackage[]{subfig}
\usepackage{amsmath}
\usepackage{amsthm}
\usepackage{amsfonts}
\usepackage{amssymb}
\usepackage{bbm}%
\usepackage{multirow}%
 \newcommand{\titleeng}{Randomized Dynamical Decoupling Strategies and Improved One-Way Key Rates for Quantum Cryptography}
 \newcommand{\einreichdate}{29.01.09}
 \newcommand{\einreichdatelang}{29. Januar 2009}
\usepackage[dvips,
            pdftitle={\titleeng},
            pdfsubject={Dissertation},
            pdfauthor={Oliver Kern},
            pdfpagelabels,
                 bookmarks,
                 bookmarksopen = false, %
                 bookmarksopenlevel = {1},
                 bookmarksnumbered = true,
                 breaklinks = true,
                 linktocpage = true,
                 colorlinks = false,
                 linkcolor = red,
                 urlcolor  = magenta,
                 citecolor = blue,
                 anchorcolor = green,
                 hyperindex = false,
                 hyperfigures = false
                 ]{hyperref}
\usepackage{graphicx}
\usepackage{pstricks,pst-node}
\usepackage[vcentermath]{youngtab}
\usepackage{rotating}%
\usepackage{svgcolor}%
\usepackage[NoDate]{currvita}%

\typearea[current]{last}

\DeclareGraphicsExtensions{.eps} %
\newrgbcolor{mygray}{0.93 0.93 0.93}

\theoremstyle{plain}
\newtheorem{lem}{Lemma}[section] %
\newtheorem{cor}[lem]{Corollary} %
\newtheorem{thm}[lem]{Theorem}   %
\theoremstyle{definition}
\newtheorem{defi}{Definition}[section]
\theoremstyle{remark}
\newtheorem*{rem}{Remark}

\newcommand{\id}{\ensuremath{\mathcal{I}}}
\newcommand{\opl}[2]{\ensuremath{#1^{\vphantom{\dagger}}_{\!#2}}}
\newcommand{\opdl}[2]{\ensuremath{#1^\dagger_{\!#2}}} %
\newcommand{\optl}[2]{\ensuremath{#1^T_{\!#2}}} %
\newcommand{\opsl}[2]{\ensuremath{#1^\ast_{\!#2}}} %

\newcommand{\ketl}[2]{\ensuremath{\vert #1 \rangle^{\vphantom{\dagger}}_{\scriptstyle\!#2}}}
\newcommand{\bral}[2]{\ensuremath{ \protect{\vphantom{\rangle^\dagger}}_{\scriptstyle{#2\!}} \langle #1\vert}}
\newcommand{\braketl}[3]{\ensuremath{ \protect{\vphantom{\rangle^\dagger}}_{\scriptstyle{#3\!}} \langle #1\vert #2 \rangle^{\vphantom{\dagger}}_{\scriptstyle\!#3}}}
\newcommand{\ketbral}[3]{\ketl{#1}{#3}\bral{#2}{#3}}
\newcommand{\tr}{\operatorname{tr}}   %
\newcommand{\ket}[1]{\ensuremath{\vert#1\rangle}}
\newcommand{\bra}[1]{\ensuremath{\langle #1|}}
\newcommand{\braket}[2]{\ensuremath{\langle #1\vert#2\rangle}}
\newcommand{\ketbra}[2]{\ensuremath{\vert#1\rangle\!\langle #2\vert}}
\newcommand{\wt}{\operatorname{wt}}   %
\newcommand{\dt}{\operatorname{dist}} %
\newcommand{\vspan}{\operatorname{span}}
\newcommand{\sthat}{\ensuremath{\: \vert\: }}
\newcommand{\XZ}{\ensuremath{X\!Z}}

\newcommand{\comment}[1]{} %

\def\clap#1{\hbox to 0pt{\hss#1\hss}}  %

\def\mathclap{\mathpalette\mathclapinternal}

\def\mathclapinternal#1#2{%
\clap{$\mathsurround=0pt#1{#2}$}}

\begin{document}
\frontmatter

\title{Randomized Dynamical Decoupling Strategies and\\
Improved One-Way Key Rates for Quantum Cryptography\\
\vspace{3\baselineskip}
\includegraphics[scale=0.25]{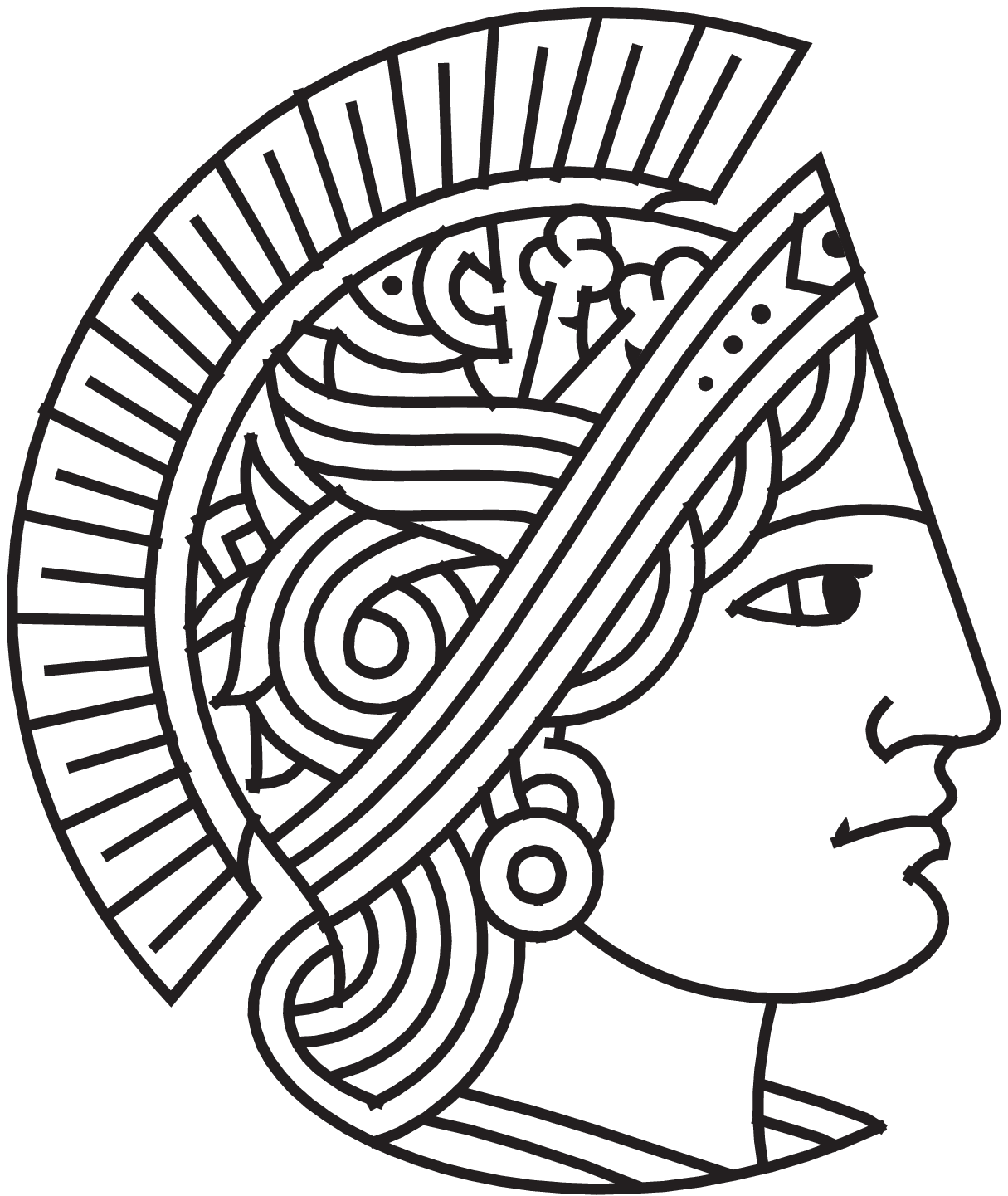}}

\selectlanguage{german}
\author{\\
\\
Vom Fachbereich Physik\\
der Technischen Universit\"at Darmstadt\\
zur Erlangung des Grades\\
eines Doktors der Naturwissenschaften\\
(Dr. rer. nat.)\\
\\
genehmigte Dissertation von\\
Dipl.-Phys. Oliver Kern\\
aus Mainz}

\date{Darmstadt 2009\\ D17\vspace{-1.5cm}} %

\lowertitleback{
\makebox[38mm][l]{Referent:} Prof. Dr. G. Alber\\
\makebox[38mm][l]{Korreferent:} Prof. Dr. J. Berges\\
\\
\makebox[38mm][l]{Tag der Einreichung:} \einreichdate \\
\makebox[38mm][l]{Tag der Pr\"ufung:} 25.05.09
}
\maketitle%
\addtolength{\oddsidemargin}{-7mm}%
\addtolength{\topmargin}{8mm}

\selectlanguage{american}
\begin{center}
\begin{minipage}{0.85\textwidth}\setlength{\parindent}{1em}
\chapter*{\centering \titleeng}
\section*{\centering Abstract}

The present thesis deals with various methods of quantum error correction.
It is divided into two parts.
In the first part, dynamical decoupling methods are considered which have the task of suppressing the influence of residual imperfections in a quantum memory.
Such imperfections might be given by couplings between the finite dimensional quantum systems (qudits) constituting the quantum memory, for instance.
The suppression is achieved by altering the dynamics of an imperfect quantum memory with the help of a sequence of local unitary operations applied to the qudits.
Whereas up to now the operations of such decoupling sequences have been constructed in a deterministic fashion, strategies are developed in this thesis which construct the operations by random selection from a suitable set.
Formulas are derived which estimate the average performance of such strategies.
As it turns out, randomized decoupling strategies offer advantages and disadvantages over deterministic ones.
It is possible to benefit from the advantages of both kind of strategies by designing combined strategies.
Furthermore, it is investigated if and how the discussed decoupling strategies can be employed to protect a quantum computation running on the quantum memory.
It is shown that a purely randomized decoupling strategy may be used by applying the decoupling operations
and adjusted gates of the quantum algorithm in an alternating fashion.
Again this method can be enhanced by the means of deterministic methods in order to obtain a combined decoupling method for quantum computations analogously to the combining strategies for quantum memories.

The second part of the thesis deals with quantum error-correcting codes and protocols for quantum key distribution.
The focus is on the BB84 and the 6-state protocol making use of only one-way communication during the error correction and privacy amplification steps.
It is shown that by adding additional errors to the preliminary key (a process called noisy preprocessing) followed by the use of a structured block code, higher secure key rates may be obtained.
For the BB84 protocol it is shown that iterating the combined preprocessing leads to an even higher gain.
In order to speed up the numerical evaluation of the key rates, results of representation theory come into play.
If a coherent version of the protocol is considered, the block code used in the preprocessing stage becomes a concatenated stabilizer code which is obtained by concatenating an outer random code with an inner deterministic one.
This concatenated stabilizer code is used to compute an improved lower bound on the quantum capacity of a certain quantum channel (the so-called qubit depolarizing channel).

\end{minipage}
\end{center}
\cleardoublepage
\selectlanguage{german}
\begin{center}
\begin{minipage}{0.85\textwidth}\setlength{\parindent}{1em}
\chapter*{\centering Zufallsbasierte dynamische Entkopplungsmethoden und verbesserte Schl"usselraten f"ur die Quantenkryptographie mit Einwegkommunikation}
\section*{\centering Zusammenfassung}
Die vorliegende Arbeit befa"st sich mit verschiedenen Methoden der Quantenfehlerkorrektur.
Sie ist in zwei Teile gegliedert.
Im ersten Teil werden dynamische Entkopplungsmethoden betrachtet, welche die Aufgabe haben, den Einflu"s verbleibender Unvollkommenheiten in einem Quantenspeicher zu unterdr"ucken.
Solche Unvollkommenheiten sind z.\,B. gegeben durch Kopplungen zwischen den einzelnen endlichdimensionalen Quantensystemen (Qudits), welche zusammen den Quantenspeicher bilden.
Um die Unterdr"uckung zu realisieren, wird die Dynamik eines fehlerbehafteten Quantenspeichers mit Hilfe einer Sequenz von lokalen unit"aren Operationen, die auf die einzelnen Qudits angewandt werden, modifiziert.
W"ahrend die Operationen einer solchen Entkopplungssequenz bislang deterministisch ausgew"ahlt wurden, werden in dieser Arbeit Strategien entwickelt, welche die Operationen durch zuf"allige Auswahl aus einer geeigneten Menge bestimmen.
Es werden Formeln hergeleitet, welche die mittlere Leistung solcher Strategien absch"atzen.
Dabei zeigt sich, da"s die zufallsbasierten dynamische Entkopplungsstrategien gegen"uber den deterministischen Vor- und Nachteile bieten.
Es ist m"oglich von den Vorteilen beider Arten von Strategien zu profitieren, indem man geeignete kombinierte Strategien entwickelt.
Weiterhin wird untersucht, inwiefern sich die diskutierten Entkopplungsstrategien einsetzen lassen, um eine auf dem Quantenspeicher laufende Quantenrechnung zu sch"utzen.
Es wird gezeigt, da"s sich eine rein zufallsbasierte Entkopplungsmethode verwenden l"a"st, indem speziell angepa"ste Gatter des zu rechnenden Quantenalgorithmus und Entkopplungsoperationen abwechselnd angewandt werden.
Diese Methode l"a"st sich wiederum mittels deterministischer Verfahren erweitern um analog zu den kombinierten Enkopplungsstrategien f"ur Quantenspeicher kombinierte Entkopplungsmethoden f"ur Quantenrechner zu erhalten.

Im zweiten Teil der Arbeit geht es um quantenfehlerkorrigierende Codes und quantenkryptographische Protokolle.
Es wird das BB84- und das 6-State-Protokoll zur sicheren Schl"usselverteilung unter Verwendung von Einwegkommunikation w"ahrend der Fehlerkorrektur und Privatsp"arhenverst"arkung betrachtet.
Es wird gezeigt, da"s durch das nachtr"agliche Hinzuf"ugen von Fehlern im vorl"aufigen Schl"ussel (\glqq noisy preprocessing\grqq{}) in Verbindung mit der Nutzung eines bestimmten Blockcodes h"ohere Schl"usselraten erzielt werden k"onnen.
F"ur das BB84-Protokoll wird weiter gezeigt, da"s sich die erzielten Vorteile verst"arken lassen, falls das kombinierte \glqq preprocessing\grqq{} iterativ verwendet wird.
Die numerische Berechnung der jeweiligen Schl"usselraten wird dabei durch das Verwenden von Resultaten der Darstellungstheorie beschleunigt.
Bei einer koh"arenten Betrachtung der Protokolle entspricht der verwendete Blockcode einem verketteten Stabilizer-Code, bei dem ein "au"serer zuf"alliger Code mit einem deterministischen inneren Code verkettet wird.
Mittels dieses verketteten quantenfehlerkorrigierenden Codes wird eine verbesserte untere Schranke f"ur die Quantenkapazit"at eines bestimmten Quantenkanals (genannt \glqq qubit depolarizing channel\grqq{}) berechnet.

\end{minipage}
\end{center}
\selectlanguage{american}
\cleardoublepage

\tableofcontents

\mainmatter

\chapter{Introduction and Preliminaries}%

\section{Introduction and Outline}

Quantum mechanics is a theory which appears rather counterintuitive:
For instance, certain variables of a quantum mechanical system like position and momentum cannot both be determined with arbitrary accuracy, particles might penetrate a barrier (tunnel effect), and cats might be dead and alive at the same time \cite{Schr35}.
Quantum information theory tries to generalize classical information theory to the quantum world by considering a quantum mechanical two-level system (a qubit) as basic information carrier.
It turns out that the properties of quantum information, i.\,e. the information encoded in a quantum system, are in strong contrast to the properties we now about classical information:
While classical information can be copied perfectly (resulting in the enormous success of file sharing networks), quantum information, in general, cannot be copied (no cloning theorem \cite{Dieks82,WZ82}).
Although it cannot be duplicated, quantum information may be teleported \cite{BBC93} by using distributed entangled states as a resource.
During the last two and a half decades the idea emerged to use quantum mechanics to implement technical applications which might not exist in a purely classical world.
Two particularly important concepts are quantum computing (promising faster computation) and quantum cryptography (promising unconditionally secure communication).

The first example of a quantum algorithm that is more efficient than any possible classical algorithm is the Deutsch-Jozsa algorithm \cite{DJ92}.
Given a black box quantum computer known as an oracle that implements a binary function which is either constant or balanced, the algorithm is able to determine if the function is constant or balanced by using the oracle only once.
By contrast, a classical computer might have to use a corresponding classical oracle on more than half the input values in the worst case.
The reason behind this speedup is the quantum parallelism which arises from the ability of a quantum memory to exist in a coherent superposition of states.
While the Deutsch-Jozsa algorithm is merely of academic interest, the potential of quantum computing drew lots of attention in 1994 when Shor presented polynomial-time algorithms for prime factorization and discrete logarithms \cite{shorpol} due to their potential to break current cryptosystems:
Nearly all current cryptosystems can be divided into two families.
One family is based on the assumption that an efficient algorithm for prime factorization does not exist (an example is the RSA public key encryption protocol~\cite{RSA}),
while the other one is based on the assumption that computation of the discrete logarithm is hard (examples are the Diffie-Hellman key distribution protocol \cite{DiHe76} and the Elgamal public key encryption protocol \cite{Elg85}).
The security of such cryptosystems is then provided by the fact that an eavesdropper with limited computational power is unable to solve these hard problems.
With Shor's algorithms the only reason that current cryptosystems can still be considered save is the tremendous difficulty to build a working quantum computer.
While the factorization of the number $15=3\times 5$ has been demonstrated on an nuclear magnetic resonance (NMR) quantum computer consisting of 7 qubits \cite{VSB01}, factorization of a 1024 bit number requires about 2000 qubits~\cite{PZ03} and is completely out of range of current technology.

The main obstacle in the realization of a quantum computer is a process called decoherence which arises from an interaction of the quantum information carriers with the environment and which destroys any coherent superpositions on a very short time scale.
But even if the quantum computer could be perfectly isolated, it still has to be accessible to perform manipulations (quantum gates) with very high accuracy. In addition, any imperfections and interactions between the finite-dimensional quantum information carriers (qudits) constituting the quantum memory of the quantum computer tend also to shorten the time scale of reliable quantum computation.
A way out might be the use of quantum error-correcting techniques:
One technique is to employ quantum error-correcting codes, introduced by Shor in 1995~\cite{shor95qec}.
By encoding the quantum information in a subspace of the total available space, they allow a recovery step involving a syndrome measurement to reverse certain decoherence processes.
Another technique is called dynamical decoupling \cite{Za99,VKL99}.
It alters the dynamics of a quantum memory by applying a series of local unitary operations to the qudits.
If the couplings to the environment effectively cancel out in the resulting dynamics, decoherence is suppressed.
This technique is inspired by refocusing techniques in NMR spectroscopy \cite{nmrbook}.

While on the one hand quantum mechanics seems to question the security of established classical cryptosystems,
on the other hand it provides key distribution protocols whose security does not rely on any unproven assumption but is guaranteed by the validity of quantum mechanics itself.
The idea of quantum key distribution (QKD) is due to Bennett and Brassard who, inspired by a paper of Wiesner written in the late 60's and not accepted for publication until 1983 \cite{Wie83}, invented the first QKD protocol (now called BB84 protocol) in 1984 \cite{BB84}.
To establish a secret key between two distant parties connected via a quantum channel and an authenticated classical channel, a QKD protocol demands one party (usually called Alice) to send non-orthogonal quantum states to the other party (usually called Bob).
The no cloning theorem \cite{Dieks82,WZ82} prevents an eavesdropper with full access to the quantum channel from copying these states in a perfect manner.
Hence any action of an eavesdropper leaves traces in the transmitted states which can be recognized by Alice and Bob by comparing measurement and preparation data of a subset of randomly chosen check states.
Thereby, one takes the conservative point of view that any noise in the channel is caused by an eavesdropper.
As long as the action of an eavesdropper seems harmless enough (i.\,e. as long as the detected error rate is low enough), Alice and Bob should be able to generate a random, correct and secure key from their raw data.
A security proof for a QKD protocol typically gives a lower bound on the length of the secret key that can be obtained from the raw data for a given error rate.
While it is still in question whether it will ever be possible to build large-scale quantum computers, the first generation of QKD systems is already available commercially \cite{QKDhardware}.
The reason that a QKD system is so much easier to build than a quantum computer is that it shares only the need to prepare and measure quantum systems but does not need to store and manipulate them.
Since the key rates of such devices are typically low, it is important to find security proofs which allow the secure key generation rate for a given error rate to be as high as possible.

The goal of this thesis is to develop improved dynamical decoupling techniques in order to contribute to the field of quantum computing and to provide improved secret key generation rates for quantum key distribution protocols in order to contribute to the field of quantum cryptography.
To achieve this goal, the thesis deals with different kinds of quantum error correction:
Part I of the thesis studies the potential of randomized dynamical decoupling strategies which are able to stabilize a quantum memory and even a running quantum computation against residual imperfections and interactions.
Part II of the thesis considers quantum error-correcting codes which are used to compute improved lower bounds on the capacity of the qubit depolarizing channel.
Furthermore, these codes are used to obtain improved secret key rates for variants of QKD protocols like the BB84 protocol \cite{BB84} and the 6-state protocol~\cite{Bru98}.
The later protocol is a natural extension of BB84, which makes use of four different quantum states, and makes use of two additional quantum states.
A more detailed introduction and outline is given in the following subsections.

\subsection{Part I: Random Decoupling}

In order to use a quantum system as a quantum memory, or even more demanding, to use it for quantum computations, we must be able to apply some kind of control.
Let us assume that the quantum system is a quantum register formed by a set of qudits.
Usually the experimentally easiest kind of control is to apply single qudit gates realized by a local control Hamiltonian.
Dynamical control of a local Hamiltonian allows the time evolution of a quantum system to be modified.
A well known example is given by the refocusing techniques used to manipulate nuclear spin Hamiltonians \cite{nmrbook}.
There are various possible controls tasks:
For example, for a closed quantum system, we might want to simulate a time evolution according to a Hamiltonian which is different from the system Hamiltonian \cite{WRJB02,BDNB04}.
In particular, we might want to simulate a vanishing Hamiltonian, a task we call decoupling from now on.
For an open quantum system, we might try to suppress decoherence by simulating vanishing couplings with the environment \cite{Za99,VKL99}, or we might try to generate at least a noiseless subsystem \cite{Za00,VKL00}.
In the simplest control scenario, the so-called bang-bang control scenario,
the local control Hamiltonian generates a set of pulses belonging to a suitable control scheme
instantaneously.
Then, for all control tasks, the fundamental deterministic control strategy is to apply the pulses belonging to the control scheme in a cyclic manner over and over again.
The control scheme is designed in such a way that, in lowest order average Hamiltonian theory (AHT) \cite{nmrbook}, the resulting dynamics corresponds to the Hamiltonian to be simulated.
Assuming the pulses to be ideal, the finite time interval $\Delta t$ in between subsequent pulses is the only obstacle preventing a control task to be achieved in a perfect manner.

One of the goals of part I of this thesis is to devise and analyze improved control strategies which lead to a better performance for a fixed time interval $\Delta t$, or in other words, which lead to a suppression of the residual higher order terms in AHT.
Let us focus for now on the control task of decoupling a closed quantum system.
In the context of decoupling, a control scheme is said to be a decoupling scheme.
As we will see, the fundamental deterministic control strategy leads to an average fidelity decay which is quadratic in time.
Thereby, the strength of the decay is determined by the strength of the system Hamiltonian,
by the size of the decoupling scheme, and by the time interval $\Delta t$ in between subsequent pulses.
For example, an improved strategy commonly used by the NMR community is a symmetrized version of the fundamental strategy:
In spite of doubling the size of the decoupling scheme, it leads to a decrease of the strength of the fidelity decay but keeps its quadratic-in-time nature.
An interesting result, first observed in the author's diploma thesis \cite{DiplKern}
(see also \cite{parec}), is that a control strategy based on random selection of the elements of a decoupling scheme leads to a fidelity decay which is only linear in time.
Subsequently, randomized decoupling was proposed for open quantum systems by Viola and Knill \cite{VK05}, who confirmed the linear-in-time decay by constructing a strict lower bound for the worst case fidelity \cite{VK05,V05}.
Meanwhile control strategies combining the advantages of purely deterministic and randomized strategies have been devised by the author \cite{combi} and by Santos and Viola \cite{SV06,VS06}.

Since, in general, the pulses of a decoupling sequence interfere with the application of an additional Hamiltonian implementing a quantum gate, the protection of a running quantum computation against imperfections of the quantum memory is not straightforward \cite{VLK99}.
Another goal of this thesis is to study how the devised decoupling strategies might be used in order to protect quantum computations.
In the bang-bang control scenario, one option for deterministic strategies is to apply quantum gates instantaneously in between completed decoupling cycles.
Under the more realistic assumption that quantum gates (especially two-qudit gates) are generated within a finite time interval by the means of bounded controls, more advanced techniques are required in order to combine decoupling and computation.
For instance, the dynamically corrected gate (DCG) of Khodjasteh and Viola \cite{KhVi08} combines a single decoupling cycle with the generation of a quantum gate.
It turns out that the decoupling pulses of a randomized decoupling strategy can be alternated with especially adjusted quantum gates, a method which was called Pauli random error correction (\textsf{PAREC}) by the author and collaborators \cite{parec}.
In order to benefit from the advantages of both methods, DCGs might be combined with the \textsf{PAREC} method.
Another scenario arises if the two-qudit gates of a quantum computer are generated by the couplings between adjacent qudits.\pagebreak{} In this case a selective decoupling method is used which switches off all but the desired coupling. The fundamental selective decoupling strategy can be improved by combining it with a randomized decoupling strategy.

\subsubsection{Outline}

\paragraph{Chapter \ref{chap:dyncontrol}: Dynamical Decoupling.}
Chapter \ref{chap:dyncontrol} deals with dynamical decoupling strategies for quantum memories in the bang-bang control scenario.
In order to improve the fundamental deterministic decoupling strategy, new randomized strategies are considered.
The performance of these strategies is analyzed by (i) deriving formulas expressing the average fidelity and (ii) by considering the variance of the fidelity.
The chapter closes with a numerical simulation of any strategy on a quantum memory perturbed by Heisenberg interactions.
The idea of the embedded decoupling strategy was published in~\cite{combi}:
\begin{quotation}\noindent
O.~Kern and G.~Alber.\\
Controlling Quantum Systems by Embedded Dynamical Decoupling Schemes.\\
\emph{Phys. Rev. Lett.}, \textbf{95}(25), 250501 (2005). {arXiv}:quant-ph/0506038v1.
\end{quotation}

\paragraph{Chapter \ref{chap:compu}: Decoupling and Computation.}
This chapter focuses on the fundamental problem of combining dynamical decoupling and quantum computation.
Here, we allow the quantum gates as well as the decoupling pulses to be generated within a finite time interval.
After presenting an overview of known results, the \textsf{PAREC} method is proposed, which is based on alternating the decoupling pulses of a randomized decoupling strategy with specially adjusted quantum gates forming the quantum algorithm.
We derive a formula for the fidelity decay of a quantum computation perturbed by static imperfections with and without the \textsf{PAREC} method.
It is shown that the \textsf{PAREC} method is a realization of an idea of Prosen and \u{Z}nidari\u{c} \cite{ProZni01}, who proposed to stabilize a quantum computation against static imperfections by increasing the decay of the correlation function measuring the fidelity decay.
Eventually, we consider the dynamically corrected gates (Euler-DCGs) of Khodjasteh and Viola \cite{KhVi08} which correspond to an implementation of a deterministic decoupling strategy for the purpose of computation.
We propose to implement the \textsf{PAREC} method by using only Euler-DCGs in order to benefit from the advantages of both methods.
Some of the results of this chapter have already been published.
The \textsf{PAREC} method together with numerical evidence was already devised in the author's diploma thesis \cite{DiplKern} and has been published in~\cite{parec}:
\begin{quotation}\noindent
O.~Kern, G.~Alber, and D.~L. Shepelyansky.\\
Quantum error correction of coherent errors by randomization.\\
\emph{Eur. Phys. J. D}, \textbf{32}(1), 153--156 (2005). {arXiv}:quant-ph/0407262v1.
\end{quotation}
The comparison of the \textsf{PAREC} method with the idea of Prosen and \u{Z}nidari\u{c} together with a formula for the average fidelity for the special case of instantaneous gates and decoupling pulses was given in~\cite{GKAJ08}:
\begin{quotation}\noindent
D.~Geberth, O.~Kern, G.~Alber, and I.~Jex.\\
Stabilization of quantum information by combined dynamical decoupling and detected-jump error correction.\\
\emph{Eur. Phys. J. D}, \textbf{46}(2), 381--394 (2008). {arXiv}:0712.1480v1.
\end{quotation}

\paragraph{Chapter \ref{chap:decrec}: Selective Recoupling and Randomized Decoupling.}
Instead of implementing a two-qudit quantum gate with the help of an external gate Hamiltonian, a quantum computer might use existing inter-qudit couplings.
Now, a non-operation is implemented by using a decoupling scheme which effectively switches off all couplings.
To implement a certain two-qudit gate, a selective decoupling (or selective recoupling) scheme is employed which removes all but the desired coupling.
By drawing on a particular example, this chapter shows how a selective decoupling strategy can be improved by devising a combined selective decoupling strategy involving randomized decoupling.
While a corresponding combined decoupling strategy can be devised quite easily, the non-vanishing lowest order AHT term of the selective decoupling strategy makes things a bit more difficult.
This chapter is a slightly enhanced version of~\cite{recoup}:
\begin{quotation}\noindent
O.~Kern and G.~Alber.\\
Stabilizing selective recoupling schemes by randomization.\\
\emph{Phys. Rev. A}, \textbf{73}(6), 062302 (2006). {arXiv}:quant-ph/0602167v1. 
\end{quotation}

\paragraph{Appendix \ref{chap:oatables} and \ref{chap:qmaps}:}
Chapter \ref{chap:oatables} of the appendix contains some examples of difference schemes and orthogonal arrays. This data can be used to obtain decoupling schemes as explained in section \ref{sec:decschemes}.
Chapter \ref{chap:qmaps} explains how certain quantum maps can be implemented as quantum algorithms.
Such quantum algorithms are used in chapters \ref{chap:compu} and \ref{chap:decrec} as test algorithms for the numerical simulations of the \textsf{PAREC} method and the improved selective decoupling method, respectively.
More detailed information on quantum maps and their implementation on a quantum computer can be found in the author's diploma thesis \cite{DiplKern}.

\subsection{Part II: Codes and Cryptography}

One of the fundamental theorems in classical information theory is Shannon's noisy channel coding theorem~\cite{Sh48}.
If classical information is to be transmitted over a classical noisy channel, Shannon's theorem assures that the transmission can be performed error-free as long as the transmission rate is below a maximum rate.
This maximum rate is called the capacity of the channel.
To achieve an error-free transmission, error-correcting codes have to be employed.
It turns out that the full capacity of a channel can be achieved by using randomly constructed block codes.
In quantum information theory, the analogous theorem is the quantum noisy channel coding theorem which states that quantum information can be sent reliably over a noisy quantum channel as long as the transmission rate is below the quantum capacity of the channel.
Quantum information which is to be sent over a noisy quantum channel has to be encoded using quantum error-correcting codes.
Surprisingly, it turns out that in contrast to the classical case, randomly constructed quantum codes do not achieve the full capacity of a quantum channel:
By considering a concatenated quantum code obtained by encoding the information encoded by a random code one more time with a so-called cat code, Shor and Smolin showed that error-free transmission over the so-called qubit depolarizing channel becomes possible at a higher rate than achievable by the random code alone \cite{ShSm96,DiSS98}.
In this thesis we extend these calculations to cat codes of larger size and obtain improved lower bounds on the capacity of the qubit depolarizing channel.

The quantum capacity of a noisy quantum channel has a close connection with the security of a quantum key distribution protocol.
If the parties Alice and Bob are able to determine how the quantum channel (i.\,e. the eavesdropper) acts on the quantum states sent from Alice to Bob, they might use a quantum error-correcting code to transmit these states error-free, i.\,e. in such a way that the eavesdropper does not learn anything about them.
A security proof of the BB84 protocol following this idea was given by Lo and Chau \cite{LC99}.
Unfortunately, in order to implement such a protocol, Alice and Bob have to be able to manipulate quantum states during the encoding and the recovery step.
By making use of the special structure of a certain class of quantum codes, the so-called CSS codes \cite{CS96,St96}, Shor and Preskill showed that a protocol based on encoding the states is equivalent to the original prepare and measure protocol \cite{SP00}.
Hence, results on the achievable transmission rate over a certain type of quantum channels (so-called memoryless Pauli channels) with the help of CSS codes can be used to prove the security of certain QKD protocols up to a certain error rate.
As discussed in the previous paragraph, randomly constructed CSS codes give a lower bound on the obtainable secure key rate, but concatenation of such codes with deterministic ones leads to even better bounds.

Another way to improve the secret key rates is to add noise to the raw key bits before they are processed into the final key.
Such a procedure is known as local randomization or noisy preprocessing and was discovered by Renner et al. \cite{KGR05,KGR05b}.
A security proof of a QKD protocol involving noisy preprocessing is not so straightforward as the Lo-Chau or Shor-Preskill proof.
The difficulty lies in the fact that a security proof based on perfect quantum error correction assures that Alice and Bob could in principle share perfectly entangled states, which are sufficient but not necessary for the generation of a secure key \cite{H3O03}.
A more sophisticated security proof involving CSS codes and noisy preprocessing was given by Renes and Smith \cite{RS07}.
Recently it was shown by the same authors for BB84 that by combining both methods --- noisy preprocessing and the use of the concatenated cat code --- even higher secure key rates can be obtained \cite{SRS06}.
In this thesis it will be shown that these results can also be applied to the 6-state protocol.
Furthermore, an iterated version of the combined preprocessing protocol is considered.
In order to evaluate the formulas expressing the secure key rates efficiently, results from representation theory have to be used.
In this context a matlab program was developed which calculates the Schur basis of the Hilbert space of $n$ qudits of dimension~$q$.

\subsubsection{Outline}

\paragraph{Chapter \ref{chap:cecc}: Classical Error Correction}
This chapter provides an introduction to the theory of classical error-correcting codes.
The main focus is on linear codes. A linear code with $q^k$ codewords of length $n$ is a $k$ dimensional subspace of the space $\mathbb{F}_q^n$ containing all strings of length $n$ with entries from the field $\mathbb{F}_q$.
Shannon's noisy coding theorem is proven for the binary symmetric channel by using random linear codes and typical set decoding.

\paragraph{Chapter \ref{chap:qecc}: Quantum Error-Correcting Codes}
We introduce the theory of quantum error-correcting codes.
An important class of quantum codes are the so-called stabilizer codes which might be viewed as the linear codes of quantum error correction.
A stabilizer code encoding $k$ qudits of dimension $q$ into $n$ is characterized by a $k$ dimensional self-orthogonal subspace (called stabilizer) in the space $\mathbb{F}_q^{2n}$ with respect to a symplectic inner product.
We explain how a unitary encoding of such a code corresponds to an extension of a basis of the stabilizer to a hyperbolic basis of $\mathbb{F}_q^{2n}$.
Then we specialize in the class of CSS codes, which form a subclass of stabilizer codes with a direct connection to classical linear codes, and show how the description of an encoding can be simplified.
Finally, we discuss the concatenation of two stabilizer codes.

\paragraph{Chapter \ref{chap:qcapa}: Quantum Channel Capacity}
While Shannon's noisy coding theorem is one of the fundamental theorems of classical information theory, this chapter deals with the quantum analog of Shannon's noisy coding theorem.
For a certain class of channels --- so-called memoryless Pauli channels --- coding theorems are proven which provide lower bounds on the capacity.
These theorems use (i) random stabilizer codes, (ii) random CSS codes, and (iii) random stabilizer codes concatenated with deterministic inner ones for encoding and joint-typical set decoding to implement the recovery operation.
The last theorem is used in connection with a specific deterministic inner code --- a so-called cat code --- to obtain better lower bounds on the capacity of the qubit depolarizing channel.

\paragraph{Chapter \ref{chap:crypto}: Quantum Cryptography}
This chapter shows how the results of the combined preprocessing step for BB84 \cite{SRS06} can be applied to the 6-state protocol.
We make use of the detailed analysis of the concatenated cat code provided by chapters \ref{chap:qecc} and \ref{chap:qcapa}, and employ the security proof of Renner \cite[corollary 6.5.2]{phdrenner}.
In addition, for the BB84 protocol, an iterative version of this preprocessing scheme is considered.
It is explained how the secret key rates can be efficiently evaluated by using insights from representation theory.
The chapter is an enhanced version of the following article~\cite{KR08}:
\begin{quotation}\noindent
O.~Kern and J.~M. Renes.\\
Improved one-way rates for BB84 and 6-state protocols.\\
\emph{Quant. Inf. \& Comp.}, \textbf{8}(8/9), 0756--0772 (2008). {arXiv}:0712.1494v2.
\end{quotation}

\paragraph{Appendix \ref{chap:techres} and \ref{chap:schur}:}
Chapter \ref{chap:techres} of the appendix contains some technical results mainly concerning error-correcting codes.
Chapter \ref{chap:schur} explains the eigenfunction method \cite{CPW02} which can be used to obtain a computer program calculating the Schur transform.
The Schur transform is a unitary transformation relating the standard computational basis of $n$ qudits of dimension $q$ with the Schur basis associated with the representation theory of the symmetric group $\textsf{S}_n$ and the general linear group $\textsf{GL}_q$.

\section{Preliminaries}\label{sec:preliminaries}

The understanding of this thesis requires the knowledge of basic quantum mechanics and basic representation theory. In addition, the theory of error-correcting codes comes into play in part II.
This section provides a brief overview of the necessary fundamentals of classical information theory, quantum mechanics and representation theory.
An introduction to error-correcting codes will be given in chapter~\ref{chap:cecc}.
An overview over classical information theory can be found in the book of MacKay \cite{MacKay}.
For an introduction to quantum mechanics we refer to the two books of Cohen-Tannoudji et al. \cite{QMTannoudji}.
A comprehensive introduction to quantum computation and quantum information can be found in the book of Nielsen and Chuang \cite{nielsenchuang}, which contains also a brief introduction to quantum mechanics and classical information theory.
In addition we refer to the lecture notes of Preskill \cite{Preskill}.
An introduction to group representation theory can be found in the book of Tung \cite{Tung85}.

\subsection{Probabilities and Entropy}

A discrete random variable $X$ is characterized by a set of outcomes
$A=(a_1,\dots,a_s)$ ($s=\vert A\vert$) together with an associated probability distribution
$P=(p_1,\dots,p_s)$ such that $X$ takes the values $a_i \in A$ with probability $\Pr(X=a_i)=p_i$.
The probabilities $p_i$ are non-negative numbers which sum up to one.
The uncertainty of the outcome a random variable is characterized by the Shannon entropy of its probability distribution.
\begin{defi}[Shannon entropy]
The $q$-ary Shannon entropy of a discrete probability distribution $P=(p_1,\dots,p_s)$ is defined as
\begin{equation}
H_{s[\log_q]}(P) = - \sum_{i=1}^s p_i \log_q p_i,
\end{equation}
where the value of $0 \log_q 0$ is taken to be $0$, which is consistent with the limit $\lim_{p\to0} p\log_q p = 0$.
\end{defi}
\noindent
Alternatively, we might say that the entropy of the random variable $X$ is given by
\begin{equation}
H_{s[\log_q]}(X) = - \sum_{a\in A} \Pr(a) \log_q \Pr(a).
\end{equation}
If the logarithm is taken to the base $2$ the entropy is expressed in bits.
Otherwise we stress such a fact by denoting the base $b$ as $H_{[\log_b]}$.
If a discrete probability distribution $P$ consists of $s$ elements, we write $H_s(P)$ to indicate the number of summands.
For $s=2$ it is sufficient to denote the first element of a probability distribution $P=(p,1-p)$, i.\,e. we write $H_2(p) \equiv  H_2(P) = H_2(p,1-p)$.

Let us consider an additional random variable $Y$ which is characterized by the set of outcomes
$B=(b_1,\dots,b_r)$ ($r=\vert B\vert$) and the probability distribution $Q=(q_1,\dots,q_r)$.
Then the conditional entropy of $X$ given $Y$ is defined as
\begin{equation}
\begin{split}
H_{[\log_q]}(X|Y)  &= \sum_{b\in B}  \underbrace{H_{[\log_q]}(X|Y=b)}_{-\sum_{a\in A}\,\Pr(a|b)\,\log_q\,\Pr(a|b)} \!\!\!\!\!\!\!\!\cdot \Pr(b)\\
 &= -\sum_{b\in B, a\in A} \Pr(a,b)\log_q\,\Pr(a|b).
\end{split}
\end{equation}
If $X$ and $Y$ are independent random variables, i.\,e. if $\Pr(a_i,b_j)=\Pr(a_i)\Pr(b_j)=p_i q_j$, it follows that $H_{[\log_q]}(X|Y) = H_{[\log_q]}(X)$.
For general $X$ and $Y$ the relation $H_{[\log_q]}(X|Y) = H_{[\log_q]}(X,Y) - H_{[\log_q]}(Y)$ can be shown to hold, where $H_{[\log_q]}(X,Y)$ denotes the joint entropy of $X$ and $Y$:
\begin{equation}
 H_{[\log_q]}(X,Y)  = -\sum_{b\in B, a\in A} \Pr(a,b)\log_q\,\Pr(a,b).
\end{equation}

\begin{defi}[Mutual information]
The mutual information of two discrete random variables $X$ and $Y$ is defined as
\begin{equation}\label{eq:mutualinfodefi}
 I_{[\log_q]}(X:Y) = H_{[\log_q]}(X) + H_{[\log_q]}(Y) - H_{[\log_q]}(X,Y).
\end{equation}
\end{defi}
\noindent
It is easy to verify the relations
$ I_{[\log_q]}(X:Y) = H_{[\log_q]}(X) - H_{[\log_q]}(X|Y) = H_{[\log_q]}(Y) - H_{[\log_q]}(Y|X) = I_{[\log_q]}(Y:X)$.
Hence the mutual information measures how much the uncertainty of $X$ is reduced when $Y$ is known (and vice versa).
The mutual information is always non-negative and $0$ if and only if $X$ and $Y$ are independent variables.

For prime $q$ the Galois field $\mathbb{F}_q$ contains the numbers $0,1,\dots,q-1$ and addition and multiplication are performed modulo $q$.
The vector space $\mathbb{F}_q^n$ contains the $q^n$ vectors $(0,0,\dots,0),(0,0,\dots,1),\dots,(q-1,q-1,\dots,q-1)$ of length $n$ with entries from $\mathbb{F}_q$.
Note that $\mathbb{F}_q^n$ forms a group with respect to addition modulo $q$.
\begin{defi}
The Hamming distance $\dt( \vec{x},\vec{y} )$
between two vectors $\vec{x},\vec{y} \in \mathbb{F}_q^n$ is defined as the number of places in which the two vectors differ.
The Hamming weight $\wt(\vec{x})$ of a vector $\vec{x} \in \mathbb{F}_q^n$ is defined as the Hamming distance between $\vec{x}$ and the null vector $\vec{0}=(0,\dots,0)$.
\end{defi}

We close this subsection proving the Chernoff bound for binomial distributions which will be used frequently in part II of the thesis to obtain asymptotic bounds. The proof is taken from the book~\cite{Roman}.
\begin{lem}[Chernoff bound]\label{lem:chernovbound}
Let $Y$ be a random variable which follows a $(n,p)$ binomial distribution, i.\,e. $\Pr(Y=k)=\binom{n}{k}p^k(1-p)^{n-k}$. Then, for any $\lambda < p$ such that $n\lambda \in \mathbb{N}_0$,
\begin{equation}
 \Pr(Y \leq n\lambda)
 \leq
 \Bigl(\frac{p}{\lambda}\Bigr)^{\lambda n} \Bigl( \frac{1-p}{1-\lambda} \Bigr)^{n(1-\lambda)}.
\end{equation}
\end{lem}
\begin{proof}
Let us define the random variable $X=e^{tY}$ with $t<0$.
Since $X$ takes only positive values, the Markov bound applies:
\begin{equation}
 \Pr(X\geq a) \leq \langle X \rangle / a.
\end{equation}
It follows that the probability of $Y$ taking on a value less than $b$ is upper bounded by
\begin{equation}
 \Pr(Y\leq b) = \Pr( X \geq e^{tb}) \leq  \langle X \rangle / e^{tb}.
\end{equation}
Plugging the expectation value of $X$,
\begin{equation}
\langle X \rangle = \sum_{k=0}^n \binom{n}{k}p^k(1-p)^{n-k} \cdot e^{tk} = (pe^t+1-p)^n,
\end{equation}
into the upper bound for $\Pr(Y\leq b)$ leads to
\begin{equation}
 \Pr(Y\leq b) = \sum_{k=0}^b \binom{n}{k}p^k(1-p)^{n-k} \leq (pe^t+1-p)^n\cdot e^{-tb}.
\end{equation}
Let us set $\lambda=b/n$ now.
The right hand side is minimized for $e^t = \frac{1-p}{p}\frac{\lambda}{1-\lambda}$ which lies in $[0,1]$ if $\lambda<p$.
\end{proof}
\noindent
For $p=1/2$ the Chernoff bound leads to the tail inequality (see e.\,g. \cite[section 3.5]{Welsh}):
\begin{cor}[Tail inequality]\label{cor:tail}
For any $\lambda$, with $0\leq \lambda < 1/2$ and $n\lambda \in \mathbb{N}_0$,
\begin{equation}
 \sum_{k=0}^{\lambda n} \binom{n}{k} %
 \leq
 \lambda^{-\lambda n} ( 1-\lambda )^{-n(1-\lambda)} = 2^{nH_2(\lambda)}.
\end{equation}
\end{cor}

\subsection{Quantum Mechanics}

The state of a quantum mechanical system $S$ is represented by a density operator $\rho$ which is a non-negative operator of trace one acting on the associated Hilbert space $\mathcal{H}_S$ of the system.
We denote the set of operators as $\mathcal{L}(\mathcal{H}_S)$ and the subset of
density operators
as $\mathcal{S}(\mathcal{H}_S)$.
A state $\rho$ is said to be a pure state if $\rho = \ket{\psi}\bra{\psi}$ for some $\ket{\psi}\in\mathcal{H}_S$ such that $\langle \psi \vert \psi \rangle=1$.

\subsubsection{Time Evolution and Measurements}
The time evolution of a pure quantum state is specified by the Schr\"odinger equation,
\begin{equation}
 i\hbar \frac{d}{dt} \ket{\psi(t)} = H(t) \ket{\psi(t)},
\end{equation}
where $H(t) \in\mathcal{L}(\mathcal{H}_S)$ denotes the self-adjoint Hamiltonian of the system.
Correspondingly, the time evolution of a general quantum state $\rho$ is described by the von Neumann equation,
\begin{equation}
 i\hbar \frac{d}{dt} \rho(t) = [ H(t) ,\rho(t) ],
\end{equation}
where the brackets denote a commutator, i.\,e. $[A,B]=AB-BA$.
As a consequence, the time evolution operator of a closed quantum system is unitary,
\begin{equation}
 \rho(t) = U(t,0) \rho(0) U^\dagger(t,0),
\end{equation}
with
\begin{equation}
 U(t,0) = \mathcal{T} \exp\Bigl(-i \int_0^t H(t') dt' \Bigr),
\end{equation}
where $\mathcal{T}$ denotes the Dyson time-ordering operator.

If the quantum system $S$ forms a part of a larger quantum system, $S$ is said to be an open quantum system.
Then the resulting time evolution of the open system alone is not necessarily unitary anymore, but is given by a trace-preserving completely positive map (tpcp-map) $\mathcal{E} : \mathcal{S}(\mathcal{H}_S) \rightarrow \mathcal{S}(\mathcal{H}_S)$.
Any tpcp-map $\mathcal{E}$ can be represented in terms of an operator sum decomposition $\{ E_\mu \}$ such that $\sum_\mu E_\mu^\dagger E_\mu = \mathcal{I}$ and
\begin{equation}
 \mathcal{E} : \rho \mapsto \mathcal{E}(\rho) = \sum_\mu E_\mu \rho E_\mu^\dagger,
\end{equation}
where $\mathcal{I}$ denotes the identity operator.

A von Neumann measurement is characterized by a self-adjoint measurement operator $M$ with spectral decomposition $M= \sum_\mu m_\mu P_\mu$, where the $m_\mu$ denote distinct measurement values and the $P_\mu$ denote orthogonal projections ($\sum_\mu P_\mu = \mathcal{I}$).
If we perform a measurement of $M$ on the state $\rho$, we obtain the result $\mu$ with probability $p_\mu=\tr(P_\mu \rho)$. Conditioned on the measurement result the state changes from $\rho$ to $P_\mu \rho P_\mu / \tr(P_\mu \rho)$.
A more general measurement is specified by a positive operator valued measure (POVM), which consists of a set $\{F_\mu \}$ of positive operators such that $\sum_\mu F_\mu = \mathcal{I}$.
In this case the probability of getting the result $\mu$ is given by $p_\mu=\tr( F_\mu \rho)$.

\subsubsection{Entropy and Quantum Mutual Information}

\begin{defi}[von Neumann entropy]
The von Neumann entropy of a quantum state $\rho \in \mathcal{S}(\mathcal{H}_S)$ is defined by
\begin{equation}
S_{[\log_q]}(\rho) = - \tr \bigl( \rho \log_q \rho \bigr).
\end{equation}
\end{defi}
For a bipartite quantum system $AB$ the joint von Neumann entropy of the state $\rho_{AB} \in \mathcal{S}(\mathcal{H}_A\otimes \mathcal{H}_B)$ is defined by
\begin{equation}
S_{[\log_q]}(A,B)\equiv
S_{[\log_q]}(\rho_{AB}) = - \tr \bigl( \rho_{AB} \log_q \rho_{AB} \bigr).
\end{equation}
By analogy with the Shannon entropies the conditional entropy of system $A$ given system $B$ is defined by
\begin{equation}
S_{[\log_q]}(A\vert B) = S_{[\log_q]}(A,B) - S_{[\log_q]}(B),
\end{equation}
where $S_{[\log_q]}(B)\equiv S_{[\log_q]}(\rho_B)$ denotes the entropy of the reduced state $\rho_B=\tr_B( \rho_{AB})$ ($\tr_B$ denotes the partial trace with respect to system $B$).
In contrast to the conditional Shannon entropy, the conditional von Neumann entropy might become negative.

\begin{defi}[Quantum mutual information]
The quantum mutual information of a bipartite quantum system $AB$ in the state $\rho_{AB} \in \mathcal{S}(\mathcal{H}_A\otimes \mathcal{H}_B)$ is defined by
\begin{equation}\label{eq:qmutualinfodefi}
 I_{[\log_q]}(A:B) = S_{[\log_q]}(A) + S_{[\log_q]}(B) - S_{[\log_q]}(A,B).
\end{equation}
\end{defi}
\noindent
As it is the case for the classical mutual information, the relation
$ I_{[\log_q]}(A:B) = S_{[\log_q]}(A) - S_{[\log_q]}(A|B) = S_{[\log_q]}(B) - S_{[\log_q]}(B|A) = I_{[\log_q]}(B:A)$ holds.
The quantum mutual information is always non-negative.

\subsubsection{Quantum Registers}
A two-dimensional quantum mechanical system is called a qubit.
Finite-dimensional quantum mechanical systems of higher dimension are called qudits.
Let $\mathcal{H}_q = \mathbb{C}^q$ denote the Hilbert space of a qudit of dimension $q$,
and fix an orthonormal basis $\{\ket{0},\dots,\ket{q-1}\}$ of $\mathcal{H}_q$.
A quantum register consisting of $n$ qudits of dimension $q$ is defined on the Hilbert space $\mathcal{H}_q^{\otimes n}$.
An orthonormal basis of $\mathcal{H}_q^{\otimes n}$ is given by the set of $n$-fold product states of the one-qudit basis states,
\begin{equation}
 \mathcal{H}_q^{\otimes n} = \vspan\bigl\{ \ket{i_1,i_2,\dots,i_n} \bigr\},
\end{equation}
with $0\leq i_j<q$ for $j\in\{1,2,\dots,n\}$ and $\ket{i_1,i_2,\dots,i_n} = \ket{i_1}\otimes\ket{i_2}\otimes\dots\otimes\ket{i_n}$.
A short hand notation for the basis states is given by $\ket{i_1,i_2,\dots,i_n} = \ket{\vec{i}}$ with $\vec{i}\in\mathbb{F}_q^n$.

\subsubsection{Pauli Operators}

We consider qudits of of prime dimensions.
The Pauli $X$ and $Z$ operators acting on $\mathcal{H}_q$ are defined by\footnote{Some authors use the definition $X\ket{i}= \ket{i-1\!\pmod{q}}$. See, for instance, \cite{Ha03Fi}.}
\begin{subequations}
\begin{align}
 X\ket{i} &= \ket{i+1\!\!\!\pmod{q}}\\
 Z\ket{i} &= \omega^i \ket{i},
\end{align}
\end{subequations}
where $\omega = \exp(2\pi i/q)$ is a complex primitive $q$-th root of unity. It follows that $ZX=\omega XZ$.

\begin{defi}
For any vector
$\vec{a} = (\vec{a}^x,\vec{a}^z) = (a^x_1,\dots,a^x_n,  a^z_1,\dots,a^z_n) \in \mathbb{F}_q^{2n}$, let the Pauli operator $\XZ(\vec{a})$ acting on $\mathcal{H}_q^{\otimes n}$ be defined by
\begin{equation}
 \XZ(\vec{a}) = \begin{cases}
     i^{a^x_1a^z_1}X^{a^x_1}Z^{a^z_1}\otimes\dots\otimes i^{a^x_na^z_n}X^{a^x_n}Z^{a^z_n} &\text{ for } q=2\\
     X^{a^x_1}Z^{a^z_1}\otimes\dots\otimes X^{a^x_n}Z^{a^z_n} &\text{ for } q\geq 3
               \end{cases},
\end{equation}
so that the eigenvalues of $\XZ(\vec{a})$ are powers of $\omega$.
\end{defi}
\begin{rem}
If we write the operator $\XZ(\vec{a})$ as $\XZ((\vec{a}^x,\vec{a}^z))$ for some
$\vec{a}=(\vec{a}^x,\vec{a}^z) \in \mathbb{F}_q^{2n}$, we will use the shorthand notation $\XZ(\vec{a}^x,\vec{a}^z)$ omitting the braces of $\vec{a}=(\vec{a}^x,\vec{a}^z)$.
For instance, the identity operator $\mathcal{I}$ is given by the operator $\XZ(\vec{0},\vec{0})$ with $\vec{0}=(0,\dots,0)\in\mathbb{F}_q^n$.
\end{rem}
\noindent
If we represent the qubit Pauli operators in the $\{ \ket{0}, \ket{1} \}$-basis, we obtain the well known Pauli matrices,
\begin{align}
 \XZ(0,0) &= \begin{pmatrix} 1&0\\ 0&1 \end{pmatrix} &
 \XZ(1,0) &= \begin{pmatrix} 0&1\\ 1&0 \end{pmatrix} &
 \XZ(1,1) &= \begin{pmatrix} 0&-i\\ i&0 \end{pmatrix} &
 \XZ(0,1) &= \begin{pmatrix} 1&0\\ 0&-1 \end{pmatrix},
\end{align}
which are also denoted as $\mathcal{I},X,Y$ and $Z$.
Hence, the qubit Pauli operators are Hermitian.
For $q\geq 3$ we obtain
\begin{equation}\label{eq:xzaxzb}
 \XZ(\vec{a}) \cdot \XZ(\vec{b}) = \omega^{\sum_i a^z_ib^x_i } \XZ(\vec{a}+\vec{b}),
\end{equation}
while for $q=2$ this expression holds up to some powers of $i$.
As a consequence, $\XZ(\cdot)$ gives rise to a unitary projective representation of $\mathbb{F}_q^{2n}$,
which by itself forms a group under addition modulo $q$:
\begin{equation}\label{eq:rayrep}
 \XZ(\cdot):\ \mathbb{F}_q^{2n}\ni \vec{a} \mapsto \XZ(\vec{a})\in \mathfrak{P}_q^n.
\end{equation}
The full Pauli group is given by
\begin{equation}
 \mathfrak{P}_q^n =\begin{cases}
     \{ \mu \XZ(\vec{a}) \ \vert\ \mu\in\{\pm 1,\pm i\}, \vec{a}\in\mathbb{F}_q^{2n} \} & ,q=2\\
     \{ \omega^j \XZ(\vec{a}) \ \vert\ j\in\mathbb{F}_q, \vec{a}\in\mathbb{F}_q^{2n} \} & ,q\geq 3
                 \end{cases}.
\end{equation}
Its order is $4\cdot 4^n$ for qubits and $q\cdot q^{2n}$ in general ($q\geq 3$).
If two elements of the Pauli group are identical up to a phase $\omega^p$, $p\in\mathbb{F}_q$,
(or some power of $i$ for $q=2$ respectively), we write $\XZ(\vec{a}) \sim \omega^p \XZ(\vec{a})$.
We denote the set containing all $n$-fold tensor products of Pauli operators as
\begin{equation}
 \mathcal{P}_q^n = \{ \XZ(\vec{a}) \ \vert\  \vec{a}\in\mathbb{F}_q^{2n} \}.
\end{equation}
Note that $\vert \mathcal{P}_q^n\vert = q^{2n}$ while $\vert \mathfrak{P}_q^n \vert = q\cdot q^{2n}$ (for $q\geq 3$).

\begin{defi}\label{defi:symip}
The symplectic inner product between elements $\vec{a}$ and $\vec{b}$ of $\mathbb{F}_q^{2n}$ is defined as
\begin{equation}
(\vec{a},\vec{b})_{sp} = \sum_{i=1}^n a^z_i b^x_i - a^x_i b^z_i \pmod{q}.
\end{equation}
\end{defi}

\begin{rem}
With the help of the inner product defined above, the order of a product of two Pauli operators $\XZ(\vec{a})$ and $\XZ(\vec{b})$ can be inverted,
\begin{equation}\label{eq:syminvertorder}
 \XZ(\vec{a}) \cdot \XZ(\vec{b})  = \omega^{(\vec{a},\vec{b})_{sp}}  \XZ(\vec{b}) \cdot \XZ(\vec{a}).
\end{equation}
Two operators commute if and only if the symplectic inner product between $\vec{a}$ and $\vec{b}$ vanishes.
\end{rem}

\subsubsection{Bell States}

\begin{defi}[Bell states]\label{defi:bellstates}
Let $\mathcal{H}_q$ denote the Hilbert space of a qudit of dimension $q$ and let $\mathcal{H}_A=\mathcal{H}_q^{\otimes n}$, $\mathcal{H}_B=\mathcal{H}_q^{\otimes n}$.
Then the states
\begin{equation}
 \ketl{\Phi_{\vec{x}}}{AB} = \frac{1}{\sqrt{q^n}} \sum_{\vec{j}\in\mathbb{F}_q^n }
 \ketl{\vec{j}}{A} \otimes \opl{\XZ(\vec{x})}{B} \ketl{\vec{j}}{B},
 \quad \vec{x}\in\mathbb{F}_q^{2n},
\end{equation}
are called Bell states.
They are maximally entangled and form an orthonormal basis of $\mathcal{H}_A \otimes \mathcal{H}_B$.
\end{defi}

\subsection{Representation Theory}\label{subsec:reptheo}

This subsection provides a brief overview of the basics of the representation theory of finite groups.
Representation theory will be relevant for decoupling in part I (if the elements of a decoupling scheme form a projective representation of an underlying group) and as a tool for the evaluation of the secure key rates of the quantum key distribution protocols in part II.

We consider a finite group $G$ of order $n_G$, i.\,e. $G$ contains $n_G = \vert G\vert$ elements.
If the elements of $G$ commute with one another, the group is called an abelian group.
\begin{defi}
An element $b\in G$ is said to be conjugate to an element $a\in G$ if there exists $u\in G$ such that
$b = u a u^{-1}$.
Elements conjugate to one another form a conjugacy class.
\end{defi}
\noindent
Since conjugacy is an equivalence relation, each element of $G$ belongs to one and only one of the classes.
If we denote the number of classes by $n_\zeta$ and the number of elements in class $i$ by $n_i$, we have $\sum_{i=1}^{n_\zeta} n_i = n_G$.
A class containing the inverse of all elements in the class is called ambivalent.
If a group is abelian, each element forms a class by itself.

\begin{defi}
A representation (rep) $R$ of $G$ is a group homomorphism from $G$ to a group $R(G)$ of operators on a vector space $\mathcal{V}$,
\begin{equation}
 R : G\ni a \mapsto R(a) = R_a \in \mathcal{L}(\mathcal{V}).
\end{equation}
From the definition of a group homomorphism we have $R_{ab} = R_a\cdot R_b$ for all $a,b\in G$.
The dimension $d=\dim(\mathcal{V})$ of $\mathcal{V}$ is called the dimension of the rep.
\end{defi}
\begin{rem}
If $\mathcal{V}$ is the vector space over the field $\mathbb{C}$, a map from $G$ to a set $R(G)$ of operators on $\mathcal{V}$ satisfying
\begin{equation}
 R_{ab} = r(a,b) \cdot R_a\cdot R_b,
\end{equation}
with $r(a,b)\in\mathbb{C}$ for all $a,b\in G$, is called a projective representation.
\end{rem}
\noindent
We will always assume that the vector space $\mathcal{V}$ is an inner product space over the field $\mathbb{C}$.
Let us fix an orthonormal basis $\{ \ket{j} \}_{j=0\dots d-1}$ of~$\mathcal{V}$.
Then,
\begin{equation}
 R_a \ket{j} = \sum_{i=0}^{d-1} D_{ij}(a) \ket{i},
\end{equation}
with $D_{ij}(a) = \bra{i} R_a \ket{j}$,
and we obtain
\begin{equation*}
R_a R_b \ket{j} = R_a \sum_{i=0}^{d-1} D_{ij}(b) \ket{i}
                = \sum_{k,i=0}^{d-1} D_{ki}(a) D_{ij}(b) \ket{k}
                = R_{ab} \ket{j} = \sum_{k=0}^{d-1} D_{kj}(ab) \ket{k}.
\end{equation*}
Since the $\{ \ket{j} \}$ form a basis, it follows that $D_{kj}(ab) = \sum_i D_{ki}(a) D_{ij}(b)$ or $D(ab) = D(a)\cdot D(b)$.
Hence, the group of matrices $D(G) = \{ D(a) \sthat a\in G\}$ forms a matrix representation of $G$.
If $R(G)$ is a representation of $G$ on a vector space $\mathcal{V}$, and $A$ is a non-singular operator on $\mathcal{V}$, then it is obvious
that $R'(G) = A R(G) A^{-1}$ also forms a representation of $G$ on $\mathcal{V}$.
In this case $R(G)$ and $R'(G)$ are related by a similarity transformation.
\begin{defi}
Two representations of a group $G$ on a vector space $\mathcal{V}$ which are related by a similarity transformation are said to be equivalent representations.
\end{defi}

\begin{defi}
If the group representation space is an inner product space and if the operators $R_g$ are unitary for all $g\in G$,
then the representation $R(G)$ is called a unitary representation.
\end{defi}
\begin{rem}
It can be shown that every representation of a finite group on an inner product space is equivalent to a unitary representation
(see e.\,g. \cite[theorem 3.3]{Tung85}).
In the following we consider only unitary representations.
\end{rem}

\begin{defi}
Let $R(G)$ be a representation of $G$ on a vector space $\mathcal{V}$.
A subspace $\mathcal{V}_1$ of $\mathcal{V}$ is called invariant subspace of $\mathcal{V}$ with respect to $R(G)$ if
$R_g \ket{ \varphi } \in \mathcal{V}_1$ for all $g\in G$ and for all $\ket{ \varphi }\in \mathcal{V}_1$.
\end{defi}
\begin{rem}%
If a space $\mathcal{V}_1$ is an invariant subspace of a representation $R(G)$ on $\mathcal{V}$, then $\mathcal{V}_1$ itself is a representation space.
\end{rem}
\begin{thm}\label{thm:commuteseigenspace}
If an operator $A$ commutes with all operators $R_g$ of a rep $R(G)$, then the eigenspace $\mathcal{V}_\lambda$ of $A$ is a representation space of $G$.
\end{thm}
\begin{proof}
We show that $\mathcal{V}_\lambda$ is an invariant subspace of $R(G)$ on $\mathcal{V}$.
Let $\ket{ \phi_\lambda } \in \mathcal{V}_\lambda$ so that $A \ket{ \phi_\lambda } = \lambda \ket{ \phi_\lambda }$.
Then, $ A R_g \ket{ \phi_\lambda } = R_g A \ket{ \phi_\lambda } = \lambda R_g \ket{ \phi_\lambda }$ and it follows that $R_g \ket{ \phi_\lambda } \in \mathcal{V}_\lambda$ for all $\ket{ \phi_\lambda } \in \mathcal{V}_\lambda$ and all $R_g\in R(G)$.
\end{proof}
\begin{defi}
A representation $R(G)$ on $\mathcal{V}$ is irreducible if there is no non-trivial invariant subspace in $\mathcal{V}$ with respect to $R(G)$ (we may also say that the representation space is irreducible).
Otherwise the representation is reducible.
\end{defi}

\noindent
Since we consider only unitary representations, reducible always means fully reducible:
Let $\mathcal{V}_1$ be an invariant subspace of the representation space $\mathcal{V}$, and let $\mathcal{V}_2$ be the space orthogonal to $\mathcal{V}_1$, i.\,e. $\mathcal{V}=\mathcal{V}_1\oplus \mathcal{V}_2$.
Then, since
$\langle R_g v_2 \vert v_1\rangle = \langle v_2 \vert R_g^\dagger v_1 \rangle
= \langle v_2 \vert R_{g^{-1}} v_1 \rangle = 0$ for all $\ket{v_1}\in \mathcal{V}_1$, all $\ket{v_2}\in \mathcal{V}_2$ and all $g\in G$, it follows that $\mathcal{V}_2$ remains invariant, too.
In other words, the operators $R_g$ of a reducible representation $R(G)$ become block-diagonal for a proper choice of basis.
For instance, if the representation space $\mathcal{V}$ decomposes into two irreducible invariant subspaces $\mathcal{V}=\mathcal{V}_1\oplus \mathcal{V}_2$ of dimension $d_1$ and $d_2=d-d_1$, we write
$R(G) = D^{(1)}(G) \oplus D^{(2)}(G)$ and
\begin{equation}
 R_g \mapsto  D(g) = \begin{pmatrix}
                        D^{(1)}(g) & 0 \\
                        0      & D^{(2)}(g)
                     \end{pmatrix},
\end{equation}
where $D^{(1)}(g)$ is a $d_1\times d_1$ matrix and $D^{(2)}(g)$ is a $d_2\times d_2$ matrix.
In general we obtain the relation
\begin{equation}
 R(G) =  \bigoplus_{\nu\in\mathcal{J}} \tau_\nu \cdot D^{(\nu)}(G),
\end{equation}
where $\nu$ labels inequivalent irreducible representations and $\tau_\nu$ denotes the number of times a certain irreducible representation $\nu$ occurs. The dimension of the irrep $D^{(\nu)}(G)$ is denoted by $d_\nu$.
Hence there exists an orthonormal basis
\begin{equation}\label{eq:nulmbasis}
\bigl\{ \ket{ \nu \ l_\nu \  m_\nu } \sthat \nu\in\mathcal{J},\ l_\nu=1\dots \tau_\nu,\ m_\nu=1\dots d_\nu  \bigr\},
\end{equation}
in which the operators $R_g$ are block-diagonal, i.\,e.
\begin{equation}
R_g \ket{ \nu \ l_\nu \  m_\nu } =
D^{(\nu)}(g) \ket{ \nu \ l_\nu \  m_\nu } =
 \sum_{m'_\nu=1}^{d_\nu} D_{m'_\nu m_\nu}^{(\nu)}(g)  \ket{ \nu \ l_\nu \ m'_\nu }.
\end{equation}
We label the subspace of the representation space $\mathcal{V}$ which is spanned by the set of basis vectors with fixed $\nu$ by $\mathcal{V}_\nu$,
\begin{equation}
 \mathcal{V}_\nu = \vspan \bigl\{ \ket{\nu \ l_\nu \ m_\nu} \sthat l_\nu=1\dots \tau_\nu, m_\nu=1\dots d_\nu \bigr\}.
\end{equation}
Since $\mathcal{V}_\nu$ has the form of a tensor space ($\ket{\nu \ l_\nu \ m_\nu} = \ket{l_\nu}\otimes\ket{m_\nu}$), we write $\mathcal{V}_\nu = \mathcal{C}_\nu \otimes \mathcal{D}_\nu$, where the dimension of $\mathcal{C}_\nu$ is given by $\tau_\nu$ and the dimension of $\mathcal{D}_\nu$ is given by $d_\nu$.
The representation space $\mathcal{V}$ decomposes into a direct sum of orthogonal subspaces,
\begin{equation}
 \mathcal{V} = \bigoplus_{\nu\in\mathcal{J}} \mathcal{V}_\nu = \bigoplus_{\nu\in\mathcal{J}} \mathcal{C}_\nu \otimes \mathcal{D}_\nu.
\end{equation}

If we restrict an irreducible representation (irrep) $D^{(\nu)}(G)$ of a group $G$ to elements of a subgroup $G_s \subset G$, we obtain a subduced representation denoted as $D^{(\nu)}(G) \downarrow G_s$.
A subduced rep is in general reducible and can be decomposed into a direct sum of irreps of $G_s$,
\begin{equation}
 D^{(\nu)}(G) \downarrow G_s  =  \bigoplus_\mu \tau^{(\nu)}_\mu \cdot D^{(\mu)}(G_s),
\end{equation}
where $\tau^{(\nu)}_\mu$ denotes the number of times the irrep $D^{(\mu)}(G_s)$ occurs in $D^{(\nu)}(G) \downarrow G_s$.
If $\tau^{(\nu)}_\mu \leq 1$ for all possible $\nu$ and $\mu$, then $G_s$ is called a canonical subgroup of $G$.
A canonical subgroup chain is a group chain $G\supset G_1 \supset G_2 \dots \supset G_n$ such that $G_{i+1}$ is a canonical subgroup of $G_i$ ($i=0,\dots,n-1$ with $G\equiv G_0$) and $G_n$ is abelian.

\begin{thm}[Schur's lemma i]\label{thm:schuri}
Let $A$ be an operator commuting with all operators of a rep $R(G)$ of $G$ on $\mathcal{V}$,
and let $\mathcal{V}_\nu \subseteq \mathcal{V}$ be an irreducible rep space of $G$ and an invariant subspace of $A$.
Then $\mathcal{V}_\nu$ is necessarily an eigenspace of $A$.
\end{thm}
\begin{proof}%
Let us assume that the invariant subspace $\mathcal{V}_\nu$ of $A$ decomposes into two eigenspaces of $A$,  $\mathcal{V}_\nu = \mathcal{V}_{\nu,1}\oplus \mathcal{V}_{\nu,2}$. According to theorem \ref{thm:commuteseigenspace}, each of these spaces would be a representation space, which is in contradiction to $\mathcal{V}_\nu$ being an irreducible rep space.
Hence, the only possibility is that $A \mathcal{V}_\nu = \nu \mathcal{V}_\nu$.
\end{proof}
\begin{rem}[i]
The representative of an operator $A$ in $\mathcal{V}_\nu$ is a multiple of the identity:
Let a basis of $\mathcal{V}_\nu$ be given by %
$\{ \ket{ i  } \}_{i=0,\dots,d_\nu-1}$.
Then the matrix representative of $A$ in $\mathcal{V}_\nu$ is given by
$D^{(\nu)}_{ij}(A) = \bra{i} A \ket{j} = \nu \delta_{ij}$.
If $\mathcal{V}_\nu = \mathcal{V}$ we obtain the result that the only operator commuting with all operators of an irrep $R(G)$ is a multiple of the identity.
\end{rem}
\begin{rem}[ii]
A direct consequence of Schur's lemma is that an irrep of an abelian group must be of dimension one.
\end{rem}

\begin{thm}[Schur's lemma ii]\label{thm:schurii}
Let $D^{(\mu)}(G)$ and $D^{(\nu)}(G)$ be two irreps of $G$ on the spaces $\mathcal{V}_\mu$ and $\mathcal{V}_\nu$ respectively, and let $A$ be a linear transformation from $\mathcal{V}_\nu$ to $\mathcal{V}_\mu$ which satisfies $A D^{(\nu)}(g) = D^{(\mu)}(g) A$ for all $g\in G$.
Then, either $A=0$, or $\mathcal{V}_\mu$ and $\mathcal{V}_\nu$ are isomorphic and $D^{(\mu)}(G) = A D^{(\nu)}(G) A^{-1}$, i.\,e. the irreps $\mu$ and $\nu$ are equivalent.
\end{thm}
\begin{proof}
It is easy to verify that the range of $A$ is an invariant subspace of $\mathcal{V}_\mu$ with respect to $D^{(\mu)}(G)$. Since $D^{(\mu)}(G)$ is irreducible it follows that either the range is $0$ (which implies $A=0$) or the range is $\mathcal{V}_\mu$.
Similarly, the null space of $A$ in $\mathcal{V}_\nu$ is an invariant subspace of  $\mathcal{V}_\nu$ with respect to $D^{(\nu)}(G)$.
Since $D^{(\nu)}(G)$ is irreducible it follows that either the null space is equal to $\mathcal{V}_\nu$ (implying $A=0$) or the null space is $0$ (implying that $A$ is a one-to-one mapping).
Hence $A$ is either an isomorphism between $\mathcal{V}_\mu$ and $\mathcal{V}_\nu$ or it vanishes.
\end{proof}

The second part of Schur's lemma can be used to prove the orthonormality of irreducible representation matrices.
\begin{thm}[Orthonormality of irreducible representation matrices]
Let $D^{(\nu)}(G)$ and $D^{(\mu)}(G)$ denote two inequivalent irreducible representations of $G$, and let the dimension of the $\mu$ representation be given by $d_\mu$.
Then the following orthonormality condition holds,
\begin{equation}
 \frac{d_\mu}{n_G} \sum_{g\in G} D^{\dagger(\mu)}_{ki}(g) \ D^{(\nu)}_{jl}(g) = \delta_{\mu\nu} \delta_{ij} \delta_{kl},
\end{equation}
with $D^{\dagger(\mu)}_{ki}(g)$ denoting the complex conjugate of the matrix element $D^{(\mu)}_{ik}(g)$.
\end{thm}

\begin{defi}[Group algebra]
The group algebra $\mathbb{C}G$ is defined as the complex vector space spanned by the group elements, i.\,e. any element $a$ in $\mathbb{C}G$ can be written as $a=\sum_{g\in G} a_g g$ with $a_g\in\mathbb{C}$.
For two elements $a,b$ in $\mathbb{C}G$ the product
\begin{equation}
 a b = \Bigl( \sum_{g\in G} a_g g \Bigr)\Bigl( \sum_{h\in G} b_h h \Bigr)
 = \sum_{g,h\in G} a_g b_h \, gh = \sum_{g' \in G} \Bigl( \sum_{h\in G} a_{g'h^{-1}}b_h \Bigr)   g'
\end{equation}
turns $\mathbb{C}G$ into an algebra.
\end{defi}%
\noindent
Any representation $R$ of $G$ extends by linearity to a representation of the elements in $\mathbb{C}G$.
Let $\mathcal{A} = R(\mathbb{C}G)$ denote the algebra generated by $R$, and let its commutant $\mathcal{A}'$ be defined as the set of elements that commutes with all the elements in $\mathcal{A}$,
$ \mathcal{A}' = \{ V\in\mathcal{L}(\mathcal{V}) \sthat VA=AV \text{ for all } A\in\mathcal{A} \}$.
The following theorem follows from the orthonormality of irreducible representation matrices and the first part of Schur's lemma.
\begin{thm}\label{thm:algebrabdiag}
In the $\{ \ket{ \nu \ l_\nu \ m_\nu } \}$-basis (as defined in equation \eqref{eq:nulmbasis}) corresponding to the representation $R$, $\mathcal{A} = R(\mathbb{C}G)$ and $\mathcal{A}'$ take the form
\begin{align}
 \mathcal{A}  &\cong \bigoplus_{\nu\in\mathcal{J}} \mathbbm{1}_{\tau_\nu} \otimes \operatorname{Mat}(d_\nu\times d_\nu,\mathbb{C})  \\
 \mathcal{A}' &\cong \bigoplus_{\nu\in\mathcal{J}} \operatorname{Mat}(\tau_\nu\times \tau_\nu,\mathbb{C}) \otimes \mathbbm{1}_{d_\nu},
\end{align}
where $\mathbbm{1}_n$ denotes an $n\times n$ dimensional identity matrix and $\operatorname{Mat}(n\times n,\mathbb{C})$ denotes the set of $n\times n$ matrices with entries in $\mathbb{C}$.
\end{thm}

\begin{rem}
In part I of this thesis we are sometimes going to deal with projective representations $R(G)$ of $G$ on $\mathcal{V}$. In this case we assume that the set of unitary matrices $\{ R_g=R(g) \sthat g\in G\}$ generates a finite group $\hat{G}$ larger than $G$ and consider the ordinary irreducible representations of $\hat{G}$.
If we define the center of $\hat{G}$ by $Z(\hat{G}) = \{z \in \hat{G} \sthat gz = zg \text{ for all } g \in \hat{G} \}$ then the quotient group $\hat{G} / Z(\hat{G})$ is isomorphic to the original group $G$.
\end{rem}

Let us close this subsection revisiting the set $\mathcal{P}_q^n$ of $n$-fold tensor products of Pauli operators.
This set is an example of a so-called nice error basis.
Such a basis was defined by Knill in \cite{Knill96a} as follows:
\begin{defi}\label{def:nicerrorb}
Let $G$ be a group of order $\vert G\vert = d^2$ and let its identity element be denoted by $e$.
A nice error basis is a set $\mathcal{E} = \{D(g) \in \textsf{U}_d \sthat g\in G\}$ of unitary $d\times d$ matrices such that
(i) $D(e)$ is given by the identity matrix,
(ii) $\tr ( D(g) ) / d = \delta_{g,e}$ for all $g\in G$,
and (iii) $D(g)D(h) = \alpha(g,h) D(gh)$ for all $g,h\in G$, where $\alpha(g,h)$ is a function from $G\times G$ to $\mathbb{C}\setminus \{0\}$.
\end{defi}
\noindent
A consequence of conditions (i) and (iii) is that the map $G\ni g \mapsto D(g) \in \textsf{U}_d$ defines a projective representation of $G$ on a $d$-dimensional Hilbert space $\mathcal{H}$.
It follows from condition (ii) that the matrices $D(g)$ are pairwise orthogonal with respect to the trace inner product $\langle A,B\rangle = \tr( A^\dagger B)/d$.
Hence they form a basis for the operators acting on $\mathcal{H}$ and the projective representation of $G$ on $\mathcal{H}$ must be irreducible.
Since the matrices are unitary we have $\vert \det D(g) \vert = 1$ for all $g\in G$ and it follows from (iii) that $\vert \alpha(g,h)\vert = 1$.
The group $G$ is also called the index group.
It is easy to verify that the set $\mathcal{P}_q^n$ of Pauli operators with the index group given by $\mathbb{F}_q^{2n}$ fulfills the definition of a nice error basis with $\alpha(\vec{g},\vec{h}) = \omega^{\sum_i g^z_i h^x_i }$ for $\vec{g}=(\vec{g}^x,\vec{g}^z), \vec{h}=(\vec{h}^x,\vec{h}^z) \in\mathbb{F}_q^{2n}$ (compare with \eqref{eq:xzaxzb}).

Finally, let us define the notation of several groups we are going to encounter.
The symmetric group $\textsf{S}_n$ on the finite set $\{1,2,\dots,n\}$ consists of all permutations of the set and has order $n!$.
The general linear group of degree $q$ over the field $\mathbb{C}$ is the group of $q\times q$ invertible matrices with entries from $\mathbb{C}$.
It is denoted by $\textsf{GL}_q = \textsf{GL}(q,\mathbb{C})$.
Subgroups of $\textsf{GL}_q$ are the unitary group $\textsf{U}_q$ containing unitary matrices, the special unitary group $\textsf{SU}_q$ containing unitary matrices with unit determinant, and the 3-dimensional rotation group which is the special orthogonal group of degree 3 over the field $\mathbb{R}$ and is denoted by $\textsf{SO}_3 = \textsf{SO}(3,\mathbb{R})$.

\part{Random Decoupling}

\chapter{Dynamical Decoupling}\label{chap:dyncontrol}

This chapter deals with dynamical decoupling strategies in the bang-bang control scenario.
After giving an introduction to dynamical control theory and average Hamiltonian theory (AHT), we present an overview over known construction methods for dynamical decoupling schemes.
The main focus is then on improved decoupling strategies which are based on a fixed decoupling scheme.
The performance of these strategies is analyzed by deriving formulas for the average fidelity decay.
For any randomized strategy, in addition, the variance of the fidelity is studied.
With the help of a numerical simulation of a quantum memory perturbed by Heisenberg interactions, these formulas are validated and conclusions concerning a general guideline for optimal decoupling are drawn.

We start by presenting the necessary framework in section \ref{sec:dynctrl}.
The overview over known construction methods for efficient decoupling schemes will then be given in section \ref{sec:decschemes}.
Improved control strategies based on a given decoupling scheme are explored in section \ref{sec:cstrategies}.
Finally, we present the results of the numerical simulation in section \ref{sec:decexample}.

\section{Dynamical Control of Quantum Systems}\label{sec:dynctrl}
Let $S$ be a quantum system defined on a finite $d$-dimensional Hilbert space $\mathcal{H}_S$ and let its dynamics be generated by the system Hamiltonian $H_0 \in \mathcal{L}(\mathcal{H}_S)$.
Typically the quantum system $S$ under consideration will be a quantum register consisting of $n$ qudits of dimension $q$ so that $\mathcal{H}_S=\mathcal{H}_q^{\otimes n}$ and the system Hamiltonian describes some static imperfections.
We assume that we are able to apply a certain set of local control operations which are realized by the time-dependent control Hamiltonian $H_c(t) \in \mathcal{L}(\mathcal{H}_S)$.
Local means that $H_c$ is a sum over one qudit Hamiltonians, i.\,e. $H_c(t)=\sum_{i=1}^{n} h^{(i)}_i(t)\otimes\mathcal{I}_{\{1,\dots,n\}\setminus \{i\}}$ with some time-dependent $h^{(i)}(t)\in\mathcal{L}(\mathcal{H}_q)$.
In turn the total Hamiltonian is given by
\begin{equation}
 H(t) = H_0  +  H_c(t)
\end{equation}
and according to the Schr\"odinger equation our system evolves in time as
\begin{equation}\label{eq::utot}
U(t) = \mathcal{T} \exp \Bigl( -i\int_0^t H(t') dt'/\hbar \Bigr),
\end{equation}
where $\mathcal{T}$ denotes the Dyson time-ordering operator.
Analogous to \eqref{eq::utot} let us denote the time evolution due to $H_c(t)$ alone by $U_c(t)$, i.\,e.
\begin{equation}\label{eq:Uct}
U_c(t) = \mathcal{T} \exp \Bigl( -i\int_0^t H_c(t') dt'/\hbar \Bigr).
\end{equation}
We now define the toggled frame as the frame that continuously follows the applied control, $\tilde{U}(t) = U_c^\dagger(t) U(t)$.
The time evolution in the toggled frame is determined by the
Schr\"odinger equation
\begin{equation}
i\hbar  \frac{d\tilde{U}(t)}{dt} = \tilde{H}(t) \tilde{U}(t),
\end{equation}
where the toggled frame Hamiltonian is given by
\begin{equation}\label{eq:toframe-h0}
 \tilde{H}(t)=U_c^\dagger(t) H_0 U_c(t).
\end{equation}
Dynamical control of $H_c(t)$ and in turn of $U_c(t)$ allows us to modify the time evolution in the toggled frame.
There are different possible control tasks.
If we deal with a quantum memory for example, we may want to freeze the evolution by demanding $\tilde{U}(t) \approx \mathcal{I}$ in order to preserve the stored data.
Another goal is the simulation of other Hamiltonians (see e.\,g. \cite{WRJB02,BDNB04}), i.\,e. we would like the system to evolve as $\tilde{U}(t) \approx \exp( -i H_0' t/\hbar)$ with $H_0' \neq H_0$.
The former of these tasks is called decoupling.

\subsection{Bang-Bang Control}

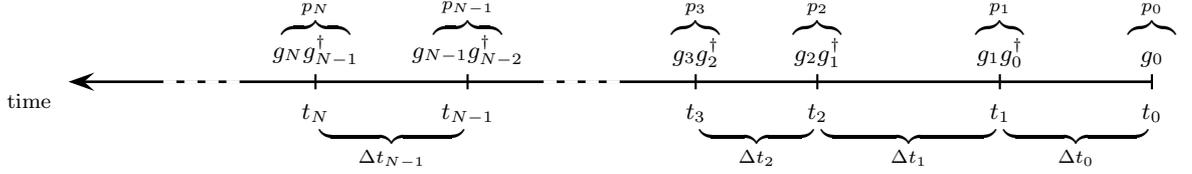
\begin{figure}\centering
\begin{pspicture}(-0.3,0.4)(15.3,-2.2)
\psline[linewidth=1pt,arrowsize=7pt]{<-}(0.75,-1.)(15,-1.)
\rput[t](0.23,-1.15){\footnotesize time}
\psline(15,-1.1)(15,-0.9)\rput[t](15,-1.3){\small $t_0$}\rput[b](15,-0.8){\small $g_0$}
\rput[b](15,-0.4){$\overbrace{\makebox(.4,0){}}^{p_0}$}
\rput[t](14,-1.6){$\underbrace{\makebox(1.9,0){}}_{\Delta t_0}$}
\psline(13,-1.1)(13,-0.9)\rput[t](13,-1.3){\small $t_1$}\rput[b](13,-0.8){\small $g_1 g_0^\dagger$}
\rput[b](13,-0.4){$\overbrace{\makebox(.4,0){}}^{p_1}$}
\rput[t](11.8,-1.6){$\underbrace{\makebox(2.3,0){}}_{\Delta t_1}$}
\psline(10.6,-1.1)(10.6,-0.9)\rput[t](10.6,-1.3){\small $t_2$}\rput[b](10.6,-0.8){\small $g_2 g_1^\dagger$}
\rput[b](10.6,-0.4){$\overbrace{\makebox(.4,0){}}^{p_2}$}
\rput[t](9.8,-1.6){$\underbrace{\makebox(1.5,0){}}_{\Delta t_2}$}
\psline(09,-1.1)(09,-0.9)\rput[t](09,-1.3){\small $t_3$}\rput[b](09,-0.8){\small $g_3 g_2^\dagger$}
\rput[b](09,-0.4){$\overbrace{\makebox(.4,0){}}^{p_3}$}
\psline[linewidth=1.2pt,linestyle=dashed,linecolor=white](8,-1.)(7,-1.)
\psline(6,-1.1)(6,-0.9)\rput[t](6,-1.3){\small $t_{N-1}$}\rput[b](6,-0.8){\small $g_{N-1} g_{N-2}^\dagger$}
\rput[b](6,-0.4){$\overbrace{\makebox(.9,0){}}^{p_{N-1}}$}
\rput[t](5,-1.6){$\underbrace{\makebox(1.9,0){}}_{\Delta t_{N-1}}$}
\psline(4,-1.1)(4,-0.9)\rput[t](4,-1.3){\small $t_N$}\rput[b](4,-0.8){\small $g_N g_{N-1}^\dagger$}
\rput[b](4,-0.4){$\overbrace{\makebox(.9,0){}}^{p_N}$}
\psline[linewidth=1.2pt,linestyle=dashed,linecolor=white](3,-1.)(2,-1.)
\end{pspicture}
\caption[Schematic representation of bang-bang control]{Schematic representation of bang-bang control. At time $t_i$ the pulse $p_i$ is applied instantaneously.\label{fig::bbcontrol}}
\end{figure}

In the quantum bang-bang control scenario \cite{VL98} it is assumed that we are able to apply a strong control $H_c(t)$ over a very short time interval.
In this case the resulting control action can be described as a quasi-instantaneous application of unitary pulses $p_i$ at times $t_i=\sum_{k=0}^{i-1} \Delta t_k$, $i\in \mathbb{N}_0$.
Since the control is assumed to be local, these pulses are of the form
$p_i = u^{(1,i)}_1\otimes u^{(2,i)}_2 \otimes \dots \otimes u^{(n,i)}_n$, where $u^{(a,i)}_c$ denotes the unitary $u^{(a,i)}\in \textsf{U}_q$ being applied to the $c$-th qudit.
After a time $t_N$ we obtain the total time evolution
\begin{equation}
 U(t_N) = p_N \, f_{\Delta t_{N-1}} \, \dots \, p_2 \, f_{\Delta t_1} \, p_1 \, f_{\Delta t_0} \, p_0,
\end{equation}
as depicted in figure \ref{fig::bbcontrol}.
Here, $f_{\Delta t_j} = \exp( -i H_0 \Delta t_j/\hbar )$ denotes free evolution due to $H_0$ over the time interval $\Delta t_j$.
Defining $g_i=p_i \dots p_1p_0$ we note that this evolution can be written as
\begin{equation}\label{eq:bbcontrolevo}
 U(t_N) = g_N (g_{N-1}^\dagger f_{\Delta t_{N-1}} g_{N-1}) \dots (g_1^\dagger f_{\Delta t_1} g_1) (g_0^\dagger f_{\Delta t_0} g_0).
\end{equation}
The time evolution operator $U_c$ at time $t_i+s$ with $s\in[0,\Delta t_i)$ is given by
$U_c(t_i+s) = g_i$, i.\,e. $U_c$ jumps from $g_{i-1}$ to $g_i=(g_ig_{i-1}^\dagger)g_{i-1} \equiv p_i g_{i-1}$ at time $t_i$.
Since $g_j^\dagger f_{\Delta t_j} g_j = \exp(-i g_j^\dagger H_0 g_j \Delta t_j/\hbar)$, let us define the toggled frame Hamiltonians $\tilde{H}_i = g_i^\dagger H_0 g_i$.
After switching to the toggled frame $\tilde{U}(t_N) = U_c^\dagger(t_N) U(t_N)$, the time evolution of equation \eqref{eq:bbcontrolevo} becomes
\begin{equation}
 \tilde{U}(t_N) = \exp(-i \tilde{H}_{N-1} \Delta t_{N-1}/\hbar) \dots \exp(-i \tilde{H}_1 \Delta t_1/\hbar) \exp(-i \tilde{H}_0 \Delta t_0/\hbar).
\end{equation}
To keep the notation as simple as possible, we set $\hbar=1$ for the remaining chapters.

\subsection{Average Hamiltonian Theory}
A convenient tool which is commonly used to analyze the resulting dynamics of a dynamical control scheme in the toggled frame is the average Hamiltonian theory (AHT) \cite{nmrbook}.
Let the time evolution in the toggled frame be generated by the time-dependent toggling frame Hamiltonian $\tilde{H}(t)$ of equation \eqref{eq:toframe-h0}.
After a time $t$ this results in the time evolution operator
\begin{equation}\label{eq:tUt1}
 \tilde{U}(t) = \mathcal{T} \exp\Bigl( -i \int_0^t \tilde{H}(t') dt' \Bigr),
\end{equation}
which can be written in terms of an average Hamiltonian $\overline{H}$ (which depends on $t$) as
\begin{equation}\label{eq:tUt2}
 \tilde{U}(t) = \exp\Bigl( -i \overline{H} t \Bigr).
\end{equation}
AHT expresses this average Hamiltonian as an infinite series of self-adjoint operators called Magnus expansion,
\begin{equation}\label{eq:magnusexp}
\overline{H} = \overline{H}^{(0)}+\overline{H}^{(1)}+\overline{H}^{(2)} +\dots,
\end{equation}
the first three terms of which are given by
\begin{subequations}\label{eq:ahts}
\begin{align}
\overline{H}^{(0)} &= \frac{1}{t} \int_0^{t}dt_1 \tilde{H}(t_1) \label{eq:ahts0}\\
\overline{H}^{(1)} &= -\frac{i}{2t} \int_0^{t}dt_2 \int_0^{t_2}dt_1 [\tilde{H}(t_2), \tilde{H}(t_1)] \label{eq:ahts1}\\
\overline{H}^{(2)} &= -\frac{1}{6t} \int_0^{t}dt_3 \int_0^{t_3}dt_2 \int_0^{t_2}dt_1
 \Bigl( [\tilde{H}(t_3),[\tilde{H}(t_2),\tilde{H}(t_1)]]+
 [[\tilde{H}(t_3),\tilde{H}(t_2)],\tilde{H}(t_1)] \Bigr). \label{eq:ahts2}
\end{align}
\end{subequations}
To obtain these expressions, we write \eqref{eq:tUt1} as an infinite series,
\begin{align}
 \tilde{U}(t) &=  \mathcal{I}  -it \sum_{n=0}^\infty \frac{(-i)^n}{t}\int_0^t dt_{n+1} \int_0^{t_{n+1}} dt_n \dots \int_0^{t_2} dt_1  \tilde{H}(t_{n+1}) \tilde{H}(t_n)\dots \tilde{H}(t_1) \nonumber\\
&\equiv \mathcal{I}  -it \sum_{n=0}^\infty h_n,
\end{align}
and expand \eqref{eq:tUt2} as
\begin{equation}
 \tilde{U}(t) = \mathcal{I} + \sum_{n=1}^\infty \frac{(-it)^n}{n!}\bigl( \overline{H}^{(0)}+\overline{H}^{(1)}+\overline{H}^{(2)} +\dots \bigr)^n.
\end{equation}
By noting that both $h_j$ and $\overline{H}^{(j)}$ are of order $j+1$ in $\tilde{H}$ and by comparing expressions of the same order in the last two equations, we obtain the expressions $h_0=\overline{H}^{(0)}$, $h_1=\overline{H}^{(1)}-it(\overline{H}^{(0)})^2/2$, et cetera, which eventually lead to \eqref{eq:ahts}.
In the bang-bang scenario at the time $t=t_N$ the Hamiltonians \eqref{eq:ahts0}--\eqref{eq:ahts2} become
\begin{subequations}\label{eq:bbh0h1h2}
\begin{align}
\overline{H}^{(0)} &= \frac{1}{t_N} \sum_{j=0}^{N-1} \tilde{H}_j \Delta t_j \label{eq:bbh0}\\
\overline{H}^{(1)} &= -\frac{i}{2t_N} \sum_{i>j=0}^{N-1} [\tilde{H}_i, \tilde{H}_j]\Delta t_i\Delta t_j \label{eq:bbh1}\\
\overline{H}^{(2)} &= -\frac{1}{6t_N} \sum_{i\geq j\geq k=0}^{N-1}
 \Bigl( [\tilde{H}_i,[\tilde{H}_j,\tilde{H}_k]]+
 [[\tilde{H}_i,\tilde{H}_j],\tilde{H}_k] \Bigr)\Delta t_i\Delta t_j\Delta t_k \times
\begin{cases}
1/2 & \text{ if } i=j \text{ or } j=k \\
1 & \text{ else }
\end{cases}.\label{eq:bbh2}
\end{align}
\end{subequations}

Finally, we state a theorem that will be used later on in this chapter to improve the performance of dynamical control schemes.
A proof of this theorem can be found in \cite{Bu81}.
\begin{thm}\label{thm:symtoggledh}
If the toggled frame Hamiltonian is symmetric in time, i.\,e. if $\tilde{H}(t-t') = \tilde{H}(t')$ for $t'\in[0,t]$,
all odd orders in the Magnus expansion \eqref{eq:magnusexp} of $\tilde{U}(t)=\exp( -i \overline{H} t)$ vanish, i.\,e. $\overline{H}^{(k)}=0$ for $k=1,3,5,\dots$ .
\end{thm}

\subsection{The Fundamental Control Strategy}\label{subsec:fundamental}

To achieve a certain control task like the simulation of a Hamiltonian $H_0'$, we make use of the simple structure of the zeroth order term $\overline{H}^{(0)}$ in the bang-bang setting.
\begin{defi}\label{def:controlscheme}
A set of unitaries $\{ g_j \}_{j=0}^{n_c-1}$ and relative times $\{ \Delta t_j \}_{j=0}^{n_c-1}$ such that
\begin{equation}\label{eq:zerothorderoverlineh}
 \overline{H}^{(0)} = \frac{1}{t_c} \sum_{j=0}^{n_c-1} \tilde{H}_j \Delta t_j
\equiv \frac{1}{t_c} \sum_{j=0}^{n_c-1} g_j^\dagger H_0 g_j \Delta t_j  = H_0' + \mathfrak{c}\cdot\frac{1}{d}\mathcal{I},
\end{equation}
where $t_c=\sum_{j=0}^{n_c-1}\Delta t_j$, $\mathfrak{c}=\tr(H_0)-\tr(H_0')$, and $d=\dim(\mathcal{H}_S)$, is called a control scheme of length $n_c$ for the simulation of the Hamiltonian $H_0'$ with the system Hamiltonian $H_0$.
\end{defi}
\begin{rem}
Without loss of generality, we usually assume all of the involved Hamiltonians to be traceless. In this case we have a vanishing constant $\mathfrak{c}=0$.
\end{rem}
\noindent
If we would like to achieve decoupling we set $H_0'\equiv 0$.
In this case a control scheme $\{ g_j,\Delta t_j \}_{j=0}^{n_c-1}$ is called a decoupling scheme.
An overview over various decoupling schemes for different types of $H_0$ is given in section \ref{sec:decschemes}.
It turns out that most of the times it is sufficient to consider control schemes with constant relative time intervals, i.\,e. $\Delta t_j=\Delta t$ for all $j\in\{0,\dots,n_c-1\}$.
In the following we will always be dealing with such schemes.

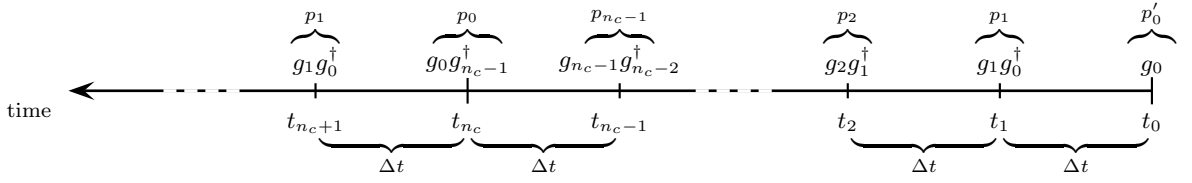
\begin{figure}\centering
\begin{pspicture}(-0.3,0.4)(15.3,-2.2)
\psline[linewidth=1pt,arrowsize=7pt]{<-}(0.75,-1.)(15,-1.)
\rput[t](0.23,-1.15){\footnotesize time}
\psline(15,-1.18)(15,-0.82)\rput[t](15,-1.3){\small $t_0$}\rput[b](15,-0.8){\small $g_0 $}
\rput[b](15,-0.4){$\overbrace{\makebox(.4,0){}}^{p'_0}$}
\rput[t](14,-1.6){$\underbrace{\makebox(1.9,0){}}_{\Delta t}$}
\psline(13,-1.1)(13,-0.9)\rput[t](13,-1.3){\small $t_1$}\rput[b](13,-0.8){\small $g_1 g_0^\dagger$}
\rput[b](13,-0.4){$\overbrace{\makebox(.4,0){}}^{p_1}$}
\rput[t](12,-1.6){$\underbrace{\makebox(1.9,0){}}_{\Delta t}$}
\psline(11,-1.1)(11,-0.9)\rput[t](11,-1.3){\small $t_2$}\rput[b](11,-0.8){\small $g_2 g_1^\dagger$}
\rput[b](11,-0.4){$\overbrace{\makebox(.4,0){}}^{p_2}$}
\psline[linewidth=1.2pt,linestyle=dashed,linecolor=white](10,-1.)(9,-1.)
\psline(8,-1.1)(8,-0.9)\rput[t](8,-1.3){\small $t_{n_c-1}$}\rput[b](8,-0.8){\small $g_{n_c-1} g_{n_c-2}^\dagger$}
\rput[b](8,-0.4){$\overbrace{\makebox(.9,0){}}^{p_{n_c-1}}$}
\rput[t](7,-1.6){$\underbrace{\makebox(1.9,0){}}_{\Delta t}$}
\psline(6,-1.17)(6,-0.83)\rput[t](6,-1.3){\small $t_{n_c}$}\rput[b](6,-0.8){\small $g_0 g_{n_c-1}^\dagger$}
\rput[b](6,-0.4){$\overbrace{\makebox(.9,0){}}^{p_0}$}
\rput[t](5,-1.6){$\underbrace{\makebox(1.9,0){}}_{\Delta t}$}
\psline(4,-1.1)(4,-0.9)\rput[t](4,-1.3){\small $t_{n_c+1}$}\rput[b](4,-0.8){\small $g_1 g_0^\dagger$}
\rput[b](4,-0.4){$\overbrace{\makebox(.4,0){}}^{p_1}$}
\psline[linewidth=1.2pt,linestyle=dashed,linecolor=white](3,-1.)(2,-1.)
\end{pspicture}
\caption[Cyclic decoupling (\textsf{PDD})]{Schematic representation of the cyclic (or periodic) control strategy (\textsf{PDD}).\label{fig::pdd}}
\end{figure}

The most basic control strategy is called cyclic (or periodic) dynamical decoupling\footnote{We call it a decoupling strategy even so it might be used for the purpose of simulating some Hamiltonian.} (\textsf{PDD}).
It consists of repeating the pulse sequence $p_0,\dots,p_{n_c-1}$,
with $p_j=g_j g_{j-1}^\dagger$  for $j=1,\dots,n_c-1$ and $p_0 = g_0 g_{n_c-1}^\dagger$ (with the exception that the first $p_0$ is simply given by $g_0$)
constructed using the elements of a control scheme $\{ g_j \}_{j=0}^{n_c-1}$ satisfying
\begin{equation}\label{eq:fctrlstrath0}
 \frac{1}{n_c} \sum_{j=0}^{n_c-1} g_j^\dagger H_0 g_j = H_0'+ \mathfrak{c}\cdot\frac{1}{d}\mathcal{I},
\end{equation}
over and over again (compare with figure \ref{fig::pdd}):
At the time $t_j = j \cdot \Delta t$, $j\in \mathbb{N}_0$, the pulse $p_{j\!\mod n_c}$ is applied.
As a result, the time evolution in the toggled frame after a time
$T = m\cdot t_c$, $m\in\mathbb{N}$, $t_c=n_c\Delta t$, is given by
\begin{equation}\label{eq:Utf_Tmtc}
 \tilde{U}( T=m\cdot t_c ) =
\Bigl(
 \exp(-i \tilde{H}_{n_c-1} \Delta t) \dots \exp(-i \tilde{H}_1 \Delta t) \exp(-i \tilde{H}_0 \Delta t)
\Bigr)^m = \exp(-i \overline{H} t_c \cdot m),
\end{equation}
where the zeroth order term in the Magnus expansion of $\overline{H}$ is given by \eqref{eq:fctrlstrath0}.
In the limit of $m\rightarrow\infty$ and $\Delta t\rightarrow 0$ with $T=m\cdot n_c \Delta t$ held constant, the influence of the higher order terms in the Magnus expansion decreases and \textsf{PDD} achieves its task perfectly: $\lim_{\Delta t\to 0}\tilde{U}(T) = \exp\bigl(-i \overline{H}^{(0)} T\bigr) = \exp(-i H_0' T )\cdot e^{-i T\mathfrak{c}/d}$.
In a realistic experiment we do not achieve this limit.
Therefore it is important to (i) quantify the error caused by the higher order terms and (ii) devise control strategies which keep the error for finite $\Delta t$ as small as possible.
In fact the main focus of the first part of this thesis is on (ii) and is dealt with in section \ref{sec:cstrategies}.
We proceed with (i) in the next subsection.

\subsection{Performance Measure}\label{subsec:performance}

If the control task is the simulation of a Hamiltonian $H_0'$,
the goal of a dynamical control strategy is to achieve a time evolution $\tilde{U}(T)$ in the toggled frame which is (up to a global phase) as close to $\tilde{U}_\text{id}(T) = \exp(- i H_0' T)$ as possible.
To quantify this closeness we define the pure state fidelity
\begin{equation}
 F_{\ket{\psi}} (T) = \bigl\vert \bra{\psi} \tilde{U}_\text{id}^\dagger(T) \, \tilde{U}(T) \ket{\psi} \bigr\vert^2.
\end{equation}
As long as $F_{\ket{\psi}}(T)$ stays close to one, we know that our control strategy was successful (at least if the quantum system was in the initial state $\ket{\psi}$).
To drop the dependence on $\ket{\psi}$, we might consider the worst case fidelity,
\begin{equation}\label{eq:Fw}
 F_w (T) = \min_{\ket{\psi}\in\mathcal{H}_S} \bigl\vert \bra{\psi} \tilde{U}_\text{id}^\dagger(T) \, \tilde{U}(T) \ket{\psi} \bigr\vert^2,
\end{equation}
as it was done in \cite{VK05} for the purpose of finding a lower bound,
or we might consider the average fidelity
\begin{equation}\label{eq:FaUU}
 F_a (T) = \int  \bigl\vert \bra{\psi} \tilde{U}_\text{id}^\dagger(T) \, \tilde{U}(T) \ket{\psi} \bigr\vert^2  \, d\psi.
\end{equation}
Here, the integration involved in the definition of the average fidelity has to be  performed over the uniform (Haar) measure on the relevant quantum state space with the normalization $\int d\psi=1$.

More generally, for a trace-preserving quantum operation $\mathcal{E}$ (i.\,e. a trace-preserving completely positive map), the average fidelity is defined as
\begin{equation}\label{eq:genAfidelity}
F_a(\mathcal{E}) = \int  \bra{\psi} \mathcal{E}(\ketbra{\psi}{\psi}) \ket{\psi} d\psi.
\end{equation}
Let $\ket{\Phi}$ be a maximally entangled state (e.\,g. a Bell state) between the quantum system under consideration and an ancilla system of the same dimension $d=\dim(\mathcal{H}_S)$.
Then the entanglement fidelity is defined as
\begin{equation}\label{eq:genEfideilty}
F_e(\mathcal{E}) = \bra{\Phi} (\mathcal{I}\otimes\mathcal{E})(\ketbra{\Phi}{\Phi}) \ket{\Phi},
\end{equation}
where $\mathcal{I}$ denotes the identity operation acting on the ancilla system.
The entanglement fidelity measures the degree to which the entanglement of quantum state is preserved by a quantum operation $\mathcal{E}$.
Apparently, it is independent of the choice of the maximally entangled state since any two maximally entangled states are related by a unitary acting only on the ancilla.
Both fidelity measures are not independent but are related by \cite{H399,Nielsen02} %
\begin{equation}\label{eq:fidcomp}
F_a(\mathcal{E}) = \frac{d F_e(\mathcal{E}) + 1}{ d+1 } = F_e(\mathcal{E}) + \mathcal{O}\left(\frac{1- F_e(\mathcal{E})}{d}\right).
\end{equation}
Thus, in the case of a quantum system which consists of a large number of qudits, i.\,e. $d=q^n\gg 1$, the difference between both measures tends to zero.
If we set $\mathcal{E}(\rho) = \tilde{U}_\text{id}^\dagger(T)  \tilde{U}(T) \,\rho\, \tilde{U}^\dagger(T) \tilde{U}_\text{id}(T)$, we obtain
\begin{equation}\label{eq:FeUU}
 F_e(\mathcal{E}) = F_e(T) =
 \Bigl\vert \frac{1}{d} \tr\bigl( \tilde{U}_\text{id}^\dagger(T) \tilde{U}(T) \bigr)\Bigr\vert^2.
\end{equation}
Typically, the evaluation of the entanglement fidelity is much simpler than the direct evaluation of the average fidelity \eqref{eq:FaUU}.
Therefore, in view of its close relationship to the average fidelity our subsequent discussion will mainly concentrate on the behavior of the entanglement fidelity.

Let us now consider the control task of decoupling, i.\,e. $\tilde{U}_\text{id}(T) = \mathcal{I}$, and let us estimate the entanglement fidelity given by \eqref{eq:FeUU} for the \textsf{PDD} control strategy.
The resulting fidelity has to be compared with the fidelity which is obtained in the absence of any decoupling. Without loss in generality, we assume that $\tr(H_0)=0$.

\subsubsection{No Decoupling (\textsf{none})}
Let us start examining the decay of the entanglement fidelity \eqref{eq:FeUU} in the absence of any decoupling.
In this case the time evolution due to the control alone is trivial, $U_c(t)=\mathcal{I}$, and the time evolution in the toggled frame coincides with the time evolution in the Schr\"odinger picture, i.\,e. $\tilde{U}(T) = U(T) = \exp(-i H_0 T)$.
In order to derive a series expansion of the fidelity, we write the system Hamiltonian $H_0$ as $\lambda H_0$ and expand in $\lambda$ (setting $\lambda=1$ in the end).
Such a series expansion up to fourth order in $\lambda$
leads to
\begin{align}
 F_e^\textsf{none}(T) &=
 \Bigl\vert \frac{1}{d} \tr\bigl( \tilde{U}(T) \bigr)\Bigr\vert^2 \nonumber\\
&= 1 - \frac{1}{d}\tr(H_0^2) T^2 + \Bigl(\frac{1}{2} \bigl(\frac{1}{d}\tr(H_0^2)\bigr)^2 + \frac{1}{6d}\tr(H_0^4)\Bigr)\frac{1}{2} T^4 + \mathcal{O}(\lambda^6 T^6).
\end{align}
Hence, for sufficiently small times, the fidelity decay is quadratic in time and its strength is determined by the trace of the square of the system Hamiltonian $H_0$.
By comparison with numerical simulations for various $H_0$, we found that a good approximation of $F_e^\textsf{none}(T)$ valid for $0\leq T \lesssim \sqrt{2}/\sqrt{ \tr(H_0^2)/d }$, or in other words as long as $F_e^\textsf{none}(T)\gtrsim 0.1$, is given by the simple expression
\begin{equation}\label{eq:FeappNone}
 F_{e\ \text{app}}^\textsf{none}(T) = \exp\Bigl( - \frac{1}{d}\tr(H_0^2) T^2 \Bigr).
\end{equation}
Viola and Knill \cite[theorem 3]{VK05} gave a strict lower bound on the worst case fidelity \eqref{eq:Fw} for \textsf{PDD} by using the matrix norm $\Vert A \Vert_2 = \max \vert \operatorname{eig}( \sqrt{A^\dagger A} )\vert$ and setting $\kappa=\Vert H_0\Vert_2$.
Analogous to this bound, a corresponding lower bound in the absence of decoupling is given by
\begin{equation}
 F_w^\textsf{none} (T) = \min_{\ket{\psi}\in\mathcal{H}_S} \bigl\vert \bra{\psi}  \tilde{U}(T) \ket{\psi} \bigr\vert^2 > 1 -  \kappa^2 T^2 + \mathcal{O}\bigl(\kappa^3 T^3 \bigr).
\end{equation}

\subsubsection{The \textsf{PDD} Fidelity}
By using a suitable control scheme $\{ g_j \}_{j=0}^{n_c-1}$, we have $\overline{H}^{(0)}=0$ and
$\tilde{U}(T) = \exp\bigl( -i \sum_{j=1}^\infty \overline{H}^{(j)} T \bigr)$ for $T=m\cdot t_c$ with $m\in\mathbb{N}$ and $t_c=n_c\Delta t$ (compare with \eqref{eq:Utf_Tmtc}).
Writing $H_0$ as $\lambda H_0$, we obtain
\begin{align}
 F_e^\textsf{PDD}(T) &= \Bigl\vert \frac{1}{d} \tr\bigl( \tilde{U}(T) \bigr)\Bigr\vert^2
         = 1 - \frac{1}{d}\tr\Bigl( \bigl(\sum_{j=1}^\infty \overline{H}^{(j)}\bigr)^2 \Bigr) T^2 + \dots \nonumber\\
&= 1 - \frac{1}{d}\tr\bigl( \bigl(\overline{H}^{(1)}\bigr)^2 \bigr) T^2 + \mathcal{O}( \lambda^5 t_c^3 T^2).
\end{align}
To evaluate this short time estimation, we have to calculate $\overline{H}^{(1)}$.
A rough estimate based on the fact that $\overline{H}^{(1)}$ is a sum over $\mathcal{O}(n_c^2)$ terms of the form $\tilde{H}_i\tilde{H}_j$ leads to
$\overline{H}^{(1)} = \mathcal{O}( \lambda^2 t_c)$.
As before, we argue that a good approximation of $F_e^\textsf{PDD}(T)$ is given by
\begin{equation}
 F_{e\ \text{app}}^\textsf{PDD}(T) = \exp\Bigl( - \frac{1}{d}\tr\bigl( \bigl(\overline{H}^{(1)}\bigr)^2 \bigr) T^2 \Bigr),
\end{equation}
as long as the fidelity has not become too small, i.\,e. for times $T$ such that $F_e^\textsf{PDD}(T) \gtrsim 0.1$.
A strict lower bound on the worst case fidelity was given by Viola and Knill \cite[theorem 3]{VK05}:
\begin{equation}
 F_w^\textsf{PDD} (T)  > 1 -  \kappa^4 t_c^2 T^2 + \mathcal{O}\bigl(\kappa^5 t_c^3 T^2\bigr).
\end{equation}

\subsection{Open Quantum Systems}\label{subsec:opens}
Up to this point we considered a closed quantum system $S$ and the task of dynamical decoupling was the removal of inter-qudit couplings.
In a real-world scenario, there will always be an interaction of the system with its surrounding environment $E$.
As a result, entanglement between the system and the environment may arise causing the quantum system to evolve in a non-unitary way and to undergo a process called decoherence.
Zanardi \cite{Za99} and Viola et\,al. \cite{VKL99} proposed that dynamical decoupling techniques may be applied to decouple such systems from their environment.
This subsection summarizes the main idea.

In this subsection we consider $S$ to be an open system, i.\,e. to be part of a larger closed system formed by $S$ and $E$ together.
Then the total system is defined on the Hilbert space $\mathcal{H}_{SE} = \mathcal{H}_S\otimes \mathcal{H}_E$, where $\mathcal{H}_S$ and $\mathcal{H}_E$ denote the system and environment Hilbert space.
The Hamiltonian of the total system is given by the sum of the Hamiltonian $H_0$ of the system $S$ and the Hamiltonian $H_E$ of the environment,
plus additional terms describing the couplings of the system with the environment,
\begin{equation}\label{eq:hamopen}
  H_{0,SE} = H_0 \otimes \mathcal{I}_E + \mathcal{I}_S\otimes H_E + \sum_\alpha S_\alpha\otimes E_\alpha.
\end{equation}
Here, the $E_\alpha$'s are supposed to be linearly independent and, without loss of generality, the coupling operators $S_\alpha$ are assumed to be traceless.
We proceed as in the case of a closed system:
By applying a time-dependent local control $H_c(t)$ on the system $S$, the total Hamiltonian becomes time dependent,
\begin{equation}\label{eq:HSEt}
  H_{SE}(t) = H_{0,SE} + H_c(t)\otimes \mathcal{I}_E,
\end{equation}
and we switch to the toggled frame defined by $\tilde{U}(t) = U^\dagger_c(t)\otimes \mathcal{I}_E \cdot U_{SE}(t)$, where $U_{SE}(t)$ denotes the time evolution operator of the combined system evolving according to \eqref{eq:HSEt}, and $U_c(t)$ is defined as in \eqref{eq:Uct} as the time evolution operator of the system evolving according to $H_c(t)$ alone.
The time evolution in the toggled frame is determined by the toggled frame Hamiltonian
\begin{equation}
 \tilde{H}_{SE}(t) = 
U^\dagger_c(t) H_0 U_c(t) \otimes \mathcal{I}_E + \mathcal{I}_S\otimes H_E + \sum_\alpha U^\dagger_c(t)S_\alpha U_c(t) \otimes E_\alpha.
\end{equation}
A decoupling scheme $\{ g_j \}_{j=0}^{n_c-1}$ that applies to all the coupling operators $S_\alpha$ satisfies
\begin{equation}
 \overline{S}_\alpha^{(0)} = \frac{1}{n_c} \sum_{j=0}^{n_c} g^\dagger_j S_\alpha g_j = \mathfrak{c}_\alpha\cdot\frac{1}{d}\mathcal{I}, \text{ with }\mathfrak{c}_\alpha=\tr(S_\alpha),
\end{equation}
for all $\alpha$.
If we use such a scheme in connection with the periodic dynamical decoupling (\textsf{PDD}) control strategy,
we achieve the desired decoupling from the environment in lowest order AHT:
\begin{equation}
 \overline{H}^{(0)}_{SE} = \overline{H}_0^{(0)} \otimes \mathcal{I}_E + \mathcal{I}_S\otimes \bigl( H_E + \sum_\alpha \frac{\mathfrak{c}_\alpha}{d} E_\alpha \bigr)
 \equiv \overline{H}_0^{(0)} \otimes \mathcal{I}_E + \mathcal{I}_S\otimes H_E'.
\end{equation}
As it was discussed before, in the fast control limit, i.\,e. for $\Delta t \rightarrow 0$ and $m\rightarrow \infty$ with the total time $T=m \cdot n_c\Delta t$ held constant, lowest order AHT becomes exact and we obtain
\begin{equation}
 \tilde{U}(T) = \exp\bigl( - i \overline{H}^{(0)}_{SE} T \bigr) = \exp\bigl(-i \overline{H}_0^{(0)} T\bigr) \otimes \exp\bigl( -i H_E' T\bigr).
\end{equation}
For quantum memories the decoupling scheme should also satisfy $\overline{H}_0^{(0)} = \mathfrak{c}\cdot\frac{1}{d}\mathcal{I}$, so that (up to a global phase determined by $\mathfrak{c}$) $\tilde{U}(T) = \mathcal{I}_S \otimes \exp\bigl( -i H_E' T\bigr)$.
\subsection{Noiseless Subsystems}\label{subsec:nlsubsys}

Dynamical decoupling was defined as a dynamical control setting in which the time evolution of a quantum system is made to freeze.
This is achieved by applying a decoupling scheme for the system Hamiltonian $H_0$ in a way specified by a certain control strategy (as for example \textsf{PDD}).
As a result, the average Hamiltonian in the toggled frame vanishes.
As discussed in the preceding subsection, for open quantum systems in principle the same method can be applied, provided that the decoupling scheme also applies to the coupling operators which are responsible for the interaction with the environment.
A less demanding goal is the dynamical generation of a noiseless subsystem \cite{Za00, VKL00}.
Instead of trying to protect the whole quantum system, control schemes are applied in order to preserve parts of the system. Information can then safely be stored by encoding it into such a part.

Let $G = \{ \mathfrak{g}_j \}_{j=0}^{n_G-1}$ be a finite group of order $n_G$,
and let $R : \mathfrak{g}_j \mapsto R(\mathfrak{g}_j)=g_j \in \textsf{U}_d$ be a unitary representation of $G$ on the $d$-dimensional system Hilbert space $\mathcal{H}_S$ (for our qudit quantum register $\mathcal{H}_S = \mathcal{H}_q^{\otimes n}$ and $d=q^n$).
As explained in the introduction in subsection \ref{subsec:reptheo},
the representation $R$ decomposes into a sum of irreps of $G$,
\begin{equation}
 R(G) =  \bigoplus_{\nu\in\mathcal{J}} \tau_\nu \cdot D^{(\nu)}(G),
\end{equation}
where the multiplicity of the irrep labeled by $\nu$ is denoted as $\tau_\nu$ and the dimension of the irrep $D^{(\nu)}(G)$ is denoted by $d_\nu$.
Since the representation space of $R$ is the system Hilbert space $\mathcal{H}_S$, any of the results of subsection \ref{subsec:reptheo} concerning the representation space apply to $\mathcal{H}_S$:
There exists an orthonormal basis
\begin{equation}
\bigl\{ \ket{ \nu \ l_\nu \  m_\nu } \sthat \nu\in\mathcal{J},\ l_\nu=1\dots \tau_\nu,\ m_\nu=1\dots d_\nu  \bigr\},
\end{equation}
in which the operators $g_j$ are block-diagonal, i.\,e.
\begin{equation}
g_j  \ket{ \nu \ l_\nu \  m_\nu } =
D^{(\nu)}(g_j) \ket{ \nu \ l_\nu \  m_\nu } =
 \sum_{m'_\nu=1}^{d_\nu} D_{m'_\nu m_\nu}^{(\nu)}(g_j)  \ket{ \nu \ l_\nu \ m'_\nu }.
\end{equation}
The subspace of $\mathcal{H}_S$ which is spanned by the set of basis vectors with fixed $\nu$ is labeled by $\mathcal{H}_\nu$,
\begin{equation}
  \mathcal{H}_\nu = \vspan \bigl\{ \ket{\nu \ l_\nu \ m_\nu} \sthat l_\nu=1\dots \tau_\nu, m_\nu=1\dots d_\nu \bigr\},
\end{equation}
and has the form of a tensor space ($\ket{\nu \ l_\nu \ m_\nu} = \ket{l_\nu}\otimes\ket{m_\nu}$), i.\,e. we write $\mathcal{H}_\nu = \mathcal{C}_\nu \otimes \mathcal{D}_\nu$, where the dimension of $\mathcal{C}_\nu$ is given by $\tau_\nu$ and the dimension of $\mathcal{D}_\nu$ is given by $d_\nu$.
The Hilbert space decomposes as
\begin{equation}
 \mathcal{H}_S = \bigoplus_{\nu\in\mathcal{J}} \mathcal{H}_\nu = \bigoplus_{\nu\in\mathcal{J}} \mathcal{C}_\nu \otimes \mathcal{D}_\nu.
\end{equation}
Let $\mathcal{A} = R(\mathbb{C}G)$ denote the group algebra generated by $R$, and let its commutant $\mathcal{A}'$ be defined as the set of elements that commute with all the elements in $\mathcal{A}$,
$\mathcal{A}' = \{ V\in\mathcal{L}(\mathcal{H}_S) \sthat VA=AV \text{ for all } A\in\mathcal{A} \}$.
According to theorem \ref{thm:algebrabdiag} the elements of $\mathcal{A}$ and $\mathcal{A}'$ become block-diagonal in the $\{ \ket{ \nu\ l_\nu\ m_\nu } \}$-basis,
\begin{align}
 \mathcal{A}  &\cong \bigoplus_{\nu\in\mathcal{J}} \mathbbm{1}_{\tau_\nu} \otimes \operatorname{Mat}(d_\nu\times d_\nu,\mathbb{C})  \\
 \mathcal{A}' &\cong \bigoplus_{\nu\in\mathcal{J}} \operatorname{Mat}(\tau_\nu\times \tau_\nu,\mathbb{C}) \otimes \mathbbm{1}_{d_\nu},
\end{align}
where $\mathbbm{1}_n$ denotes an $n\times n$ dimensional identity matrix and $\operatorname{Mat}(n\times n,\mathbb{C})$ denotes the set of $n\times n$ matrices with entries in $\mathbb{C}$.

We start by describing the idea of a noiseless subsystem \cite{ZaRa97a,ZaRa97b,LCW98}.
Let us imagine that the quantum system $S$ under consideration is open and its Hamiltonian is given by equation \eqref{eq:hamopen}.
If $H_0$ and the coupling operators $S_\alpha$ are elements of $\mathcal{A}$, we have
\begin{equation}
 H_0 = \bigoplus_{\nu\in\mathcal{J}}  \mathcal{I}_{\mathcal{C}_\nu} \otimes D^{(\nu)}(H_0),
\end{equation}
and corresponding expressions for the $S_\alpha$.
It follows that information encoded in the $\mathcal{C}_\nu$-part of the subspace $\mathcal{H}_\nu$ remains unchanged over time:
Let the information be described by $\rho= \sum_{i,j=1}^{\tau_\nu} \rho_{ij} \ket{i}\bra{j}$ and let it be encoded in $\mathcal{H}_\nu$ as
\begin{equation}
\rho_{\mathcal{C}_\nu} \, \otimes \, \sigma_{\mathcal{D}_\nu} = \sum_{i,j=1}^{\tau_\nu} \sum_{k,l=1}^{d_\nu} \rho_{ij}\sigma_{kl} \ket{\nu i_\nu k_\nu}\bra{\nu j_\nu l_\nu}
\end{equation}
for some arbitrary $\sigma = \sum_{k,l=1}^{d_\nu} \sigma_{kl} \ket{k}\bra{l}$.
Denoting the time evolution operator of the total system as $U_{SE}(t)$ and assuming that the environment is initially not entangled with the system, we obtain
\begin{align}
 \bigl( (\rho_{\mathcal{C}_\nu}\otimes\sigma_{\mathcal{D}_\nu})_S \otimes \tau_E \bigr) (t) &= U_{SE}(t) \ \bigl( (\rho_{\mathcal{C}_\nu}\otimes\sigma_{\mathcal{D}_\nu})_S \otimes \tau_E \bigr) \ U^\dagger_{SE}(t) \nonumber\\
&=\exp\Bigl(-it \, \mathcal{I}_{\mathcal{C}_\nu} \otimes \bigl( D^{(\nu)}(H_0) \otimes \mathcal{I}_E + \mathcal{I}_{\mathcal{D}_\nu} \otimes H_E + \sum_\alpha D^{(\nu)}(S_\alpha) \otimes E_\alpha \bigr) \Bigr) \times \nonumber\\
& \qquad\qquad\qquad\qquad\qquad\qquad\qquad\qquad\qquad \rho_{\mathcal{C}_\nu} \otimes \sigma_{\mathcal{D}_\nu} \otimes \tau_E \ \exp\bigl(+it \dots \bigr) \nonumber\\
&= \rho_{\mathcal{C}_\nu} \otimes U_{\mathcal{D}_\nu E}(t) \, (\sigma_{\mathcal{D}_\nu} \otimes \tau_E) \,  U^\dagger_{\mathcal{D}_\nu  E}(t).
\end{align}
Hence the $\{\mathcal{C}_\nu\}_{\nu\in\mathcal{J}}$ are indeed noiseless (or decoherence-free) subsystems.
In the special case that $d_\nu=1$, $\mathcal{C}_\nu$ is a noiseless subspace.

Unfortunately, the interactions of a typical quantum system hardly allow the existence of large noiseless subsystems.
Hence Zanardi and Viola et\,al. \cite{Za00, VKL00} came up with the idea to modify the interactions in terms of dynamical control, such that the resulting symmetrized dynamics allows for larger noiseless subsystems.
Let a control scheme $\mathcal{G} = \{ g_j \}_{j=0}^{n_c-1}$ of length $n_c=n_G$ be defined by a unitary projective representation $R$ of a group $G = \{ \mathfrak{g}_j \}_{j=0}^{n_G-1}$ acting on the system Hilbert space $\mathcal{H}_S$, i.\,e. $g_j = R(\mathfrak{g}_j)$.
As a result of the applied control scheme (let us assume here for simplicity that we use the \textsf{PDD} control strategy in the fast control limit), the operators $H_0$ and $S_\alpha$ become
\begin{equation}
 \Pi_\mathcal{G}(X) = \frac{1}{n_c} \sum_{j=0}^{n_c-1} g_j^\dagger X g_j,
\end{equation}
with $X\in\{H_0,S_\alpha\}$.
Since $\Pi_\mathcal{G}(X)$ commutes with any element $g_j$ of the control scheme, it follows that $\Pi_\mathcal{G}(X)$ is in $\mathcal{A}'$.
As a result, the subsystems $\{\mathcal{D}_\nu\}_{\nu\in\mathcal{J}}$ are noiseless.
The standard decoupling scenario ($\Pi_\mathcal{G}(X) = \mathfrak{c}_X\cdot\mathcal{I}$ for all $X\in\{H_0,S_\alpha\}$) is included as a special case:
If the representation is irreducible, the set $\mathcal{J}$ consist of only one element $\nu$ and we have $\tau_\nu=1$ and $d_\nu=\dim(\mathcal{H}_S)$.

\subsection{Bounded Controls}\label{subsec:boundedctrls}
\begin{figure}\centering
\scalebox{0.91}{%
\begin{pspicture}(-2.0,0.4)(15.3,-2.2)
\psline[linewidth=1pt,arrowsize=7pt]{<-}(-1.50,-1.)(15,-1.)
\rput[t](-2.,-1.15){\footnotesize time}
\psline(15,-1.1)(15,-0.9)\rput[t](15,-1.3){\small $t_0$}%
\rput[B](14.6,-0.4){\gray \small $H_Y$}
\psline[linewidth=3pt,arrowsize=7pt, linecolor=gray]{->}(14.975,-1.)(14,-1.)
\rput[t](14,-1.6){$\underbrace{\makebox(1.9,0){}}_{\Delta t}$}
\psline(13,-1.1)(13,-0.9)\rput[t](13,-1.3){\small $t_1$}\rput[b](14.1,-0.8){\small $Y\mathcal{I}^\dagger$}
\rput[B](12.6,-0.4){\gray \small $H_X$}
\psline[linewidth=3pt,arrowsize=7pt, linecolor=gray]{->}(12.975,-1.)(12,-1.)
\rput[t](12.5,-1.6){$\underbrace{\makebox(0.9,0){}}_{\tau_p}$}
\psline(11,-1.1)(11,-0.9)\rput[t](11,-1.3){\small $t_2$}\rput[b](12.1,-0.8){\small $ZY^\dagger$}
\rput[B](10.6,-0.4){\gray \small $H_Y$}
\psline[linewidth=3pt,arrowsize=7pt, linecolor=gray]{->}(10.975,-1.)(10,-1.)
\psline(9,-1.1)(9,-0.9)\rput[t](9,-1.3){\small $t_3$}\rput[b](10.1,-0.8){\small $XZ^\dagger$}
\rput[B](8.6,-0.4){\gray \small $H_Y$}
\psline[linewidth=3pt,arrowsize=7pt, linecolor=gray]{->}(8.975,-1.)(8,-1.)
\psline(7,-1.1)(7,-0.9)\rput[t](7,-1.3){\small $t_4$}\rput[b](8.1,-0.8){\small $ZX^\dagger$}
\rput[B](6.6,-0.4){\gray \small $H_X$}
\psline[linewidth=3pt,arrowsize=7pt, linecolor=gray]{->}(6.975,-1.)(6,-1.)
\psline(5,-1.1)(5,-0.9)\rput[t](5,-1.3){\small $t_5$}\rput[b](6.1,-0.8){\small $YZ^\dagger$}
\rput[B](4.6,-0.4){\gray \small $H_Y$}
\psline[linewidth=3pt,arrowsize=7pt, linecolor=gray]{->}(4.975,-1.)(4,-1.)
\psline(3,-1.1)(3,-0.9)\rput[t](3,-1.3){\small $t_6$}\rput[b](4.1,-0.8){\small $\mathcal{I}Y^\dagger$}
\rput[B](2.6,-0.4){\gray \small $H_X$}
\psline[linewidth=3pt,arrowsize=7pt, linecolor=gray]{->}(2.975,-1.)(2,-1.)
\rput[t](1.5,-1.6){$\underbrace{\makebox(0.9,0){}}_{\Delta t-\tau_p}$}
\psline(1,-1.1)(1,-0.9)\rput[t](1,-1.3){\small $t_7$}\rput[b](2.1,-0.8){\small $X\mathcal{I}^\dagger$}
\rput[B](0.6,-0.4){\gray \small $H_X$}
\psline[linewidth=3pt,arrowsize=7pt, linecolor=gray]{->}(0.975,-1.)(0,-1.)
\psline(-1,-1.1)(-1,-0.9)\rput[t](-1,-1.3){\small $t_8$}\rput[b](0.1,-0.8){\small $\mathcal{I}X^\dagger$}
\end{pspicture}}%
\caption[Euler decoupling]{Schematic representation of an Euler decoupling cycle based on the decoupling set $\mathcal{G}=\{\mathcal{I},X,Y,Z\}$ and the generators $\Gamma=\{X,Y\}$.
The above cycle of length $t_c = \vert\mathcal{G}\vert\cdot \vert\Gamma\vert\cdot \Delta t = 8\Delta t$ is based on the Eulerian cycle on the Cayley graph of $\mathcal{G}$ with respect to $\Gamma$ shown in figure \ref{fig::digraph}.
It is repeated over and over again.
$H_X$ denotes a potentially time-dependent control Hamiltonian which generates the generator $X$, i.\,e. up to a phase we have $X=\mathcal{T}\exp\bigl(-i \int_0^{\tau_p} H_X(t') dt' \bigr)$. $H_Y$ is defined analogously.
As a result, the applied control generates the gates denoted in the second line.\label{fig::euler}}
\end{figure}
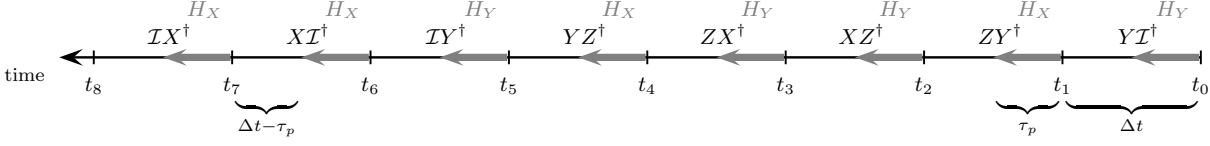
\begin{figure}
\begin{minipage}{.5\textwidth}
\centering
\scalebox{1}{
\begin{pspicture}(0bp,0bp)(120bp,134bp)
\psset{linewidth=1bp}

  \pstVerb{2 setlinejoin} %
\psset{linecolor=black}
  \psset{linecolor=mediumblue}
  \psbezier[arrows=->](41bp,119bp)(47bp,121bp)(53bp,123bp)(60bp,124bp)(65bp,125bp)(71bp,124bp)(86bp,120bp)
  \psset{linecolor=black}
  \rput(63bp,129bp){\small $7$}
  \psset{linecolor=red}
  \psbezier[arrows=->](19bp,95bp)(17bp,89bp)(15bp,82bp)(14bp,76bp)(13bp,66bp)(13bp,55bp)(15bp,36bp)
  \psset{linecolor=black}
  \rput(22bp,65bp){\small $1$}
  \psset{linecolor=red}
  \psbezier[arrows=->](21bp,36bp)(22bp,42bp)(23bp,48bp)(23bp,54bp)(24bp,64bp)(24bp,74bp)(24bp,94bp)
  \psset{linecolor=black}
  \rput(29bp,65bp){\small $6$}
  \psset{linecolor=mediumblue}
  \psbezier[arrows=->](83bp,8bp)(75bp,4bp)(65bp,1bp)(55bp,3bp)(51bp,4bp)(47bp,5bp)(34bp,10bp)
  \psset{linecolor=black}
  \rput(58bp,8bp){\small $5$}
  \psset{linecolor=mediumblue}
  \psbezier[arrows=->](36bp,18bp)(46bp,18bp)(59bp,18bp)(81bp,18bp)
  \psset{linecolor=black}
  \rput(58bp,23bp){\small $2$}
  \psset{linecolor=red}
  \psbezier[arrows=->](100bp,36bp)(100bp,50bp)(101bp,69bp)(101bp,94bp)
  \psset{linecolor=black}
  \rput(104bp,65bp){\small $3$}
  \psset{linecolor=red}
  \psbezier[arrows=->](93bp,96bp)(90bp,90bp)(87bp,83bp)(85bp,76bp)(83bp,66bp)(84bp,63bp)(85bp,54bp)(86bp,51bp)(87bp,47bp)(91bp,34bp)
  \psset{linecolor=black}
  \rput(91bp,65bp){\small $4$}
  \psset{linecolor=mediumblue}
  \psbezier[arrows=->](84bp,112bp)(74bp,112bp)(63bp,112bp)(42bp,112bp)
  \psset{linecolor=black}
  \rput(63bp,117bp){\small $8$}
{%
  \psset{linecolor=black}
  \psellipse[](24bp,112bp)(18bp,18bp)
  \rput(24bp,112bp){$I$}
}%
{%
  \psset{linecolor=black}
  \psellipse[](102bp,112bp)(18bp,18bp)
  \rput(102bp,112bp){$X$}
}%
{%
  \psset{linecolor=black}
  \psellipse[](99bp,18bp)(18bp,18bp)
  \rput(99bp,18bp){$Z$}
}%
{%
  \psset{linecolor=black}
  \psellipse[](18bp,18bp)(18bp,18bp)
  \rput(18bp,18bp){$Y$}
}%
\end{pspicture}}%
\end{minipage}%
\begin{minipage}{.5\textwidth}
\caption[Directed graph]{Eulerian cycle on the Cayley graph of $\mathcal{G}=\{\mathcal{I},X,Y,Z\}$ with respect to the generators $\Gamma=\{X,Y\}$. The edges colored by $X$ are depicted in blue, those colored by $Y$ are shown in red.\label{fig::digraph}}
\end{minipage}
\end{figure}
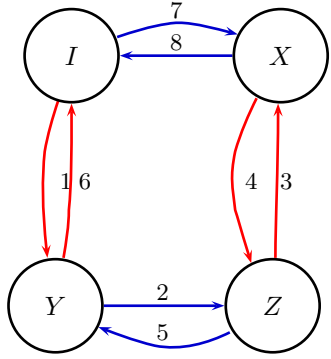

The current chapter of this thesis deals with dynamical decoupling in the bang-bang control scenario, i.\,e. we assume a strong control Hamiltonian such that any applied control pulse may be considered as being applied instantaneously.
Of course such a scenario is an idealization. This subsection discusses the effects of bounded controls.

In order to analyze the effects of bounded controls, let us assume we apply the control scheme $\{ g_j \}_{j=0}^{n_c-1}$ of length $n_c$ using the fundamental control strategy (also called periodic dynamical decoupling), i.\,e. we repeat the pulse sequence $p_0,\dots,p_{n_c-1}$, with $p_j=g_{j+1} g_j^\dagger$ for $j=0,\dots,n_c-1$ and $g_{n_c}=g_0$, over and over again\footnote{In subsection \ref{subsec:fundamental} the original definition of $p_j$ was $p_j=g_j g_{j-1}^\dagger$. Here it is changed it to $p_j=g_{j+1} g_j^\dagger$ in order to close the basic cycle with $g_0$ instead of $g_{n_c-1}$.}.
But instead of applying the pulses $p_{j\!\mod n_c}$ instantaneously at times $j\cdot \Delta t$, $j\in\mathbb{N}_0$,
we now assume that each pulse is generated by switching on a possibly time-dependent control Hamiltonian $H_j(t')$ for a time $\tau_p < \Delta t$ during the time interval $[j\cdot \Delta t, j\cdot \Delta t+\tau_p]$ such that
$p_j = p_j(\tau_p) = \mathcal{T} \exp\bigl( -i \int_0^{\tau_p} H_j(t') dt'  \bigr)$.
As a result, after $m\in \mathbb{N}$ such cycles of length $t_c=n_c\Delta t$, the time evolution operator in the toggled frame is given by $\tilde{U}(T=mt_c) = \exp\bigl(-i \overline{H} T\bigr)$, where in lowest order AHT $\overline{H}$ is given by equation \eqref{eq:ahts0}:
\begin{align}\label{eq:bctrlohneeuler}
 \overline{H}^{(0)} &= \frac{1}{t_c} \int_0^{t_c} dt_1 \tilde{H}(t_1) \nonumber\\
&= \frac{1}{n_c} \sum_{j=0}^{n_c-1} g_j^\dagger \biggl( \frac{1}{\Delta t}
 \int_0^{\tau_p} p_j^\dagger(t')  \ H_0 \ p_j(t') dt' + H_0 \cdot(1-\tau_p/\Delta t)  \biggr)  g_j.
\end{align}
For $\tau_p \rightarrow 0$ this expression reduces to the corresponding expression \eqref{eq:fctrlstrath0} of the bang-bang scenario.
If the control scheme $\{ g_j \}_{j=0}^{n_c-1}$ is for the simulation of the Hamiltonian $H_0'$ with the system Hamiltonian $H_0$, this means that for $\tau_p=0$ we would get
\begin{equation}\label{eq:bbctrlcond_eulersection}
 \overline{H}^{(0)} = \frac{1}{n_c} \sum_{j=0}^{n_c-1} g_j^\dagger H_0 g_j = H_0'+ \mathfrak{c}\cdot\frac{1}{d}\mathcal{I},
\end{equation}
with $\mathfrak{c}=\tr(H_0)-\tr(H_0')$.
For finite $\tau_p$ the first term within the braces in equation \eqref{eq:bctrlohneeuler} depends on $j$ and prevents the bang-bang control condition from above to be fulfilled.

If the elements of the control scheme $\mathcal{G} = \{ g_j \}_{j=0}^{n_c-1}$ are defined by a unitary projective representation $R$ of a group $G = \{ \mathfrak{g}_j \}_{j=0}^{n_c-1}$ acting on the system Hilbert space $\mathcal{H}_S$, i.\,e. if we have $g_j = R(\mathfrak{g}_j)$, this problem may be circumvented by using the so-called Eulerian decoupling proposed by Viola and Knill \cite{VK03}.
Before we describe their idea, we have to make some definitions.
First, let $\mathcal{A} = R(\mathbb{C}G)$ denote the corresponding group algebra, and let its commutant $\mathcal{A}'$ be defined as the set of elements that commutes with all the elements in $\mathcal{A}$.
Second, the Cayley graph of $\mathcal{G}$ with respect of to a set of generators is defined as follows:
\begin{defi}[Cayley graph]
Let $\mathcal{G} = \{ g_j \}_{j=0}^{n_c-1}$ be a finite group of order $n_c$, and let $\Gamma = \{ p_i \}_{i=1}^{\vert\Gamma\vert}$ be a generating set.
Then the Cayley graph of $\mathcal{G}$ with respect to $\Gamma$ is defined as the directed multigraph whose edges are colored by the generators $p_i\in \Gamma$, such that vertex $g_j$ is joined to vertex $g_k$ by an edge of color $p_i$ if and only if $g_k = p_i g_j$ (or $p_i = g_kg_j^\dagger$).
\end{defi}
\noindent
Last, an Eulerian path in the Cayley graph is defined as a path which uses each edge exactly once.
The proposal of Viola and Knill is now to replace the basic \textsf{PDD} cycle of length $n_c = \vert\mathcal{G}\vert$ by a cycle corresponding to an Eulerian path of length $n_c\cdot \vert\Gamma\vert$.
As a consequence, instead of \eqref{eq:bctrlohneeuler}, we obtain in lowest order AHT
\begin{equation}\label{eq:H0bctrlecycle}
 \overline{H}^{(0)} =
\frac{1}{n_c} \sum_{j=0}^{n_c-1} g_j^\dagger \biggl(
\frac{1}{\vert\Gamma\vert} \sum_{i=1}^{\vert\Gamma\vert} \frac{1}{\Delta t}
 \Bigl( \int_0^{\tau_p} p_i^\dagger(t')  \ H_0 \ p_i(t') dt' + H_0\cdot(\Delta t-\tau_p) \Bigr)
\biggr)  g_j,
\end{equation}
where, as before, $p_i(t) = \mathcal{T} \exp\bigl( -i \int_0^t H_i(t') dt'  \bigr)$ is generated using a possibly time-dependent control Hamiltonian $H_i$ ($i=1,\dots,\vert\Gamma\vert$).
By using the definitions
\begin{align}
\Pi_\mathcal{G} (X) &= \frac{1}{n_c} \sum_{j=0}^{n_c-1} g_j^\dagger X g_j \\
 F_\Gamma (X) &= \frac{1}{\vert\Gamma\vert} \sum_{i=1}^{\vert\Gamma\vert} \frac{1}{\tau_p}
 \int_0^{\tau_p} p_i^\dagger(t')  \ X \ p_i(t') dt',\label{eq:FGamma}
\end{align}
this expression can be written as
\begin{equation}
 \overline{H}^{(0)} =
\Pi_\mathcal{G}\bigl(  F_\Gamma(H_0)  \bigr) \cdot \frac{\tau_p}{\Delta t} +
\Pi_\mathcal{G}\bigl( H_0 \bigr) \cdot(\Delta t-\tau_p)/\Delta t.
\end{equation}
Due to the following theorem this is equal to $\Pi_\mathcal{G}\bigl( H_0 \bigr)$ and we arrive at the standard control condition \eqref{eq:bbctrlcond_eulersection} of the bang-bang scenario.

\begin{thm}[\cite{VK03}]\label{thm:vk03}
Let $X$ be any time-independent operator acting on the system Hilbert space $\mathcal{H}_S$.
If the control Hamiltonians $H_i(t)$ are in the group algebra $\mathcal{A}=R(\mathbb{C}G)$ for all $t\in[0,\tau_p]$ and all $i\in\{1,\dots,\vert\Gamma\vert\}$, then $\Pi_\mathcal{G}\bigl(  F_\Gamma(X)  \bigr) = \Pi_\mathcal{G}\bigl(  X  \bigr)$. 
\end{thm}
\begin{proof}
If $H_i(t)\in\mathcal{A}$ it follows that $p_i(t)\in\mathcal{A}$ for all $t\in[0,\tau_p]$ and all $i\in\{1,\dots\vert\Gamma\vert\}$.
Hence, $F_\Gamma(Y)=Y$ for any time-independent operator $Y\in\mathcal{A}'$.
We are now going to show that $Q(X)=\Pi_\mathcal{G}\bigl(  F_\Gamma(X)  \bigr)$ is a projector.
First, we note that $Q^2(X)=\Pi_\mathcal{G}\bigl(  F_\Gamma\bigl(  \Pi_\mathcal{G}\bigl(  F_\Gamma(X)  \bigr) \bigr)  \bigr) =
\Pi_\mathcal{G}\bigl(    \Pi_\mathcal{G}\bigl(  F_\Gamma(X)  \bigr)  \bigr)$, which follows from $F_\Gamma(Y)=Y$ for $Y\in\mathcal{A}'$. By using the fact that $\Pi_\mathcal{G}$ is a projector, we find that $Q^2(X)=Q(X)$.
Since the range of $Q$ is in $\mathcal{A}'$, we have $Q=\Pi_\mathcal{G}$ if and only if $Q$ acts on $\mathcal{A}'$ as the identity. Let $Y\in\mathcal{A}'$, then $Q(Y)=\Pi_\mathcal{G}(Y)=Y$.
\end{proof}
\noindent
As an example we consider the decoupling scheme for one qubit given by the Pauli group $\mathcal{G} = \{\mathcal{I},X,Y,Z\}$.
As a set of generators we choose $\Gamma=\{X,Y\}$.
The  Cayley graph of $\mathcal{G}$ with respect to $\Gamma$ is shown in figure \ref{fig::digraph}.
An Eulerian path is obtained by following the numbers $1,\dots,8$.
The decoupling cycle corresponding to this path is depicted in figure \ref{fig::euler}.

The above method increases the length of a basic \textsf{PDD} cycle by a factor $\vert\Gamma\vert$.
For local system Hamiltonians shorter decoupling schemes may be devised using Eulerian orthogonal arrays \cite{Woc04}.
For a geometric perspective on the theory of decoupling with bounded controls we refer to \cite{Ch06}.

\section{Decoupling Schemes}\label{sec:decschemes}

A decoupling scheme for the system Hamiltonian $H_0$ acting on the system Hilbert space $\mathcal{H}_S$ was defined in definition \ref{def:controlscheme} as a set of unitaries $\{ g_j \}_{j=0}^{n_c-1}$ and relative times $\{ \Delta t_j \}_{j=0}^{n_c-1}$ such that
\begin{equation}\label{eq:deccondi}
\frac{1}{t_c} \sum_{j=0}^{n_c-1} g_j^\dagger H_0 g_j \Delta t_j  = \tr(H_0)\cdot\frac{1}{d}\mathcal{I},
\end{equation}
where $t_c=\sum_{j=0}^{n_c-1}\Delta t_j$ and $d=\dim(\mathcal{H}_S)$.
In this section we give an overview over known decoupling schemes for different types of system Hamiltonians.
All these schemes work with constant relative time intervals, i.\,e. $\Delta t_j=\Delta t$ for all $j\in\{0,\dots,n_c-1\}$.
Since the quantum system under consideration forms a quantum register consisting of $n$ qudits of dimension $q$ we have $\mathcal{H}_S = \mathcal{H}_q^{\otimes n}$ and the local control assumption requires the unitaries $g_j$ to be of the form
$g_j = g^{(1,j)}_1\otimes g^{(2,j)}_2 \otimes \dots \otimes g^{(n,j)}_n$, where $g^{(i,j)}_k$ denotes the unitary $g^{(i,j)} \in \textsf{U}_q$ being applied to the $k$-th qudit.

\subsection{General Hamiltonians}

We start with decoupling schemes which apply to all traceless Hamiltonians $H_0$ acting on $\mathcal{H}_S$.
\begin{defi}
An annihilator is a decoupling scheme $\{ g_j, \Delta t_j\}_{j=0}^{n_c-1}$ satisfying
\begin{equation}
\frac{1}{t_c} \sum_{j=0}^{n_c-1} g_j^\dagger H_0 g_j \Delta t_j  = 0,
\end{equation}
for all traceless system Hamiltonians $H_0$.
\end{defi}
\noindent
It was shown in \cite{WRJB02} %
that an annihilator has to contain at least $n_c = \dim(\mathcal{H}_S)^2$ elements $g_j$ and that the relative times for such a minimal annihilator have to be equal, i.\,e. $\Delta t_j=\Delta t$ for all $j\in\{0,\dots,n_c-1\}$.
Annihilators can be found using the following group-theoretic averaging procedure \cite{Za99,VKL99}.
\begin{thm}\label{thm:irrepgroupaverage}
Let $G = \{ \mathfrak{g}_j \}_{j=0}^{n_c-1}$ be a finite group of order $n_c$, and let $R : \mathfrak{g}_j \mapsto g_j \in \textsf{U}_d$ be an irreducible representation of $G$ on a $d$-dimensional Hilbert space $\mathcal{H}_S$.
Then, for any $H_0 \in \mathcal{L}(\mathcal{H}_S)$,
\begin{equation}
 \Pi_{R(G)}(H_0) \equiv \frac{1}{n_c} \sum_{j=0}^{n_c-1} g_j^\dagger H_0 g_j = \tr(H_0) \cdot \frac{1}{d} \mathcal{I}.
\end{equation}
\end{thm}
\begin{proof}
First we note that the left hand side of the above equation commutes with all the unitaries $g_j$.
Since the $g_j$ form an irreducible representation, Schur's lemma (theorem \ref{thm:schuri}) tells us that the only operator commuting with all the $g_j$ is a multiple of the identity.
The correct factor is obtained by taking the trace on both sides of the equation.
\end{proof}
\noindent
This theorem was shown in \cite{WRJB02} to hold for irreducible projective representations as well.
Since, by definition, any nice error basis (see definition \ref{def:nicerrorb}) forms an irreducible projective representation, it can be used as an annihilator.
A particular example for a nice error basis --- and hence for an annihilator --- for $\mathcal{H}_S = \mathcal{H}_q^{\otimes n}$ is the set of Pauli operators,
\begin{equation}\label{eq:pauliops:gendecscheme}
 \mathcal{P}_q^n = \{ \XZ(\vec{a}) \ \vert\  \vec{a}\in\mathbb{F}_q^{2n} \},
\end{equation}
as defined in section \ref{sec:preliminaries}.

Decoupling according to theorem \ref{thm:irrepgroupaverage} corresponds to the special case of a dynamical generated noiseless subsystem (subsection \ref{subsec:nlsubsys}) which is identical with the whole system.

\subsection{Local Hamiltonians}\label{subsec:localhams}

Let us first define a map mapping an operator of the form
$A = A^{(1)}_1\otimes \dots \otimes A^{(s)}_s$ acting on $\mathcal{H}_q^{\otimes s}$ to an operator acting on $\mathcal{H}_q^{\otimes n}$ with $n\geq s$ via
\begin{equation}
 A \mapsto \bigl[ A \bigr]_{(k_1,k_2,\dots,k_s)} = A^{(1)}_{k_1}\otimes \dots \otimes A^{(s)}_{k_s}\otimes \mathcal{I}_{\{1,2,\dots,n\}\setminus \{k_1,\dots, k_s\}},
\end{equation}
for any $1\leq k_1<k_2<\dots < k_s \leq n$.
Here, the index $i$ in $A^{(j)}_i$ indicates that the operator $A^{(j)} \in\mathcal{L}(\mathcal{H}_q)$ acts on the $i$-th qudit.
Using this kind of notation, a $t$-local Hamiltonian is defined as follows:
\begin{equation}\label{eq:tlocalh}
 H_0 = \sum_{s=1}^t
 \
 \sum_{k_1 = 1}^{n-s+1}
 \sum_{k_2 = k_1+1}^{n-s+2}
 \dots
 \sum_{k_s = k_{s-1}+1}^n
 \
 \sum_{\vec{a}\in\mathbb{F}_q^{2s}\setminus\{\vec{0}\}}
 J^{k_1\dots k_s}_{\vec{a}} \bigl[ \XZ(\vec{a}) \bigr]_{(k_1\dots k_s)}.
\end{equation}
Since the Pauli operators form an operator basis, any Hamiltonian that couples no more than $t$ of the qudits can be written as in \eqref{eq:tlocalh}.
Decoupling schemes for $t$-local qubit Hamiltonians ($q=2$) have been devised by Leung \cite{Leu02} in terms of Hadamard matrices and by Stollsteimer and Mahler \cite{SM01} using orthogonal arrays \cite{OABook}.
The orthogonal array approach was generalized to qudits by Wocjan et\,al. in \cite{WRJB02b}.
Eventually it was shown by R\"{o}tteler and Wocjan \cite{RW04} that both methods are equivalent.
We proceed explaining the generalized orthogonal array approach.
\begin{defi}[Orthogonal arrays]
Let $\mathcal{A}$ be an alphabet containing $a$ symbols.
An orthogonal array $OA_\lambda(n_c,n,t,a)$ with $a$ levels, strength $t$ and index $\lambda$ is an $n\times n_c$ matrix $M = (m_{ij})$ with entries from $\mathcal{A}$ if any $s\times n_c$ sub-matrix (obtained from $M$ by selecting $s$ rows) contains any possible $s$-tuple of elements from $\mathcal{A}$ exactly $\lambda$ times as a column.
\end{defi}
\noindent
Let $\{ u_i \}_{i=0}^{q^2-1}$ denote an annihilator for the one qudit Hilbert space $\mathcal{H}_q$
(for example we could choose the set of Pauli operators, i.\,e. $\{ u_i \}_{i=0}^{q^2-1} = \mathcal{P}_q$).
Given an $OA_\lambda(n_c,n,t,q^2)$ with $q^2$ levels, a control scheme $\{ g_j \}_{j=0}^{n_c-1}$ can be obtained as follows:
The $j$-th unitary $g_j$ is constructed using the $(j+1)$-th column of the orthogonal array as $g_j = u_{m_{1,j+1}} \otimes u_{m_{2,j+1}} \otimes \dots \otimes u_{m_{n,j+1}}$.
The following theorem due to Wocjan and R\"{o}tteler \cite{WRJB02b,RW04} shows that such a control scheme is in fact a decoupling scheme for any $t$-local Hamiltonian.
\begin{thm}
A control scheme $\{ g_j \}_{j=0}^{n_c-1}$ constructed from an $OA_\lambda(n_c,n,t,q^2)$ with $q^2$ levels and strength $t$ as described above, is a decoupling scheme for all $t$-local Hamiltonians acting on $\mathcal{H}_S=\mathcal{H}_q^{\otimes n}$.
\end{thm}
\begin{proof}
The annihilator $\{ u_i \}_{i=0}^{q^2-1}$ for the one-qudit Hilbert space $\mathcal{H}_q$ consists of the elements of a nice error basis for operators acting on $\mathcal{H}_q$.
Hence the collection of all $s$-fold tensor products of the $u_i$'s forms a nice error basis for $\mathcal{H}_q^{\otimes s}$ and we obtain
\begin{equation}
\frac{1}{q^{2s}} \sum_{i_1,\dots,i_s=0}^{q^2-1} \bigl( u_{i_1}^\dagger \dots u_{i_s}^\dagger \bigr) H \bigl( u_{i_1} \dots u_{i_s} \bigr) = 0
\end{equation}
for all traceless Hamiltonians $H$ acting on $\mathcal{H}_q^{\otimes s}$.
Let us pick now the term characterized by $(k_1\dots k_s)$ and $\vec{a}$ from the $t$-local $H_0$ given by \eqref{eq:tlocalh}.
For the control scheme $\{ g_j \}_{j=0}^{n_c-1}$ constructed from the $OA$ we obtain
\begin{align}
&\phantom{\mathrel{=}} \quad
 \frac{1}{n_c}\sum_{j=0}^{n_c-1} g_j^\dagger \ J^{k_1\dots k_s}_{\vec{a}} \bigl[ \XZ(\vec{a}) \bigr]_{(k_1\dots k_s)} \ g_j \nonumber\\
&=
\frac{1}{n_c}\sum_{j=1}^{n_c}
\bigl(u_{m_{1,j}}^\dagger \otimes \dots \otimes u_{m_{n,j}}^\dagger\bigr)
\ J^{k_1\dots k_s}_{\vec{a}} \bigl[ \XZ(\vec{a}) \bigr]_{(k_1\dots k_s)} \
\bigl(u_{m_{1,j}} \otimes  \dots \otimes u_{m_{n,j}}\bigr) \nonumber\\
&=\biggl[ J^{k_1\dots k_s}_{\vec{a}}
\frac{1}{n_c}\sum_{j=1}^{n_c}
\bigl(u_{m_{k_1,j}}^\dagger \otimes \dots \otimes u_{m_{k_s,j}}^\dagger\bigr)
\  \XZ(\vec{a})  \
\bigl(u_{m_{k_1,j}} \otimes  \dots \otimes u_{m_{k_s,j}}\bigr)
\biggr]_{(k_1\dots k_s)} \nonumber\\
&=\biggl[ J^{k_1\dots k_s}_{\vec{a}}
\frac{1}{q^{2s}}\sum_{i_1,\dots,i_s=0}^{q^2-1}
\bigl(u_{i_1}^\dagger \otimes \dots \otimes u_{i_s}^\dagger\bigr)
\  \XZ(\vec{a})  \
\bigl(u_{i_1} \otimes  \dots \otimes u_{i_s}\bigr)
\biggr]_{(k_1\dots k_s)} = 0.
\end{align}
The last line is obtained by noting that the $OA$ contains each possible $s$-tuple (with $s\leq t$) with entries in $\mathbb{F}_q^2$ equally often.
\end{proof}
\begin{rem}
A decoupling scheme $\{ g_j \}_{j=0}^{n_c-1}$ for a $t$-local Hamiltonian acting on $\mathcal{H}_q^{\otimes n}$ based on an orthogonal array $OA(n_c,n,t,q^2)$ can be extended to a decoupling scheme $\{ g'_j \}_{j=0}^{n_c-1}$ for a $t$-local Hamiltonian acting on $\mathcal{H}_q^{\otimes n+1}$ by setting
$g'_j = (g_j)_{\{1\dots n\}} \otimes \mathcal{I}_{n+1}$,
as long as there are no local terms in the Hamiltonian which act only the $(n+1)$-th qudit.
\end{rem}

Physical interactions are typically described by $2$-local Hamiltonians.
Hence orthogonal arrays of strength two are of special importance.
Using a construction method based on Hamming codes \cite[chapter 5.3]{OABook},
orthogonal arrays $OA\bigl( s^i, (s^i-1)/(s-1), 2, s\bigr)$, with $s$ being a prime power (here $s=q^2$) and $i\geq 2$, can be obtained.
It follows that any $2$-local Hamiltonian acting on up to $n$ qudits of dimension $q$ can be decoupled using a decoupling scheme $\{ g_j \}_{j=0}^{n_c-1}$ of length $n_c$, where an upper bound on $n_c$ is given by $n_c \leq n(s-1)s+2s-s^2$.
Even though this bound is far from optimal (orthogonal arrays exist which cannot be obtained by the Hamming code method), it shows that the length of a decoupling scheme scales linearly with the number of qudits.
In the appendix \ref{app:oatables} we list the orthogonal arrays $OA(16,5,2,4)$, $OA(32,9,2,4)$ and $OA(48,13,2,4)$, which can be used to decouple up to $5$, $9$ and $13$ qubits, respectively.

\subsubsection{Diagonal Couplings}

Let us consider now the special case of an $n$-qubit Hamiltonian $H_0$ involving only bipartite couplings,
\begin{equation}
 H_0 =
 \sum_{k_1 = 1}^{n-1}
 \sum_{k_2 = k_1+1}^n
 \
 \sum_{\vec{a}\in\mathbb{F}_2^2\setminus\{\vec{0}\}}
 J^{k_1,k_2}_{\vec{a}} \bigl[ \XZ(\vec{a})\otimes\XZ(\vec{a})  \bigr]_{(k_1,k_2)}.
\end{equation}
These kind of couplings are called diagonal couplings, since the coefficient matrix $J_{\vec{a}}$ is diagonal when compared with the one for the general case
$\bigl( \sum_{\vec{a},\vec{c}} J_{\vec{a},\vec{c}} \XZ(\vec{a}) \otimes \XZ(\vec{c}) \bigr)$.
It was shown by Stollsteimer and Mahler \cite{SM01} that such Hamiltonians can be decoupled using decoupling schemes constructed from difference schemes \cite[chapter 6]{OABook}.
The advantage over corresponding decoupling schemes using orthogonal arrays is the shorter length of such schemes.
We generalize this approach to the qudit case.
For qudits of dimension $q\geq 3$ let us consider the Hamiltonian
\begin{equation}\label{eq:2localdiagh}
 H_0 =
 \sum_{k_1 = 1}^{n-1}
 \sum_{k_2 = k_1+1}^n
 \
 \sum_{\vec{a}\in\mathbb{F}_q^{2}\setminus\{\vec{0}\}}
 J^{k_1,k_2}_{\vec{a}} \bigl[ \XZ(\vec{a})\otimes\XZ^\dagger(\vec{a}) \bigr]_{(k_1,k_2)}.
\end{equation}
For $H_0$ to be Hermitian, the coefficients must satisfy $J^{k_1,k_2}_{\vec{a}} = J^{k_1,k_2}_{-\vec{a}}$ since
$\XZ(-\vec{a})\otimes\XZ^\dagger(-\vec{a}) = \XZ(\vec{a})^\dagger\otimes\XZ(\vec{a})$. It follows that the Hamiltonian is symmetric with respect to $k_1$ and $k_2$.
\begin{rem}
Note that interactions of the form \eqref{eq:2localdiagh} might be of interest for quantum computation, since the swap gate,
$U_\text{SWAP} \ket{\phi}\otimes\ket{\psi} = \ket{\psi}\otimes\ket{\phi}$, which can be written as
\begin{equation}
U_\text{SWAP} =
\frac{1}{q^2}
 \sum_{\vec{a}\in\mathbb{F}_q^{2}}
  \XZ(\vec{a}) \otimes \XZ^\dagger(\vec{a}),
\end{equation}
can be generated (up to a global phase) as $U_\text{SWAP} = \exp\bigl(-i H_\text{SWAP} \pi/2 \bigr)$ by the interaction
\begin{equation}
H_\text{SWAP} =
\frac{1}{q}
 \sum_{\vec{a}\in\mathbb{F}_q^{2}\setminus\{\vec{0}\}}
  \XZ(\vec{a})\otimes\XZ^\dagger(\vec{a})
\end{equation}
which is of the form \eqref{eq:2localdiagh}. In the qubit case the square root swap gate --- $\exp\bigl(-i H_\text{SWAP} \pi/4 \bigr)$ --- in connection with all single qubit gates forms a universal set of gates.
\end{rem}

\begin{defi}[Difference schemes]
A difference scheme $D(n_c,n,s)$ based on a finite abelian group $(\mathcal{A},+)$ of order $s$ is an $n\times n_c$ matrix $M=(m_{ij})$ such that for all $1\leq i<j \leq n$, the vector difference between the $i$-th and the $j$-th row contains each element of $\mathcal{A}$ equally often.
\end{defi}
\noindent
Necessarily $n_c$ is a multiple of $s$. It can be shown that if a difference scheme $D(n_c,n,s)$ exists, then $n\leq n_c$ \cite[chapter 6]{OABook}.

Let the set of Pauli operators be given by $\mathcal{P}_q = \{ \XZ(\vec{a}) \sthat \vec{a}\in\mathbb{F}_q^2 \}$.
Given a $D(n_c,n,q^2)$ based on $\mathbb{F}_q^2$, a control scheme $\{ g_j \}_{j=0}^{n_c-1}$ can be constructed as follows:
The $j$-th unitary $g_j$ is constructed using the $(j+1)$-th column of the difference scheme as
$g_j = \XZ(m_{1,j+1})\otimes \XZ(m_{2,j+1}) \otimes \dots \otimes \XZ(m_{n,j+1})$.
\begin{thm}\label{thm:diffschemedec}
A control scheme $\{ g_j \}_{j=0}^{n_c-1}$ constructed from a $D(n_c,n,q^2)$ as described above, is a decoupling scheme for all $n$-qudit Hamiltonians involving diagonal qudit-qudit couplings as in \eqref{eq:2localdiagh}.
\end{thm}
\begin{proof}
Let us pick a single term characterized by $(k_1,k_2)$ and $\vec{a}$ from $H_0$ in \eqref{eq:2localdiagh}.
For the control scheme $\{ g_j \}_{j=0}^{n_c-1}$ constructed from a $D(n_c,n,q^2)=(m_{ij})$, $i=1\dots n$, $j=1\dots n_c$, $m_{ij}\in\mathbb{F}_q^2$, we obtain
\begin{align}
&\phantom{\mathrel{=}} \quad
 \frac{1}{n_c}\sum_{j=0}^{n_c-1} g_j^\dagger \ J^{k_1,k_2}_{\vec{a}} \bigl[ \XZ(\vec{a})\otimes \XZ^\dagger(\vec{a}) \bigr]_{(k_1,k_2)} \ g_j \nonumber\\
&=
\frac{1}{n_c}\sum_{j=1}^{n_c}
\XZ^\dagger(m_{1,j})\otimes  \dots \otimes \XZ^\dagger(m_{n,j})
\ J^{k_1,k_2}_{\vec{a}} \bigl[ \XZ(\vec{a})\otimes \XZ^\dagger(\vec{a}) \bigr]_{(k_1,k_2)} \
\XZ(m_{1,j})\otimes \dots \otimes \XZ(m_{n,j}) \nonumber\\
&=\biggl[
J^{k_1,k_2}_{\vec{a}} \frac{1}{n_c} \sum_{j=1}^{n_c}
\bigl(\XZ^\dagger(m_{k_1,j}) \otimes \XZ^\dagger(m_{k_2,j})\bigr)
\bigl(\XZ(\vec{a})\otimes \XZ(\vec{a})^\dagger \bigr)
\bigl(\XZ(m_{k_1,j}) \otimes  \XZ(m_{k_2,j})\bigr)
\biggr]_{(k_1,k_2)}. \nonumber\\
\intertext{Using the symplectic inner product as in \eqref{eq:syminvertorder}, the order of the Pauli operators can be inverted leading to}
&=\biggl[ J^{k_1,k_2}_{\vec{a}} \XZ(\vec{a})\otimes \XZ^\dagger(\vec{a}) \biggr]_{(k_1,k_2)}
\frac{1}{n_c}\sum_{j=1}^{n_c} %
\omega^{(\vec{a},m_{k_1,j}-m_{k_2,j})_{sp}}
\nonumber\\
&=\biggl[ J^{k_1,k_2}_{\vec{a}} \XZ(\vec{a})\otimes \XZ^\dagger(\vec{a}) \biggr]_{(k_1,k_2)}
\frac{1}{q^2}\sum_{\vec{d}\in\mathbb{F}_q^2} \omega^{(\vec{a},\vec{d})_{sp}} = 0.
\end{align}
The last line is obtained by noting that in the difference scheme $(m_{ij})$, with $m_{ij}\in\mathbb{F}_q^2$, the vector difference between row $k_1$ and $k_2$ contains each element in $\mathbb{F}_q^2$ exactly $n_c/s$ times.
As it can be seen from the last two lines, the position of the dagger operator is not important: The decoupling scheme also eliminates couplings of the form $\XZ(\vec{a})^\dagger\otimes \XZ(\vec{a})$.
\end{proof}

Construction methods for difference schemes $D(q^m,q^m,q^2)$ with $q$ prime and $m\geq 2$ are known (see for example \cite[chapter 6.1]{OABook}).
It follows that any Hamiltonian with diagonal couplings between up to $n$ qudits of dimension $q$ can be decoupled using a decoupling scheme $\{ g_j \}_{j=0}^{n_c-1}$ of length $n_c$, where an upper bound on $n_c$ is given by $n_c \leq n q - q$.
This bound is of the order $\mathcal{O}(n q)$ and has to be compared with the bound for orthogonal arrays which was $\mathcal{O}(n q^4)$.
In the appendix \ref{app:dstables} we list difference schemes $D(4\lambda,4\lambda,4)$ for $\lambda\in\{1,2,3,4\}$, which can be used in order to decouple up to $4\lambda$ qubits, respectively.

There exist diagonal couplings for which even shorter decoupling schemes can be devised.
A famous example are dipolar inter-qubit couplings,
\begin{equation}
 H_0 =
 \sum_{k_1 = 1}^{n-1}
 \sum_{k_2 = k_1+1}^n
 \
 J^{k_1,k_2} \bigl[ 2Z\otimes Z - X\otimes X - Y\otimes Y\bigr]_{(k_1,k_2)},
\end{equation}
for which a decoupling scheme of constant length $n_c=3$ is given by the set $\{g_j\}_{j=1}^3$ \cite{WHH68}
of non-selective $\pi/2$ pulses,
\begin{equation}
g_j = \exp\Bigl( -\frac{i}{2} \alpha_j \frac{\pi}{2} \Bigr)^{\otimes n}, \quad \text{ with } \alpha_j=X,Y,Z \text{ for } j=1,2,3.
\end{equation}
The $\pi/2$ pulses convert the diagonal terms in the Hamiltonian $H_0$ in a cyclic manner, thereby achieving the decoupling condition $\tilde{H}_1+\tilde{H}_2+\tilde{H}_3=0$ with $\tilde{H}_j=g_j^\dagger H_0 g_j$.

\subsection{Selective Decoupling}\label{subsec:seldec}
In the preceding subsection, among others, decoupling schemes for general and diagonal Hamiltonians involving only bipartite inter-qudit couplings have been presented. These schemes turn off all qudit-qudit couplings in
\begin{equation}
 H_0 =
 \sum_{k_1 = 1}^{n-1}
 \sum_{k_2 = k_1+1}^n
 \
 \sum_{\vec{a},\vec{b}\in\mathbb{F}_q^2\setminus\{\vec{0}\}}
 J^{k_1,k_s}_{\vec{a},\vec{b}} \bigl[ \XZ(\vec{a})\otimes\XZ(\vec{b}) \bigr]_{(k_1,k_2)},
\end{equation}
or its diagonal counterpart \eqref{eq:2localdiagh}.
Under certain circumstances we might want to keep one (or more than one) particular coupling alive, i.\,e. we want to simulate the Hamiltonian
\begin{equation}
 H_0' = \sum_{\vec{a},\vec{b}\in\mathbb{F}_q^2\setminus\{\vec{0}\}}
 J^{k_1,k_s}_{\vec{a},\vec{b}} \bigl[ \XZ(\vec{a})\otimes\XZ(\vec{b}) \bigr]_{(k_1,k_2)},
\end{equation}
for some fixed pair $(k_1, k_2)$ with $1\leq k_1<k_2\leq n$.
This control task is called selective decoupling.
An example for such a scenario is a quantum computer in which the two qudit gates are generated by the qudit-qudit couplings.
A control scheme $\{ g'_j \}_{j=0}^{n_c-1}$ for the simulation of $H_0'$ can easily be obtained from the corresponding decoupling scheme $\{ g_j \}_{j=0}^{n_c-1}$ as follows \cite{SM01}:
Let $g_j$ be given by $g^{(1,j)}_1 \otimes g^{(2,j)}_2 \otimes \dots \otimes g^{(n,j)}_n$, where $g^{(i,j)}_k$ denotes the unitary $g^{(i,j)}$ being applied to the $k$-th qudit.
We set $g'_j=g_j$ and apply the following modifications:
\begin{itemize}
\item For general couplings, the decoupling scheme was constructed with the help of an orthogonal array.
 To keep the $(k_1,k_2)$-coupling, we replace the unitaries $g^{(k_1,j)}_{k_1}$ and $g^{(k_2,j)}_{k_2}$ by $\mathcal{I}_{k_1}$ and $\mathcal{I}_{k_2}$.
\item For diagonal couplings, the decoupling scheme was constructed with the help of a difference scheme.
To keep the $(k_1,k_2)$-coupling, we replace $g^{(k_2,j)}_{k_2}$ by $g^{(k_1,j)}_{k_2}$ (or vice versa $g^{(k_1,j)}_{k_1}$ by $g^{(k_2,j)}_{k_1}$).
\end{itemize}

\subsection{Nearest-Neighbor Couplings}

A general $2$-local $n$-qudit Hamiltonian $H_0$ involves couplings between up to $n(n-1)/2$ pairs.
If the only inter-qudit couplings involved in $H_0$ are nearest-neighbor couplings and the qudits are arranged on a linear chain, i.\,e. if
\begin{equation}\label{eq:2localnnh}
 H_0 =
 \
 \sum_{k = 1}^{n-1}
 \
 \sum_{\vec{a},\vec{c}\in\mathbb{F}_q^2\setminus\{\vec{0}\}}
 J^{k,k+1}_{\vec{a},\vec{c}} \bigl[ \XZ(\vec{a})\otimes \XZ(\vec{c}) \bigr]_{(k,k+1)},
\end{equation}
far shorter decoupling schemes can be devised as the ones discussed in the preceding subsection.
Let $\{ u(j) \}_{j=0}^{q^2-1}$ denote an annihilator for the one qudit Hilbert space $\mathcal{H}_q$.
A decoupling scheme $\{ g_j \}_{j=0}^{n_c-1}$ of constant length $n_c=q^2$ can be constructed by letting the elements of the annihilator act on the even numbered qudits, i.\,e. by setting
$g_j = \mathcal{I}_1 \otimes u(j)_2 \otimes \mathcal{I}_3 \otimes u(j)_4 \otimes \dots $ for all $j\in \{0,\dots,q^2-1\}$.
\begin{thm}
An $n$-qudit Hamiltonian $H_0$ involving only nearest-neighbor couplings as in \eqref{eq:2localnnh} can be decoupled using a decoupling scheme of constant length $n_c=q^2$ as it is described above.
\end{thm}
\begin{proof}
Let us pick a term in \eqref{eq:2localnnh} with odd $k$ (for even $k$ the proof goes analogously). Then,
\begin{align}
&\phantom{\mathrel{=}} \quad
 \frac{1}{n_c}\sum_{j=0}^{n_c-1} g_j^\dagger \ J^{k,k+1}_{\vec{a},\vec{c}} \bigl[ \XZ(\vec{a})\otimes \XZ(\vec{c}) \bigr]_{(k,k+1)} \ g_j \nonumber\\
&=
\frac{1}{n_c}\sum_{j=0}^{n_c-1 }
\bigl(\mathcal{I}_1\otimes u(j)_2^\dagger \otimes \mathcal{I}_3 \otimes u(j)_4^\dagger\dots \bigr)
\ J^{k,k+1}_{\vec{a},\vec{c}} \bigl[ \XZ(\vec{a})\otimes \XZ(\vec{c}) \bigr]_{(k,k+1)} \
\bigl(\mathcal{I}_1\otimes u(j)_2 \otimes \mathcal{I}_3 \otimes u(j)_4\dots \bigr) \nonumber\\
&=\biggl[ J^{k,k+1}_{\vec{a},\vec{c}} \ \XZ(\vec{a})\otimes \Bigl(
\frac{1}{q^2}\sum_{j=0}^{q^2-1}
  u(j)^\dagger \XZ(\vec{c}) u(j) \Bigr)
\biggr]_{(k,k+1)} = 0.
\end{align}
The last step is due to the fact that the set $\{ u(j) \}_{j=0}^{q^2-1}$ forms an annihilator and $\XZ(\vec{c})$ is traceless for $\vec{c}\in\mathbb{F}_q^2\setminus\{\vec{0}\}$.
\end{proof}

\section{Control Strategies}\label{sec:cstrategies}

Dynamical control over a local Hamiltonian allows the time evolution of a quantum system to be modified.
In the bang-bang scenario, a control scheme consisting of a set of unitaries generated by the local Hamiltonian, is used to achieve a certain control task.
For example, for a closed quantum system, we might want to simulate a time evolution according to a Hamiltonian which is different from the system Hamiltonian.
In particular, the simulation of a vanishing Hamiltonian is called decoupling.
For an open system, we might try to generate a noiseless subsystem (see subsection \ref{subsec:nlsubsys}).
For all these tasks, the fundamental control strategy (as discussed in subsection \ref{subsec:fundamental}) is to apply the pulses determined by the control scheme with the help of the local control Hamiltonian over and over again.
Assuming the pulses to be ideal, the finite time interval in between subsequent pulses is the only obstacle preventing a control task to be achieved in a perfect manner.
For the task of decoupling, it was shown in subsection \ref{subsec:performance}, that the fundamental control strategy (\textsf{PDD}) leads to an average fidelity decay which is quadratic in time.
The strength of the decay is determined by (i) the strength of the system Hamiltonian, (ii) by the length of the decoupling scheme, and (iii) by the time interval $\Delta t$ in between subsequent pulses.

In this section, we consider control strategies which improve the average fidelity decay of a given decoupling scheme for a fixed time interval $\Delta t$.
The standard technique used by the nuclear magnetic resonance (NMR) community is a symmetrized version of the \textsf{PDD} strategy,
which leads to a decrease of the strength of the decay, but keeps its quadratic-in-time nature.
In the author's diploma thesis \cite{DiplKern} it was observed that a control strategy based on a random selection of the elements of a decoupling scheme leads to a fidelity decay which is only linear in time.
Subsequently, randomized decoupling was proposed for open quantum systems by Viola and Knill \cite{VK05}.
The linear-in-time decay was confirmed by constructing a lower bound on the worst case fidelity \cite{VK05,V05}.
Control strategies combining the advantages of purely deterministic and randomized strategies have been devised by the author \cite{combi} and by Santos and Viola \cite{SV06,VS06}, and have been explored numerically for open \cite{SV05} and closed systems \cite{SV08}.
We start presenting the deterministic strategies in subsection \ref{subsec:detstrat},
and proceed with the randomized strategies in subsection \ref{subsec:randstrat}.
For most of the strategies we calculate a short time expansion of the average fidelity decay, which allows us to discuss the advantages and disadvantages of a certain strategy.
Even though we focus on decoupling, the control strategies discussed in this section are applicable to other control tasks as well.
We label the strategies using the abbreviations introduced by Santos and Viola in \cite{SV06,VS06,SV08}.

As in the preceding chapters, let $S$ be a closed quantum system defined on a finite-dimensional Hilbert space $\mathcal{H}_S$ of dimension $d=\dim(\mathcal{H}_S)$, and let its Hamiltonian be given by $H_0$ acting on $\mathcal{H}_S$.
Without loss of generality $H_0$ is assumed to be traceless, i.\,e. $\tr(H_0)=0$.
Occasionally, we write $H_0$ as $\lambda H_0$ and use powers of $\lambda$ to indicate the dependence on $H_0$.
We assume that a certain decoupling scheme $\{ g_j \}_{j=0}^{n_c-1}$ of length $n_c$ for $H_0$ is given.

\subsection{Deterministic Strategies}\label{subsec:detstrat}

\subsubsection{Periodic Dynamical Decoupling (\textsf{PDD})}\label{subsubsec:pdd}
The fundamental decoupling strategy, as described in subsection \ref{subsec:fundamental}, is called periodic dynamical decoupling.
At the time $t_i = i \cdot \Delta t$, $i\in \mathbb{N}_0$, the local control Hamiltonian is used to generate the pulse $p_{i\!\mod n_c}$, where $p_j=g_j g_{j-1}^\dagger$ (for $j=0\dots n_c-1$) is defined in terms of the elements $g_j$ of the decoupling scheme by setting $g_{-1}=g_{n_c-1}$ with the exception that the first $p_0$ is simply given by $p_0'=g_0$ (compare with figure \ref{fig::pdd}).
As a result, the time evolution in the toggled frame after the time
$T = m\cdot t_c$ with $m\in\mathbb{N}$ and $t_c=n_c\Delta t$ is given by
\begin{equation}\label{eq:pddmcycles}
 \tilde{U}( T=m\cdot t_c ) =
\Bigl(
 \exp(-i \tilde{H}_{n_c-1} \Delta t) \dots \exp(-i \tilde{H}_1 \Delta t) \exp(-i \tilde{H}_0 \Delta t)
\Bigr)^m = \exp(-i \overline{H} t_c \cdot m),
\end{equation}
with $\tilde{H}_j = g^\dagger_j H_0 g_j$.
The zeroth order term in the Magnus expansion of $\overline{H}$ vanishes by definition of the decoupling scheme,
\begin{equation}\label{eq:zerothorderoverlineh-inpdd}
  \overline{H}^{(0)} = \frac{1}{n_c} \sum_{j=0}^{n_c-1} g_j^\dagger H_0 g_j  = 0,
\end{equation}
and, as it was shown in subsection \ref{subsec:performance}, the decay of the entanglement fidelity,
\begin{align}
 F_e^\textsf{PDD}(T) &= 1 - \frac{1}{d}\tr\bigl( \bigl(\overline{H}^{(1)}\bigr)^2 \bigr) T^2 + \mathcal{O}( \lambda^5 t_c^3 T^2), \label{eq:fepdd} \\
 F_{e\ \text{app}}^\textsf{PDD}(T) &= \exp\Bigl( - \frac{1}{d}\tr\bigl( \bigl(\overline{H}^{(1)}\bigr)^2 \bigr) T^2 \Bigr), \label{eq:feapppdd}
\end{align}
is in lowest order only due to the first order term in $\overline{H}$, which is given by \eqref{eq:bbh1}:
\begin{equation}\label{eq:firstorderoverlineh}
 \overline{H}^{(1)} = -\frac{i}{2n_c} \sum_{i>j=0}^{n_c-1} [\tilde{H}_i, \tilde{H}_j]\Delta t
 = \mathcal{O}( \lambda^2 t_c).
\end{equation}
A strict lower bound on the worst case fidelity \eqref{eq:Fw} was given in \cite{VK05} by using the matrix norm $\Vert A \Vert_2 = \max \vert \operatorname{eig}( \sqrt{A^\dagger A} )\vert$ and setting $\kappa=\Vert H_0\Vert_2$,
\begin{equation}\label{eq:fwpdd}
 F_w^\textsf{PDD} (T) > 1 -  \kappa^4 t_c^2 T^2 + \mathcal{O}\bigl(\kappa^5 t_c^3 T^2\bigr).
\end{equation}

In summary, the fidelity decay using \textsf{PDD} is of the order $\mathcal{O}(\lambda^4 t_c^2 T^2)$ and is caused mainly by the first order term \eqref{eq:firstorderoverlineh} in the Magnus expansion of a single \textsf{PDD} cycle of length $t_c=n_c\Delta t$.
Suppose we cannot decrease the time interval in between pulses below a certain value $\Delta t$.
Then, to optimize the fidelity decay of the \textsf{PDD} strategy, we have to find a decoupling scheme as small as possible (i.\,e. we minimize $n_c$).
The performance of a minimal decoupling scheme may be optimized further by noting that the first order term \eqref{eq:firstorderoverlineh} depends on the order of the elements $g_j$ in the decoupling scheme:
There are $n_c!$ possibilities and the term $\tr\bigl( \bigl(\overline{H}^{(1)}\bigr)^2 \bigr)/d$ becomes minimal for the new decoupling scheme $\{g'_j\}$ specified by $g'_j = g_{\pi(j)}$, where $\pi \in \textsf{S}_{n_c}$ denotes a particular permutation of $0,1,\dots,n_c-1$.
We might also say that $\pi$ denotes a particular path which traverses the elements of the decoupling scheme.
Unfortunately, such an optimal path is hard to find, depends on $H_0$, and the improvement might be relatively small.

\subsubsection{Symmetric Dynamical Decoupling (\textsf{SDD})}
The decoupling technique commonly used by the NMR community is a symmetrized version of the \textsf{PDD} strategy.
We call it symmetric dynamical decoupling.
Let us construct a symmetrized decoupling scheme $\{ g'_j \}_{j=0}^{n'_c-1}$ of length $n_c'=2n_c$ from the given decoupling scheme $\{ g_j \}_{j=0}^{n_c-1}$ of length $n_c$ as follows:
\begin{equation}
 g_j' = \begin{cases}
g_j & \text{ for } j=0,\dots,n_c-1 \\
g_{2n_c-1-j} & \text{ for } j=n_c,\dots,2n_c-1
\end{cases}.
\end{equation}
The \textsf{SDD} strategy is to apply the new scheme using the \textsf{PDD} strategy.
As a consequence, the time evolution of a single \textsf{SDD} cycle of length $t_c'=n_c'\Delta t$ in the toggled frame is given by
\begin{multline}
 \tilde{U}( t_c' ) =
 \exp(-i \tilde{H}_0 \Delta t) \exp(-i \tilde{H}_1 \Delta t) \dots \exp(-i \tilde{H}_{n_c-1} \Delta t) \times \\
 \exp(-i \tilde{H}_{n_c-1} \Delta t) \dots \exp(-i \tilde{H}_1 \Delta t) \exp(-i \tilde{H}_0 \Delta t) = \exp(-i \overline{H} t_c').
\end{multline}
Each cycle is symmetric in time, and according to theorem \ref{thm:symtoggledh}, all odd orders in the Magnus expansion of $\overline{H}$ vanish.
Hence, any resulting error is generated mainly by the second order term \eqref{eq:bbh2},
\begin{multline}%
 \overline{H}^{(2)} = -\frac{1}{6n'_c} \sum_{i\geq j\geq k=0}^{n'_c-1}
 \Bigl( [\tilde{H}_{f(i)},[\tilde{H}_{f(j)},\tilde{H}_{f(i)}]]+ \\
 [[\tilde{H}_{f(i)},\tilde{H}_{f(j)}],\tilde{H}_{f(k)}] \Bigr) \Delta t^2\times
\begin{cases}
1/2 & \text{ if } i=j \text{ or } j=k \\
1 & \text{ else }
\end{cases},
\end{multline}
where $f(i)=i$ for $i\in\{0,\dots,n_c-1\}$ and $f(i)=2n_c-1-i$ for $i\in\{n_c,\dots,2n_c-1\}$.
The above expression can be simplified as explained by the following lemma.
\begin{lem}\label{lem:h2sddh2pdd}
The second-order term in the Magnus expansion of a single \textsf{SDD} cycle as given by the above equation is equal to the second-order term in the Magnus expansion of the corresponding \textsf{PDD} cycle, i.\,e.
\begin{equation}\label{eq:2ndorderoverlineh}
 \overline{H}^{(2)} = -\frac{1}{6n_c} \sum_{i\geq j\geq k=0}^{n_c-1}
 \Bigl( [\tilde{H}_i,[\tilde{H}_j,\tilde{H}_k]]+
 [[\tilde{H}_i,\tilde{H}_j],\tilde{H}_k] \Bigr) \Delta t^2\times
\begin{cases}
1/2 & \text{ if } i=j \text{ or } j=k \\
1 & \text{ else }
\end{cases}.
\end{equation}
\end{lem}
\begin{proof}
Let us divide the interval $[0,t_c']$ into the two subintervals $[0,t_c]$ and $[t_c,t_c']$.
If we calculate the average Hamiltonian for each of these subintervals, we obtain vanishing zeroth-order terms of the form of equation
\eqref{eq:zerothorderoverlineh-inpdd}.
The results presented in \cite[section IV.D]{Bu81} state that in such a case the second-order term of the entire interval is given by the sum of the second-order terms of the subintervals, divided by two.
The proof is finished by noting that the second-order term of each of the subintervals is given by \eqref{eq:2ndorderoverlineh}.
\end{proof}

\noindent
Analogously to equations \eqref{eq:fepdd}, \eqref{eq:feapppdd} and \eqref{eq:fwpdd}, we obtain the expressions
\begin{align}
 F_e^\textsf{SDD}(T) &= 1 - \frac{1}{d}\tr\bigl( \bigl(\overline{H}^{(2)}\bigr)^2 \bigr) T^2 + \mathcal{O}( \lambda^8 t_c'^6 T^2), \label{eq:FeSDD}\\
 F_{e\ \text{app}}^\textsf{SDD}(T) &= \exp\Bigl( - \frac{1}{d}\tr\bigl( \bigl(\overline{H}^{(2)}\bigr)^2 \bigr) T^2 \Bigr), \label{eq:FappSDD}\\
 F_w^\textsf{SDD} (T) &> 1 -  \kappa^6 t_c^4 T^2 + \mathcal{O}\bigl(\kappa^8 t_c'^6 T^2\bigr), \vphantom{\frac{1}{d}}
\end{align}
and the estimate $\overline{H}^{(2)} = \mathcal{O}(\lambda^3 t_c^2 )$.

In summary, the fidelity decay using \textsf{SDD} is of the order $\mathcal{O}(\lambda^6 t_c^4 T^2)$ and is caused mainly by the second order term \eqref{eq:2ndorderoverlineh} in the Magnus expansion of a single \textsf{SDD} cycle of length $t'_c=2n_c\Delta t$.
For $\kappa t_c < 1$ this is an improvement over \textsf{PDD} in the sense that the strength of the \textsf{SDD} decay ($\mathcal{O}(\lambda^6 t_c^4)$) is smaller than the corresponding \textsf{PDD} strength ($\mathcal{O}(\lambda^4 t_c^2)$).
As it was the case for \textsf{PDD}, the performance of \textsf{SDD} might be optimized further by choosing an optimal path $\pi\in \textsf{S}_{n_c}$ traversing the elements $g_j$ of the underlying decoupling scheme, i.\,e. an order of the elements such that $\tr\bigl(\bigl(\overline{H}^{(2)}\bigr)^2\bigr)/d$ is minimal.

\subsubsection{Higher Order Decoupling}
A natural question is whether the \textsf{SDD} approach can be generalized to suppress even higher order terms in the Magnus expansion.
For a given decoupling scheme of length $n_c$, we have the set $\{ \tilde{H}_j \}_{j=0}^{n_c-1}$ of toggled frame Hamiltonians.
Is there a set of indices $\{ j(i) \}_{i=1}^N$, $j(i)\in\{0,\dots,n_c-1\}$, and relative times $\{ \Delta t_i \}_{i=1}^N$ of length $N$ such that the sequence
\begin{equation}
 \tilde{U}\bigl( T=\sum_{i=1}^N \Delta t_i \bigr) = \exp(-i \tilde{H}_{j(N)} \Delta t_N) \dots \exp(-i \tilde{H}_{j(1)} \Delta t_1) = \exp(-i \overline{H} T )
\end{equation}
has vanishing zeroth, first and second order terms in the Magnus expansion ?
(\textsf{SDD} is obtained for $N=2n_c$, $\Delta t_i = \Delta t$, $T=2n_c\Delta t$ and
$j(i)=\{ i \text{ for } i=0\dots n_c-1 \text{ and } 2n_c-1-i \text{ for } i=n_c\dots 2n_c-1 \}$.
It leads to a vanishing zeroth and first order term.)
According to \eqref{eq:bbh2}, the second order Magnus term is of third order in $H_0$.
Sets $\{ j(i) \}_{i=1}^N$ and $\{ \Delta t_i \}_{i=1}^N$ of length $N$ satisfying $\overline{H} \sim \mathcal{O}(H_0)^m$ can be found using a Trotter-Suzuki decomposition \cite{Su91}, but according to the non-existence theorem of positive decompositions (ibd.), they always involve negative times $\Delta t_i$ for $m\geq 4$.
This fact forbids general higher order decoupling according to some simple rule (see also the comment in \cite[section V]{KL06}).
(Nevertheless, there exist specific examples for which second order decoupling is achievable by repetition of a decoupling scheme traversing a series of different paths, see for example the '\textsf{H2}' scheme in \cite{SV08}.)

\subsubsection{Concatenated Dynamical Decoupling (\textsf{CDD})}
When using the \textsf{PDD} strategy, the time evolution of a single cycle in the toggled frame is given by \eqref{eq:pddmcycles},
\begin{equation}
 \tilde{U}( t_c ) =
 \exp(-i \tilde{H}_{n_c-1} \Delta t) \dots \exp(-i \tilde{H}_1 \Delta t) \exp(-i \tilde{H}_0 \Delta t)
= \exp(-i \overline{H} t_c ).
\end{equation}
Khodjasteh and Lidar proposed a concatenated dynamical decoupling strategy \cite{KL05}, %
which tries to fight the remaining higher order terms in the Magnus expansion of $\overline{H}$ as follows:
As a first step, the basic \textsf{PDD} cycle is embedded into an additional one,
\begin{equation}
\tilde{U}( n_c\cdot t_c ) =
g_{n_c-1}^\dagger \tilde{U}(t_c) g_{n_c-1} \cdot \hdots \cdot g_1^\dagger \tilde{U}(t_c) g_1 \cdot g_0^\dagger \tilde{U}(t_c) g_0,
\end{equation}
leading to a cycle of length $n_c^2$.
We may now either repeat this cycle over and over again (called periodic concatenated level 2 decoupling (\textsf{PCDD2})), or iterate the embedding process one more time to obtain a cycle of length $n_c^3$.
After $k$ recursive embeddings, one obtains a cycle of length $n_c^k$.
Periodic decoupling with such a cycle is called periodic concatenated level $k$ decoupling (\textsf{PCDDk}) \cite{SV06,VS06,SV08}.
The $\textsf{CDD}$ strategy is to repeat the embedding process ad infinitum.

In order to achieve a good performance with $\textsf{CDD}$, the underlying decoupling scheme should be able to suppress the correlations in the remaining effective Hamiltonian of the $k$-th embedded cycle for increasing $k$.
Since these correlations increase with $k$, we expect $\textsf{CDD}$ to work best when the decoupling scheme is an annihilator of short length $n_c$.
Due to the fact that the length of a minimal annihilator is equal to the dimension of the system Hilbert space, it will be hard to meet this criterion.
In fact, \textsf{CDD} was proposed to decouple a single qubit from its environment \cite{KL05}, in which case an annihilator of length four is given by the Pauli operators $\mathcal{I},X,Y$, and $Z$.

\subsection{Randomized Strategies}\label{subsec:randstrat}

\subsubsection{Naive Random Decoupling (\textsf{NRD})}
The simplest randomized control strategy is to apply the pulses $p_i$ at times $t_i=i\Delta t$, $i\in\mathbb{N}_0$, where $p_i = g_{r(i)} g^\dagger_{r(i-1)}$ is constructed by picking the elements of the decoupling scheme at random: The indices $r(i)\in\{0,\dots,n_c-1\}$ are chosen independently according to a uniform distribution.
As a result, after a time $T=t_N$ the time evolution operator in the toggled frame is given by
\begin{equation}\label{eq:tildeUTNRD}
 \tilde{U}(T=t_N) = \exp(-i \tilde{H}_{r(N-1)} \Delta t) \dots \exp(-i \tilde{H}_{r(1)} \Delta t) \exp(-i \tilde{H}_{r(0)} \Delta t).
\end{equation}
The resulting decay of the entanglement fidelity \eqref{eq:FeUU} (corresponding to the average state fidelity) depends on the particular choice of indices.
To obtain a general statement, we take the average over all random realizations (denoted by $\mathbb{E}$), i.\,e. we define
\begin{equation}\label{eq:FeNRDdefi}
 F_e^\textsf{NRD}(T)  = \mathbb{E}  \Bigl\vert \frac{1}{d} \tr\bigl( \tilde{U}(T) \bigr) \Bigr\vert^2
\end{equation}
as the relevant performance measure.
\begin{thm}\label{thm:nrdf}
In lowest order, the average \textsf{NRD} fidelity \eqref{eq:FeNRDdefi} is given by
\begin{equation}\label{eq:FeNRD}
F_e^\textsf{NRD}(T) = 1 - \frac{1}{d} \tr \bigl(H_0^2\bigr) \Delta t T + \mathcal{O}(\lambda^4\Delta t^2 T).
\end{equation}
\end{thm}
\begin{proof}
Writing $H_0$ as $\lambda H_0$, we calculate the fidelity \eqref{eq:FeNRDdefi} up to fourth order in $\lambda$.
This allows any result to be used later on to obtain the variance of the fidelity.
We start by expanding each of the products in \eqref{eq:tildeUTNRD} as
\begin{equation}
 \exp(-i \tilde{H}_{r(s)} \Delta t ) = \mathcal{I} - i \tilde{H}_{r(s)} \Delta t -\frac{1}{2} \tilde{H}_{r(s)}^2 \Delta t^2 + \frac{i}{6} \tilde{H}_{r(s)}^3 \Delta t^3 + \frac{1}{24} \tilde{H}_{r(s)}^4 \Delta t^4 + \mathcal{O}(\lambda^5),
\end{equation}
with $0 \leq s \leq N-1$. Taking the trace leads to
\begin{multline}
\frac{1}{d} \tr\bigl( \tilde{U}(T) \bigr) = 1
- \frac{1}{2} \sum_s \frac{1}{d} \tr \bigl(  \tilde{H}_{r(s)}^2 \bigr) \Delta t^2
- \sum_{s>u} \frac{1}{d} \tr \bigl(  \tilde{H}_{r(s)} \tilde{H}_{r(u)} \bigr) \Delta t^2
+\frac{i}{6} \sum_s \frac{1}{d} \tr \bigl(  \tilde{H}_{r(s)}^3 \bigr) \Delta t^3 \\
+\frac{i}{2} \sum_{s>u} \frac{1}{d} \tr \bigl(\tilde{H}_{r(s)}\tilde{H}_{r(u)}^2 +  \tilde{H}_{r(s)}^2 \tilde{H}_{r(u)} \bigr) \Delta t^3
+i \sum_{s>u>v} \frac{1}{d} \tr \bigl(\tilde{H}_{r(s)}\tilde{H}_{r(u)}\tilde{H}_{r(v)}\bigr) \Delta t^3 + \dots + \mathcal{O}(\lambda^5).
\end{multline}
The fidelity is obtained by averaging the absolute square of the above expression over all random realizations.
With the help of the decoupling condition \eqref{eq:zerothorderoverlineh} for traceless $H_0$,
\begin{equation}
\frac{1}{n_c} \sum_{j=0}^{n_c-1} \tilde{H}_j  = 0,
\end{equation}
and due to the independence of the random selections, we obtain
\begin{multline}\label{eq:proofaveragenrd}
 \mathbb{E}  \Bigl\vert \frac{1}{d} \tr\bigl( \tilde{U}(T) \bigr) \Bigr\vert^2 =
 1 - \frac{1}{d} \tr \bigl(H_0^2\bigr) \Delta t T
+\frac{1}{4} \Bigl( \frac{1}{d} \tr \bigl(H_0^2\bigr) \Delta t T \Bigr)^2
+\frac{1}{12} \frac{1}{d} \tr \bigl(H_0^4\bigr) \Delta t^3 T \\
+\frac{1}{2}\frac{1}{n_c}\sum_{j=0}^{n_c-1} \frac{1}{n_c}\sum_{j'=0}^{n_c-1} \Bigl( \frac{1}{d} \tr \bigl( \tilde{H}_j \tilde{H}_{j'} \bigr)  \Bigr)^2 \Delta t^2 T(T-\Delta t)\\
+\frac{1}{4} \frac{1}{d}\tr\Bigl( \Bigl( \frac{1}{n_c}\sum_{j=0}^{n_c-1} \tilde{H}^2_j \Bigr)^2 \Bigr) \Delta t^2 T(T-\Delta t)
+ \mathcal{O}(\lambda^5). \qedhere
\end{multline}
\end{proof}

\begin{rem}
As it turns out by looking at various numeric examples, a good approximation of \eqref{eq:FeNRDdefi},
valid for all times $T\geq 0$ and in lowest order identical to \eqref{eq:FeNRD}, is given by
\begin{equation}\label{eq:FeappNRD}
F_{e\ \text{app}}^\textsf{NRD}(T) = \exp\Bigl( - \frac{1}{d} \tr \bigl(H_0^2\bigr) \Delta t T \Bigr).
\end{equation}
\end{rem}

A strict lower bound on the average worst case fidelity was given in \cite{VK05},
\begin{equation}
F_w^\textsf{NRD} (T) =
\mathbb{E} \min_{\ket{\psi}\in\mathcal{H}_S} \bigl\vert \bra{\psi}  \tilde{U}(T) \ket{\psi} \bigr\vert^2
> 1 -  4 \kappa^2 \Delta t T + \mathcal{O}\bigl(\kappa^3 \Delta t^2 T \bigr),
\end{equation}
with $\kappa = \Vert H_0\Vert_2$.
The bound remains valid for time-dependent system Hamiltonians $H_0(t)$
if $\Vert H_0(t) \Vert_2 < \kappa$ for $0\leq t\leq T$
and the decoupling condition \eqref{eq:deccondi} is satisfied for $0\leq t\leq T$.
For an appropriate redefinition of $\kappa$, the bound applies to open quantum systems as well \cite{VK05,V05}.

In summary, \textsf{NRD} offers some interesting advantages over deterministic strategies like \textsf{PDD} and \textsf{SDD}:
The fidelity decay ($\mathcal{O}(\lambda^2\Delta t T)$) is only linear in time, while it is quadratic in time for \textsf{PDD} and \textsf{SDD}.
The strength of the decay does not depend on the length $n_c$ of the underlying decoupling scheme.
As a consequence, it is always possible to choose an annihilator as decoupling scheme.
Since the lower bound guarantees a linear decay also for time dependent Hamiltonians, it is possible to apply \textsf{NRD} even if the system Hamiltonian is completely unknown.
An additional advantage over \textsf{PDD} is that the \textsf{NRD} strategy remains applicable if we use bounded control instead of bang-bang control (a fact that turns out in subsection \ref{subsec:parec_ana}), while the \textsf{PDD} cycles have to be replaced by the longer Euler cycles of subsection \ref{subsec:boundedctrls}.
A disadvantage is the higher strength of the decay ($\mathcal{O}(\lambda^2\Delta t)$) compared to \textsf{PDD} and \textsf{SDD} ($\mathcal{O}(\lambda^4 t_c^2)$ and $\mathcal{O}(\lambda^6 t_c^4)$, respectively).
As pointed out in \cite{VK05}, \textsf{NRD} outperforms \textsf{PDD} if $\kappa^2 \Delta t T \cdot n_c^2 \gg 1$, i.\,e. for long times and/or long decoupling schemes.

The linear-in-time fidelity decay of \textsf{NRD} was first observed in the author's diploma thesis \cite[chapter 4.2]{DiplKern} where a quantum memory consisting of $n=10$ qubits was protected against inter-qubit couplings by using a decoupling scheme of length $n_c=4^n$ given by the set of Pauli operators $\mathcal{P}_2^n$.
As it will be shown in section \ref{sec:parec}, in contrast with any periodic strategy, \textsf{NRD} allows the protection of a quantum computation in which the quantum gates are applied in between subsequent decoupling pulses \cite{parec,GKAJ08}.
In this context, \textsf{NRD} using a decoupling scheme given by the set of Pauli operators was called Pauli random error correction (\textsf{PAREC}).

For any decoupling strategy which involves some kind of randomization,
in addition to the average fidelity,
an important quantity is its variance.
It is a measure of how close the fidelity of a single run comes to the average fidelity:
The smaller the variance, the smaller the expected difference.
\begin{thm}\label{thm:nrdv}
In lowest non-vanishing order, the variance of the \textsf{NRD} fidelity \eqref{eq:FeNRDdefi} is given by
\begin{equation}\label{eq:nrdv}
\sigma^2_\textsf{NRD} (T) =
2T(T-\Delta t) \frac{1}{n_c}\sum_{j=0}^{n_c-1} \frac{1}{n_c}\sum_{j'=0}^{n_c-1} \Bigl( \frac{1}{d} \tr \bigl( \tilde{H}_j \tilde{H}_{j'} \bigr)  \Bigr)^2 \Delta t^2 + \mathcal{O}(\lambda^6).
\end{equation}
\end{thm}
\begin{proof}
We calculate the quantity
\begin{equation}\label{eq:nrdvproof}
 \sigma^2_\textsf{NRD} (T) =
\mathbb{E} \Bigl( \Bigl\vert \frac{1}{d} \tr\bigl( \tilde{U}(T) \bigr) \Bigr\vert^2 \Bigr)^2 -
\Bigl( \mathbb{E} \Bigl\vert \frac{1}{d} \tr\bigl( \tilde{U}(T) \bigr) \Bigr\vert^2 \Bigr)^2
\end{equation}
up to fourth order in $\lambda$ as it was done in the proof of theorem \ref{thm:nrdf}.
The term whose square is subtracted on the right hand side is given by \eqref{eq:proofaveragenrd}.
\end{proof}
\begin{rem}[i]
Equation \eqref{eq:nrdv} can be upper and lower bounded as follows:
Using the fact that $\langle A,B\rangle = \tr(A^\dagger B)$ denotes the Hilbert-Schmidt inner product,
the Cauchy-Schwarz inequality,
$\vert \langle A,B\rangle \vert^2 \leq \langle A,A\rangle \cdot \langle B,B\rangle$, in connection with $T(T-\Delta t)< T^2$ leads to an upper bound.
Since the averaging is performed over a non-negative expression, we obtain a lower bound by picking the elements where $j=j'$. Altogether,
\begin{equation}\label{eq:nrdvbounds}
\frac{2T(T-\Delta t)}{n_c} \Bigl( \frac{1}{d} \tr \bigl( H_0^2 \bigr) \Delta t \Bigr)^2 \leq \sigma^2_\textsf{NRD} (T) \leq 2 \Bigl( \frac{1}{d} \tr \bigl( H_0^2 \bigr) \Delta t T\Bigr)^2.
\end{equation}
\end{rem}
\begin{rem}[ii]
If the elements $g_j$ of the decoupling set $\{g_j\}_{j=0}^{n_c-1}$ form a group, equation \eqref{eq:nrdv} simplifies to
\begin{equation}
\sigma^2_\textsf{NRD} (T) =
2T(T-\Delta t) \frac{1}{n_c}\sum_{j=0}^{n_c-1}  \Bigl( \frac{1}{d} \tr \bigl( H_0 \tilde{H}_j \bigr)  \Bigr)^2 \Delta t^2 + \mathcal{O}(\lambda^6).
\end{equation}
\end{rem}
\noindent
While the average \textsf{NRD} fidelity does not depend on the length of the decoupling scheme,
equation \eqref{eq:nrdv} in connection with the lower bound in \eqref{eq:nrdvbounds} leads to the conclusion that its variance actually becomes smaller, the greater the length of the decoupling scheme.
We expect the variance to become minimal if the underlying decoupling scheme is an annihilator.
This feature is in strong contrast to \textsf{PDD} and \textsf{SDD} where smaller decoupling schemes increase the performance.

\subsubsection{Embedded Decoupling (\textsf{EMD})}
In order to combine the advantages of the \textsf{PDD} and the \textsf{NRD} strategy, the following embedded dynamical decoupling strategy has been devised by the author in \cite{combi}.
Let $\tilde{U}(t_c)$ denote the time evolution operator of a single \textsf{PDD} cycle in the toggled frame (compare with \eqref{eq:pddmcycles}),
\begin{equation}\label{eq:pddcycleinemdcontext}
 \tilde{U}( t_c ) =
 \exp(-i \tilde{H}_{n_c-1} \Delta t) \dots \exp(-i \tilde{H}_1 \Delta t) \exp(-i \tilde{H}_0 \Delta t)
= \exp(-i \overline{H} t_c ).
\end{equation}
By definition of the decoupling scheme, the zeroth order term in $\overline{H}$ vanishes and we have the residual Hamiltonian $\overline{H} = \overline{H}^{(1)}+\overline{H}^{(2)}+\dots$, with $\overline{H}^{(1)}$ given by \eqref{eq:firstorderoverlineh}.
Let us now take a second decoupling scheme $\{\gamma_j\}_{j=0}^{\nu_c-1}$ eliminating the residual Hamiltonian.
The embedded decoupling strategy is to apply the \textsf{NRD} strategy at times $i \cdot t_c$, $i\in\mathbb{N}_0$, using the second decoupling set to suppress the residual Hamiltonian of the \textsf{PDD} cycles.
As a result, after a time $T=N\cdot t_c$, $N\in \mathbb{N}_0$, we obtain the following time evolution,
\begin{align}
 \tilde{U}( T= N\cdot t_c ) &=
  \gamma_{r(N-1)}^\dagger \tilde{U}( t_c ) \gamma_{r(N-1)} \ \dots \
  \gamma_{r(1)}^\dagger \tilde{U}( t_c ) \gamma_{r(1)} \
  \gamma_{r(0)}^\dagger \tilde{U}( t_c ) \gamma_{r(0)} \nonumber\\
&=
\exp(-i \tilde{\overline{H}}_{r(N-1)} \Delta t) \dots \exp(-i \tilde{\overline{H}}_{r(1)} \Delta t) \exp(-i \tilde{\overline{H}}_{r(0)} \Delta t),
\end{align}
where $\tilde{\overline{H}}_{r(i)} = \gamma_{r(i)}^\dagger \, \overline{H} \, \gamma_{r(i)}$ for $i=\{0,1,\dots, N-1\}$, and $r(i)\in\{0,1,\dots,\nu_c-1\}$.
Typically, we choose the second decoupling set to be an annihilator given by the set of Pauli operators, i.\,e. $\{\gamma_j\}_{j=0}^{\nu_c-1} = \mathcal{P}_q^n$.
To analyze the performance of \textsf{EMD}, we can simply adopt the results obtained for \textsf{NRD} if we apply the substitutions $H_0 \mapsto \overline{H}$ and $\Delta t \mapsto t_c$.
In particular, to obtain the lowest order results, it suffices to replace $H_0$ with $\overline{H}^{(1)}$.
Hence, we obtain the entanglement fidelity
\begin{align}
F_e^\textsf{EMD}(T) &= 1 - \frac{1}{d} \tr\bigl( \bigl(\overline{H}^{(1)}\bigr)^2 \bigr) t_c T + \mathcal{O}((\lambda^2t_c)^4 t_c^2 T) , \label{eq:FeEMD} \\
F_{e\ \text{app}}^\textsf{EMD}(T) &= \exp\Bigl( - \frac{1}{d} \tr\bigl( \bigl(\overline{H}^{(1)}\bigr)^2  \bigr) t_c T \Bigr), \label{eq:FeappEMD}
\end{align}
the worst case fidelity
\begin{equation}
F_w^\textsf{EMD} (T) > 1 -  4 \kappa^4 t_c^3 T + \mathcal{O}\bigl(\kappa^6 t_c^5 T \bigr),
\end{equation}
and the variance
\begin{equation}\label{eq:emdv}
\sigma^2_\textsf{EMD} (T) =
2T(T-t_c) \frac{1}{\nu_c}\sum_{j=0}^{\nu_c-1} \frac{1}{\nu_c}\sum_{j'=0}^{\nu_c-1} \Bigl( \frac{1}{d} \tr \bigl( \tilde{\overline{H}}^{(1)}_j \tilde{\overline{H}}^{(1)}_{j'} \bigr)  \Bigr)^2 t_c^2 + \mathcal{O}(\lambda^{12}),
\end{equation}
with $\tilde{\overline{H}}^{(1)}_j = \gamma^\dagger_j \overline{H}^{(1)} \gamma_j$.
The fidelity decay is of order $\mathcal{O}(\lambda^4 t_c^3 T)$ and does indeed combine the advantage of the linear-in-time decay of \textsf{NRD} with the stronger suppression of \textsf{PDD}.
As it was discussed in the \textsf{PDD} paragraph, the performance of \textsf{PDD} depends slightly on the order of the elements in the decoupling scheme, or in other words, on the path which traverses the elements during a cycle. To eliminate this dependence and to achieve an average performance, we might choose a random path for each basic cycle (compare with the \textsf{RPD} strategy). We label such an embedded strategy involving the additional path randomization \textsf{EMDr}.
An overview over the dependencies of the average fidelity decay for different control strategies can be found in table \ref{tab:overviewdecay}.

\begin{table}\centering
\begin{tabular}{c|c}
strategy & decay \\
\hline
\textsf{none} & $\mathcal{O}(\lambda^2 T^2)$\\
\textsf{NRD} & $\mathcal{O}(\lambda^2 \Delta t T)$\\
\textsf{PDD} & $\mathcal{O}(\lambda^4 t_c^2 T^2)$\\
\textsf{EMD},\textsf{EMDr},\textsf{RPD} & $\mathcal{O}(\lambda^4 t_c^3 T)$\\
\textsf{SDD} & $\mathcal{O}(\lambda^6 t_c^4 T^2)$\\
\textsf{ESDD},\textsf{ESDDr},\textsf{SRPD}& $\mathcal{O}(\lambda^6 t_c^5 T)$
\end{tabular}
\caption[Average fidelity decay of various control strategies]{The average fidelity decay of various control strategies using an underlying decoupling scheme of length $n_c$ and a pulse distance in time $\Delta t$ ($t_c=n_c\Delta t$) to suppress the system Hamiltonian $\lambda H_0$.
Note that the strength of the decay of \textsf{NRD} does not depend on the length $n_c$.
}\label{tab:overviewdecay}
\end{table}

\subsubsection{Embedded Symmetric Decoupling (\textsf{ESDD})}
The embedded decoupling strategy described in the preceeding paragraph can naturally be extended to an underlying \textsf{SDD} scheme, as it was done implicitly in \cite{recoup} (see chapter \ref{chap:decrec}).
We call the resulting decoupling strategy embedded symmetric dynamical decoupling.
For a single \textsf{SDD} cycle, equation \eqref{eq:pddcycleinemdcontext} becomes
\begin{multline}\label{eq:sddcycleinsemdcontext}
 \tilde{U}( t_c' ) =
 \exp(-i \tilde{H}_0 \Delta t) \exp(-i \tilde{H}_1 \Delta t) \dots \exp(-i \tilde{H}_{n_c-1} \Delta t) \times \\
 \exp(-i \tilde{H}_{n_c-1} \Delta t) \dots \exp(-i \tilde{H}_1 \Delta t) \exp(-i \tilde{H}_0 \Delta t) = \exp(-i \overline{H} t_c'),
\end{multline}
with $t_c'=n_c'\Delta t$ and $n_c'=2n_c$, and the Magnus expansion of the residual Hamiltonian $\overline{H}$ contains only terms of second and higher order, i.\,e. $\overline{H} = \overline{H}^{(2)}+\overline{H}^{(4)}+\dots$, with $\overline{H}^{(2)}$ given by \eqref{eq:2ndorderoverlineh}.
As it was done in the analysis of the performance of \textsf{EMD}, we can simply adopt the results obtained for \textsf{NRD} if we apply the substitutions $H_0 \mapsto \overline{H}$ and $\Delta t \mapsto t_c'$ in the corresponding expressions. To obtain the lowest order results, it suffices to replace $H_0$ with $\overline{H}^{(2)}$, and we obtain the average fidelity
\begin{align}
F_e^\textsf{ESDD}(T) &= 1 - \frac{1}{d} \tr\bigl( \bigl(\overline{H}^{(2)}\bigr)^2 \bigr) t_c' T + \mathcal{O}((\lambda^3t_c'^2)^4 t_c'^2 T), \label{eq:FeESDD} \\
F_{e\ \text{app}}^\textsf{ESDD}(T) &= \exp\Bigl( - \frac{1}{d} \tr\bigl( \bigl(\overline{H}^{(2)}\bigr)^2  \bigr) t_c' T \Bigr), \label{eq:FeappESDD}
\end{align}
the worst case fidelity
\begin{equation}
F_w^\textsf{ESDD} (T) > 1 -  4 \kappa^6 t_c'^5 T + \mathcal{O}\bigl(\kappa^9 t_c'^8 T \bigr),
\end{equation}
and the variance
\begin{equation}\label{eq:esddv}
\sigma^2_\textsf{ESDD} (T) =
2T(T-t_c') \frac{1}{\nu_c}\sum_{j=0}^{\nu_c-1} \frac{1}{\nu_c}\sum_{j'=0}^{\nu_c-1} \Bigl( \frac{1}{d} \tr \bigl( \tilde{\overline{H}}^{(2)}_j \tilde{\overline{H}}^{(2)}_{j'} \bigr)  \Bigr)^2 t_c'^2 + \mathcal{O}(\lambda^{18}).
\end{equation}
As in the \textsf{EMD} case, we might bring the decoupling elements after each cycle into a new random order. We label such a strategy involving this additional randomization by \textsf{ESDDr} (to be compared with \textsf{SRPD}).

\subsubsection{Random Path Decoupling (\textsf{RPD})}
Another approach to combine the advantages of the deterministic and randomized strategies is called random path decoupling.
It was proposed by Viola and Knill \cite{VK05} and explored by Santos and Viola in \cite{SV06,VS06}.
While the performance of \textsf{RPD} was conjectured to be comparable with \textsf{EMD} \cite{VS06}, we are going to prove this conjecture.
The \textsf{RPD} strategy is basically to apply \textsf{PDD}, but now each \textsf{PDD} cycle is constructed from a randomly reordered decoupling scheme.
In other words, each \textsf{PDD} cycle traverses the elements of the decoupling scheme according to a random path.
The time evolution operator of such a \textsf{PDD} cycle is given by
\begin{equation}
\tilde{U}_\pi( t_c ) =
 \exp(-i \tilde{H}_{\pi(n_c-1)} \Delta t) \dots \exp(-i \tilde{H}_{\pi(1)} \Delta t) \exp(-i \tilde{H}_{\pi(0)} \Delta t) = \exp(-i \overline{H}_\pi t_c ),
\end{equation}
where $\pi \in \textsf{S}_{n_c}$ denotes a randomly chosen permutation of the elements of the decoupling scheme.
The reordering obviously does not affect the zeroth order term in the Magnus expansion of $\overline{H}_\pi = \overline{H}^{(0)}_\pi + \overline{H}^{(1)}_\pi + \dots$, which is still given by \eqref{eq:zerothorderoverlineh-inpdd} (i.\,e. $\overline{H}^{(0)}_\pi = 0$ for all $\pi$), but the first order term,
\begin{equation}%
 \overline{H}^{(1)}_\pi = -\frac{i}{2n_c} \sum_{i>j=0}^{n_c-1} [\tilde{H}_{\pi(i)}, \tilde{H}_{\pi(j)}]\Delta t
\end{equation}
depends on $\pi$.
\begin{lem}\label{lem:rpdh1}
The average of $ \overline{H}^{(1)}_\pi$ taken over all $\pi\in \textsf{S}_{n_c}$ vanishes, i.\,e. we have
\begin{equation}
 \Bigl\langle \overline{H}^{(1)}_\pi \Bigr\rangle_{\pi\in \textsf{S}_{n_c}} =
\frac{1}{n_c!}\sum_{\pi\in \textsf{S}_{n_c}} \overline{H}^{(1)}_\pi  = 0.
\end{equation}
\end{lem}
\begin{proof}
This result is a simple consequence of the fact that $[\tilde{H}_i, \tilde{H}_j] = - [\tilde{H}_j, \tilde{H}_i]$.
\end{proof}
\noindent
According to the above lemma, we are in a situation similar to \textsf{EMD}, where the residual Hamiltonian $\overline{H}^{(1)}+\overline{H}^{(2)}+\dots$ of a fixed \textsf{PDD} cycle
is eliminated on average by the additional pulses generated by random selection from the second decoupling scheme.
While \textsf{EMD} achieves the suppression perfectly in the sense that the average taken over all toggled residual Hamiltonians vanishes,
it is unclear whether \textsf{RPD} achieves annihilation of the second- and higher-order terms in the residual Hamiltonian as well.
(It will be shown in the next paragraph that annihilation is still achieved for the second-order term.)
Therefore, we expect \textsf{RPD} to perform slightly worse than \textsf{EMD} (or \textsf{EMDr} if we eliminate the influence of the order of the decoupling elements).
In fact \textsf{RPD} is equivalent to \textsf{EMDr}, if we replace each element of the second decoupling scheme by the identity.
Nevertheless, \textsf{RPD} offers the advantage that no second decoupling scheme is involved.
Hence, all the applied pulses are of the form $g_j g_i^\dagger$ for some $i,j\in\{0,\dots,n_c-1\}$.

\subsubsection{Symmetric Random Path Decoupling (\textsf{SRPD})}
The \textsf{RPD} strategy of the preceding paragraph can be improved by symmetrizing the randomly traversed \textsf{PDD} cycles as it was done by the \textsf{SDD} strategy.
The resulting strategy is called symmetric random path decoupling \cite{SV06,VS06}.
Using \textsf{SRPD}, a basic random cycle of length $n_c'=2n_c$ is given by
\begin{multline}
 \tilde{U}_\pi( t_c' ) =
 \exp(-i \tilde{H}_{\pi(0)} \Delta t) \exp(-i \tilde{H}_{\pi(1)} \Delta t) \dots \exp(-i \tilde{H}_{\pi(n_c-1)} \Delta t) \times \\
 \exp(-i \tilde{H}_{\pi(n_c-1)} \Delta t) \dots \exp(-i \tilde{H}_{\pi(1)} \Delta t) \exp(-i \tilde{H}_{\pi(0)} \Delta t) = \exp(-i \overline{H}_\pi t_c'),
\end{multline}
with $\pi \in \textsf{S}_{n_c}$, and by using lemma \ref{lem:h2sddh2pdd}, the lowest non-vanishing term in the Magnus expansion of $\overline{H}_\pi$ is given by
\begin{equation}
 \overline{H}^{(2)}_\pi = -\frac{1}{6n_c} \sum_{i\geq j\geq k=0}^{n_c-1}
 \Bigl( [\tilde{H}_{\pi(i)},[\tilde{H}_{\pi(j)},\tilde{H}_{\pi(k)}]]+
 [[\tilde{H}_{\pi(i)},\tilde{H}_{\pi(j)}],\tilde{H}_{\pi(k)}] \Bigr) \Delta t^2\times
\begin{cases}
1/2 & \text{ if } i=j \text{ or } j=k \\
1 & \text{ else }
\end{cases}.
\end{equation}
\begin{lem}\label{lem:rpdh2}
The average of the above expression taken over all permutations $\pi\in \textsf{S}_{n_c}$ vanishes, i.\,e. we have
\begin{equation}\label{eq:h2averagedoverPi}
 \Bigl\langle \overline{H}^{(2)}_\pi \Bigr\rangle_{\pi\in \textsf{S}_{n_c}} = 0.
\end{equation}
\end{lem}
\begin{proof}
We start with the observation that all the terms in the sum forming $\overline{H}^{(2)}_\pi$ with $i=j$ or $j=k$ add up to zero:
\begin{multline}
  \sum_{i > j =0}^{n_c-1} \Bigl( [\tilde{H}_{\pi(i)},[\tilde{H}_{\pi(i)},\tilde{H}_{\pi(j)}]]+
 [[\tilde{H}_{\pi(i)},\tilde{H}_{\pi(j)}],\tilde{H}_{\pi(j)}] \Bigr) =\\
 \sum_{i \neq j =0}^{n_c-1} \Bigl( \tilde{H}_{\pi(i)}\tilde{H}_{\pi(i)}\tilde{H}_{\pi(j)} -2\tilde{H}_{\pi(i)}\tilde{H}_{\pi(j)}\tilde{H}_{\pi(i)} + \tilde{H}_{\pi(j)}\tilde{H}_{\pi(i)}\tilde{H}_{\pi(i)} \Bigr) = 0.
\end{multline}
The last identity follows if we extend the sum by the terms $i=j$ and use the fact that $\overline{H}^{(0)}=0$.
Hence, the average over all permutations can be taken over the simpler expression
\begin{equation}
 \overline{H}^{(2)}_\pi = -\frac{1}{6n_c} \sum_{i> j> k=0}^{n_c-1}
 \Bigl( [\tilde{H}_{\pi(i)},[\tilde{H}_{\pi(j)},\tilde{H}_{\pi(k)}]]+
 [[\tilde{H}_{\pi(i)},\tilde{H}_{\pi(j)}],\tilde{H}_{\pi(k)}] \Bigr) \Delta t^2,
\end{equation}
and as in the proof of lemma \ref{lem:rpdh1}, the property $[\tilde{H}_{\pi(i)},\tilde{H}_{\pi(j)}] = - [\tilde{H}_{\pi(j)},\tilde{H}_{\pi(i)}]$ leads to the vanishing mean.
\end{proof}
\noindent
Since it remains unclear whether \textsf{SRPD} eliminates the remaining higher order terms in the Magnus expansion as well, we expect it to perform slightly worse than an average \textsf{ESDD} or \textsf{ESDDr}, respectively.

\section{Example}\label{sec:decexample}

In the preceding section various decoupling strategies and their advantages have been discussed.
We are now going to examine the performance of these strategies by means of numerical simulations.
Results on the entanglement fidelity obtained numerically are compared with the corresponding formulas which have been derived in the preceding section.
We start by presenting the model, a quantum register perturbed by Heisenberg couplings, in subsection \ref{subsec:model}.
Then, in subsection \ref{subsec:varnrd}, we focus on the variance of the naive random decoupling \textsf{NRD} strategy using different decoupling sets.
In subsection~\ref{subsec:compstra}, we compare different strategies in order to identify the best one.
Finally, we conclude in subsection~\ref{subsec:exconcl} with a general guideline for a good decoupling strategy.

\subsection{The Model}\label{subsec:model}

We choose the same model Hamiltonian as in \cite{SV06}, i.\,e. we consider $n=8$ qubits with Heisenberg couplings arranged on a linear chain,
\begin{equation}\label{eq:decmodel}
 H_0 =
 \sum_{k_1 = 1}^{n-1}
 \sum_{k_2 = k_1+1}^n
 \
 J^{k_1,k_2} \bigl[  X\otimes X + Y\otimes Y + Z\otimes Z \bigr]_{(k_1,k_2)},
\end{equation}
where the coupling strength between qubits $k_1$ and $k_2$ decays cubically with their separation distance, i.\,e. $J^{k_1,k_2} = J\cdot \vert k_1 - k_2\vert^{-3}$.
We construct a decoupling scheme $\{ g_j \}_{j=0}^{n_c-1}$ of length $n_c=8$ for $H_0$ by using the difference scheme $D(8,8,4)$ listed in table \ref{ds:8} in a way explained in theorem \ref{thm:diffschemedec}.
Another decoupling scheme for $H_0$ is given by the annihilator $\{ \gamma_j \}_{j=0}^{\nu_c-1}$ of length $\nu_c = 4^8$ consisting of Pauli operators, i.\,e. $\gamma_j = \XZ( j )$ with $j\in\mathbb{F}_2^{2\cdot 8}$.
\subsection{The Naive Random Strategy}\label{subsec:varnrd}

\begin{figure}
\centering
\subfloat[Entanglement fidelity.]{\label{fig:vnrd1a}
\includegraphics[scale=0.9]{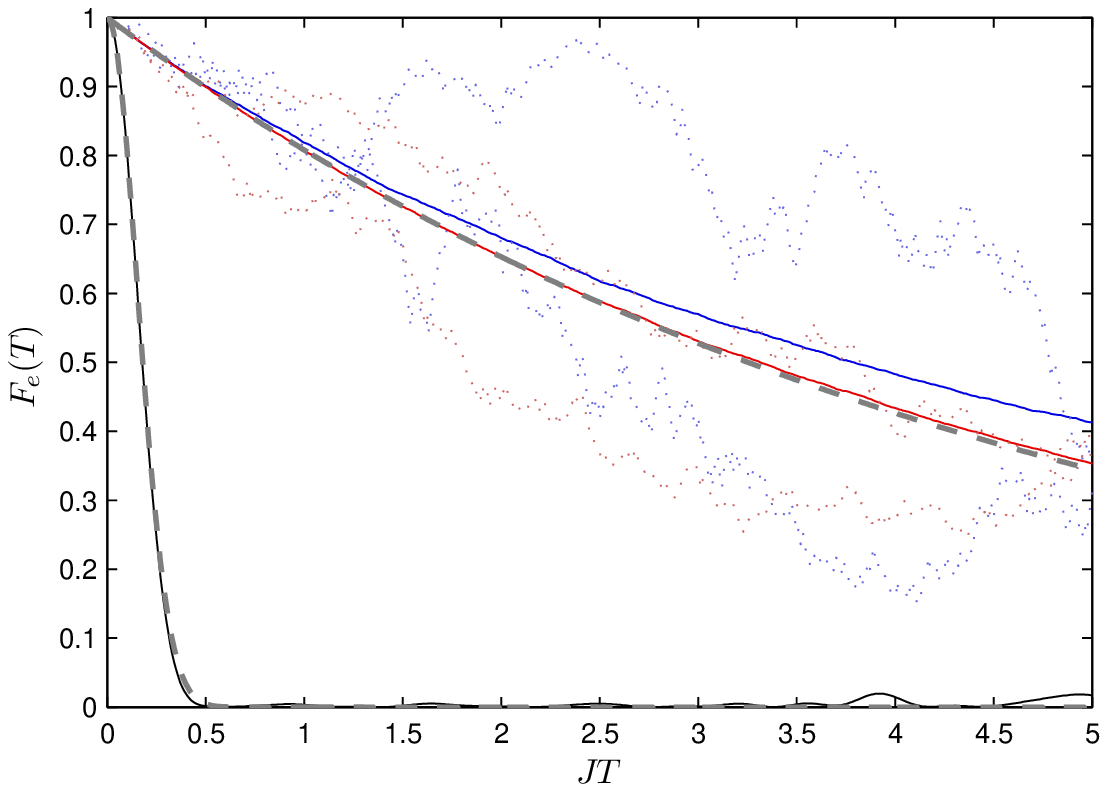}}\\
\subfloat[Entanglement fidelity and its root mean square.]{\label{fig:vnrd1b}
\includegraphics[scale=0.9]{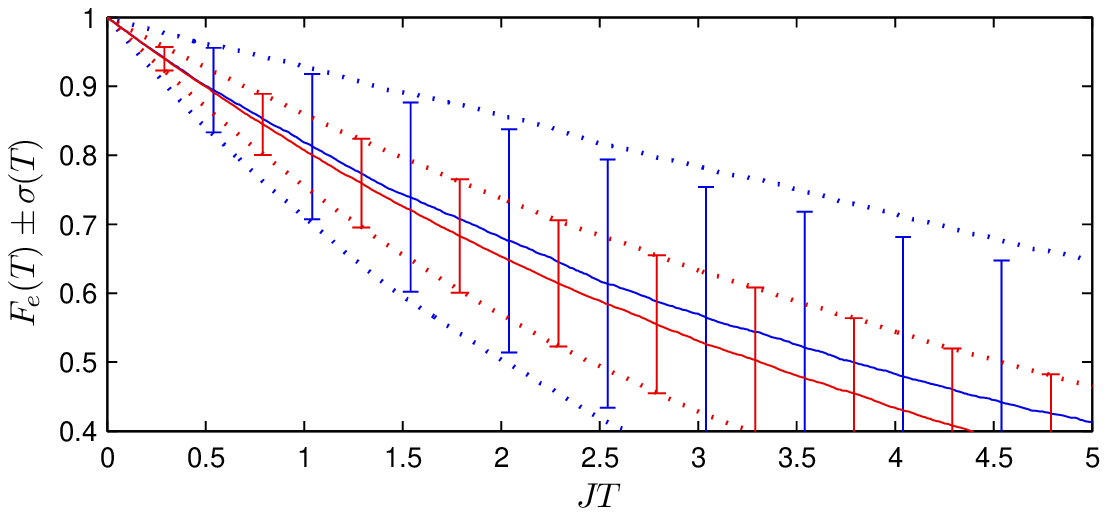}}
\caption[Entanglement fidelity]{
The entanglement fidelity of a quantum register with $n=8$ qubits, perturbed by the Hamiltonian given in \eqref{eq:decmodel}.
The time interval between adjacent decoupling pulses is $\Delta t=0.01 J^{-1}$.
The \textsf{NRD} fidelities are averaged over $1500$ runs.\\
\textbf{(a)}
Without decoupling (\textit{solid line, black}),
with \textsf{NRD} using the set $\{ g_j\}_{j=0}^7$ (\textit{solid line, blue}),
and \textsf{NRD} using the set $\{ \gamma_j\}_{j=0}^{4^8-1}$ (\textit{solid line, red}).
For both of the \textsf{NRD} strategies two individual runs are shown (\textit{dotted lines}).
The \textit{dashed lines} indicate the estimations given by \eqref{eq:FeappNone} and \eqref{eq:FeappNRD}, respectively.\\
\textbf{(b)}
In addition to the two \textsf{NRD} fidelities $F_{e\ \text{num}}^\textsf{NRD}(T)$ (\textit{solid lines}), we indicate the intervals
$F_{e\ \text{num}}^\textsf{NRD}(T) \pm \sigma^\text{num}_\textsf{NRD}(T)$ (\textit{error bars}) and
$F_{e\ \text{num}}^\textsf{NRD}(T) \pm \sigma^\text{app}_\textsf{NRD}(T)$ (\textit{dotted lines}), where $\sigma^\text{num}_\textsf{NRD}(T)$ denotes the standard deviation of the numerical fidelity and $\sigma^\text{app}_\textsf{NRD}(T)$ the corresponding estimation given by \eqref{eq:nrdvapp}.
}\label{fig:vnrd1}
\end{figure}

We performed a numerical simulation of model \eqref{eq:decmodel} over the time $0\leq T \leq 5J^{-1}$.
The resulting entanglement fidelity without decoupling, $F_e^\textsf{none}(T)$, drops down to zero after the time $\approx 0.5 J^{-1}$ and is shown in figure \ref{fig:vnrd1a} (\textit{black, solid line}).
It is in excellent agreement with our estimation $F_{e\ \text{app}}^\textsf{none}(T)$ given by \eqref{eq:FeappNone} (\textit{dashed line}).
In addition, figure \ref{fig:vnrd1a} shows the numerically obtained \textsf{NRD} fidelity
$F_{e\ \text{num}}^\textsf{NRD}(T)$ when using the small decoupling set $\{ g_j \}_{j=0}^{n_c-1}$ of length $n_c=8$ with a pulse distance in time of $\Delta t=0.01J^{-1}$ (\textit{blue, solid line}).
The index $\text{num}$ in $F_{e\ \text{num}}^\textsf{NRD}(T)$ indicates that the quantity differs from the definition of $F_e^\textsf{NRD}(T)$ in equation \eqref{eq:FeNRDdefi} with respect to the average over the random pulse realizations: The latter quantity was defined by averaging over all realizations, while $F_{e\ \text{num}}^\textsf{NRD}(T)$ is averaged over a random subset of simulated runs.
The \textsf{NRD} fidelity based on the small decoupling set (\textit{blue, solid line}) is compared with the corresponding \textsf{NRD} fidelity based on the annihilator $\{ \gamma_j \}_{j=0}^{\nu_c-1}$ of length $\nu_c = 4^8$ (\textit{red, solid line}).
Both fidelities have been obtained by averaging over $1500$ single runs with independent random pulse realizations.
As predicted by our short time expansion \eqref{eq:FeNRD}, both fidelities are identical for short times.
In the region where higher orders become relevant, \textsf{NRD} based on the small decoupling set performs slightly better.
Our estimation $F_{e\ \text{app}}^\textsf{NRD}(T)$ \eqref{eq:FeappNRD} (\textit{dashed line}) is in excellent agreement with the \textsf{NRD} fidelity using the annihilator (\textit{red, solid line}).
To evaluate the estimations $F_{e\ \text{app}}^\textsf{none}(T)$ \eqref{eq:FeappNone} and $F_{e\ \text{app}}^\textsf{NRD}(T)$ \eqref{eq:FeappNRD}, we need the quantity $\tr( H_0^2)/d \approx 21.30 J^2$.

We are now going to study the variance of the \textsf{NRD} fidelities.
In figure \ref{fig:vnrd1b} we indicate the value of the quantity $\sigma^{2\ \text{num}}_\textsf{NRD} (T)$,
which is defined as in \eqref{eq:nrdvproof} with the average over all random realizations (denoted by $\mathbb{E}$) being replaced by the average over the subset of simulated random realizations,
by plotting $F_{e\ \text{num}}^\textsf{NRD}(T) \pm \sigma^{\ \text{num}}_\textsf{NRD} (T)$ (\textit{error bars}) in addition to $F_{e\ \text{num}}^\textsf{NRD}(T)$ (\textit{solid line}).
As in figure \ref{fig:vnrd1a}, the plots corresponding to the \textsf{NRD} strategy based on the small decoupling set are depicted in \textit{blue}, while plots corresponding to the \textsf{NRD} strategy based on the annihilator are depicted in \textit{red}.
It can be seen that the variance is smaller with the annihilator as the underlying decoupling set.
A short time estimation for the variance $\sigma^2_\textsf{NRD} (T)$ was calculated in equation \eqref{eq:nrdv}. Evaluating this expression for the two different decoupling sets leads to
\begin{equation}
 \sigma^2_\textsf{NRD} (T) \approx 2 T^2 \Delta t^2 \times
\begin{cases}
92.47 J^4 & \text{,for } \{ g_j \}_{j=0}^7 \\
21.00 J^4 & \text{,for } \{ \gamma_j\}_{j=0}^{4^8-1}
\end{cases} \quad + \mathcal{O}(J^6) .
\end{equation}
As it turns out, this expression overestimates the variance for longer times.
Hence, we propose the following estimation,
\begin{equation}\label{eq:nrdvapp}
\sigma^{2\ \text{app}}_\textsf{NRD} (T) =
2T(T-\Delta t) \mathbb{E}_j \mathbb{E}_{j'} \Bigl( \frac{1}{d} \tr \bigl( \tilde{H}_j \tilde{H}_{j'} \bigr)  \Bigr)^2 \Delta t^2 \times
\exp\Bigl( - 2\frac{1}{d} \tr \bigl(H_0^2\bigr) \Delta t T \Bigr),
\end{equation}
which we expect to deliver a good approximation for all relevant times.
Here, $\mathbb{E}_j$ ($\mathbb{E}_{j'}$) denotes the average taken over all elements of the underlying decoupling set,
i.\,e. $\tilde{H}_j=g_j^\dagger H_0 g_j$ for the small set $\{g_j\}_{j=0}^{n_c-1}$ of length $n_c=8$ and $\tilde{H}_j=\gamma_j^\dagger H_0 \gamma_j$ for the annihilator $\{\gamma_j\}_{j=0}^{\nu_c-1}$ of length $\nu_c=4^8$.
Note that for short times the exponential can be neglected and this estimation is identical to the (exact) short time expression \eqref{eq:nrdv}.
We put \eqref{eq:nrdvapp} to the test by plotting the quantities
$F_{e\ \text{num}}^\textsf{NRD}(T) \pm \sigma^{\ \text{app}}_\textsf{NRD} (T)$ for both \textsf{NRD} cases (\textit{dotted lines} in figure \ref{fig:vnrd1b}).
As expected, the estimation \eqref{eq:nrdvapp} is excellent for short times.
For longer times it remains excellent when using the annihilator, but slightly overestimates the variance when the small decoupling set is involved.

A decoupling strategy like \textsf{NRD} will be of interest only as long as the resulting fidelity is reasonably high.
In this range, it doesn't make any difference from what kind of decoupling set the elements for \textsf{NRD} are chosen: all choices lead essentially to the same performance.
But since one is interested in a reliable result, one might prefer an annihilator like the set of Pauli operators to constitute the underlying decoupling set, because of the smaller variance.

\subsection{Comparison of Strategies}\label{subsec:compstra}

We are now going to compare the long-time performance of different decoupling strategies.
For this purpose we simulated the time evolution of model \eqref{eq:decmodel} up to the time $T=100J^{-1}$.
All decoupling strategies apply their pulses at times $t_i=i\cdot \Delta t$, $i\in\{0,1,\dots,2000\}$, with $\Delta t=0.05J^{-1}$.
Each of the entanglement fidelities of the randomized strategies is averaged over 100 individual runs with independent random selections.

For this setting, a simulation of various decoupling strategies up to the time $T=50J^{-1}$ has already been done by Santos and Viola in \cite{SV06} (even though for a slightly different underlying decoupling scheme).
We extend this research by taking a closer look on the influence of the traversing path of the decoupling elements and by comparing the obtained fidelities with their corresponding estimations, which have been obtained in section \ref{sec:cstrategies}.
In addition, we study the variance of the randomized schemes and analyze the performance of the \textsf{EMDr} and \textsf{ESDDr} strategies, which have not been considered in~\cite{SV06}.

\subsubsection{Influence of the Traversing Path}
\begin{figure}\centering
\subfloat[Ea][Entanglement fidelity of \textsf{PDD} and \textsf{SDD} for three different permutations of the decoupling scheme.]{\label{fig:compstrata}
\includegraphics[scale=0.88]{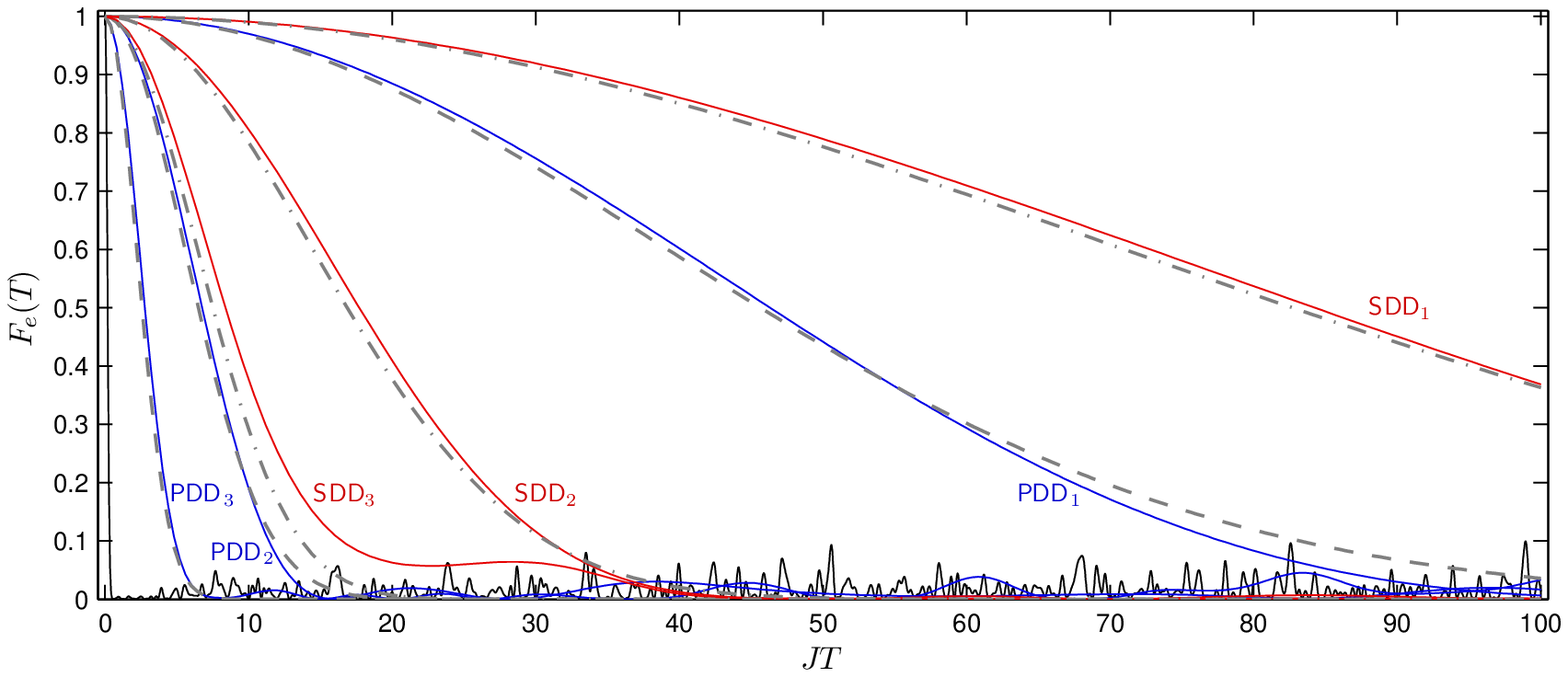}}\\
\subfloat[Eb][Entanglement fidelity of various deterministic and randomized strategies.
The upper part shows an enlarged representation of the high fidelity range \protect{$[0.975,1]$}.]{\label{fig:compstratb}
\includegraphics[scale=0.88, trim=1.5mm 0 0 0]{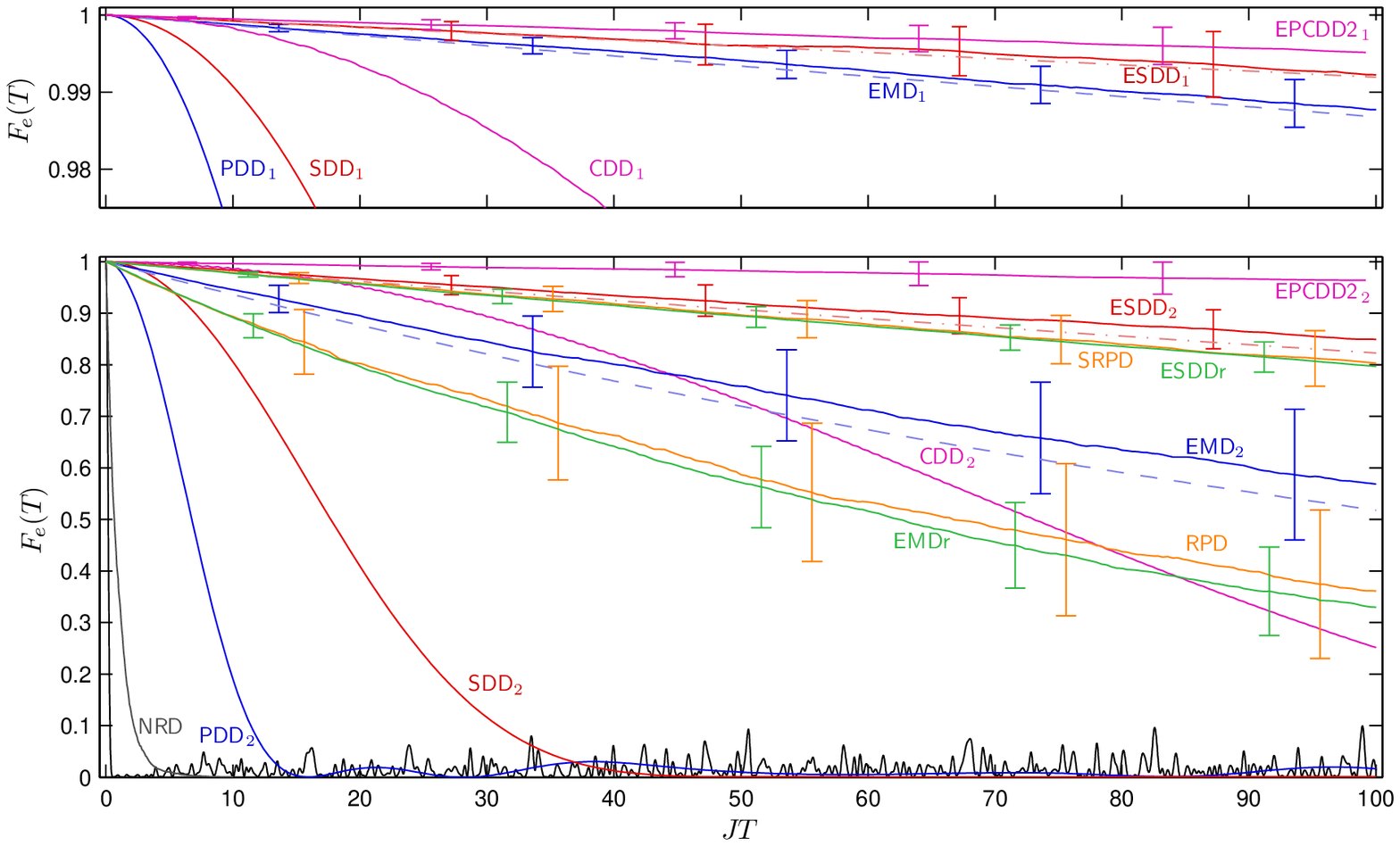}} %
\caption[Entanglement fidelity]{
The entanglement fidelity of a quantum register with $n=8$ qubits, perturbed by the Hamiltonian given in \eqref{eq:decmodel}.
The time interval between adjacent decoupling pulses is $\Delta t=0.05 J^{-1}$.
All randomized fidelities are averaged over $100$ individual runs.\\
\textbf{(a)}
Without decoupling (\textit{solid line, black}),
with \textsf{PDD} using the decoupling set $\{ g_j\}_{j=0}^7$ for three different traversing paths (\textit{blue, solid lines}), the corresponding estimations (\textit{dashed lines}),
the corresponding \textsf{SDD} fidelities (\textit{red, solid lines}) and their estimations (\textit{dashed-dotted lines}).\\
\textbf{(b)}
Strategies using the standard path (labeled as 2):
$\textsf{PDD}_2$ (\textit{blue}), $\textsf{SDD}_2$ (\textit{red}), $\textsf{CDD}_2$ (\textit{purple})
$\textsf{EMD}_2$ (\textit{blue}), $\textsf{ESDD}_2$ (\textit{red}), and $\textsf{EPCDD2}_2$ (\textit{purple}).
Fully randomized strategies:
\textsf{NRD} (\textit{gray}),
\textsf{RPD} (\textit{orange}), \textsf{EMDr} (\textit{green}),
\textsf{SRPD} (\textit{orange}), and \textsf{ESDDr} (\textit{green}).
The upper part shows strategies using the optimal path (labeled as 1):
$\textsf{PDD}_1$ (\textit{blue}), $\textsf{SDD}_1$ (\textit{red}), $\textsf{CDD}_1$ (\textit{purple})
$\textsf{EMD}_1$ (\textit{blue}), $\textsf{ESDD}_1$ (\textit{red}), and $\textsf{EPCDD2}_1$ (\textit{purple}).
In addition the estimations \eqref{eq:FeappEMDi} and \eqref{eq:FeappESSDi} for $\textsf{EMD}_i$  and $\textsf{ESDD}_i$ are shown (\textit{dashed lines}).
The standard deviation of the randomized strategies is indicated by error bars.
}\label{fig:compstrat}
\end{figure}
Let us start with an examination of the performance of the fundamental decoupling strategy (\textsf{PDD}) based on the decoupling scheme $\{g_j\}_{j=0}^{n_c-1}$ constructed using the difference scheme $D(8,8,4)$ listed in table \ref{ds:8}.
As it was discussed in the \textsf{PDD} paragraph in subsection \ref{subsec:detstrat}, the resulting fidelity decay is mainly due to the first order term in the Magnus expansion of a single decoupling cycle of length $t_c=n_c\Delta t$, and we proposed the estimate \eqref{eq:feapppdd}
\begin{equation}
 F_{e\ \text{app}}^\textsf{PDD}(T) = \exp\Bigl( - \frac{1}{d}\tr\bigl( \bigl(\overline{H}^{(1)}\bigr)^2 \bigr) T^2 \Bigr).
\end{equation}
Since, with exception of the vanishing zeroth-order term,
all orders in the Magnus expansion depend on the order of the elements in the decoupling scheme,
the performance of $\textsf{PDD}$ may be optimized by finding the permutation $\pi\in \textsf{S}_{n_c}$ which minimizes $\tr\bigl( \bigl(\overline{H}_\pi^{(1)}\bigr)^2 \bigr)$, or in other words by finding an optimal traversing path for the elements of the decoupling scheme.
We calculated the latter quantity for all permutations and found that it lies in the range
$0.09252 J^4 \Delta t^2 \leq \tr\bigl( \bigl(\overline{H}_\pi^{(1)}\bigr)^2 \bigr)/d \leq 36.963 J^4 \Delta t^2$.
Permutations corresponding to these extremal values are shown in table \ref{tab:h1h2}.
\begin{table}\centering
\begin{tabular}{cccc}
$\textsf{PDD}_i$ & traversing path & $\tr\bigl( \bigl(\overline{H}_i^{(1)}\bigr)^2 \bigr)/d$ & $\tr\bigl( \bigl(\overline{H}_i^{(2)}\bigr)^2 \bigr)/d$ \\
\hline
$\textsf{PDD}_1$& $g_0, g_2, g_4, g_7, g_1, g_3, g_5, g_6$ & $0.09252 J^4\Delta t^2$ & $ 16.2032 J^6\Delta t^4$ \\
$\textsf{PDD}_2$& $g_0, g_1, g_2, g_3, g_4, g_5, g_6, g_7$ & $5.5994  J^4\Delta t^2$ & $389.5980 J^6\Delta t^4$ \\
$\textsf{PDD}_3$& $g_0, g_1, g_6, g_5, g_2, g_3, g_4, g_7$ & $36.963  J^4\Delta t^2$ & $1971.425 J^6\Delta t^4$
\end{tabular}
\caption{The trace of the squared first- and second-order terms of the residual Hamiltonian of a \textsf{PDD} cycle as a function of the order of the decoupling elements. From top to bottom: optimal order (i.\,e. the order which minimizes $\tr\bigl( \bigl(\overline{H}_i^{(1)}\bigr)^2 \bigr)/d$), standard order (close to average performance), and worst order.}\label{tab:h1h2}
\end{table}
We label the \textsf{PDD} strategy based on the optimal path as $\textsf{PDD}_1$,
the one corresponding to the standard path as $\textsf{PDD}_2$, and the worst one as $\textsf{PDD}_3$.
The resulting fidelities $F_e^{\textsf{PDD}_i}(T)$, $i\in\{1,2,3\}$, are compared in figure \ref{fig:compstrata} (\textit{blue, solid lines}).
As it is to be expected from the estimation $F_{e\ \text{app}}^\textsf{PDD}(T)$, we have
$F_e^{\textsf{PDD}_1}(T) > F_e^{\textsf{PDD}_2}(T) > F_e^{\textsf{PDD}_3}(T)$.
In figure \ref{fig:compstrata} we also depicted the improved estimations
\begin{align}
 F_{e\ \text{app}}^{\textsf{PDD}_i}(T) &=
\exp\Bigl( - \frac{1}{d}\Bigl(
\tr\bigl( \bigl(\overline{H}_i^{(1)} + \overline{H}_i^{(2)} \bigr)^2 \bigr)
\Bigr) T^2 \Bigr) \nonumber\\
&=
 \exp\Bigl( - \frac{1}{d}\Bigl(
\tr\bigl( \bigl(\overline{H}_i^{(1)}\bigr)^2 \bigr) + \tr\bigl( \bigl(\overline{H}_i^{(2)}\bigr)^2 \bigr)
\Bigr) T^2 \Bigr),
\end{align}
with the first and second-order Magnus term of the $i$-th path given in table \ref{tab:h1h2}, as \textit{dashed} lines. It can be seen that they are quite close to the actual curves $F_e^{\textsf{PDD}_i}(T)$.

The better deterministic control strategy is \textsf{SDD} which achieves a vanishing first-order Magnus term by doubling the length of a single decoupling cycle.
Hence, the expected fidelities of the three traversing paths are given by \eqref{eq:FappSDD},
\begin{equation}%
F_{e\ \text{app}}^{\textsf{SDD}_i}(T) = \exp\Bigl( - \frac{1}{d}\tr\bigl( \bigl(\overline{H}_i^{(2)}\bigr)^2 \bigr) T^2 \Bigr).
\end{equation}
They are shown in figure \ref{fig:compstrata} as \textit{dashed-dotted} lines, and are in good agreement with the actual \textsf{SDD} fidelities $F_e^{\textsf{SDD}_i}(T)$ (\textit{red, solid lines}).
In principle $F_{e\ \text{app}}^{\textsf{SDD}_1}(T)$ is not necessarily the best \textsf{SDD} fidelity since we minimized the quantity $\tr\bigl( \bigl(\overline{H}_\pi^{(1)}\bigr)^2\bigr)$ which is now vanishing.
Hence, in order to obtain the optimal \textsf{SDD} fidelity we should search for the permutation $\pi$ which minimizes $\tr\bigl( \bigl(\overline{H}_\pi^{(2)}\bigr)^2 \bigr)$.
Although we did not perform this search (due to computational limitations), we expect the optimal \textsf{SDD} fidelity to be quite close to $F_e^{\textsf{SDD}_1}(T)$.

The last remaining deterministic strategy we are going to consider is \textsf{CDD}.
It turns out that for the model and decoupling scheme under consideration, \textsf{CDD} leads to the same fidelity as \textsf{PCDD2} repeating a \textsf{PCDD2} cycle of length $n_c^2\Delta t$.
This is a result of the fact that the residual Hamiltonian of such a cycle cannot be eliminated by the decoupling scheme which was designed to eliminate the system Hamiltonian $H_0$.
Again, the fidelity depends on the traversing path of the underlying \textsf{PDD} cycle.
We show $\textsf{CDD}_i$ for the optimal \textsf{PDD} path ($i=1$) and the standard path ($i=2$) in figure \ref{fig:compstratb} (\textit{purple}).
It can be seen that the $\textsf{CDD}_i$ fidelity surpasses the $\textsf{SDD}_i$ fidelity.
Since, for the model under consideration, the performance of $\textsf{CDD}_i$ is equal to the performance of $\textsf{PCDD2}_i$, this means that periodic dynamical decoupling using a single $\textsf{PCDD2}_i$ cycle of length $n_c^2\Delta t$ is superior than periodic dynamical decoupling based on a $\textsf{SDD}_i$ cycle of length $2n_c\Delta t$.
Hence, according to the estimating formulas for periodic decoupling strategies, the trace of the square of the residual Hamiltonian of a $\textsf{PCDD2}_i$ cycle has to be smaller than the one of a $\textsf{SDD}_i$ cycle.

The randomized strategies which depend on the traversing path are \textsf{EMD} and \textsf{ESDD}, for which the estimations \eqref{eq:FeappEMD} and \eqref{eq:FeappESDD} have been proposed:
\begin{align}
 F_{e\ \text{app}}^{\textsf{EMD}_i}(T) &=
 \exp\Bigl( - \frac{1}{d}\Bigl(
\tr\bigl( \bigl(\overline{H}_i^{(1)}\bigr)^2 \bigr) + \tr\bigl( \bigl(\overline{H}_i^{(2)}\bigr)^2 \bigr)
\Bigr) T\cdot n_c\Delta t \Bigr) \label{eq:FeappEMDi}\\
 F_{e\ \text{app}}^{\textsf{ESDD}_i}(T) &=
 \exp\Bigl( - \frac{1}{d} \tr\bigl( \bigl(\overline{H}_i^{(2)}\bigr)^2 \bigr) T\cdot 2n_c\Delta t \Bigr). \label{eq:FeappESSDi}
\end{align}
The improvement over \textsf{PDD} and \textsf{SDD} is the conversion of the quadratic decay into a linear-in-time one.
We show the fidelities $F_{e\ \text{num}}^{\textsf{EMD}_i}(T)$ (\textit{blue}) and
$F_{e\ \text{num}}^{\textsf{ESDD}_i}(T)$ (\textit{red}) for $i=1,2$ in the lower and upper part of \ref{fig:compstratb}, respectively.
The corresponding approximations $F_{e\ \text{app}}^{\textsf{EMD}_i}(T)$ and $F_{e\ \text{app}}^{\textsf{ESDD}_i}(T)$ are also shown (\textit{dashed lines}).
Analogous to \textsf{ESDD}, we might as well embed the $\textsf{PCDD2}_i$ cycles into a naive random decoupling scheme based on an annihilator.
We label the resulting strategy \textsf{EPCDD2} for embedded periodic concatenated second level dynamical decoupling.
In figure \ref{fig:compstratb}, $F_{e\ \text{num}}^{\textsf{EPCDD2}_i}(T)$ is depicted for $i=1,2$ (\textit{purple}).
As to be expected from the result that the $\textsf{PCDD2}_i$ fidelity surpasses the $\textsf{SDD}_i$ fidelity, $\textsf{EPCDD2}_i$ is superior to $\textsf{ESDD}_i$.
In fact, the best decoupling strategy we found for our model is $\textsf{EPCDD2}_1$ for the optimized traversing path.
It has to be compared with the best previously known strategy in \cite{SV06}, which was \textsf{SRPD} (\textsf{SRPD} will be discussed in the next paragraph) and which achieves a fidelity of $\approx 0.8$ at $T=100J^{-1}$, while $\textsf{EPCDD2}_1$ manages to sustain the fidelity nearly perfectly.
The standard deviation of the fidelity of each randomized decoupling strategy is indicated in figure \ref{fig:compstratb} by error bars.

\subsubsection{Fully Randomized Strategies}

Randomized decoupling strategies which do not involve a fixed traversing path through the elements of the decoupling set are \textsf{NRD}, \textsf{RPD}, and \textsf{EMDr} as well as their symmetrized counterparts \textsf{SRPD} and \textsf{ESDDr}.
We refer to these strategies as being fully randomized.
The \textsf{NRD} fidelity based on the set of Pauli operators performs quite poor, as it can be seen from the \textit{gray} curve in figure \ref{fig:compstratb}.
This fact can be understood by looking at the estimation given by \eqref{eq:FeappNRD},
\begin{equation}%
F_{e\ \text{app}}^\textsf{NRD}(T) = \exp\Bigl( - \frac{1}{d} \tr \bigl(H_0^2\bigr) T \Delta t \Bigr).
\end{equation}
Even though the fidelity decay is linear in time, the value of $\tr \bigl(H_0^2\bigr)/d \approx 21.30J^2$ is huge compared to the worst (i.\,e. largest) first-order term $\tr\bigl( \bigl(\overline{H}_3^{(1)}\bigr)^2 \bigr) \approx 36.963J^4\Delta t^2=0.0924J^2$ relevant for \textsf{PDD}.
A higher suppression of $H_0$ is obtained by using the random path decoupling (\textsf{RPD}) strategy, which chooses the traversing path through $\{ g_j \}_{j=0}^7$ for each successively applied \textsf{PDD} cycle of length $n_c\Delta t=8\Delta t$ at random.
While the \textsf{EMD} fidelity depends on the particular choice of a fixed path, \textsf{RPD} delivers an average \textsf{EMD} fidelity, i.\,e. we propose that a good approximation is given by
\begin{equation}%
F_{e\ \text{app}}^\textsf{RPD}(T) = \exp\Bigl( - \frac{1}{d} \mathbb{E}_\pi
\Bigl( \tr\bigl( \bigl(\overline{H}_\pi^{(1)}\bigr)^2 \bigr) + \tr\bigl( \bigl(\overline{H}_\pi^{(2)}\bigr)^2 \bigr) \Bigr)
T \cdot n_c\Delta t \Bigr),
\end{equation}
where $\mathbb{E}_\pi$ denotes the average over all permutations $\pi\in \textsf{S}_{n_c}$.
The numerically obtained fidelity $F_{e\ \text{num}}^\textsf{RPD}(T)$ is depicted in figure \ref{fig:compstratb} in \textit{orange}.
The symmetrized counterpart of \textsf{RPD} is \textsf{SRPD} and makes use of random \textsf{SDD} cycles of length $2n_c\Delta t$.
As a result, \textsf{SRPD} removes the first-order Magnus terms and leads to the improved fidelity
\begin{equation}%
F_{e\ \text{app}}^\textsf{SRPD}(T) = \exp\Bigl( - \frac{1}{d} \mathbb{E}_\pi \tr\bigl( \bigl(\overline{H}_\pi^{(2)}\bigr)^2 \bigr) T \cdot 2n_c\Delta t \Bigr).
\end{equation}
$F_{e\ \text{num}}^\textsf{SRPD}(T)$ is also shown in figure \ref{fig:compstratb} in \textit{orange}.
From \textsf{RPD} and \textsf{SRPD} we obtain the strategies \textsf{EMDr} and \textsf{ESDDr} by plugging in additional pulses in between subsequent \textsf{PDD} or \textsf{SDD} cycles, where these additional pulses are constructed by random selection from a second decoupling set (typically an annihilator given by the set of Pauli operators).
Since the average over the residual Hamiltonian of the underlying cycles vanishes for the random path strategies even if we do not apply this additional embedding\footnote{This might not be true for terms of third and higher order in the Magnus expansion of a basic cycle.}, we expect the resulting fidelity to be effectively identical with the one of \textsf{RPD} and \textsf{SRPD}.
This fact is confirmed by the data shown in figure \ref{fig:compstratb}, although a bit surprisingly the \textsf{EMDr} and \textsf{ESDDr} fidelities appear to be slightly worse.
Nevertheless, the \textsf{EMDr} and \textsf{ESDDr} fidelities shown in figure \ref{fig:compstratb} (\textit{green}) indicate an advantage:
The square root of the variance indicated by the length of the error bars is approximately only half the size as the corresponding quantity for \textsf{RPD} and \textsf{SRPD}.
This feature might be important in practice, since it is a priori unknown whether a particular single run of a randomized strategy delivers a fidelity above or below average.

\subsection{Conclusions}\label{subsec:exconcl}

The general guideline for the construction of a good decoupling strategy for a system Hamiltonian $H_0$ turned out to be the following:
\begin{itemize}
\item
We start by looking for a deterministic strategy, for which the average Hamiltonian $\overline{H} = \overline{H}^{(0)} + \overline{H}^{(1)} + \overline{H}^{(2)} + \dots $ of a basic decoupling cycle gets as small as possible.
Such a strategy is usually based on a decoupling scheme of length $n_c$ for $H_0$, which satisfies the decoupling condition $\overline{H}^{(0)}=0$.
In order to minimize the residual Hamiltonian, the length $n_c$ should be as small as possible (since we have $\overline{H}^{(i)} = \mathcal{O}( (H_0)^{i+1} (n_c\Delta t)^i )$).
The standard trick to improve a given decoupling scheme is to make it symmetric in time.
Even though the length of such a symmetrized scheme is twice the length of the basic decoupling scheme, this leads to a vanishing first-order term $\overline{H}^{(1)}$.
In addition we saw that the residual Hamiltonian depends on the order of the elements of the decoupling scheme.
By finding an optimal order, the remaining quantity $\overline{H}^{(1)}$ (or for $\overline{H}^{(1)}=0$ the quantity $\overline{H}^{(2)}$) can be minimized.
For our example, the basic decoupling scheme was based on a difference scheme of length $n_c=8$ and the best deterministic decoupling strategy we found was the \textsf{PCDD2} cycle of length $n_c^2$ for an order of the decoupling elements which minimized the quantity $\tr\bigl( \bigl(\overline{H}^{(1)}\bigr)^2 \bigr)$.
\item
The second step is to suppress the residual Hamiltonian.
In principle we could use the same guideline that was used in the first step for the suppression of $H_0$, but because of the complicated structure of the typically highly correlated residual Hamiltonian, a small decoupling scheme usually does not exist.
Instead we have to use an annihilator like the set of Pauli operators.
Because of the large length of this second decoupling scheme (which is equal to the square of the dimension of the system Hilbert space), now the method of choice is naive random decoupling.
Hence, we end up with an embedded decoupling scheme.
For our example, the best result was obtained for \textsf{EPCDD2}, while the second best result was obtained for \textsf{ESDD} (in both cases for an optimal order of the decoupling elements).
\end{itemize}
While it might be hard to find a deterministic strategy which surpasses \textsf{SDD} for a given decoupling scheme, the \textsf{SDD} strategy can always be applied.
If we are not able to determine a good order of the decoupling elements, we might ensure at least an average performance by using the symmetric random path strategy (\textsf{SRPD}) instead of embedding the \textsf{SDD} strategy.
The variance of \textsf{SRPD} can then be minimized by an additional embedding of the basic \textsf{SRPD} cycles in a naive random decoupling strategy based on an annihilator (leading to \textsf{ESDDr}).
In addition, \textsf{SRPD} is the method of choice if we cannot afford the second decoupling scheme, i.\,e. if we are restricted to apply only pulses of the form $g_i g_j^\dagger$, with $g_i$ being an element of the basic decoupling scheme $\{ g_j \}_{j=0}^{n_c-1}$ for $H_0$.

Let us close this chapter by giving a small outlook.
According to the results presented in the last subsection, \textsf{NRD} alone seems to be a rather poor choice for decoupling. Nevertheless it holds many useful features:
For example, it can be applied even if the system Hamiltonian is time dependent.
Even more important, in chapter \ref{chap:compu} \textsf{NRD} turns out to be applicable even if the decoupling pulses have to be implemented using bounded controls, and in addition, it turns out to be able to stabilize quantum computations.

So far only the control task of decoupling has been considered.
We expect similar results for the task of simulating a non-vanishing Hamiltonian.
For example, the potential of \textsf{ESDD} for the simulation of a two qubit gate Hamiltonian in the context of a selective decoupling scheme will be explored in chapter~\ref{chap:decrec}.

The assumption that the decoupling pulses can be applied in a perfect manner is a strong idealization.
In practice, each pulse will be non-ideal and we have to distinguish between systematic and random pulse errors.
An important question is how such errors affect the performance of a given decoupling strategy.
First results concerning this question have been obtained by Santos in Viola with the help of numerical simulations \cite{SV08}.
In addition, the question arises whether decoupling sequences might be designed that are stable against pulse imperfections.
For instance, an Eulerian decoupling cycle (as discussed in subsection \ref{subsec:boundedctrls}) projects any systematic errors of the decoupling pulses (which are elements of the group algebra $\mathcal{A}=R(\mathbb{C}G)$) into the commutant $\mathcal{A}'$ and an additional subsystem encoding might protect against these residual errors \cite{VK03}.
\begin{rem}
The latter fact can be seen by looking at equation \eqref{eq:H0bctrlecycle} in which the effect of systematic pulse errors is reflected by replacing the left $H_0$ by $H_0+H^\text{err}_i(t')$ where
$H^\text{err}_j(t')$ specifies the error of the pulse
\begin{equation*}
p^\text{err}_j = p^\text{err}_j(\tau_p) = \mathcal{T}
\exp\Bigl( -i \int_0^{\tau_p} \bigl(H_j(t')+H^\text{err}_j(t')\bigr) dt'  \Bigr),
\end{equation*}
while the corresponding ideal pulse is given by
\begin{equation*}
p_j = p_j(\tau_p) = \mathcal{T}
\exp\Bigl( -i \int_0^{\tau_p} H_j(t') dt'  \Bigr).
\end{equation*}
\end{rem}

\chapter{Decoupling and Computation}\label{chap:compu}%

In chapter \ref{chap:dyncontrol} we studied dynamical decoupling methods which were designed to suppress the influence of imperfections in a quantum memory.
A more demanding goal is to use these methods to protect a running quantum computation, which consists of a sequence of one- and two-qudit quantum gates.
While we assumed in chapter \ref{chap:dyncontrol} that the decoupling pulses are applied quasi-instantaneously using a \textit{strong} local control Hamiltonian (with the exception of subsection \ref{subsec:boundedctrls}), we are going to assume that the experimentally more demanding quantum gates (especially the two-qudit quantum gates) are realized by applying a \textit{weak} gate Hamiltonian over a finite time interval $\tau_g$ larger than the time interval $\Delta t$ in between subsequent decoupling pulses.
As a consequence, in general, the applied decoupling scheme also alters the gate Hamiltonians.
Solutions for this fundamental problem have been discussed by Viola et\,al. in \cite{VLK99}.
In particular, by using a subsystem encoding it becomes possible to achieve universal control via a set of gate Hamiltonians which commute with the decoupling pulses, and hence remain unaffected.
For example, the hybrid decoupling and computing scheme analyzed in \cite{KhLi08} by Khodjasteh and Lidar is based on the above approach.
Even more general, we might assume that the decoupling pulses are realized over a finite time interval as well.
In this case the dynamically corrected gates based on an Eulerian decoupling cycle (Euler-DCGs) proposed recently by Khodjasteh and Viola \cite{KhVi08} are able to achieve simultaneous computation and decoupling:
An Euler-DCG is generated by extending an Eulerian path in the Cayley graph of the Eulerian decoupling strategy \cite{VK03} described in subsection \ref{subsec:boundedctrls},
by applying a corresponding gate Hamiltonian after completing the path.
In addition, in order to get a vanishing lowest order average Hamiltonian, a gate leading to the same error as the gate Hamiltonian, but implementing the identity, is applied after visiting each of the non-identity vertices in the Cayley graph for the last time.

In this chapter we consider the most general setting, i.\,e. we consider decoupling pulses which are generated by applying a local control Hamiltonian for a time $\tau_p$ and quantum gates which are generated by applying a two-qudit gate Hamiltonian for a time $\tau_g$.
We are going to show that a quantum computation can be stabilized against static imperfections by executing the quantum gates in between subsequent decoupling pulses.
This is in contrast with the Euler-DCGs of Khodjasteh and Viola \cite{KhVi08}, where a quantum gate is effectively implemented only in between completed cycles.
Thereby, our decoupling pulses are constructed by random selection from an annihilator as the set of Pauli operators, or in other words by using the naive random decoupling (\textsf{NRD}) strategy presented in the preceding chapter.
Our method has been published in \cite{parec}, where we devised the acronym Pauli random error correction (\textsf{PAREC}), and provided numerical evidence of its error suppressing properties.
We derive a formula for the fidelity decay of a stabilized quantum computation (for the special case of instantaneous gates and pulses we derived such a formula in \cite{GKAJ08}).
A numerical simulation of the \textsf{PAREC} method is performed for the quantum computation of a quantum map running on a quantum computer perturbed by Heisenberg couplings.
The \textsf{PAREC} method is compared with an idea of Prosen and \u{Z}nidari\u{c} \cite{ProZni01}, who proposed to stabilize a quantum computation against static imperfections by increasing the decay of the correlation function measuring the fidelity decay.
It turns out that our approach does exactly that, i.\,e. it leads to an ultimate decay of correlations.
Eventually, we consider the Euler-DCGs of Khodjasteh and Viola \cite{KhVi08}.
By implementing each quantum gate as an Euler-DCG, a deterministic decoupling method for quantum computations is obtained.
We propose to implement the \textsf{PAREC} method by using only Euler-DCGs in order to benefit from the advantages of both methods.

\pagebreak
Another scenario in which the decoupling strategies of the preceding chapter may be used to improve the performance of a quantum computation is given if the quantum gates are implemented using a selective decoupling scheme. It will be dealt with in chapter \ref{chap:decrec}.

We start by presenting an overview of known results on the fundamental problem of combining quantum computation and dynamical decoupling
in section \ref{sec:decqlogic}.
The \textsf{PAREC} method based on the randomized decoupling strategy is presented, analyzed and simulated in section \ref{sec:parec}.
In section \ref{sec:incrcor}, we compare the \textsf{PAREC} method with the idea of Prosen and \u{Z}nidari\u{c} \cite{ProZni01}, who proposed to increase the correlation decay.
Eventually, we present the Euler-DCGs of Khodjasteh and Viola \cite{KhVi08} in section \ref{sec:enEuler} and show how they might be combined with the \textsf{PAREC} method.

\section{Decoupling and Quantum Logic}\label{sec:decqlogic}
Let us consider a quantum register $S$ defined on a $d$-dimensional Hilbert space $\mathcal{H}_S$.
Typically the register consists of $n$ qudits of dimension $q$ such that $d=q^n$.
For the sake of simplicity, we assume $S$ to be a closed system perturbed by static imperfections modeled by the system Hamiltonian $H_0$ acting on $\mathcal{H}_S$.
(It is straightforward to extend any of the forthcoming results to the case where $S$ is an open system coupled to an environment $E$ via a set of coupling operators as in subsections \ref{subsec:opens} and \ref{subsec:nlsubsys}).
In this section we assume that the decoupling pulses are applied quasi-instantaneously by using a strong local control Hamiltonian, or in other words, by using bang-bang control, but all results are also applicable if the Euler decoupling method (\cite{VK03}, subsection \ref{subsec:boundedctrls}) for bounded strength control is applied.
The fundamental control strategy, called periodic dynamic decoupling (\textsf{PDD}, subsection \ref{subsubsec:pdd}), repeats a basic control cycle traversing all the elements of a control scheme $\mathcal{G} = \{ g_j \}_{j=0}^{n_c-1}$ over and over again.
The length $t_c=n_c\Delta t$ of such a basic cycle is determined by the number $n_c$ of elements in the control scheme and by the time $\Delta t$ in between subsequent pulses.
Let us assume now, that we would like to generate a certain two-qudit quantum gate by applying a possibly time-dependent gate Hamiltonian $H_g(t)$ for a time $\tau_g = m\cdot t_c$, $m\in \mathbb{N}$.
Then, the total Hamiltonian is given by the sum of the Hamiltonians describing the static imperfections ($H_0$), the quantum gate ($H_g(t)$), and the decoupling pulses ($H_c(t)$),
\begin{equation}
 H(t) = H_0 + H_g(t) + H_c(t),
\end{equation}
for $t\in[0,\tau_g]$.
As in section \ref{sec:dynctrl}, we switch to the toggled frame $\tilde{U}(t) = U_c^\dagger(t) \cdot U(t)$.
As a result of the control, we obtain (in lowest order AHT) the effective total Hamiltonian
\begin{equation}\label{eq:hgaltered}
 \overline{ H }^{(0)} = \Pi_\mathcal{G} ( H_0 ) + \Pi_\mathcal{G} ( H_g ),
\end{equation}
where we assumed for simplicity that the gate Hamiltonian remains constant over the time interval $\tau_g$, and where we used the definition
\begin{equation}\label{eq:defi:pigsec21}
 \Pi_\mathcal{G}(X) = \frac{1}{n_c} \sum_{j=0}^{n_c-1} g_j^\dagger X g_j,
\end{equation}
for any operator $X$ acting on $\mathcal{H}_S$.
Hence, any gate Hamiltonian gets altered by the applied decoupling scheme. In particular, a time-independent gate Hamiltonian $H_g$ becomes $\Pi_\mathcal{G}(H_g)$.
We are now going to discuss solutions to this problem.
For the remaining section, let us assume that the elements of the control scheme $\mathcal{G} = \{ g_j \}_{j=0}^{n_c-1}$
are defined by a unitary projective representation $R$ of a group $G = \{ \mathfrak{g}_j \}_{j=0}^{n_c-1}$ acting on the system Hilbert space $\mathcal{H}_S$, i.\,e. we assume that $g_j = R(\mathfrak{g}_j)$.
We will call $G$ the underlying index group. Assuming that the elements in $\mathcal{G}$ generate a larger but finite group $\hat{G}$, we consider the ordinary irreducible representations of $\hat{G}$.
As in subsections \ref{subsec:nlsubsys} and \ref{subsec:boundedctrls} we denote the corresponding group algebra $R(\mathbb{C}G)$ by $\mathcal{A}$ and its commutant by $\mathcal{A}'$.

\subsection{Universal Computation on a Subsystem}

As discussed in subsection \ref{subsec:nlsubsys}, the Hilbert space of the quantum register decomposes with respect to the irreps $\mathcal{J}$ of $\mathcal{G}$,
\begin{equation}
 \mathcal{H}_S = \bigoplus_{\nu\in\mathcal{J}} \mathcal{H}_\nu = \bigoplus_{\nu\in\mathcal{J}} \mathcal{C}_\nu \otimes \mathcal{D}_\nu,
\end{equation}
where $\tau_\nu=\dim(\mathcal{C}_\nu)$ denotes the degeneracy and $d_\nu=\dim(\mathcal{D}_\nu)$ denotes the dimension of the irrep $\nu\in\mathcal{J}$.
Since, for any operator $X$ acting on $\mathcal{H}_S$, $\Pi_\mathcal{G}(X)$ commutes with all the group elements, it follows that $\Pi_\mathcal{G}(X)$ is in $\mathcal{A}'$.
Hence, the subsystems $\{\mathcal{D}_\nu\}_{\nu\in\mathcal{J}}$ are dynamically generated noiseless subsystems (\cite{Za00, VKL00}, subsection \ref{subsec:nlsubsys}).
In order to generate a universal set of gates acting on subsystem $\mathcal{D}_\nu$, we have to apply gate Hamiltonians which belong to the group algebra $\mathcal{A}$.
Unfortunately, according to equation \eqref{eq:hgaltered}, this is impracticable since such a Hamiltonian gets projected onto $\mathcal{A}'$.
A very elegant solution appears for the case that $\Pi_\mathcal{G}(H_0) \in \mathcal{A}' \cap \mathcal{A} = \bigoplus_{\nu\in\mathcal{J}} \lambda_\nu \mathcal{I}_\nu$, with $\lambda_\nu\in\mathbb{C}$ and $\mathcal{I}_\nu$ denoting the identity acting on $\mathcal{H}_\nu$:
In this case we might use one of the subsystems $\{\mathcal{C}_\nu\}_{\nu\in\mathcal{J}}$ as a noiseless subsystem and generate the corresponding quantum gates using a gate Hamiltonian belonging to $\mathcal{A}'$.
Any Hamiltonian belonging to $\mathcal{A}'$ remains unaffected by the action of $\Pi_\mathcal{G}$ \cite{Za00, VKL00}.
The method becomes infeasible if $\mathcal{G}$ acts irreducible on $\mathcal{H}_S$. Then, the set $\mathcal{J}$ contains only one element $\nu$ with $d_\nu=\dim(\mathcal{H}_S)$ and $\tau_\nu=1$.

In the above scenario, universal control is achieved via a set of gate Hamiltonians which commute with the decoupling pulses, and hence remain unaffected.
For instance, the hybrid decoupling and computing scheme analyzed in \cite{KhLi08} by Khodjasteh and Lidar is based on the assumption that the computational operations commute with the decoupling pulses.

\subsection{Universal Computation using Multiple Decoupling Schemes}

By using a decoupling scheme $\mathcal{G}=\{ g_j \}_{j=0}^{n_c-1}$ defined by a unitary projective representation $R$,
any time-independent gate Hamiltonian $H_g$ gets projected onto the commutant $\mathcal{A}'$ of the group algebra $\mathcal{A}$ via $\Pi_\mathcal{G}(H_g)$ (compare with \eqref{eq:hgaltered}).
Hence, the only applicable gate Hamiltonians are those which belong to $\mathcal{A}'$.
If an additional decoupling group $\tilde{\mathcal{G}}=\{ \tilde{g}_j \}_{j=0}^{\tilde{n}_c-1}$, with group algebra $\tilde{\mathcal{A}}$ and commutant $\tilde{\mathcal{A}}'$, is available, it becomes also possible to apply any gate Hamiltonian belonging to $\tilde{\mathcal{A}}'$.
Let $A\in \mathcal{A}'$ and let $B\in \tilde{\mathcal{A}}'$.
It was recognized by Viola et\,al. in \cite{VLK99}, that by applying $A$ and $B$ interchangeably, any gate $U_g=e^L$ could be created, where $L$ belongs to the Lie algebra generated by $iA$ and $iB$ under commutation.
Additional decoupling groups $\tilde{\mathcal{G}}$ might be generated by employing the following trick:
We apply the additional bang-bang pulses $P$ and $P^\dagger$ at the beginning and the end of a single $\mathcal{G}$-decoupling cycle, respectively.
As a result, the time evolution of a single \textsf{PDD} cycle is changed from
\begin{equation}
 \tilde{U}(t_c) = \exp\bigr(-i g_{n_c-1}^\dagger (H_0+H_g) g_{n_c-1} \Delta t\bigl)
\dots
\exp\bigr(-i g_1^\dagger (H_0+H_g) g_1 \Delta t\bigl)\exp\bigr(-i g_0^\dagger (H_0+H_g) g_0 \Delta t\bigl)
\end{equation}
(compare with \eqref{eq:pddmcycles}) to $P^\dagger \tilde{U}(t_c) P$, and lowest order AHT leads to $H_g \mapsto \Pi_{\tilde{\mathcal{G}}}( P^\dagger H_g P) \in \tilde{\mathcal{A}}'$ with $\tilde{\mathcal{G}} = P^\dagger \mathcal{G} P$.
The decoupling of $H_0 \mapsto \Pi_\mathcal{G}(H_0) = \lambda\cdot\mathcal{I}$ (with $\lambda\in\mathbb{R}$) remains unaffected since $\Pi_{\tilde{\mathcal{G}}}( P^\dagger H_0 P) = P^\dagger \Pi_\mathcal{G}(H_0) P = P^\dagger \lambda\mathcal{I} P = \lambda\cdot\mathcal{I}$.
Note that for $\tilde{\mathcal{A}}' \neq \mathcal{A}'$, $P$ must not be in $\mathcal{A}$.
If, in addition to $\mathcal{G}$, a large enough set of bang-bang pulses $P \notin \mathcal{A}$ is available, it might become feasible to construct a universal set of gates~\cite{VLK99}.
Again, the method becomes infeasible if $\mathcal{G}$ acts irreducible on $\mathcal{H}_S$:
Then, $\mathcal{A}' = \lambda \mathcal{I}$, with $\lambda \in\mathbb{C}$, generates only a trivial action.

\subsection{Gates via Fast Switching}

In the previous two subsections we assumed that a gate Hamiltonian $H_g$ was switched on over a period corresponding to an integer number of decoupling cycles, each of which is of length $t_c = n_c \cdot \Delta t$.
As a consequence, in lowest order AHT, $H_g$ became projected onto $\Pi_\mathcal{G}(H_g)$.
Let us now assume that we are able to switch $H_g$ on and off for shorter periods $\Delta t$,
a scenario which is called 'weak strength/fast switching' in \cite{VLK99}.
If $H_g$ is switched on only during the interval $\Delta t$ corresponding to the identity element $g_0\in\mathcal{G}$, lowest order AHT leads to
\begin{equation}
 \overline{H}^{(0)} = \Pi_\mathcal{G}(H_0) + \frac{1}{\vert\mathcal{G}\vert} H_g.
\end{equation}
Now any quantum gate $U_g=\exp( -i H_g \cdot m t_c)$ with $m\in\mathbb{N}$ could be generated by repeating such a cycle an integer number of times.
If we are also able to switch on the Hamiltonians $g_j H_g g^\dagger_j$ during the $j$-th part of the cycle (for $j=1,\dots,n_c-1$), the factor $1 / \vert\mathcal{G}\vert$ in the above equation vanishes \cite{VLK99}.
Note that this method works even if the control scheme $\mathcal{G}$ is not related to an underlying index group.

\subsection{Dynamically Corrected Gates}\label{subsec:dcgs} %
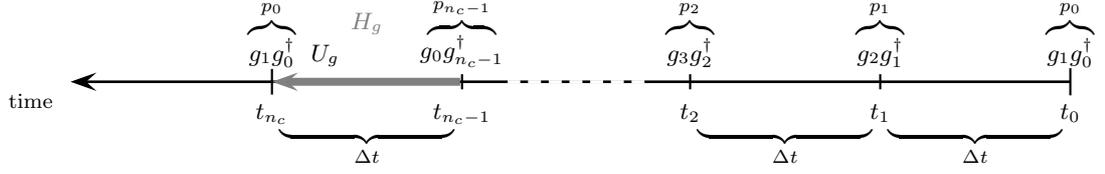
\begin{figure}\centering
\scalebox{1}{
\begin{pspicture}(1.1,0.4)(15.3,-2.2)
\psline[linewidth=1pt,arrowsize=7pt]{<-}(1.85,-1.)(15,-1.)
\rput[t](1.33,-1.15){\footnotesize time}
\psline(15,-1.18)(15,-0.82)\rput[t](15,-1.3){\small $t_0$}\rput[b](15,-0.8){\small $g_1 g_0^\dagger$}
\rput[b](15,-0.4){$\overbrace{\makebox(.4,0){}}^{p_0}$}
\rput[t](13.75,-1.6){$\underbrace{\makebox(2.3,0){}}_{\Delta t}$}
\psline(12.5,-1.1)(12.5,-0.9)\rput[t](12.5,-1.3){\small $t_1$}\rput[b](12.5,-0.8){\small $g_2 g_1^\dagger$}
\rput[b](12.5,-0.4){$\overbrace{\makebox(.4,0){}}^{p_1}$}
\rput[t](11.25,-1.6){$\underbrace{\makebox(2.3,0){}}_{\Delta t}$}
\psline(10,-1.1)(10,-0.9)\rput[t](10,-1.3){\small $t_2$}\rput[b](10,-0.8){\small $g_3 g_2^\dagger$}
\rput[b](10,-0.4){$\overbrace{\makebox(.4,0){}}^{p_2}$}
\psline[linewidth=1.2pt,linestyle=dashed,linecolor=white](9.4,-1.)(7.6,-1.)
\psline(7,-1.1)(7,-0.9)\rput[t](7,-1.3){\small $t_{n_c-1}$}\rput[b](7,-0.8){\small $g_0 g_{n_c-1}^\dagger$}
\rput[b](7,-0.4){$\overbrace{\makebox(.9,0){}}^{p_{n_c-1}}$}
\rput[t](5.75,-1.6){$\underbrace{\makebox(2.3,0){}}_{\Delta t}$}
\rput[B](5.75,-0.3){\gray \small $H_g$}\rput[b](5.2,-0.8){\small $U_g$}
\psline[linewidth=3pt,arrowsize=7pt, linecolor=gray]{->}(6.975,-1.)(4.5,-1.)
\psline(4.5,-1.17)(4.5,-0.83)\rput[t](4.5,-1.3){\small $t_{n_c}$}\rput[b](4.5,-0.8){\small $g_1 g_0^\dagger$}
\rput[b](4.5,-0.4){$\overbrace{\makebox(.4,0){}}^{p_0}$}
\end{pspicture}}%
\caption[DCGates (a)]{Schematic representation of a \textsf{PDD} cycle, which tries to implement a quantum gate $U_g=\mathcal{T} \exp\bigl(-i \int_0^{\tau_g} H_g(t') dt' \bigr)$ with $\tau_g = \Delta t$ by switching on the gate Hamiltonian $H_g$ during the period where the control visits the identity element $g_0$ of the control scheme $\{g_j\}_{j=0}^{n_c-1}$.\label{fig::dcgA}}
\end{figure}

In this subsection we present an idea due to Khodjasteh and Viola \cite{KhVi08}, who proposed to combine decoupling and computation by constructing dynamically corrected gates (DCGs)\footnote{In \cite{KhVi08} the idea of dynamically corrected gates was presented in the context of Eulerian decoupling using bounded controls; here we consider the simpler case of instantaneous decoupling pulses.}.
We consider a decoupling scheme $\mathcal{G} = \{ g_j \}_{j=0}^{n_c-1}$ of length $n_c$, where $g_0=\mathcal{I}$ denotes the identity element.
The basic \textsf{PDD} cycle of length $t_c=n_c\cdot \Delta t$ is constructed by traversing the elements of the decoupling scheme in the order $g_1,g_2,\dots,g_{n_c-1},g_0$, i.\,e. we close the cycle by visiting the identity element.
If the gates implementing a quantum computation could be generated instantaneously, they could simply be executed in between subsequent cycles without introducing any errors.
Instead, we assume that a quantum gate $U_g$ has to be generated by switching on a time-dependent gate Hamiltonian $H_g(t)$ for a time $\tau_g=\Delta t$: $U_g \equiv U_g(\tau_g)$ with $U_g(t) = \mathcal{T} \exp\bigl(-i \int_0^t H_g(t') dt' \bigr)$ for $t\in[0,\tau_g]$.
In order to combine a decoupling cycle with the generation of a quantum gate $U_g$, we apply the corresponding gate Hamiltonian during the last part of the cycle, in which the control visits the identity element. A schematic representation is given in figure \ref{fig::dcgA}.
As a consequence, the time evolution of such a cycle is given by
\begin{equation}
 \tilde{U}(t_c) =
U_g \cdot \mathcal{T}\exp\Bigr(-i \int_0^{\Delta t}\!\! U_g^\dagger(t') H_0 U_g(t') dt'\Bigl) \cdot
\exp\bigr(-i g_{n_c-1}^\dagger H_0 g_{n_c-1} \Delta t\bigl)
\dots
\exp\bigr(-i g_1^\dagger H_0 g_1 \Delta t\bigl),
\end{equation}
and in lowest order AHT the average Hamiltonian of such a cycle is given by
\begin{equation}
 \overline{H}^{(0)} = \frac{1}{n_c\Delta t}\biggl( g_0^\dagger \underbrace{\int_0^{\Delta t}\!\! U_g^\dagger(t') H_0 U_g(t') dt'}_{\Phi_g}  g_0 + \sum_{j=1}^{n_c-1} g_j^\dagger H_0 g_j \Delta t \biggr).
\end{equation}
Because of the lowest order gate error $\Phi_g$, we do not obtain the usual result $\overline{H}^{(0)} = \Pi_\mathcal{G}(H_0)$.
The idea of Khodjasteh and Viola \cite{KhVi08} is now to produce the same error during all the non-identity steps of the decoupling cycle. As a result, the lowest order average Hamiltonian of such a cycle would be given by
\begin{equation}\label{eq:deccondiDCGs}
 \overline{H}^{(0)} = \frac{1}{n_c\Delta t} \sum_{j=0}^{n_c-1} g_j^\dagger \Phi_g g_j
 = \Pi_\mathcal{G}\Bigl(  \frac{1}{\Delta t}\int_0^{\Delta t}\!\! U_g^\dagger(t') H_0 U_g(t') dt' \Bigr).
\end{equation}
The above expression leads to a trivial time evolution, if we demand a decoupling scheme which satisfies $\Pi_\mathcal{G}\bigl( \Phi_g ) = \lambda\cdot\mathcal{I}$, with $\lambda\in\mathbb{C}$ (this point will be further discussed in subsection \ref{subsec:eulerdcg} dealing with Euler-DCGs).

We close this subsection by showing how these additional errors could be generated.
Khodjasteh and Viola \cite{KhVi08} proposed the following trick:
Let us assume that the quantum gate $U_g \equiv U_g(\tau_g) = \exp( -i H_g\tau_g )$ is generated using a fixed gate Hamiltonian $H_g$ whose strength is modulated by a time-dependent pulse shape $f(t)$ such that $\int_0^1 f(t') dt' = 1$:
\begin{equation}
  U_g(t) = \exp\Bigl( -i H_g\tau_g \cdot \frac{1}{\tau_g} \int_0^t f(t'/\tau_g) dt' \Bigr).
\end{equation}
Assuming that $f(t)=0$ for $t\notin[0,1]$ we could generate an identity gate $\mathcal{I} \equiv U_I(\tau_g)$ by using the following pulse shape:
\begin{equation}\label{eq:UI}
  U_I(t) = \exp\Bigl( -i H_g\tau_g \cdot \frac{2}{\tau_g} \int_0^t \bigl( f(2t'/\tau_g) - f(2-2t'/\tau_g)\bigr) dt' \Bigr).
\end{equation}
Calculating the lowest order error $\Phi_I$ of such an identity gate,
\begin{equation}
\Phi_I = \int_0^{\Delta t}\!\! U_I^\dagger(t') H_0 U_I(t') dt',
\end{equation}
is straightforward and shows that indeed $\Phi_I = \Phi_g$.
Hence, in order to generate the additional errors, we have to implement these identity gates by switching on the Hamiltonian in the exponent of \eqref{eq:UI} during the first $n_c-1$ steps of the decoupling cycle.

\section{Pauli Random Error Correction}\label{sec:parec}

The methods for quantum computation in the presence of decoupling, which have been discussed in the preceding section, all have some drawbacks:
The first two proposals, subsystem-encoding and multiple decoupling schemes, become infeasible if the decoupling group acts irreducible on the system Hilbert space.
The fast-switching method demands the ability to switch a gate Hamiltonian on and off quickly, and in addition, weakens the interaction strength of any applied gate Hamiltonian by a factor in inverse proportion to the size of the decoupling set.
Eventually, dynamically corrected quantum gates demand a decoupling set which satisfies the decoupling condition for perturbations which have been twisted by the gate errors \eqref{eq:deccondiDCGs}, and in addition, demands the generation of additional identity-gates mirroring the gate errors.

We are now going to present a method which uses naive random decoupling (\textsf{NRD}, subsection \ref{subsec:randstrat}) to stabilize arbitrary quantum algorithms against static imperfections
(like inter-qudit couplings, for instance) in a rather simple way.
While any method based on deterministic decoupling strategies,
like for instance the method of dynamically corrected gates (\cite{KhVi08}, subsection \ref{subsec:dcgs}),
is only allowed to implement quantum gates in between completed decoupling cycles, random decoupling allows the quantum gates to be implemented in between subsequent decoupling pulses.
It will be shown that, as it is the case for \textsf{NRD} in the absence of any computation,
the fidelity decay caused by static imperfections will be slowed down to a linear-in-time one.
Our method was proposed for the first time in the author's diploma thesis \cite{DiplKern} and subsequently in \cite{parec}, where the acronym Pauli random error correction (\textsf{PAREC}) was devised.
In these publications, all pulses and gates were assumed to be of the bang-bang kind, and only numerical evidence of the resulting linear-in-time decay was provided.
We derived a formula for the resulting fidelity decay in \cite{GKAJ08}.
In this section, we consider the more general case of bounded controls generating finite decoupling pulses of duration $\tau_p$ and finite quantum gates of duration $\tau_g$.

We start with a detailed description of the \textsf{PAREC} method in subsection \ref{subsec:parec}.
To evaluate the stabilizing properties of \textsf{PAREC}, we have to compare a stabilized computation with an unprotected one.
Before we proceed with an analysis of the fidelity decay of an unprotected quantum computation in subsection \ref{subsec:par_ohnekorrektur}, we derive a general second order expansion of the entanglement fidelity of a perturbed quantum algorithm in subsection \ref{subsec:parec_expans}.
The fidelity decay of a stabilized computation is analyzed in subsection \ref{subsec:parec_ana}.
Eventually, in subsection \ref{subsec:par_ex}, we present the results of a numerical simulation of a protected and an unprotected quantum algorithm, which allow us to put the derived fidelity formulas to the test.

\subsection{Implementation}\label{subsec:parec}
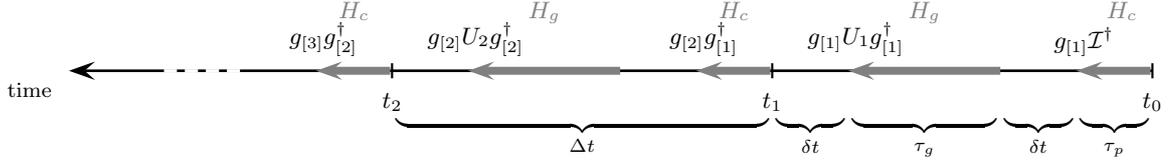
\begin{figure}\centering
\scalebox{1}{
\begin{pspicture}(-0.3,0.4)(15.3,-2.2)
\psline[linewidth=1pt,arrowsize=7pt]{<-}(0.75,-1.)(15,-1.)
\rput[t](0.23,-1.15){\footnotesize time}
\psline(15,-1.1)(15,-0.9)\rput[t](15,-1.3){\small $t_0$}
\rput[B](14.6,-0.3){\gray \small $H_c$}\rput[b](14.1,-0.8){\small $g_{[1]}\mathcal{I}^\dagger$}
\psline[linewidth=3pt,arrowsize=7pt, linecolor=gray]{->}(14.975,-1.)(14,-1.)
\rput[t](14.5,-1.6){$\underbrace{\makebox(0.9,0){}}_{\tau_p}$}
\rput[t](13.5,-1.6){$\underbrace{\makebox(0.9,0){}}_{\delta t}$}
\rput[t](10.5,-1.6){$\underbrace{\makebox(0.9,0){}}_{\delta t}$}
\rput[B](12,-0.3){\gray \small $H_g$}\rput[b](11.1,-0.8){\small $g_{[1]}U_1g_{[1]}^\dagger$}
\psline[linewidth=3pt,arrowsize=7pt, linecolor=gray]{->}(13,-1.)(11,-1.)
\rput[t](12,-1.6){$\underbrace{\makebox(1.9,0){}}_{\tau_g}$}
\psline(10,-1.1)(10,-0.9)\rput[t](10,-1.3){\small $t_1$}
\rput[B](9.5,-0.3){\gray \small $H_c$}\rput[b](9.1,-0.8){\small $g_{[2]}g_{[1]}^\dagger$}
\psline[linewidth=3pt,arrowsize=7pt, linecolor=gray]{->}(9.975,-1.)(9,-1.)
\rput[B](7,-0.3){\gray \small $H_g$}\rput[b](6.1,-0.8){\small $g_{[2]}U_2g_{[2]}^\dagger$}
\psline[linewidth=3pt,arrowsize=7pt, linecolor=gray]{->}(8,-1.)(6,-1.)
\rput[t](7.5,-1.6){$\underbrace{\makebox(4.9,0){}}_{\Delta t}$}
\psline(5,-1.1)(5,-0.9)\rput[t](5,-1.3){\small $t_2$}
\rput[B](4.5,-0.3){\gray \small $H_c$}\rput[b](4.1,-0.8){\small $g_{[3]}g_{[2]}^\dagger$}
\psline[linewidth=3pt,arrowsize=7pt, linecolor=gray]{->}(4.975,-1.)(4,-1.)
\psline[linewidth=1.2pt,linestyle=dashed,linecolor=white](3,-1.)(2,-1.)
\end{pspicture}}%
\caption[PAREC]{Schematic representation of the \textsf{PAREC} method. The gate sequence of the original quantum algorithm $U_{QA} = \dots \cdot U_3\cdot U_2\cdot U_1$ is replaced by an alternating sequence of randomly chosen decoupling pulses $g_{[i+1]} g^\dagger_{[i]}$ of duration $\tau_p$ generated by the local control Hamiltonian $H_c(t)$, and twisted quantum gates $g_{[i]} U_i g_{[i]}^\dagger$ of duration $\tau_g$ generated by a gate Hamiltonian $H_g(t)$.\label{fig::parecscheme}}
\end{figure}

Let us consider $n_a \in \mathbb{N}$ iterations of a quantum algorithm given by the ideal unitary transformation
$U_{QA} = U_{n_g}\cdots U_3\cdot U_2\cdot U_1$, where $U_i$, $i=1,\dots,n_g$, denotes an elementary one- or two-qudit quantum gate.
In the \textsf{PAREC} method before each quantum gate $U_i$ of the $\tau$-th iteration ($\tau=1,...,n_a$) of the unitary transformation $U_{QA}$, a unitary of the form $g_{[\tau,i]} g_{[\tau,i-1]}^\dagger$ is applied.
Here, the unitaries $g_{[\tau,i]}$ (with $g_{[\tau,0]} = g_{[\tau-1,n_g]}$ and $g_{[1,0]} = \mathcal{I}$) are drawn at random from a decoupling set $\mathcal{G} = \{ g_j \}_{j=0}^{n_c-1}$, i.\,e. the index $[\tau,i]$ is in $\{0,1,\dots,n_c-1\}$ for all $\tau=1,...,n_a$ and all $i=1,\dots,n_g$.
Simultaneously the changes on the quantum algorithm due to these random unitary gates have to be compensated by replacing each elementary quantum gate $U_i$ of the $\tau$-th iteration of the original algorithm by
$U^{(\tau)}_i = g_{[\tau,i]}  U_i  g_{[\tau,i]}^\dagger$.
The locality of the control assures that any quantum gate acting on $m$ qudits remains an $m$-qudit gate:
With $g_{[\tau,i]} = g_{[\tau,i],1} \otimes g_{[\tau,i],2} \otimes \dots \otimes g_{[\tau,i],n} \in \textsf{U}_q^{\otimes n}$
it follows that
\begin{align}
U^{(\tau)}_i &= g_{[\tau,i]}  \cdot U_i  \cdot g_{[\tau,i]}^\dagger \nonumber \\
   &= \bigl( g_{[\tau,i],k_1}\otimes g_{[\tau,i],k_2} \cdot  U_i  \cdot g_{[\tau,i],k_1}^\dagger \otimes g_{[\tau,i],k_2}^\dagger \bigr) \otimes \mathcal{I}_{\{1,\dots,n\}\setminus \{k_1,k_2\}},
\end{align}
for any $m=2$ qudit gate $U_i$ acting on qudits $k_1$ and $k_2$, for instance.
Furthermore, after the last quantum gate $U^{(n_a)}_{n_g}$ a final unitary gate $g_{[n_a,n_g]}^\dagger$ is applied.
As a result each iteration of a unitary transformation $U_{QA}$ is replaced by $2n_g$ unitary quantum gates so that after $n_a$ iterations one obtains the result
\begin{equation}\label{eq::uparec}
\begin{split}
U_{QA}^{n_a} &= U_{QA}  \dots U_{QA} \cdot U_{QA} \\
&= g_{[n_a,n_g]}^\dagger \bigl(
U^{(n_a)}_{n_g} \cdot \hdots \cdot g_{[n_a,3]}g_{[n_a,2]}^\dagger
\cdot U^{(n_a)}_2 \cdot g_{[n_a,2]}g_{[n_a,1]}^\dagger \cdot U^{(n_a)}_1 \cdot g_{[n_a,1]}g_{[n_a-1,n_g]}^\dagger
\bigr) \cdot \\
&\qquad\qquad \dots
\bigl( U^{(2)}_{n_g} \cdot \hdots \cdot g_{[2,3]}g_{[2,2]}^\dagger \cdot U^{(2)}_2 \cdot g_{[2,2]}g_{[2,1]}^\dagger \cdot U^{(2)}_1 \cdot g_{[2,1]}g_{[1,n_g]}^\dagger \bigr) \cdot\\
&\qquad\qquad\qquad\qquad
\bigl( U^{(1)}_{n_g} \cdot \hdots \cdot g_{[1,3]}g_{[1,2]}^\dagger \cdot U^{(1)}_2 \cdot g_{[1,2]}g_{[1,1]}^\dagger \cdot U^{(1)}_1 \cdot g_{[1,1]}\mathcal{I}^\dagger \bigr) \\
&\equiv g_{[n_a,n_g]}^\dagger \bigl( V^{(n_a)}_{2n_g} \dots V^{(n_a)}_2 V^{(n_a)}_1 \bigr) \dots
\bigl(V^{(2)}_{2n_g} \dots V^{(2)}_2 V^{(2)}_1\bigr)
\bigl(V^{(1)}_{2n_g} \dots V^{(1)}_2 V^{(1)}_1\bigr)
\end{split}
\end{equation}
with $V^{(\tau)}_{2k} = U^{(\tau)}_k$ and $V^{(\tau)}_{2k-1} = g_{[\tau,k]}g_{[\tau,k-1]}^\dagger$  for $k=1,...,n_g$.
A particular \textsf{PAREC} implementation of the quantum Fourier transform (QFT) is schematically represented in figure \ref{fig::qft} for the special case of $n=4$ qubits and $\mathcal{G} = \mathcal{P}^n_2$ given by the set of Pauli operators \eqref{eq:pauliops:gendecscheme}.
Definitely, this random application of decoupling elements together with the associated change of elementary quantum gates does not affect any quantum algorithm.

In this section, we consider the general case of bounded controls, i.\,e. we assume that the decoupling pulses $V^{(\tau)}_{2k-1} \equiv V^{(\tau)}_{2k-1}(\tau_p)$ are generated by switching on a local control Hamiltonian $H_c$ for a time $\tau_p$,
\begin{equation}
 V^{(\tau)}_{2k-1}(t) = \mathcal{T} \exp\Bigl( -i \int_0^t H_c(t') dt' \Bigr), \text{ for } t\in[0,\tau_p],
\end{equation}
and the quantum gates $V^{(\tau)}_{2k} \equiv V^{(\tau)}_{2k}(\tau_g)$ are generated by switching on a gate Hamiltonian $H_g$ for a time $\tau_g$,
\begin{equation}
 V^{(\tau)}_{2k}(t) = \mathcal{T} \exp\Bigl( -i \int_0^t H_g(t') dt' \Bigr), \text{ for } t\in[0,\tau_g].
\end{equation}
The situation is depicted in figure \ref{fig::parecscheme}, where we consider a single iteration of $U_{QA}$.
The time in between subsequent decoupling pulses is denoted as usual as $\Delta t$.
As a consequence, the time $\delta t$ of free evolution in between gates and pulses is given by $\delta t = (\Delta t -\tau_g-\tau_p)/2$.

While equation \eqref{eq::uparec} denotes the ideal time evolution of an iterated quantum algorithm $U_{QA}^{n_a}$ employing the \textsf{PAREC} method, the total time evolution in the presence of static imperfections described by a Hamiltonian $H_0$ is given by
\begin{multline}\label{eq:parec_total}
 U_{QA\ \text{perturbed}}^{n_a} =
\bigl( G_{n_g}^{(n_a)} \dots G_2^{(n_a)} G_1^{(n_a)} \bigr) \cdot \hdots \cdot
\bigl( G_{n_g}^{(2)} \dots G_2^{(2)} G_1^{(2)} \bigr) \cdot \\
\bigl( G_{n_g}^{(1)} \dots G_2^{(1)} G_1^{(1)} \bigr) \cdot \mathcal{T} \exp\Bigl(-i \int_0^{\tau_p} V_1^{(1)\dagger}(t') H_0 V_1^{(1)}(t') dt' \Bigr),
\end{multline}
where we used the abbreviations
\begin{multline}
G_k^{(\tau)} = g^\dagger_{[\tau,k]} \cdot
\mathcal{T} \exp\Bigl(-i \int_0^{\tau_p} V_{2k+1}^{(\tau)\dagger}(t') H_0 V_{2k+1}^{(\tau)}(t') dt'\Bigr) \cdot
\exp(-i H_0\delta t) \cdot \\
\underbrace{g_{[\tau,k]} \cdot U_k \cdot g^\dagger_{[\tau,k]}}_{V_{2k}^{(\tau)}(\tau_g)} \cdot
\mathcal{T} \exp\Bigl(-i \int_0^{\tau_g} V_{2k}^{(\tau)\dagger}(t') H_0 V_{2k}^{(\tau)}(t') dt'\Bigr) \cdot
\exp(-i H_0\delta t) \cdot g_{[\tau,k]}.
\end{multline}
In other words, to obtain the total time evolution, the quantum gate $U_k$, $k=1,\dots,n_g$, in the $\tau$-th iteration of the ideal quantum algorithm $U_{QA}$ is replaced by the gate
\begin{equation}\label{eq:Gktau}
G_k^{(\tau)} =
g^\dagger_{[\tau,k]} \exp( - i H_{kl}^\tau ) g_{[\tau,k]} \cdot  U_k \cdot
g^\dagger_{[\tau,k]} \exp( - i H_{kr}^\tau ) g_{[\tau,k]},
\end{equation}
where in lowest order AHT the average Hamiltonians $H_{kl}^\tau$ and $H_{kr}^\tau$ are given by
\begin{subequations}\label{eq:Hkptau}
\begin{align}
H_{kl}^\tau &= \int_0^{\tau_p} V_{2k+1}^{(\tau)\dagger}(t') H_0 V_{2k+1}^{(\tau)}(t') dt' + H_0\delta t \label{eq:Hkltau}\\ \text{ and }
H_{kr}^\tau &= \int_0^{\tau_g} V_{2k}^{(\tau)\dagger}(t') H_0 V_{2k}^{(\tau)}(t') dt' + H_0\delta t, \label{eq:Hkrtau}
\end{align}
\end{subequations}
respectively. If the decoupling pulses and the quantum gates are applied in the bang-bang limit ($\tau_p\rightarrow 0$, $\tau_g\rightarrow 0$), we obtain the simpler and exact expressions $H_{kl}^\tau = H_{kr}^\tau = H_0\Delta t/2$.

\subsection{Expansion of the Entanglement Fidelity}\label{subsec:parec_expans}

In the following we are mainly interested in the entanglement fidelity comparing a unitary operation $U$ and its slightly perturbed version $U_\delta$.
Thus the relevant quantum operation $\mathcal{E}$ involves a single unitary Kraus operator $K$ which is given by $K = U^\dagger \cdot U_\delta$.
On the basis of \eqref{eq:fidcomp} in the case of high dimensional quantum systems
the average fidelity is approximately given by the entanglement fidelity \eqref{eq:FeUU}
\begin{equation}
F_e( \mathcal{E} ) = \left \vert \frac{1}{d} \tr\bigl( U^\dagger U_\delta \bigr) \right \vert^2
\end{equation}
which is determined by the absolute square of a fidelity amplitude
\begin{equation}\label{eq:fidamp}
 A_e = \frac{1}{d} \tr \bigl( U^\dagger U_\delta \bigr).
\end{equation}

In this subsection a perturbative short-time approximation of the fidelity amplitude
is derived, which will be used at several occasions in the current and the following section.
Let us consider $n_a$ iterations of a quantum algorithm given by the ideal unitary transformation
$U_{QA} = U_{n_g}\cdots U_3\cdot U_2\cdot U_1$, i.\,e. we set $U^\dagger = U_{QA}^{-n_a}$ in \eqref{eq:fidamp}.
We make the general assumption that the ideal time evolution is perturbed, where the $j$-th quantum gate of the $\tau$-th iteration of $U_{QA}$ is replaced by the perturbed unitary quantum gate
\begin{equation}\label{eq:replaceUj}
U_j \mapsto  \exp( - i \delta\!H_{jl}^\tau )  U_j  \exp( - i \delta\!H_{jr}^\tau ).
\end{equation}
The index $\tau$ in \eqref{eq:replaceUj} takes into account that perturbations may be different
in successive iterations of the unitary transformation $U_{QA}$.
\begin{lem}\label{lem:2orderfeqa}
A second order expansion of the fidelity amplitude $A_e$ \eqref{eq:fidamp} after $n_a$ iterations of the perturbed quantum algorithm 
with respect to $\delta\!H_{jl}^\tau$ and $\delta\!H_{jr}^\tau$ is given by
\begin{equation}\label{eq::secorderQA}
\begin{split}
A_e(n_a) &= 1 - \sum_{p=l,r}\sum_{\tau=1}^{n_a}\sum_{j=1}^{n_g} \frac{1}{d}\Bigl[
i\tr \bigl( \delta\!H_{jp}^\tau \bigr)  + \frac{1}{2}\tr \bigl( (\delta\!H_{jp}^\tau)^2 \bigr) 
\Bigr]\\
&\mathrel{\phantom{=}}-\sum_{\tau=1}^{n_a}\sum_{j=2}^{n_g}\sum_{k=1}^{j-1}\frac{1}{d}\Bigl[
 \tr \bigl( \delta\!H_{jl}^\tau(j)  \delta\!H_{kl}^\tau(k) \bigr)
 +\tr \bigl( \delta\!H_{jl}^\tau(j)  \delta\!H_{kr}^\tau(k-1) \bigr)\\
&\qquad\qquad\qquad +\tr \bigl( \delta\!H_{jr}^\tau(j-1)  \delta\!H_{kl}^\tau(k) \bigr)
  +\tr \bigl( \delta\!H_{jr}^\tau(j-1)  \delta\!H_{kr}^\tau(k-1) \bigr)  \Bigr]\\
&\mathrel{\phantom{=}}-\sum_{\tau=1}^{n_a}\sum_{j=1}^{n_g}\frac{1}{d} \tr \bigl( U^\dagger_j \delta\!H_{jl}^\tau U_j \delta\!H_{jr}^\tau \bigr)\\
&\mathrel{\phantom{=}}-\sum_{\tau_1=2}^{n_a}\sum_{\tau_2=1}^{\tau_1-1}\sum_{j,k=1}^{n_g}\frac{1}{d}\Bigl[
  \tr \bigl( U^{\tau_2-\tau_1} \delta\!H_{jl}^{\tau_1}(j) U^{\tau_1-\tau_2}
\delta\!H_{kl}^{\tau_2}(k) \bigr) \\
&\qquad\qquad
 +\tr \bigl( U^{\tau_2-\tau_1} \delta\!H_{jl}^{\tau_1}(j) U^{\tau_1-\tau_2} \delta\!H_{kr}^{\tau_2}(k-1) \bigr)
 +\tr \bigl( U^{\tau_2-\tau_1} \delta\!H_{jr}^{\tau_1}(j-1) U^{\tau_1-\tau_2} \delta\!H_{kl}^{\tau_2}(k) \bigr) \\
&\qquad\qquad
 +\tr \bigl( U^{\tau_2-\tau_1} \delta\!H_{jr}^{\tau_1}(j-1) U^{\tau_1-\tau_2} \delta\!H_{kr}^{\tau_2}(k-1) \bigr) \Bigr] + \mathcal{O}\bigl( (\delta\!H)^3 \bigr),
\end{split}
\end{equation}
with the abbreviation
\begin{equation}
\delta\!H_{jp}^\tau (i) = U_1^\dagger U_2^\dagger \dots U_i^\dagger \cdot \delta\!H_{jp}^\tau \cdot U_i \dots U_2 U_1
\equiv  U_{1\dots i}^\dagger  \, \delta\!H_{jp}^\tau \, U_{i\dots 1}.
\end{equation}
The terms linear in the perturbing Hamiltonians $\delta\!H_{jp}^\tau$ vanish if all Hamiltonians involved are traceless. %
\end{lem}
\begin{proof}
To obtain the expansion, all terms of the form $\exp( - i \delta\!H_{jl}^\tau )$ and $ \exp( - i \delta\!H_{jr}^\tau )$ are expanded as $ \exp( - i \delta\!H_{jr}^\tau ) = \mathcal{I} - i \delta\!H_{jr}^\tau - \frac{1}{2} \bigl( \delta\!H_{jr}^\tau \bigr)^2 + \dots $.
\end{proof}
\begin{rem}
Note that all the terms of \eqref{eq::secorderQA} involving $\tr \bigl( \cdot \bigr)$ terms
are real valued so that up to second order the fidelity $F_e(n_a) = \vert A_e(n_a) \vert^2$ is simply obtained
by multiplying all these terms of $A_e(n_a)$ with a factor of magnitude two.
\end{rem}
\subsection{Fidelity Decay of Unprotected Computations}\label{subsec:par_ohnekorrektur}

Before we are going to derive a formula for the entanglement fidelity of a quantum computation in the presence of static imperfections which is protected by the \textsf{PAREC} method, we have to examine the corresponding fidelity decay of an unprotected computation.
Typically, the fundamental unitary transformation $U_{QA}$ constituting a quantum algorithm can be decomposed into a sequence of $n_g$ elementary one- and two-qudit quantum gates, i.e.
\begin{equation}\label{eq:Uqa}
U_{QA} = U_{n_g} \cdot \dots \cdot U_3 \cdot U_2 \cdot U_1.
\end{equation}
Let us assume in our subsequent discussion that the quantum algorithm under consideration involves $n_a$ iterations of such a fundamental unitary transformation $U_{QA}$.
Such quantum algorithms appear in the context of search algorithms, for example \cite{grov1}.
Furthermore, let us focus our attention on the case of static imperfection in which the perturbing influence on such a quantum algorithm arises from a fixed and time-independent Hamiltonian coupling $H_0$ between the qudits constituting the quantum information processor.
Without loss in generality, $H_0$ is taken to be traceless throughout the remaining section.
We assume that an elementary quantum gate $U_g$ is generated by switching on a possibly time-dependent gate Hamiltonian $H_g$ for a time $\tau_g$, i.\,e. we have $U_g \equiv U_g(\tau_g)$ with
\begin{equation}
 U_g(t) = \mathcal{T} \exp\bigl(-i \int_0^t H_g(t') dt' \bigr)
\end{equation}
for $t\in[0,\tau_g]$.
Instead, because of the imperfections, after the time $\tau_g$ we obtain the perturbed evolution
\begin{equation}
 U_g' = \mathcal{T} \exp\bigl(-i \int_0^{\tau_g} (H_g(t')+H_0) dt' \bigr)
 = U_g \cdot \mathcal{T} \exp\bigl(-i \int_0^{\tau_g} U_g^\dagger(t')H_0 U_g(t') dt' \bigr)
\end{equation}
Let us assume in addition, that subsequent quantum gates are performed after time intervals of duration $\Delta t \geq \tau_g$, i.\,e. in between subsequent gates there is also a period $\Delta t-\tau_g$ of free evolution during which the inter-qudit couplings perturb the quantum algorithm.
Hence, in order to describe the perturbed quantum algorithm, we replace each elementary quantum gate $U_j$ in \eqref{eq:Uqa} by
\begin{align}
 U_j &\mapsto U_j \cdot \mathcal{T} \exp\bigl(-i \int_0^{\tau_g} U_j^\dagger(t')H_0 U_j(t') dt' \bigr) \cdot \exp\bigl(-i H_0 (\Delta t-\tau_g) \bigr), \\
&\equiv U_j \cdot \exp\bigl(-i \delta\!H_j \bigr), \label{eq:efperturbHjr}
\end{align}
where (in lowest order AHT) the Hamiltonian $\delta\!H_j$ is given by
\begin{equation}
 \delta\!H_j = \overline{H}^{(0)} \Delta t = \int_0^{\tau_g} U_j^\dagger(t')H_0 U_j(t') dt' + H_0 \cdot (\Delta t-\tau_g).
\end{equation}
The total time $T$ taken by the $n_a$ iterations of the quantum algorithm $U_{QA}$ is $T = n_a \cdot  n_g \Delta t$.
Equation \eqref{eq:efperturbHjr} allows us to use the second order expansion of the fidelity amplitude which was derived in the preceding subsection:
By setting $\delta\!H_{jl}^\tau = 0$ and $\delta\!H_{jr}^\tau = \delta\!H_j$, equation \eqref{eq::secorderQA} reduces to
\begin{multline}\label{eq::fstat}
F_e(n_a) = \vert A(n_a) \vert^2 = 1 - n_a \sum_{j,k=1}^{n_g} \frac{1}{d} \tr\bigl(
U_{1\dots j-1}^\dagger  \delta\!H_j  U_{j-1\dots 1} \cdot
U_{1\dots k-1}^\dagger  \delta\!H_k  U_{k-1\dots 1} \bigr) \\
 -2 \sum_{\tau=1}^{n_a-1} (n_a-\tau) \sum_{j,k=1}^{n_g} \frac{1}{d} \tr \bigl(
U_{QA}^{-\tau} \cdot U_{1\dots j-1}^\dagger  \delta\!H_j  U_{j-1\dots 1} \cdot
U_{QA}^\tau \cdot U_{1\dots k-1}^\dagger  \delta\!H_k  U_{k-1\dots 1} \bigr) + \mathcal{O}\bigl( H_0^3 \bigr).
\end{multline}
Here, the first term in the sum of \eqref{eq::fstat} describes the influence of perturbations occurring in the same iteration $\tau\in\{1,\dots,n_a\}$ and the second double sum describes their influence in different iterations.

Let us switch now to the simpler scenario of instantaneously applied gates.
By letting $\tau_g \rightarrow 0$, we find that the effective perturbation $\delta\!H_j = H_0 \Delta t$ becomes the same for all quantum gates.
In this case, the short-time behavior of the entanglement fidelity $F_e(n_a)$ has been studied in detail by Frahm et\,al. \cite{shep144}.
In particular, these authors demonstrated that whenever an ideal unitary transformation of a quantum map $U_{QA}$ can be modeled by a random matrix after $n_a$ iterations
the corresponding decay of the entanglement fidelity is given by
\begin{equation}\label{eq::frahm}
F_e^\textsf{QMap} (n_a) = 1 - \frac{n_a}{t_a} - \frac{2}{d \sigma} \frac{n_a^2}{t_a} + \mathcal{O}\bigl( H_0^3 \bigr),
\end{equation}
where $\sigma$ denotes the relative fraction of the chaotic component of the phase space of this map and $t_a$ is defined by
\begin{equation}
 \frac{1}{t_a} = \sum_{j,k=1}^{n_g} \frac{1}{d} \tr\bigl(
U_{1\dots j-1}^\dagger  H_0  U_{j-1\dots 1} \cdot
U_{1\dots k-1}^\dagger  H_0  U_{k-1\dots 1} \bigr) \Delta t^2 = \alpha \cdot n_g^2 \frac{1}{d}\tr( H_0^2 ) \Delta t^2,
\end{equation}
with $\alpha \leq 1$.
Furthermore, numerical studies indicate that the behavior of higher order terms is such that the fidelity decay becomes approximately  exponential, i.\,e.
\begin{equation}\label{eq:FeappQMap}
F_{e\ \text{app}}^\textsf{QMap}(n_a) = \exp \Bigl(
- \frac{n_a}{t_a} - \frac{2}{d \sigma} \frac{n_a^2}{t_a}
 \Bigr).
\end{equation}
While this formula was derived considering instantaneously applied quantum gates ($\tau_g=0$), it should remain valid for finite $\tau_g\in[0,\Delta t]$ as well.
The fidelity in the above expression has to be compared with the fidelity of a quantum memory after the time $T = n_a \cdot n_g \Delta t$, which was derived in subsection \ref{subsec:performance}:
\begin{equation}
 F_{e\ \text{app}}^\textsf{none}(T = n_a \cdot n_g \Delta t ) =
 \exp\Bigl( - \frac{1}{d}\tr(H_0^2) T^2 \Bigr) = \exp\Bigl( - \frac{n_a^2}{t_a} \Bigr) \text{ with } \alpha=1.
\end{equation}
It can be seen that the application of a quantum map slows down the quadratic fidelity decay by a factor $2/(d \sigma)$.
Hence, the more chaotic the quantum map ($\sigma \rightarrow 1$), the slower is the fidelity decay.
This is essentially the observation of Prosen and \u{Z}nidari\u{c} \cite{ProZni01}, who proposed to stabilize a quantum algorithm $U_{QA}$ against static imperfections by devising more chaotic gate decompositions (see section \ref{sec:incrcor}).

\subsection{Fidelity Decay of Protected Computations}\label{subsec:parec_ana}

The goal of this subsection is to derive a formula for the entanglement fidelity of a quantum computation which is perturbed by static imperfections, and protected using the \textsf{PAREC} method.
It will be shown that the quadratic time dependence of the resulting fidelity decay \eqref{eq::frahm} of an unprotected computation will be converted into a linear one.

As we showed in subsection \ref{subsec:parec}, the total time evolution in the presence of static imperfections of $n_a$ iterations of a quantum algorithm $U_{QA} = U_{n_g} \dots U_2 U_1$ which is stabilized using the \textsf{PAREC} method, is obtained by replacing the $k$-th quantum gate ($k=1,\dots,n_g$) of the $\tau$-th iteration by the gate \eqref{eq:Gktau}
\begin{equation}
G_k^{(\tau)} =
g^\dagger_{[\tau,k]} \exp( - i H_{kl}^\tau ) g_{[\tau,k]} \cdot  U_k \cdot
g^\dagger_{[\tau,k]} \exp( - i H_{kr}^\tau ) g_{[\tau,k]}.
\end{equation}
Hence, by setting
\begin{subequations}
\begin{align}
\delta\!H^\tau_{kl} &= g^\dagger_{[\tau,k]} \cdot H_{kl}^\tau \cdot g_{[\tau,k]} \\ \text{ and }
\delta\!H^\tau_{kr} &= g^\dagger_{[\tau,k]} \cdot H_{kr}^\tau \cdot g_{[\tau,k]},
\end{align}
\end{subequations}
equation \eqref{eq::secorderQA} yields the second order expansion of the entanglement fidelity between the total time evolution \eqref{eq:parec_total} and the ideal time evolution $U_{QA}^{n_a}$.
(We neglect the first term in \eqref{eq:parec_total} which describes the time evolution of the first decoupling pulse.)
We proceed by calculating the quantities
$\mathbb{E} \, \delta\!H^\tau_{kl}$ and
$\mathbb{E} \, \delta\!H^\tau_{kr}$, where $\mathbb{E}$ denotes the average taken over all random selections $g_{[\tau,k]}$ from the decoupling set $\mathcal{G} = \{ g_j \}_{j=0}^{n_c-1}$.

According to \eqref{eq:Hkltau}, $H_{kl}^\tau$ depends on the random index $[\tau,k]$, because the time integral of the integrand $V_{2k+1}^{(\tau)\dagger}(t') H_0 V_{2k+1}^{(\tau)}(t')$ involves the unitary $V_{2k+1}^{(\tau)}(t')$ generating the pulse $V_{2k+1}^{(\tau)}(\tau_p) = g_{[\tau,k+1]} g^\dagger_{[\tau,k]}$.
If the elements of the decoupling set $\mathcal{G}$ form a projective representation $R$ of a group $G=\{\mathfrak{g}_j\}_{j=0}^{n_c-1}$ (i.\,e. if $g_j=R(\mathfrak{g}_j)$), the pulse $g_{[\tau,k+1]} g^\dagger_{[\tau,k]}$ corresponds to a random member $g_{j'}$ of the group.
In the following we make this assumption and are going to use the notation $\Pi_\mathcal{G}(X)$, which was introduced in \eqref{eq:defi:pigsec21} as the projection of the operator $X$ onto the commutant $\mathcal{A}'$ of the group algebra $\mathcal{A}=R(\mathbb{C}G)$.
Hence, the average becomes
\begin{align}
\mathbb{E} \, \delta\!H^\tau_{kl} &= \Pi_\mathcal{G} \biggl(
\frac{1}{n_c}\sum_{j'=0}^{n_c-1} \int_0^{\tau_p} g_{j'}^\dagger(t') H_0 g_{j'}(t') dt' \biggr) + \Pi_\mathcal{G}(H_0) \cdot \delta t \\
&= \Pi_\mathcal{G} (H_0) \cdot (\tau_p + \delta t), \label{eq:finitetpgroup}
\end{align}
where $g_{j'}(t) = \mathcal{T}\exp\bigl( \int_0^t H_c(t') dt' \bigr)$ for $t\in[0,\tau_p]$ denotes the unitary generating the pulse $g_{j'} \equiv g_{j'}(\tau_p)$.
The last identity is obtained analogously to the proof of theorem \ref{thm:vk03} by demanding that the control Hamiltonian $H_c(t')$ generating $g_{j'}(t)$ is within the group algebra $\mathcal{A}$ for all $t'\in[0,\tau_p]$ and for all $j\in\{0,1,\dots,n_c-1\}$.

In order to calculate $\mathbb{E} \, \delta\!H^\tau_{kr}$, we note that according to \eqref{eq:Hkrtau}, $H_{kr}^\tau$ depends on the random index $[\tau,k]$ because the time integral of $V_{2k}^{(\tau)\dagger}(t') H_0 V_{2k}^{(\tau)}(t')$ involves the unitary $V_{2k}^{(\tau)}(t')$ generating the twisted quantum gate
$V_{2k}^{(\tau)}(\tau_g) = g_{[\tau,k]} \cdot  U_k \cdot g^\dagger_{[\tau,k]}$.
Let us assume now that the quantum gate $U_k = \exp\bigl(-i K \int_0^{\tau_g} f(t') dt' \bigr)$ is generated by a gate Hamiltonian $K$, shaped by a pulse form $f(t)$ such that $\int_0^{\tau_g} f(t) dt = 1$.
The corresponding twisted gate could now be generated by the altered gate Hamiltonian
$K'_{[\tau,k]}= g_{[\tau,k]} \cdot  K \cdot g^\dagger_{[\tau,k]}$, i.\,e.
$g_{[\tau,k]} \cdot  U_k \cdot g^\dagger_{[\tau,k]} = \exp\bigl(-i K'_{[\tau,k]} \int_0^{\tau_g} f(t') dt' \bigr)$.
Then,
\begin{align}
\mathbb{E} \, \delta\!H^\tau_{kr} &= \frac{1}{n_c}\sum_{j=0}^{n_c-1} g_j^\dagger \biggl(
\int_0^{\tau_g}  \exp\Bigl(+i K'_j \int_0^{t'} f(t'') dt'' \Bigr)\, H_0\, \exp\Bigl(-i K'_j \int_0^{t'} f(t'') dt'' \Bigr) dt'
+ H_0\delta t \biggr) g_j \nonumber\\
&= \int_0^{\tau_g} \exp\Bigl(+i K \int_0^{t'} f(t'') dt'' \Bigr)\, \Pi_\mathcal{G}(H_0) \, \exp\Bigl(-i K \int_0^{t'} f(t'') dt'' \Bigr) dt'
+ \Pi_\mathcal{G}( H_0 ) \cdot \delta t \\
&= \Pi_\mathcal{G}( H_0 ) \cdot (\tau_g + \delta t),
\end{align}
where the last step is obtained provided that the action of $\Pi_\mathcal{G}(H_0)$ is trivial.

We are now going to use the results of the preceding two paragraphs on $\mathbb{E} \, \delta\!H^\tau_{lr}$ and $\mathbb{E} \, \delta\!H^\tau_{kr}$ to calculate the average $\mathbb{E}$ of the second order expansion of the entanglement fidelity given by equation \eqref{eq::secorderQA}.
For a traceless Hamiltonian $H_0$ a suitable decoupling scheme $\mathcal{G}$ leads to $\Pi_\mathcal{G}( H_0 ) = 0$ and we obtain the expectation value of the amplitude
\begin{equation}\label{eq:EAeGeneral}
\begin{split}
\mathbb{E} A_e(n_a) = 1 &- \frac{1}{2} \sum_{\tau=1}^{n_a}\sum_{j=1}^{n_g}
\Bigl( \frac{1}{d}\tr \bigl( (H_{jl}^\tau)^2 \bigr) + \frac{1}{d}\tr \bigl( (H_{jr}^\tau)^2 \bigr) \Bigr)\\
&-\sum_{\tau=1}^{n_a}\sum_{j=1}^{n_g} \mathbb{E} \frac{1}{d} \tr \bigl( U^\dagger_j \delta\!H_{jl}^\tau U_j \delta\!H_{jr}^\tau \bigr)
+ \mathcal{O}\bigl( (\delta\!H)^3 \bigr).
\end{split}
\end{equation}
In order to derive a simple expression for the fidelity, we are now going to consider the limit in which the pulses and gates are generated instantaneously ($\tau_p,\tau_g\rightarrow 0$), but we stress that the crucial step in the derivation of our fidelity formula was performed for the general case of finite pulses.
In the bang-bang limit, we have $H_{kl}^\tau = H_{kr}^\tau = H_0\Delta t/2$ and \eqref{eq:EAeGeneral} simplifies to
\begin{align}
\mathbb{E} A_e(n_a) &= 1 -  \frac{n_a}{4} \frac{1}{d} \Bigl(
 n_g \tr \bigl( H_0^2 \bigr) + \sum_{j=1}^{n_g}\frac{1}{n_c}\sum_{i=0}^{n_c-1}  \tr \bigl( U^\dagger_j \, g_i^\dagger H_0 g_i \, U_j \, g_i^\dagger H_0 g_i \bigr) \Bigr) \Delta t^2 + \mathcal{O}\bigl( H_0^3 \bigr), \label{eq:AeParec}\\
 &\geq  1 - \frac{n_a n_g}{2} \frac{1}{d} \tr\bigl( H_0^2 \bigr)\Delta t^2 + \mathcal{O}\bigl( H_0^3 \bigr).
\end{align}
The last inequality can be obtained by recalling that $\tr( A^\dagger B )$ constitutes
a Hermitian inner product for which the Cauchy-Schwarz inequality applies.
We proved the following theorem:
\begin{thm}\label{thm:fidparec}
Let a quantum computation consist of $n_a$ iterations of a quantum algorithm $U_{QA}$ consisting of $n_g$ quantum gates.
The entanglement fidelity between an ideal computation and a non-ideal computation protected by the \textsf{PAREC} method is (on average) given by
\begin{equation}\label{eq:FeParec}
F_e^\textsf{PAREC}(n_a) \geq 1 - n_a n_g \frac{1}{d} \tr\bigl( H_0^2 \bigr)\Delta t^2 + \mathcal{O}\bigl( H_0^3 \bigr),
\end{equation}
where the Hamiltonian $H_0$ describes the imperfections of the quantum computer, and $\Delta t$ denotes the time in between subsequent quantum gates (compare with figure \ref{fig::parecscheme}).
\end{thm}
\begin{rem}
In subsection \ref{subsec:randstrat} we derived a short time expansion of the entanglement fidelity $F_e^\textsf{NRD}(T)$ \eqref{eq:FeNRD} of a quantum memory protected by \textsf{NRD} and argued that a good approximation (valid for all times $T$) is given by the exponential $F_{e\ \text{app}}^\textsf{NRD}(T)$ given by \eqref{eq:FeappNRD}.
Analogously, we propose that for all numbers of iterations $n_a$ a good approximation of the \textsf{PAREC} fidelity is given by
\begin{equation}\label{eq:FeappParec}
F_{e\ \text{app}}^\textsf{PAREC}(n_a) = \exp\Bigl( -  n_a n_g \frac{1}{d} \tr\bigl( H_0^2 \bigr) \Delta t^2  \Bigr).
\end{equation}
\end{rem}

As it turned out in this subsection, the decoupling scheme $\mathcal{G} = \{ g_j \}_{j=0}^{n_c-1}$ employed by \textsf{PAREC} has to satisfy the decoupling condition $\Pi_\mathcal{G}(H_0) = \lambda \cdot \mathcal{I}$, with $\lambda \in \mathbb{R}$.
In addition, for a finite pulse width $\tau_p$, the elements of the decoupling scheme should also form a  group (or at least a projective representation of a group) in order to arrive at \eqref{eq:finitetpgroup}.
Since the order $n_c$ of the decoupling group does not enter in the formula for the resulting fidelity decay,
it is always possible to choose $\mathcal{G}$ to be an annihilator, such as the set $\mathcal{P}^n_q$ of Pauli operators.
Equation \eqref{eq:FeParec} explicitly exhibits the dependence of the entanglement fidelity decay on the number $n_g$ of elementary quantum gates and the strictly linear dependence on the numbers of iterations of the unitary transformation~$U_{QA}$.

Several straightforward improvements of the basic relation \eqref{eq:FeParec} are possible.
For example, it is also possible to apply the random decoupling pulses not before each elementary quantum gate but less often.
One random decoupling pulse between each iteration of a quantum algorithm, for example, is already
enough to get rid of the terms of \eqref{eq::fstat} quadratic in $n_a$.
In this case \eqref{eq:FeParec} is replaced by the inequality
\begin{equation}
F_e^\textsf{PAREC}(n_a) \geq 1 - n_a n_g^2 \frac{1}{d} \tr\bigl( H_0^2 \bigr)\Delta t^2 + \mathcal{O}\bigl( H_0^3 \bigr),
\end{equation}
at the expense that the term linear in $n_a$ has a coefficient quadratic in the number of elementary quantum gates per iteration $n_g$.

In order to determine the decay of the average entanglement fidelity of a quantum memory stabilized by \textsf{NRD} we use \eqref{eq:FeParec} and specialize to the case of $n_a$ iterations of a quantum algorithm consisting of $n_g$ identity gates.
Denoting the total interaction time between the qudits of the quantum memory by $T = n_a n_g \Delta t$ one obtains the result
\begin{equation}\label{eq:FePvsNRDi}
F_e^\textsf{PAREC}(n_a) \geq 1 -  n_a n_g \frac{1}{d} \tr\bigl( H_0^2 \bigr) \Delta t^2 +\mathcal{O}\bigl( H_0^3 \bigr)  = 1 - \frac{1}{d} \tr\bigl( H_0^2 \bigr) \Delta t T +\mathcal{O}\bigl( H_0^3 \bigr),
\end{equation}
which is identical to the average \textsf{NRD} fidelity \eqref{eq:FeNRD},
\begin{equation}\label{eq:FePvsNRDii}
F_e^\textsf{NRD}(T) = 1 - \frac{1}{d} \tr \bigl(H_0^2\bigr) \Delta t T + \dots,
\end{equation}
of theorem \ref{thm:nrdf} derived in subsection \ref{subsec:randstrat} by considering bang-bang control.
Since we derived the \textsf{PAREC} fidelity by considering bounded controls
(generating the decoupling pulses within a finite time interval $\tau_p$),
this fact indicates that the \textsf{NRD} strategy remains applicable even if only bounded controls are available.

\subsection{Numerical Example}\label{subsec:par_ex}

\begin{figure}\centering
\subfloat[Entanglement fidelity of a $n=8$ qubit quantum computation.]{\label{fig:thesispA}
\includegraphics[scale=0.9]{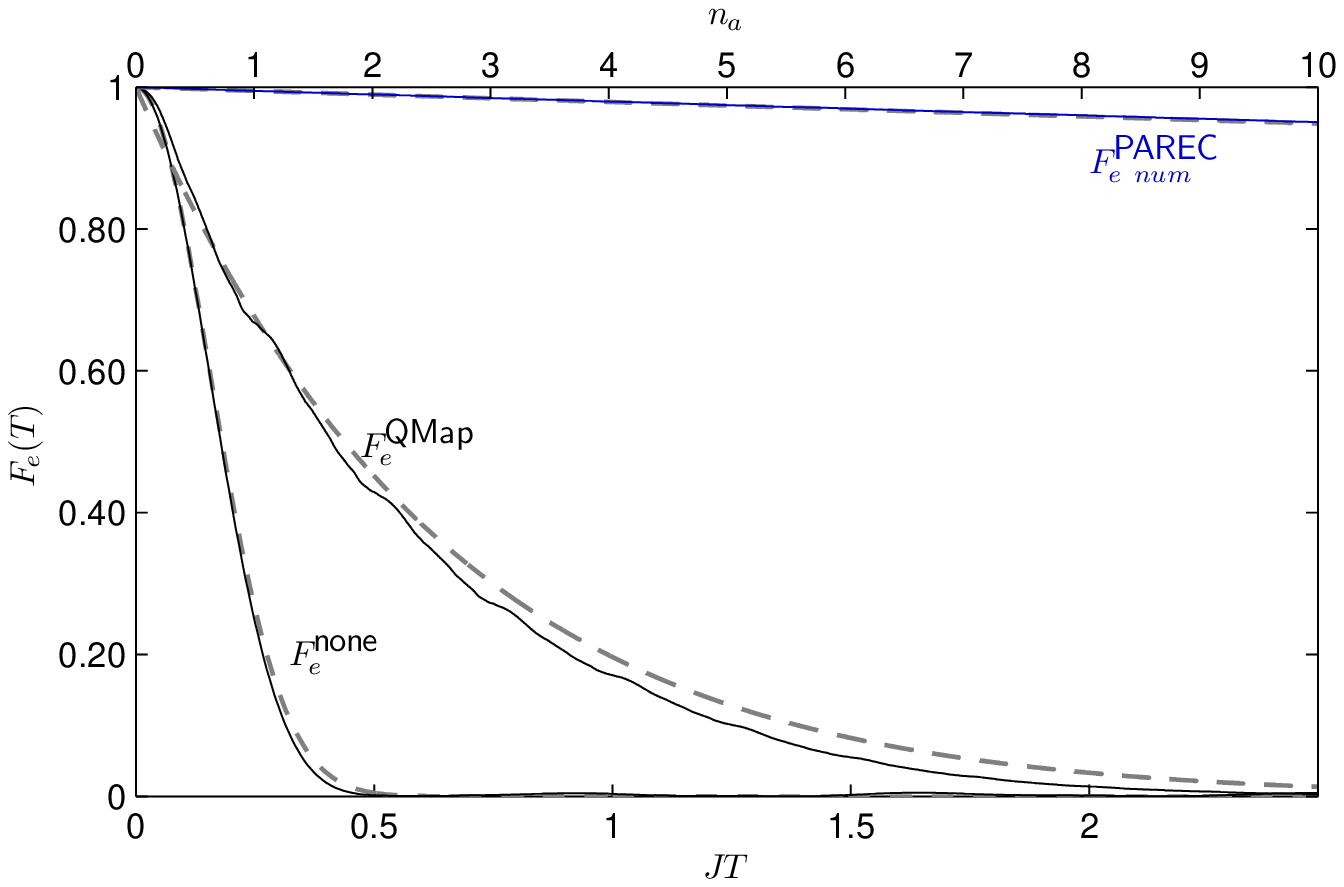}}\\
\subfloat[Enlarged part of figure \ref{fig:thesispA}.]{\label{fig:thesispB}
\includegraphics[scale=0.9]{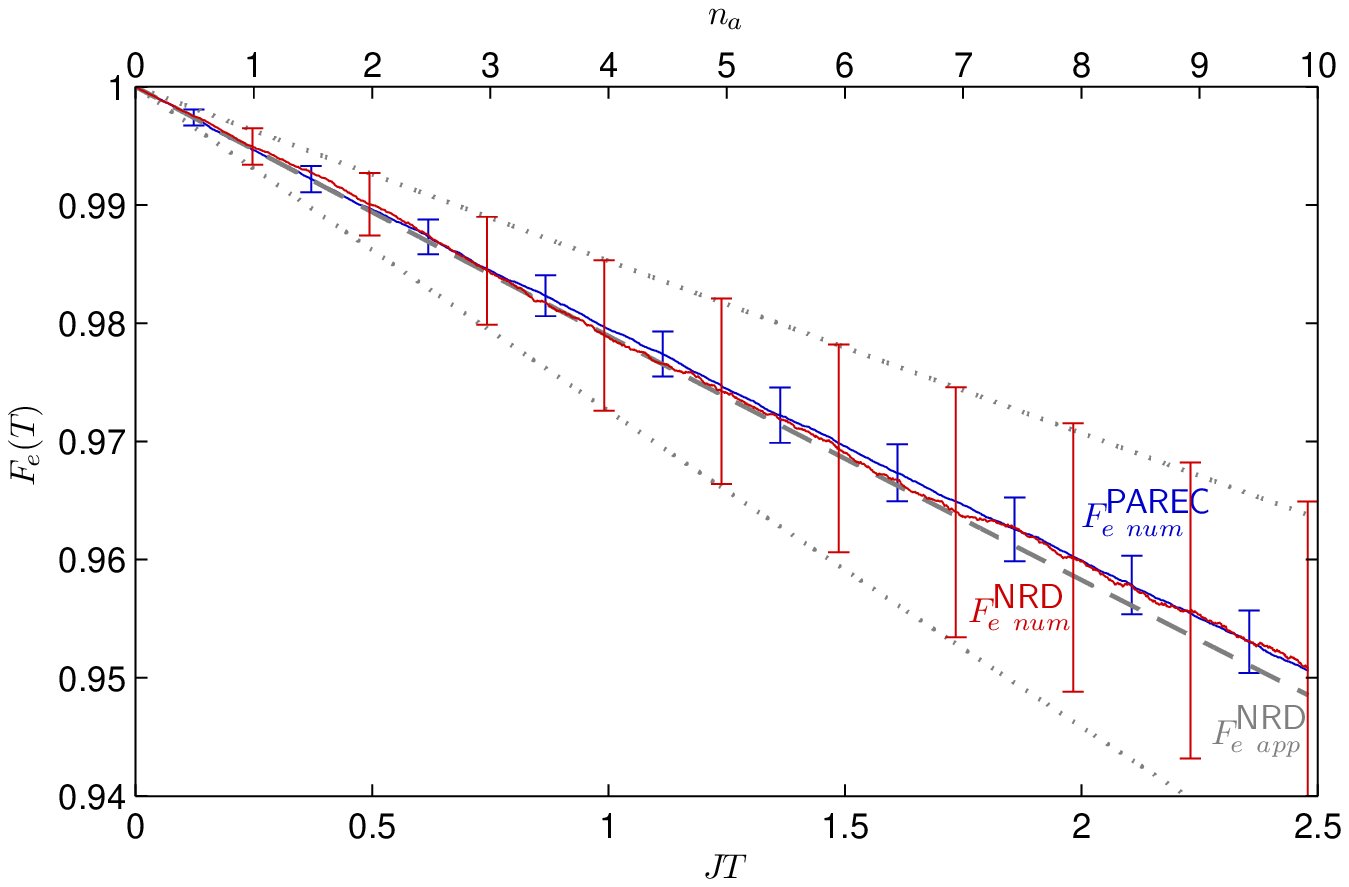}}
\caption[Entanglement fidelity of PAREC]{%
Entanglement fidelity of a $n=8$ qubit quantum computation perturbed by the imperfections given in \eqref{eq:decmodel:PAREC}. Each of the $n_a=10$ iterations of the quantum algorithm consists of $n_g=248$ quantum gates. The time in between subsequent gates is given by $\Delta t=0.001 J^{-1}$.\\
(\textit{a}) The fidelity $F_e^\textsf{QMap}$ of an unprotected computation,
the fidelity $F_e^\textsf{none}$ of a quantum memory, the fidelity $F_{e\ \text{num}}^\textsf{PAREC}$ (\textit{blue}) of the stabilized computation, and the corresponding estimations $F_{e\ \text{app}}^\textsf{QMap}$ \eqref{eq:FeappQMap}, $F_{e\ \text{app}}^\textsf{none}$ \eqref{eq:FeappNone}, and $F_{e\ \text{app}}^\textsf{PAREC}$ \eqref{eq:FeappParec} (\textit{dashed lines}).\\
(\textit{b})
The fidelity $F_{e\ \text{num}}^\textsf{PAREC}$ (\textit{blue}) of the stabilized quantum computation, the fidelity $F_{e\ \text{num}}^\textsf{NRD}$ (\textit{red}) of a quantum memory stabilized by \textsf{NRD}, and its estimation $F_{e\ \text{app}}^\textsf{NRD}$ \eqref{eq:FeappNRD} (\textit{dashed line}).
In addition, the standard deviation of the \textsf{PAREC} and the \textsf{NRD} fidelity is indicated by error bars.
The estimate $\sigma^{\ \text{app}}_\textsf{NRD}$ \eqref{eq:nrdvapp} of the \textsf{NRD} standard deviation is indicated by $F_{e\ \text{app}}^\textsf{NRD} \pm \sigma^{\ \text{app}}_\textsf{NRD}$ (\textit{dotted lines}).\label{fig:thesisparec}}
\end{figure}

We close the discussion of the \textsf{PAREC} method with a numerical simulation.
Let us consider a quantum computer with $n = 8$ qubits arranged on a linear chain, which are perturbed by Heisenberg couplings,
\begin{equation}\label{eq:decmodel:PAREC}
 H_0 =
 \sum_{k_1 = 1}^{n-1}
 \sum_{k_2 = k_1+1}^n
 \
 J^{k_1,k_2} \bigl[  X\otimes X + Y\otimes Y + Z\otimes Z \bigr]_{(k_1,k_2)},
\end{equation}
where the coupling strength between qubits $k_1$ and $k_2$ decays cubically with their separation distance, i.\,e. $J^{k_1,k_2} = J\cdot \vert k_1 - k_2\vert^{-3}$.
Note that these are the same imperfections as assumed for the numerical simulations of the decoupling strategies in section \ref{sec:decexample}.
As a quantum algorithm we consider multiple iterations of the quantum tent map,
\begin{equation}
 U_{QA} = \exp\Bigl( -\frac{i}{2} m^2 T \Bigr) \exp\Bigl( -i k V( q ) \Bigr),
\end{equation}
with parameters $T=2\pi/2^n$ and $kT=1.7$.
A definition of the operators $m$ and $q$ and the tent-map potential $V$ can be found in appendix \ref{sec:gatesqmaps}.
It is also explained in the appendix that each iteration of the tent map can be decomposed into $n_g = \frac{9}{2}n^2-\frac{11}{2}n + 4$ elementary one- and two-qubit quantum gates, which for $n=8$ leads to $n_g=248$.
We assume that the gates and pulses are performed instantaneously, and that the time interval $\Delta t$ in between subsequent quantum gates is given by $\Delta t = 0.001 J^{-1}$.
The simulations cover $n_a=10$ iterations.
Hence, the total run time of the quantum computation is given by $T=10 n_g \Delta t = 2.48 J^{-1}$.
The results of our simulations are presented in figure \ref{fig:thesisparec}.

In figure \ref{fig:thesispA}, we compare the fidelity $F_e^\textsf{QMap}$ of the unprotected quantum computation with the corresponding fidelity $F_e^\textsf{none}$ of an unprotected quantum memory.
The corresponding estimations $F_{e\ \text{app}}^\textsf{QMap}$ given by \eqref{eq:FeappQMap} with $\alpha = 0.294$ and $F_{e\ \text{app}}^\textsf{none}$ given by \eqref{eq:FeappNone} are also shown (\textit{dashed lines}).
It can be seen that the quantum computation itself leads to a slow down of the fidelity decay.
If the computation is stabilized using the \textsf{PAREC} method, the resulting fidelity $F_{e\ \text{num}}^\textsf{PAREC}$ (\textit{blue}) is significantly improved and in good agreement with the predicted fidelity $F_{e\ \text{app}}^\textsf{PAREC}$ of equation \eqref{eq:FeappParec} (\textit{dashed line}).
(The index \text{num} indicates the fact that the fidelity is obtained numerically by averaging over a subset of $70$ random pulse realizations.)

Figure \ref{fig:thesispB} shows an enlarged part of the high fidelity region.
In addition to $F_{e\ \text{num}}^\textsf{PAREC}$ (\textit{blue}), the fidelity $F_{e\ \text{num}}^\textsf{NRD}$ (\textit{red}) of a quantum memory protected by the naive random decoupling strategy (\textsf{NRD}) is shown together with its corresponding estimation $F_{e\ \text{app}}^\textsf{NRD}$ (\textit{dashed}) given by equation \eqref{eq:FeappNRD}.
The memory protected via \textsf{NRD} corresponds to a trivial quantum computation (all the quantum gates are identity gates) which is protected by the \textsf{PAREC} method.
As predicted by equations \eqref{eq:FePvsNRDi} and \eqref{eq:FePvsNRDii}, all three fidelities are quite close to each other.
Let us focus now on the variance of $F_e^\textsf{PAREC}$ and $F_e^\textsf{NRD}$.
An estimation of the latter quantity $\sigma_\textsf{NRD}^2$ was proposed in subsection \ref{subsec:varnrd} to be given by \eqref{eq:nrdvapp}.
This estimation is indicated by the two \textit{dotted lines} representing
$F_{e\ \text{app}}^\textsf{NRD} \pm \sigma_\textsf{NRD}^{\ \text{app}}$.
It is in good agreement with the actual standard deviation $\sigma_\textsf{NRD}^{\ \text{num}}$ indicated by the error bars (\textit{red}).
An interesting observation is that the standard deviation $\sigma_\textsf{PAREC}^{\ \text{num}}$ of the fidelity $F_{e\ \text{num}}^\textsf{PAREC}$ of the stabilized computation (indicated by the \textit{blue} error bars) is considerably smaller.

\section{Stabilizing Computations by Increasing the Correlation Decay}\label{sec:incrcor}

The fidelity decay of an unprotected quantum computation in the presence of static imperfections depends on the decomposition of the quantum algorithm into elementary one- and two-qudit gates (subsection \ref{subsec:par_ohnekorrektur}).
The first non-trivial term in a short-time expansion of the fidelity is called correlation function.
The larger the value of this correlation function, the faster the decay of the fidelity.
Based on this observation, Prosen and \u{Z}nidari\u{c} \cite{Prosen02,ProZni01} proposed to stabilize quantum algorithms by rewriting them in such a way, that the new gate decomposition leads to an increased decay of the correlation function.
For a particular type of imperfections, they demonstrated their idea by designing an alternative gate decomposition for the quantum Fourier transform \cite{ProZni01}.
An open question is how to find good gate decompositions for general algorithms and general imperfections.
In this section, we are going to demonstrate that the \textsf{PAREC} method of the preceding section provides a solution to this question:
By viewing the random decoupling pulses as additional quantum gates, \textsf{PAREC} translates an arbitrary quantum algorithm consisting of $n_g$ quantum gates into one containing twice as much gates.
This new gate decomposition leads (on average) to an ultimate decay of the correlation function.

We start in the first subsection with a summary of the main results of \cite{ProZni01}.
The second subsection explains how the \textsf{PAREC} method wipes out the correlations.
As in subsection \ref{subsec:par_ohnekorrektur} we consider $n_a$ iterations of quantum algorithm $U_{QA} = U_{n_g} \dots U_2 U_1$.
To keep things as simple as possible, we assume that the gates (quantum gates and decoupling pulses) are applied instantaneously ($\tau_g,\tau_p \rightarrow 0$).

\subsection{Fidelity and Correlation Decay}\label{subsec::prosen}

In special cases in which an ideal unitary transformation $U_{QA}$ is not decomposed into elementary gates
we may simplify \eqref{eq::fstat} by taking $n_g=1$ thus obtaining the fidelity decay
\begin{equation}\label{eq::Prosen1}
F_e(n_a) =  1 - \sum_{\tau=-(n_a-1)}^{n_a-1} ( n_a - |\tau| ) \frac{1}{d} \tr\bigl( U_{QA}^{-\tau} H_0 U_{QA}^\tau H_0 \bigr)\Delta t^2 + \mathcal{O}\bigl( H_0^3 \bigr).
\end{equation}
This expression has been studied previously by Prosen \cite{Prosen02}.
It indicates that the faster the decay of the correlation function
$\tr\bigl( U_{QA}^{-\tau} H_0 U_{QA}^\tau H_0 \bigr)$
the slower the decay of the fidelity.
According to an original proposal by Prosen and \u{Z}nidari\u{c} \cite{ProZni01} this characteristic feature of the fidelity decay can be exploited for stabilizing a quantum algorithm against static imperfections.
This aspect was investigated in detail by these authors for the special case of $n_a=1$.
In this case \eqref{eq::fstat} reduces to the simpler form
\begin{equation}\label{eq:defiCjk}
F_e =  1 - \sum_{j,k=1}^{n_g} \underbrace{ \frac{1}{d} \tr\bigl(
U_{1\dots j-1}^\dagger  H_0  U_{j-1\dots 1} \cdot U_{1\dots k-1}^\dagger  H_0  U_{k-1\dots 1}
\bigr)\Delta t^2 }_{C(j,k)} + \mathcal{O}\bigl( H_0^3 \bigr).
\end{equation}
Prosen and \u{Z}nidari\u{c} based their error suppression method on the idea to rewrite a quantum algorithm $U_{QA}$ in such a way that for the new gate decomposition the sum over the off-diagonal elements of the correlation matrix $C(j,k)$ becomes smaller than for the original gate sequence (thereby using possibly even a larger number of quantum gates).
They considered as an example perturbations of the form $H_0\Delta t = V \delta$ with $V$ being represented by a $d$-dimensional matrix randomly chosen from the Gaussian unitary ensemble (GUE).
Thus, on average the matrix elements of $V$ fulfill
the condition $\langle V_{jk} V_{lm} \rangle = \delta_{jm} \delta_{kl} / d$.
With this kind of imperfections on average the correlation function becomes
\begin{equation}\label{eq::cfkt}
\langle C(j,k)\rangle =
\left( \Bigl\vert \frac{1}{d} \tr\bigl( U_{j-1}\dots U_2 U_1 \cdot U^\dagger_1 U^\dagger_2\dots U^\dagger_{k-1} \bigr) \Bigr\vert^2 -\frac{1}{d^2}\right) \delta^2.
\end{equation}
The $1/d^2$-term comes from the fact that according to our assumption of traceless perturbing Hamiltonians
also our matrices $V$ have to be chosen traceless.
(In the case of a non-traceless perturbation $V$ this restriction can be achieved by the replacement
$V \mapsto V - \mathcal{I}\cdot \tr(V)/d$).
It should be mentioned that this latter $1/d^2$-term was not taken into account in reference \cite{ProZni01} so that these authors investigated the quantity
$\bigl\vert \frac{1}{d} \tr\bigl( U^{\phantom{\dagger}}_{j-1\dots 1} \cdot U^\dagger_{1\dots k-1} \bigr) \bigr\vert^2 \delta^2$.

\begin{figure*}
\centerline{\includegraphics[width=.48\columnwidth]{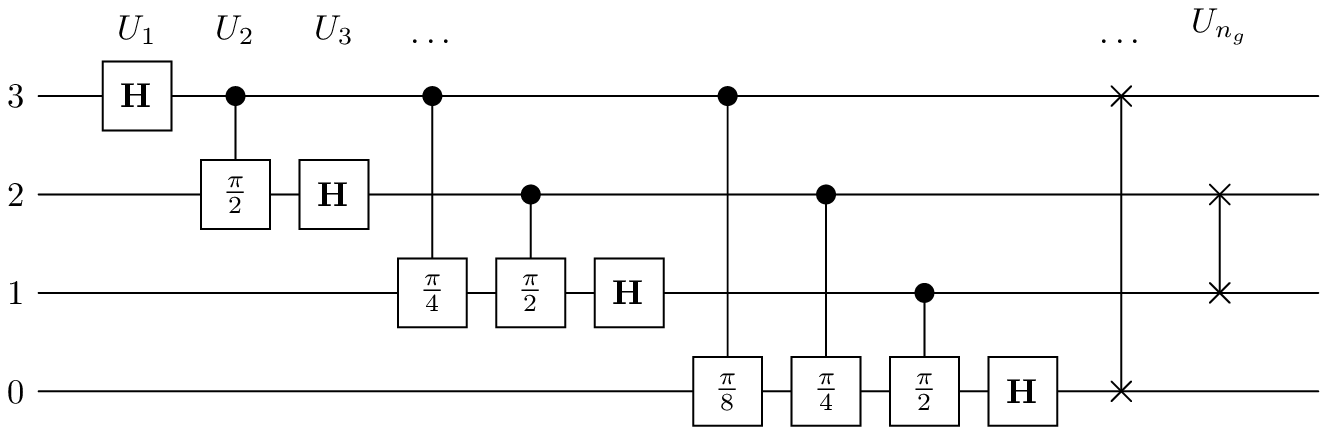}\hfill
\includegraphics[width=.48\columnwidth]{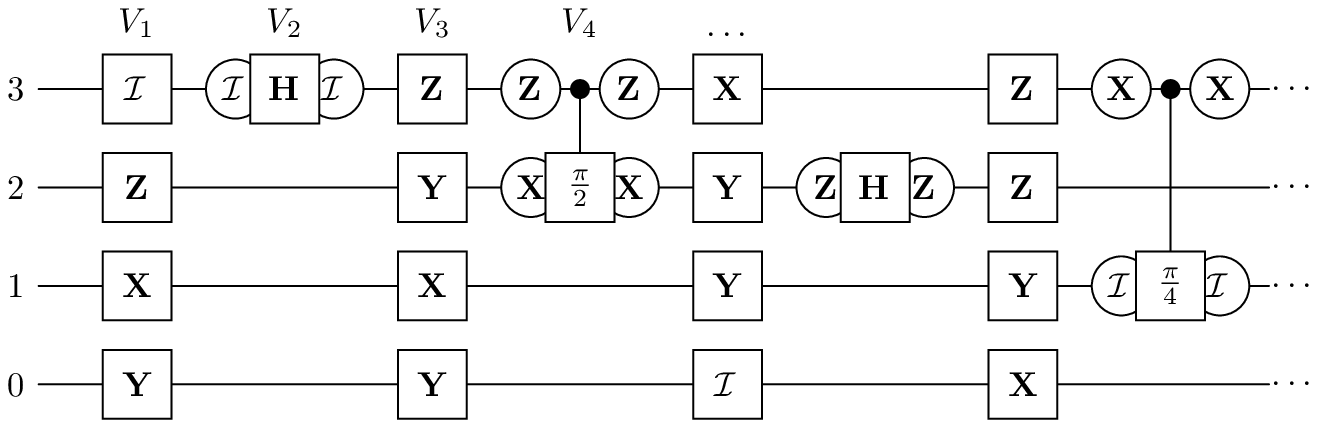}%
}
\caption{Quantum circuit of the quantum Fourier transform for
$n=4$ qubits \textit{(left)}.  The first four gates of the same circuit
involving the \textsf{PAREC} method \textit{(right)}.\label{fig::qft}}
\end{figure*}
In order to demonstrate their idea, Prosen and \u{Z}nidari\u{c} considered the quantum Fourier transformation (QFT) as an example.
Typically, this unitary transformation $U_{QA}$ is decomposed into $n_g = \lfloor n(n+2)/2 \rfloor$ quantum gates which involve Hadamard operations, controlled-phase gates, and swap gates.
(compare with the left-hand side of figure \ref{fig::qft}, see also subsection \ref{subsec:AppQFT} of the appendix).
Instead, Prosen and \u{Z}nidari\u{c} used a different decomposition involving $n_g'=\lfloor n(2n+1)/2 \rfloor$ quantum gates.
In figure \ref{fig::prosen} the correlation matrix $\langle C(j,k) \rangle$ is depicted for both gate decompositions.
Compared to the conventional gate decomposition (left) the off-diagonal elements of this correlation matrix are suppressed significantly by this new gate decomposition (middle).
Diagonal values are always constant, i.e. $\langle C(j,j)\rangle/\delta^2 + 1/d^2 = 1$.

Though of interest this proposal of Prosen and \u{Z}nidari\u{c} leaves important questions unanswered. 
How can such an improved gate sequence be found for an arbitrary quantum algorithm ?
How can this be achieved for repeated iterations of a unitary quantum map ?
Is it possible to suppress all off-diagonal elements of the correlation function perfectly ?
All these questions can be addressed and solved in a rather straightforward way utilizing \textsf{NRD} decoupling as described in the preceding section.

\subsection{Destroying Correlations with the \textsf{PAREC} Method}
\begin{figure*}
\begin{minipage}[b]{0.21\textwidth}
\includegraphics[scale=1.32, trim = 0 41 0 0,clip=true]{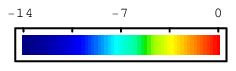}\\
\includegraphics[scale=1.3]{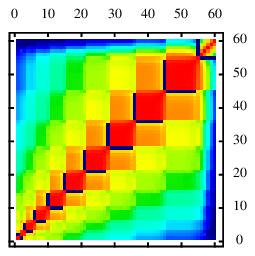}%
\end{minipage}%
\begin{minipage}[b]{0.79\textwidth}
\hfill\includegraphics[scale=1.3]{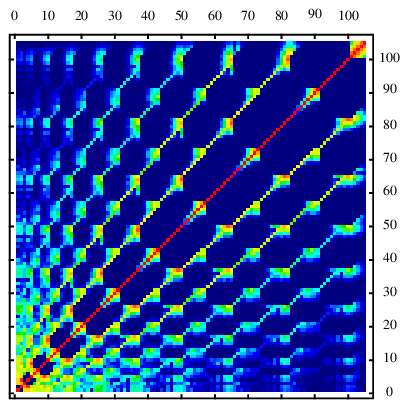}%
\includegraphics[scale=1.3]{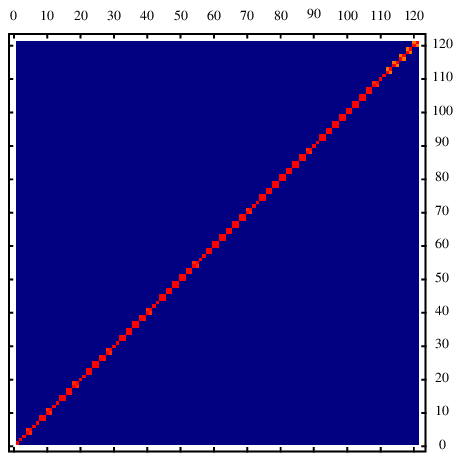}%
\end{minipage}
\caption{$\ln \left[ \langle C(j,k) \rangle/\delta^2 + 1/d^2\right] $ for the QFT with $n=10$ qubits using the usual gate decomposition with $n_g=60$ gates \textit{(left)},
the decomposition by Prosen using $n_g'=105$ gates \textit{(middle)} and
$\ln \left[ \mathbb{E} \langle C(j,k) \rangle/\delta^2 + 1/d^2  \right]$  using the \textsf{PAREC} method with $n_g'=2n_g+1=121$ gates \textit{(right)}.\label{fig::prosen}}
\end{figure*}

In this subsection it is explicitly shown that the \textsf{PAREC} method is capable of canceling the off-diagonal terms of the correlation function $\langle C(j,k)\rangle$ \eqref{eq::cfkt} perfectly.
According to equation \eqref{eq::uparec}, the \textsf{PAREC} method translates a quantum algorithm $U_{QA}$ consisting of $n_g$ quantum gates, into one containing $n_g' = 2n_g + 1$ quantum gates.
Let us consider the stabilizing properties of the \textsf{PAREC} method with respect to static imperfections which can be characterized by traceless perturbing Hamiltonians of the form
$H_0\Delta t \equiv \frac{1}{2} \cdot \bigl(V - \mathcal{I}\cdot\tr(V)/d \bigr) \cdot \delta $
with $V$ chosen randomly from the Gaussian unitary ensemble (GUE).
The strength of the interaction is reduced by the factor $1/2$ so that the situation is equivalent to the one depicted in figure \ref{fig::parecscheme}, where $\Delta t$ denotes the time interval in between 'real' subsequent quantum gates (not counting the decoupling pulses as gates).
These perturbations describe physical situations in which in each individual realization of a quantum
algorithm the inter-qudit Hamiltonian perturbing the dynamics of the qudits of the quantum information processor  is time independent but random.
To eliminate such GUE-governed static imperfections we have to choose an annihilator, such as the set of Pauli operators $\mathcal{P}^n_q$, as a decoupling set $\mathcal{G} = \{ g_j \}_{j=0}^{n_c-1}$.
As a result the fidelity averaged over all possible random gates reduces to the expression
\begin{equation}
\mathbb{E} \langle F_e \rangle = 1 - \frac{1}{4}\sum_{j,k=1}^{n_g'} \mathbb{E} \langle C(j,k) \rangle + \mathcal{O}\bigl( H_0^3 \bigr),
\end{equation}
where the factor $1/4$ is a consequence of the reduced interaction strength $\delta/2$.
In view of the statistical independence of subsequent Pauli operations
almost all off-diagonal terms of the correlation function vanish, i.e.
\begin{equation}\label{step}
\mathbb{E} \langle C(j,k) \rangle  = \delta^2 \cdot \begin{cases}
1-\frac{1}{d^2}  &,\text{if $j=k$}\\
\bigl\vert \frac{1}{d} \tr U_{(j-1)/2} \bigr\vert^2 -\frac{1}{d^2} &, \text{if $j$ odd and $j=k+1$}\\
\bigl\vert \frac{1}{d} \tr U_{(k-1)/2} \bigr\vert^2 -\frac{1}{d^2} &, \text{if $k$ odd and $k=j+1$}\\
0 &, \text{else}.
\end{cases}
\end{equation}
Here, it has been taken into account that for all unitary matrices $U$ the relation
\begin{equation}
\mathbb{E} \Bigl\vert \frac{1}{d} \tr\bigl(g U \bigr) \Bigr\vert^2 \equiv
\frac{1}{d^2} \sum_{j=0}^{n_c-1} \Bigl\vert \frac{1}{d}\tr\bigl( g_j U \bigr) \Bigr\vert^2 = \frac{1}{d^2}
\end{equation}
holds since the average is performed over all unitary random Pauli gates  $g_j\in\mathcal{G} = \mathcal{P}_q^n$ which are elements of an orthonormal unitary error basis.
As a result the expectation value of the entanglement fidelity becomes
\begin{align}
\mathbb{E} \langle F_e \rangle &= 1 - (2 n_g+1) \frac{\delta^2}{4} (1-d^{-2})
-2 \frac{\delta^2}{4} \sum_{j=1}^{n_g} \Bigl( \Bigl\vert \frac{1}{d} \tr U_j \Bigr\vert^2 -d^{-2} \Bigr)
+ \mathcal{O}(\delta^3) \nonumber\\
&\geq 1 - n_g \delta^2 (1-d^{-2}) + \mathcal{O}(\delta^3).
\end{align}
Alternatively this expression can also be derived by averaging \eqref{eq:FeParec} over all elements of the GUE after substituting the relevant perturbing Hamiltonian
$H_0\Delta t \equiv \bigl(V - \mathcal{I}\cdot\tr(V)/d \bigr) \cdot \delta $ and setting $n_a=1$.
For the special case of a quantum Fourier transform (QFT) the resulting values of $\mathbb{E} \langle C(j,k) \rangle$ are shown on the right-hand side of figure \ref{fig::prosen}.
In this figure they are also compared to the corresponding values resulting from the improved QFT proposed by Prosen.

The \textsf{PAREC} method works not only for general algorithms, but also for general imperfections:
In the general case of a traceless Hamiltonian $H_0$, the quantity $C(j,k)$ defined in \eqref{eq:defiCjk} becomes on average (compare with \eqref{eq:AeParec})
\begin{equation}\label{eq:generalECjk}
\mathbb{E} C(j,k)   = \Delta t^2 \cdot \begin{cases}
\frac{1}{d}\tr\bigl( H_0^2 \bigr)  &,\text{if $j=k$}\\
\mathbb{E}_i \frac{1}{d}\tr\bigl( U^\dagger_{(j-1)/2} \, g_i^\dagger H_0 g_i \,  U_{(j-1)/2} \, g_i^\dagger H_0 g_i \bigr) &, \text{if $j$ odd and $j=k+1$}\\
\mathbb{E}_i \frac{1}{d}\tr\bigl( U^\dagger_{(k-1)/2} \, g_i^\dagger H_0 g_i \, U_{(k-1)/2} \, g_i^\dagger H_0 g_i \bigr) &, \text{if $k$ odd and $k=j+1$}\\
0 &, \text{else},
\end{cases}
\end{equation}
for any decoupling scheme $\mathcal{G} = \{g_j\}_{j=0}^{n_c-1}$ satisfying the standard decoupling condition
\begin{equation}
 \frac{1}{n_c} \sum_{j=0}^{n_c-1} g_j^\dagger H_0 g_j = 0.
\end{equation}

\section{Stabilizing Computations using Dynamically Corrected Gates}\label{sec:enEuler}

The \textsf{PAREC} method of section \ref{sec:parec} combines quantum computation with the naive random decoupling (\textsf{NRD}) strategy of subsection \ref{subsec:randstrat}.
Unfortunately, the suppression potential of \textsf{NRD} is rather low.
If the imperfections of a quantum computer are described by a Hamiltonian $\lambda H_0$, the decay of the entanglement fidelity after the time $T$ is of the order $\mathcal{O}\bigl( \lambda^2 \Delta t T \bigr)$, where $\Delta t$ denotes the time interval in between the application of subsequent decoupling pulses.
On the other hand, periodic dynamical decoupling (\textsf{PDD}, subsection \ref{subsec:detstrat}) is able to achieve a decay of the order $\mathcal{O}\bigl( \lambda^4 (\Delta t n_c T)^2 \bigr)$.
Even though the quadratic time dependence of \textsf{PDD} is inferior to the linear one of \textsf{NRD}, the fact that the imperfection strength $\lambda$ enters in the fourth power is a serious advantage.
In subsection \ref{subsec:dcgs}, we discussed the dynamically corrected gate (DCG) of Khodjasteh and Viola \cite{KhVi08}, which combines a single \textsf{PDD} cycle with the generation of a quantum gate.
By implementing each gate constituting a quantum algorithm as a DCG, a complete quantum computation might be stabilized against imperfections.
In contrast to the \textsf{PAREC} method, that way the resulting fidelity decay of the stabilized algorithm would benefit from the \textsf{PDD} characteristics.
This chapter considers the general case of decoupling pulses being generated by turning on a bounded control Hamiltonian for a time $\tau_p>0$.
Hence, in place of \textsf{PDD}, the Eulerian decoupling strategy (\cite{VK03}, subsection \ref{subsec:boundedctrls}) has to be applied.
In subsection \ref{subsec:eulerdcg}, we consider a generalization of the DCG approach of subsection \ref{subsec:dcgs} from \textsf{PDD} to Eulerian decoupling.
(In fact the original DCG proposal of Khodjasteh and Viola \cite{KhVi08} was for Eulerian decoupling.)
We compare the error suppression potential of the \textsf{PAREC} method and the Euler-DCG method for quantum algorithms.
By embedding \textsf{PDD} cycles within \textsf{NRD}, the embedded decoupling (\textsf{EMD}) strategy was devised in subsection \ref{subsec:randstrat}, which combines the advantages of both strategies in order to protect a quantum memory.
Motivated by this idea, we propose to combine Euler-DCGs with the \textsf{PAREC} method in order to protect quantum computations in subsection \ref{subsec:DCGPAREC}.

\subsection{Dynamically Corrected Gates (Euler-DCGs)}\label{subsec:eulerdcg}

\begin{figure}\centering
\scalebox{0.91}{%
\begin{pspicture}(-2.0,0.4)(15.3,-2.2)
\psline[linewidth=1pt,arrowsize=7pt]{<-}(-1.50,-1.)(15,-1.)
\rput[t](-2.,-1.15){\footnotesize time}
\psline(15,-1.1)(15,-0.9)\rput[t](15,-1.3){\small $t_0$}
\rput[B](14.6,-0.4){\gray \small $H_Y$}
\psline[linewidth=3pt,arrowsize=7pt, linecolor=gray]{->}(14.975,-1.)(14,-1.)
\rput[t](14.55,-1.6){$\underbrace{\makebox(0.9,0){}}_{\tau_p}$}
\psline(14,-1.1)(14,-0.9)%
\rput[b](14.1,-0.8){\small $Y\mathcal{I}^\dagger$}
\rput[B](13.6,-0.4){\gray \small $H_X$}
\psline[linewidth=3pt,arrowsize=7pt, linecolor=gray]{->}(13.975,-1.)(13,-1.)
\rput[t](13.55,-1.6){$\underbrace{\makebox(0.9,0){}}_{\tau_p}$}
\psline(13,-1.1)(13,-0.9)%
\rput[b](13.1,-0.8){\small $ZY^\dagger$}
\rput[B](12.6,-0.4){\gray \small $H_Y$}
\psline[linewidth=3pt,arrowsize=7pt, linecolor=gray]{->}(12.975,-1.)(12,-1.)
\psline(12,-1.1)(12,-0.9)%
\rput[b](12.1,-0.8){\small $XZ^\dagger$}
\rput[B](11.6,-0.4){\gray \small $H_Y$}
\psline[linewidth=3pt,arrowsize=7pt, linecolor=gray]{->}(11.975,-1.)(11,-1.)
\psline(11,-1.1)(11,-0.9)%
\rput[b](11.1,-0.8){\small $ZX^\dagger$}
\rput[B](10.1,-0.4){\gray \small $H_I$}
\psline[linewidth=3pt,arrowsize=7pt, linecolor=gray]{->}(10.975,-1.)(9,-1.)
\rput[t](10.05,-1.6){$\underbrace{\makebox(1.8,0){}}_{\tau_g}$}
\psline(9,-1.1)(9,-0.9)%
\rput[B](8.6,-0.4){\gray \small $H_X$}
\psline[linewidth=3pt,arrowsize=7pt, linecolor=gray]{->}(8.975,-1.)(8,-1.)
\psline(8,-1.1)(8,-0.9)%
\rput[b](8.1,-0.8){\small $YZ^\dagger$}
\rput[B](7.1,-0.4){\gray \small $H_I$}
\psline[linewidth=3pt,arrowsize=7pt, linecolor=gray]{->}(7.975,-1.)(6,-1.)
\psline(6,-1.1)(6,-0.9)%
\rput[B](5.6,-0.4){\gray \small $H_Y$}
\psline[linewidth=3pt,arrowsize=7pt, linecolor=gray]{->}(5.975,-1.)(5,-1.)
\rput[t](5.55,-1.6){$\underbrace{\makebox(0.9,0){}}_{\tau_p}$}
\psline(5,-1.1)(5,-0.9)%
\rput[b](5.1,-0.8){\small $IY^\dagger$}
\rput[B](4.6,-0.4){\gray \small $H_X$}
\psline[linewidth=3pt,arrowsize=7pt, linecolor=gray]{->}(4.975,-1.)(4,-1.)
\psline(4,-1.1)(4,-0.9)%
\rput[b](4.1,-0.8){\small $XI^\dagger$}
\rput[B](3.1,-0.4){\gray \small $H_I$}
\psline[linewidth=3pt,arrowsize=7pt, linecolor=gray]{->}(3.975,-1.)(2,-1.)
\psline(2,-1.1)(2,-0.9)%
\rput[B](1.6,-0.4){\gray \small $H_X$}
\psline[linewidth=3pt,arrowsize=7pt, linecolor=gray]{->}(1.975,-1.)(1,-1.)
\rput[t](1.55,-1.6){$\underbrace{\makebox(0.9,0){}}_{\tau_p}$}
\psline(1,-1.1)(1,-0.9)%
\rput[b](1.1,-0.8){\small $IX^\dagger$}
\rput[B](0.1,-0.4){\gray \small $H_g$}
\psline[linewidth=3pt,arrowsize=7pt, linecolor=gray]{->}(0.975,-1.)(-1,-1.)
\rput[t](0.05,-1.6){$\underbrace{\makebox(1.8,0){}}_{\tau_g}$}
\psline(-1,-1.1)(-1,-0.9)\rput[t](-1,-1.3){\small $t_g$}\rput[b](-0.9,-0.8){\small $U_g$}
\end{pspicture}}%
\caption[Euler-DCG]{Schematic representation of an Euler-DCG based on the decoupling set $\mathcal{G}=\{\mathcal{I},X,Y,Z\}$ and the generators $\Gamma=\{X,Y\}$.
The above cycle of length
$t_g = \vert\mathcal{G}\vert \cdot \vert\Gamma\vert \cdot \tau_p + \vert\mathcal{G}\vert \cdot \tau_g$ is based on the Eulerian path in the Cayley graph of $\mathcal{G}$ with respect to $\Gamma$ shown in figure \ref{fig::digraphDCG}.
$H_X$ denotes a potentially time-dependent control Hamiltonian which generates the generator $X$, i.\,e. up to a phase we have $X=\mathcal{T}\exp\bigl(-i \int_0^{\tau_p} H_X(t') dt' \bigr)$.
$H_Y$ is defined analogously.
Furthermore, $H_g$ denotes the Hamiltonian generating the quantum gate
$U_g = \mathcal{T}\exp\bigl(-i \int_0^{\tau_g} H_g(t') dt' \bigr)$ and $H_I$ denotes the Hamiltonian mirroring the error of $H_g$, but implementing the identity.
The gates generated by the applied Hamiltonians are denoted in the second line.\label{fig::eulerDCG}}
\end{figure}
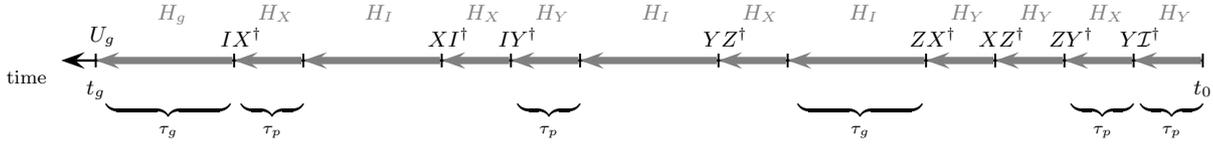
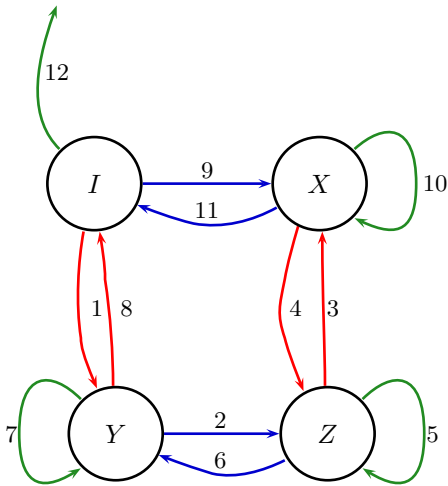
\begin{figure}
\begin{minipage}{.5\textwidth}\centering
\scalebox{1}{
\begin{pspicture}(0bp,0bp)(168bp,170bp) %
\psset{linewidth=1bp}

  \pstVerb{2 setlinejoin} %
\psset{linecolor=black}
  \psset{linecolor=mediumblue}
  \psbezier[arrows=->](52bp,117bp)(63bp,117bp)(77bp,117bp)(100bp,117bp)
  \psset{linecolor=black}
  \rput(76bp,122bp){\small $9$}
  \psset{linecolor=red}
  \psbezier[arrows=->](30bp,99bp)(28bp,88bp)(27bp,72bp)(29bp,59bp)(29bp,56bp)(30bp,52bp)(35bp,40bp)
  \psset{linecolor=black}
  \rput(35bp,70bp){\small $1$}
  \psset{linecolor=red}
  \psbezier[arrows=->](41bp,41bp)(41bp,53bp)(40bp,68bp)(38bp,81bp)(38bp,84bp)(38bp,86bp)(36bp,99bp)
  \psset{linecolor=black}
  \rput(46bp,70bp){\small $8$}
  \psset{linecolor=forestgreen}
  \psbezier[arrows=->](134bp,36bp)(145bp,47bp)(157bp,47bp)(157bp,23bp)(157bp,5bp)(150bp,0bp)(134bp,10bp)
  \psset{linecolor=black}
  \rput(160bp,23bp){\small $5$}
  \psset{linecolor=mediumblue}
  \psbezier[arrows=->](105bp,13bp)(97bp,9bp)(87bp,6bp)(78bp,8bp)(75bp,9bp)(71bp,10bp)(58bp,15bp)
  \psset{linecolor=black}
  \rput(81bp,13bp){\small $6$}
  \psset{linecolor=mediumblue}
  \psbezier[arrows=->](60bp,23bp)(70bp,23bp)(82bp,23bp)(103bp,23bp)
  \psset{linecolor=black}
  \rput(81bp,28bp){\small $2$}
  \psset{linecolor=red}
  \psbezier[arrows=->](120bp,41bp)(120bp,55bp)(119bp,74bp)(119bp,99bp)
  \psset{linecolor=black}
  \rput(123bp,70bp){\small $3$}
  \psset{linecolor=forestgreen}
  \psbezier[arrows=->](131bp,130bp)(142bp,140bp)(154bp,140bp)(154bp,117bp)(154bp,100bp)(148bp,96bp)(131bp,104bp)
  \psset{linecolor=black}
  \rput(161bp,117bp){\small $10$}
  \psset{linecolor=forestgreen}
  \psbezier[arrows=->](29bp,36bp)(18bp,47bp)(6bp,47bp)(6bp,23bp)(6bp,5bp)(13bp,0bp)(29bp,10bp)
  \psset{linecolor=black}
  \rput(3bp,23bp){\small $7$}
  \psset{linecolor=red}
  \psbezier[arrows=->](110bp,101bp)(108bp,94bp)(106bp,87bp)(105bp,81bp)(103bp,71bp)(102bp,68bp)(105bp,59bp)(106bp,55bp)(107bp,52bp)(112bp,39bp)
  \psset{linecolor=black}
  \rput(109bp,70bp){\small $4$}
  \psset{linecolor=mediumblue}
  \psbezier[arrows=->](102bp,108bp)(93bp,103bp)(82bp,100bp)(71bp,102bp)(67bp,103bp)(63bp,104bp)(50bp,109bp)
  \psset{linecolor=black}
  \rput(76bp,107bp){\small $11$}
  \psset{linecolor=forestgreen}
  \psbezier[arrows=->](21bp,130bp)(10bp,141bp)(11bp,159bp)(20bp,184bp)
  \psset{linecolor=black}
  \rput(20bp,159bp){\small $12$}
{%
  \psset{linecolor=black}
  \psellipse[](34bp,117bp)(18bp,18bp)
  \rput(34bp,117bp){$I$}
}%
{%
  \psset{linecolor=black}
  \psellipse[](118bp,117bp)(18bp,18bp)
  \rput(118bp,117bp){$X$}
}%
{%
  \psset{linecolor=black}
  \psellipse[](121bp,23bp)(18bp,18bp)
  \rput(121bp,23bp){$Z$}
}%
{%
  \psset{linecolor=black}
  \psellipse[](42bp,23bp)(18bp,18bp)
  \rput(42bp,23bp){$Y$}
}%
\end{pspicture}}%
\end{minipage}%
\begin{minipage}{.5\textwidth}
\caption[Directed graph 2]{Eulerian path in the Cayley graph of $\mathcal{G}=\{\mathcal{I},X,Y,Z\}$ with respect to the generators $\Gamma=\{X,Y\}$. The edges colored by $X$ are depicted in blue, those colored by $Y$ are shown in red.
After a vertex is visited for the last time, a loop (depicted in green) is applied (with the exception that the final loop of the vertex assigned to the identity element is not closed).\label{fig::digraphDCG}}
\end{minipage}
\end{figure}

Subsection \ref{subsec:dcgs} dealt with a dynamically corrected gate (DCG) combining a single \textsf{PDD} cycle with the generation of a quantum gate $U_g$.
Thereby, the decoupling pulses constituting the \textsf{PDD} cycle were assumed to be implemented instantaneously (i.\,e. in the bang-bang fashion), while the quantum gate $U_g$ was assumed to be generated within the finite time $\tau_g$ using bounded controls: $U_g \equiv U_g(\tau_g)$ with $U_g(t) = \mathcal{T} \exp\bigl(-i \int_0^t H_g(t') dt' \bigr)$ for $t\in[0,\tau_g]$.
Let us assume now that the decoupling pulses have to be generated using bounded controls as well.
The standard decoupling condition demands that the action of the lowest-order average Hamiltonian of a basic decoupling cycle is trivial. If the decoupling scheme is given by the set $\mathcal{G} = \{ g_j \}_{j=0}^{n_c-1}$, the decoupling condition for a \textsf{PDD} cycle becomes (compare with equation~\eqref{eq:zerothorderoverlineh-inpdd})
\begin{equation}
\Pi_\mathcal{G}(H_0) \equiv \frac{1}{n_c} \sum_{j=0}^{n_c-1} g_j^\dagger H_0 g_j = \tr(H_0) \cdot \frac{1}{d}\mathcal{I}.
\end{equation}
As we know from subsection \ref{subsec:boundedctrls}, in order to maintain the decoupling condition from above for finite pulses of duration $\tau_p$, the \textsf{PDD} cycle of length $t_c=n_c\Delta t$ has to be replaced by an Eulerian cycle.
To construct an Eulerian cycle, the elements of the decoupling scheme $\mathcal{G}$ have to form a group (strictly speaking a projective representation $R$ of a group is sufficient).
After choosing a subset of generators $\Gamma$, an Eulerian cycle is obtained by choosing an Eulerian path in the Cayley graph of $\mathcal{G}$ with respect to~$\Gamma$~\cite{VK03}.

We are now going to show how an Eulerian decoupling cycle has to be modified in order to generate a dynamically corrected gate.
As discussed in subsection \ref{subsec:dcgs}, a DCG generates the quantum gate $U_g$ within the last step of a \textsf{PDD} cycle visiting the identity element.
The error produced by generating the gate has to be mirrored during all the remaining steps of the \textsf{PDD} cycle.
Since an Eulerian cycle visits each element $g_j\in\mathcal{G}$ exactly $\Gamma$ times, this means that we have to implement the identity-gates mirroring the gate error only once, say after an element is visited for the last time.
As a result, the duration of an Euler-DCG is given by $t_g = n_c \vert\Gamma\vert \tau_p + n_c \tau_g$ (compared with the scenario depicted in figure \ref{fig::euler}, we set $\Delta t=\tau_p$ for simplicity).
Hence the zeroth-order average Hamiltonian of an Euler-DCG implementing $U_g$ is given by
\begin{align}
\overline{H}^{(0)} &= \frac{1}{t_g} \sum_{j=0}^{n_c-1} g_j^\dagger \Bigl(
F_\Gamma(H_0)\cdot\vert\Gamma\vert\tau_p + \int_0^{\tau_g} U_g^\dagger(t') H_0 U_g(t') dt'  \Bigr) g_j \nonumber\\
&=\frac{1}{\vert\Gamma\vert\tau_p+\tau_g} \Bigl( \Pi_\mathcal{G}(H_0)\cdot \vert\Gamma\vert\tau_p +\Pi_\mathcal{G}\bigl( \int_0^{\tau_g} U_g^\dagger(t') H_0 U_g(t') dt' \bigr) \Bigr), \label{eq:EDCGcondi}
\end{align}
where we used definition \eqref{eq:FGamma} and theorem \ref{thm:vk03} from subsection \ref{subsec:boundedctrls}.
If, in addition, the gate Hamiltonian generating $U_g(t')$ is an element of the group algebra $\mathcal{A}=R(\mathbb{C}G)$, analogous to theorem \ref{thm:vk03} we finally arrive at $\overline{H}^{(0)}=\Pi_\mathcal{G}(H_0)$, i.\,e we recover the standard decoupling condition that the action of $\Pi_\mathcal{G}(H_0)$ has to be trivial.
(Note that in order to achieve universal quantum computation, not all the gate Hamiltonians are allowed to be in $\mathcal{A}$. Hence, finding a decoupling scheme satisfying \eqref{eq:EDCGcondi} is not trivial.
The problem might be solved by a suitable subsystem encoding, for example. Another possibility would be to employ multiple decoupling groups with different group algebras.)
To illustrate the method, figure \ref{fig::eulerDCG} shows an Euler-DCG corresponding to the Eulerian path in the Cayley graph of $\mathcal{G}=\{\mathcal{I},X,Y,Z\}$ with respect to $\Gamma=\{X,Y\}$ which is depicted in figure \ref{fig::euler}.

\subsubsection{Fidelity of Protected Computations}

We are now going to analyze the entanglement fidelity of a quantum computation whose gates are all realized by Euler-DCGs.
The analysis is performed as in subsection \ref{subsec:par_ohnekorrektur}.
The computation consists of $n_a$ iterations of a quantum algorithm $U_{QA} = U_{n_g}  \dots  U_2 \cdot U_1$ which is decomposed into $n_g$ elementary quantum gates.
In order to describe the time evolution of the perturbed algorithm, in subsection \ref{subsec:par_ohnekorrektur} the $j$-th gate $U_j$ of the ideal algorithm was replaced by the perturbed gate $U_j\cdot \exp\bigl(-i \delta H_j \bigr)$ \eqref{eq:efperturbHjr}, where in the case of instantaneously applied gates $\delta H_j$ was given by $\delta H_j = H_0\Delta t$.
Now each gate is realized as Euler-DCG.
If the underlying Eulerian cycle is based on a decoupling scheme $\mathcal{G}$ of length $n_c = \vert\mathcal{G}\vert$ together with a set of generators $\Gamma\subset\mathcal{G}$, the corresponding gate time is now given by $t_g = n_c \vert\Gamma\vert \tau_p + n_c\tau_g$.
Hence the gate error is mainly due to the first-order correction following the zeroth-order term \eqref{eq:EDCGcondi} describing an Euler-DCG, i.\,e. we have $\delta H_j = \overline{H}^{(1)} = \mathcal{O}\bigl( (H_0)^2 \cdot t_g \bigr)$.
If the quantum algorithm $U_{QA}$ describes a quantum map, according to equation \eqref{eq:FeappQMap} we expect the resulting fidelity to behave as
\begin{equation}%
F_{e\ \text{app}}^\textsf{Euler-DCG}(n_a) = \exp \Bigl(
- \frac{n_a}{t_a} - \frac{2}{d \sigma} \frac{n_a^2}{t_a}
 \Bigr),
\end{equation}
where $t_a$ is now of the order $1/t_a = \mathcal{O}\bigl( n_g^2\cdot (H_0)^4 \cdot t_g^2 \bigr)$.

\subsection{Combining the \textsf{PAREC} Method with Euler-DCGs}\label{subsec:DCGPAREC}

The \textsf{PAREC} method can be understood as translating a $n_g$ gate decomposition $U_{QA}= U_{n_g} \dots U_1 \cdot U_1$ of a quantum algorithm $U_{QA}$ into a new gate decomposition containing twice as much gates.
If we consider $n_a$ iterations of the quantum algorithm, we obtain \eqref{eq::uparec}:
\begin{equation}\label{eq:uparecC}
U_{QA}^{n_a} = g_{[n_a,n_g]}^\dagger \bigl( V^{(n_a)}_{2n_g} \dots V^{(n_a)}_2 V^{(n_a)}_1 \bigr) \dots
\bigl(V^{(2)}_{2n_g} \dots V^{(2)}_2 V^{(2)}_1\bigr)
\bigl(V^{(1)}_{2n_g} \dots V^{(1)}_2 V^{(1)}_1\bigr),
\end{equation}
with $V^{(\tau)}_{2k} = g_{[\tau,k]}  \cdot U_k  \cdot g_{[\tau,k]}^\dagger$ and $V^{(\tau)}_{2k-1} = g_{[\tau,k]}g_{[\tau,k-1]}^\dagger$ for $k=1,2,...,n_g$ and $\tau=1,2,\dots,n_a$.
In subsection \ref{subsec:parec_ana} we showed that a formula for the fidelity decay of a \textsf{PAREC} computation is given by~\eqref{eq:FeappParec},
\begin{equation}
F_{e\ \text{app}}^\textsf{PAREC}(n_a) = \exp\Bigl( -  n_a n_g \frac{1}{d} \tr\bigl( H_0^2 \bigr) \Delta t^2  \Bigr),
\end{equation}
where we considered the simplified scenario in which each pulse $V^{(\tau)}_{2k-1}$ and each gate $V^{(\tau)}_{2k}$ is generated instantaneously, and where $V^{(\tau)}_{2k}$ is separated from $V^{(\tau)}_{2k-1}$ by the time interval $\Delta t/2$.

In order to combine the \textsf{PAREC} method with the use of Euler-DCGs we simply propose to implement each of the $2n_g+1$ gates in equation \eqref{eq:uparecC} as an Euler-DCG.
As a consequence, each pulse and each gate now takes up the time $t_g=n_c \vert\Gamma\vert \tau_p + n_c\tau_g$ instead of $\Delta t/2$. In addition, the error of a pulse and/or gate is now characterized by $\overline{H}^{(1)}$ instead of $H_0$, where $\overline{H}^{(1)} = \mathcal{O}\bigl( (H_0)^2 \cdot t_g \bigr)$ denotes the first-order correction following the zeroth-order term \eqref{eq:EDCGcondi} in the Magnus expansion of the average Hamiltonian of the Euler-DCG.
Hence we expect the fidelity of the combined stabilization method to be given by
\begin{equation}
F_{e\ \text{app}}^\textsf{PAREC+Euler-DCG}(n_a) = \exp\Bigl( -  n_a n_g \frac{1}{d} \tr\bigl( \bigl(\overline{H}^{(1)}\bigr)^2 \bigr) 4t_g^2  \Bigr).
\end{equation}
As it was the case for the embedded dynamical decoupling strategy (\textsf{EMD}) which was obtained by embedding periodic dynamical decoupling (\textsf{PDD}) into naive random decoupling (\textsf{NRD}), the combined stabilization method for computations allows us to benefit from the advantages of both underlying methods:
The strong suppression $\mathcal{O}\bigl( (H_0)^4 \bigr)$ of the Euler-DCGs and the linear decay $\mathcal{O}\bigl( n_a \bigr)$ of the \textsf{PAREC} method.

\chapter{Selective Recoupling and Randomized Decoupling}\label{chap:decrec}

In chapter \ref{chap:dyncontrol} we considered decoupling strategies which, with the help of instantaneously applied pulses (bang-bang pulses), suppressed the action of a system Hamiltonian describing static imperfections of a quantum memory, for instance.
The performance of the fundamental decoupling strategy --- called periodic dynamical decoupling (\textsf{PDD}) --- was significantly improved by embedding it into a naive random decoupling strategy (\textsf{NRD}).
As a result we obtained the so-called embedded decoupling strategy (\textsf{EMD}), which combines the advantages of both underlying strategies (strong suppression and linear fidelity decay).
In an analogous fashion, by embedding the symmetrized decoupling strategy (\textsf{SDD}), we obtained embedded symmetric decoupling (\textsf{ESDD}).
We are now going to show how to embed a symmetric recoupling scheme.
In contrast to a decoupling scheme, a recoupling scheme leads to a non-vanishing zeroth-order average Hamiltonian describing the desired recoupling.
Hence, we have to be careful not to affect this zeroth-order term when trying to eliminate residual higher order terms.

As a specific example, let us consider the recently proposed recoupling scheme for dipole-coupled nuclear spins in a crystalline solid \cite{YLMY04}.
While in all previously proposed similar schemes \cite{JK99,LCYY00,SM01,Leu02} the evolution-time overhead
grows linearly with the number of spins, this particular scheme leads to an evolution-time overhead which is
independent of the number of spins involved.
Thus, it appears to be well suited for the stabilization of quantum information processors against unwanted inter-qubit interactions.
This recoupling scheme uses particular combinations of fast broadband and slower selective radio-frequency fields to turn off all couplings except those between two particularly selected ensembles of spins.
Thereby, spins within each ensemble representing a particular logical qubit are decoupled \cite{LGYY02}.
Furthermore, cross-couplings between selected ensembles are avoided by requiring that qubit couplings have to be much stronger than any other couplings within each ensemble.
Unwanted couplings are suppressed up to second-order average Hamiltonian theory with the help of time-symmetric pulse sequences.
Despite many advantages in this recoupling scheme the residual higher-order interactions accumulate coherently thus leading to a quadratic-in-time decay of the fidelity of any quantum state
(compare with subsection \ref{subsec:par_ohnekorrektur}).
This restricts the achievable time scales of reliable quantum computation significantly.

In this chapter it is demonstrated that the performance of this recoupling scheme can be improved significantly by embedding it into a stochastic decoupling scheme (\textsf{NRD}).
In contrast to a deterministic scheme which repetitively applies a certain sequence of pulses (compare with subsection \ref{subsec:detstrat}), the corresponding stochastic scheme selects its pulses randomly (compare with subsection \ref{subsec:randstrat}).
Stochastic schemes are advantageous whenever the set these pulses are chosen from is large.
In the case of an annihilator like the set of Pauli operators, for example, this set grows exponentially with the number of qubits.
By a suitable embedding of the recoupling scheme into a \textsf{NRD} scheme based on Pauli operators, the coherent accumulation of higher-order residual interactions can be destroyed to a large extent so that the fidelity decay of any quantum state is slowed down significantly to an almost linear-in-time one.
As a result, reliable quantum computation can be performed on significantly longer time scales.
The results presented in this chapter have been published in \cite{recoup}.

This chapter is organized as follows:
The basic ideas underlying the recently proposed deterministic recoupling scheme of reference \cite{YLMY04}
are summarized briefly in section \ref{sec:rec} for the sake of completeness.
In section \ref{sec:rec2} a simple restricted embedded decoupling scheme is introduced.
Though it already leads to first improvements in comparison with the deterministic selective recoupling scheme of reference \cite{YLMY04}, its error suppressing properties can still be improved significantly by an additional simple symmetrization procedure.
We analyze the stabilization properties of this symmetrized embedded recoupling scheme for a unitary two-qubit swap gate.
In section \ref{sec:saw} its stabilizing properties are investigated by applying it to the iterated quantum algorithm of the quantum sawtooth map \cite{shep129}.

\section{Deterministic Selective Recoupling of Qubits}\label{sec:rec}

In this section the basic ideas underlying the recently proposed recoupling scheme of reference \cite{YLMY04} are summarized.
In particular, the form and magnitude of the residual higher-order interaction is discussed which cannot be suppressed by the suggested pulse sequences.

Let us consider $n$ nuclear spin-$1/2$ systems in a crystalline solid which are interacting with an external static magnetic field in $z$-direction.
In the rotating wave approximation their Hamiltonian is given by \cite[chapter IV section II A]{nmrAbragam}
\begin{equation}\label{eq:sysham}
H_0 = \underbrace{- \sum_{k=0}^{n-1} \frac{\hbar \omega_k}{2} Z_k}_{H_Z} + \underbrace{\sum_{k=0}^{n-2}\sum_{l=k+1}^{n-1} \frac{J_{kl}}{4} \bigl( 2Z_kZ_l - X_kX_l - Y_kY_l \bigr) }_{H_D}
\end{equation}
with the Pauli spin operators $X$, $Y$, and $Z$.
Thereby, the Larmor frequencies $\omega_k$ of the first term characterize the interaction strengths of these spins with the external magnetic field.
Using a magnetic field gradient the $\omega_k$ are adjusted in such a way that the spins can be addressed individually.
The second term of the Hamiltonian \eqref{eq:sysham} describes the dipole-dipole interaction of the nuclear spins with the coupling strength $J_{kl}$ between spins $k$ and $l$ being inversely proportional to the cubic power of their distance.
To keep the notation as simple as possible, we set $\hbar=1$ for the remaining chapter.

\subsection{Decoupling}

If these nuclear spins are used as qubits of a quantum memory, for example, one has to protect them
against the perturbing influence of the interaction Hamiltonian \eqref{eq:sysham}.
In the framework of a deterministic decoupling scheme (chapter \ref{chap:dyncontrol}) this may be achieved by
an appropriate sequence of fast electromagnetic pulses.
For $\alpha\in\{X,Y,Z\}$, let us define a global $\pi/2$-pulse as
\begin{equation}\label{eq:Palpha}
 P_\alpha = \bigotimes_{k=0}^{n-1} \exp\bigl( -i \alpha_k \pi/4 \bigr)
 = P^\dagger_{\bar{\alpha}}.
\end{equation}
Analogously, a global $\pi$-pulse is defined as $\alpha^{\otimes n}$.
A decoupling scheme for the Zeeman term $H_Z$ is given by the set $\{ \mathcal{I}, X^{\otimes n}\}$, for instance. Hence, in order to suppress $H_Z$, a series of fast global $X^{\otimes n}$-pulses is applied, leaving the dipole-dipole coupling term $H_D$ invariant.
This latter term can be suppressed by the well known WHH scheme $\{ \mathcal{I},P_x,  P_{\bar{y}}P_x\}$ (\cite{WHH68}, subsection \ref{subsec:localhams}).
Using the symmetric dynamical decoupling (\textsf{SDD}) strategy, the WHH pulse sequence consists of four fast $\pi/2$-pulses applied at times $\Delta t$, $2\Delta t$, $4\Delta t$ and $5\Delta t$.
Thus, the resulting unitary time evolution after this pulse sequence, i.\,e. at time $t_c = 6\Delta t$, is given by
\begin{equation}
\begin{split}
U(t_c) &=
\exp(-iH_D\Delta t) P_{\bar{x}} \exp(-iH_D\Delta t) P_y \exp(-iH_D2\Delta t) P_{\bar{y}} \exp(-iH_D\Delta t) P_x \exp(-iH_D\Delta t) \\
&\equiv  \exp(-i\tilde{H}_6\Delta t)\dots\exp(-i\tilde{H}_2\Delta t)\exp(-i\tilde{H}_1\Delta t),\\
\end{split}
\end{equation}
with the interaction-picture (toggled) Hamiltonians
$\tilde{H}_1 = \tilde{H}_6 = H_D$,
$\tilde{H}_2 = \tilde{H}_5 = P_{\bar{x}} \hat{H}_D \hat{P}_x$
and
$\tilde{H}_3 = \tilde{H}_4 = P_{\bar{x}}P_y H_D P_{\bar{y}} P_x$.
As a consequence, in zeroth-order average Hamiltonian theory (AHT) the time-averaged Hamiltonian vanishes, i.\,e.
\begin{equation}
 \overline{H}_D^{(0)} = \frac{1}{6} \sum_{j=1}^6 \tilde{H}_j = 0.
\end{equation}
Due to the time reversal symmetry of the WHH pulse sequence, i.\,e.
$\tilde{H}(t) = \tilde{H}(t_c-t)$, in AHT all odd higher-order Hamiltonians vanish (theorem \ref{thm:symtoggledh}): $\overline{H}_D^{(2i+1)}=0$ for $i\in \mathbb{N}_0$.

\subsection{Selective Recoupling}

\begin{figure}
\includegraphics[width=\textwidth]{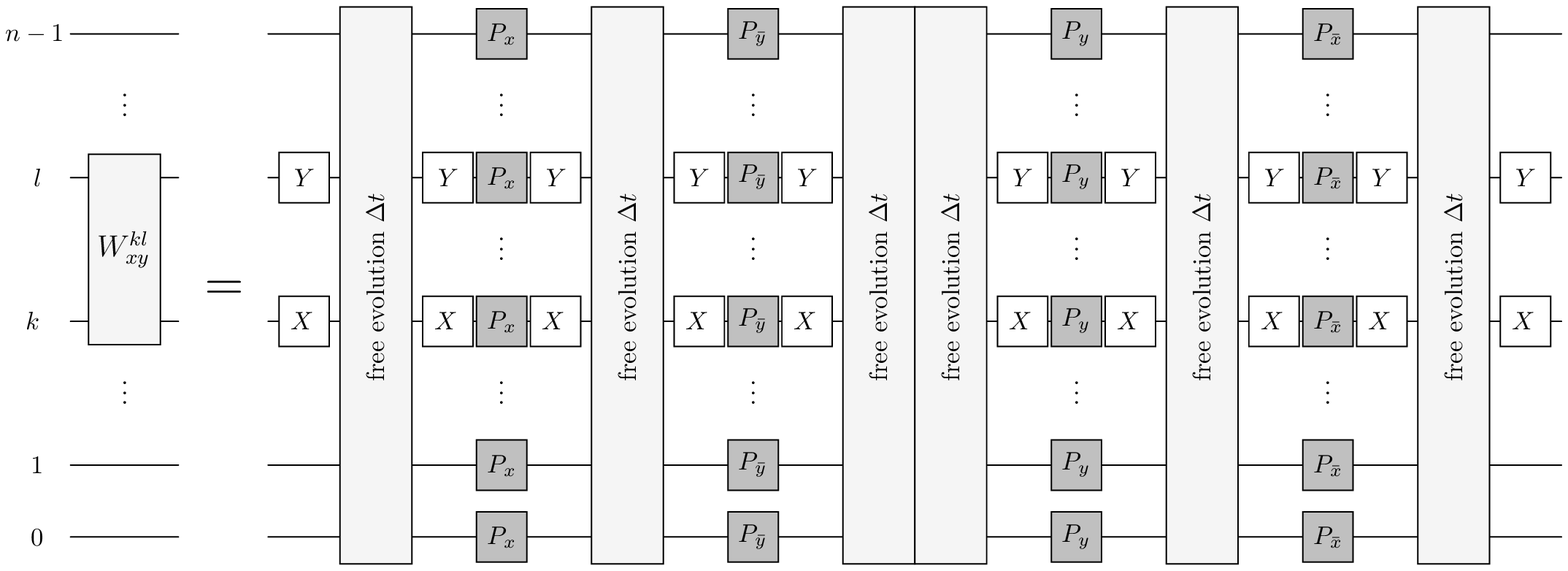}
\caption{Schematic representation of the unitary $W_{xy}^{kl}$ quantum gate acting on qubits $k$ and $l$:
\textit{Free evolution} indicates time evolution according to the Hamiltonian $H_D$ over a time interval of duration $\Delta t$.\label{fig:wxy}}
\end{figure}

\begin{figure}
\includegraphics[width=\textwidth]{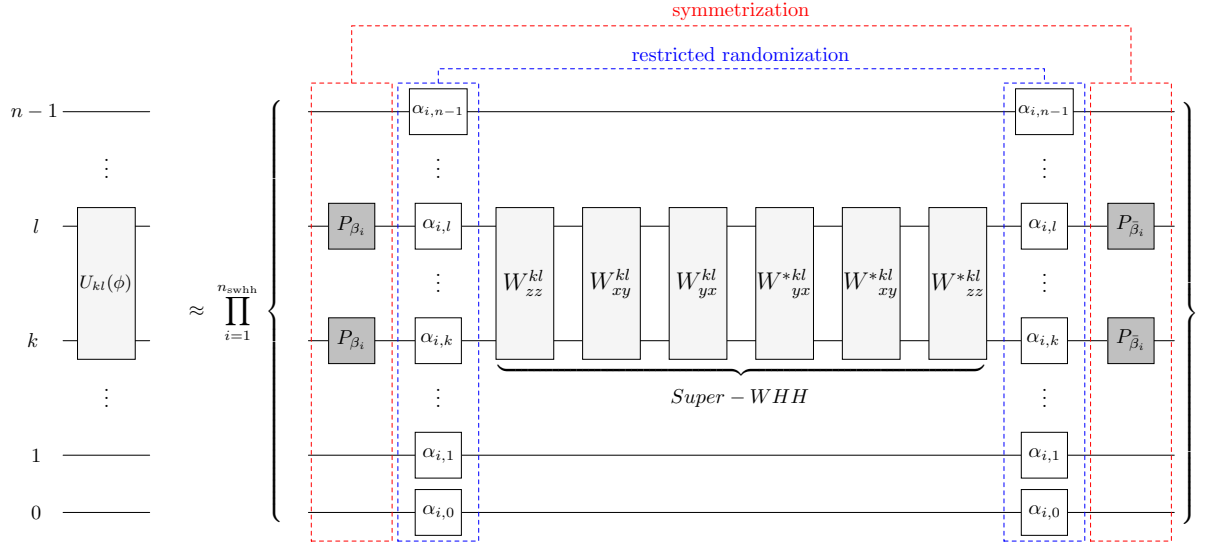}
\caption{Schematic representation of the two-qubit gate $U_{kl}(\phi)$ obtained by recoupling qubits $k$ and $l$ according to equation \eqref{eq:uphi}:
The $W^*$ gates are obtained by reversing the order of the broadband pulses suppressing the Zeeman term.
Residual second-order terms of AHT can be eliminated by the restricted randomization step accomplished by random selective $\pi$-pulses $\alpha_{i,j} \in \{\mathcal{I},X,Y,Z\}$.
Thereby $\alpha_{i,k}$ has to be equal to $\alpha_{i,l}$ to ensure that the wanted gate action is not disturbed.
Still remaining terms are symmetrized by random $\pi/2$-pulses
$P_{\beta_i} = \exp\bigl( -i \beta_i \pi/4 \bigr)$, $\beta_i \in \{X,Y,Z\}$.
Either condition \eqref{eq:0bedingung} or condition \eqref{eq:2bedingung} has to be fulfilled depending on whether the original Super-WHH or the symmetrized Super-WHH sequence is used.\label{fig:uphi}}
\end{figure}

If these nuclear spins are used as qubits of a quantum information processor one also has to implement universal sets of unitary quantum gates.
In particular, one needs to be able to implement two-qubit entanglement gates, such as controlled-phase gates.
This can be accomplished by recoupling qubits selectively with the help of a Super-WHH pulse sequence as proposed in reference \cite{YLMY04}.
Such a Super-WHH sequence recoupling qubits $k$ and $l$ consists of three WHH sequences applied to the toggled Hamiltonians
\begin{subequations}
\begin{align}
\tilde{H}_{zz}^{kl} &= Z_kZ_l H_D Z_kZ_l,\\
\tilde{H}_{xy}^{kl} &= X_kY_l H_D X_kY_l,\\
\text{ and } \tilde{H}_{yx}^{kl} &= Y_kX_l H_D Y_kX_l,
\end{align}
\end{subequations}
respectively.
Correspondingly, there are 18 time periods of duration $\Delta t$ during which the time evolution is described by the double-toggled Hamiltonians
$\tilde{\tilde{H}}_1 = \tilde{H}_{zz}^{kl}$,
$\tilde{\tilde{H}}_2 = \hat{P}_{\bar{x}} \tilde{H}_{zz}^{kl} \hat{P}_x$, et cetera.
The appropriate WHH pulse sequence of the $\tilde{H}_{xy}^{kl}$ Hamiltonian, for example,
is illustrated in figure \ref{fig:wxy}, where \textit{free evolution} denotes the time evolution according to the Hamiltonian $H_D$ over a time interval of duration $\Delta t$.
The quantum gates resulting from these WHH sequences are denoted by $W_{xy}^{kl}$, $W_{zz}^{kl}$, and $W_{yx}^{kl}$, respectively.
The Super-WHH sequence is finally obtained from a combination of these latter quantum gates preceeded
by the corresponding time reversed sequence (compare with the inner part of figure \ref{fig:uphi}).
As a consequence \cite{YLMY04}, this Super-WHH sequence yields the average Hamiltonian $\overline{H}_D = \overline{H}_D^{(0)} + \overline{H}_D^{(1)} + \overline{H}_D^{(2)} + \dots$, with
\begin{equation}\label{eq:hr}
\overline{H}_D^{(0)} = \frac{1}{t_c} \sum_{j=1}^{n_c} \tilde{\tilde{H}}_j \Delta t=
J^{(0)}_{kl} \bigl( X_kX_l + Y_kY_l + Z_kZ_l \bigr),
\end{equation}
$n_c=36$, $t_c=n_c\Delta t$, and with the renormalized zeroth-order recoupling strength $J^{(0)}_{kl} = (J_{kl}/4)\times(8/9)$.
Due to the time reversal symmetry of the Super-WHH sequence, in AHT all odd-valued higher order Hamiltonians vanish, i.\,e. $\overline{H}_D^{(1)} = \overline{H}_D^{(3)} = 0$, et cetera.
Note that in contrast to the selective decoupling schemes of subsection \ref{subsec:seldec}, the selective recoupling scheme presented above changes the form of the selected coupling (from $2Z_kZ_l - X_kX_l - Y_kY_l$ in \eqref{eq:sysham} to $X_kX_l + Y_kY_l + Z_kZ_l$ in \eqref{eq:hr}).

With the help of the zeroth-order recoupled Hamiltonian $\overline{H}_D^{(0)}$ of equation \eqref{eq:hr} one can approximate unitary two-qubit quantum gates of the form
\begin{equation}\label{eq:uphi}
U_{kl}(\phi) = \exp\bigl( -i ( X_kX_l + Y_kY_l + Z_kZ_l ) \phi \bigr) \equiv \exp\bigl( -i H_g^{kl} \phi \bigr).
\end{equation}
Thereby, for a particular value of the phase $\phi$ one has to adjust the time $\Delta t$ between two successive pulses of a WHH sequence and the number of times $n_\text{swhh}$ a Super-WHH sequence has to be applied according to the relation
\begin{equation}\label{eq:0bedingung}
J^{(0)}_{kl} \cdot n_\text{swhh} n_c\Delta t = \phi %
\end{equation}
(compare with figure \ref{fig:uphi}).
However, because of the residual higher-order interactions which have not been canceled by the Super-WHH pulse sequence, this implementation of a two-qubit quantum gate is only approximate.
The error resulting from these residual higher-order interactions is dominated by the second-order term of AHT which is given by \eqref{eq:bbh2},
\begin{equation}
\overline{H}_D^{(2)} = -\frac{1}{6t_c} \sum_{i\geq j\geq k=1}^{n_c}
 \Bigl( [\tilde{\tilde{H}}_i,[\tilde{\tilde{H}}_j,\tilde{\tilde{H}}_k]]+
 [[\tilde{\tilde{H}}_i,\tilde{\tilde{H}}_j],\tilde{\tilde{H}}_k] \Bigr)\Delta t^3 \times
\begin{cases}
1/2 & \text{ if } i=j \text{ or } j=k\\
1 & \text{ else }
\end{cases}.
\end{equation}
Therefore, the lowest-order correction to the recoupled Hamiltonian of equation \eqref{eq:hr} is given by
\begin{equation}\label{eq:hd2}
\begin{split}
\overline{H}_D^{(2)} = \sum_a^{\neq k,l} \Bigl[ &
  X_kX_l \Bigl( -322 J_{al}^2 J_{ak} +446 J_{ak}^2 J_{al}
+3628 J_{al} J_{ak} J_{kl} -2906 J_{ak}^2 J_{kl} -1370 J_{al}^2 J_{kl} \Bigr) + \\
 &Y_kY_l \Bigl( +308 J_{al}^2 J_{ak} +308 J_{ak}^2 J_{al} +3208 J_{al} J_{ak} J_{kl}
 -2588 J_{ak}^2 J_{kl} -2588 J_{al}^2 J_{kl} \Bigr) +\\
 &Z_kZ_l \Bigl( +446 J_{al}^2 J_{ak} -322 J_{ak}^2 J_{al} +3580 J_{al} J_{ak} J_{kl}
 -1922 J_{ak}^2 J_{kl} -3458 J_{al}^2 J_{kl} \Bigr) \\
& \Bigr] \Delta t^2/1728 + \dots \, .
\end{split}
\end{equation}
Thereby, only terms of the form
$\alpha_k \beta_l =
\alpha_k\otimes\beta_l\otimes\mathcal{I}_{\{0,1,\dots,n-1\}\setminus\{k,l\}}$ with $\alpha,\beta \in \{X,Y,Z\}$ are indicated as all other terms are irrelevant for our subsequent discussion.
As a consequence, the gate Hamiltonian resulting from recoupling qubits $k$ and $l$ by a Super-WHH sequence is of the form
\begin{align}\label{eq:hrklp1}
H'^{kl}_g=\overline{H}_D &= \overline{H}_D^{(0)} + \overline{H}_D^{(2)} + \overline{H}_D^{(4)} + \dots \nonumber\\
          &=  J^{(0)}_{kl} \bigl( X_kX_l + Y_kY_l + Z_kZ_l \bigr) + \mathcal{O}\bigl(J (J\Delta t)^2 \bigr).
\end{align}
To estimate the resulting error affecting the unitary gate $U'_{kl}(\phi)$ generated by $H'^{kl}_g$
we study the entanglement fidelity given by \eqref{eq:FeUU},
\begin{equation}
 F_e = \Bigl\vert \frac{1}{d}\tr\Bigl( U^\dagger_{kl}(\phi) \cdot U'_{kl}(\phi)    \Bigr) \Bigr\vert^2,
\end{equation}
comparing the action of $U'_{kl}(\phi)$ with the action of the ideal gate $U_{kl}(\phi)$ generated by $H^{kl}_g$.
A short time expansion of $F_e$ can be derived by using the following lemma.

\begin{lem}
Let $x$ and $y$ denote Hermitian operators, and let the unitaries $U$ and $U'$ be defined as $U=\exp(-i x t)$ and $U'=\exp\bigl(-i (x+y) t\bigr)$, respectively.
Then a series expansion of $U^\dagger\cdot U'$ is given by
\begin{equation}
U^\dagger\cdot U' =
\mathcal{I} -iyt +\frac{1}{2}[x,y]t^2  -\frac{1}{2}y^2 +\frac{i}{6}[x,[x,y]]t^3 -\frac{i}{6}y[x,y]t^3 -\frac{i}{3}[x,y]yt^3 +\mathcal{O}(t^4).
\end{equation}
\end{lem}

\noindent
By setting $x=H^{kl}_g$, $y=\overline{H}_D^{(2)} + \overline{H}_D^{(4)} + \dots$,
and the gate-time $t \equiv n_\text{swhh} n_c \Delta t = \phi/J^{(0)}_{kl}$ according to condition \eqref{eq:0bedingung}, we obtain the expression
\begin{align}
 F_e = \Bigl\vert \frac{1}{d}\tr\Bigl( U^\dagger_{kl}(\phi) \cdot U'_{kl}(\phi)    \Bigr) \Bigr\vert^2
&= 1 - \frac{1}{d}\tr\Bigl(  \bigl(\overline{H}_D^{(2)}\bigr)^2 \Bigr)t^2 + \mathcal{O}\bigl(t^4\bigr) \\
&= 1 -\mathcal{O}\bigl( (J^3\Delta t^2  \cdot n_\text{swhh} n_c \Delta t)^2 \bigr)
= 1 -\mathcal{O}\bigl(  \phi^6/(n_\text{swhh} n_c)^4  \bigr). \label{eq:FeUphi}
\end{align}
For a fixed phase $\phi$, the strength of the fidelity decay of $U'_{kl}(\phi)$ is inversely proportional to the fourth power of the number of Super-WHH iterations.

\section{Embedded Selective Recoupling}\label{sec:rec2}

The selective recoupling scheme of the preceding section applies the symmetric dynamical decoupling (\textsf{SDD}, see subsection \ref{subsec:detstrat}) strategy in order to get a vanishing first-order term in the Magnus expansion of the average Hamiltonian describing the time evolution of a single recoupling cycle.
However, in contrast to \textsf{SDD} the zeroth-order AHT term does not vanish and describes the desired recoupling.
In this section we are going to show how the recoupling scheme can be embedded into a naive random decoupling (\textsf{NRD}, see subsection \ref{subsec:randstrat}) scheme.
By embedding \textsf{SDD} into \textsf{NRD}, we devised the embedded symmetric decoupling (\textsf{ESDD}, subsection \ref{subsec:randstrat}) strategy combining the advantages of both underlying strategies.
Now, however, we have to prevent the \textsf{NRD} pulses from averaging out the desired recoupling action, i.\,e. they should merely suppress the remaining second (and higher) order AHT term(s) and leave the zeroth-order term unaffected.
As a consequence, we are not able to suppress the remaining terms entirely.
Fortunately, the non-suppressible part can be cast into the form of the desired recoupling, thereby simply renormalizing the effective recoupling strength.

\subsection{Embedding the Selective Recoupling Scheme}

The residual interaction described by the Hamiltonian \eqref{eq:hd2} can be suppressed significantly
by embedding the recoupling scheme of section \ref{sec:rec} into a naive random decoupling (\textsf{NRD}) scheme based on an annihilator as the set of Pauli operators $\mathcal{P}_2^n$.
For this purpose we choose at random an $n$-fold tensor product of Pauli-matrices
$\alpha_{i,0} \otimes \alpha_{i,1} \otimes \dots \otimes \alpha_{i,n-1}$, with
$\alpha_{i,j} \in \mathcal{P}_2=\{\mathcal{I},X,Y,Z\}$ for $j=0,1,\dots,n-1$, and apply it before and after the $i$-th Super-WHH sequence.
This way each deterministic Super-WHH sequence is embedded within two statistically independent random Pauli operations.
In contrast to a usual dynamical decoupling scenario (\cite{combi}, chapter \ref{chap:dyncontrol}) in our case we have to choose the Pauli-matrices in such a way that they leave the ideally recoupled gate Hamiltonian $H^{kl}_g$ of equation \eqref{eq:uphi} invariant.
This can be achieved by imposing the restriction that the randomly chosen statistically independent Pauli spin operators have to be identical for qubits $k$ and $l$ for each Super-WHH sequence, i.\,e. $\alpha_{i,k} = \alpha_{i,l}$ for all $i\in\{1,\dots,n_\text{swhh}\}$.
This restriction assures that terms of the form $\alpha_k \alpha_l$ in $H^{kl}_g$ remain invariant (compare with figure \ref{fig:uphi}).
Since $\overline{H}_D^{(2)}$ contains no terms of the form $\alpha_k \beta_l$ with $\alpha \neq \beta$
(compare with equation \eqref{eq:hd2}) the Pauli-matrices for qubits $k$ and $l$ can always be omitted, i.\,e. chosen to be the identity, $\alpha_{i,k}=\alpha_{i,l}=\mathcal{I}$.

The only terms of the Hamiltonian $\overline{H}_D^{(2)}$ which cannot be eliminated by this constrained randomization method are the ones containing terms of the form
$\alpha_k \alpha_l$ ($\alpha \in \{X,Y,Z\}$) which are shown in equation \eqref{eq:hd2}.
However, by an additional symmetrization these terms can be made rotationally invariant so that they can be cast into the form of equation \eqref{eq:hr}.
Thus, for a given value of $\phi$ these terms lead to a renormalization of the values of the required gate parameters $\Delta t$ and $n_\text{swhh}$.
This rotational symmetrization can be achieved by selective $\pi/2$-pulses as defined in equation \eqref{eq:Palpha}.
For this purpose one chooses one of the three unitary transformations $\{ P_{\beta_i,k} P_{\beta_i,l} \}_{\beta_i \in \{X,Y,Z\}}$ acting on qubits $k$ and $l$ at random and applies it before and the corresponding inverse transformation after the $i$-th Super-WHH sequence (compare with Fig. \ref{fig:uphi}).
This way the coefficients of the $\alpha_k \alpha_l$-terms are permuted in the relevant toggled Hamiltonians.
As a consequence one obtains the statistically and rotationally averaged second-order contribution
\begin{equation}
\mathbb{E} \overline{H}_D^{(2)} =
\bigl(X_kX_l + Y_kY_l + Z_kZ_l\bigr) J^{(2)}_{kl} \Delta t^2
\end{equation}
with
\begin{equation}\label{eq:Jkl2}
J^{(2)}_{kl} = \sum_a^{\neq k,l} \Bigl(  \frac{1}{12} ( J_{al}^2 J_{ak} + J_{ak}^2 J_{al} )
+\frac{217}{108} J_{al} J_{ak} J_{kl} -\frac{103}{72} ( J_{ak}^2 J_{kl} +  J_{al}^2 J_{kl} ) \Bigr).
\end{equation}
Here, $\mathbb{E}$ denotes the average taken over the $\alpha_{i,j} \in \mathcal{P}_2=\{\mathcal{I},X,Y,Z\}$ and the $\beta_i \in \{X,Y,Z\}$.
By this combined randomization and symmetrization method the improved recoupled Hamiltonian 
\begin{align}\label{eq:hrklp2}
H''^{kl}_g &= \mathbb{E} \overline{H}_D^{(0)} + \mathbb{E} \overline{H}_D^{(2)} + \mathbb{E} \overline{H}_D^{(4)} + \dots \nonumber\\
 &= \Bigl( J^{(0)}_{kl} + J^{(2)}_{kl} \Delta t^2 + \mathcal{O}\bigl( J(J\Delta t)^4 \bigr) \Bigr)
\times \bigl( X_kX_l + Y_kY_l + Z_kZ_l \bigr)
\end{align}
is obtained.
In contrast to $H'^{kl}_g$ given by equation \eqref{eq:hrklp1}, now
the effective recoupling strength is renormalized and  the residual error is suppressed up to fourth order in the small coupling parameter $J\Delta t \ll 1$.
Thus, in order to implement a $U_{kl}(\phi)$-gate, for example, we now have to choose the renormalized characteristic parameter $\Delta t'$ in such a way that the condition
\begin{equation}\label{eq:2bedingung}
\bigl( J^{(0)}_{kl} + J^{(2)}_{kl} \Delta t'^2 \bigr) n_\text{swhh} n_c \Delta t' = \phi
\end{equation}
is fulfilled. As a result, in general the required time of free evolution $\Delta t'$ depends on the chosen qubit pair $(k,l)$.

\subsection{Performance of a Recoupled Quantum Gate}\label{sec:pi4gate}

In this section the stabilizing properties of selective recoupling by the
embedded symmetric dynamical decoupling (\textsf{ESDD}) method of the preceding section is investigated for a unitary phase gate as described by equation \eqref{eq:uphi}.
As shown in equation \eqref{eq:FeUphi} the fidelity of a unitary phase gate $U_{kl}(\phi)$ which is realized by recoupling qubits $k$ and $l$ with the help of the average Hamiltonian of equation \eqref{eq:hrklp1} (i.\,e. by applying the \textsf{SDD} strategy)
deviates from unity by terms of the order of $\mathcal{O}\left(\phi^6/(n_\text{swhh}n_c)^4\right)$.
Here, $n_\text{swhh}$ denotes the number of required iterations of the Super-WHH sequence which is related to the time $\Delta t$ of the intermediate free evolution and the phase $\phi$ as determined by relation \eqref{eq:0bedingung}.

In order to estimate the improvement achievable with the help of the embedded recoupling scheme, let us recall our result for the non-embedded original scheme \eqref{eq:FeUphi}:
\begin{align}
 F_e = \Bigl\vert \frac{1}{d}\tr\Bigl( U^\dagger_{kl}(\phi) \cdot U'_{kl}(\phi)    \Bigr) \Bigr\vert^2
= 1 - \frac{1}{d}\tr\Bigl(  \bigl(\overline{H}_D^{(2)}\bigr)^2 \Bigr)t^2 + \dots
= 1 -\mathcal{O}\bigl(  \phi^6/(n_\text{swhh} n_c)^4  \bigr).
\end{align}
This expression is of the same form as the short time expansion of the fidelity of a quantum memory protected using the \textsf{SDD} strategy \eqref{eq:FeSDD}.
As we found out in subsection \ref{subsec:randstrat}, the corresponding \textsf{ESDD} fidelity \eqref{eq:FeESDD} is obtained by replacing one power of the total time $t$ by the time of a basic cycle.
Applying these results to the recoupling case, this means that we have to replace one power of the total time $t=n_\text{swhh}\cdot n_c\Delta t$ in the preceding equation by the time $n_c\Delta t$ taken by a single Super-WHH cycle.
As a result we obtain the estimation
\begin{align}
 F_e = \mathbb{E} \Bigl\vert \frac{1}{d}\tr\Bigl( U^\dagger_{kl}(\phi) \cdot U'_{kl}(\phi)    \Bigr) \Bigr\vert^2
&= 1 - \frac{1}{d}\tr\Bigl(  \bigl(\overline{H}_D^{(2)}\bigr)^2 \Bigr) n_\text{swhh}n_c\Delta t \cdot n_c\Delta t + \dots \nonumber\\
&= 1 -\mathcal{O}\bigl(  \phi^6/(n_\text{swhh}^5 n_c^4)  \bigr), \label{eq:FeUphiESDD}
\end{align}
where we used condition \eqref{eq:0bedingung} to approximate the relevant condition \eqref{eq:2bedingung}.

\begin{figure}\centering
 \psset{unit=0.75cm}
 \begin{psmatrix}[colsep=1.8,rowsep=1.8,mnode=circle]
  [name=N3] \large 3 & [name=N2] \large 2 & [name=N1] \large 1 & [name=N0] \large 0
 \end{psmatrix}
 \ncline[]{N0}{N1}\naput{$J$}\ncline[]{N0}{N1}
 \ncline[]{N1}{N2}\naput{$J$}\ncline[]{N1}{N2}
 \ncline[]{N2}{N3}\naput{$J$}\ncline[]{N2}{N3}
 \vspace{3mm}\\
 \includegraphics{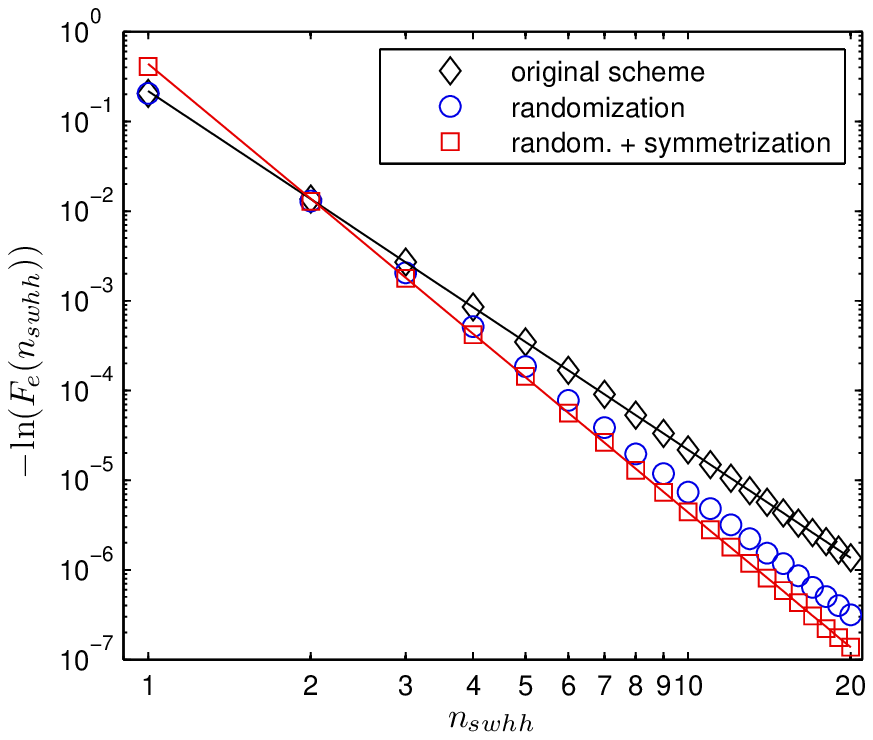}
\caption{The entanglement fidelity (\textit{bottom}) of the $U_{12}(\pi/4)$-gate on a linear four-qubit chain (\textit{top}) as a function of the number of repetitions $n_\text{swhh}$ of the Super-WHH sequence:
the original Super-WHH sequence (diamonds),
the unsymmetrized embedded scheme (circles),
and the complete embedded scheme with adapted pulse interval $\Delta t'$ according to \eqref{eq:2bedingung} (squares).
The solid lines represent the fitting functions $\exp(-c/n_\text{swhh}^4)$ and $\exp(-c_\text{ESDD}/n_\text{swhh}^5)$ with $c = 0.22$ and $c_\text{ESDD} = 0.44$.}
\label{fig:pi4gate}
\end{figure}

In figure \ref{fig:pi4gate} (\textit{bottom}) the entanglement fidelity $F_e$
of a unitary $U_{12}(\pi/4)$-gate and its dependence on the number of performed Super-WHH sequences $n_\text{swhh}$ is depicted.
In these numerical simulations this unitary quantum gate is realized by recoupling of the two central qubits $1$ and $2$ of a linear four-qubit chain (containing the qubits $0,1,2,$ and $3$).
The coupling strength is assumed to be constant for adjacent qubits and to be vanishing between all other qubits (compare with Fig. \ref{fig:pi4gate} (\textit{top})).
Apart from an irrelevant global phase this unitary $U(\pi/4)$-gate is nothing but a \textsf{SWAP}-gate (compare with figure \ref{fig:gates}).
The statistical averaging was performed over 100 runs with statistically independent realizations of the random pulses involved.
Figure \ref{fig:pi4gate} (\textit{bottom}) demonstrates that the fidelity (diamonds) resulting from non-embedded original Super-WHH pulse sequences can be fitted well by a function of the form $\exp(-c/n_\text{swhh}^4)$ with $c \approx 0.22$.
This is consistent with the simple estimate \eqref{eq:FeUphi}.
Using a recoupling scheme based on the embedded procedure discussed in section \ref{sec:rec2} while choosing $\Delta t$ according to condition \eqref{eq:2bedingung}, we notice that the resulting fidelity (squares) is fitted well by a function of the form $\exp(-c_\text{ESDD}/n_\text{swhh}^5)$ with $c_\text{ESDD} \approx 0.44$, which confirms our estimate \eqref{eq:FeUphiESDD}.
If symmetrization is omitted an intermediate behavior is obtained (circles).

\section{Numerical Simulation of a Quantum Algorithm}\label{sec:saw}

In this section the question is explored how much can be gained by stabilizing an iterative quantum algorithm by the embedded recoupling scheme of section \ref{sec:rec2}.
Using the embedded recoupling scheme to implement a quantum algorithm is reminiscent of the \textsf{PAREC}-method of section \ref{sec:parec} in the sense that each period of imperfect evolution is suppressed using naive random decoupling (\textsf{NRD}).
Hence, in addition to the improvement which is achieved for a single recoupled quantum gate, we expect the fidelity decay of a quantum algorithm using the embedded recoupling scheme to be linear in time instead of quadratic in time.

\subsection{Quantum Computation with a Recoupled Quantum Gate}\label{sec:qc}

For purposes of quantum computation one needs to know how to perform two-qubit entanglement gates,
such as the controlled-not gate (\textsf{CNOT}-gate) or the controlled-phase gate ($\textsf{CP}(\varphi)$-gate), on the basis of the recoupled Hamiltonian $H_g^{kl}$ \eqref{eq:uphi}.
Definitely, such quantum gates can be performed only between qubits $k$ and $l$ which are coupled, i.\,e. for which $J_{kl} \neq 0$.
\begin{figure}
 \hfill\includegraphics[scale=0.81]{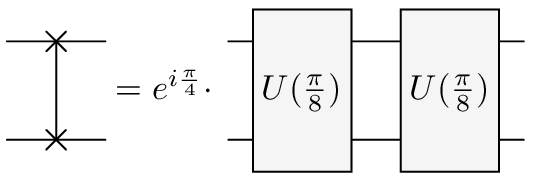}%
 \hfill\includegraphics[scale=0.81]{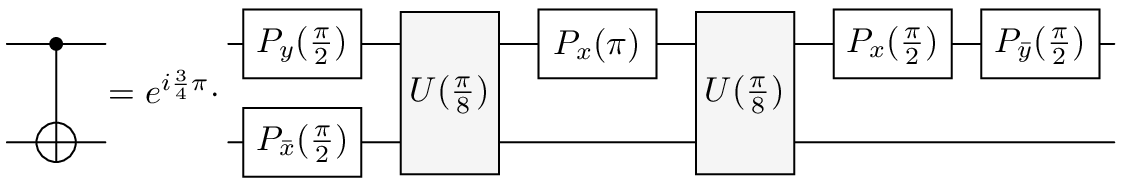}\hspace*{\fill}%
 \vspace{1cm}
 \includegraphics[scale=0.81]{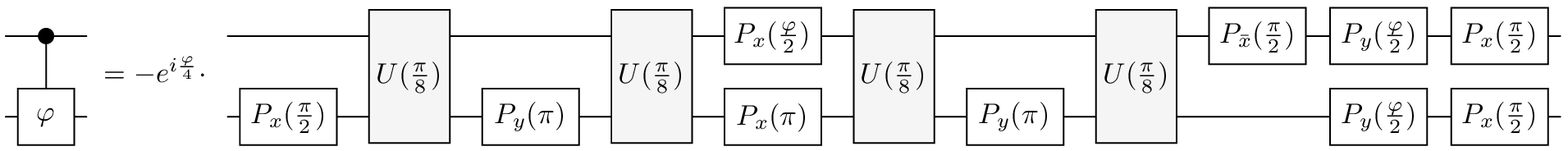}
\caption{Quantum circuits implementing the \textsf{SWAP}, the \textsf{CNOT}, and the controlled-phase gate $\textsf{CP}(\varphi)$ by using a $U(\pi/8)$ gate generated by Super-WHH recoupling.
The single qubit gate $P_\alpha(\varphi)$ is defined as
$P_\alpha(\varphi) = \exp\bigl(-i \alpha \varphi/2 \bigr) = P_{\bar{\alpha}}^\dagger(\varphi)$ for $\alpha \in \{X,Y,Z\}$.\label{fig:gates}}
\end{figure}
Therfore, in order to be able to entangle any two qubits of a quantum computer it is necessary to swap qubit pairs with vanishing coupling constants to neighboring positions.
Fortunately, such a unitary swapping gate can be realized easily by the unitary phase gate of equation \eqref{eq:uphi} because $\textsf{SWAP}_{kl} = U_{kl}(\pi/4)$.
Throughout the rest of this section we will use the quantum phase gate $U_{kl}(\pi/8)$
as a basic building block for all two-qubit quantum gates.
Thus, the quantum $\textsf{SWAP}_{kl}$-gate consists of the repeated application of two such gates.
For the realization of other two-qubit quantum gates repeated applications of this $U_{kl}(\pi/8)$-gate in
combination with single-qubit gates are required.
In figure \ref{fig:gates} basic gate decompositions are depicted for the \textsf{CNOT}-gate, the $\textsf{CP}(\varphi)$, and for the \textsf{SWAP}-gate.
A description of these gates can be found in appendix \ref{sec:qgates}.
These decompositions will be used in the next section for the simulation of a quantum algorithm.
The $U_{kl}(\pi/8)$-gate itself can be generated approximately either by repeated application of the original or of the embedded Super-WHH recoupling sequence using either condition \eqref{eq:0bedingung} or relation \eqref{eq:2bedingung} for the determination of the free evolution time $\Delta t$ between successive fast pulses.

\subsection{Lattice Model of a Quantum Computer}\label{subsec:latticem}

\begin{figure}
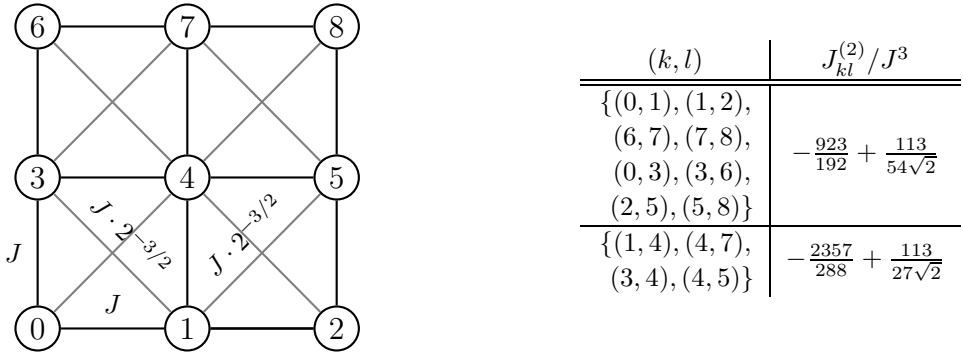

\hfill\begin{minipage}[t]{.45\textwidth}\centering
 \vspace{0pt}
 \psset{unit=0.75cm}
 \begin{psmatrix}[colsep=1.8,rowsep=1.8,mnode=circle]
  [name=N6] \large 6 & [name=N7] \large 7 & [name=N8] \large 8 \\[0pt]
  [name=N3] \large 3 & [name=N4] \large 4 & [name=N5] \large 5 \\[0pt]
  [name=N0] \large 0 & [name=N1] \large 1 & [name=N2] \large 2
 \end{psmatrix}
 \ncline[]{N0}{N1}\naput{$J$}\ncline[]{N1}{N2}
 \ncline[]{N3}{N4}\ncline[]{N4}{N5}
 \ncline[]{N6}{N7}\ncline[]{N7}{N8}
 \ncline[]{N0}{N3}\naput{$J$}\ncline[]{N1}{N2}\ncline[]{N3}{N6}
 \ncline[]{N1}{N4}\ncline[]{N4}{N7}
 \ncline[]{N2}{N5}\ncline[]{N5}{N8}
 \psset{nrot=:U}
 \ncline[linecolor=gray]{N0}{N4}\ncline[linecolor=gray]{N3}{N1}\naput{$J\cdot 2^{-3/2}$}
 \ncline[linecolor=gray]{N1}{N5}\naput{$J\cdot 2^{-3/2}$}\ncline[linecolor=gray]{N4}{N2}
 \ncline[linecolor=gray]{N3}{N7}\ncline[linecolor=gray]{N6}{N4}
 \ncline[linecolor=gray]{N4}{N8}\ncline[linecolor=gray]{N7}{N5}
\end{minipage}\hfill%
\begin{minipage}[t]{.45\textwidth}\centering
 \vspace{0pt}\vspace{0.5cm}
 \begin{tabular}{c|c}
  $(k,l)$ & $ J^{(2)}_{kl} / J^3$ \\
  \hline
  \hline
  $\{(0,1),(1,2),$ & \multirow{4}{*}{$-\frac{923}{192}+\frac{113}{54 \sqrt{2}}$} \\
  $\hphantom{\{}(6,7),(7,8),$ & \\
  $\hphantom{\{}(0,3),(3,6),$ & \\
  $\hphantom{\{}(2,5),(5,8)\}$ & \\
  \hline
  $\{(1,4),(4,7),$ & \multirow{2}{*}{$-\frac{2357}{288}+\frac{113}{27 \sqrt{2}}$}\\
  $\hphantom{\{}(3,4),(4,5)\}$  &  \\
 \end{tabular}
\end{minipage}\hspace*{\fill}%
\caption{\textit{Left}:
The qubits of the nine-qubit quantum information processor are arranged on a lattice.
The lines connecting qubits $i$ and $j$ indicate the values of the coupling constants $J_{ij}$;
\textit{right}: The table shows the two different values of the coupling constants $J^{(2)}_{kl}$ for each qubit pair $(k,l)$.\label{fig:qubits}}
\end{figure}

For the subsequent numerical simulations of a quantum algorithm we consider a quantum information processor
consisting of $n = 9$ qubits which are arranged on a lattice as indicated in figure \ref{fig:qubits}.
The coupling constants of vertical or horizontal qubit pairs are assumed to be equal while the coupling constants of diagonal neighbors are smaller by a factor of $2^{-3/2}$ due to the larger distance between them.
Non-neighboring qubits are assumed to be uncoupled.
According to relation \eqref{eq:2bedingung} this implies that in the embedded recoupling scheme two different time intervals $\Delta t$ are required for the free evolutions.
The values of the coupling strengths $J^{(2)}_{kl}$ \eqref{eq:Jkl2} for the 9-qubit lattice used in our subsequent simulation are apparent from the table of figure \ref{fig:qubits}.

In the following it is assumed that a quantum algorithm is performed on this quantum information processor according to the following rules:
\begin{enumerate}
\item Single-qubit gates are performed instantaneously and perfectly.
(Even though in the setting of \cite{YLMY04} selective gates are generated slowly using weak pulses, reference \cite{YLMY04} describes a way of implementing them in such a way that the inter-qubit couplings are decoupled during the application time. Hence, in good approximation, they might be viewed as being applied instantaneously.)
\item Two-qubit gates between vertical or horizontal neighboring qubits are performed by repeated applications of the unitary $U_{kl}(\pi/8)$-gate in combination with single-qubit gates as illustrated in figure \ref{fig:gates}.
The $U_{kl}(\pi/8)$-gate itself is generated by applying Super-WHH sequences $n_\text{swhh}$ times as indicated in figure \ref{fig:uphi}.
\item If the target qubits of a two-qubit gate are not vertical or horizontal neighbors
they are moved into such positions by applying a sequence of \textsf{SWAP}-gates according to the following simple strategy
\footnote{A better but more complicated strategy would be to minimize the number of \textsf{SWAP}-gates.
Note that due to the simple strategy used in this paper the first few iterations of a quantum algorithm take different amounts of computation time because the initial positions of the logical qubits are varying and so does the number of \textsf{SWAP}-gates.}:
If the vertical position of the qubits is the same, move the lower qubit to the upper one.
Otherwise, move the lower one to the same horizontal position and afterwards move the left one as far as necessary to the right.
\item A Super-WHH sequence is always applied in such a way that the qubit whose physical position has the smaller label (compare with figure \ref{fig:qubits}) is  qubit $k$ in $W_{xy}^{kl}$, i.\,e. it is transformed by the $X$ transformations.
The gate sequence of the $\textsf{CP}(\varphi)$-gate (compare with figure \ref{fig:gates}) is applied in such a way that the first single-qubit gate is applied always to the qubit with the smaller label.
\end{enumerate}

\subsection{The Quantum Algorithm}

In order to investigate the stabilizing properties of the embedded recoupling scheme the quantum algorithm of the quantum sawtooth map \cite{shep129} is simulated according to the rules of the preceding subsection.
One iteration of the quantum sawtooth map transforms an initial $n$-qubit quantum state $\ket{\Psi(0)}$ to the quantum state
\begin{equation}
\ket{\Psi(1)} = \exp\Bigl( -\frac{i}{2} m^2 T \Bigr) \exp\Bigl( -i k V( q ) \Bigr) \ket{\Psi(0)}
\end{equation}
with the sawtooth potential $V(q) = -\frac{1}{2} (q-\pi)^2$ ($0\leq q <2\pi$) and the (dimensionless) momentum operator $m$ whose eigenstates form the computational basis,
$ m \ket{ i } = i\ket{ i }$ for $i=0,1\dots,2^n-1$.
The position operator $q$ is related to the momentum operator via the quantum Fourier transform (QFT):
\begin{equation}
 q = U_\text{QFT}^{-1} \cdot \frac{2\pi}{d} m \cdot U_\text{QFT}.
\end{equation}
Initially the nine-qubit quantum information processor is prepared in the momentum eigenstate $\ket{\Psi(0)} = \ket{ 100110011 }$.
The (dimensionless) parameters of the sawtooth map are assumed to have the same values as in the previous simulations of reference \cite{shep152}, i.\,e. $T=2\pi/2^n$ and $kT = -0.5$.
Therefore, in Husimi functions\footnote{The definition of a Husimi function is given in appendix \ref{sec:csandhf}}, such as the ones presented in figure \ref{fig:husimi}, the dynamics of the sawtooth map are restricted to a phase-space cell of size $2\pi \times 2\pi$ and its corresponding classical dynamics are integrable.
In these Husimi functions the initial state corresponds to a horizontal line slightly above the middle.

Our gate decomposition of the quantum algorithm of this sawtooth map
consists of $n_g= 2 n^2 + 2n$ quantum gates. A detailed description can be found in appendix \ref{sec:gatesqmaps}.
In particular, $2\times n(n+1)/2$ quantum gates originate from the two quantum Fourier transforms
after which the inversion of the qubit positions is taken care of by relabeling instead of swapping.

\subsection{Numerical Results}

\begin{figure}\centering
\includegraphics[scale=0.88]{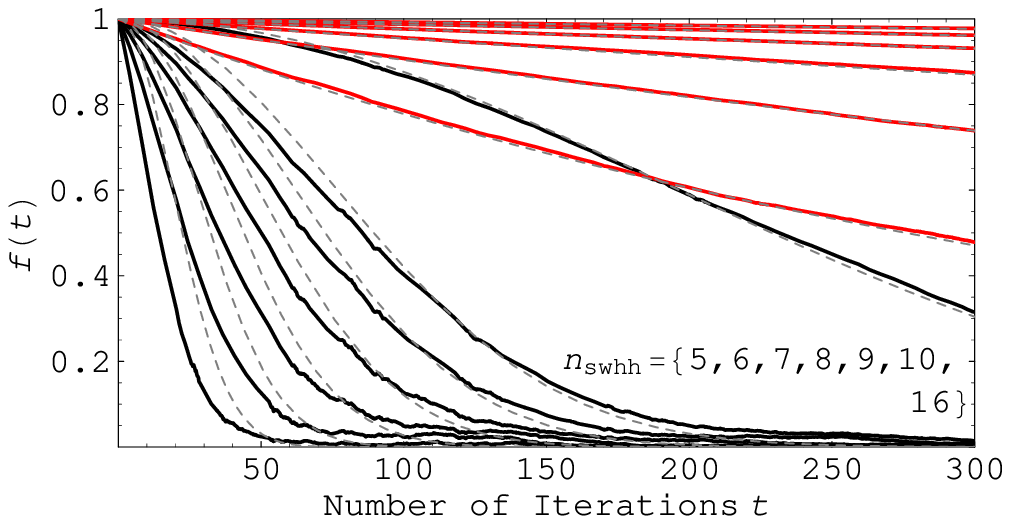}\vspace{4mm}\\
\noindent\includegraphics[scale=0.88]{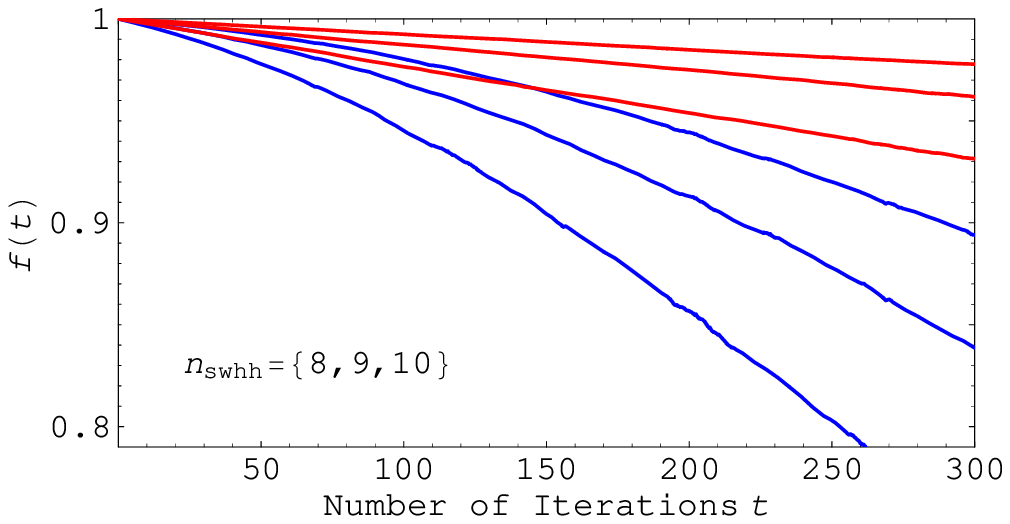}
\caption{
\textit{Upper plot:}
Fidelity plots of the quantum sawtooth map implemented with the original recoupling scheme of \citeauthor{YLMY04}\cite{YLMY04} (lower plots, black)
and the corresponding plots of the embedded recoupling scheme (upper plots, red):
Dashed curves show the fidelity estimations according to equations \eqref{eq:fidohne} and \eqref{eq:fidmit}.
\textit{Lower plot:}
Fidelity plots of the embedded recoupling scheme (upper plots, red) and the embedded but unsymmetrized scheme (lower plots, blue).\label{fig:fid}}
\end{figure}
\begin{figure}
 \begin{minipage}[b]{0.5\columnwidth}
  \vspace{0pt}\centering
  \includegraphics[scale=0.85]{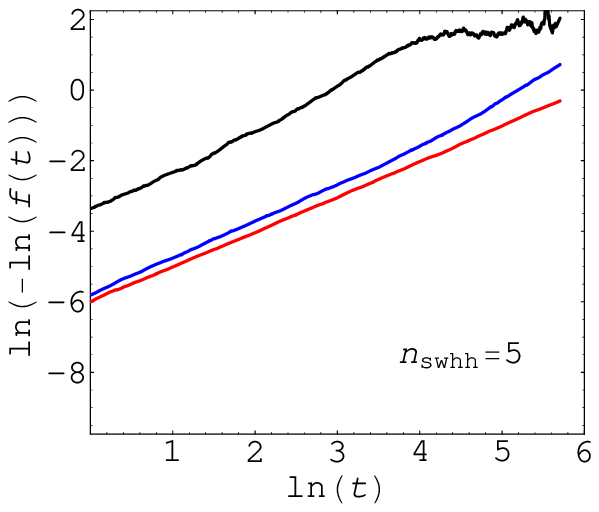}
 \end{minipage}%
 \begin{minipage}[b]{0.5\columnwidth}
  \vspace{0pt}\centering
  \includegraphics[scale=0.85]{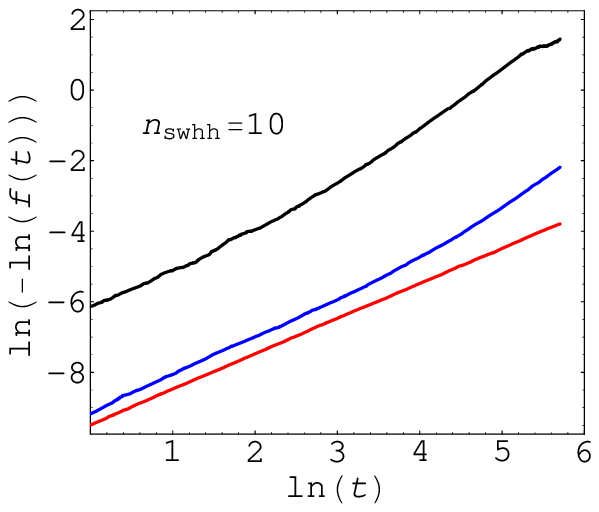}
 \end{minipage}%
\caption{Logarithmic fidelity plots of the quantum sawtooth map with $n_\text{swhh}=5$ \textit{(left)} and $n_\text{swhh}=10$ \textit{(right)}:
the original scheme of \citeauthor{YLMY04}\cite{YLMY04} (upper curve, black),
the embedded scheme (lowest curve, red),
and the embedded scheme without symmetrization (middle curve, blue).\label{fig:lnln}}
\end{figure}

\begin{figure}
 \begin{minipage}[b]{0.333\columnwidth}
  \vspace{0pt}\centering
  \includegraphics[scale=0.5]{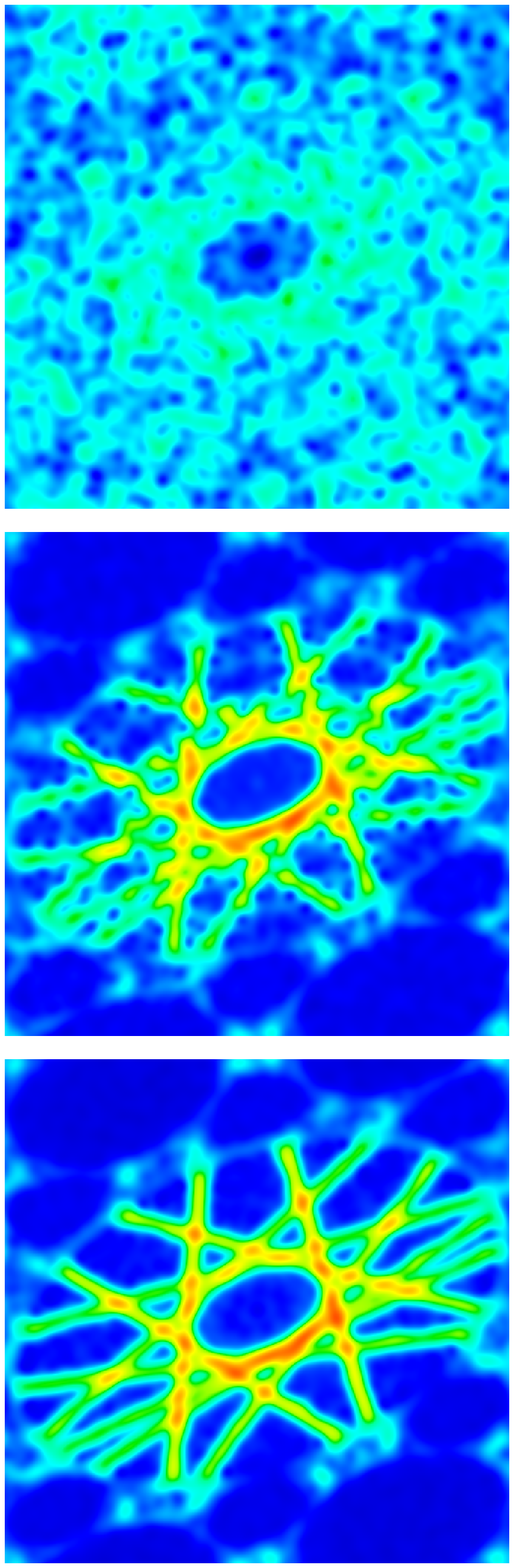}\\
  $n_\text{swhh}=6$
 \end{minipage}%
 \begin{minipage}[b]{0.333\columnwidth}
  \vspace{0pt}\centering
  \includegraphics[scale=0.5]{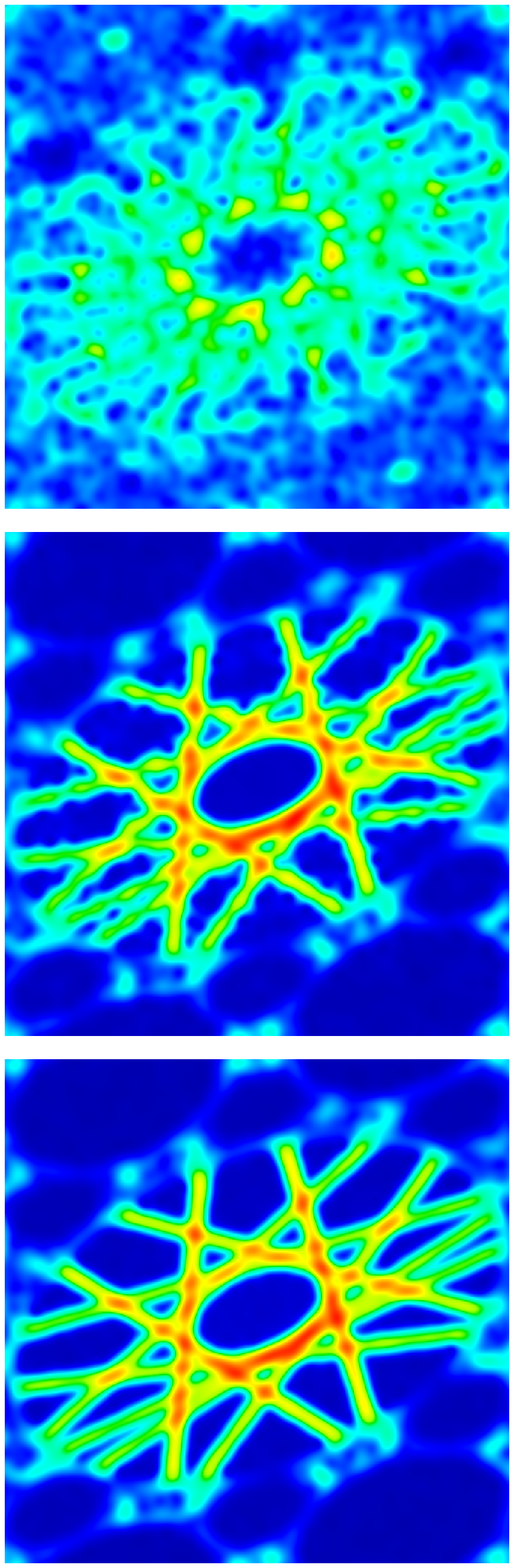}\\
  $n_\text{swhh}=8$
 \end{minipage}%
 \begin{minipage}[b]{0.333\columnwidth}
  \vspace{0pt}\centering
  \includegraphics[scale=0.5]{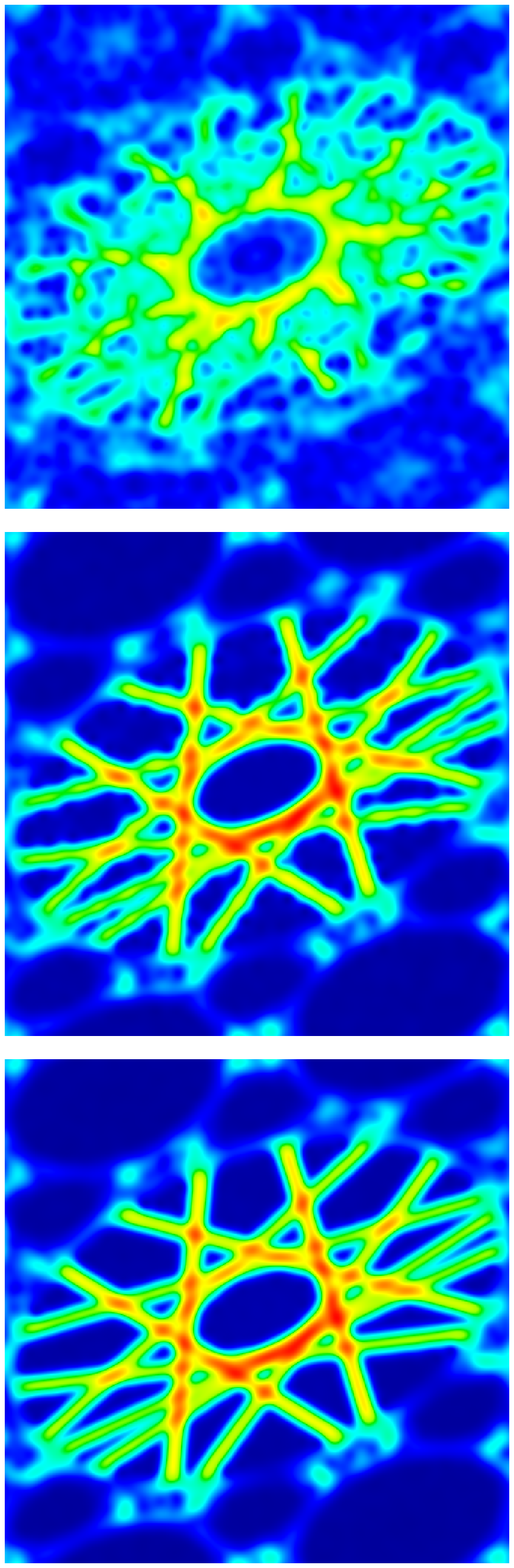}\\
  $n_\text{swhh}=10$
 \end{minipage}%
\caption{Husimi functions of the quantum states resulting from the quantum sawtooth map:
\textit{(Upper row)} The original scheme of \citeauthor{YLMY04}\cite{YLMY04},
\textit{(Middle row)} the embedded but unsymmetrized scheme,
\textit{(Lower row)} the embedded scheme.
The $U(\pi/8)$-gates used in the computations consist of $n_\text{swhh} = \{6,8,10\}$ Super-WHH sequences (from left to right). These functions are averaged over $290 \leq t \leq 299$ numbers $t$ of iterations of the sawtooth map.\label{fig:husimi}}
\end{figure}

In figures \ref{fig:fid}, \ref{fig:lnln}, and \ref{fig:husimi} results of our numerical simulations of the pure state fidelity
\begin{equation}
f(t) = \bigl\vert  \braket{\Psi(t)}{\Psi_\text{ideal}(t)}  \bigr\vert^2
\end{equation}
are presented for different numbers of repetitions $n_\text{swhh} \in \{5,6,\dots,10,16\}$ of the Super-WHH sequences.
For each value of $n_\text{swhh}$ we calculated the fidelity of the quantum state $\ket{\Psi(t)}$ of the quantum sawtooth map for up to $t=300$ iterations as well as the corresponding Husimi functions.

The quadratic-in-time fidelity decay of the original recoupling scheme is clearly apparent from figures \ref{fig:fid} and \ref{fig:lnln}. (The corresponding fidelities are plotted in \textit{black}).
This decay is caused by the coherent accumulation of errors due to the second-order AHT-term of the Super-WHH sequences involved in the realizations of the unitary $U_{kl}(\pi/8)$-gates.
The situation is somewhat reminiscent of the situation analyzed in subsection \ref{subsec:par_ohnekorrektur}, where it was assumed that each gate is preceded by a static imperfection.
(Here the imperfection depends on the index pair $(k,l)$.)
The $t$-dependence of the fidelity can be fitted by the function
\begin{equation}\label{eq:fidohne}
f(t) = \exp\bigl( - c \cdot t^2 / n_\text{swhh}^4 \bigr)
\end{equation}
with $c \approx 0.87$
(compare with the seven lowest dashed lines of the upper picture of figure \ref{fig:fid}).
According to equation \eqref{eq:FeappQMap} describing the behavior of the entanglement fidelity in the presence of static imperfections (subsection \ref{subsec:par_ohnekorrektur}), there should also be a linear contribution in the exponent of \eqref{eq:fidohne} which dominates the fidelity decay for small numbers of iterations.
Neglecting this linear contribution is the reason for the slightly imperfect overlap of our fitted fidelities with the corresponding numerical results.

Using the embedded Super-WHH sequence together with the appropriately chosen free evolution times given by equation \eqref{eq:2bedingung}, it is possible to get an almost linear-in-time fidelity decay at least on time scales where errors of the order of $\mathcal{O}\bigl( J(J\Delta t)^4 \bigr)$ are negligible (compare with Figs. \ref{fig:fid} and \ref{fig:lnln} (\textit{red} plots)).
In these cases the fidelity decay can be fitted by the function
\begin{equation}\label{eq:fidmit}
f(t) = \exp \bigl( - c_\textsf{ESDD} \cdot t / n_\text{swhh}^5 \bigr)
\end{equation}
with $c_\textsf{ESDD} \approx 7.85$ (compare with the six upper dashed lines of the upper picture of figure \ref{fig:fid} which are almost indistinguishable from the corresponding full curves).
The use of the embedded Super-WHH sequence does not only improve the action of a single $U_{kl}(\pi/8)$-gate but also prevents the residual imperfections to accumulate during the subsequent application of multiple gates.
It can therefore be seen as a variant of the \textsf{PAREC}-method of section \ref{sec:parec}.

Simulations based on the embedded recoupling scheme without the symmetrization step are shown in figure \ref{fig:fid} (lower part, \textit{blue} plots) and figure \ref{fig:lnln} (\textit{blue}).
The fidelity decay is suppressed significantly but on the time scale of these plots it is still quadratic in time.
This originates from the fact that terms of the Hamiltonian of equation \eqref{eq:hd2}
of the form $\alpha_k \alpha_l$, $\alpha \in \{X,Y,Z\}$, are not eliminated by the restricted randomization.

\section{Conclusions}\label{sec:conclude}

We showed how a selective recoupling scheme can be embedded into a stochastic decoupling scheme in such a way
that the desired coupling remains conserved and that, in addition, the coherent accumulation of higher-order errors is suppressed significantly.
While we focused on a specific example, the same general idea applies to other recoupling schemes as well.
Even if computation times of a quantum information processor are so long that the residual higher-order interaction term of equation \eqref{eq:hrklp2} of the order of $\mathcal{O}\bigl( J(J\Delta t)^4 \bigr)$ is no longer negligible, it is possible to suppress also these errors significantly by a suitable adjustment of the free evolution time $\Delta t$ involved in the realization of the relevant two-qubit gates ($U(\phi)$-gates).
In generalization of the procedure discussed in section \ref{sec:rec2} (compare with condition \eqref{eq:2bedingung}) this can be achieved either by explicitly calculating the fourth-order contribution of AHT and by solving the corresponding implicit equation of fifth order for $\Delta t$ involving renormalized coupling strengths or, alternatively, adjusting the value of $\Delta t$ so that the resulting fidelity decay is as small as possible.

Basic properties of our embedded scheme were analyzed for a single two-qubit gate.
In particular, it was demonstrated that our proposed embedded symmetrized recoupling scheme results in an improvement of the scaling of the error of a swapping gate with $n_\text{swhh}^{-5}$ instead of $n_\text{swhh}^{-4}$.
Here, $n_\text{swhh}$ denotes the number of repetitions of an embedded Super-WHH sequence which are required for the realization of the phase gate.
Therefore, in our embedded recoupling scheme fewer numbers of repetitions of Super-WHH sequences are necessary for achieving a particular degree of error suppression.
Typically, this also implies fewer pulses which are required for performing a quantum computation with a particular error tolerance.
This aspect is apparent from the upper plot of figure \ref{fig:fid} where  at $t\approx 70$ iterations the fidelity of the original recoupling scheme with $n_\text{swhh}=16$ is the same as the one of the embedded symmetrized recoupling scheme with $n_\text{swhh}=6$.

While the original Super-WHH sequence makes use of selective pulses on two of the qubits at the same time, our embedded scheme also makes use of simultaneous selective pulses on all qubits.
Since a selective pulse addressing a qubit with Larmor frequency $\omega_k$ induces erroneous rotations of qubits with nearby Larmor frequencies it may become important to use correction techniques as described in reference \cite{StVaCh00}.

\part{Codes and Cryptography}
\chapter{Classical Error Correction}\label{chap:cecc}

One of the fundamental quests of classical information theory is to transmit information reliably over a noisy channel.
Addressed by Shannon in 1948 \cite{Sh48}, his famous noisy coding theorem associates to each channel a non-negative number $C$, the so-called capacity of the channel,
and assures that for any rate below $C$, reliable information transmission over the channel is possible with the help of error-correcting codes.
This chapter serves to provide the background on classical error correction which is necessary for the understanding of the forthcoming chapters on quantum error correction.
For a more complete introduction to coding theory we refer to the books of MacKay \cite{MacKay} and Welsh~\cite{Welsh}.

After defining the capacity of discrete memoryless channels in section \ref{sec:dmcs}, we take a closer look at error-correcting codes in section \ref{sec:ecc}.
Linear codes form an important subclass of codes and are treated separately in section \ref{sec:lincodes}.
Eventually, we show in section \ref{sec:randlincodes} that picking a linear code at random allows us to transmit information over the binary symmetric channel at a rate arbitrary close to the capacity,
i.\,e. we prove a special case of the noisy coding theorem.

\section{Capacity of Discrete Memoryless Channels}\label{sec:dmcs}

In classical information theory, a discrete memoryless channel is a simple model of a noisy channel used for information tranmission.
It takes as input a symbol $a_i$ from a certain input alphabet $\Sigma_1=\{a_1,\dots,a_s\}$ and outputs a symbol $b_j$ from a certain output alphabet $\Sigma_2=\{b_1,\dots,b_r\}$ according to a fixed conditional probability distribution $p_{ji} = \Pr(b_j \vert a_i)$.
The $r\times s$ dimensional matrix $p_{ji}$ is called channel matrix.
Most of the time we will consider channels where the input alphabet as well as the output alphabet is the set $\mathbb{F}_q$ containing the numbers from $0$ to $q-1$.

The binary symmetric channel (BSC) is the most simple discrete memoryless channel.
It is defined on the binary alphabet $\mathbb{F}_2$ and its channel matrix is given by $\Pr(a\vert a)=1-p$ and $\Pr(a\oplus 1\vert a)=p$ with $a\in\mathbb{F}_2$.
It is therefore completely specified by a single parameter $p\in[0,1]$.

An $n$-fold extension of a discrete memoryless channel corresponds to $n$ uses of the channel.
Such an extended channel takes as input a string $\vec{a}=(a_{i_1},\dots,a_{i_n}) \in \Sigma_1^n$ and outputs a string $\vec{b} = (b_{j_1},\dots,b_{j_n}) \in \Sigma_2^n$ according to the conditional probability distribution
$\Pr(\vec{b} \vert \vec{a}) = \Pr(b_{j_1}\vert a_{i_1}) \cdot \Pr(b_{j_2}\vert a_{i_2})\dots \Pr(b_{j_n}\vert a_{i_n}) $.

\begin{defi}[Capacity of discrete memoryless channels]
Consider a discrete memoryless channel $\mathcal{\chi}$ with input alphabet $\Sigma_1=\{a_1,\dots,a_s\}$,
output alphabet $\Sigma_2=\{b_1,\dots,b_r\}$ and channel matrix $p_{ji} = \Pr(b_j \vert a_i)$.
Let $P=\{p_1,\dots, p_s\}$ be the probability distribution of a source $S$ outputting symbol $a_i \in \Sigma_1$, i.\,e. $\Pr(a_i) = p_i$.
Then the joint probability $\Pr(b_j,a_i)$ of the channel outputting symbol $b_j \in \Sigma_2$ and getting the input $a_i$ is given by $\Pr(b_j,a_i) = \Pr(b_j\vert a_i)\Pr(a_i) = p_{ji} p_i$.
The total probability of receiving output $b_j$ is given by $q_j = \Pr(b_j) =\sum_{i=1}^s \Pr(b_j,a_i)$.
The \textit{capacity} $C(\mathcal{\chi})$ of the channel $\mathcal{\chi}$ is defined as the mutual information between the source $S$ and the receiver $R$, maximized over all input probability distributions $P$:
\begin{equation}
 C(\mathcal{\chi}) = \max_P I(S:R).
\end{equation}
(The mutual information $I(S:R)$ was defined in equation \eqref{eq:mutualinfodefi} as $H(S)+H(R) -H(S,R)$, where
$H(S)=-\sum_i p_i \log_2 p_i$ denotes the Shannon entropy of the source,
$H(R)=-\sum_j q_j \log_2 q_j$ the Shannon entropy of the receiver,
and $H(S,R)=-\sum_{ij} \Pr(b_j,a_i) \log_2 \Pr(b_j,a_i)$ the joint entropy of source and receiver.)
\end{defi}

\begin{rem}
It is straightforward to calculate the capacity of the binary symmetric channel.
The mutual information is maximal for a uniform input distribution and we get
\begin{equation}\label{eq:capabsc}
  C(\text{BSC}_p) = 1 - H_2(p).
\end{equation}
Naturally, the capacity of the $n$-fold extension of the BSC is $n$ times its single capacity since the mutual information is additive.
\end{rem}

\section{Error Correction}\label{sec:ecc}

Let us assume now that the input alphabet and the output alphabet of the noisy channel under consideration are both given by the set $\mathbb{F}_q$.
When a string $\vec{x}\in\mathbb{F}_q^n$ is sent over the channel, the output will be a string $\vec{y}\in\mathbb{F}_q^n$ which is altered by the noise in the channel.
What we would like to do is to deduce the original input $\vec{x}$ from the received string $\vec{y}$.
Such a task becomes feasible only if we restrict the set of possible input strings.
\begin{defi}
A $q$-ary error-correcting code $\mathcal{C}$ of length $n$ is a subset $\mathcal{C} \subset \mathbb{F}_q^n$ of all possible $q$-ary strings of length $n$.
The members $\vec{x}\in\mathcal{C}$ of a code are called codewords.
\end{defi}
\noindent
The next step is to choose a decoding rule $\mathcal{D}$, which tells us which output strings have to be mapped to which codewords.
The optimal decoding rule $\mathcal{D}_\text{opt}$ decodes an output $\vec{y}$ as the codeword $\vec{x}\in\mathcal{C}$, which has the highest probability $\Pr(\vec{x}\vert\vec{y})$ of being sent through the channel conditioned on the event that $\vec{y}$ was received,
\begin{equation}
 \mathcal{D}_\text{opt} (\vec{y}) = \vec{x}\in\mathcal{C} \text{ s.\,t. } \Pr(\vec{x}\vert\vec{y}) \text{ is maximal.}
\end{equation}
The probabilities $\Pr(\vec{x})$ of having codeword $\vec{x}$ as input must be known to implement such a decoder, since
\begin{equation}
 \Pr(\vec{x}\vert\vec{y}) = \frac{ \Pr(\vec{y} \vert \vec{x}) \Pr(\vec{x}) }{ \sum_{\vec{x}'} \Pr(\vec{y} \vert \vec{x}') \Pr(\vec{x}') },
\end{equation}
where $ \Pr(\vec{y} \vert \vec{x}) = \prod_i \Pr(y_i \vert x_i)$ is specified by the channel matrix $\Pr(y_i \vert x_i)$.
Hence, usually a so called maximum likelihood decoder is used, which decodes $\vec{y}$ to the codeword that maximizes $\Pr(\vec{y} \vert \vec{x})$,
\begin{equation}
 \mathcal{D}_\text{mlk} (\vec{y}) = \vec{x}\in\mathcal{C} \text{ s.\,t. } \Pr(\vec{y}\vert\vec{x}) \text{ is maximal.}
\end{equation}
For the binary symmetric channel with $p<1/2$, the maximum likelihood decoder is equivalent to a minimum distance decoder $\mathcal{D}_\text{min}$ which decodes $\vec{y}$ as the codeword $\vec{x}$ that has minimum Hamming distance to $\vec{y}$,
\begin{equation}
 \mathcal{D}_\text{min} (\vec{y}) = \vec{x}\in\mathcal{C} \text{ s.\,t. } \dt(\vec{x},\vec{y}) \text{ is minimal.}
\end{equation}
\noindent
For a given noisy channel, code $\mathcal{C}$ and decoding rule $\mathcal{D}$, the average error probability is given by
\begin{equation}\label{eq:avdecerror}
p_\text{error}
=
\sum_{\vec{x}_\text{in}\in\mathcal{C}} \Pr(\vec{x}_\text{in}) \Pr(\vec{x}_\text{out} \neq \vec{x}_\text{in} \vert \vec{x}_\text{in})
=
\sum_{\vec{x}_\text{in}\in\mathcal{C}} \Pr(\vec{x}_\text{in}) \sum_{\vec{y} : \mathcal{D}(\vec{y})\neq \vec{x}_\text{in} } \Pr(\vec{y} \vert \vec{x}_\text{in} ),
\end{equation}
where $\Pr(\vec{x}_\text{in})$ denotes the probability of having $\vec{x}_\text{in}$ as input string.
In order to communicate reliably over the channel, we have to find a code $\mathcal{C}$ and decoder $\mathcal{D}$ such that this error probability, or even better the maximum error probability
\begin{equation}
 p_\text{error}^\ast = \max_{\vec{x}_\text{in}\in\mathcal{C}} \Pr(\vec{x}_\text{out} \neq \vec{x}_\text{in} \vert \vec{x}_\text{in}),
\end{equation}
becomes very small.

The next subsection examines the conditions under which perfect error correction ($p_\text{error}^\ast=0$) becomes possible.
Afterwards, the succeeding subsection deals with Shannon's noisy coding theorem, which tells us under which conditions error correction is possible if we allow some small probability of error ($p_\text{error}^\ast < \varepsilon$).

\subsection{Perfect Error Correction}
If a code $\mathcal{C}$ has the property that its codewords are very distinct, it may become possible to reconstruct the originally sent codeword $\vec{x}\in\mathcal{C}$ from the received $\vec{y}$ in a perfect manner (at least as long as not to many errors occur).
To formulate this idea precisely, we need the following definition.
\begin{defi}
The minimum distance $d$ of an error-correcting code $\mathcal{C}$ is defined as the minimum Hamming distance between different codewords $\vec{x},\vec{y}\in\mathcal{C}$:
\begin{equation}
  d(\mathcal{C}) = \min_{\vec{x},\vec{y}\in\mathcal{C} \text{ s.\,t. } \vec{x}\neq \vec{y} } \dt( \vec{x},\vec{y} ) .
\end{equation}
\end{defi}
\begin{lem}\label{lem:eccd=2e+1}
Given a $q$-ary error-correcting code $\mathcal{C}$ of length $n$ with minimum distance $d\geq 2e+1$,
information can be sent reliably over a noisy channel as long as the channel does not introduce more than $e$ errors. The transmission rate is given by $\log_q (\vert\mathcal{C}\vert) / n$.
\end{lem}
\begin{proof}
To deduce the originally sent codeword $\vec{x}$, we use minimum distance decoding.
Since the $e$-spheres $S_e(\vec{x})=\{ \vec{s}\in\mathbb{F}_q^n \vert \dt(\vec{x},\vec{s})\leq e \}$ around distinct codewords of a code with distance greater than $2e$ do not overlap,
the original codeword can be recovered from the received $\vec{y}$ as long as no more than $e$ errors are made by the channel.
\end{proof}
\noindent
At which rate can we encode information if we want to protect it perfectly against $e$ errors, i.\,e if we demand a distance $d=2e+1$ ?
A lower bound on this rate is given by the Gilbert Varshamov bound.
\begin{thm}[see e.\,g. chapter 4.2 in \cite{Welsh}]\label{thm:generalgv}
Gilbert Varshamov lower bound for $q$-ary codes.
A lower bound on the maximum number of codewords $A_q(n,d)$ of a $q$-ary code of length $n$ with minimum distance $d$ is given by
\begin{equation}
A_q(n,d) \geq q^n / \left( \sum_{i=0}^{d-1} \binom{n}{i} (q-1)^i \right).
\end{equation}
\end{thm}
\begin{proof}
Suppose $\mathcal{C}$ is a code of length $n$ with minimum distance $d$ and maximum number of codewords.
There can be no vector in $\mathbb{F}_q^n \setminus \mathcal{C}$ that has distance greater than $d$ from all the codewords of $\mathcal{C}$.
All $q^n$ vectors have to be included in the $d-1$ spheres around the $A_q$ codewords.
An upper bound on the number of vectors contained in these spheres is given by
\begin{equation*}
A_q(n,d) \cdot  \sum_{i=0}^{d-1} \binom{n}{i} (q-1)^i . \qedhere
\end{equation*}
\end{proof}
\begin{cor}\label{cor:gvasymp}
 For large $n$ the Gilbert Varshamov lower bound becomes
\begin{equation}
  \frac{ \log_q A_q(d,n) }{n} \geq 1 - H_{q [\log_q]} \Bigl(1-\frac{d}{n},\frac{d/n}{q-1},\dots,\frac{d/n}{q-1} \Bigr).
\end{equation}
\end{cor}
\begin{proof}
Setting $\lambda=(d-1)/n$ and writing the sum over $i$ as
\begin{equation}
\sum_{i=0}^{d-1} \binom{n}{i} (q-1)^i = \sum_{i=0}^{\lambda n} \binom{n}{i} \Bigl(\frac{q-1}{q}\Bigr)^i \Bigl(\frac{1}{q}\Bigr)^{n-i} \cdot q^n,
\end{equation}
the Chernoff bound \ref{lem:chernovbound} can be applied to obtain the upper bound
\begin{equation}
\exp_q \Bigl( n H_{q [\log_q]} \bigl (1-\lambda, \lambda/(q-1), \dots, \lambda/(q-1) \bigr) \Bigr)
\end{equation}
if $\lambda = (d-1)/n < (q-1)/q$. The proof is completed noting that $H_{q [\log_q]}$ is monotonically increasing in~$\lambda$.
\end{proof}

\subsection{Shannon's Noisy Coding Theorem}

Here we state Shannon's noisy coding theorem for discrete memoryless channels (see e.\,g. \cite[chapter 10]{MacKay} or \cite[section 3.5]{Welsh}).
\begin{thm}[Shannon's noisy coding theorem]\label{thm:shannonnoisyc}
For any $\varepsilon > 0$ and $R$ smaller than the channel capacity $C$,
there exists (for large enough $n$) a code $\mathcal{C}$ of length $n$ and rate not smaller than $R$,
together with a decoding rule $\mathcal{D}$,
such that the maximum probability $p_\text{error}^\ast$ of getting a decoding error is smaller than $\varepsilon$.
\end{thm}
\noindent
We will give a proof for the special case of the binary symmetric channel in section \ref{sec:randlincodes}.
It can also be shown that transmission at rates above the capacity becomes an impossible task if we continue to demand an arbitrary low error rate, see e.\,g. \cite[section 3.6]{Welsh}.

\section{Linear Codes}\label{sec:lincodes}

\begin{defi}
A linear $q$-ary error-correcting code $\mathcal{C}$ of length $n$ is a subspace of $\mathbb{F}_q^n$.
If $\mathcal{C}$ is a $k$-dimensional subspace, we say $\mathcal{C}$ is an $[n,k]_q$ code or denote it as $\mathcal{C}_{[n,k]_q}$.
If its minimum distance $d$ is known, we say it is an $[n,k,d]_q$ code.
\end{defi}

\begin{rem}
The minimum distance $d$ of an $[n,k]_q$ code $\mathcal{C}$ is the minimum weight of its nonzero codewords since $d(\mathcal{C}) = \min_{\vec{x}\neq\vec{y} \in \mathcal{C}} \dt(\vec{x},\vec{y}) = \min_{\vec{x}\neq\vec{0} \in \mathcal{C}} \wt(\vec{x})$.
\end{rem}

\noindent
If $\dim(\mathcal{C})=k$, $\mathcal{C}$ consists of $q^k$ codewords which are linear combinations of $k$ linearly independent generating elements $\vec{g}_i \in \mathbb{F}_q^n$ ($i=1,\dots,k$).
The $k\times n$ matrix $G$ whose rows are the $\vec{g}_i$ is called generator matrix.
The $k$ row vectors of the generator matrix $G$ can be extended to form a basis of $\mathbb{F}_q^n$ by adding $n-k$ additional linearly independent vectors $\vec{g}_j$ ($j=k+1,\dots,n$).
Each element $\vec{x}$ in $\mathbb{F}_q^n$ can then be expressed as a linear combination of the
$\vec{g}_i$: $\vec{x}=\sum_{i=1}^n u_i \vec{g}_i$, $u_i\in\mathbb{F}_q$.
The string $(u_{k+1},\dots,u_n)\in\mathbb{F}_q^{n-k}$ is called the syndrome.
For a given string $\vec{x}$, the syndrome can easily be calculated by matrix multiplication with an $(n-k)\times n$ dimensional parity check matrix $H$ whose rows $\vec{h}_i$ ($i=1,\dots,n-k$) satisfy $\vec{h}_i \cdot \vec{g}_j = 0$ for $j=1,\dots,k$ and $\vec{h}_i \cdot \vec{g}_j = \delta_{i,j-k}$ for $j=k+1,\dots,n$.
There is a one-to-one correspondence between the cosets of $\mathcal{C}$ in $\mathbb{F}_q^n$ and the syndromes.

When a string $\vec{y}$ is received over a noisy channel, the set of possible errors is given by
$\{ \vec{e}=\vec{y}-\vec{x} \vert \vec{x} \in \mathcal{C} \}$.
If $\mathcal{C}$ is a linear code, $\vec{x}\in\mathcal{C}$ implies that $-\vec{x}$ is also a member of $\mathcal{C}$.
This means that the set of possible errors is given by the coset of $\mathcal{C}$ in $\mathbb{F}_q^n$ which contains $\vec{y}$ and which can be identified unambiguously by the syndrome $\vec{s}=H\vec{y}^T$ of the received string $\vec{y}$.
Certain decoders are able to make use of this fact to speed up the decoding process to some extent.
The minimum distance decoder $\mathcal{D}_\text{min}$ for example has to find the element $\vec{e}_0$ of minimum weight in the coset of $\mathcal{C}$ which contains $\vec{y}$.
For all coset members $\vec{y}'\in \vec{y}+\mathcal{C}$ the result of the decoder is given by
$\mathcal{D}_\text{min}(\vec{y}') = \vec{y}'-\vec{e}_0$.
Therefore, knowledge of the syndrome $\vec{s}=H\vec{y}^T$ of the received $\vec{y}$ allows the use of a look-up table $\vec{e}_0(\vec{s})$ (which has to be calculated only once in the beginning) to find the required minimum-weight-element $\vec{e}_0$.

In the last section we gave a lower bound (Gilbert Varshamov bound) on the rate of codes with minimum distance $d$.
For linear codes we can find a better lower bound (which is sometimes called Varshamov bound) by taking into account the structure of such codes.
The following lemma establishes a relation between the parity check matrix $H$ and the minimum distance $d$.
It is then used to prove the lower bound given in the following theorem which is a generalization of \cite[theorem 12 from chapter 1, \textsection 10]{MacWS} or \cite[problem 21 chapter 4]{Welsh} to $q$-ary codes.
\begin{lem}[Theorem 10 from chapter 1, \textsection 10 in \cite{MacWS}]\label{lem:paritycd}
If $H$ is the $(n-k)\times n$-dimensional parity check matrix of an $[n,k]_q$ code, then the code has minimum distance $d$ iff every $d-1$ columns of $H$ are linearly independent and some $d$ columns are linear dependent.
\end{lem}
\begin{proof}
Some $d$ columns of $H$ are linear dependent
$\Leftrightarrow$ $H \vec{x}^T = 0$ for some $\vec{x}$ with weight $d$
$\Leftrightarrow$ There is a codeword $\vec{x}$ of weight $d$.
The same chain applies to the $d-1$ linearly independent columns of $H$ with the result that there are no codewords of weight less than $d$.
\end{proof}
\begin{thm}[Varshamov lower bound for linear $q$-ary codes]\label{thm:gvbest}%
An $[n,k,d]_q$ code exists provided that
\begin{equation}
\sum_{j=0}^{d-2} \binom{n-1}{j} (q-1)^j < q^{n-k}.
\end{equation}
\end{thm}
\begin{proof}
We construct an $(n-k)\times n$ dimensional parity check matrix $H$ such that all $d-1$ columns are linearly independent and use lemma \ref{lem:paritycd}.
The first column can by any nonzero $n-k$ column vector.
Suppose we have chosen $i$ columns such that all $d-1$ columns are linearly independent.
We can add another column and keep this property if the number of distinct linear combinations of $d-2$ or fewer of these columns is less than $q^{n-k}$.
This number is
\begin{equation*}
\sum_{j=0}^{d-2} \binom{i}{j} (q-1)^j. \qedhere
\end{equation*}
\end{proof}

\begin{rem}
As an example we calculate the above bound for a binary code of length $n=11$ and distance $d=3$ and get $A_2(11,3) \geq 128$.
The Gilbert Varshamov bound for general codes given in theorem \ref{thm:generalgv} assures us only that $A_2(11,3) \geq 31$.
\end{rem}
\begin{rem}
The asymptotic version of the above bound coincides with the asymptotic version of the bound for general codes given in corollary \ref{cor:gvasymp} if we replace $\log_q A_q(n,d)$ by $k$.
\end{rem}

We close this section by giving the definition of the dual code $\mathcal{C}^\perp$ of a code $\mathcal{C}$.
Dual codes are helpful in connection with quantum CSS codes as will become clear in section \ref{sec:csscodes}.
\begin{defi}
The dual code $\mathcal{C}^\perp$ of a $q$-ary code $\mathcal{C}$ of length $n$ is defined using the ordinary inner product of vectors modulo $q$,
\begin{equation}
 \mathcal{C}^\perp = \{\vec{x}\in\mathbb{F}_q^n \sthat \forall \vec{c}\in\mathcal{C},\ \vec{x}\cdot\vec{c}=0 \pmod q\}.
\end{equation}
\end{defi}
\begin{rem}
If $\mathcal{C}$ is an $[n,k]_q$ code, its dual code $\mathcal{C}^\perp$ is an $[n,n-k]_q$ code.
\end{rem}

\section{Random Linear Codes and the Binary Symmetric Channel}\label{sec:randlincodes}

In this section it is shown that a random linear code can --- at least in principle --- be used to communicate reliably over a binary symmetric channel at a rate arbitrary close to its capacity.
To achieve this goal, we do not demand perfect error correction as it was done in lemma \ref{lem:eccd=2e+1}, but we demand only a small maximum probability $p_\text{error}^\ast$ of getting a decoding error.
Since we are going to use a typical set decoder $\mathcal{D}_\text{typ}$,
we first need to define typical sets and discuss their relevant asymptotic properties in subsections \ref{subsec:typsets} and \ref{subsec:jtypsets}.
Then, in subsection \ref{subsec:randomlincodes}, it is shown that taking the average over all linear $[n,k]_q$ codes leads to an arbitrary small error probability $p_\text{error}^\ast$ (for large enough $n$) which proves a special case of Shannon's noisy coding theorem.

\subsection{Typical Sets}\label{subsec:typsets}

This subsection deals with typical sequences \cite[section 2.6]{HK02}.
The asymptotic properties of a set of typical sequences allows such a set to be used to construct so-called typical-set decoders.

A discrete random variable $X$ is characterized by a set of possible outcomes
$A=(a_1,\dots,a_s)$, $s=\vert A\vert$, together with an associated probability distribution
$P=(p_1,\dots,p_s)$ such that $X$ takes on the values $a_i \in A$ with probability $\Pr(X=a_i)=P(a_i)=p_i$.
The outcome of an ensemble of $n$ independent and identically distributed (iid) random variables $X^n=(X_1,\dots,X_n)$ is a sequence $\vec{x}=(x_1,\dots,x_n) \in A^n$ where the probability of getting outcome $\vec{x}$ is given by $P^n(\vec{x})=P(x_1)P(x_2)\dots P(x_n)$.
If the ensemble is large, the output sequence will contain about $p_1\cdot n$ times the symbol $a_1\in A$, about $p_2\cdot n$ times the symbol $a_2\in A$, etc.,
which motivates the definition of a subset of typical sequences:

\begin{defi}\label{def:typset}
The set $T_\delta^n(X)$ of strongly $\delta$-typical sequences is defined as the collection of strings in~$A^n$ whose relative frequency distribution of the symbols $A$ is close to the probability distribution~$P$:
\begin{equation}
 T_\delta^n (X) = \Bigl\{ \vec{x}\in A^n \text{ s.\,t. for every } x\in A,
 \bigl\vert N(x\vert\vec{x}) - n P(x) \bigr\vert
 < \frac{\delta nP(x)}{\log_q\vert A\vert}   \Bigr\},
\end{equation}
where $N(x\vert\vec{x})$ denotes the number of times the letter $x\in A$ occurs in $\vec{x}$
(i.\,e. $N(x\vert\vec{x}) = \vert\{ i \sthat x_i=x \}\vert$)
and the logarithm is taken with respect to the base $q$.
\end{defi}
\begin{thm}[Asymptotic equipartition property of $T_\delta^n(X)$]\label{thm:proptypset} $\,$\\
 $(a)$ For any length $n$ and any $\vec{x}\in T_\delta^n(X)$,
 \begin{equation}
  \Bigl\vert \frac{1}{n}\log_q P^n(\vec{x}) + H_{[\log_q]}(X) \Bigr\vert \leq \delta,
 \end{equation}
 or in other words, for all $\vec{x}\in T_\delta^n(X)$,
 $\exp_q\bigl(-n(H_{[\log_q]}(X)-\delta)\bigr) \geq P^n(\vec{x}) \geq \exp_q\bigl(-n(H_{[\log_q]}(X)+\delta)\bigr)$.\\
 $(b)$ For any $\Delta > 0$ and for $n$ sufficiently large,
 \begin{equation}
  \Pr( \vec{x}\in T_\delta^n (X)  ) = \sum_{\vec{x}\in T_\delta^n (X)} P^n(\vec{x}) \geq 1 - \Delta.
 \end{equation}
 $(c)$ For any $\Delta > 0$ and for $n$ sufficiently large,
 the cardinality of $T_\delta^n(X)$ is bounded by
 \begin{equation}
   ( 1 - \Delta ) \exp_q\bigl(n (H_{[\log_q]}(X)-\delta) \bigr)   \,\leq\,
                     \vert T_\delta^n(X)\vert                     \,\leq\,
                 \exp_q\bigl(n (H_{[\log_q]}(X)+\delta) \bigr).
 \end{equation}
\end{thm}
\begin{proof}[Proof of (a)]
 \begin{align*}
  \Bigl\vert \frac{1}{n}\log_q P^n(\vec{x}) + H_{[\log_q]}(X) \Bigr\vert &=
  \Bigl\vert \sum_{x\in A} \frac{N(x\vert\vec{x})}{n} \log_q P(x) - \sum_{x\in A} P(x) \log_q P(x) \Bigr\vert \\
  &\leq \sum_{x\in A} \frac{1}{n} \bigl\vert N(x\vert\vec{x})  - n P(x)\bigr\vert (-\log_q P(x)) \\
  &\leq \sum_{x\in A} \frac{\delta P(x)}{\log_q \vert A\vert} (-\log_q P(x)) && \text{by def.}\\
  &= \delta \cdot H_{[\log_q]}(X) / \log_q \vert A\vert     \leq \delta \qedhere
 \end{align*}
\end{proof}
\begin{proof}[Proof of (b)]
For each $x\in A$, let $F_x$ be the event that $X^n=(X_1,\dots, X_n)$ takes on a value $\vec{x}\in A^n$ that does not satisfy
\begin{equation*}
 \bigl\vert N(x\vert\vec{x}) - n P(x) \bigr\vert  < \frac{\delta nP(x)}{\log_q\vert A\vert}.
\end{equation*}
Chebyshev's inequality tells us that
\begin{equation*}
 \Pr(F_x) = \Pr\Bigl( \Bigl\vert \frac{N(x\vert\vec{x})}{n} - P(x)\Bigr\vert \geq \frac{\delta P(x)}{\log_q \vert A\vert} \Bigr) \leq
 \frac{P(x)(1-P(x))}{n} \cdot \Bigl( \frac{\log_q\vert A\vert}{\delta P(x)} \Bigr)^2.
\end{equation*}
If a sequence $\vec{x}$ is not in $T^n_\delta(X)$, it follows that at least one of the events $\{F_x\}_{x\in A}$ occurs and by the union bound we have
\begin{equation*}
  \Pr( \vec{x} \notin T_\delta^n(X) ) \leq \sum_{x\in A} \Pr(F_x) \leq \vert A\vert
  \frac{ (\log_q\vert A\vert)^2 }{n \delta^2} \max_{x\in A} \frac{1-P(x)}{P(x)},
\end{equation*}
which is smaller than any $\Delta>0$ for sufficiently large $n$.
\end{proof}
\begin{proof}[Proof of (c)]
We prove the upper bound using (a),
\begin{equation*}
 \vert T_\delta^n(X) \vert \cdot \exp_q\bigl(-n (H_{[\log_q]}(X)+\delta) \bigr) \leq
 \sum_{\vec{x}\in T_\delta^n(X)} P^n(\vec{x}) \leq 1.
\end{equation*}
The lower bound follows from (b) and (a),
\begin{equation*}
 1 - \Delta \leq \sum_{\vec{x}\in T_\delta^n(X)} P^n(\vec{x}) \leq  \vert T_\delta^n(X) \vert \cdot \exp_q\bigl( -n (H_{[\log_q]}(X)-\delta) \bigr). \qedhere
\end{equation*}
\end{proof}

\begin{rem}
The set $\tilde{T}_\delta^n(X)$ of weakly $\delta$-typical sequences is defined as the collection of strings in $A^n$ satisfying property (a) of theorem \ref{thm:proptypset},
\begin{equation}
 \tilde{T}_\delta^n (X) = \Bigl\{ \vec{x}\in A^n \text{ s.\,t. }
 \Bigl\vert \frac{1}{n}\log_q P^n(\vec{x}) + H_{[\log_q]}(X) \Bigr\vert < \delta
  \Bigr\}.
\end{equation}
It is possible to show that the set of weakly typical sequences also satisfies the remaining asymptotic equipartition properties (b) and (c).
Therefore it would be sufficient to use weakly typical sets for the purpose of typical set decoding.
But since we will need the strongly typical set later on in this thesis to construct conditional typical sets for the purpose of decoding certain random quantum codes, we decided to work with strongly typical sets right from the start.
In the following, when we speak of typical sets or sequences we always mean strongly typical.
\end{rem}

\subsection{Joint Typical Sets}\label{subsec:jtypsets}
In the context of random quantum codes, occasionally we'll have to work with the conditional typical sets \cite[section 2.6]{HK02} corresponding to a certain joint typical set.
We present the necessary material here, since it fits in this section dealing with typical sets in general.
\begin{defi}\label{def:jtypset}
Let the joint probability distribution of two random variables $X$ and $Y$ taking on values in the finite alphabets $A$ and $B$ be given by $\{ P(x,y) \sthat x\in A \text{ and } y\in B \}$.
The set of jointly strongly $\delta$-typical sequences
$(\vec{x},\vec{y}) = ( (x_1,y_1),\dots,(x_n,y_n) )$  ($\vec{x}\in A^n \text{ and } \vec{y}\in B^n$)
of length $n$ is defined by
\begin{equation}
  T_\delta^n(XY) = \Bigl\{ (\vec{x},\vec{y})
 \text{ s.\,t. for all } x \in A \text{ and } y \in B,
 \bigl\vert N(xy\vert\vec{x}\vec{y}) - n P(x,y)\bigr\vert \leq \frac{\delta n P(x,y)}{\log_q \vert A \times B\vert}  \Bigr\},
\end{equation}
where $N(xy\vert\vec{x}\vec{y}) = \vert\{ i \sthat (x_i,y_i)=(x,y) \}\vert$.
For a given joint typical set $T_\delta^n(XY)$, we define the set of typical $X$-sequences as
\begin{equation}
 T'^n_\delta(X) = \{ \vec{x}\in A^n \sthat (\vec{x},\vec{y}) \in T_\delta^n(XY) \text{ for some } \vec{y}\in B^n \},
\end{equation}
and we define the conditional typical set for a given $\vec{x}\in A^n$ as
\begin{equation}
 T_\delta^n(Y\vert\vec{x}) = \{ \vec{y}\in B^n \sthat (\vec{x},\vec{y}) \in T_\delta^n(XY) \}.
\end{equation}
\end{defi}
\begin{rem}
Any $\vec{x}\in T'^n_\delta(X)$ also belongs to $T_\delta^n(X)$.
\textit{Proof.} For all $x\in A$ we have
\begin{equation*}
 \bigl\vert N(x\vert\vec{x}) - n P(x) \bigr\vert   =
 \Bigl\vert \sum_{y\in B} \bigl( N(xy\vert\vec{x}\vec{y}) - n P(x,y) \bigr) \Bigr\vert  \leq
 \sum_{y\in B} \bigl\vert N(xy\vert\vec{x}\vec{y}) - n P(x,y) \bigr\vert,
\end{equation*}
which holds for any $\vec{y}\in B^n$. By the definition of $T'^n_\delta(X)$,
\begin{equation*}
 \sum_{y\in B} \bigl\vert N(xy\vert\vec{x}\vec{y}) - n P(x,y) \bigr\vert \leq
 \sum_{y\in B} \frac{ \delta n P(x,y) }{\log_q \vert A \times B\vert} =
 \frac{ \delta n P(x) }{\log_q \vert A \times B\vert} \leq
 \frac{ \delta n P(x) }{\log_q \vert A \vert}. \qed
\end{equation*}
\end{rem}
\begin{thm}[Asymptotic equipartition property of $T_\delta^n(XY)$]\label{thm:propjtypset}
$(a)$ For any $(\vec{x},\vec{y})\in T_\delta^n(XY)$,
\begin{subequations}
\begin{align}
\Bigl\vert \frac{1}{n} \log_q P^n(\vec{x},\vec{y}) + H_{[\log_q]}(XY) \Bigr\vert &\leq \delta,  \\
\Bigl\vert \frac{1}{n} \log_q P^n(\vec{x}) + H_{[\log_q]}(X) \Bigr\vert &\leq \delta,  \\
\Bigl\vert \frac{1}{n} \log_q P^n(\vec{y}\vert \vec{x}) + H_{[\log_q]}(Y\vert X) \Bigr\vert &\leq 2\delta.
\end{align}
\end{subequations}
$(b)$ For any $\Delta > 0$, and $n$ sufficiently large,
\begin{subequations}
\begin{align}
 \Pr\bigl( (\vec{x},\vec{y})\in T_\delta^n(XY) \bigr) &\geq 1 -\Delta, \\
 \Pr\bigl( \vec{x}\in T'^n_\delta(X) \bigr) &\geq 1 -\Delta.
\end{align}
\end{subequations}
$(c)$ For any $\Delta > 0$, $\vec{x}\in T'^n_\delta(X)$, and $n$ sufficiently large,
\begin{subequations}
\begin{gather}
\begin{array}{rcccl}
(1-\Delta)\exp_q\bigl( n(H_{[\log_q]}(XY)-\delta) \bigr) &\leq& \vert T_\delta^n(XY)\vert &\leq& \exp_q\bigl( n(H_{[\log_q]}(XY)+\delta) \bigr),
\end{array}\\
\begin{array}{rcccl}
(1-\Delta)\exp_q\bigl( n(H_{[\log_q]}(X)-\delta) \bigr) &\leq& \vert T'^n_\delta(X)\vert &\leq& \exp_q\bigl( n(H_{[\log_q]}(X)+\delta) \bigr),
\end{array}\\
\begin{array}{rcccl}
 \phantom{(1-\Delta)\exp_q\bigl( n(H_{[\log_q]}(Y\vert X)-2\delta) \bigr)} &\phantom{\leq}& \vert T_\delta^n(Y\vert \vec{x})\vert &\leq& \exp_q\bigl( n(H_{[\log_q]}(Y\vert X)+2\delta) \bigr).
\end{array}
\end{gather}
\end{subequations}
\end{thm}
\begin{proof}[Proof of (a)]
The proof of the first inequality is nearly identical to the proof of part (a) of theorem \ref{thm:proptypset}.
To prove the second inequality, note that it was shown in the above remark that
$(\vec{x},\vec{y})\in T_\delta^n(XY)$ implies $\vec{x}\in T_\delta^n(X)$.
The last inequality is proven by applying the first two inequalities to the expression
$P^n(\vec{y}\vert\vec{x}) = P^n(\vec{x},\vec{y}) / P^n(\vec{x})$.
\end{proof}
\begin{proof}[Proof of (b)]
The proof of the first part is nearly identical to the proof of part (b) of theorem \ref{thm:proptypset}.
For the proof of the second part, we note that
\begin{align*}
 \Pr\bigl( (\vec{x},\vec{y})\in T_\delta^n(XY) \bigr) &=
 \sum_{ (\vec{x},\vec{y})\in T_\delta^n(XY) } P^n(\vec{x},\vec{y}) =
 \sum_{ \vec{x} } \sum_{ \vec{y} }^{\text{s.\,t. } (\vec{x},\vec{y})\in T_\delta^n(XY) } P^n(\vec{x},\vec{y}) \\
 &\leq  \sum_{ \vec{x}\in T'^n_\delta(X) } \sum_{ \vec{y} } P^n(\vec{x},\vec{y}) =
        \sum_{ \vec{x}\in T'^n_\delta(X) } P^n(\vec{x}) =
        \Pr\bigl( \vec{x}\in T'^n_\delta(X) \bigr). \qedhere
\end{align*}
\end{proof}
\begin{proof}[Proof of (c)]
The proof goes as the proof of part (c) of theorem \ref{thm:proptypset}, using the results of part (a) and (b).
For $\vert T_\delta^n(Y\vert \vec{x})\vert$ only an upper bound can be proved, since the corresponding statement in (b) which is needed to prove the lower bound does not hold.
\end{proof}

\subsection{Random Coding}\label{subsec:randomlincodes}

We are now going to prove a special case of Shannon's noisy coding theorem (theorem \ref{thm:shannonnoisyc}).
We consider the binary symmetric channel with error probability $p$, the capacity of which was shown to be $1-H_2(p)$ in equation \eqref{eq:capabsc}.
\begin{thm}
Let $\text{BSC}_p$ be the binary symmetric channel with error probability $p$ and let $\varepsilon>0$.
Then, as long as
\begin{equation}
 \frac{k}{n} < C(\text{BSC}_p) = 1 - H_2(p),
\end{equation}
and for large enough $n$, there exists an $[n,k]_q$ code $\mathcal{C}$,
together with a decoder $\mathcal{D}$,
such that the maximum probability $p_\text{error}^\ast$ of getting a decoding error is smaller than $\varepsilon$.
\end{thm}
\begin{proof}
A binary linear $[n,k]_2$ code $\mathcal{C}$ is a $k$-dimensional subspace of $\mathbb{F}_2^n$.
Hence it is completely specified by an $(n-k)\times n$ dimensional parity check matrix $H$ such that $H\cdot \vec{x}^T=0$ for all $\vec{x}\in\mathcal{C}$.
If we want to use such a code to send information over a $n$-fold extension of the binary symmetric channel with bit flip probability $p$ ($\text{BSC}_p^n$), we need to specify the decoding algorithm.
Let $X$ be a random variable representing the error of $\text{BSC}_p$, i.\,e. $X$ takes on the values $A=\{0,1\}$ with probability $P=\{1-p,p\}$.
We are going to use a typical set decoder $\mathcal{D}_\text{typ}$ which calculates the syndrome $H\cdot\vec{y}^T$ of the received vector $\vec{y}\in\mathbb{F}_2^n$, and checks whether there is exactly one error vector $\vec{e}$ within the typical set
$T_\delta^n(X)$ such that $H\cdot\vec{e}^T = H\cdot\vec{y}^T$.
If this is the case, the decoder outputs $\vec{x}_\text{out}=\vec{y}-\vec{e}$, otherwise it produces a decoding error.

We are now going to determine an upper bound on the maximum decoding error probability $p_\text{error}^\ast$.
Since the error produced by the $\text{BSC}_p^n$ does not depend on its input $\vec{x}_\text{in}$,
the probability $\Pr(\vec{x}_\text{out} \neq \vec{x}_\text{in} \vert \vec{x}_\text{in})$ of getting a decoding error does not depend on the input $\vec{x}_\text{in}$, either.
Hence,
\begin{equation}
 p_\text{error}^\ast = \max_{\vec{x}_\text{in}\in\mathcal{C}}
  \Pr(\vec{x}_\text{out} \neq \vec{x}_\text{in} \vert \vec{x}_\text{in}) =
  \Pr(\vec{x}_\text{out} \neq \vec{x}_\text{in} ).
\end{equation}
To estimate the decoding error probability $\Pr(\vec{x}_\text{out} \neq \vec{x}_\text{in} )$, we have to sum over all possible errors produced by the $\text{BSC}_p^n$:
\begin{align}
 \Pr(\vec{x}_\text{out} \neq \vec{x}_\text{in} ) &=
 \sum_{\vec{e}\in\mathbb{F}_2^n} p^e(1-p)^{n-e} \cdot \begin{cases}
    1 & \text{if typical set decoding for $\vec{e}$ fails}\\
    0 & \text{else}
\end{cases} \nonumber\\
&\equiv
 \sum_{\vec{e}\in\mathbb{F}_2^n} p^e(1-p)^{n-e} \cdot
 \mathbbm{1}\!\left[ \text{typical set decoding for $\vec{e}$ fails}\right].
\end{align}
Here we denoted by $e = \wt( \vec{e} )$ the number of 1s in $\vec{e}$, a notation we shall use throughout.
The function $\mathbbm{1}[x]$ returns $1$ if the boolean expression $x$ is true and $0$ if it is false.
We split up this sum into a sum over typical errors and a sum over the remaining ones.
The later can be upper bounded by theorem \ref{thm:proptypset}b leading to
\begin{equation}
 \Pr(\vec{x}_\text{out} \neq \vec{x}_\text{in} )
  \leq \Delta +
  \sum_{\vec{e}\in T_\delta^n(X)} p^e(1-p)^{n-e} \cdot
  \mathbbm{1}\!\left[ \text{typical set decoding for $\vec{e}$ fails} \right].
\end{equation}
The sum over the typical errors can be upper bounded by
\begin{equation}
  \sum_{\vec{e}\in T_\delta^n(X)} p^e(1-p)^{n-e} \cdot
 \sum_{ \vec{e}'\in T_\delta^n(X)}^{ \text{s.\,t. } \vec{e}'\neq\vec{e} }
 \mathbbm{1}\!\left[ H\cdot(\vec{e}-\vec{e}')^T=\vec{0}^T \right].
\end{equation}
Now we take the average of $p_\text{error}^\ast$ over all linear codes.
Let
\begin{equation}
 A_{n,k,q} = \{ \mathcal{C}\subseteq\mathbb{F}_q^n \sthat \mathcal{C}\text{ is an } [n,k]_q \text{-code} \}
\end{equation}
denote the set containing all $[n,k]_q$ codes and let
\begin{equation}
A_{n,k,q} (\vec{x}) = \{ \mathcal{C}\in A_{n,k,q} \sthat \vec{x} \in \mathcal{C} \}
\end{equation}
be the subset of codes which contain a certain nonzero codeword $\vec{x}\in\mathbb{F}_q^n$.
In the following we need an upper bound for the quantity
$ \vert A_{n,k,2}(\vec{x}) \vert / \vert A_{n,k,2} \vert $.
It is proved in corollary \ref{cor:numberlincodes} in appendix \ref{sec:app:lincodes} that such a bound is given by
\begin{equation}\label{eq:AnkBoundFromApp}
 \frac{ \vert A_{n,k,q} (\vec{x}) \vert }{ \vert A_{n,k,q} \vert } = \frac{ q^k-1 }{q^n-1} \leq \frac{1}{q^{n-k}}.
\end{equation}
With the help of the above bound we obtain
\begin{align}
 \Bigl\langle p_\text{error}^\ast \Bigr\rangle_{\mathcal{C}\in A_{n,k,2}}
 &\leq \Delta +
 \sum_{\vec{e}\in T_\delta^n(X)} p^e(1-p)^{n-e} \cdot
 \sum_{ \vec{e}'\in T_\delta^n(X) }^{ \text{s.\,t. } \vec{e}'\neq\vec{e} } (1/2)^{n-k} && \text{by \eqref{eq:AnkBoundFromApp}}\nonumber\\
 &\leq\Delta + (\vert T_\delta^n(X) \vert -1) 2^{k-n} \nonumber\\
 &\leq\Delta + 2^{n(H_2(p)+\delta)-n+k} && \text{by theorem \ref{thm:proptypset}c.}
\end{align}
This quantity becomes arbitrarily small for large enough $n$ as long as
\begin{equation}
 \frac{k}{n} < 1 - H_2(p) -\delta.
\end{equation}
Since the above statement holds for any $\delta$, we are free to choose $\delta$ as small as we like.
Hence, for any $\varepsilon > 0$ and any rate $R$ below the channel capacity $C(\text{BSC}_p)=1 - H_2(p)$,
there exists (for large enough $n$) a linear code $\mathcal{C}$ of length $n$ and rate not smaller than $R$,
such that the maximum probability $p_\text{error}^\ast$ of getting a decoding error is smaller than $\varepsilon$.
\end{proof}

\begin{rem}
The achievable rate for reliable transmission over the $\text{BSC}_p$ as proven above is given by
$ 1 - H_{2[\log_2]}(p) $.
Demanding perfect error correction of up to $np$ errors,
the Gilbert-Varshamov bound in corollary \ref{cor:gvasymp} assures the existence of codes with a rate of at least $ 1 - H_{2[\log_2]}(2p) $.
Let the maximum value of tolerable noise $p_\text{max}$ of the $\text{BSC}_p$ be defined as the value of $p$ for which the transmission rate becomes zero.
By comparing the two rates we find that permitting a small decoding error probability results in a value of $p_\text{max}$ twice as high as in the case of perfect error correction.
\end{rem}

\chapter{Quantum Error-Correcting Codes}\label{chap:qecc}

To be of any practical use, a quantum memory has to be accessible from the outside to allow for measurements and the manipulation of the stored data.
Therefore, it can never be isolated perfectly from the environment and has to be treated as an open quantum system, i.\,e. as part of a larger quantum system.
In such a system, the most general state evolution is not unitary anymore,
but is given by a trace preserving completely positive map (tpcp-map)
$\mathcal{A} : \mathcal{S}(\mathcal{H}) \rightarrow \mathcal{S}(\mathcal{H})$
between density operators on a Hilbert space $\mathcal{H}$ describing the system.
Whereas unitary evolution is --- at least in theory --- always reversible, an error described by a tpcp-map can in general not be reversed, i.\,e. there exists no tpcp-map $\mathcal{R}$ such that $\mathcal{R}(\mathcal{A}(\rho))=\rho$ for any $\rho \in \mathcal{S}(\mathcal{H})$.
To be able to perform quantum error correction, we therefore have to demand less.
The trick is to restrict our attention to a subspace $\mathcal{C}$, called quantum code, of the Hilbert space of the quantum memory which has to be protected.
The question is whether it is possible to undo an error described by a tpcp-map $\mathcal{A}$ at least on such a subspace $\mathcal{C}$.

In section \ref{sec:suffnecforrec} we present the necessary and sufficient conditions a quantum code has to fulfill in order to be able to recover from a given set of errors.
An important family of quantum codes is given by the so-called stabilizer codes, which are discussed in section \ref{sec:stabilizercodes}.
CSS codes form a subclass of stabilizer codes and are treated separately in section \ref{sec:csscodes}.
By encoding a quantum register which is already encoded by some 'outer' stabilizer code a second time,
this time using some other 'inner' stabilizer code, one obtains a so-called concatenated code as we will discuss in section \ref{sec:concat}.

\section{Reversibility of Quantum Operations}\label{sec:suffnecforrec}

\begin{defi}
A quantum error-correcting code $\mathcal{C}$ is a subspace of the Hilbert space of a quantum memory which we would like to preserve.
For instance, a code which protects $k$ qubits might encode them into a $2^k$ dimensional subspace $\mathcal{C}$ of the Hilbert space $\mathcal{H} = \mathcal{H}_2^{\otimes n}$ of $n$ physical qubits.
\end{defi}

Is it possible to undo a quantum error described by a tpcp-map $\mathcal{A}$ on such a subspace $\mathcal{C}$, i.\,e. does there exist a recovery operation described by a tpcp-map $\mathcal{R}$ such that
\begin{equation}
 \mathcal{R}\bigl( \mathcal{A}(\rho)  \bigr) = \rho
\end{equation}
for all $\rho \in \mathcal{S}(\mathcal{C})$ ?
The necessary and sufficient condition a quantum code has to fulfill to allow for the recovery from a tpcp-map $\mathcal{A}$ was found by Knill and Laflamme \cite{KnillLa97}:

\begin{thm}[\cite{KnillLa97,NCSB97}]
Let $\{A_\mu\}$ be the operators in an operator sum representation of a tpcp-map~$\mathcal{A} : \mathcal{S}(\mathcal{H}) \rightarrow \mathcal{S}(\mathcal{H})$,
\begin{equation}
 \mathcal{A}: \rho\mapsto \mathcal{A}(\rho)=\sum_\mu A_\mu \rho A^\dagger_\mu.
\end{equation}
Then a necessary and sufficient condition for reversibility of $\mathcal{A}$ on a quantum code $\mathcal{C}$ is given by
\begin{equation}\label{eq:kl}
 \Pi_\mathcal{C} A^\dagger_\mu A_\nu \Pi_\mathcal{C} = \Pi_\mathcal{C} C_{\mu\nu},
\end{equation}
where $\Pi_\mathcal{C}$ denotes the projection on the code space and $C_{\mu\nu}$ is a %
Hermitian matrix.
\end{thm}

\begin{rem}
The operator sum representation is not unique.
But since different representations $\{A_\mu\}$,$\{B_\nu\}$ of a certain tpcp-map are related as $A_\mu=\sum_\nu u_{\mu\nu} B_\nu$ with unitary $u_{\mu\nu}$, the criterion given above does not depend on the representation.
\end{rem}

Let us introduce the set containing all $n$-fold tensor products of Pauli operators, %
\begin{equation}
 \mathcal{P}_q^n = \{ \XZ(\vec{a}) \ \vert\  \vec{a}\in\mathbb{F}_q^{2n} \},
\end{equation}
as defined in section \ref{sec:preliminaries},
as a basis for quantum errors acting on a quantum memory consisting of $n$ qudits of dimension $q$.
\begin{lem}
If we consider a subset $\mathcal{E} \subseteq \mathcal{P}_q^n$ of such error operators and Knill and Laflamme's condition is satisfied for all errors $E_a \in \mathcal{E}$, i.\,e.
\begin{equation}\label{eq:kl:forpauli}
 \Pi_\mathcal{C} E^\dagger_a E_b \Pi_\mathcal{C} = \Pi_\mathcal{C} C_{ab} \quad \text{ for all } E_{a,b} \in \mathcal{E},
\end{equation}
then the quantum code $\mathcal{C}$ allows for the correction of all tpcp-maps whose operator sum representation contains only elements which can be written as linear combinations of the $E_a \in \mathcal{E}$.
\end{lem}
\begin{proof}
If the elements of a operator sum representation $\{A_\mu\}$ of $\mathcal{A}$ can be written as $A_\mu=\sum_i a_{\mu i}E_i$ with $E_i \in \mathcal{E}$ and \eqref{eq:kl:forpauli} is satisfied, then equation \eqref{eq:kl} is satisfied, too:
\begin{equation}
 \Pi_\mathcal{C} A^\dagger_\mu A_\nu \Pi_\mathcal{C} =
 \sum_{ij} a^\ast_{\mu i}  a_{\nu j} \Pi_\mathcal{C}  E^\dagger_i  E_j \Pi_\mathcal{C} =
 \Pi_\mathcal{C} \sum_{ij} a^\ast_{\mu i}  a_{\nu j} C_{ij} =
 \Pi_\mathcal{C} C'_{\mu\nu}.
\end{equation}
\end{proof}

\begin{defi}\label{defi:degencode}
For a given set $\mathcal{E} \subseteq \mathcal{P}_q^n$, a code is said to be degenerate if the matrix $C_{ab}$ in \eqref{eq:kl:forpauli} is singular.
\end{defi}

\begin{defi}
A code is said to correct $t$ errors if \eqref{eq:kl:forpauli} is satisfied for the set $\mathcal{E}$ containing all Pauli operators which are composed of at least $n-t$ $\id$'s.
If we define the weight of a Pauli operator as the number of qudits on which it acts non-trivially,
the statement can be reformulated as follows:
A code is said to correct $t$ errors if \eqref{eq:kl:forpauli} is satisfied for the set
$\mathcal{E} = \{ E_i\in\mathcal{P}_q^n \vert \wt(E_i) \leq t \}$.
\end{defi}

\begin{defi}\label{defi:distanceqc}
A quantum code is said to have minimum distance $d$ if it detects all errors in $\mathcal{E} = \{ E_i\in\mathcal{P}_q^n \vert \wt(E_i) \leq d-1 \}$, i.\,e. if
$\Pi_\mathcal{C} E_i \Pi_\mathcal{C} = \alpha_i \Pi_\mathcal{C}$ for all $E_i\in \mathcal{E}$ with $\alpha_i\in\mathbb{C}$.
\end{defi}

\begin{rem}
A quantum code with distance $d\geq 2t+1$ corrects $t$ errors because \eqref{eq:kl:forpauli} will be satisfied for the set $\mathcal{E} = \{ E_i\in\mathcal{P}_q^n \vert \wt(E_i) \leq t \}$.
\end{rem}

\section{Stabilizer Codes}\label{sec:stabilizercodes}

The stabilizer code formalism has been developed mainly by Gottesman in \cite{Go96,PhdGottesman}.
It has been generalized to handle quantum systems of dimension higher than two in \cite{Gott98,Rains99}.
This section deals with quantum systems of dimension $q$ (prime), but in principle $q$ could also be a power of a prime.

The stabilizer formalism proposes the common eigenspaces of an abelian subgroup of the Pauli group $\mathfrak{P}_q^n$ as codespaces.
Since $\mathfrak{P}_q^n$ and the space $\mathbb{F}_q^{2n}$, which forms a group under addition modulo $q$, are related by the ray representation \eqref{eq:rayrep}, stabilizer codes can be described in two equivalent ways.
We will focus mainly on the description in the $\mathbb{F}_q^{2n}$ picture.

\subsection{Stabilizers and Codespaces}

\begin{defi}
A stabilizer is a self-orthogonal subspace $L\subset \mathbb{F}_q^{2n}$ with respect to the symplectic inner product,
i.\,e. $L\subseteq L^\perp$ where
$L^\perp = \{ \vec{x}\in\mathbb{F}_q^{2n} \ \vert\ \forall\vec{l}\in L, (\vec{x},\vec{l})_{sp}=0 \}$.
Equivalently, using the $\XZ(\cdot)$ representation,
a stabilizer $S=\{ \omega^k \XZ(\vec{l}) \ \vert\  \vec{l}\in L, k\in\mathbb{F}_q  \}$\footnote{If $q=2$, $\omega^k$ should be replaced by $\mu\in\{\pm1,\pm i\}$} is an abelian subgroup of the Pauli group~$\mathfrak{P}_q^n$.
\end{defi}
\begin{rem}
An $(n-k)$-dimensional self-orthogonal subspace $L\subset\mathbb{F}_q^{2n}$ can always be specified by $n-k$ linearly independent generating elements,
e.\,g. $L=\vspan \{\vec{g}_1,\dots,\vec{g}_{n-k} \}$ with $\vec{g}_i=(\vec{g}_i^x,\vec{g}_i^x)\in\mathbb{F}_q^{2n}$ for $i=1,\dots,n-k$.
\end{rem}
\begin{lem}[see e.\,g. \cite{PhdGottesman} or \cite{nielsenchuang}]
A commutative subgroup $S \subset \mathfrak{P}_q^n$ corresponding to an $(n-k)$-dim\-ensional self-orthogonal subspace $L\subset \mathbb{F}_q^{2n}$ divides the Hilbert space $\mathcal{H}_q^{\otimes n}$ into $q^{n-k}$ common eigenspaces of dimension $q^k$.
\end{lem}
\begin{proof}
The construction of a basis of such a $q^k$-dimensional eigenspace in the next subsection implies the proof.
\end{proof}
\begin{defi}
The $q^k$-dimensional eigenspaces corresponding to an $(n-k)$-dimensional stabilizer $L\subseteq L^\perp\subseteq \mathbb{F}_q^{2n}$ can be labeled by a vector $\vec{s}\in\mathbb{F}_q^{n-k}$.
They are defined to be the corresponding stabilizer codes $\mathcal{C}(L,\vec{s})$.
We will use the notation $[[n,k]]_q$ code to denote an $(n-k)$-dimensional stabilizer code $L$,
or strictly speaking, to denote the collection of all code spaces $\mathcal{C}(L,\vec{s})$ corresponding to a specific stabilizer $L$ of dimension $n-k$.
If the distance $d$ of an $[[n,k]]_q$ code is known, we say the code is an $[[n,k,d]]_q$ code.
\end{defi}
\begin{rem}
We will see below that all these code spaces are equivalent in the sense that they have identical error correcting properties.
\end{rem}

\subsection{Encoding Operations}\label{sec:stabenc}

\begin{lem}[see e.\,g. \cite{Ha02, Ha03Fi, Matsu05}]
For a given set $\{\vec{g}_1,\dots,\vec{g}_{n-k} \}$ of generating elements of some self-orthogonal $(n-k)$-dimensional subspace $L\subseteq L^\perp\subseteq \mathbb{F}_q^{2n}$, it is always possible to find vectors $\{\vec{g}_{n-k+1},\dots,\vec{g}_n\}$ and $\{\vec{h}_1,\dots,\vec{h}_n\}$ such that
\begin{align}\label{eq:stab:commutatorrelations}
 (\vec{g}_i,\vec{h}_j)_{sp}&=\delta_{ij},  &
 (\vec{g}_i,\vec{g}_j)_{sp}&=0,  &
 (\vec{h}_i,\vec{h}_j)_{sp}&=0.
\end{align}
Vectors $\{\vec{g}_1,\dots,\vec{g}_n, \vec{h}_1,\dots,\vec{h}_n\}$ satisfying the above conditions are said to form a hyperbolic basis of~$\mathbb{F}_q^{2n}$.
\end{lem}
\begin{rem}
Note that $L^\perp=\vspan\{\vec{g}_1,\dots,\vec{g}_n,\vec{h}_{n-k+1},\dots,\vec{h}_n\}$. $L^\perp$ is called the normalizer.
\end{rem}

\begin{figure}
\begin{minipage}[c]{0.45\textwidth}
 \centering  %
 \begin{pspicture}(-0.17,-0.17)(5.70,5.44)%
 \scalebox{0.85}{%
   \rput[origin=c]{-90}(1.5,6){$\rotatebox[origin=c]{90}{$2n$}\left\{\makebox(0,1.6){}\right.$}
   \psframe[linecolor=gray](-0.1,-0.1)(3.1,1.6)
   \psframe[linecolor=gray](-0.1,1.9)(3.1,5.6)
   \psframe[fillstyle=solid,fillcolor=lightgray,linestyle=none](0,0)(1.5,0.5)
   \psframe[fillstyle=solid,fillcolor=white,linestyle=none](1.5,0)(3,0.5)
   \psframe(0,0)(3,0.5)
   \rput(1.5,0.25){$\vec{g}_n$}
   \rput(1.5,0.8){$\vdots$}
   \psframe[fillstyle=solid,fillcolor=lightgray,linestyle=none](0,1)(1.5,1.5)
   \psframe[fillstyle=solid,fillcolor=white,linestyle=none](1.5,1)(3,1.5)
   \psframe(0,1)(3,1.5)
   \rput(1.5,1.25){$\vec{g}_{n-k+1}$}
   \psframe[fillstyle=solid,fillcolor=lightgray,linestyle=none](0,2)(1.5,2.5)
   \psframe[fillstyle=solid,fillcolor=white,linestyle=none](1.5,2)(3,2.5)
   \psframe(0,2)(3,2.5)
   \rput(1.5,2.25){$\vec{g}_{n-k}$}
   \rput(1.5,3.8){$\vdots$}
   \psframe[fillstyle=solid,fillcolor=lightgray,linestyle=none](0,5)(1.5,5.5)
   \psframe[fillstyle=solid,fillcolor=white,linestyle=none](1.5,5)(3,5.5)
   \psframe(0,5)(3,5.5)
   \rput(1.5,5.25){$\vec{g}_1$}

   \psframe[linecolor=gray](3.4,-0.1)(6.6,1.6)
   \psframe[linecolor=gray](3.4,1.9)(6.6,5.6)
   \psframe[fillstyle=solid,fillcolor=lightgray,linestyle=none](3.5,0)(5,0.5)
   \psframe[fillstyle=solid,fillcolor=white,linestyle=none](5,0)(6.5,0.5)
   \psframe(3.5,0)(6.5,0.5)
   \rput(5,0.25){$\vec{h}_n$}
   \rput(5,0.8){$\vdots$}
   \psframe[fillstyle=solid,fillcolor=lightgray,linestyle=none](3.5,1)(5,1.5)
   \psframe[fillstyle=solid,fillcolor=white,linestyle=none](5,1)(6.5,1.5)
   \psframe(3.5,1)(6.5,1.5)
   \rput(5,1.25){$\vec{h}_{n-k+1}$}
   \psframe[fillstyle=solid,fillcolor=lightgray,linestyle=none](3.5,2)(5,2.5)
   \psframe[fillstyle=solid,fillcolor=white,linestyle=none](5,2)(6.5,2.5)
   \psframe(3.5,2)(6.5,2.5)
   \rput(5,2.25){$\vec{h}_{n-k}$}
   \rput(5,3.8){$\vdots$}
   \psframe[fillstyle=solid,fillcolor=lightgray,linestyle=none](3.5,5)(5,5.5)
   \psframe[fillstyle=solid,fillcolor=white,linestyle=none](5,5)(6.5,5.5)
   \psframe(3.5,5)(6.5,5.5)
   \rput(5,5.25){$\vec{h}_1$}
 }%
 \end{pspicture}
\end{minipage}%
\begin{minipage}[c]{0.52\textwidth}
 \caption[Stabilizer code]{A stabilizer code is specified by the generating elements $\vec{g}_i =(\vec{g}^x_i,\vec{g}^z_i)\in \mathbb{F}_q^{2n}$ of a self-orthogonal subspace
$L = \vspan \{\vec{g}_1,\dots,\vec{g}_{n-k} \} \subseteq L^\perp$.
Any extension of these vectors to a hyperbolic basis
$\mathbb{F}_q^{2n} = \vspan \{\vec{g}_1,\dots,\vec{g}_n,\vec{h}_1,\dots,\vec{h}_n\}$ with
$(\vec{g}_i,\vec{h}_j)_{sp}=\delta_{ij}$, $(\vec{g}_i,\vec{g}_j)_{sp}=0$ and $(\vec{h}_i,\vec{h}_j)_{sp}=0$, specifies a specific encoding.}%
 \label{fig:stab-to-css:stab}%
\end{minipage}
\end{figure}
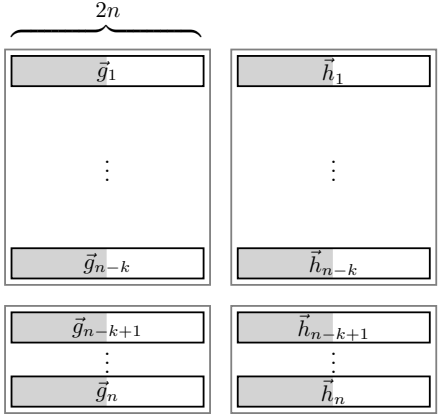

We are now going to show that such an extension of the generating set of a stabilizer
to a hyperbolic basis
together with a set of phase factors to be defined below, completely specifies a unitary encoding operation.
Let us define the operators
\begin{align}\label{eq:defilogicalxandzops}
  \overline{Z}_i &= \theta_z(i) \XZ(\vec{g}_i) &
  \overline{X}_i &= \theta_x(i) \XZ(\vec{h}_i)
\end{align}
using some fixed set
$\{ \theta_\alpha(i) \in \{\omega^r\vert r\in\mathbb{F}_q\} \}_{\alpha\in\{x,z\}, i\in\{1\dots n\}}$
of phase factors, and let us define the abbreviations
\begin{subequations}
\begin{align}
 \overline{X}^{\vec{u}} &= \prod_{i=1}^n \overline{X}_i^{u_i}, & X^{\vec{u}} &= X^{u_1}\otimes \dots \otimes X^{u_n} = \XZ(\vec{u},\vec{0}), \\
 \overline{Z}^{\vec{v}} &= \prod_{i=1}^n \overline{Z}_i^{v_i}, & Z^{\vec{v}} &= Z^{v_1}\otimes \dots \otimes Z^{v_n} = \XZ(\vec{0},\vec{v}).
\end{align}
\end{subequations}
Since the $\{ \overline{Z}_i \}_{i\in\{1,\dots,n\}}$ commute with each other, there has to be a non-empty common eigenspace with eigenvalue list $(\lambda_1,\dots, \lambda_n)$.
Let us define $\overline{\ket{0,\dots,0}}$ as a normalized vector in this eigenspace.
By applying the operator $\overline{X}^{\vec{u}}$ to both sides of the eigenequation
\begin{equation}
 \overline{Z}_i \overline{\ket{0,\dots,0}} = \lambda_i \overline{\ket{0,\dots,0}},
\end{equation}
and by making use of the fact that $\overline{Z}_i \overline{X}_i = \omega \overline{X}_i \overline{Z}_i$, we find that the state $\overline{X}^{\vec{u}} \overline{\ket{0,\dots,0}}$ is an eigenstate of the
$\{ \overline{Z}_i \}_{i\in\{1,\dots,n\}}$ with eigenvalue list $(\lambda_1 \omega^{u_1},\dots ,\lambda_n \omega^{u_n})$.
Hence there have to exist at least $q^n$ different eigenspaces, each of which must be of dimension one.
In the following we will always chose $\overline{\ket{0,\dots,0}}$ as the common eigenvector with eigenvalue list $(\lambda_1,\dots, \lambda_n)=(1,\dots,1)$.
The encoding operator $U_\text{enc}$ is defined as the unitary which maps the states $\ket{ \vec{u} }$ of the computational basis onto the states $ \overline{\ket{\vec{u}}} = \overline{X}^{\vec{u}} \overline{\ket{0,\dots,0}}$,
\begin{equation}\label{eq:stabencoderdefi}
 U_\text{enc} : X^{\vec{u}} \ket{0,\dots,0}=\ket{u_1,\dots,u_n}  \mapsto
           \overline{X}^{\vec{u}} \overline{\ket{0,\dots,0}} = \overline{\ket{u_1,\dots,u_n}}.
\end{equation}
It is straightforward to show that i)
\begin{equation}
\overline{Z}^{\vec{v}} \overline{\ket{l_1,\dots,l_n}} =  \omega^{\vec{v}\cdot\vec{l}} \overline{\ket{l_1,\dots,l_n}}
\end{equation}
and that ii)
\begin{align}\label{eq:stabenc_uxu}
 U_\text{enc} X^{\vec{u}} U^\dagger_\text{enc} &= \overline{X}^{\vec{u}} &
 U_\text{enc} Z^{\vec{v}} U^\dagger_\text{enc} &= \overline{Z}^{\vec{v}}.
\end{align}
Because of equation \eqref{eq:stabenc_uxu}, we will call the operators
$\{ \overline{X}_i,\overline{Z}_i \}_{i\in\{1,\dots,n\}}$
defined in \eqref{eq:defilogicalxandzops} encoded $X$- and $Z$-operators.
\begin{rem}
The Clifford group consists of all operators which map Pauli operators to Pauli operators.
It follows from equation \eqref{eq:stabenc_uxu} that $U_\text{enc}$ is an element of the Clifford group.
\end{rem}
The codespace with label (sometimes called syndrome) $(s_1,\dots,s_{n-k})$
is the common eigenspace of (the generators of) the stabilizer
$\{ \overline{Z}_i \}_{i\in\{1,\dots,n-k\}}$, with eigenvalue list $(\omega^{s_1},\dots,\omega^{s_{n-k}})$ and can be written as
\begin{equation}
 \mathcal{C}(L,\vec{s}) = \vspan\{ \overline{ \ket{s_1,\dots,s_{n-k},c_1,\dots,c_k} } \ \vert\ (c_1,\dots,c_k)\in\mathbb{F}_q^k \}.
\end{equation}
An encoded quantum state is given by
\begin{equation}
 \mathcal{C}(L,\vec{s}) \ni \ket{\psi}_{\vec{s}} =
 \sum_{c_1,\dots,c_k}
 \alpha_{c_1,\dots,c_k} \overline{ \ket{s_1,\dots,s_{n-k},c_1,\dots,c_k} }, \text{ with } \alpha_{c_1,\dots,c_k}\in\mathbb{C}.
\end{equation}
Operators $\{\overline{Z}_i,\overline{X}_i\}_{i\in\{n-k+1,\dots,n\}}$ manipulate the encoded state, i.\,e. they perform logical $Z_{i-n+k}$ and $X_{i-n+k}$ operations on the $(i-n+k)$-th encoded qudit.

For a given hyperbolic basis $\{\vec{g}_1,\dots,\vec{g}_n, \vec{h}_1,\dots,\vec{h}_n\}$ of~$\mathbb{F}_q^{2n}$, any vector $\vec{a}\in\mathbb{F}_q^{2n}$ can be expressed as linear combination of the basis elements,
\begin{equation}\label{eq:stabdecompoa}
\begin{split}
 \vec{a} &= ( a_1^x,\dots,a_n^x, a_1^z,\dots,a_n^z ) \\
 &= \sum_{i=1}^{n-k} \bigl( s_i \vec{h}_i + n_i \vec{g}_i \bigr)
 + \sum_{i=n-k+1}^n \bigl(
 l_{i-(n-k)}^x \vec{h}_i + l_{i-(n-k)}^z \vec{g}_i \bigr),
\end{split}
\end{equation}
where $s_i = (\vec{g}_i,\vec{a})_{sp}$, et cetera.
Together with equation \eqref{eq:xzaxzb} we obtain the following lemma.
\begin{lem}\label{lem:xzdecompostab}
Any Pauli operator $\XZ(\vec{a}\in\mathbb{F}_q^{2n}) \in \mathcal{P}_q^n$ can be expressed (up to a phase) as product of some powers of the operators $\XZ(\vec{g}_i),XZ(\vec{h}_i)$,
\begin{align}
 XZ(\vec{a})
 &\sim
 \prod_{i=1}^{n-k} \bigl( \XZ(\vec{h}_i)^{s_i} \XZ(\vec{g}_i)^{n_i} \bigr)
 \prod_{i=1}^k \bigl( \XZ(\vec{h}_{i+n-k})^{l_i^x} \XZ(\vec{g}_{i+n-k})^{l_i^z} \bigr),\\
\intertext{
or by using the operators $\overline{Z}_i,\overline{X}_i$ defined in \eqref{eq:defilogicalxandzops},
}
 &\sim
 \prod_{i=1}^{n-k} \bigl( \overline{X}_i^{s_i} \overline{Z}_i^{n_i} \bigr)
 \prod_{i=1}^k \bigl( \overline{X}_{i+n-k}^{l_i^x} \overline{Z}_{i+n-k}^{l_i^z} \bigr)
 =
  \overline{X}^{(\vec{s},\vec{l}^x)}  \overline{Z}^{(\vec{n},\vec{l}^z)},
\end{align}
where the strings $\vec{s},\vec{n}\in\mathbb{F}_q^{n-k}$ and $\vec{l}^x,\vec{l}^z\in\mathbb{F}_q^k$ are defined in \eqref{eq:stabdecompoa}.
\end{lem}

\subsection{Correctable Errors}

For which sets of errors $\mathcal{E}\subseteq \mathcal{P}_q^n$ is Knill and Laflamme's condition for reversibility satisfied on the codespaces $\mathcal{C}(L,\vec{s})$ of a stabilizer code,
or in other words, what are the errors that can be corrected ?
As we will see, neither does the answer depend on the label $\vec{s}$ of the codespace we have chosen to encode some information, nor does it depend on the encoding operation $U_\text{enc}$.

\begin{lem}[see e.\,g. \cite{PhdGottesman}]\label{lem:gott:stab:correctable}
Let $\Pi_{\mathcal{C}(L,\vec{s})}$ be the projector on the codespace $\mathcal{C}(L,\vec{s})$.
Then equation \eqref{eq:kl:forpauli} with the substitution
$\Pi_{\mathcal{C}} \mapsto \Pi_{\mathcal{C}(L,\vec{s})}$ will be satisfied for $\mathcal{E}\subseteq \mathcal{P}_q^n$ iff for each $E_a, E_b \in\mathcal{E}$ one of the following holds:
\begin{itemize}
 \item $E_a^\dagger E_b$ is an element of the stabilizer $S$.
 \item There exists an element in $S$ that does not commute with $E_a^\dagger E_b$.
\end{itemize}
\end{lem}
\begin{proof}
We are going to show that if one of the above conditions is satisfied for each $E_a, E_b \in\mathcal{E}$,
equation \eqref{eq:kl:forpauli} will be satisfied, too.
If not, i.\,e. if there exists $E_a^\dagger E_b \notin S$ and there doesn't exist any element in $S$ that does not commute with $E_a^\dagger E_b$, then equation \eqref{eq:kl:forpauli} cannot be satisfied.
The first point is equivalent to
$\Pi_{\mathcal{C}(L,\vec{s})}  E_a^\dagger E_b \Pi_{\mathcal{C}(L,\vec{s})} =
 \Pi_{\mathcal{C}(L,\vec{s})} C_{ab}$, with $C_{ab}=\omega^k$, $k\in\mathbb{F}_q$.
Regarding the second point, let $M$ be the non-commuting element in $S$ and let its eigenvalue of the eigenspace $\mathcal{C}(L,\vec{s})$ be $m$, $M \Pi_{\mathcal{C}(L,\vec{s})} = m \Pi_{\mathcal{C}(L,\vec{s})}$.
Then,
\begin{equation*}
m \Pi_{\mathcal{C}(L,\vec{s})}  E_a^\dagger E_b \Pi_{\mathcal{C}(L,\vec{s})} =
   \Pi_{\mathcal{C}(L,\vec{s})}  E_a^\dagger E_b M \Pi_{\mathcal{C}(L,\vec{s})} =
 \omega^x \Pi_{\mathcal{C}(L,\vec{s})} M E_a^\dagger E_b \Pi_{\mathcal{C}(L,\vec{s})} =
 \omega^x m \Pi_{\mathcal{C}(L,\vec{s})} E_a^\dagger E_b \Pi_{\mathcal{C}(L,\vec{s})},
\end{equation*}
for some $x\neq 0 \in\mathbb{F}_q$ and it follows that $\eqref{eq:kl:forpauli}$ is fulfilled with $C_{ab}=0$.
The remaining possibility is that $E_a^\dagger E_b \notin S$, and  there doesn't exists any element in $S$ that does not commute with $E_a^\dagger E_b$.
It follows that $E_a^\dagger E_b$ commutes with the stabilizer, but is not in the stabilizer itself.
Hence, it performs a logical operation on the encoded data and equation \eqref{eq:kl:forpauli} cannot be satisfied.
\end{proof}

To visualize the structure of the correctable error sets $\mathcal{E}\subseteq \mathcal{P}_q^n$,
let us first define three quotient groups together with their corresponding transversals (generating sets for the coset decompositions):
\begin{itemize}
 \item The cosets of $L^\perp$ in $\mathbb{F}_q^{2n}$ ($\mathbb{F}_q^{2n}/L^\perp$).
 Let a transversal of this decomposition be given by
 $G=\{ \vec{f}_\alpha \}$, i.\,e.     %
 $\vec{f}_\alpha L^\perp   \cap  \vec{f}_\beta L^\perp = \emptyset $ if $\alpha\neq\beta$ and
 $\cup  \vec{f}_\alpha L^\perp = \mathbb{F}_q^{2n}$.
 (Note that in the case under consideration $\vec{f}_\alpha L^\perp = \{ \vec{f}_\alpha \times \vec{l} \ \vert\ \vec{l}\in L^\perp\}$, where the group multiplication rule $\times$ is addition modulo $q$.)
 There are $\vert G\vert = q^{2n}/q^{n+k}=q^{n-k}$ such cosets.\newline
 [For a specific encoding specified by a hyperbolic basis
  $\{\vec{g}_1,\dots,\vec{g}_n,\vec{h}_1\dots,\vec{h}_n\}$ we could choose $G=\vspan\{ \vec{h}_1,\dots,\vec{h}_{n-k} \}$, for instance.
  Each of these cosets might be labeled unambiguously by a syndrome vector $\vec{s}\in\mathbb{F}_q^{n-k}$
  such that $s_i = (\vec{g}_i, \vec{v})_{sp}$, where $\vec{v}$ is an arbitrary vector in the corresponding coset.]

 \item The cosets of $L$ in $L^\perp$ ($L^\perp/L$).
 Let a corresponding transversal be given by $G^\perp=\{ l^\perp_\alpha \}$.
 There are $\vert G^\perp \vert = q^{n+k}/q^{n-k}=q^{2k}$ such cosets.\newline
 [For a specific encoding we could choose $G^\perp=\vspan\{ \vec{g}_{n-k+1},\dots,\vec{g}_n,\vec{h}_{n-k+1},\dots,\vec{h}_n \}$, for instance.
  Each of these cosets might be labeled by a logical error vector $\vec{l}=(\vec{l}^x,\vec{l}^z)\in\mathbb{F}_q^{2k}$ such that $l^x_i=(\vec{g}_{i+n-k},\vec{v})_{sp}$ and $l^z_i=(\vec{v},\vec{h}_{i+n-k})_{sp}$, where $\vec{v}$ is an arbitrary vector in the corresponding coset.]

 \item The cosets of $L$ in $\mathbb{F}_q^{2n}$ ($\mathbb{F}_q^{2n}/L$).
 A corresponding transversal might be obtained by taking the direct product $G\otimes G^\perp$.
 There are
 $\vert G\otimes G^\perp \vert = %
 q^{n-k}q^{2k}=q^{n+k}$
 such cosets.\newline
 [For a specific encoding we could choose $G\otimes G^\perp=\vspan\{ \vec{g}_{n-k+1},\dots,\vec{g}_n,\vec{h}_1,\dots,\vec{h}_n \}$, for instance.
  Each of these cosets can be labeled by a vector $(\vec{s},\vec{l})\in\mathbb{F}_q^{n+k}$.]
\end{itemize}
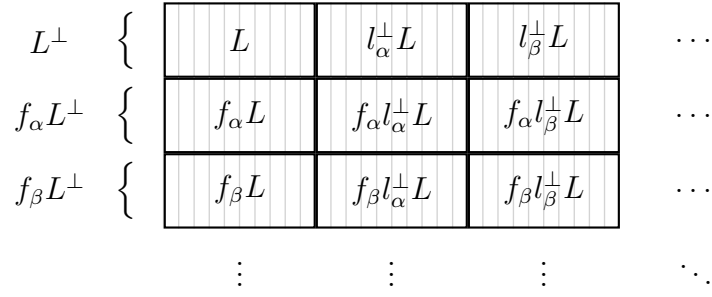
\begin{figure}
\centering
\begin{pspicture}(0,0)(10,-4)%
\multirput(2,0)(0.2,0){30}{\psline[linewidth=0.5pt,linecolor=lightgray](0,0)(0,-1)}
\psframe(2,0)(4,-1)\rput(3,-0.5){\large{$L$}}
\psframe(4,0)(6,-1)\rput(5,-0.5){\large{$l^\perp_\alpha L$}}
\psframe(6,0)(8,-1)\rput(7,-0.5){\large{$l^\perp_\beta L$}}
\rput(9,-0.5){\large{$\dots$}}
\multirput(2,-1)(0.2,0){30}{\psline[linewidth=0.5pt,linecolor=lightgray](0,0)(0,-1)}
\psframe(2,-1)(4,-2)\rput(3,-1.5){\large{$f_\alpha L$}}
\psframe(4,-1)(6,-2)\rput(5,-1.5){\large{$f_\alpha l^\perp_\alpha L$}}
\psframe(6,-1)(8,-2)\rput(7,-1.5){\large{$f_\alpha l^\perp_\beta L$}}
\rput(9,-1.5){\large{$\dots$}}
\multirput(2,-2)(0.2,0){30}{\psline[linewidth=0.5pt,linecolor=lightgray](0,0)(0,-1)}
\psframe(2,-2)(4,-3)\rput(3,-2.5){\large{$f_\beta L$}}
\psframe(4,-2)(6,-3)\rput(5,-2.5){\large{$f_\beta l^\perp_\alpha L$}}
\psframe(6,-2)(8,-3)\rput(7,-2.5){\large{$f_\beta l^\perp_\beta L$}}
\rput(9,-2.5){\large{$\dots$}}
\rput(3,-3.5){\large{$\vdots$}}
\rput(5,-3.5){\large{$\vdots$}}
\rput(7,-3.5){\large{$\vdots$}}
\rput(9,-3.5){\large{$\ddots$}}
\rput(.5,-0.5){\large{$L^\perp $}}
\rput(.5,-1.5){\large{$f_\alpha L^\perp$}}
\rput(.5,-2.5){\large{$f_\beta L^\perp$}}
\rput(1.5,-0.5){\large{$\Bigl\{$}}
\rput(1.5,-1.5){\large{$\Bigl\{$}}
\rput(1.5,-2.5){\large{$\Bigl\{$}}
\end{pspicture}%
\caption[$\mathbb{F}_q^{2n}$ arranged in cosets $\mathbb{F}_q^{2n}/L$]{
All elements of $\mathbb{F}_q^{2n}$ (gray boxes) are arranged in cosets $\mathbb{F}_q^{2n}/L$ (black boxes) generated by some stabilizer $L$.
A corresponding stabilizer code corrects a subset $\mathcal{E}\subseteq\mathbb{F}_q^{2n}$ iff in each row of the diagram no more than one black box is populated by members of $\mathcal{E}$.}%
\label{fig:stabilizer:cosets}%
\end{figure}
Obviously an error set $\mathcal{E}\subseteq\mathcal{P}_q^n$ can equivalently be expressed as a subset $\mathcal{E}_\mathbb{F}\subseteq\mathbb{F}_q^{2n}$ s.\,t. $\mathcal{E}= \{ \XZ(\vec{e}) \ \vert\ \vec{e}\in\mathcal{E}_\mathbb{F} \}$.
In the following we will use the same notation $\mathcal{E}$ for both sets.

\begin{cor}\label{cor:stab:correctable}
Using the coset language, a subset $\mathcal{E} \subseteq \mathbb{F}_q^{2n}$ can be corrected by a stabilizer $L$,
iff in each of the cosets of $L^\perp$ in $\mathbb{F}_q^{2n}$, no more than one of the $\mathbb{F}_q^{2n}/L$-cosets includes elements of $\mathcal{E}$ (compare with figure \ref{fig:stabilizer:cosets}).
\end{cor}
\begin{proof}
If a $\mathbb{F}_q^{2n}/L$-coset includes some elements $\vec{a},\vec{b} \in\mathcal{E}$ it follows that $-\vec{a}+\vec{b} \in L$ and the first of the two conditions in lemma \ref{lem:gott:stab:correctable} is satisfied.
If there is no more than one $\mathbb{F}_q^{2n}/L$-coset populated within a $\mathbb{F}_q^{2n}/L^\perp$-coset,
it follows that for all $\vec{a},\vec{b} \in\mathcal{E}$ s.\,t. $-\vec{a}+\vec{b} \notin L$,
$-\vec{a}+\vec{b} \notin L^\perp$ which is equivalent to $-\vec{a}+\vec{b} \in \mathbb{F}_q^{2n}\setminus L^\perp$, and there exists an element $\vec{g}\in L$ s.\,t. $(\vec{g}, -\vec{a}+\vec{b})_{sp}\neq 0$ and the second condition in lemma \ref{lem:gott:stab:correctable} is satisfied.
\end{proof}
\begin{rem}
Since errors in different $\mathbb{F}_q^{2n}/L^\perp$-cosets lead to different syndromes when the stabilizer is measured, and errors in different $\mathbb{F}_q^{2n}/L$-cosets generate different encoded operations, the corollary makes the following intuitive statement:
All errors having the same syndrome must act in the same way on the encoded information.
Otherwise, knowing the syndrome wouldn't be enough.
\end{rem}
\begin{lem}
A stabilizer code is degenerate if and only if more than one element in the error set $\mathcal{E}\subseteq\mathbb{F}_q^{2n}$ belongs to the same coset of $L$ in $\mathbb{F}_q^{2n}$.
\end{lem}
\begin{proof}
If the code corrects $\mathcal{E}$, the condition
\begin{equation}
 \Pi_{\mathcal{C}(L,\vec{\mathfrak{s}})}  \XZ(\vec{a})^\dagger \XZ(\vec{b}) \Pi_{\mathcal{C}(L,\vec{\mathfrak{s}})} =
\Pi_{\mathcal{C}(L,\vec{\mathfrak{s}})} C_{\vec{a},\vec{b}}
\end{equation}
is satisfied for all $\vec{a},\vec{b}\in\mathcal{E}$ and for all code spaces $\mathcal{C}(L,\vec{\mathfrak{s}})$.
According to definition \ref{defi:degencode}, in order to determine whether or not the code is degenerate, we have to determine whether or not $C_{\vec{a},\vec{b}}$ is singular.
We do this by examining the eigenvalues of $C_{\vec{a},\vec{b}}$.
Note that each element $\vec{a}$ in $\mathcal{E}$ can be decomposed as in \eqref{eq:stabdecompoa},
\begin{equation}
 \vec{a} = \sum_{i=1}^{n-k} \bigl( s_i \vec{h}_i + n_i \vec{g}_i \bigr)
 + \sum_{i=n-k+1}^n \bigl(
 l_{i-(n-k)}^x \vec{h}_i + l_{i-(n-k)}^z \vec{g}_i \bigr),
\end{equation}
and the corresponding Pauli operator $\XZ(\vec{a})$ can be written as
$\overline{X}^{(\vec{s},\vec{l}^x)}\overline{Z}^{(\vec{n},\vec{l}^z)}$ (see lemma \ref{lem:xzdecompostab}).
Let us sort the elements of $\mathcal{E}$ according to their syndrome $\vec{s}=(s_1,\dots,s_{n-k})$.
Then it is clear that $(C_{\vec{a},\vec{b}})$ becomes block-diagonal since
$\Pi_{\mathcal{C}(L,\vec{\mathfrak{s}})}  \XZ(\vec{a})^\dagger \XZ(\vec{b}) \Pi_{\mathcal{C}(L,\vec{\mathfrak{s}})} = 0$ for $\vec{s}(\vec{a}) \neq \vec{s}(\vec{b})$.
We restrict our attention to one of these blocks, i.\,e. we consider only elements of $\mathcal{E}$ with the same syndrome $\vec{s}$.
Corollary \ref{cor:stab:correctable} tells us that there is only one coset of $L$ in the coset of $L^\perp$ in $\mathbb{F}_q^{2n}$ characterized by $\vec{s}$ which is populated with members of $\mathcal{E}$.
Let us assume first that $\mathcal{E}$ contains all $q^{n-k}$ members of this particular coset of $L$.
The Pauli operators of these members are given by the set
$\{ \overline{X}^{(\vec{s},\vec{l}^x)}\overline{Z}^{(\vec{n},\vec{l}^z)} \}_{\vec{n}\in\mathbb{F}_q^{n-k}}$
and the matrix elements $c_{\vec{n},\vec{m}}$ of the block are given by
\begin{equation}
\begin{split}
 c_{\vec{n},\vec{m}} \Pi_{\mathcal{C}(L,\vec{\mathfrak{s}})} &=
 \Pi_{\mathcal{C}(L,\vec{\mathfrak{s}})}
 (\overline{X}^{(\vec{s},\vec{l}^x)}\overline{Z}^{(\vec{n},\vec{l}^z)})^\dagger
 (\overline{X}^{(\vec{s},\vec{l}^x)}\overline{Z}^{(\vec{m},\vec{l}^z)})
 \Pi_{\mathcal{C}(L,\vec{\mathfrak{s}})} = \\
&=\Pi_{\mathcal{C}(L,\vec{\mathfrak{s}})}
 \overline{Z}^{-\vec{n}} \overline{Z}^{\vec{m}}
 \Pi_{\mathcal{C}(L,\vec{\mathfrak{s}})} =
 \omega^{(\vec{m}-\vec{n})\cdot\vec{\mathfrak{s}}}\Pi_{\mathcal{C}(L,\vec{\mathfrak{s}})}.
\end{split}
\end{equation}
Using the fact that $\sum_{\vec{m}\in\mathbb{F}_q^{n-k}} \omega^{\vec{m}\cdot\vec{v}} /q^{n-k} = \delta_{\vec{v},\vec{0}}$ we find that the unitary
\begin{equation}
u = \sum_{\vec{i},\vec{n}\in\mathbb{F}_q^{n-k}} u_{\vec{i},\vec{n}} \ketbra{\vec{i}}{\vec{n}}
  = \sum_{\vec{i},\vec{n}\in\mathbb{F}_q^{n-k}} \omega^{\vec{i}\cdot\vec{n}}/\sqrt{q^{n-k}} \ketbra{\vec{i}}{\vec{n}}
\end{equation}
diagonalizes $c = \sum_{\vec{n},\vec{m}\in\mathbb{F}_q^{n-k}} c_{\vec{n},\vec{m}} \ketbra{\vec{n}}{\vec{m}}$,
\begin{equation}
  ucu^\dagger =
\sum_{\vec{i},\vec{j}\in\mathbb{F}_q^{n-k}} \ketbra{\vec{i}}{\vec{j}}
\sum_{\vec{n},\vec{m}\in\mathbb{F}_q^{n-k}}
 u_{\vec{i},\vec{n}} c_{\vec{n},\vec{m}} u^\dagger_{\vec{m},\vec{j}} =
 q^{n-k} \ketbra{\vec{\mathfrak{s}}}{\vec{\mathfrak{s}}} .
\end{equation}
Hence the eigenvalues of the block $c$ are $(q^{n-k},0,\dots,0)$ which makes the block singular.
Inverting the diagonalization leads to $c = q^{n-k} \ketbra{\psi}{\psi}$ with
$\ket{\psi} = u^\dagger\ket{\vec{\mathfrak{s}}} = \sum_{\vec{i}\in\mathbb{F}_q^{n-k}} \omega^{-\vec{\mathfrak{s}}\vec{i}}/\sqrt{q^{n-k}}\ket{\vec{i}}$.
If the set $\mathcal{E}$ contains only a subset $S$ of the $q^{n-k}$ members of the coset of $L$,
we have to consider the operator $c\vert_S=\sum_{\vec{n},\vec{m}\in S} c_{\vec{n},\vec{m}} \ketbra{\vec{n}}{\vec{m}}$.
Since $c\vert_S$ can be written as $c\vert_S = q^{n-k} \ketbra{\psi_S}{\psi_S}$ with $\ket{\psi_S} = \sum_{\vec{i}\in S} \omega^{-\vec{\mathfrak{s}}\vec{i}}/\sqrt{q^{n-k}}\ket{\vec{i}}$, normalization of $\ket{\psi_S}$ leads to $\ket{\tilde{\psi}_S} = \sqrt{q^{n-k}}/\sqrt{\vert S\vert} \cdot \ket{\psi_S}$ and we obtain $c\vert_S = \vert S\vert \ketbra{\tilde{\psi}_S}{\tilde{\psi}_S}$.
Hence the eigenvalues of the block $c\vert_S$ are $(\vert S\vert,0,\dots,0)$ and again the block is singular.
The only possibility to obtain a non-singular block is that at most one member of the coset of $L$ in $\mathbb{F}_q^{2n}$ is in $\mathcal{E}$, i.\,e. $\vert S\vert =1$. Hence, if the code is non-degenerated,  $C_{\vec{a},\vec{b}}$ is the identity matrix.
\end{proof}

If we want to correct the set $\mathcal{E} = \{ E_i\in\mathcal{P}_q^n \vert \wt(E_i) \leq t \}$ containing error operators of weight $\leq t$, the stabilizer code has to be at least of distance $d\geq 2t+1$.
Let us first give a simple rule to calculate the distance $d$ of a given stabilizer code.
\begin{cor}\label{cor:diststab}
The distance $d$ of a stabilizer code is the minimum weight\footnote{Here the weight of an element $\vec{e}\in\mathbb{F}_q^{2n}$ is defined as the weight of $\XZ(\vec{e})$.} of the elements in $L^\perp\setminus L$.
\end{cor}
\begin{proof}
It follows from lemma \ref{lem:gott:stab:correctable} and definition \ref{defi:distanceqc} that for a stabilizer code of distance $d$, each error operator $E\in\mathcal{P}_q^n$ of weight less than $d$ is either in $S$ or does not commute with some $M\in S$.
This statement is equivalent to each of the following statements and to the corollary itself:
Each $\vec{e}\in\mathbb{F}_q^{2n}$ of weight less than $d$ is in $L\cup (\mathbb{F}_q^{2n}\setminus L^\perp)$;
In $L^\perp\setminus L$ is no element of weight less than $d$;
\end{proof}
\noindent
Now we state a quantum Gilbert Varshamov lower bound on the rate of $q$-ary stabilizer codes of distance~$d$.
\begin{thm}[Gilbert Varshamov bound for stabilizer codes \cite{FeMa04}]\label{thm:qgv}
Suppose $n>k\geq 2$, $d\geq 2$ and $n=k \pmod 2$.
Then there exists a stabilizer code
of distance $d$ encoding $k$ qudits into $n$, provided that
\begin{equation}
 \frac{q^{n-k+2}-1}{q^2-1} > \sum_{i=1}^{d-1} (q^2-1)^{i-1} \binom{n}{i}.
\end{equation}
\end{thm}
\noindent
Since the proof is more sophisticated, we refer to the original work \cite{FeMa04}.
A weaker bound is given in \cite{KKKS06} (and \cite{MaUy02} for the binary case).
Note that the bound found for the binary case in \cite{EM96} and \cite[chapter 7.1]{PhdGottesman} has been criticized (see e.\,g. \cite{HNO03}).
An asymptotic version of the above bound was known previously \cite{AsKn01}.
\begin{cor}[Asymptotic GV for stabilizer codes \cite{AsKn01}]For large $n$,
there exist stabilizer codes of distance $d$ encoding $k$ qudits into $n$, such that
\begin{equation}
 \frac{k}{n} \geq 1 - 2H_{q^2[\log_{q^2}]}\Bigl( 1-\frac{d}{n},\frac{d/n}{q^2-1},\dots, \frac{d/n}{q^2-1} \Bigr).
\end{equation}
\end{cor}
\begin{proof}
Using the Chernoff bound \ref{lem:chernovbound} (as it was done in proving the asymptotic limit of theorem \ref{thm:generalgv}), this corollary follows from theorem \ref{thm:qgv}.
\end{proof}
\begin{rem}
For qubits ($q=2$) the asymptotic bound becomes \cite{CRSS97} \cite[chapter 7.14]{Preskill}
\begin{equation}
 \frac{k}{n} \geq 1 - H_2\Bigl(\frac{d}{n}\Bigr) - \frac{d}{n}\log_2 3.
\end{equation}
\end{rem}

\subsection{Recovery Operation}\label{subsec:stabrecover}

For stabilizer codes Knill and Laflamme's criterion for reversibility of a quantum operation $\mathcal{A}$ on a codespace $\mathcal{C}$ leads to lemma \ref{lem:gott:stab:correctable} and corollary \ref{cor:stab:correctable}, telling us what kind of error subsets $\mathcal{E}\subseteq\mathcal{P}_q^n$ might be corrected by a certain stabilizer code.
We are now going to write down the recovery operation which achieves the desired correction of such an error subset.

As discussed in the last subsection, the cosets of $L^\perp$ in $\mathbb{F}_q^{2n}$ can be labeled by a syndrome vector $\vec{s}\in\mathbb{F}_q^{n-k}$ such that $s_i = (\vec{g}_i, \vec{v})_{sp}$, where $\vec{v}$ is an arbitrary member of the corresponding coset.
Let us construct a set of coset representatives (a transversal) $J_0$ by choosing a vector $\vec{J}_0(\vec{s})$ from each coset $\vec{s}$ of $L^\perp$ in $\mathbb{F}_q^{2n}$,
$J_0=\{ \vec{J}_0(\vec{s}) \vert \vec{s}\in\mathbb{F}_q^{n-k} \}$.
The error subset $J=J_0+L \subseteq\mathbb{F}_q^{2n}$\footnote{$A+B=\{a+b\vert a\in A,b\in B\}$}
can obviously be corrected:
Using figure \ref{fig:stabilizer:cosets}, $J_0$ by construction has the property that each of the rows in the figure contains exactly one of its elements.
Now we can easily write down the recovery operation which reverses all quantum operations $\mathcal{A}$ with support on $J$ on the codespace $\mathcal{C}(L,\vec{s}_0)$:
\begin{equation}
 \mathcal{R}_{\vec{s}_0}^{(J)} \bigl( \mathcal{A}(\rho) \bigr) =
 \sum_{\vec{t}\in\mathbb{F}_q^{n-k}}
  \XZ^\dagger(\vec{J}_0(\vec{t})) \, \Pi_{\mathcal{C}(L,\vec{s}_0+\vec{t})}
    \ \mathcal{A}(\rho)\
  \Pi_{\mathcal{C}(L,\vec{s}_0+\vec{t})} \, \XZ(\vec{J}_0(\vec{t})) = \rho \in\mathcal{S}(\mathcal{C}(L,\vec{s}_0)).
\end{equation}
To generate this tpcp-map, we could first measure the syndrome $\vec{s}_0+\vec{t}$, thereby projecting onto the codespace $\mathcal{C}(L,\vec{s}_0+\vec{t})$.
Afterwards we apply the Pauli-operator $\XZ^\dagger(\vec{J}_0(\vec{t}))$ to go back into the original codespace $\mathcal{C}(L,\vec{s}_0)$ and to undo the remaining logical error.
Note that a stabilizer code correcting $J_0$ is non-degenerate, but becomes degenerate when correcting $J$.

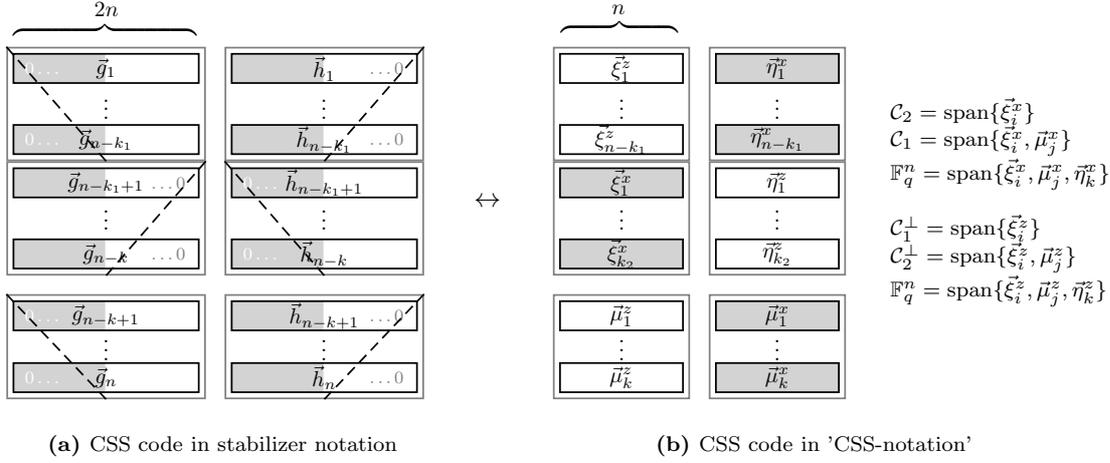
\begin{figure}
\subfloat[CSS code in stabilizer notation]{\label{fig:stab-to-css:cssa}%
\begin{minipage}[c]{0.42\textwidth}
 \centering
 \begin{pspicture}(-0.17,-0.37)(5.70,5.44)%
 \scalebox{0.82}{%
   \rput[origin=c]{-90}(1.5,6){$\rotatebox[origin=c]{90}{$2n$}\left\{\makebox(0,1.6){}\right.$}
   \psframe[linecolor=gray](-0.1,-0.1)(3.1,1.6)
   \psframe[linecolor=gray](-0.1,1.9)(3.1,3.75)\psframe[linecolor=gray](-0.1,3.75)(3.1,5.6)
   \psframe[fillstyle=solid,fillcolor=lightgray,linestyle=none](0,0)(1.5,0.5)
   \psframe[fillstyle=solid,fillcolor=white,linestyle=none](1.5,0)(3,0.5)
   \psframe(0,0)(3,0.5)
   \rput(1.5,0.25){$\vec{g}_n$}\rput(0.5,0.25){\white{\footnotesize{$0\dots$}}}
   \rput(1.5,0.8){$\vdots$}
   \psframe[fillstyle=solid,fillcolor=lightgray,linestyle=none](0,1)(1.5,1.5)
   \psframe[fillstyle=solid,fillcolor=white,linestyle=none](1.5,1)(3,1.5)
   \psframe(0,1)(3,1.5)
   \rput(1.5,1.25){$\vec{g}_{n-k+1}$}\rput(0.5,1.25){\white{\footnotesize{$0\dots$}}}
  \psline[linestyle=dashed](-0.1,1.6)(1.5,-0.1)
   \psframe[fillstyle=solid,fillcolor=lightgray,linestyle=none](0,2)(1.5,2.5)
   \psframe[fillstyle=solid,fillcolor=white,linestyle=none](1.5,2)(3,2.5)
   \psframe(0,2)(3,2.5)
   \rput(1.5,2.25){$\vec{g}_{n-k}$}\rput(2.5,2.25){\gray{\footnotesize{$\dots 0$}}}
   \rput(1.5,2.9){$\vdots$}
   \psframe[fillstyle=solid,fillcolor=lightgray,linestyle=none](0,3.15)(1.5,3.65)
   \psframe[fillstyle=solid,fillcolor=white,linestyle=none](1.5,3.15)(3,3.65)
   \psframe(0,3.15)(3,3.65)
   \rput(1.5,3.4){$\vec{g}_{n-k_1+1}$}\rput(2.5,3.4){\gray{\footnotesize{$\dots 0$}}}
  \psline[linestyle=dashed](1.5,1.9)(3.1,3.75)
   \psframe[fillstyle=solid,fillcolor=lightgray,linestyle=none](0,3.85)(1.5,4.35)
   \psframe[fillstyle=solid,fillcolor=white,linestyle=none](1.5,3.85)(3,4.35)
   \psframe(0,3.85)(3,4.35)
   \rput(1.5,4.1){$\vec{g}_{n-k_1}$}\rput(0.5,4.1){\white{\footnotesize{$0\dots$}}}
   \rput(1.5,4.7){$\vdots$}
   \psframe[fillstyle=solid,fillcolor=lightgray,linestyle=none](0,5)(1.5,5.5)
   \psframe[fillstyle=solid,fillcolor=white,linestyle=none](1.5,5)(3,5.5)
   \psframe(0,5)(3,5.5)
   \rput(1.5,5.25){$\vec{g}_1$}\rput(0.5,5.25){\white{\footnotesize{$0\dots$}}}
  \psline[linestyle=dashed](-0.1,5.6)(1.5,3.75)

   \psframe[linecolor=gray](3.4,-0.1)(6.6,1.6)
   \psframe[linecolor=gray](3.4,1.9)(6.6,3.75)\psframe[linecolor=gray](3.4,3.75)(6.6,5.6)
   \psframe[fillstyle=solid,fillcolor=lightgray,linestyle=none](3.5,0)(5,0.5)
   \psframe[fillstyle=solid,fillcolor=white,linestyle=none](5,0)(6.5,0.5)
   \psframe(3.5,0)(6.5,0.5)
   \rput(5,0.25){$\vec{h}_n$}\rput(6,0.25){\gray{\footnotesize{$\dots 0$}}}
   \rput(5,0.8){$\vdots$}
   \psframe[fillstyle=solid,fillcolor=lightgray,linestyle=none](3.5,1)(5,1.5)
   \psframe[fillstyle=solid,fillcolor=white,linestyle=none](5,1)(6.5,1.5)
   \psframe(3.5,1)(6.5,1.5)
   \rput(5,1.25){$\vec{h}_{n-k+1}$}\rput(6,1.25){\gray{\footnotesize{$\dots 0$}}}
  \psline[linestyle=dashed](5,-0.1)(6.6,1.6)
   \psframe[fillstyle=solid,fillcolor=lightgray,linestyle=none](3.5,2)(5,2.5)
   \psframe[fillstyle=solid,fillcolor=white,linestyle=none](5,2)(6.5,2.5)
   \psframe(3.5,2)(6.5,2.5)
   \rput(5,2.25){$\vec{h}_{n-k}$}\rput(4,2.25){\white{\footnotesize{$0\dots$}}}
   \rput(5,2.9){$\vdots$}
   \psframe[fillstyle=solid,fillcolor=lightgray,linestyle=none](3.5,3.15)(5,3.65)
   \psframe[fillstyle=solid,fillcolor=white,linestyle=none](5,3.15)(6.5,3.65)
   \psframe(3.5,3.15)(6.5,3.65)
   \rput(5,3.4){$\vec{h}_{n-k_1+1}$}\rput(4,3.4){\white{\footnotesize{$0\dots$}}}
  \psline[linestyle=dashed](5,1.9)(3.4,3.75)
   \psframe[fillstyle=solid,fillcolor=lightgray,linestyle=none](3.5,3.85)(5,4.35)
   \psframe[fillstyle=solid,fillcolor=white,linestyle=none](5,3.85)(6.5,4.35)
   \psframe(3.5,3.85)(6.5,4.35)
   \rput(5,4.1){$\vec{h}_{n-k_1}$}\rput(6,4.1){\gray{\footnotesize{$\dots 0$}}}
   \rput(5,4.7){$\vdots$}
   \psframe[fillstyle=solid,fillcolor=lightgray,linestyle=none](3.5,5)(5,5.5)
   \psframe[fillstyle=solid,fillcolor=white,linestyle=none](5,5)(6.5,5.5)
   \psframe(3.5,5)(6.5,5.5)
   \rput(5,5.25){$\vec{h}_1$}\rput(6,5.25){\gray{\footnotesize{$\dots 0$}}}
  \psline[linestyle=dashed](5,3.75)(6.6,5.6)
 }%
 \end{pspicture}
\end{minipage}}%
\begin{minipage}[c]{0.03\textwidth}
$\leftrightarrow$
\end{minipage}%
\subfloat[CSS code in 'CSS-notation']{\label{fig:stab-to-css:cssb}
\begin{minipage}[c]{0.31\textwidth}
 \centering
 \begin{pspicture}(-0.17,-0.37)(4.00,5.44)%
\scalebox{0.82}{%
   \rput[origin=c]{-90}(1,6){$\rotatebox[origin=c]{90}{$n$}\left\{\makebox(0,1.1){}\right.$}
   \psframe[linecolor=gray](-0.1,-0.1)(2.1,1.6)
   \psframe[linecolor=gray](-0.1,1.9)(2.1,3.75)\psframe[linecolor=gray](-0.1,3.75)(2.1,5.6)
   \psframe[fillstyle=solid,fillcolor=white](0,0)(2,0.5)
   \rput(1,0.25){$\vec{\mu}^z_k$}
   \rput(1,0.8){$\vdots$}
   \psframe[fillstyle=solid,fillcolor=white](0,1)(2,1.5)
   \rput(1,1.25){$\vec{\mu}^z_1$}
   \psframe[fillstyle=solid,fillcolor=lightgray](0,2)(2,2.5)
   \rput(1,2.25){$\vec{\xi}^x_{k_2}$}
   \rput(1,2.9){$\vdots$}
   \psframe[fillstyle=solid,fillcolor=lightgray](0,3.15)(2,3.65)
   \rput(1,3.4){$\vec{\xi}^x_1$}
   \psframe[fillstyle=solid,fillcolor=white](0,3.85)(2,4.35)
   \rput(1,4.1){$\vec{\xi}^z_{n-k_1}$}
   \rput(1,4.7){$\vdots$}
   \psframe[fillstyle=solid,fillcolor=white](0,5)(2,5.5)
   \rput(1,5.25){$\vec{\xi}^z_1$}

   \psframe[linecolor=gray](2.4,-0.1)(4.6,1.6)
   \psframe[linecolor=gray](2.4,1.9)(4.6,3.75)\psframe[linecolor=gray](2.4,3.75)(4.6,5.6)
   \psframe[fillstyle=solid,fillcolor=lightgray](2.5,0)(4.5,0.5)
   \rput(3.5,0.25){$\vec{\mu}^x_k$}
   \rput(3.5,0.8){$\vdots$}
   \psframe[fillstyle=solid,fillcolor=lightgray](2.5,1)(4.5,1.5)
   \rput(3.5,1.25){$\vec{\mu}^x_1$}
   \psframe[fillstyle=solid,fillcolor=white](2.5,2)(4.5,2.5)
   \rput(3.5,2.25){$\vec{\eta}^z_{k_2}$}
   \rput(3.5,2.9){$\vdots$}
   \psframe[fillstyle=solid,fillcolor=white](2.5,3.15)(4.5,3.65)
   \rput(3.5,3.4){$\vec{\eta}^z_1$}
   \psframe[fillstyle=solid,fillcolor=lightgray](2.5,3.85)(4.5,4.35)
   \rput(3.5,4.1){$\vec{\eta}^x_{n-k_1}$}
   \rput(3.5,4.7){$\vdots$}
   \psframe[fillstyle=solid,fillcolor=lightgray](2.5,5)(4.5,5.5)
   \rput(3.5,5.25){$\vec{\eta}^x_1$}
 }%
 \end{pspicture}
\end{minipage}%
\begin{minipage}[c]{0.19\textwidth}\footnotesize
$\mathcal{C}_2=\vspan\{\vec{\xi}^x_i\}$\\
$\mathcal{C}_1=\vspan\{\vec{\xi}^x_i,\vec{\mu}^x_j\}$\\
$\mathbb{F}_q^n=\vspan\{\vec{\xi}^x_i,\vec{\mu}^x_j,\vec{\eta}^x_k\}$\\

$\mathcal{C}_1^\perp=\vspan\{\vec{\xi}^z_i\}$\\
$\mathcal{C}_2^\perp=\vspan\{\vec{\xi}^z_i,\vec{\mu}^z_j\}$\\
$\mathbb{F}_q^n=\vspan\{\vec{\xi}^z_i,\vec{\mu}^z_j,\vec{\eta}^z_k\}$
\end{minipage}}%
\caption[CSS codes]{(a)~CSS codes form a subclass of stabilizer codes:
For the first $n-k_1$ generating elements $\vec{g}=(g_1^x,\dots,g_n^x, g_1^z,\dots,g^z_n)$, the $x$-part of the vector is $0$, while for the next $k_2$ generating elements, the $z$-part is $0$ ($k=k_1-k_2$).
An extension to a hyperbolic basis can be chosen which shows an analogous structure.
(b)~Since each of the vectors in $\mathbb{F}_q^{2n}$ becomes effectively a vector in $\mathbb{F}_q^n$, we refer to these $n$-dit vectors as indicated in the figure.
Using the definition of a CSS code by the means of two classical codes $\mathcal{C}_2 \subseteq \mathcal{C}_1$, the relations between these codes and the $n$-dit vectors is shown on the right.
}%
\label{fig:stab-to-css:css}%
\end{figure}

\section{CSS Codes}\label{sec:csscodes}

CSS codes are constructed from two classical linear codes $\mathcal{C}_1$ and $\mathcal{C}_2$ such that $\mathcal{C}_2 \subseteq \mathcal{C}_1$.
They have been developed independently by Calderbank, Shor and Steane \cite{CS96,St96} in 1996.
Since CSS codes also form a subclass of stabilizer codes, we will start the description of these codes from this point of view, and establish the connection with the classical codes later on in this section.

\begin{defi}
CSS codes form a subclass of stabilizer codes in which the generating elements of the stabilizer
$L = \vspan \{\vec{g}_1,\dots,\vec{g}_{n-k}\}$, $\vec{g}_i\in\mathbb{F}_q^{2n}$,
have either a vanishing $x$-part
($\vec{g}=(0,\dots,0, g_1^z,\dots,g^z_n)$, $z$-type $\vec{g}$)
or a vanishing $z$-part
($\vec{g}=(g_1^x,\dots,g_n^x, 0,\dots,0)$, $x$-type $\vec{g}$).
Setting $k=k_1-k_2$, we will use the convention that the first $n-k_1$ generating elements are $z$-type vectors, while the next $k_2$ generating elements are $x$-type vectors.
\end{defi}

\subsection{Encoding Operations}\label{subsec::cssencod}
As discussed in the last section, an encoding for a stabilizer code $L$ is specified by an extension of the generating elements $\{\vec{g}_1,\dots,\vec{g}_{n-k}\}$ of $L$ to a hyperbolic basis $\{\vec{g}_1,\dots,\vec{g}_n,\vec{h}_1,\dots,\vec{h}_n\}$ spanning $\mathbb{F}_q^{2n}$.
The elements of a hyperbolic basis obey relations \eqref{eq:stab:commutatorrelations},
i.\,e. vanishing symplectic inner products between any two $\vec{g}$'s and any two $\vec{h}$'s,
and non-vanishing inter inner product: $(\vec{g}_i,\vec{h}_j)_{sp}=\delta_{ij}$ (compare with figure \ref{fig:stab-to-css:stab}).
According to their definition, the generating elements of CSS codes are of $x$-type and $z$-type only.
Considering possible extensions to hyperbolic bases for such codes, it turns out that it is always possible to find extensions which have
the same $x$-type/$z$-type structure.
For example the first $n-k_1$ vectors $\{\vec{h}_1,\dots,\vec{h}_{n-k_1}\}$ have to be $x$-type vectors in order to fulfill $(\vec{g}_i,\vec{h}_j)_{sp}=\delta_{ij}$, since the first $n-k_1$ generating elements are $z$-type vectors.
The detailed form of such extensions is shown in figure \ref{fig:stab-to-css:cssa}.
This means that a CSS code plus an encoding is effectively specified by $2n$ vectors in $\mathbb{F}_q^n$.
Each of these $n$-dit vectors is given a unique notation as indicated in figure \ref{fig:stab-to-css:cssb}, e.\,g. the first $n-k_1$ generating elements $\vec{g}_i\in \mathbb{F}_q^{2n}$ (which are $z$-type vectors) are denoted as $\vec{\xi}^z_i\in\mathbb{F}_q^n$ now ($\vec{g}_i=(\vec{0},\vec{\xi}^z_i)$).
The basis $\{\vec{g}_1,\dots,\vec{g}_n  ;\ \vec{h}_1,\dots,\vec{h}_n\}$ becomes
\begin{equation}\label{eq:hypercssbasis}
 \{\vec{\xi}^z_1,\dots,\vec{\xi}^z_{n-k_1},
   \vec{\xi}^x_1,\dots,\vec{\xi}^x_{k_2}, \vec{\mu}^z_1,\dots,\vec{\mu}^z_k  ;\
   \vec{\eta}^x_1,\dots,\vec{\eta}^x_{n-k_1},
   \vec{\eta}^z_1,\dots,\vec{\eta}^z_{k_2}, \vec{\mu}^x_1,\dots,\vec{\mu}^x_k   \}
\end{equation}
in the new notation.
Since both notations are equivalent, occasionally we will use them simultaneously.
The three relations \eqref{eq:stab:commutatorrelations} a hyperbolic basis has to fulfill, translate into nine relations the $n$-dit vectors \eqref{eq:hypercssbasis} have to fulfill.
Regarding the $n$-dit vectors as row-vectors, we can put these nine relations into one single equation:
\begin{equation}\label{eq:css:nineconditions}
\left(
\begin{array}{c}
 \vec{\xi}^z_1 \\
 \vdots\\
 \vec{\eta}^z_1\\
 \vdots\\
 \vec{\mu}^z_1\\
 \vdots\\
\end{array}
\right)\cdot\left(
 (\vec{\eta}^x_1)^T \cdots
 (\vec{\xi}^x_1)^T \cdots
 (\vec{\mu}^x_1)^T
\right)=\left(
\scalebox{0.7}{
\text{\scriptsize{$
\begin{array}{c@{}c@{}cc@{}c@{}cc@{}c@{}c}
\makebox(11,11){1}&\makebox(11,11){0}&\makebox(11,11){$\cdots$}&
\makebox(11,11){0}&\makebox(11,11){0}&\makebox(11,11){$\cdots$}&
\makebox(11,11){0}&\makebox(11,11){0}&\makebox(11,11){$\cdots$}\\
\makebox(11,11){0}&\makebox(11,11){1}&&
\makebox(11,11){0}&\makebox(11,11){0}&&
\makebox(11,11){0}&\makebox(11,11){0}&\\
\makebox(11,11){$\vdots$}& &\makebox(11,11){$\ddots$}&
\makebox(11,11){$\vdots$}& &\makebox(11,11){$\ddots$}&
\makebox(11,11){$\vdots$}& &\makebox(11,11){$\ddots$}\\
\\
\makebox(11,11){0}&\makebox(11,11){0}&\makebox(11,11){$\cdots$}&
\makebox(11,11){1}&\makebox(11,11){0}&\makebox(11,11){$\cdots$}&
\makebox(11,11){0}&\makebox(11,11){0}&\makebox(11,11){$\cdots$}\\
\makebox(11,11){0}&\makebox(11,11){0}&&
\makebox(11,11){0}&\makebox(11,11){1}&&
\makebox(11,11){0}&\makebox(11,11){0}&\\
\makebox(11,11){$\vdots$}& &\makebox(11,11){$\ddots$}&
\makebox(11,11){$\vdots$}& &\makebox(11,11){$\ddots$}&
\makebox(11,11){$\vdots$}& &\makebox(11,11){$\ddots$}\\
\\
\makebox(11,11){0}&\makebox(11,11){0}&\makebox(11,11){$\cdots$}&
\makebox(11,11){0}&\makebox(11,11){0}&\makebox(11,11){$\cdots$}&
\makebox(11,11){1}&\makebox(11,11){0}&\makebox(11,11){$\cdots$}\\
\makebox(11,11){0}&\makebox(11,11){0}&&
\makebox(11,11){0}&\makebox(11,11){0}&&
\makebox(11,11){0}&\makebox(11,11){1}&\\
\makebox(11,11){$\vdots$}& &\makebox(11,11){$\ddots$}&
\makebox(11,11){$\vdots$}& &\makebox(11,11){$\ddots$}&
\makebox(11,11){$\vdots$}& &\makebox(11,11){$\ddots$}\\
\end{array}$}}%
}
\right).
\end{equation}
It follows that the two matrices which are multiplied above, cannot be singular.
This fact is equivalent to
\begin{subequations}\label{eq:xietamuspanf}%
\begin{align}
\mathbb{F}_q^n &= \vspan\{\vec{\xi}^z_1,\dots,\vec{\xi}^z_{n-k_1}, \vec{\eta}^z_1,\dots,\vec{\eta}^z_{k_2},
        \vec{\mu}^z_1,\dots,\vec{\mu}^z_k\} \\
\text{ and } \mathbb{F}_q^n &= \vspan\{\vec{\eta}^x_1,\dots,\vec{\eta}^x_{n-k_1}, \vec{\xi}^x_1,\dots,\vec{\xi}^x_{k_2},
         \vec{\mu}^x_1,\dots,\vec{\mu}^x_k\},
\end{align}
\end{subequations}
respectively.

As it is mentioned in the beginning of this section, the original construction of CSS codes makes use of two classical codes $\mathcal{C}_2\subseteq \mathcal{C}_1$.
Let $\mathcal{C}_1$ be an $[n,k_1]_q$ code code encoding $k_1$ dits into $n$, and $\mathcal{C}_2$ be an $[n,k_2]_q$ code with $k_2\leq k_1$.
Then the CSS code which is constructed using these classical codes, plus an encoding, is specified by the two lists of vectors,
\begin{align*}
&\{\vec{\xi}^z_1,\dots,\vec{\xi}^z_{n-k_1} \,,\, \vec{\eta}^z_1,\dots,\vec{\eta}^z_{k_2} \,,\,
        \vec{\mu}^z_1,\dots,\vec{\mu}^z_k\} \text{ and } \\
&\{\vec{\eta}^x_1,\dots,\vec{\eta}^x_{n-k_1} \,,\, \vec{\xi}^x_1,\dots,\vec{\xi}^x_{k_2} \,,\,
         \vec{\mu}^x_1,\dots,\vec{\mu}^x_k\},
\end{align*}
both spanning $\mathbb{F}_q^n$ and satisfying \eqref{eq:css:nineconditions}, where
$\mathcal{C}_1^\perp=\vspan\{ \vec{\xi}^z_1,\dots,\vec{\xi}^z_{n-k_1} \}$ and
$\mathcal{C}_2=\vspan\{ \vec{\xi}^x_1,\dots,\vec{\xi}^x_{k_2} \}$.
It follows that $\mathcal{C}_2^\perp$ has to be spanned by
$\{ \vec{\xi}^z_i, \vec{\mu}^z_j\}_{i\in\{1\dots n-k_1\},j\in\{1\dots k\}}$, while $\mathcal{C}_1$ has to be spanned by
$\{ \vec{\xi}^x_i, \vec{\mu}^x_j\}_{i\in\{1\dots k_2\},j\in\{1\dots k\}}$ in order to satisfy \eqref{eq:css:nineconditions}.

Keeping in mind that a CSS code together with a corresponding encoding operation is fully specified by the two sets of $n$-dit vectors
in equation \eqref{eq:xietamuspanf} and by a set of phases
$\{ \theta_\alpha(i) \in \{\omega^r\vert r\in\mathbb{F}_q\} \}_{\alpha\in\{x,z\}, i\in\{1\dots n\}}$,
we are now going to explicitly construct the $q^k$ encoded basis states for all $q^{n-k}$ codespaces using the definition of the encoding operator $U_\text{enc}$ given in \eqref{eq:stabencoderdefi}.
First, we have to find the common eigenvector of the set $\{ \overline{Z}_i \}_{ i\in\{1,\dots,n\} }$ of encoded $Z$-operators
with eigenvalue list $(\omega^0,\dots,\omega^0)$.
Let us set the phase factors $\theta_z(\cdot)$ and $\theta_x(\cdot)$ equal to one, i.\,e. we use the encoded operators
$\overline{Z}_i = \XZ(\vec{g}_i)$ and
$\overline{X}_i = \XZ(\vec{h}_i)$ for $i\in\{1,\dots,n\}$.
The only $X$-operators in the set of encoded $Z$-operators are those constructed from elements spanning $\mathcal{C}_2$.
Hence, the state
\begin{equation}\label{eq:css000}
 \overline{\ket{0\dots 0}} = \frac{1}{\sqrt{\vert \mathcal{C}_2\vert}} \sum_{\vec{\mathfrak{v}}\in\mathcal{C}_2} \ket{\ \vec{\mathfrak{v}}\ }
\end{equation}
is certainly a common eigenstate of these operators with eigenvalue $+1$.
It is also a common $+1$ eigenstate of the $Z$-operators in $\{ \overline{Z}_i \}$, since all these operators are generated by elements of $\mathcal{C}_2^\perp$.
Applying the $\{ \overline{X}_j= \XZ(\vec{h}_j) \}_{j\in\{1,\dots,n\}}$-operators onto the state $\overline{\ket{0\dots 0}}$
constructs all encoded states:
\begin{equation}\label{eq:cssencodedstates}
\begin{split}
 \overline{ \ket{ \vec{x}, \vec{z} ,\vec{c} } }
&= \overline{X}^{(\vec{x},\vec{z},\vec{c})} \overline{\ket{0\dots 0}}\\
&= \frac{1}{\sqrt{\vert \mathcal{C}_2\vert}} \sum_{\vec{\mathfrak{v}}\in\mathcal{C}_2}
   \omega^{\vec{\mathfrak{z}}\cdot\vec{\mathfrak{v}}} \ket{\ \vec{\mathfrak{v}}+\vec{\mathfrak{c}}+\vec{\mathfrak{x}}\ },
\end{split}
\end{equation}
where the vectors $\vec{\mathfrak{x}},\vec{\mathfrak{z}}$ and $\vec{\mathfrak{c}}$ are given by
\begin{align}\label{eq:css:encodedstatesdefi}
 \vec{\mathfrak{x}} &= \sum_{i=1}^{n-k_1} x_i \vec{\eta}^x_i,  &
 \vec{\mathfrak{z}} &= \sum_{i=1}^{k_2} z_i \vec{\eta}^z_i,  &
 \text{ and }
 \vec{\mathfrak{c}} &= \sum_{i=1}^k c_i \vec{\mu}^x_i.
\end{align}
The basis of the $q^k$-dimensional code space $\mathcal{C}(L,\vec{s})$ with syndrome $\vec{s}=(\vec{x},\vec{z})$ is given by the orthonormal set of states
$\{  \overline{ \ket{ \vec{x}, \vec{z} ,\vec{c} } } \}_{ \vec{c}\in\mathbb{F}_q^k }$.

As it was mentioned in section \ref{sec:stabenc},
any vector $\vec{a}\in\mathbb{F}_q^{2n}$ can be expressed as linear combination of the basis elements
of a given hyperbolic basis $\{\vec{g}_1,\dots,\vec{g}_n, \vec{h}_1,\dots,\vec{h}_n\}$ of~$\mathbb{F}_q^{2n}$,
\begin{equation}
\begin{split}
 \vec{a} &= ( a_1^x,\dots,a_n^x, a_1^z,\dots,a_n^z ) = (\vec{a}^x,\vec{a}^z) \\
 &= \sum_{i=1}^{n-k} \bigl( s_i \vec{h}_i + n_i \vec{g}_i \bigr)
 + \sum_{i=1}^k \bigl(
 l_i^x \vec{h}_{i+n-k} + l_i^z \vec{g}_{i+n-k} \bigr),
\end{split}
\end{equation}
where $s_i = (\vec{a},\vec{g}_i)_{sp}$ etc.
Taking into account the special structure of such a basis in the CSS case
(i.\,e. the fact that $\vec{g}_1=(\vec{0},\vec{\xi}^z_1)$ etc.),
the $x$- and $z$-part of $\vec{a}$ can be decomposed separately,
\begin{subequations}\label{eq:cssdecompoa}
\begin{align}
 \vec{a}^x &=
     \sum_{i=1}^{n-k_1} s_i^x \vec{\eta}^x_i +
     \sum_{i=1}^{k_2}   n_i^z \vec{\xi}^x_i  +
     \sum_{i=1}^k       l_i^x \vec{\mu}^x_i \\
 \vec{a}^z &=
     \sum_{i=1}^{n-k_1} n_i^x \vec{\xi}^z_i +
     \sum_{i=1}^{k_2}   s_i^z \vec{\eta}^z_i  +
     \sum_{i=1}^k       l_i^z \vec{\mu}^z_i,
\end{align}
\end{subequations}
where $\vec{s}=(\vec{s}^x,\vec{s}^z)$, $\vec{n}=(\vec{n}^x,\vec{n}^z)$ and
e.\,g. $s_1 = (\vec{g}_1,\vec{a})_{sp} = \vec{\xi}^z_1 \cdot \vec{a}^x = s^x_1$, et cetera.
Analogous to lemma \ref{lem:xzdecompostab}, expression \eqref{eq:xzaxzb} gives the next lemma.
\begin{lem}\label{lem:xzdecompocss}
Any Pauli operator $\XZ(\vec{a}\in\mathbb{F}_q^{2n}) \in \mathcal{P}_q^n$ can be expressed (up to a phase) as product of some powers of the operators
$\XZ(\vec{\eta}^x_i,\vec{0}), \XZ(\vec{\xi}^x_j,\vec{0}), \XZ(\vec{\mu}^x_l,\vec{0})$ and
$\XZ(\vec{0},\vec{\xi}^z_i), \XZ(\vec{0},\vec{\eta}^z_j), \XZ(\vec{0},\vec{\mu}^z_l)$,
\begin{multline}
 \XZ(\vec{a}) \sim
  \prod_{i=1}^{n-k_1}\bigl( \XZ(\vec{\eta}^x_i,\vec{0})^{s^x_i} \XZ(\vec{0},\vec{\xi}^z_i)^{n^x_i} \bigr)
  \prod_{i=1}^{k_2}  \bigl( \XZ(\vec{0},\vec{\eta}^z_i)^{s^z_i} \XZ(\vec{\xi}^x_i,\vec{0})^{n^z_i} \bigr) \\
  \prod_{i=1}^k \bigl( \XZ(\vec{\mu}^x_i,\vec{0})^{l_i^x} \XZ(\vec{0},\vec{\mu}^z_i)^{l_i^z} \bigr),
\end{multline}
or by using the operators $\overline{Z}_i,\overline{X}_i$ as defined in \eqref{eq:defilogicalxandzops},
\begin{multline}
\sim
 \prod_{i=1}^{n-k_1}\bigl( \overline{X}_i^{s^x_i} \overline{Z}_i^{n^x_i} \bigr)
 \prod_{i=1}^{k_2}  \bigl( \overline{X}_{i+n-k_1}^{s^z_i} \overline{Z}_{i+n-k_1}^{n^z_i} \bigr)
 \prod_{i=1}^k \bigl( \overline{X}_{i+n-k}^{l_i^x} \overline{Z}_{i+n-k}^{l_i^z} \bigr)
=  \overline{X}^{(\vec{s}^x,\vec{s}^z,\vec{l}^x)}  \overline{Z}^{(\vec{n}^x,\vec{n}^z,\vec{l}^z)},
\end{multline}
where the strings $\vec{s}^x,\vec{s}^z,\vec{l}^x$ and $\vec{n}^x,\vec{n}^z,\vec{l}^z$ are defined by \eqref{eq:cssdecompoa}.
\end{lem}

\subsection{Correctable Errors}

\begin{cor}
The distance $d$ of a CSS quantum code constructed from classical codes $\mathcal{C}_2\subseteq \mathcal{C}_1$ is given by
\begin{equation}
d = \min \{ \wt(\vec{c}) \ \vert\ \vec{c}\in (\mathcal{C}_1\setminus\mathcal{C}_2)\cup(\mathcal{C}_2^\perp\setminus\mathcal{C}_1^\perp) \}.
\end{equation}
\end{cor}
\begin{proof}
According to corollary \ref{cor:diststab}, the distance of a stabilizer code is the weight of the lightest element in $L^\perp\setminus L$.
As can be seen in figure \ref{fig:stab-to-css:cssb}, the weight of the lightest non-zero element in $L^\perp$ is the minimum distance of $\mathcal{C}_1$ and $\mathcal{C}_2^\perp$ since
$\mathcal{C}_1=\vspan\{\vec{\xi}^x_i,\vec{\mu}^x_j\}$,
$\mathcal{C}_2^\perp=\vspan\{\vec{\xi}^z_i,\vec{\mu}^z_j\}$ and
$L^\perp = \vspan\{ (\vec{a},\vec{0})  ,   (\vec{0},\vec{b}) \sthat
                     \vec{a}\in\mathcal{C}_1, \vec{b}\in \mathcal{C}_2^\perp \}$.
It remains to subtract
$L = \vspan\{ (\vec{a},\vec{0}) , (\vec{0},\vec{b}) \sthat
               \vec{a}\in\mathcal{C}_2, \vec{b}\in\mathcal{C}_1^\perp  \}$.
\end{proof}

\begin{thm}
There exist CSS codes of distance $d$ encoding $k$ qudits into $n$ such that (for large enough $n$)
\begin{equation}\label{eq:cssgvrate}
 \frac{k}{n} \geq 1 - 2H_{q[\log_q]}\Bigl( 1-\frac{d}{n}, \frac{d/n}{q-1},\dots,\frac{d/n}{q-1} \Bigr).
\end{equation}
\end{thm}
\begin{proof}
In chapter \ref{sec:app:goodselfortho}, a Gilbert-Varshamov lower bound for self-orthogonal codes is established.
It guarantees the existence of $[n,n-\textsf{k},d]_q$ codes $\mathcal{C}^\perp$ of rate
\begin{equation}
 \frac{n-\textsf{k}}{n}  \geq 1 - H_{q[\log_q]}\Bigl( 1-\frac{d}{n}, \frac{d/n}{q-1},\dots, \frac{d/n}{q-1} \Bigr)
\end{equation}
such that $\mathcal{C}\subseteq\mathcal{C}^\perp$.
A CSS-code constructed from such a code encodes $k=k_1-k_2=(n-\textsf{k})-\textsf{k}$ qudits into $n$.
Hence its rate is given by \eqref{eq:cssgvrate}.
\end{proof}

\noindent
For CSS codes a transversal $J_0$ for the cosets of $L^\perp$ in $\mathbb{F}_q^{2n}$ can be specified by fixing a transversal
$\Gamma_1$ of $\mathbb{F}_q^n / \mathcal{C}_1$ and a transversal
$\Gamma_2$ of $\mathbb{F}_q^n / \mathcal{C}_2^\perp$.
Then,
\begin{equation}
 J_0 = \{ \XZ(\vec{a}^x,\vec{a}^z) \sthat \vec{a}^x\in\Gamma_1 , \: \vec{a}^z\in\Gamma_2 \},
\end{equation}
and the correctable error set $J = J_0 + L$ is given by
\begin{equation}
 J = \{ \XZ(\vec{a}^x,\vec{a}^z) \sthat \vec{a}^x\in\Gamma_1+\mathcal{C}_2 , \: \vec{a}^z\in\Gamma_2+\mathcal{C}_1^\perp \}.
\end{equation}

\section{Concatenated Codes}\label{sec:concat}

If a quantum register corresponding to a certain set of qudits is encoded using a stabilizer code, the resulting qudits may be encoded once more using some other stabilizer code.
Equivalently, such a twofold encoding process may be considered as a single one, encoding the initial register only once using a so-called concatenated stabilizer code.
We will call the code which is used first the outer code
and the code used for the second encoding the inner code%
\footnote{Some authors label the codes the other way round making the first code the inner code and the second code the outer code.}.
This section examines how such a concatenated code is obtained from its two subcodes.

\subsection{The Outer Code}
The stabilizer code used in a twofold encoding process to encode the qudits before the second encoding is applied is called the outer code.
Let the outer code encode $K$ qudits into $N$ and let its stabilizer $L^\text{out}$ be spanned by
$\{ \vec{G}_1,\dots,\vec{G}_{N-K} \}$.
As discussed in section \ref{sec:stabenc}, any extension
\begin{equation*}
\{\vec{G}_{N-K+1},\dots,\vec{G}_N, \vec{H}_1,\dots,\vec{H}_N\}
\end{equation*}
of the generating elements of $L^\text{out}$ to a hyperbolic basis of $\mathbb{F}_q^{2N}$
together with a set of phases
$\{ \Theta_\alpha(i) \in \{\omega^r\vert r\in\mathbb{F}_q\} \}_{\alpha\in\{x,z\}, i\in\{1\dots N\}}$
defines a unitary encoding operation $U^\text{out}$ as follows:
\begin{equation}\label{eq:stabenc_u_out}
 U^\text{out} \ket{ \beta_1 \dots \beta_N } = \overline{\ket{ \beta_1 \dots \beta_N }}_{\text{out}}
 = \overline{X}^{\vec{\beta}}_{\text{out}} \overline{\ket{ 0 \dots 0 }}_{\text{out}},
\end{equation}
where $\overline{\ket{0\dots0}}_{\text{out}}$ is defined as the common eigenvector of the operators
$\{ \overline{Z}_{\text{out},i} \}_{i\in 1\dots N}$ with all the eigenvalues equal to $\omega^0$, and the $\overline{X}_{\text{out},i}$ and $\overline{Z}_{\text{out},i}$ are defined as
\begin{subequations}\label{eq:stabenc_xz_out}
\begin{align}
\overline{X}^{\vec{\beta}}_\text{out} &= \prod_i \overline{X}_{\text{out},i}^{\beta_i}  &
\overline{X}_{\text{out},i} &= \Theta_x(i) \XZ(\vec{H}_i) \\
\overline{Z}^{\vec{\beta}}_\text{out} &= \prod_i \overline{Z}_{\text{out},i}^{\beta_i}  &
\overline{Z}_{\text{out},i} &= \Theta_z(i) \XZ(\vec{G}_i).
\end{align}
\end{subequations}
The encoding operator $U^\text{out}$ as defined above has the property of mapping the Pauli operators $X^{\vec{u}}$ and $Z^{\vec{v}}$ onto their encoded versions $\overline{X}^{\vec{u}}_\text{out}$ and $\overline{Z}^{\vec{v}}_\text{out}$ (see \eqref{eq:stabenc_uxu}):
\begin{align}\label{eq:stabenc_uxu_out}
 U^\text{out} X^{\vec{u}} U^{\text{out}\dagger} &= \overline{X}^{\vec{u}}_\text{out} &
 U^\text{out} Z^{\vec{v}} U^{\text{out}\dagger} &= \overline{Z}^{\vec{v}}_\text{out}.
\end{align}

\subsection{The Inner Code}
Imagine we would like to encode the $N$ qudits resulting from the application of $U^\text{out}$
once more, this time using a stabilizer code encoding $k$ qudits into $n$.
We will call the stabilizer code used for such a second level encoding the inner code.
Then the $N$ qudits have to be partitioned into groups of size $k$ (we assume that $N$ is divisible by $k$),
and the encoding operation $U^\text{in}$ of the inner code has to be applied to all of these groups.
Let the stabilizer of the inner code be $L^\text{in} = \vspan \{\vec{g}_1,\dots,\vec{g}_{n-k} \}$.
As it is the case for the outer code,
any extension of these vectors to a hyperbolic basis
$\{\vec{g}_1,\dots,\vec{g}_n,\vec{h}_1,\dots,\vec{h}_n\}$ of $\mathbb{F}_q^{2n}$
together with a set of phases
$\{ \theta_\alpha(i) \in \{\omega^r\vert r\in\mathbb{F}_q\} \}_{\alpha\in\{x,z\}, i\in\{1\dots n\}}$
specifies an encoding operator $U^\text{in}$.
The set of expressions \eqref{eq:stabenc_u_out}, \eqref{eq:stabenc_xz_out} and \eqref{eq:stabenc_uxu_out} applies if the token 'out' is replaced by 'in'.

\subsection{The Concatenated Code}
As a result of such a two step encoding procedure, $K$ qudits have been encoded into $\mathfrak{n} = N/k\times n$.
We are interested in the unitary encoder $U^\text{con}$ of the concatenated code.
For given encoding operations $U^\text{out}$ and $U^\text{in}$ derived from corresponding hyperbolic bases as described above,
can we construct a corresponding hyperbolic basis, let's say $\{\vec{\mathfrak{g}}_1,\dots,\vec{\mathfrak{g}}_{\mathfrak{n}} , \vec{\mathfrak{h}}_1,\dots,\vec{\mathfrak{h}}_{\mathfrak{n}}\}$, of $\mathbb{F}_q^{2\mathfrak{n}}$ that specifies $U^\text{con}$~?
Let us denote the initial state of the $N$ qudits which are going to be encoded first by $U^\text{out}$ by
$\ket{ \beta_1^1\dots\beta^1_k \,,\, \dots \,,\, \beta_1^{N/k}\dots\beta^{N/k}_k }$ and let us label the first group of $k$ qudits by $B_1$, the second group by $B_2$, et cetera.
Before the inner encoding is applied, additional $n-k$ qudits have to be added to each of the groups $B_i$.
Let us label the $n-k$ qudits added to $B_i$ by $A_i$ and let them be in the state $\ket{ \alpha^i_1\dots \alpha^i_{n-k} }$.
Then, the inner encoding operator $U^\text{in}$ is applied to each of the sets $A_i \cup B_i$
and the total encoding procedure can be viewed as applying the single operator
\begin{equation}
U^\text{con}_{AB} =
[ U^\text{in}_{A_1B_1} \otimes \dots \otimes U^\text{in}_{A_{N/k}B_{N/k}}] \cdot U^\text{out}_B,
\qquad (A=\cup_i A_i, \, B=\cup_j B_j),
\end{equation}
describing the encoding of the concatenated code, to the state
\begin{equation}
\bigl\vert \,
\underbrace{\alpha^1_1\dots \alpha^1_{n-k}}_{A_1}, \underbrace{\beta_1^1\dots \beta^1_k}_{B_1} \,;\,
\underbrace{\alpha^2_1\dots \alpha^2_{n-k}}_{A_2}, \underbrace{\beta_1^2\dots \beta^2_k}_{B_2} \,;\,
\dots \,;\,
\underbrace{\alpha^{N/k}_1\dots \alpha^{N/k}_{n-k}}_{A_{N/k}}, \underbrace{\beta^{N/k}_1\dots \beta^{N/k}_k}_{B_{N/k}} \,\bigr\rangle.
\end{equation}
\begin{figure}
\centering
\begin{pspicture}(-0.45,0)(9.675,7.65)
\scalebox{0.9}{
 \psframe[fillstyle=solid,fillcolor=white](3,0.3)(4,8.2)
 \rput(3.5,4.25){$U^{\text{out}}_B$}
 \psline(2.0,0.75)(3,0.75)\rput(1.5,0.75){$\ket{\Psi_{\!K}}_{\!K}$}
\rput(1.5,1.85){$\vdots$}
 \psline(2.0,2.75)(3,2.75)\rput(1.5,2.75){$\ket{\Psi_1}_1$}
 \psline(2.0,3.75)(3,3.75)\rput(1.5,3.75){$\ket{0}_{\!N-K}$}
\rput(1.5,5.35){$\vdots$}
 \psline(2.0,6.75)(3,6.75)\rput(1.5,6.75){$\ket{0}_{2}$}
 \psline(2.0,7.75)(3,7.75)\rput(1.5,7.75){$\ket{0}_{1}$}

 \rput(0.8,5.75){\gray{$\left\{ \makebox(0,2.5){} \right.$}}\rput(0.1,5.75){\gray{$N-K$}}
 \rput(0.8,1.75){\gray{$\left\{ \makebox(0,1.4){} \right.$}}\rput(0.1,1.75){\gray{$K$}}

 \rput(9.6,4.25){\gray{$\left. \makebox(0,4.1){} \right\}$}}\rput(10.25,4.25){\gray{$\frac{N}{k}\cdot n$}}

 \rput(5.5,0.9){$\vdots$}\rput(7.5,0.9){$\vdots$}\rput(8.00,0.8){$B_{N/k}$}
 \rput(5.5,1.8){$\vdots$}\rput(7.5,1.8){$\vdots$}\rput(8.00,1.7){$A_{N/k}$}
 \psline(4.0,0.5)(8,0.5)
 \psline(4.0,1.1)(8,1.1)
   \rput(5.0,1.4){$\ket{0}$}\psline(5.5,1.4)(8,1.4)
   \rput(5.0,2.0){$\ket{0}$}\psline(5.5,2.0)(8,2.0)
 \psframe[fillstyle=solid,fillcolor=white](6,0.3)(7,2.2)
 \rput{90}(6.5,1.25){$U^\text{in}_{A_{N/k}B_{N/k}}$}

 \rput(5.5,4.9){$\vdots$}\rput(7.5,4.9){$\vdots$}\rput(8.00,4.8){$B_{2}$}
 \rput(5.5,5.8){$\vdots$}\rput(7.5,5.8){$\vdots$}\rput(8.00,5.7){$A_{2}$}
 \psline(4.0,4.5)(8,4.5)
 \psline(4.0,5.1)(8,5.1)
   \rput(5.0,5.4){$\ket{0}$}\psline(5.5,5.4)(8,5.4)
   \rput(5.0,6.0){$\ket{0}$}\psline(5.5,6.0)(8,6.0)
 \psframe[fillstyle=solid,fillcolor=white](6,4.3)(7,6.2)
 \rput{90}(6.5,5.25){$U^\text{in}_{A_2B_2}$}

 \rput(5.5,6.9){$\vdots$}\rput(7.5,6.9){$\vdots$}\rput(8.00,6.8){$B_{1}$}
 \rput(5.5,7.8){$\vdots$}\rput(7.5,7.8){$\vdots$}\rput(8.00,7.7){$A_{1}$}
 \psline(4.0,6.5)(8,6.5)
 \psline(4.0,7.1)(8,7.1)
   \rput(5.0,7.4){$\ket{0}$}\psline(5.5,7.4)(8,7.4)
   \rput(5.0,8.0){$\ket{0}$}\psline(5.5,8.0)(8,8.0)
 \psframe[fillstyle=solid,fillcolor=white](6,6.3)(7,8.2)
 \rput{90}(6.5,7.25){$U^\text{in}_{A_1B_1}$}

 \rput(8.4,7.7){\gray{$\left. \makebox(0,0.45){} \right\}$}}\rput(9.0,7.7){\gray{$n-k$}}
 \rput(8.4,6.8){\gray{$\left. \makebox(0,0.45){} \right\}$}}\rput(9.0,6.8){\gray{$k$}}
}%
\end{pspicture}
\caption[Quantum circuit of the encoder of a concatenated code]{\label{fig:concated-encoder-qc}
Quantum circuit of the encoder of a concatenated quantum code.
First the outer code is applied which encodes $K$ data qudits into $N$ qudits after adding the state $\ket{0}^{\otimes N-K}$.
Then the inner code encodes $k$ qudits into $n$ after adding $N/k \times (n-k)$ additional qudits prepared as $\ket{0}$ (we assume that $N/k$ is an integer).
Altogether the concatenated code encodes $K$ logical qudits into $\mathfrak{n} = N/k\cdot n$ physical qudits.
}
\end{figure}
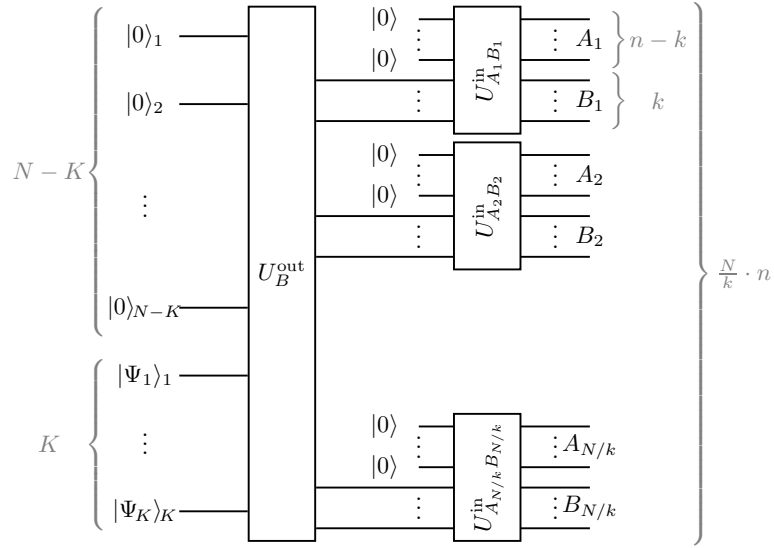%
A quantum circuit depicting the situation
(for $\ket{\beta^1_1,\dots,\beta^{N/k}_k} = \ket{0\dots 0,\Psi_1,\dots,\Psi_K}$ and $\ket{\alpha^i_j}=\ket{0}$) is presented in figure \ref{fig:concated-encoder-qc}.

We are now going to determine the elements of the hyperbolic basis
$\{\vec{\mathfrak{g}}_1,\dots,\vec{\mathfrak{g}}_{\mathfrak{n}} ,
  \vec{\mathfrak{h}}_1,\dots,\vec{\mathfrak{h}}_{\mathfrak{n}}\}$ of $\mathbb{F}_q^{2 \mathfrak{n}}$
that specifies $U^\text{con}$ by calculating the operators
$\{ \overline{X}_{\text{con},i} \sim \XZ(\vec{\mathfrak{h}}_i),
  \overline{Z}_{\text{con},i} \sim \XZ(\vec{\mathfrak{g}}_i) \}_{i\in 1\dots \mathfrak{n}}$
using \eqref{eq:stabenc_uxu_out} and the corresponding expressions for the inner and the concatenated code.
Before we proceed, let us define a map with parameter $j\in\{1,\dots,N/k\}$ mapping a string
$\vec{a}=(\vec{a}^x,\vec{a}^z) \in \mathbb{F}_q^{2n}$ to
a string $\vec{\mathfrak{a}} = (\vec{\mathfrak{a}}^x, \vec{\mathfrak{a}}^z ) \in \mathbb{F}_q^{2\mathfrak{n}}$
by
\begin{equation}
  \mathbb{F}_q^{2n} \ni \ \vec{a} \mapsto  \vec{a}^{(j)} = \vec{\mathfrak{a}} \ \in \mathbb{F}_q^{2\mathfrak{n}},
\end{equation}
where $\vec{\mathfrak{a}}^x = ( \vec{0},\dots,\vec{0},\vec{a}^x,\vec{0},\dots,\vec{0} )$ contains $\vec{a}^x$ in position $j$ and $\vec{\mathfrak{a}}^z$ is defined analogously.
Let $i\in\{1,\dots,n-k\}$, $j\in\{1,\dots,N/k\}$ and let the entries of $\vec{u}\in\mathbb{F}_q^{n-k}$ be given by $u_s=\delta_{s,i}$.
Then,
\begin{align*}
 \overline{X}_{\text{con},(j-1)(n-k)+i}  &= U^\text{con} \,X_{A_j}^{\vec{u}}\, U^{\text{con}\dagger} \\
&= [U^\text{in}_{A_1B_1} \otimes \dots ]\cdot U^\text{out}_B  \,X_{A_j}^{\vec{u}}\,
   U^{\text{out}\dagger}_B \cdot [U^\text{in}_{A_1B_1} \otimes \dots]^\dagger \\
&= [U^\text{in}_{A_1B_1} \otimes \dots ]   \,X_{A_j}^{\vec{u}}\,
   [U^\text{in}_{A_1B_1} \otimes \dots]^\dagger \\
&= U^\text{in}_{A_jB_j} \,X_{A_j}^{\vec{u}}\,  U^{\text{in}\dagger}_{A_jB_j} \otimes \id_{AB\setminus \{A_jB_j\}} \\
&= \theta_x(i) \XZ(\vec{h}_i)_{A_jB_j} \otimes \id_{AB\setminus \{A_jB_j\}},
\end{align*}
and essentially the same calculation for $\overline{Z}_{\text{con},(j-1)(n-k)+i}$ leads to the conclusion that
\begin{subequations}\label{eq:concstab:first}
\begin{align}
 \vec{\mathfrak{h}}_{ (j-1)(n-k)+i } &= \vec{h}_i^{(j)}, \\
 \vec{\mathfrak{g}}_{ (j-1)(n-k)+i } &= \vec{g}_i^{(j)}.
\end{align}
\end{subequations}
To determine the remaining $2N$ elements of the hyperbolic basis
let $i\in\{1,\dots,N\}$ and let the entries of $\vec{u}\in\mathbb{F}_q^N$ be given by $u_s=\delta_{s,i}$.
Denoting the entries of $\vec{H}_i = (\vec{H}_i^x, \vec{H}_i^z)$ as
\begin{align*}
\vec{H}_i^x &=
\bigl( (\vec{H}_i^x)^1_1\dots(\vec{H}_i^x)^1_k , \dots , (\vec{H}_i^x)^{N/k}_1\dots(\vec{H}_i^x)^{N/k}_k \bigr) \\
\vec{H}_i^z &=
\bigl( (\vec{H}_i^z)^1_1\dots(\vec{H}_i^z)^1_k , \dots , (\vec{H}_i^z)^{N/k}_1\dots(\vec{H}_i^z)^{N/k}_k \bigr),
\end{align*}
we obtain
\begin{align*}
 \overline{X}_{\text{con},\mathfrak{n}-N+i}  &= U^\text{con} \,X_B^{\vec{u}}\, U^{\text{con}\dagger} \\
&=[U^\text{in}_{A_1B_1} \otimes \dots ]\cdot U^\text{out}_B  \,X_B^{\vec{u}}\,
  U^{\text{out}\dagger}_B \cdot [U^\text{in}_{A_1B_1} \otimes \dots]^\dagger \\
&=[U^\text{in}_{A_1B_1} \otimes \dots ]
  \, \Theta_x(i) \XZ(\vec{H}_i)_B\,
  [U^\text{in}_{A_1B_1} \otimes \dots]^\dagger \\
&=\Theta_x(i) \bigotimes_{j=1}^{N/k}
  U^\text{in}_{A_jB_j} \,\XZ\bigl((\vec{H}_i^x)^j_1\dots(\vec{H}_i^x)^j_k , (\vec{H}_i^z)^j_1\dots(\vec{H}_i^z)^j_k \bigr)_{B_j}\,  U^{\text{in}\dagger}_{A_jB_j} \\
&=\Theta_x(i) \bigotimes_{j=1}^{N/k}
  \Bigl(\prod_{s=1}^k
  \bigl[ \theta_x(n-k+s)\XZ(\vec{h}_{n-k+s})\bigr]^{(\vec{H}_i^x)^j_s} \!\cdot\!
  \bigl[ \theta_z(n-k+s)\XZ(\vec{g}_{n-k+s})\bigr]^{(\vec{H}_i^z)^j_s}
  \Bigr)_{\!\!A_jB_j} \\
&\sim \bigotimes_{j=1}^{N/k}
  X\!Z\Bigl(\sum_{s=1}^k \bigl(
   (\vec{H}_i^x)^j_s \cdot \vec{h}_{n-k+s} + (\vec{H}_i^z)^j_s \cdot \vec{g}_{n-k+s} \bigr)
  \Bigr)_{\!A_jB_j} ,
\end{align*}
and again essentially the same calculation for $\overline{Z}_{\text{con},\mathfrak{n}-N+i}$ leads to the conclusion that
\begin{subequations}\label{eq:concstab:lastN}
\begin{align}
 \vec{\mathfrak{h}}_{ \mathfrak{n}-N+i } &=  \sum_{j=1}^{N/k} \sum_{s=1}^k \Bigl(
   (\vec{H}_i^x)^j_s \cdot \vec{h}_{n-k+s}^{(j)} + (\vec{H}_i^z)^j_s \cdot \vec{g}_{n-k+s}^{(j)} \Bigr),  \\
 \vec{\mathfrak{g}}_{ \mathfrak{n}-N+i } &=  \sum_{j=1}^{N/k} \sum_{s=1}^k \Bigl(
   (\vec{G}_i^x)^j_s \cdot \vec{h}_{n-k+s}^{(j)} + (\vec{G}_i^z)^j_s \cdot \vec{g}_{n-k+s}^{(j)} \Bigr).
\end{align}
\end{subequations}

\begin{figure}%
\begin{minipage}{0.25\textwidth}\centering
 \subfloat[inner code]{
 \scalebox{0.8}{
 \begin{pspicture}(-0.1,-0.1)(3.1,6.5)
   \rput[origin=c]{-90}(1.5,6){$\rotatebox[origin=c]{90}{$2n$}\left\{\makebox(0,1.6){}\right.$}
   \psframe[linecolor=gray](-0.1,-0.1)(3.1,1.6)
   \psframe[linecolor=gray](-0.1,1.9)(3.1,5.6)
   \psframe[fillstyle=solid,fillcolor=lightgray,linestyle=none](0,0)(1.5,0.5)
   \psframe[fillstyle=solid,fillcolor=white,linestyle=none](1.5,0)(3,0.5)
   \psframe(0,0)(3,0.5)   \rput(1.5,0.25){$\vec{g}_n$}
   \rput(1.5,0.8){$\vdots$}
   \psframe[fillstyle=solid,fillcolor=lightgray,linestyle=none](0,1)(1.5,1.5)
   \psframe[fillstyle=solid,fillcolor=white,linestyle=none](1.5,1)(3,1.5)
   \psframe(0,1)(3,1.5)   \rput(1.5,1.25){$\vec{g}_{n-k+1}$}
   \psframe[fillstyle=solid,fillcolor=lightgray,linestyle=none](0,2)(1.5,2.5)
   \psframe[fillstyle=solid,fillcolor=white,linestyle=none](1.5,2)(3,2.5)
   \psframe(0,2)(3,2.5)   \rput(1.5,2.25){$\vec{g}_{n-k}$}
   \rput(1.5,3.8){$\vdots$}
   \psframe[fillstyle=solid,fillcolor=lightgray,linestyle=none](0,4.5)(1.5,5.0)
   \psframe[fillstyle=solid,fillcolor=white,linestyle=none](1.5,4.5)(3,5.0)
   \psframe(0,4.5)(3,5.0)   \rput(0.75,4.75){$\vec{g}_2^x$}\rput(2.25,4.75){$\vec{g}_2^z$}
   \psframe[fillstyle=solid,fillcolor=lightgray,linestyle=none](0,5)(1.5,5.5)
   \psframe[fillstyle=solid,fillcolor=white,linestyle=none](1.5,5)(3,5.5)
   \psframe(0,5)(3,5.5)   \rput(1.5,5.25){$\vec{g}_1$}
  \end{pspicture}}}\\
  \subfloat[outer code]{
  \scalebox{0.8}{
  \begin{pspicture}(-0.1,-0.1)(3.1,6.5)
   \rput[origin=c]{-90}(1.5,6){$\rotatebox[origin=c]{90}{$2N$}\left\{\makebox(0,1.6){}\right.$}
   \psframe[linecolor=gray](-0.1,-0.1)(3.1,1.6)
   \psframe[linecolor=gray](-0.1,1.9)(3.1,5.6)
   \psframe[fillstyle=solid,fillcolor=lightgray,linestyle=none](0,0)(1.5,0.5)
   \psframe[fillstyle=solid,fillcolor=white,linestyle=none](1.5,0)(3,0.5)
   \psframe(0,0)(3,0.5)   \rput(1.5,0.25){$\vec{G}_N$}
   \rput(1.5,0.8){$\vdots$}
   \psframe[fillstyle=solid,fillcolor=lightgray,linestyle=none](0,1)(1.5,1.5)
   \psframe[fillstyle=solid,fillcolor=white,linestyle=none](1.5,1)(3,1.5)
   \psframe(0,1)(3,1.5)   \rput(1.5,1.25){$\vec{G}_{N-K+1}$}
   \psframe[fillstyle=solid,fillcolor=lightgray,linestyle=none](0,2)(1.5,2.5)
   \psframe[fillstyle=solid,fillcolor=white,linestyle=none](1.5,2)(3,2.5)
   \psframe(0,2)(3,2.5)   \rput(1.5,2.25){$\vec{G}_{N-K}$}
   \rput(1.5,3.8){$\vdots$}
   \psframe[fillstyle=solid,fillcolor=lightgray,linestyle=none](0,5)(1.5,5.5)
   \psframe[fillstyle=solid,fillcolor=white,linestyle=none](1.5,5)(3,5.5)
   \psframe(0,5)(3,5.5)   \rput(1.5,5.25){$\vec{G}_1$}
 \end{pspicture}}}
 \end{minipage}%
 \begin{minipage}{0.74\textwidth}%
  \subfloat[concatenated code]{
  \scalebox{0.8}{
 \begin{pspicture}(0.0,-0.1)(0.,14.5)%
 \rput[origin=c]{-90}(6.0,14){$\rotatebox[origin=c]{90}{$2\mathfrak{n}$}\left\{\makebox(0,6){}\right.$}
 \rput(12.25,12.25){\gray{$\left. \makebox(0,1.25){} \right\}$}}\rput(13.0,12.25){\gray{$n-k$}}
 \rput(12.25,02.75){\gray{$\left. \makebox(0,0.8){} \right\}$}}\rput(13.0,02.75){\gray{$N-K$}}
 \rput(12.25,00.75){\gray{$\left. \makebox(0,0.8){} \right\}$}}\rput(13.0,00.75){\gray{$K$}}
 \begin{pspicture}(0.00,0.)(6.,13.5)
   \psframe[fillstyle=solid,fillcolor=lightgray](0.00,02.0)(06.0,13.5)
   \multiput(00.0,11.0)(1.5,-2.5){2}{
    \psframe(0.0,0.0)(1.5,2.5)\rput(0.75,1.35){$\vdots$}
    \psframe(0.0,0.0)(1.5,0.5)\rput(0.75,0.25){$\vec{g}_{n-k}^x$}
    \psframe(0.0,2.0)(1.5,2.5)\rput(0.75,2.25){$\vec{g}_{1}^x$}
   }%
   \rput(3.75,7.35){$\ddots$}
   \put(04.5,3.5){
    \psframe(0.0,0.0)(1.5,2.5)\rput(0.75,1.35){$\vdots$}
    \psframe(0.0,0.0)(1.5,0.5)\rput(0.75,0.25){$\vec{g}_{n-k}^x$}
    \psframe(0.0,2.0)(1.5,2.5)\rput(0.75,2.25){$\vec{g}_{1}^x$}
   }
   \multiput(0.75,2.60)(1.5,0){4}{$\vdots$}
   \psframe(0.0,2.0)(1.5,2.5)
   \psframe(1.5,2.0)(3.0,2.5)
   \rput(3.75,2.25){$\dots$}
   \psframe(4.5,2.0)(6.0,2.5)
   \psframe[fillstyle=solid,fillcolor=lightgray,linecolor=lightgray](0.06,2.06)(5.94,2.44)
   \rput(3.0,2.25){\footnotesize$\vec{\mathfrak{g}}_{ \mathfrak{n}-K }^x$}
   \psframe(0.0,3.0)(1.5,3.5)
   \psframe(1.5,3.0)(3.0,3.5)
   \rput(3.75,3.25){$\dots$}
   \psframe(4.5,3.0)(6.0,3.5)
   \psframe[fillstyle=solid,fillcolor=lightgray,linecolor=lightgray](0.06,3.06)(5.94,3.44)
   \rput(3.0,3.25){\footnotesize$\vec{\mathfrak{g}}_{ \mathfrak{n}-N+1 }^x$}

   \psframe[fillstyle=solid,fillcolor=lightgray](0.00,00.0)(06.0,01.5)
   \multiput(0.75,0.60)(1.5,0){4}{$\vdots$}
   \psframe(0.0,0.0)(1.5,0.5)
   \psframe(1.5,0.0)(3.0,0.5)
   \rput(3.75,0.25){$\dots$}
   \psframe(4.5,0.0)(6.0,0.5)
   \psframe[fillstyle=solid,fillcolor=lightgray,linecolor=lightgray](0.06,0.06)(5.94,0.44)
   \rput(3.0,0.25){\footnotesize$\vec{\mathfrak{g}}_{ \mathfrak{n} }^x$}
   \psframe(0.0,1.0)(1.5,1.5)
   \psframe(1.5,1.0)(3.0,1.5)
   \rput(3.75,1.25){$\dots$}
   \psframe(4.5,1.0)(6.0,1.5)
   \psframe[fillstyle=solid,fillcolor=lightgray,linecolor=lightgray](0.06,1.06)(5.94,1.44)
   \rput(3.0,1.25){\footnotesize$\vec{\mathfrak{g}}_{ \mathfrak{n}-K+1 }^x$}
 \end{pspicture}%
 \begin{pspicture}(0.00,0.)(6.,13.5)
   \psframe[fillstyle=solid,fillcolor=white](0.00,02.0)(06.0,13.5)
   \multiput(00.0,11.0)(1.5,-2.5){2}{
    \psframe(0.0,0.0)(1.5,2.5)\rput(0.75,1.35){$\vdots$}
    \psframe(0.0,0.0)(1.5,0.5)\rput(0.75,0.25){$\vec{g}_{n-k}^z$}
    \psframe(0.0,2.0)(1.5,2.5)\rput(0.75,2.25){$\vec{g}_{1}^z$}
   }%
   \rput(3.75,7.35){$\ddots$}
   \put(04.5,3.5){
    \psframe(0.0,0.0)(1.5,2.5)\rput(0.75,1.35){$\vdots$}
    \psframe(0.0,0.0)(1.5,0.5)\rput(0.75,0.25){$\vec{g}_{n-k}^z$}
    \psframe(0.0,2.0)(1.5,2.5)\rput(0.75,2.25){$\vec{g}_{1}^z$}
   }
   \multiput(0.75,2.60)(1.5,0){4}{$\vdots$}
   \psframe(0.0,2.0)(1.5,2.5)
   \psframe(1.5,2.0)(3.0,2.5)
   \rput(3.75,2.25){$\dots$}
   \psframe(4.5,2.0)(6.0,2.5)
   \psframe[fillstyle=solid,fillcolor=white,linecolor=white](0.06,2.06)(5.94,2.44)
   \rput(3.0,2.25){\footnotesize$\vec{\mathfrak{g}}_{ \mathfrak{n}-K }^z$}
   \psframe(0.0,3.0)(1.5,3.5)
   \psframe(1.5,3.0)(3.0,3.5)
   \rput(3.75,3.25){$\dots$}
   \psframe(4.5,3.0)(6.0,3.5)
   \psframe[fillstyle=solid,fillcolor=white,linecolor=white](0.06,3.06)(5.94,3.44)
   \rput(3.0,3.25){\footnotesize$\vec{\mathfrak{g}}_{ \mathfrak{n}-N+1 }^z$}

   \psframe[fillstyle=solid,fillcolor=white](0.00,00.0)(06.0,01.5)
   \multiput(0.75,0.60)(1.5,0){4}{$\vdots$}
   \psframe(0.0,0.0)(1.5,0.5)
   \psframe(1.5,0.0)(3.0,0.5)
   \rput(3.75,0.25){$\dots$}
   \psframe(4.5,0.0)(6.0,0.5)
   \psframe[fillstyle=solid,fillcolor=white,linecolor=white](0.06,0.06)(5.94,0.44)
   \rput(3.0,0.25){\footnotesize$\vec{\mathfrak{g}}_{ \mathfrak{n} }^z$}
   \psframe(0.0,1.0)(1.5,1.5)
   \psframe(1.5,1.0)(3.0,1.5)
   \rput(3.75,1.25){$\dots$}
   \psframe(4.5,1.0)(6.0,1.5)
   \psframe[fillstyle=solid,fillcolor=white,linecolor=white](0.06,1.06)(5.94,1.44)
   \rput(3.0,1.25){\footnotesize$\vec{\mathfrak{g}}_{ \mathfrak{n}-K+1 }^z$}
 \end{pspicture}%
 \end{pspicture}}}
 \end{minipage}
\caption[Concatenated code]{\label{fig:concatcompo}
An inner $[[n,k]]_q$ code is concatenated with an outer $[[N,K]]_q$ code resulting in an
$[[\mathfrak{n},K]]_q$ code with $\mathfrak{n}=N/k\times n$ (we assume $N$ is divisible by $k$).
If the inner code plus an encoding $U^\text{in}$ is specified by the hyperbolic basis
$\{\vec{g}_1,\dots,\vec{g}_{n}, \vec{h}_1,\dots,\vec{h}_{n}\}$ \textit{(upper left part)}
and the outer code plus an encoding $U^\text{out}$ is specified by
$\{\vec{G}_1,\dots,\vec{G}_{N}, \vec{H}_1,\dots,\vec{h}_{N}\}$ \textit{(lower left part)},
the resulting concatenated code plus an encoding $U^\text{con}$ is specified by
$\{\vec{\mathfrak{g}}_1,\dots,\vec{\mathfrak{g}}_{\mathfrak{n}}, \vec{\mathfrak{h}}_1,\dots,\vec{\mathfrak{h}}_{\mathfrak{n}}\}$ \textit{(right part)} which is related to the bases of the inner and outer code via equations
\eqref{eq:concstab:first}
(for $(\vec{\mathfrak{g}}_i,\vec{\mathfrak{h}}_i)$, $i\in\{1,\dots,\mathfrak{n}-N\}$) and
\eqref{eq:concstab:lastN}
(for $(\vec{\mathfrak{g}}_i,\vec{\mathfrak{h}}_i)$, $i\in\{\mathfrak{n}-N+1,\dots,\mathfrak{n}\}$).
In the figure, only the first half of the bases (i.\,e. $\{\vec{g}_1,\dots,\vec{g}_{n}\}$, etc.) is shown.}
\end{figure}
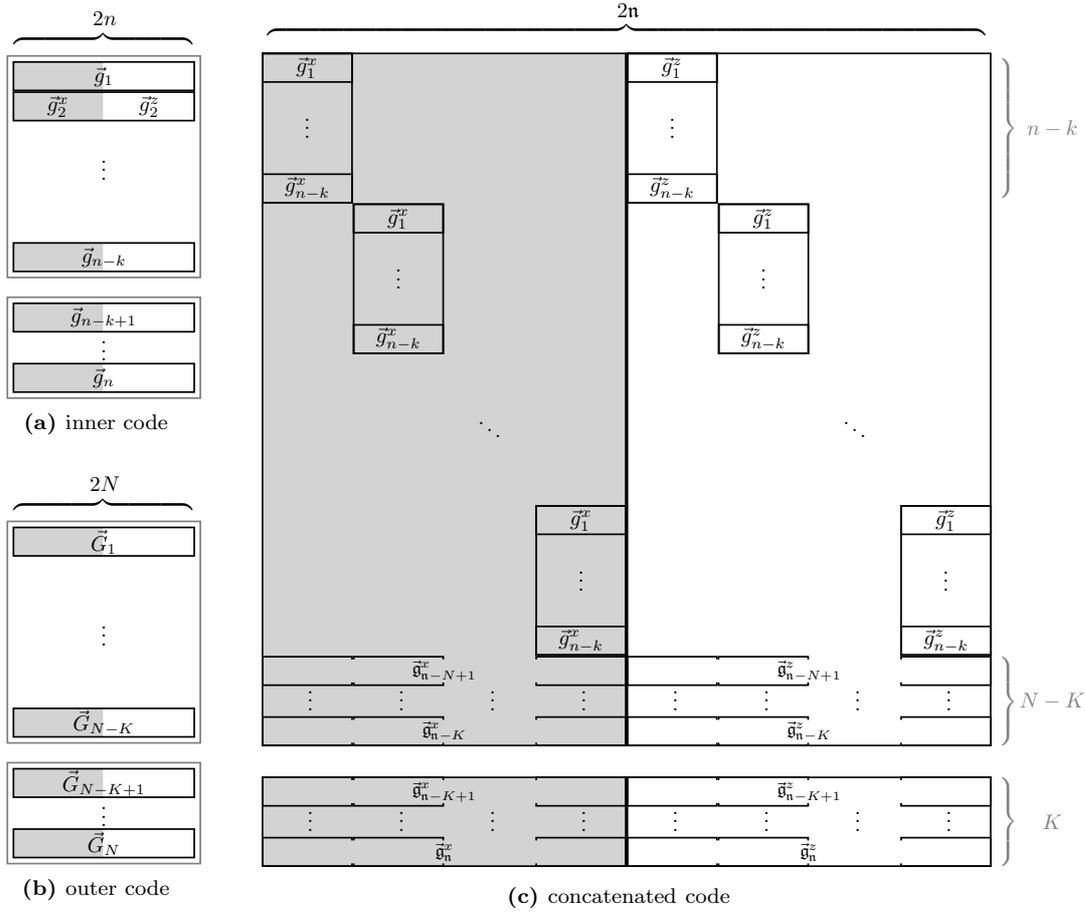

\chapter{Quantum Channel Capacity}\label{chap:qcapa}

Shannon's noisy coding theorem is one of the fundamental theorems of classical information theory.
As discussed in chapter \ref{chap:cecc},
it assigns to each channel a non-negative number $C$, called channel capacity, such that for any rate below the capacity, there exists an error correcting scheme achieving reliable transmission over the channel.
The channel capacity is given by the maximum mutual information between source and receiver and
the theorem is proven by showing that typical set decoding using random linear codes leads to an arbitrary small decoding error probability.
This chapter deals with the quantum analog of Shannon's noisy coding theorem.

It was not until Shor presented a nine qubit quantum error-correcting code in his seminal paper \cite{shor95qec},
that it was known whether there exist error correction methods for quantum information at all.
In the same paper, Shor stated that the ultimate goal would be to find a quantum analog of Shannon's noisy coding theorem,
i.\,e. to define a quantum analog of the Shannon capacity for a quantum channel,
and to find encoding schemes which approach this capacity.
About a year later the demanded quantum noisy coding theorem was proposed by Lloyd \cite{Ll96}.
As it was conjectured by Schumacher and Nielsen \cite{SchN96},
the role analogous to that played by the mutual information in the classical theory is taken by the regularized coherent information,
which corresponds to the limit of the coherent information as the number of channel uses goes to infinity.
A rigorous proof that the quantum capacity is upper bounded by the regularized coherent information was given by Barnum, Nielsen and coworkers in \cite{BNS98,BKN00},
while the converse part (the capacity is lower bounded by the regularized coherent information) was shown by
Shor himself \cite{Shor02} (unpublished) and Devetak \cite{De0304}. %

While the Shannon capacity of a classical channel is given by a formula involving a single use of the channel, the quantum capacity involves the limit as the number of channel uses goes to infinity and
cannot be expressed by a single letter formula.
Therefore, the computation of the quantum capacity for a given quantum channel remains to be a hard problem and is not feasible in general.
To obtain at least a lower bound on the quantum capacity,
one may calculate the achievable rate of the so-called one-way hashing entanglement distillation protocol by Bennett et\,al. \cite{BDSW}, which corresponds to a quantum error correcting scheme making use of random stabilizer codes
(see \cite[section 7.6]{PhdGottesman} and \cite[section 7.16.2]{Preskill} for the binary case,
and \cite{Ha01} for the general one).
The fact that the 'hashing'-rate is indeed only a lower bound on the quantum capacity was shown by Shor and Smolin in \cite{ShSm96}
(and later together with DiVincenzo in \cite{DiSS98}).
By concatenating an outer random stabilizer code with a deterministic inner one, they found that rates above the hashing rate could be achieved for very noisy depolarizing qubit channels.
This result came somewhat as a surprise since it stands in contrast to the classical case where random codes do achieve the capacity of a channel.

In section \ref{sec:qcapa} we define the quantum capacity of a noisy quantum channel and present the quantum noisy coding theorem, i.\,e. the representation of the capacity in terms of the regularized coherent information.
For the remaining part of the chapter, we restrict our attention to a certain subclass of quantum channels, so-called memoryless Pauli channels.
As it is discussed in section \ref{sec:paulich}, this kind of channels are especially easy to analyze and allow us to obtain lower bounds on the capacity of general channels.
We present the quantum coding scheme based on random stabilizer codes and corresponding to the one-way hashing protocol in section \ref{sec:randqcodes}. %
In addition we give a rigorous proof that the hashing-rate can be obtained by using only CSS codes,
a result which has been used by Lo in \cite{Lo01} to prove the security of the 6-state quantum key distribution protocol, but for which no elaborated proof can be found in the literature.
Concatenation of random codes with deterministic ones \cite{ShSm96,DiSS98} allows for rates surpassing the hashing-rate under certain circumstances.
We determine the achievable rate of such concatenated coding schemes in section \ref{sec:concrandom}.
Eventually we apply the results of the preceeding sections to calculate new lower bounds on the capacity of the qubit depolarizing channel in section \ref{sec:concexamples}.
After giving a detailed description of the deterministic inner code used by \cite{DiSS98,SmSm07}, we evaluate the achievable rate for this code for larger code sizes than it was done before in \cite{SmSm07}.

\section{Quantum Noisy Coding Theorem}\label{sec:qcapa}

A quantum channel is a trace preserving complete positive map (tpcp-map)
$\mathcal{M} : \mathcal{S}(\mathcal{H}) \rightarrow \mathcal{S}(\mathcal{H})$ map between density matrices on a Hilbert space $\mathcal{H}$.
In this thesis we are primary concerned with discrete Hilbert spaces of dimension $q$.
To send quantum information reliably over a noisy quantum channel, one might protect it by encoding it into a quantum error-correcting code $\mathcal{C} \subset \mathcal{H}^{\otimes n}$, which encodes say $k$ qudits into $n$.
The rate at which we send quantum information in this case would be given by the ratio $k/n$.
To quantify how good the protection works, we may use the minimum pure-state fidelity which is defined for a quantum channel $\mathcal{M}$ and a quantum code $\mathcal{C}$ with corresponding recovery operation $\mathcal{R}$ as
\begin{equation}
 F_p( \mathcal{C}, \mathcal{R}\mathcal{M}^{\otimes n} ) = \min_{ \ket{\psi} \in \mathcal{C} }
 \bra{\psi} \mathcal{R}\bigl(\mathcal{M}^{\otimes n} ( \ketbra{\psi}{\psi} )\bigr) \ket{\psi}.
\end{equation}
The capacity of a quantum channel for transmitting quantum information was defined by Bennett et\,al.  \cite{BDSW,BDS97,DiSS98} with the help of the minimum pure-state fidelity as follows:
\begin{defi}
The quantum capacity $Q(\mathcal{M})$ of a quantum channel $\mathcal{M} : \mathcal{S}(\mathcal{H}) \rightarrow \mathcal{S}(\mathcal{H})$ is defined as the maximum number $Q$, such that for any rate $R<Q$ and any $\varepsilon>0$, there exists a quantum code $\mathcal{C}$ with rate $k/n \geq R$, together with a recovery operation $\mathcal{R}$, such that
\begin{equation}
 F_p( \mathcal{C}, \mathcal{R}\mathcal{M}^{\otimes n} )
> 1 - \varepsilon.
\end{equation}
\end{defi}
\begin{rem}
There exist quite a lot of different definitions for the quantum capacity.
For example, the minimum pure-state fidelity might be replaced by the entanglement fidelity \cite{BKN00}.
As it turns out, all these definitions are equivalent.
For an overview see \cite{KretW04}: 'Tema con variazioni: quantum channel capacity'.
\end{rem}

The question raised by Shor in his seminal paper on quantum error correction \cite{shor95qec} was
whether there exits a quantum analog of Shannon's noisy coding theorem
relating the quantum capacity of a quantum channel to a quantity corresponding to the mutual information in the classical theory.
Such a quantum noisy coding theorem was proposed by Lloyd \cite{Ll96}.
The quantity taking the role the mutual information played in the classical case is taken by the coherent information, which is defined for a quantum channel $\mathcal{M} : \mathcal{S}(\mathcal{H}) \rightarrow \mathcal{S}(\mathcal{H})$ and a quantum state $\rho \in \mathcal{S}(\mathcal{H})$ as
\begin{equation}\label{eq:defcinfo}
 I_c (\rho, \mathcal{M} ) = S( \mathcal{M}(\rho) ) - S(\mathcal{M}\otimes\id (\ketbra{\psi}{\psi}) ),
\end{equation}
where $\ket{\psi}\in \mathcal{H}\otimes\mathcal{H}$ is a purification of $\rho$.
\begin{thm}[Quantum noisy coding theorem]
The quantum capacity $Q(\mathcal{M})$ of a quantum channel $\mathcal{M} : \mathcal{S}(\mathcal{H}) \rightarrow \mathcal{S}(\mathcal{H})$ is given by the regularized coherent information,
\begin{equation}\label{eq:regcohinfo}
 Q(\mathcal{M}) =  \lim_{n\to\infty} \frac{1}{n} \max_{\rho} I_c(\rho, \mathcal{M}^{\otimes n} ),
\end{equation}
which is obtained by taking the limit as $n$ goes to infinity of $I_c (\rho, \mathcal{M}^{\otimes n} )/n$ maximized over all density operators on $\mathcal{H}^{\otimes n}$.
\end{thm}
\noindent
It was proved rigorously by Barnum, Nielsen and coworkers in \cite{BNS98,BKN00},
that the regularized coherent information
is an upper bound on the capacity $Q(\mathcal{M})$,
while the other direction of the theorem
($Q(\mathcal{M})$ is lower bounded by the regularized coherent information)
was shown by Shor himself \cite[(unpublished)]{Shor02} and Devetak \cite{De0304}.

\section{Pauli Channels}\label{sec:paulich}

In this section we consider a special class of tpcp-maps called Pauli channels.
Pauli channels have the nice property of being easy to analyze.
In addition, any more general channel may be converted into a Pauli channel by a process called discrete twirling.
This allows lower bounds on the capacity of Pauli channels to be applicable to more general channels as well.

In the first subsection we give the definition of a Pauli channel.
The subsequent subsection explains how a general channel may be twirled to become a Pauli channel.

\subsection{Definitions}

\begin{defi}\label{def:paulichannel}
A Pauli channel
$\mathcal{A} : \mathcal{S}(\mathcal{H}) \rightarrow \mathcal{S}(\mathcal{H})$
is a tpcp-map between density operators on a $q$-dimensional Hilbert space $\mathcal{H}$ given by
\begin{equation}
 \mathcal{A} :   \rho \mapsto \mathcal{A}(\rho) = \sum_{\vec{e}\in\mathbb{F}^2_q} P_{\!\mathcal{A}}(\vec{e}) \, \XZ(\vec{e}) \rho \XZ(\vec{e})^\dagger 
\end{equation}
for some probability distribution $P_{\!\mathcal{A}}$ on $\mathbb{F}_q^2$.
\end{defi}
\noindent
If we speak of a memoryless quantum channel, we mean a channel acting identically and independently on multiple qudits.
For example, a memoryless Pauli channel
$\mathcal{A}^{\otimes n} : \mathcal{S}(\mathcal{H}^{\otimes n}) \rightarrow \mathcal{S}(\mathcal{H}^{\otimes n})$ between density operators on $\mathcal{H}^{\otimes n}$ is given by
\begin{equation}\label{eq:memlesspauli}
 \mathcal{A}^{\otimes n} :  \rho \mapsto
 \mathcal{A}^{\otimes n}(\rho) = \sum_{\vec{e}\in\mathbb{F}^{2n}_q}
 P^n_{\!\mathcal{A}}(\vec{e}) \, \XZ(\vec{e})  \rho  \XZ(\vec{e})^\dagger ,
\end{equation}
where
$P^n_{\!\mathcal{A}}(\vec{e}=(e^x_1,\dots,e^x_n,e^z_1,\dots,e^z_n)) =
 \prod_{i=1}^n P_{\!\mathcal{A}}(e^x_i,e^z_i)$.
In contrast to \eqref{eq:memlesspauli}, a general Pauli channel
$\mathcal{G} : \mathcal{S}(\mathcal{H}^{\otimes n}) \rightarrow \mathcal{S}(\mathcal{H}^{\otimes n})$ is defined by a probability distribution $P_{\!\mathcal{G}}$ on $\mathbb{F}_q^{2n}$ which is not necessarily a product distribution.

\subsection{Discrete Twirling}

We follow \cite[section 2.3--2.5]{Ha03Fi}.
First we note that there is a one-to-one map between a complete positive map
$\mathcal{M} : \mathcal{S}(\mathcal{H}^{\otimes n}) \rightarrow \mathcal{S}(\mathcal{H}^{\otimes n})$
and a non-negative operator $\rho_\mathcal{M}$ in $\mathcal{S}(\mathcal{H}^{\otimes n}\otimes \mathcal{H}^{\otimes n})$ defined by
\begin{equation}
 \rho_\mathcal{M} = [\id\otimes\mathcal{M}]( \ketbra{\Phi_{\vec{0}}}{\Phi_{\vec{0}}} ),
\end{equation}
where $\ket{\Phi_{\vec{0}}}$ denotes a Bell state (compare with definition \ref{defi:bellstates}).
Let an operator sum representation of $\mathcal{M}$ be given by $\mathcal{M}: \rho\mapsto \sum_\mu M_\mu\rho M^\dagger_\mu$.
Then, by comparing the expressions
\begin{align}
\rho_\mathcal{M} &= \sum_{\vec{y},\vec{z}\in\mathbb{F}_q^{2n}}
 \ket{\Phi_{\vec{y}}}
   \underbrace{\bra{\Phi_{\vec{y}}} \rho_\mathcal{M} \ket{\Phi_{\vec{z}}} }_{m_{\vec{y},\vec{z}}}
 \bra{\Phi_{\vec{z}}} \nonumber\\
 &=\frac{1}{q^n} \sum_{\vec{i},\vec{j}\in\mathbb{F}_q^n} \ketbral{\vec{i}}{\vec{j}}{A} \otimes
  \sum_{\vec{y},\vec{z}\in\mathbb{F}_q^{2n}} m_{\vec{y},\vec{z}}
  \opl{\XZ(\vec{y})}{B}
  \ketbral{\vec{i}}{\vec{j}}{B}
  \opdl{\XZ(\vec{z})}{B}
\intertext{and}
\rho_\mathcal{M} &= \frac{1}{q^n} \sum_{\vec{i},\vec{j}\in\mathbb{F}_q^n} \ketbral{\vec{i}}{\vec{j}}{A} \otimes \sum_\mu M_\mu \ketbral{\vec{i}}{\vec{j}}{B} M^\dagger_\mu,
\end{align}
it follows that any
$\mathcal{M} : \mathcal{S}(\mathcal{H}^{\otimes n}) \rightarrow \mathcal{S}(\mathcal{H}^{\otimes n})$ may be expressed as
\begin{equation}
 \mathcal{M}: \rho\mapsto \sum_{\vec{y},\vec{z}\in\mathbb{F}_q^{2n}}
 m_{\vec{y},\vec{z}} \,  \XZ(\vec{y}) \rho \XZ(\vec{z})^\dagger, \quad
\text{with } m_{\vec{y},\vec{z}} =
 \bra{\Phi_{\vec{y}}}
     [\id\otimes\mathcal{M}]( \ketbra{\Phi_{\vec{0}}}{\Phi_{\vec{0}}} )
 \ket{\Phi_{\vec{z}}}.
\end{equation}
Discrete twirling (\cite{Ha03Fi}, \cite{BBPSSW96e,BDSW} for the binary case) converts the state
\begin{equation}
\mathcal{S}(\mathcal{H}^{\otimes n}\otimes \mathcal{H}^{\otimes n}) \ni
\rho_\mathcal{M} = \sum_{\vec{y},\vec{z}\in\mathbb{F}_q^{2n}}
 m_{\vec{y},\vec{z}} \, \ketbra{\Phi_{\vec{y}}}{\Phi_{\vec{z}}} 
\end{equation}
into a Bell diagonal one by applying one of the bilateral rotations
$\{ \XZ(\vec{x})^\ast \otimes \XZ(\vec{x}) \sthat \vec{x}\in\mathbb{F}_q^{2n}\}$ (${}^\ast$~denoting complex conjugation) at random,
\begin{align}
\tilde{\rho}_\mathcal{M} &=
\frac{1}{q^{2n}} \sum_{\vec{x}\in\mathbb{F}_q^{2n}}
\bigl( \XZ(\vec{x})^\ast \otimes \XZ(\vec{x}) \bigr)
\rho_\mathcal{M}
\bigl( \XZ(\vec{x})^\ast \otimes \XZ(\vec{x}) \bigr)^\dagger \nonumber\\
&=
 \frac{1}{q^{2n}} \sum_{ \vec{x},\vec{y},\vec{z} \in\mathbb{F}_q^{2n} }
  m_{\vec{y},\vec{z}}
 \bigl(\id\otimes \XZ(\vec{x})\XZ(\vec{y})\XZ(\vec{x})^\dagger\bigr)
 \ketbra{\Phi_{\vec{0}}}{\Phi_{\vec{0}}}
 \bigl(\id\otimes \XZ(\vec{x})\XZ(\vec{z})\XZ(\vec{x})^\dagger\bigr)^\dagger \label{eq:twzwischen}\\
&=
 \frac{1}{q^{2n}} \sum_{\vec{y},\vec{z} \in\mathbb{F}_q^{2n} }
  m_{\vec{y},\vec{z}}
 \sum_{\vec{x}\in\mathbb{F}_q^{2n} }
  \omega^{(\vec{x},\vec{y})_{sp}-(\vec{x},\vec{z})_{sp}}
 \bigl(\id\otimes \XZ(\vec{y})\bigr)
 \ketbra{\Phi_{\vec{0}}}{\Phi_{\vec{0}}}
 \bigl(\id\otimes \XZ(\vec{z})\bigr)^\dagger \nonumber\\
&=
 \sum_{\vec{y}\in\mathbb{F}_q^{2n}}
 m_{\vec{y},\vec{y}} \, \ketbra{\Phi_{\vec{y}}}{\Phi_{\vec{y}}} .
\end{align}
To arrive at \eqref{eq:twzwischen} we made use of lemma \ref{lem:OTo1gleich1oO}.
We obtain from \eqref{eq:twzwischen} that
\begin{equation}
 \tilde{\rho}_\mathcal{M} = \bigl[ \id\otimes \frac{1}{q^{2n}}\sum_{\vec{x}\in\mathbb{F}_q^{2n}}\mathcal{N}_{\vec{x}}\mathcal{M}\mathcal{N}^\dagger_{\vec{x}} \bigr] ( \ketbra{\Phi_{\vec{0}}}{\Phi_{\vec{0}}} ),
\end{equation}
with $\mathcal{N}_{\vec{x}} : \rho \mapsto \XZ(\vec{x}) \rho \XZ(\vec{x})^\dagger$, which leads to the central theorem of this subsection.

\begin{thm}\label{thm:twirledcpmap}
Any completely positive map
$\mathcal{M} : \mathcal{S}(\mathcal{H}^{\otimes n}) \rightarrow \mathcal{S}(\mathcal{H}^{\otimes n})$
can be converted into a general Pauli channel
$\tilde{\mathcal{M}} : \mathcal{S}(\mathcal{H}^{\otimes n}) \rightarrow \mathcal{S}(\mathcal{H}^{\otimes n})$ such that
\begin{equation}
 \tilde{\mathcal{M}} : \rho \mapsto
 \frac{1}{q^{2n}} \sum_{\vec{x}\in\mathbb{F}_q^{2n}}
 \mathcal{N}_{\vec{x}} \mathcal{M} \mathcal{N}^\dagger_{\vec{x}} (\rho)
 =
 \sum_{\vec{e}\in\mathbb{F}^{2n}_q}
 P_\mathcal{M}(\vec{e}) \, \XZ(\vec{e})  \rho  \XZ(\vec{e})^\dagger,
\end{equation}
with $P_\mathcal{M}(\vec{e}) = m_{\vec{e},\vec{e}} =
 \bra{\Phi_{\vec{e}}}
     [\id\otimes\mathcal{M}]( \ketbra{\Phi_{\vec{0}}}{\Phi_{\vec{0}}} )
 \ket{\Phi_{\vec{e}}}$.
\end{thm}

To obtain a lower bound on the quantum capacity of a general memoryless channel
$\mathcal{M}^{\otimes n} : \mathcal{S}(\mathcal{H}^{\otimes n}) \rightarrow \mathcal{S}(\mathcal{H}^{\otimes n})$,
we apply twirling to convert the channel into the memoryless Pauli channel
$\tilde{\mathcal{M}}^{\otimes n} : \mathcal{S}(\mathcal{H}^{\otimes n}) \rightarrow \mathcal{S}(\mathcal{H}^{\otimes n})$.
Hence, any lower bound for $\tilde{\mathcal{M}}^{\otimes n}$ is automatically a lower bound for $\mathcal{M}^{\otimes n}$.

\begin{rem}
Let an operator sum representation of
$\mathcal{M} : \mathcal{S}(\mathcal{H}) \rightarrow \mathcal{S}(\mathcal{H})$ be given by
$\mathcal{M}: \rho\mapsto \sum_\mu M_\mu\rho M^\dagger_\mu$ with
$M_\mu = \sum_{w^x,w^z} a_{\mu,w^x,w^z} \XZ(w^x,w^z)$.
Then,
\begin{align}
 P_\mathcal{M}(\vec{e}) %
&= \bra{\Phi_{\vec{e}}}   [\id\otimes\mathcal{M}]( \ketbra{\Phi_{\vec{0}}}{\Phi_{\vec{0}}} ) \ket{\Phi_{\vec{e}}} \nonumber\\
&= \sum_\mu \vert a_{\mu,e^x,e^z} \vert^2,
\end{align}
which coincides with the definition of a probability distribution $P_\mathcal{M}$ on $\mathbb{F}_q^2$ of a general memoryless channel in \cite[section II]{Ha01b}.%
\end{rem}

\section{Lower Bounds on the Capacity of Memoryless Pauli Channels}\label{sec:randqcodes}

A lower bound on the quantum capacity of a binary memoryless Pauli channel was found by Bennett et\,al. \cite{BBPSSW96} by constructing the breeding entanglement distillation protocol.
Imagine two distant parties, say Alice and Bob, who would like to share a set of maximally entangled states, are connected only via a noisy quantum channel.
If Alice prepares a set of maximally entangled bipartite states and sends Bobs half through the channel, they end up sharing a set of imperfect maximally entangled states.
The task of an entanglement distillation protocol is now to distill a smaller set of (nearly) maximally entangled states by means of classical communication and local operations only.
Since the breeding protocol has the need for some pre-distilled maximally entangled states, a revised version of this protocol, the so-called one-way hashing protocol, was proposed in \cite{BDSW}.
Both protocols make use of one-way classical communication only and are therefore equivalent \cite{BDSW} to a scheme where Alice uses a quantum error correcting code to protect Bobs half of the smaller set of perfect states during transmission over the noisy quantum channel.

In the first subsection, the quantum error correcting scheme (generalized to qudits) corresponding to the one-way hashing entanglement distillation protocol is presented.
It corresponds to the use of a random stabilizer code (see \cite[section 7.6]{PhdGottesman} and \cite[section 7.16.2]{Preskill} for the binary case, \cite{Ha01} for the general one).
The achievable rate of this scheme is a lower bound on the quantum capacity of the memoryless Pauli channel (the quantum capacity is by definition the highest achievable rate).
In the second subsection it is shown that the same result can be achieved using random CSS codes,
which is of interest for quantum key distribution since entanglement distillation protocols based on CSS codes are reducible to prepare and measure QKD schemes (\cite{SP00,Ha06a}, subsection \ref{subsec:spproof}).
In fact this result was used by Lo in \cite{Lo01} to prove the security of the 6-state protocol.

\subsection{Random Stabilizer Codes}

In this section we prove the following theorem due to \cite[section 7.6]{PhdGottesman} and
\cite[section 7.16.2]{Preskill} (binary case) and \cite{Ha01} (general case).
\begin{thm}\label{thm:randstab}
Let $\mathcal{A} : \mathcal{S}(\mathcal{H}) \rightarrow \mathcal{S}(\mathcal{H})$ be a Pauli channel with probability distribution $P_{\!\mathcal{A}}$ on $\mathbb{F}_q^2$
and let $\varepsilon>0$.
Then, as long as
\begin{equation}\label{eq:randstabcondi}
 \frac{k}{n} < 1 - H_{q^2[\log_q]} ( P_{\!\mathcal{A}} ),
\end{equation}
and for large enough $n$, there exists a stabilizer $L$ of dimension $n-k$ such that
for any corresponding stabilizer code $\mathcal{C}_{(L,\vec{s})}$,
there exists a recovery operation
$\mathcal{R}_{(\vec{s})}$
with minimum fidelity
\begin{equation}\label{eq:randstabf}
F_p\bigl(  \mathcal{C}_{(L,\vec{s})}  , \mathcal{R}_{( \vec{s} )} \mathcal{A}^{\otimes n}  \bigr)
=
\min_{ \ket{\psi}\in\mathcal{C}_{(L,\vec{s})} }
\bra{\psi}  \mathcal{R}_{( \vec{s} )}( \mathcal{A}^{\otimes n}(\ketbra{\psi}{\psi}) )  \ket{\psi}
 > 1-\varepsilon .
\end{equation}
\end{thm}

\begin{rem}
\cite{Ha01} shows the following stronger result:
Let integers $n,k$ and $R\in\mathbb{R}$ satisfy $0\leq k \leq Rn$ and $0\leq R < 1$.
Then, the minimum fidelity of \eqref{eq:randstabf} is at least
\begin{equation}
 1-(n+1)^{2(q^2-1)} q ^ {-n E(R,P_{\!\mathcal{A}}) },
\end{equation}
where the random coding exponent $E(R, P_{\!\mathcal{A}} )$ stays positive as long as
$R < 1 - H_{q^2[\log_q]} ( P_{\!\mathcal{A}} )$.
The proof of the stronger statement is more elaborate than
the simple proof of theorem \ref{thm:randstab}, which uses typical set decoding as in section \ref{sec:randlincodes}.
\end{rem}

\noindent
Before we start with the proof of theorem \ref{thm:randstab}, we need the following lemma.

\begin{lem}[Lemma 6 of \cite{Ha01b}]\label{lem:boundstabAdivAx}
Let the set of all stabilizers of dimension $n-k$ be given by
\begin{equation}
 \textsf{A}_{n,k} = \{ L\subset F_q^{2n} \sthat L \text{ is linear }, L\subseteq L^\perp, \dim L=n-k \}
\end{equation}
and let
\begin{equation}
 \textsf{A}_{n,k}(\vec{x}) = \{ L\in \textsf{A}_{n,k} \sthat \vec{x}\in L^\perp\setminus\{\vec{0}\} \ \}.
\end{equation}
Then, $\vert \textsf{A}_{n,k}(\vec{0})\vert =0$ and
\begin{equation}
  \frac{\vert \textsf{A}_{n,k}(\vec{x}) \vert}{\vert \textsf{A}_{n,k} \vert}
= \frac{q^{n+k}-1}{q^{2n}-1} \leq \frac{1}{q^{n-k}}
\end{equation}
for any nonzero $\vec{x}\in\mathbb{F}_q^{2n}$.
\end{lem}

\begin{proof}[Proof of theorem \ref{thm:randstab}]
For fixed $n$ and $k$, we pick a stabilizer $L\in \textsf{A}_{n,k}$ and encode $k$ qudits into one of the codespaces $\mathcal{C}(L,\vec{s})$ labeled by $\vec{s}\in\mathbb{F}_q^{n-k}$.
We apply the definition \ref{def:typset} of a typical set to
the random variable $X$ taking on values $(e^x,e^z)\in \mathbb{F}_q^2$ according to the probability distribution $P_{\!\mathcal{A}} $ on~$\mathbb{F}_q^2$:
\begin{multline}
 T_\delta^n = \Bigl\{ \vec{e}=(\vec{e}^x,\vec{e}^z) \in\mathbb{F}_q^{2n} \text{ s.\,t. for every }
 (e^x,e^z)\in \mathbb{F}_q^2, \\
 \bigl\vert N((e^x,e^z)\vert\vec{e}) - n P_{\!\mathcal{A}}(e^x,e^z) \bigr\vert
 < \frac{ \delta n P_{\!\mathcal{A}}(e^x,e^z) }{ \log_q \vert \mathbb{F}_q^2 \vert }  \Bigr\},
\end{multline}
For a given stabilizer $L$, we construct a transversal $J(L)$ for the cosets of $L^\perp$ in $\mathbb{F}_q^{2n}$ according to the following rule:
If a coset contains exactly one typical vector $\vec{e} \in T_\delta^n$, then add this vector to $J(L)$, else
pick the corresponding representative at random.
A recovery operation $\mathcal{R}_{( J(L),\vec{s} )}$ which corrects an error set like $J(L)$ was defined in subsection \ref{subsec:stabrecover}. As a consequence of $J(L)$ being a transversal, our code will be non-degenerate.
The minimum fidelity of our coding scheme,
\begin{equation}
F_p\bigl(  \mathcal{C}_{(L,\vec{s})}  , \mathcal{R}_{( J(L),\vec{s} )} \mathcal{A}^{\otimes n}  \bigr)
=
\min_{ \ket{\psi}\in\mathcal{C}_{(L,\vec{s})} }
\bra{\psi}  \mathcal{R}_{( J(L),\vec{s} )}( \mathcal{A}^{\otimes n}(\ketbra{\psi}{\psi}) )  \ket{\psi},
\end{equation}
will certainly be not less than $\sum_{\vec{e} \in J(L)} P_{\!\mathcal{A}}^n(\vec{e})$, since the fidelity will be one if $\vec{e} \in J(L)$.
In other words,
\begin{align}
1-F_p\bigl(  \mathcal{C}_{(L,\vec{s})}  , \mathcal{R}_{( J(L),\vec{s} )} \mathcal{A}^{\otimes n}  \bigr)
&\leq \sum_{\vec{e} \notin J(L)} P_{\!\mathcal{A}}^n(\vec{e}) \nonumber\\
&\leq    \sum_{ \vec{e} \notin T_\delta^n } P_{\!\mathcal{A}}^n(\vec{e}) +
      \sum_{ \vec{e} \in T_\delta^n } P_{\!\mathcal{A}}^n(\vec{e}) \cdot
 \mathbbm{1}\!\left[
 \begin{array}{@{}l@{}}
   \text{\small{$\exists\, \vec{e}'\in T_\delta^n$ with $\vec{e}'\neq\vec{e}$ s.\,t.}}\\
   \text{\small{$(\vec{g}_i,\vec{e}-\vec{e}')_{sp}=0$ for $1\leq i \leq n-k$}}
 \end{array}\right].
\end{align}
The first sum is upper bounded by $\Delta \sim 1/(\delta^2 n)$ (part b of theorem \ref{thm:proptypset}) and the latter by
\begin{equation}\label{eq:ProofRandStabLater}
   \sum_{ \vec{e} \in T_\delta^n } P_{\!\mathcal{A}}^n(\vec{e})
   \sum_{ \vec{e}' \in T_\delta^n }^{\text{ s.\,t. } \vec{e}'\neq\vec{e}}
\mathbbm{1}\!\left[ (\vec{g}_i,\vec{e}-\vec{e}')_{sp}=0 \text{ for } 1\leq i \leq n-k \right].
\end{equation}
Therefore, averaging over all stabilizers $L\in\textsf{A}_{n,k}$ leads to
\begin{align*}
1-\overline{F}_p &\equiv
\bigl\langle
1-F_p\bigl(  \mathcal{C}_{(L,\vec{s})}  , \mathcal{R}_{( J(L),\vec{s} )} \mathcal{A}^{\otimes n}  \bigr)
\bigr\rangle_{L\in\textsf{A}_{n,k}} \\
&\leq  \Delta +
   \sum_{ \vec{e} \in T_\delta^n } P_{\!\mathcal{A}}^n(\vec{e})
   \sum_{ \vec{e}' \in T_\delta^n }^{\text{ s.\,t. } \vec{e}'\neq\vec{e}}
   \frac{\vert \textsf{A}_{n,k}(\vec{e}-\vec{e}') \vert}{\vert \textsf{A}_{n,k} \vert} && \text{by theorem \ref{thm:proptypset}b and \eqref{eq:ProofRandStabLater}}\\
&\leq  \Delta +  (\vert T_\delta^n \vert -1) q^{k-n} && \text{by lemma \ref{lem:boundstabAdivAx}}\\
&\leq  \Delta +  \exp_q\bigl( n(H_{q^2[\log_q]}(P_{\!\mathcal{A}})+\delta)+k-n\bigr)  && \text{by theorem \ref{thm:proptypset}c}.
\end{align*}
This quantity becomes arbitrary small for large enough $n$ as long as
\begin{equation}
 \frac{k}{n} < 1 - H_{q^2[\log_q]}(P_{\!\mathcal{A}}) - \delta.
\end{equation}
Since the above statement holds for any $\delta$, we are free to choose $\delta$ as small as we like.
So far we have shown that the fidelity $\overline{F}_p$ averaged over all stabilizers is larger than $1-\varepsilon$.
It follows that there exists at least one stabilizer $L\in \textsf{A}_{n,k}$ such that
$F_p\bigl(  \mathcal{C}_{(L,\vec{s})}  , \mathcal{R}_{( J(L),\vec{s} )} \mathcal{A}^{\otimes n}\bigr) > 1-\varepsilon$.
\end{proof}

\subsection{Random CSS Codes}\label{subsec:randcss}

In this subsection we show that using CSS codes instead of general stabilizer codes is sufficient for theorem \ref{thm:randstab} to hold, i.\,e. we prove the following theorem proposed by Lo in \cite{Lo01} to prove the security of the 6-state quantum key distribution protocol.

\begin{thm}\label{thm:randcss}
Let $\mathcal{A} : \mathcal{S}(\mathcal{H}) \rightarrow \mathcal{S}(\mathcal{H})$ be a Pauli channel with probability distribution $P_{\!\mathcal{A}}$ on $\mathbb{F}_q^2$
and let $\varepsilon>0$.
Then, as long as
\begin{equation}
 \frac{k}{n} < 1 - H_{q^2[\log_q]} ( P_{\!\mathcal{A}} ),
\end{equation}
and for large enough $n$, there exists
a pair of codes $\mathcal{C}_2 \subset \mathcal{C}_1$ such that
for any codespace $\mathcal{C}_{(L(\mathcal{C}_1,\mathcal{C}_2),\vec{s})}$ of the corresponding CSS code with stabilizer $L(\mathcal{C}_1,\mathcal{C}_2)$,
there exists a two step recovery operation $\mathcal{R}_{(\vec{s})}$,
first correcting the bit errors and then, by using the bit error syndrome to reduce the uncertainty on the phase errors, correcting the phase errors,
with minimum fidelity
\begin{equation}\label{eq:randcssf}
F_p\bigl(  \mathcal{C}_{(L(\mathcal{C}_1,\mathcal{C}_2),\vec{s})}  , \mathcal{R}_{(\vec{s})} \mathcal{A}^{\otimes n}  \bigr)
=
\min_{ \ket{\psi}\in\mathcal{C}_{(L(\mathcal{C}_1,\mathcal{C}_2),\vec{s})} }
\bra{\psi}  \mathcal{R}_{(\vec{s})}( \mathcal{A}^{\otimes n}(\ketbra{\psi}{\psi}) )  \ket{\psi}
 > 1-\varepsilon .
\end{equation}
\end{thm}

\noindent
For the proof of theorem \ref{thm:randcss} we need the joint- and conditional typical sets from subsection \ref{subsec:jtypsets},
and corollaries \ref{cor:numberlincodes} and \ref{cor:numberc2ifc1} from appendix \ref{sec:app:lincodes}.

\begin{proof}[Proof of theorem \ref{thm:randcss}]
We apply definition \ref{def:jtypset} of a set of jointly strongly $\delta$-typical sequences of length $n$ to the two random variables $X$ and $Z$
with joint probability distribution
$P_{\!\mathcal{A}} = \{
  P_{\!\mathcal{A}}(x,z) = P_{\!\mathcal{A}}(z\vert x)\cdot P_{\!\mathcal{A}}(x)
     \}_{x,z\in\mathbb{F}_q}$ and obtain (i) the joint typical set:
\begin{equation}
  T_\delta^n(XZ) = \Bigl\{ (\vec{x},\vec{z})
 \text{ s.\,t. for all } x,z\in\mathbb{F}_q,
 \bigl\vert N(xz\vert\vec{x}\vec{z}) - n P_{\!\mathcal{A}}(x,z)\bigr\vert \leq \frac{\delta n P_{\!\mathcal{A}}(x,z)}{\log_q \vert \mathbb{F}_q^4 \vert}  \Bigr\},
\end{equation}
(ii) the set of typical $X$-sequences:
\begin{equation}
 T'^n_\delta(X) = \{ \vec{x}\in\mathbb{F}_q^n \sthat (\vec{x},\vec{z}) \in T_\delta^n(XZ) \text{ for some } \vec{z}\in\mathbb{F}_q^n \},
\end{equation}
and (iii) the conditional typical set of $Z$-sequences for a given $\vec{x}\in\mathbb{F}_q^n$:
\begin{equation}
 T_\delta^n(Z\vert\vec{x}) = \{ \vec{z}\in\mathbb{F}_q^n \sthat (\vec{x},\vec{z}) \in T_\delta^n(XZ) \}.
\end{equation}
A CSS code $\mathcal{C}$ encoding $k=k_1-k_2$ qudits into $n$ is a stabilizer code whose $n-k$ dimensional stabilizer $L$ is constructed from two linear codes $\mathcal{C}_2 \subseteq \mathcal{C}_1$, where
$\mathcal{C}_1$ is an $[n,k_1]_q$ code correcting bit errors and
$\mathcal{C}_2^\perp$ is an $[n,n-k_2]_q$ code correcting phase errors.
Let us fix $n$, $k$ and an $n-k$ dimensional stabilizer $L\equiv L(\mathcal{C}_1,\mathcal{C}_2)$ and encode $k$ qudits into one of the codespaces $\mathcal{C}(L,\vec{s})$ labeled by $\vec{s}\in\mathbb{F}_q^{n-k}$.
A non-degenerate correctable error set $J(L)$ for the CSS-type stabilizer $L$ can be specified by fixing a transversal $\Gamma_1$ of $\mathbb{F}_q^n / \mathcal{C}_1$ and a
transversal $\Gamma_2$ of $\mathbb{F}_q^n / \mathcal{C}_2^\perp$,
\begin{equation}
 J(L) = \{ \XZ(\vec{a}^x,\vec{a}^z) \sthat \vec{a}^x\in\Gamma_1 , \: \vec{a}^z\in\Gamma_2 \}.
\end{equation}
Let us assume now that the actual error of the Pauli channel is in the set $T_\delta^n(XZ)$.
We split up the recovery operation for the correctable error set $J(L)$ into two parts.
In the first step, we try identify the bit error $\vec{x} \in T'^n_\delta(X)$ by using a typical set decoder:
$\Gamma_1$ is chosen in such a way that each of its coset representatives is either the only coset member which is in $T'^n_\delta(X)$, or, if there are none or multiple coset members which are in $T'^n_\delta(X)$, it is chosen at random.
By measuring the bit error syndrome $\vec{s}^x$
(i.\,e. by measuring the eigenvalue list of the Pauli operators corresponding to the first $n-k_1$ generating elements of $L$),
we identify a coset of $\mathcal{C}_1$ in $\mathbb{F}_q^n$ and conclude that the actual bit error $\vec{x}$ is the corresponding coset representative in $\Gamma_1$.
In the next step, we use the information about the bit error $\vec{x}$ to reduce the uncertainty on the remaining phase error $\vec{z}$:
Since we know that $\vec{z}$ has to be in $T_\delta^n(Z\vert\vec{x})$, we apply typical set decoding for the set $T_\delta^n(Z\vert\vec{x})$ by setting $\Gamma_2$ accordingly.
The measurement of the phase error syndrome $\vec{s}^z$
(corresponding to the eigenvalue list of the Pauli operators corresponding to the last $k_2$ generating elements of $L$),
identifies a coset of $\mathcal{C}_2^\perp$ in $\mathbb{F}_q^n$ and we conclude that the actual phase error $\vec{z}$ is the corresponding coset representative in $\Gamma_2$.
To find a lower bound on the minimum fidelity,
\begin{equation}
F_p\bigl(  \mathcal{C}_{(L,\vec{s})}  , \mathcal{R}_{( J(L),\vec{s} )} \mathcal{A}^{\otimes n}  \bigr)
=
\min_{ \ket{\psi}\in\mathcal{C}_{(L,\vec{s})} }
\bra{\psi}  \mathcal{R}_{( J(L),\vec{s} )}( \mathcal{A}^{\otimes n}(\ketbra{\psi}{\psi}) )  \ket{\psi},
\end{equation}
of our coding scheme, we note that the fidelity will certainly be greater or equal than the probability of success of the coding scheme.
In other words, one minus the fidelity will be upper bounded by the probability of failure.
We proceed by finding an upper bound on the probability of failure.
Our scheme fails if
(i) the actual error is not within the joint typical set $T_\delta^n(XZ)$,
(ii) it is in $T_\delta^n(XZ)$, but bit error correction fails because the measured coset of $\mathcal{C}_1$ in $\mathbb{F}_q^n$ contains multiple coset members which are in $T'^n_\delta(X)$,
or (iii) the actual error is in $T_\delta^n(XZ)$, bit error corrections works, but phase error correction fails because the measured coset of $\mathcal{C}_2^\perp$ in $\mathbb{F}_q^n$ contains multiple coset members which are in $T_\delta^n(Z\vert\vec{x})$.
Conditioned on the assumption that the actual error is $(\vec{x},\vec{z}) \in T_\delta^n(XZ)$,
bit error correction fails if the following boolean expression is true,
\begin{equation}
F_\text{bit}=(\exists \vec{x}'\in T'^n_\delta(X)\text{ with }\vec{x}'\neq\vec{x}\text{ s.\,t.}H_1(\vec{x}-\vec{x}')^T=\vec{0}^T), \quad ( H_1 \text{ parity check matrix of } \mathcal{C}_1),
\end{equation}
and phase error correction fails (assuming that bit error correction succeeded) if
\begin{equation}
F_\text{phase}=(\exists \vec{z}'\in T_\delta^n(Z\vert\vec{x})\text{ with }\vec{z}'\neq\vec{z}\text{ s.\,t. }H_2(\vec{z}-\vec{z}')^T=\vec{0}^T), \quad ( H_2 \text{ parity check matrix of } \mathcal{C}_2^\perp),
\end{equation}
is true.
Using these boolean expressions, we obtain
\begin{align*}
 1-F_p &\equiv 1-F_p\bigl(  \mathcal{C}_{(L,\vec{s})}  , \mathcal{R}_{( J(L),\vec{s} )} \mathcal{A}^{\otimes n}  \bigr)\\
 &\leq
         \sum_{\vec{e}\notin T_\delta^n(XZ)} P_{\!\mathcal{A}}^n(\vec{e}) +
         \sum_{\vec{e}\in T_\delta^n(XZ)} P_{\!\mathcal{A}}^n(\vec{e}) \cdot
         \mathbbm{1}\!\left[ F_\text{bit} \vee ( \neg F_\text{bit}  \wedge F_\text{phase} ) \right] \\
&\leq
         \Delta +
         \sum_{\vec{e}\in T_\delta^n(XZ)} P_{\!\mathcal{A}}^n(\vec{e}) \cdot
         \mathbbm{1}\!\left[ F_\text{bit} \vee F_\text{phase} \right] && \text{by thm \ref{thm:propjtypset}b}\\
&\leq
        \Delta +
        \sum_{\vec{e}\in T_\delta^n(XZ)} P_{\!\mathcal{A}}^n(\vec{e}) \cdot
\bigl(  \mathbbm{1}\!\left[ F_\text{bit}  \right] +
        \mathbbm{1}\!\left[ F_\text{phase}\right]   \bigr)\\
&\leq
         \Delta    +
 \,\sum_{\mathclap{\vec{e}\in T_\delta^n(XZ)}}\, P_{\!\mathcal{A}}^n(\vec{e}) \cdot \Bigl(
 \sum_{\mathclap{\vec{x}'\in T'^n_\delta(X)}}^{\smash{\vec{x}'\neq\vec{x}}}
 \mathbbm{1}\!\left[ \text{\small{$ H_1(\vec{x}-\vec{x}')^T=\vec{0}^T $}}\right]   +
 \sum_{\mathclap{\vec{z}'\in T_\delta^n(Z\vert\vec{x})}}^{\smash{\vec{z}'\neq\vec{z}}}
 \mathbbm{1}\!\left[ \text{\small{$H_2(\vec{z}-\vec{z}')^T=\vec{0}^T $}}\right] \Bigr).
\end{align*}
Now we are going to take the average of $1-F_p$ over all code pairs $(\mathcal{C}_1,\mathcal{C}_2)$ which satisfy $\mathcal{C}_2 \subseteq \mathcal{C}_1$.
Let us denote by
\begin{equation}
 A_{n,k,q} = \{ \mathcal{C}\subseteq\mathbb{F}_q^n \sthat \mathcal{C}\text{ is a } [n,k]_q \text{-code} \}
\end{equation}
the set of all $[n,k]_q$ codes.
Let $\mathcal{K}$ be an $[n,\kappa]_q$ code and let $\vec{c}$ be some nonzero codeword in $\mathbb{F}_q^n$,
then we denote by $A_{n,k,q}(\vec{c})$ the set of all $[n,k]_q$ codes which contain $\vec{c}$,
and, in an analogous fashion, we denote by $A_{n,k,q}( \mathcal{K} )$ and $A_{n,k,q}( \mathcal{K},\vec{c} )$
the set of codes which contain $\mathcal{K}$ and $\mathcal{K}\cup\vec{c}$, respectively (see section \ref{sec:app:lincodes}).
We denote the average over all codes by
$\bigl\langle \bigl\langle \cdot \bigr\rangle_{\mathcal{C}_2^\perp} \bigr\rangle_{\mathcal{C}_1}$
using the shorthand notation
$\bigl\langle \cdot \bigr\rangle_{\mathcal{C}_2^\perp}
 \equiv
 \bigl\langle \cdot \bigr\rangle_{\mathcal{C}_2^\perp \in A_{n,n-k_2,q}(\mathcal{C}_1^\perp)}$
since we are allowed to average only over those codes $\mathcal{C}_2^\perp$ which include $\mathcal{C}_1^\perp$.
With the help of corollaries \ref{cor:numberlincodes} and \ref{cor:numberc2ifc1} we obtain
\begin{align*}
\bigl\langle \bigl\langle 1-F_p \bigr\rangle_{\mathcal{C}_2^\perp} \bigr\rangle_{\mathcal{C}_1} &\leq
         \Delta    +
 \,\sum_{\mathclap{\vec{e}\in T_\delta^n(XZ)}}\, P_{\!\mathcal{A}}^n(\vec{e})
 \sum_{\mathclap{\vec{x}'\in T'^n_\delta(X)}}^{\smash{\vec{x}'\neq\vec{x}}}
 \frac{ \vert A_{n,k_1,q} (\vec{x}-\vec{x}') \vert }{ \vert A_{n,k_1,q} \vert }
  +
 \,\sum_{\mathclap{\vec{e}\in T_\delta^n(XZ)}}\, P_{\!\mathcal{A}}^n(\vec{e})
 \sum_{\mathclap{\vec{z}'\in T_\delta^n(Z\vert\vec{x})}}^{\smash{\vec{z}'\neq\vec{z}}}
\Bigl\langle
\frac{\vert A_{n,n-k_2,q}( \mathcal{C}_1^\perp,\vec{z}-\vec{z}' ) \vert}{\vert A_{n,n-k_2,q}( \mathcal{C}_1^\perp ) \vert}
\Bigr\rangle_{\mathcal{C}_1} \\
&\leq
       \Delta    +
 \,\sum_{\mathclap{\vec{e}\in T_\delta^n(XZ)}}\, P_{\!\mathcal{A}}^n(\vec{e})
 \bigl(\vert T'^n_\delta(X)\vert-1\bigr)
 q^{-n+k_1}
  +
 \,\sum_{\mathclap{\vec{e}\in T_\delta^n(XZ)}}\, P_{\!\mathcal{A}}^n(\vec{e})
 \bigl(\vert T_\delta^n(Z\vert\vec{x})\vert-1 \bigr)
 q^{-k_2}, \\
\intertext{and by part c of theorem \ref{thm:propjtypset},}
&\leq
       \Delta    +
 \exp_q\bigl( n(H_{[\log_q]}(X)+\delta) -n+k_1 \bigr)  +
 \exp_q\bigl( n(H_{[\log_q]}(Z\vert X)+2\delta) -k_2 \bigr).
\end{align*}
This quantity becomes arbitrary small for sufficiently large $n$, as long as
$n H_{[\log_q]}(X) - n + k_1 < 0$ and $n H_{[\log_q]}(Z\vert X) -k_2 < 0$, which can always be satisfied as long as
\begin{equation*}
 \frac{k_1-k_2}{n} < 1 - n H_{[\log_q]}(Z\vert X) -n H_{[\log_q]}(X) = 1 - H_{[\log_q]}(XZ).
\end{equation*}
So far we have shown that the fidelity averaged over all code pairs $(\mathcal{C}_1,\mathcal{C}_2)$ such
that $\mathcal{C}_2\subset \mathcal{C}_1$ is larger than $1-\varepsilon$,
\begin{equation*}
 \bigl\langle \bigl\langle
 F_p( \mathcal{C}_{(L(\mathcal{C}_1,\mathcal{C}_2),\vec{s})}, \mathcal{R}_{(J(L),\vec{s})} \mathcal{A}^{\otimes n}  )
 \bigr\rangle_{\mathcal{C}_2^\perp} \bigr\rangle_{\mathcal{C}_1} > 1-\varepsilon.
\end{equation*}
It follows that there exists at least one pair of codes $(\mathcal{C}_1,\mathcal{C}_2)$ such that
$F_p\bigl( \mathcal{C}_{(L(\mathcal{C}_1,\mathcal{C}_2),\vec{s})} , \mathcal{R}_{(J(L),\vec{s})} \mathcal{A}^{\otimes n}\bigr)$ is larger than $1-\varepsilon$.
\end{proof}

\section{Concatenating Random and Deterministic Codes}\label{sec:concrandom}

It was shown by Shor and Smolin in \cite{ShSm96}
(and later together with DiVincenzo in \cite{DiSS98})
that the achievable rate for reliable quantum communication over a memoryless Pauli channel $\mathcal{A} : \mathcal{S}(\mathcal{H}) \rightarrow \mathcal{S}(\mathcal{H})$ using random stabilizer codes (section \ref{sec:randqcodes}) is indeed only a lower bound on the quantum capacity of the channel:
By concatenating a certain deterministic inner code with a random outer code, they found that reliable transmission over the depolarizing channel,
a special type of Pauli channel characterized by a single noise parameter $p$,
becomes feasible for higher values of noise than allowed by random codes alone.
This result is somewhat surprising since in the classical case, random codes do achieve the capacity of discrete memoryless channels.

For a given inner code, concatenated as described above,
we determine the achievable rate for reliable quantum communication over a memoryless Pauli channel in subsection \ref{subsec:rateconc} \cite{DiSS98,Ha02}.
In the subsequent subsection \ref{subsec:ratecoinf} we show that this rate can be expressed as coherent information of a maximally mixed state in the codespace of the inner code \cite{DiSS98,Ha02}.
We apply these results to the depolarizing channel using a so-called cat code as inner code in the following section.

\subsection{Achievable Rate}\label{subsec:rateconc}

We are going to determine the achievable rate for reliable quantum communication over a memoryless Pauli channel, when using a concatenated code whose outer code is chosen at random.
Let the deterministic inner code be an $[[n,k]]_q$ code with stabilizer $L^\text{in}=\{\vec{g}_1,\dots,\vec{g}_{n-k}\}$, and let an extension to a hyperbolic basis of~$\mathbb{F}_q^{2n}$ be given by $\{\vec{g}_{n-k+1},\dots,\vec{g}_n, \vec{h}_1,\dots,\vec{h}_n\}$.
By writing a vector $\vec{a}\in\mathbb{F}_q^{2n}$ as linear combination of the basis elements of such a basis,
\begin{equation*}%
\begin{split}
 \vec{a} &= ( a_1^x,\dots,a_n^x, a_1^z,\dots,a_n^z ) \\
 &= \sum_{i=1}^{n-k} \bigl( s_i \vec{h}_i + n_i \vec{g}_i \bigr)
 + \sum_{i=n-k+1}^n \bigl(
 l_{i-(n-k)}^x \vec{h}_i + l_{i-(n-k)}^z \vec{g}_i \bigr),
\end{split}
\end{equation*}
with $s_i = (\vec{g}_i,\vec{a})_{sp}$, $n_i = (\vec{a},\vec{h}_i)_{sp}$ for $i\in\{1,\dots,n-k\}$ and
$l_{i-(n-k)}^x = (\vec{g}_i,\vec{a})_{sp}$, $l_{i-(n-k)}^z = (\vec{a},\vec{h}_i)_{sp}$ for $i\in\{n-k+1,\dots,n\}$, we derived lemma \ref{lem:xzdecompostab}, relating the corresponding Pauli operators:
\begin{equation}
 XZ(\vec{a}) \sim  \overline{X}^{(\vec{s},\vec{l}^x)}  \overline{Z}^{(\vec{n},\vec{l}^z)}.
\end{equation}
This relation allows us to rewrite the action of a memoryless Pauli channel
$\mathcal{A}^{\otimes n}$ defined by the probability distribution
$P^n_{\!\mathcal{A}}(\vec{a}=(a^x_1,\dots,a^x_n,a^z_1,\dots,a^z_n)) =
 \prod_{i=1}^n P_{\!\mathcal{A}}(a^x_i,a^z_i)$, as follows:
\begin{align}\label{eq:paulirewritten}
 \mathcal{A}^{\otimes n} :  \rho \mapsto \mathcal{A}^{\otimes n}(\rho)
&= \sum_{\vec{a}\in\mathbb{F}^{2n}_q} P^n_{\!\mathcal{A}}(\vec{a}) \, \XZ(\vec{a}) \rho  \XZ(\vec{a})^\dagger, \nonumber\\
&=\sum_{\vec{s},\vec{n}\in\mathbb{F}_q^{n-k}} \sum_{\vec{l}^x,\vec{l}^z\in\mathbb{F}_q^{k}}
P_{\!\mathcal{A}}(\vec{s},\vec{n},\vec{l}^x,\vec{l}^z) \,
\bigl(\overline{X}^{(\vec{s},\vec{l}^x)}  \overline{Z}^{(\vec{n},\vec{l}^z)}\bigr) \rho \,
\bigl(\overline{X}^{(\vec{s},\vec{l}^x)}  \overline{Z}^{(\vec{n},\vec{l}^z)}\bigr)^\dagger.
\end{align}
In addition to $P_{\!\mathcal{A}}(\vec{s},\vec{n},\vec{l}^x,\vec{l}^z) = P^n_{\!\mathcal{A}}(\vec{a})$ we define
\begin{align}\label{eq:paulislxlz}
 P_{\!\mathcal{A}}(\vec{s},\vec{l}^x,\vec{l}^z) &=
 \sum_{\mathclap{\vec{n}\in\mathbb{F}_q^{n-k}}} P_{\!\mathcal{A}}(\vec{s},\vec{n},\vec{l}^x,\vec{l}^z), &
 P_{\!\mathcal{A}}(\vec{s}) &=
 \sum_{\mathclap{\vec{l}^x,\vec{l}^z\in\mathbb{F}_q^k}} P_{\!\mathcal{A}}(\vec{s},\vec{l}^x,\vec{l}^z),
\end{align}
and the conditional probability
$P_{\!\mathcal{A}}(\vec{l}^x,\vec{l}^z\vert \vec{s}) =
  P_{\!\mathcal{A}}(\vec{s},\vec{l}^x,\vec{l}^z)/P_{\!\mathcal{A}}(\vec{s})$.
Note that $P_{\!\mathcal{A}}(\vec{s},\vec{l}^x,\vec{l}^z)$ denotes the probability of having an error in a certain coset of $L^\text{in}$ in $\mathbb{F}_q^{2n}$,
and $P_{\!\mathcal{A}}(\vec{s})$ denotes the probability of having an error in a certain coset of $L^{\text{in}\perp}$ in $\mathbb{F}_q^{2n}$.
Hence, the probability distributions given by $\{ P_{\!\mathcal{A}}(\vec{s},\vec{l}^x,\vec{l}^z) \}$ and $\{ P_{\!\mathcal{A}}(\vec{s}) \}$ do not depend on the detailed form of the hyperbolic basis (and therefore on the encoding), but depend only on the stabilizer $L^\text{in}$ itself.

Let the random outer code be an $[[N,K]]_q$ code as in section \ref{sec:concat} (with $N$ divisible by $k$).
We encode some $K$-qudit quantum state within one of the codespaces of the concatenated code and send the resulting $\mathfrak{n} = N/k\times n$ qudits through the Pauli channel $\mathcal{A}^{\otimes N/k\times n }$.
The result of a measurement of the first $N/k\times (n-k)$ operators $\overline{Z}_{\text{con},i}$, $i\in\{1,\dots,N/k\times(n-k)\}$
(which corresponds to a measurement of the $N/k$
syndromes of the inner codes)
can be expressed as $\vec{S} = (\vec{s}_1,\dots,\vec{s}_{N/k})$, $\vec{s}_j\in\mathbb{F}_q^{n-k}$ for $j=1,\dots,N/k$.
Applying definition \ref{def:jtypset} of a set of jointly strongly $\delta$-typical sequences of length $N/k$ to the two random variables $E$ and $S$ taking on values $\vec{l}=(\vec{l}^x,\vec{l}^z)\in\mathbb{F}_q^{2k}$ and $\vec{s}\in\mathbb{F}_q^{n-k}$ according to the joint probability distribution
$\{  P_{\!\mathcal{A}}( \vec{l},\vec{s} ) \}$ given by \eqref{eq:paulislxlz}, we obtain (i) the joint typical set:
\begin{equation}
 T_\delta^{N/k}(ES) = \Bigl\{ (\vec{L},\vec{S})
 \text{ s.\,t. for all } \vec{l}\in\mathbb{F}_q^{2k},\vec{s}\in\mathbb{F}_q^{n-k}, \,
\Bigl\vert N(\vec{l}\vec{s}\vert\vec{L}\vec{S}) - \frac{N}{k} P_{\!\mathcal{A}}(\vec{l},\vec{s})\Bigr\vert \leq \frac{\delta N P_{\!\mathcal{A}}(\vec{l},\vec{s})}{k \log_q \vert \mathbb{F}_q^{n+k} \vert}  \Bigr\},
\end{equation}
with $\vec{L} = (\vec{l}_1,\dots,\vec{l}_{N/k}) = (\vec{l}_1^x,\vec{l}_1^z,\dots,\vec{l}_{N/k}^x,\vec{l}_{N/k}^z) \in \mathbb{F}_q^{N/k\times 2k}$,
(ii) the set of typical $S$-sequences:
\begin{equation}
 T'^{N/k}_\delta(S) = \{ \vec{S}\in\mathbb{F}_q^{N/k\times(n-k)} \sthat (\vec{L},\vec{S}) \in T_\delta^{N/k}(ES) \text{ for some } \vec{L}\in\mathbb{F}_q^{N/k\times 2k} \},
\end{equation}
and (iii) the conditional typical set of $E$-sequences for a given $\vec{S}\in\mathbb{F}_q^{N/k\times(n-k)}$:
\begin{equation}
 T_\delta^{N/k}(E\vert\vec{S}) = \{ (\vec{L},\vec{S}) \sthat (\vec{L},\vec{S}) \in T_\delta^{N/k}(ES) \}.
\end{equation}
Let us assume now that the actual error of $\mathcal{A}^{\otimes N/k\times n}$ is in $T_\delta^{N/k}(ES)$.
This assumption is satisfied, since for $N/k$ sufficiently large, the probability of the error being in $T_\delta^{N/k}(ES)$ is larger than $1-\Delta$ for any $\Delta > 0$ (part b of theorem \ref{thm:propjtypset}).
Conditioned on the result $\vec{S}$ of the measurement described above,
the situation is equivalent to a scenario where only an $[[N,K]]_q$ code is used to protect against the Pauli channel
\begin{equation}
\mathcal{A}_\text{eff}=\bigotimes_{j=1}^{N/k} \mathcal{G}_j, \quad \text{ with }
\mathcal{G}_j : \mathcal{S}(\mathcal{H}^{\otimes k}) \rightarrow \mathcal{S}(\mathcal{H}^{\otimes k}),
\end{equation}
where $\mathcal{G}_j$ is a general Pauli channel whose probability distribution
$P_{\!\smash{\mathcal{G}_j}} = \{ P_{\!\mathcal{A}}( \vec{l}\vert\vec{s}_j  )\}$
depends on the value of $\vec{s}_j$ in $\vec{S} = (\vec{s}_1,\dots,\vec{s}_{N/k})$.
Since the errors of $\mathcal{A}_\text{eff}$ are known to be in $T_\delta^{N/k}(E\vert\vec{S})$, we are going to use corresponding typical set decoding.
It is known from the proof of theorem \ref{thm:randstab}
that taking the average over all $[[N,K]]_q$ codes results in an average minimum fidelity which is greater than $1-\varepsilon$ for any $\varepsilon>0$, as long as the exponent of
\begin{equation}
  ( T_\delta^n(E\vert\vec{S}) - 1) \cdot q^{K-N} \leq
  \exp_q\Bigl( \frac{N}{k}(H_{[\log_q]}(E\vert S)+2\delta)-(N-K)  \Bigr)
\end{equation}
is negative and $N/k$ is sufficiently large.
Since
\begin{equation}
  H_{[\log_q]}(E\vert S) = \sum_{\vec{s}\in\mathbb{F}_q^{n-k}} P_{\!\mathcal{A}}(\vec{s})
  H_{q^{2k}[\log_q]}\bigl( \{ P_{\!\mathcal{A}}(\vec{l}^x,\vec{l}^z \vert \vec{s}) \} \bigr),
\end{equation}
we have proven the following theorem due to
\cite[for $k=1$ and $q=2$]{DiSS98} and \cite{Ha02}.
\begin{thm}\label{thm:concrate}
Let $\mathcal{A} : \mathcal{S}(\mathcal{H}) \rightarrow \mathcal{S}(\mathcal{H})$ be a Pauli channel with probability distribution $P_{\!\mathcal{A}}$ on $\mathbb{F}_q^2$,
let some inner $[[n,k]]_q$ code be fixed,
and let $\varepsilon>0$.
Then there exists an outer $[[N,K]]_q$ code such that for any codespace of the corresponding concatenated $[[\mathfrak{n},K]]$ code (with $\mathfrak{n} = N/k\cdot n$), there exists a recovery operation with minimum fidelilty larger than $1-\varepsilon$,
as long as the total rate $K/\mathfrak{n}$ satisfies
\begin{equation}\label{eq:concrate}
 \frac{K}{\mathfrak{n}} %
  <
 \frac{1}{n}\Bigl( k - \sum_{\vec{s}\in\mathbb{F}_q^{n-k}} P_{\!\mathcal{A}}(\vec{s})
  H_{q^{2k}[\log_q]}\bigl(
   \{ P_{\!\mathcal{A}}(\vec{l}^x,\vec{l}^z \vert \vec{s}) \}
 \bigr)    \Bigr),
\end{equation}
where $\{ P_{\!\mathcal{A}}(\vec{s}) \}$ and $\{ P_{\!\mathcal{A}}(\vec{l}^x,\vec{l}^z \vert \vec{s}) \}$ are defined by \eqref{eq:paulislxlz}.
\end{thm}
\begin{rem}[i]
Hamada shows the stronger result that one minus the minimum fidelity is upper bounded by epsilon, where epsilon drops exponentially in $N/k$ as long as condition \eqref{eq:concrate} is satisfied \cite{Ha02}.
\end{rem}
\begin{rem}[ii]
If the deterministic inner code is a CSS code,
we might concatenate it with random outer CSS codes as in theorem \ref{thm:randcss}.
Since the resulting code will also be a CSS code, this means we could achieve the rate in equation \eqref{eq:concrate} by using only CSS codes.
This result has been used by Lo \cite{Lo01} to improve the security proof of the 6-state protocol:
While the standard security proof obtains the maximum tolerable bit error rate from the hashing rate of theorem \ref{thm:randcss}, Lo used the CSS analog of theorem \ref{thm:concrate} to obtain a maximum tolerable bit error rate given by equation \eqref{eq:concrate}.
By using an inner CSS code whose stabilizer consists entirely of $Z$-type operators, the protocol remains to be reducible to a prepare and measure scheme.
(The inner code used by Lo is the so-called cat code which is treated in detail in section \ref{sec:concexamples}.)
\end{rem}

\subsection{Achievable Rate and Coherent Information}\label{subsec:ratecoinf}

In the preceding subsection we showed that by concatenating a deterministic inner code with a random outer one, we can achieve reliable quantum communication over a memoryless Pauli channel up to a rate given by theorem \ref{thm:concrate}.
We are now going to express this rate in terms of the coherent information of a maximally mixed state defined on one of the codespaces of the inner code.
This way a relationship with the quantum capacity $Q(\mathcal{A})$ of a memoryless Pauli channel $\mathcal{A}:\mathcal{S}(\mathcal{H})\rightarrow\mathcal{S}(\mathcal{H})$ is established,
which can be expressed as regularized coherent information~\eqref{eq:regcohinfo},
\begin{equation*}
 Q(\mathcal{A}) =  \lim_{n\rightarrow\infty} \frac{1}{n} \max_{\rho} I_c(\rho, \mathcal{A}^{\otimes n} ).
\end{equation*}
We prove the following theorem due to \cite[for $k=1$ and $q=2$]{DiSS98} and \cite{Ha02}:

\begin{thm}\label{thm:concrate=cinfo}
Concatenation of a random outer $[[N,K]]_q$ code with an inner $[[n,k]]_q$ code with stabilizer $L$ allows for reliable quantum communication over a Pauli channel defined by a probability distribution $P_{\!\mathcal{A}}$ on $\mathbb{F}_q^2$ as long as the total rate $K/\mathfrak{n}$ satisfies (theorem \ref{thm:concrate})
\begin{align}
 \frac{K}{\mathfrak{n}}      &<
 \frac{1}{n}\Bigl( k - \sum_{\vec{s}\in\mathbb{F}_q^{n-k}} P_{\!\mathcal{A}}(\vec{s})
  H_{q^{2k}[\log_q]}\bigl(
   \{ P_{\!\mathcal{A}}(\vec{l}^x,\vec{l}^z \vert \vec{s}) \}
 \bigr)    \Bigr).
\intertext{This rate can be expressed as the coherent information of a maximally mixed state defined on one of the codespaces $\mathcal{C}(L,\vec{s})$ of the inner code,
}
&= \frac{1}{n} I_c\Bigl( \frac{1}{q^k}\Pi_{\mathcal{C}(L,\vec{s})}, \mathcal{A}^{\otimes n}\Bigr).
\end{align}
\end{thm}
\begin{proof}
Let $\rho$ be a maximally mixed state defined on one of the $q^{n-k}$ codespaces $\mathcal{C}(L,\vec{s})$ of the inner code,
\begin{equation*}
\rho = \frac{1}{q^k} \Pi_{\mathcal{C}(L,\vec{s}_0)}
     = \frac{1}{q^k} \sum_{\vec{c} \in \mathbb{F}_q^k }
 \overline{\ket{\vec{s}_0,\vec{c}}}\overline{\bra{\vec{s}_0,\vec{c}}},
\end{equation*}
where $\Pi_{\mathcal{C}(L,\vec{s}_0)}$ denotes the projector on codespace $\mathcal{C}(L,\vec{s}_0)$,
and let
\begin{equation*}
 \ket{\psi_{\vec{s}_0}} = \frac{1}{\sqrt{q^k}} \sum_{\vec{c} \in \mathbb{F}_q^k } \overline{\ket{\vec{s}_0,\vec{c}}} \otimes \ket{\vec{c}}
\end{equation*}
be a corresponding purification of $\rho$.
Since the coherent information
$I_c( \rho, \mathcal{A}^{\otimes n})$
is defined as the difference between
$S( \mathcal{A}^{\otimes n}(\rho) )$ and
$S\bigl(\mathcal{A}^{\otimes n}\otimes\id (\ketbra{\psi_{\vec{s}_0}}{\psi_{\vec{s}_0}}) \bigr)$ in equation \eqref{eq:defcinfo},
we proceed by calculating these quantities.
We start with the von Neumann entropy of $\mathcal{A}^{\otimes n}(\rho)$:
By making use of the channel representation in equation \eqref{eq:paulirewritten}, we obtain
\begin{equation*}
 \rho = \frac{1}{q^k} \sum_{\vec{c} \in \mathbb{F}_q^k } \overline{\ket{\vec{s}_0,\vec{c}}}\overline{\bra{\vec{s}_0,\vec{c}}} \mapsto
\mathcal{A}^{\otimes n}(\rho) = \frac{1}{q^k} \sum_{\vec{s}\in\mathbb{F}_q^{n-k}} P_{\!\mathcal{A}}(\vec{s}) \Pi_{\mathcal{C}(L,\vec{s})},
\end{equation*}
and eventually
\begin{equation}\label{eq:cinfoprf1}
  S( \mathcal{A}^{\otimes n}(\rho) ) = k + H_{q^{n-k}[\log_q]}\bigl( \{P_{\!\mathcal{A}}(\vec{s})\} \bigr).
\end{equation}
To determine the von Neumann entropy resulting from a channel application to a purification of $\rho$,
we use again the channel representation in \eqref{eq:paulirewritten} and obtain
\begin{equation*}
 \mathcal{A}^{\otimes{n}} \otimes \id ( \ketbra{\psi_{\vec{s}_0}}{\psi_{\vec{s}_0}} ) =
 \sum_{\vec{s}\in\mathbb{F}_q^{n-k}} \sum_{\vec{l}=(\vec{l}^x,\vec{l}^z)\in\mathbb{F}_q^{2k}}
 P_{\!\mathcal{A}}(\vec{s},\vec{l})
 \bigl\vert \psi_{\vec{s}_0+\vec{s},\vec{l}^x}^{\vec{l}^z} \bigr\rangle%
 \bigl\langle \psi_{\vec{s}_0+\vec{s},\vec{l}^x}^{\vec{l}^z} \bigr\vert
\end{equation*}
with
\begin{equation*}
\bigl\vert \psi_{\vec{s}_0+\vec{s},\vec{l}^x}^{\vec{l}^z} \bigr\rangle = 
\frac{1}{\sqrt{q^k}} \sum_{\vec{c} \in \mathbb{F}_q^k } \omega^{\vec{c}\cdot\vec{l}^z} \overline{\ket{\vec{s}_0+\vec{s},\vec{c}+\vec{l}^x}} \otimes \ket{\vec{c}}.
\end{equation*}
One can easily check that the set of kets
\begin{equation*}
 \bigl\{ \bigl\vert \psi_{\vec{s}_0+\vec{s},\vec{l}^x}^{\vec{l}^z} \bigr\rangle
\sthat
 \vec{s}\in\mathbb{F}_q^{n-k}, \vec{l}=(\vec{l}^x,\vec{l}^z)\in\mathbb{F}_q^{2k} \bigr\}
\end{equation*}
forms an orthonormal basis of $\mathcal{H}^{\otimes(n+k)}$.
Therefore, the von Neumann entropy of $\mathcal{A}^{\otimes{n}} \otimes \id ( \ketbra{\psi_{\vec{s}_0}}{\psi_{\vec{s}_0}} )$ is given by the corresponding Shannon entropy,
\begin{equation}\label{eq:cinfoprf2}
  S\bigl( \mathcal{A}^{\otimes{n}} \otimes \id ( \ketbra{\psi_{\vec{s}_0}}{\psi_{\vec{s}_0}} ) \bigr) = H_{q^{n+k}[\log_q]}\bigl( \{P_{\!\mathcal{A}}(\vec{s},\vec{l})\} \bigr).
\end{equation}
The proof is finished by subtracting \eqref{eq:cinfoprf2} from \eqref{eq:cinfoprf1}.
\end{proof}

\section{Concatenated Codes and the Depolarizing Channel}\label{sec:concexamples}

The depolarizing channel is a special type of Pauli channel which is characterized by a single noise parameter $p \in [0,1]$.
It can be interpreted as a quantum channel which transmits a qudit of dimension $q$ unperturbed with probability $1-\tilde{p}=1-p q^2/(q^2-1)$, while exchanging it with the completely mixed state $\id/q$ with probability $\tilde{p}$.
By concatenating certain inner codes with random outer codes,
it was shown by Shor and Smolin in \cite{ShSm96}
(and later together with DiVincenzo in \cite{DiSS98})
that for very noisy depolarizing channels,
the hashing rate given by theorem \ref{thm:randstab}
(representing the achievable rate for reliable quantum communication using random codes alone)
can be exceeded by the rate in theorem \ref{thm:concrate}
(representing the achievable rate using concatenated codes).
In this section we present these inner codes and determine the resulting rates.
For the depolarizing channel, the maximum amount of noise $p_\text{max}$ is defined as the level of noise for which the quantum capacity becomes zero.
The best lower bound on $p_\text{max}$ of the qubit depolarizing channel known so far was found in \cite{SmSm07} by using an inner $[[5\times 16,1]]_2$ code.
We improve this bound by presenting the results of numerical calculations up to an inner $[[5\times 22,1]]_2$ code.

First, we define the depolarizing channel in subsection \ref{subsec:depola}.
Then, in subsection \ref{subsec:review}, we briefly review how the action of a Pauli channel is rewritten for a fixed (inner) code as it was done in subsection \ref{subsec:rateconc}.
Subsection \ref{subsec:thecatcode} presents the so-called cat code, the inner code used in \cite{ShSm96} and \cite{DiSS98}.
The succeeding subsection \ref{subsec:theconcatcode} deals with the concatenated cat code of \cite{DiSS98} and \cite{SmSm07}.
This code results from concatenating an outer 'flipped'-type cat code with an inner 'standard' cat code and leads to the best known lower bound on the maximum tolerable noise $p_\text{max}$ of the qubit depolarizing channel.

\subsection{Depolarizing Channel}\label{subsec:depola}

\begin{defi}\label{def:depola}
The depolarizing channel
$\mathcal{D}_p : \mathcal{S}(\mathcal{H}) \rightarrow \mathcal{S}(\mathcal{H})$
is a Pauli channel between density operators on a $q$-dimensional Hilbert space $\mathcal{H}$,
whose probability distribution on $\mathbb{F}_q^2$ is characterized by a single parameter $p\in[0,1]$:
\begin{equation}
 \mathcal{D}_p :   \rho \mapsto \mathcal{D}_p(\rho) =
 (1-p)\rho +
 \sum_{\vec{e}\in\mathbb{F}^2_q}^{\vec{e}\neq (0,0)} \frac{p}{q^2-1} \, \XZ(\vec{e}) \rho \XZ(\vec{e})^\dagger.
\end{equation}
\end{defi}

\begin{rem}
The depolarizing channel $\mathcal{D}_p$ can be written as
\begin{equation}\label{eq:alternate-depola}
 \mathcal{D}_{\tilde{p}} :   \rho \mapsto \mathcal{D}_{\tilde{p}}(\rho) =
 (1-\tilde{p})\cdot \rho + \tilde{p}\cdot \frac{1}{q} \id,
\end{equation}
with  $\tilde{p} = p\cdot q^2/(q^2-1)$ by using the fact that
\begin{equation}
 \frac{1}{q^2} \sum_{\vec{e}\in\mathbb{F}^2_q} \XZ(\vec{e}) \rho \XZ(\vec{e})^\dagger = \frac{1}{q}\id,
\end{equation}
for any normalized $\rho\in\mathcal{S}(\mathcal{H})$.
\end{rem}

The highest value of $p$ up to which the quantum capacity of the depolarizing channel $\mathcal{D}_p$ remains non-zero is defined as the channels maximal tolerable level of noise $p_\text{max}$,
\begin{equation}
  Q( \mathcal{D}_{p_\text{max}} ) = 0 . %
\end{equation}
For the qubit depolarizing channel ($q=2$) we get a lower bound on $p_\text{max}$ by calculating the value of $p$ for which the hashing rate of theorems \ref{thm:randstab} and \ref{thm:randcss} becomes zero.
We obtain $p_\text{max} >  p_\text{max}^\text{hash} = 18.9290\%$.

While taking the limit as the number of channel uses goes to infinity prevents us from calculating the quantum capacity of the depolarizing channel using the regularized coherent information in equation~\eqref{eq:regcohinfo},
\begin{equation}%
 Q( \mathcal{D}_p ) =  \lim_{n\rightarrow\infty} \frac{1}{n} \max_{\rho} I_c(\rho, \mathcal{D}_p^{\otimes n} ),
\end{equation}
we are going to calculate the one-shot capacity of the qubit depolarizing channel,
\begin{equation}
 Q^{(1)}(\mathcal{D}_p) =  \max_{\rho} I_c(\rho, \mathcal{D}_p ).
\end{equation}
\begin{lem}\label{lem:oneshotdepola}
The one-shot capacity of the qubit depolarizing channel is given by
\begin{equation}
 Q^{(1)}(\mathcal{D}_p) = 1 - H_{4[\log_2]}\bigl( \{ 1-p,p/3,p/3,p/3 \}\bigr),
\end{equation}
which equals the hashing rate of theorem \ref{thm:randstab}.
\end{lem}
\begin{proof}
The representation of the depolarizing channel in \eqref{eq:alternate-depola} shows us that the depolarizing channel does not depend on the basis in which the Pauli operators $\XZ(\cdot)$ are defined.
Therefore, we can assume without restriction of any kind that the state $\rho$ which maximizes $ Q^{(1)}(\mathcal{D}_p) $ is given by
$\rho = c \ketbra{0}{0} + (1-c) \ketbra{1}{1}$, where $\{ \ket{i} \}_{i=0,1}$ is the basis of the qubit Hilbert space which defines the Pauli operators (i.\,e. $Z\ket{1}=-\ket{1}$ for example).
A purification of $\rho$ is given by $\ket{\psi} = \sqrt{c}\ket{0}\otimes\ket{0} + \sqrt{1-c}\ket{1}\otimes\ket{1}$.
Now we follow the proof given in \cite[section V]{AC97} which shows by a straightforward calculation of
\begin{equation}
 f( c,p ) = I_c(\rho, \mathcal{D}_p ) = S( \mathcal{D}_p(\rho) ) - S( [\mathcal{D}_p\otimes\id](\ketbra{\psi}{\psi}) )
\end{equation}
that for all values of $p$ the maximum of $f( c,p )$ is obtained for $c=1/2$.
\end{proof}

\subsection{Pauli Channel Representation for a CSS Code}\label{subsec:review}

In subsection \ref{subsec:rateconc} we determined the achievable rate for reliable quantum communication over a memoryless Pauli channel for a concatenated code whose outer code is chosen at random.
We repeat briefly how we rewrote the action of a Pauli channel $\mathcal{A}$ with probability distribution $\{ P_{\!\mathcal{A}}(a^x,a^z) \}$, $(a^x,a^z)\in\mathbb{F}_q^2$, for some given inner $[[n,k]]_q$ code to arrive at a channel with probability distribution $\{ P_{\!\mathcal{A}}(\vec{l},\vec{s}) \}$.
Since all inner codes considered in this section are CSS codes, this time we specialize in an inner CSS code.

As discussed in section \ref{subsec::cssencod},
an $[[n,k]]_q$ CSS code together with an encoding may be specified by two bases of~$\mathbb{F}_q^n$,
\begin{subequations}\label{eq:pcreviewbases}
\begin{align}
\mathbb{F}_q^n &= \vspan\{\vec{\xi}^z_1,\dots,\vec{\xi}^z_{n-k_1}, \vec{\eta}^z_1,\dots,\vec{\eta}^z_{k_2},
        \vec{\mu}^z_1,\dots,\vec{\mu}^z_k\} \\
\text{ and } \mathbb{F}_q^n &= \vspan\{\vec{\eta}^x_1,\dots,\vec{\eta}^x_{n-k_1}, \vec{\xi}^x_1,\dots,\vec{\xi}^x_{k_2},
         \vec{\mu}^x_1,\dots,\vec{\mu}^x_k\},
\end{align}
\end{subequations}
with $k=k_1-k_2$, fulfilling conditions \eqref{eq:css:nineconditions}.
By writing the $x$-component [$z$-component] of a vector
$\vec{a} = ( a_1^x,\dots,a_n^x, a_1^z,\dots,a_n^z ) \in\mathbb{F}_q^{2n}$
as linear combination of the basis elements of the
$\{\vec{\xi}^z_1\dots, \vec{\eta}^z_1\dots, \vec{\mu}^z_1\dots\}$
[$\{\vec{\eta}^x_1\dots, \vec{\xi}^x_1\dots, \vec{\mu}^x_1\dots\}$] basis,
we obtained~\eqref{eq:cssdecompoa},
\begin{align}\label{eq:cssdecompoaDie2te}
 \vec{a}^x &=
     \sum_{i=1}^{n-k_1} s_i^x \vec{\eta}^x_i +
     \sum_{j=1}^{k_2}   n_j^z \vec{\xi}^x_j  +
     \sum_{r=1}^k       l_r^x \vec{\mu}^x_r \\
 \vec{a}^z &=
     \sum_{i=1}^{n-k_1} n_i^x \vec{\xi}^z_i +
     \sum_{j=1}^{k_2}   s_j^z \vec{\eta}^z_j  +
     \sum_{r=1}^k       l_r^z \vec{\mu}^z_r,
\end{align}
with
$s_i^x = \vec{\xi}_i^z \cdot \vec{a}^x$,
$n_j^z = \vec{\eta}_j^z \cdot \vec{a}^x$,
$l_r^x = \vec{\mu}_r^z \cdot \vec{a}^x$
and
$n_i^x = \vec{\eta}_i^x \cdot \vec{a}^z$,
$s_j^z = \vec{\xi}_j^x \cdot \vec{a}^z$,
$l_r^z = \vec{\mu}_r^x \cdot \vec{a}^z$,
which led to lemma \ref{lem:xzdecompocss}, relating the corresponding Pauli operators:
\begin{equation}
XZ(\vec{a})
\sim
\overline{X}^{(\vec{s}^x,\vec{s}^z,\vec{l}^x)}  \overline{Z}^{(\vec{n}^x,\vec{n}^z,\vec{l}^z)}.
\end{equation}
This relation allows us to rewrite the action of a memoryless Pauli channel $\mathcal{A}^{\otimes n}$ with
probability distribution
$P^n_{\!\mathcal{A}}(\vec{a}=(a^x_1,\dots,a^x_n,a^z_1,\dots,a^z_n)) =
 \prod_{i=1}^n P_{\!\mathcal{A}}(a^x_i,a^z_i)$ as
\begin{align}
 \mathcal{A}^{\otimes n}(\rho)
&= \sum_{\vec{a}\in\mathbb{F}^{2n}_q} P^n_{\!\mathcal{A}}(\vec{a}) \, \XZ(\vec{a}) \rho  \XZ(\vec{a})^\dagger \nonumber\\
&=\sum_{\vec{s},\vec{n}\in\mathbb{F}_q^{n-k}} \sum_{\vec{l}\in\mathbb{F}_q^{2k}}
P_{\!\mathcal{A}}(\vec{s},\vec{n},\vec{l}) \,
\bigl(\overline{X}^{(\vec{s}^x,\vec{s}^z,\vec{l}^x)}  \overline{Z}^{(\vec{n}^x,\vec{n}^z,\vec{l}^z)}\bigr)
\rho \,
\bigl(\overline{X}^{(\vec{s}^x,\vec{s}^z,\vec{l}^x)}  \overline{Z}^{(\vec{n}^x,\vec{n}^z,\vec{l}^z)}\bigr)^\dagger,
\end{align}
with
$\vec{s} = (\vec{s}^x\in\mathbb{F}_q^{n-k_1}, \vec{s}^z\in\mathbb{F}_q^{k_2}) \in \mathbb{F}_q^{n-k}$,
$\vec{n} = (\vec{n}^x\in\mathbb{F}_q^{n-k_1}, \vec{n}^z\in\mathbb{F}_q^{k_2}) \in \mathbb{F}_q^{n-k}$ and
$\vec{l} = (\vec{l}^x\in\mathbb{F}_q^k,\vec{l}^z\in\mathbb{F}_q^k) \in \mathbb{F}_q^{2k}$.
In addition to
$P_{\!\mathcal{A}}(\vec{s},\vec{n},\vec{l}) = P^n_{\!\mathcal{A}}(\vec{a})$ we define the probabilities
\begin{align}\label{eq:paulislxlzDie2te}
 P_{\!\mathcal{A}}(\vec{s},\vec{l}) &=
 \sum_{\vec{n}\in\mathbb{F}_q^{n-k}} P_{\!\mathcal{A}}(\vec{s},\vec{n},\vec{l}), &
  P_{\!\mathcal{A}}(\vec{s}) &=
 \sum_{\vec{l}\in\mathbb{F}_q^{2k}} P_{\!\mathcal{A}}(\vec{s},\vec{l}), &
 P_{\!\mathcal{A}}(\vec{l}\vert\vec{s}) &= P_{\!\mathcal{A}}(\vec{s},\vec{l}) / P_{\!\mathcal{A}}(\vec{s}),
\end{align}
as in equation \eqref{eq:paulislxlz}.
The achievable rate $R$ for reliable quantum communication over a memoryless Pauli channel characterized by $\{ P_{\!\mathcal{A}}(a^x,a^z) \}$, $(a^x,a^z)\in\mathbb{F}_q^2$, is given by theorem \ref{thm:concrate}:
\begin{equation}\label{eq:rofconcthm}
 R = \frac{1}{n}\Bigl( k - \sum_{\vec{s}\in\mathbb{F}_q^{n-k}} P_{\!\mathcal{A}}(\vec{s})
  H_{q^{2k}[\log_q]}\bigl(
   \{ P_{\!\mathcal{A}}(\vec{l} \vert \vec{s}) \}
 \bigr)    \Bigr).
\end{equation}
To determine $R$ for the inner $[[n,k,]]_q$ code specified by \eqref{eq:pcreviewbases},
we obviously have to know the corresponding probability distributions
$\{ P_{\!\mathcal{A}}(\vec{s}) \}$ and $\{ P_{\!\mathcal{A}}(\vec{l} \vert \vec{s}) \}$.
In the following subsections we determine these distributions for various inner codes.

\subsection{The Cat Code}\label{subsec:thecatcode}

\begin{figure}
\begin{minipage}{0.49\textwidth}\centering
\subfloat['Standard' cat code]
['Standard' cat code. The operators on the left hand side are
$\{ \XZ(\vec{0},\vec{\xi}^z_i) \}$ ($i=1\dots m-1$ from top to bottom)
and $\XZ(\vec{0},\vec{\mu}^z)$, those on the right hand side are
$\{ \XZ(\vec{\eta}^x_i,\vec{0}) \}$ and $\XZ(\vec{\mu}^x,\vec{0})$.]{\label{fig:standard-cat-code}
\scalebox{0.7}{
\begin{pspicture}(-1.5,0.5)(7.25,-2.75)
%
\psframe[linecolor=lightgray,linestyle=dotted](-0.25,0.25)(2.75,-1.75) %
\pspolygon[linecolor=lightgray,linestyle=dashed](-0.5,0.5)(3,0.5)(3,-1.75)(6.25,-1.75)(6.25,-2.75)(-0.5,-2.75) %
\psframe[linecolor=lightgray,fillstyle=solid](0,0)(2.5,-1.5)
\rput[c](1.25,-0.75){
$\begin{array}{cccc}
 Z & Z & \id & \id \\
 Z & \id & Z & \id \\
 Z & \id & \id & Z
 \end{array}$}
\psframe[linecolor=lightgray,fillstyle=solid](0,-2)(2.5,-2.5)
\rput[c](1.25,-2.25){
$\begin{array}{cccc}
  Z & \id & \id & \id
 \end{array}$}
\psframe[linecolor=lightgray,fillstyle=solid,fillcolor=mygray](3.5,0)(6,-1.5)
\rput[c](4.75,-0.75){
$\begin{array}{cccc}
 \id & X & \id & \id \\
 \id & \id & X & \id \\
  \id & \id & \id & X
 \end{array}$}
\psframe[linecolor=lightgray,fillstyle=solid,fillcolor=mygray](3.5,-2)(6,-2.5)
\rput[c](4.75,-2.25){
$\begin{array}{cccc}
 X & X & X & X
 \end{array}$}
\rput[r](-0.05,-0.75){$m-1 \left\{ \makebox(0,0.9)[b]{}\right.$}
\rput[B](1.25,0.6){$m$}
\rput[B](1.25,0.1){\rotatebox[origin=c]{-90}{$\left\{ \makebox(0,1.35)[b]{}\right.$}}
\end{pspicture}}}%
\end{minipage}%
\begin{minipage}{0.49\textwidth}\centering
\subfloat['Flipped' cat code]
['Flipped' cat code. The operators on the left hand side are
$\{ \XZ(\vec{\xi}^x_i,\vec{0}) \}$ ($i=1\dots m-1$ from top to bottom)
and $\XZ(\vec{0},\vec{\mu}^z)$, those on the right hand side are
$\{ \XZ(\vec{0},\vec{\eta}^z_i) \}$ and $\XZ(\vec{\mu}^x,\vec{0})$.]{\label{fig:flipped-cat-code}
\scalebox{0.7}{
\begin{pspicture}(-1.5,0.5)(7.25,-3.0)
\psframe[linecolor=lightgray,linestyle=dotted](-0.25,0.25)(2.75,-1.75) %
\pspolygon[linecolor=lightgray,linestyle=dashed](-0.5,0.5)(3,0.5)(3,-1.75)(6.25,-1.75)(6.25,-2.75)(-0.5,-2.75) %
\psframe[linecolor=lightgray,fillstyle=solid,fillcolor=mygray](0,0)(2.5,-1.5)
\rput[c](1.25,-0.75){
$\begin{array}{cccc}
 X & X & \id & \id \\
 X & \id & X & \id \\
 X & \id & \id & X
 \end{array}$}
\psframe[linecolor=lightgray,fillstyle=solid](0,-2)(2.5,-2.5)
\rput[c](1.25,-2.25){
$\begin{array}{cccc}
  Z & Z & Z & Z
 \end{array}$}
\psframe[linecolor=lightgray,fillstyle=solid](3.5,0)(6,-1.5)
\rput[c](4.75,-0.75){
$\begin{array}{cccc}
 \id & Z & \id & \id \\
 \id & \id & Z & \id \\
  \id & \id & \id & Z
 \end{array}$}
\psframe[linecolor=lightgray,fillstyle=solid,fillcolor=mygray](3.5,-2)(6,-2.5)
\rput[c](4.75,-2.25){
$\begin{array}{cccc}
 X & \id & \id & \id
 \end{array}$}
\rput[r](-0.05,-0.75){$m-1 \left\{ \makebox(0,0.9)[b]{}\right.$}
\rput[B](1.25,0.6){$m$}
\rput[B](1.25,0.1){\rotatebox[origin=c]{-90}{$\left\{ \makebox(0,1.35)[b]{}\right.$}}
\end{pspicture}}}%
\end{minipage}
\caption[Encoded Pauli operators corresponding to various cat codes.]{\label{fig:cat-code}
The encoded Pauli operators corresponding to a certain encoding of
(a) the 'standard' cat code and
(b) the 'flipped' cat code, both encoding one qubit into $m=4$.
Operators corresponding to (generators of) the stabilizer are within the dotted line,
operators corresponding to (generators of) the normalizer within the dashed one.}
\end{figure}
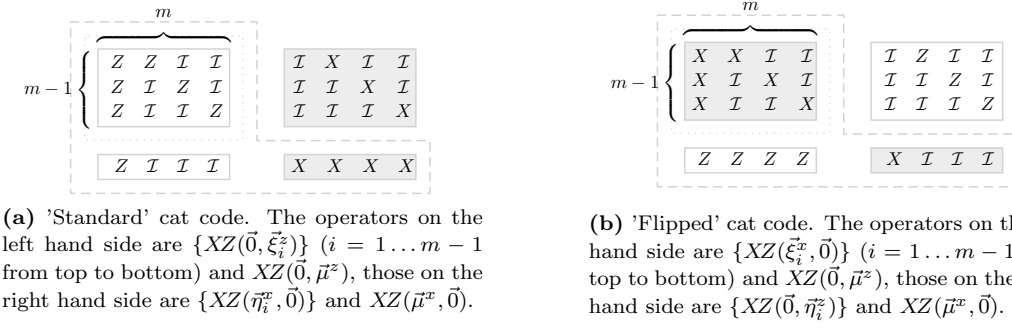

The cat code used in \cite{ShSm96,DiSS98} is an $[[m,k=1]]_2$ CSS code with $k=k_1=1$ and $k_2=0$.
It is specified by a classical code $\mathcal{C}_1^
\perp = \vspan \{ \vec{\xi}^z_1,\dots,\vec{\xi}^z_{m-k_1} \}$,
where the entries of the vector $\vec{\xi}^z_i$ are given by
$(\vec{\xi}^z_i)_j = \delta_{j,1}+\delta_{j,i+1}$ for $j\in\{1,\dots,m\}$.
The corresponding stabilizer is $L = \vspan \{ (\vec{0},\vec{\xi}^z_1),\dots,(\vec{0},\vec{\xi}^z_{m-1}) \}$.
We consider an extension to the bases
\begin{equation}
\begin{split}
            \mathbb{F}_2^n&=\{ \vec{\xi}^z_1,\dots,\vec{\xi}^z_{m-1} \,,\, \vec{\mu}^z \} \\
\text{and } \mathbb{F}_2^n&=\{ \vec{\eta}^x_1,\dots,\vec{\eta}^x_{m-1} \,,\, \vec{\mu}^x \}
\end{split}
\end{equation}
as shown in figure \ref{fig:cat-code}\subref{fig:standard-cat-code}.
To construct the encoding $U_\text{enc}$ associated with these extensions, we set the phase factors $\theta_z(\cdot)$ and $\theta_x(\cdot)$ equal to one as it was done in subsection \ref{subsec::cssencod}.
Then, the corresponding encoded states of equation \eqref{eq:cssencodedstates} become
\begin{equation}
 \overline{ \ket{ \vec{s}^x,l^x } }
= \overline{X}^{(\vec{s}^x,l^x)} \overline{\ket{0\dots 0}}\\
= \bigl\vert\ l^x\cdot\vec{\mu}^x + \sum_{i=1}^{m-1} s^x_i\cdot \vec{\eta}^x_i \ \bigr\rangle.
\end{equation}
The code is called cat code because a pure one qubit state $\alpha\ket{0} + \beta\ket{1}$ encoded in the codespace $\mathcal{C}(L,\vec{0})$ becomes a cat state,
\begin{equation}
 U_\text{enc} \ket{\vec{0}}\otimes \bigl( \alpha \ket{0} + \beta \ket{1} \bigr) = \alpha \ket{0,\dots,0,0} + \beta \ket{1,\dots,1,1}.
\end{equation}
We do not calculate the probability distributions
$\{ P_{\!\mathcal{A}}(\vec{s}) \}$ and $\{ P_{\!\mathcal{A}}(\vec{l} \vert \vec{s}) \}$ for the cat code, since they emerge as a special case of the corresponding distributions of the concatenated cat code in subsection \ref{subsec:theconcatcode} (see equation \eqref{eq:probcatcode}).

\begin{figure}
\centering
\includegraphics[scale=0.9]{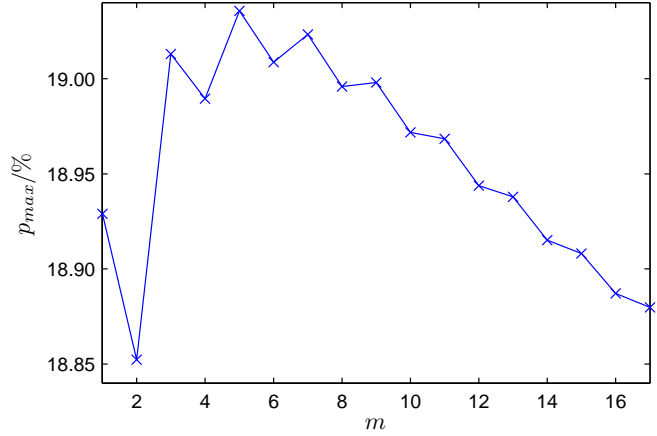}
\caption{\label{fig:pmaxcatcode}
The maximum tolerable value of noise $p$ for the qubit depolarizing channel $\mathcal{D}_p$ as a function of the size of the inner $[[m,1]]_2$ cat code.
The highest value $p_\text{max}^\text{cat}(m=5) = 19.0356\%$ is obtained for $m=5$.
}
\end{figure}

The cat code improves the hashing rate lower bound
$p_\text{max}^\text{hash} = 18.9290\%$
on the maximum tolerable level of noise $p_\text{max}$ of the qubit depolarizing channel.
By setting the rate $R_m(p)$ of \eqref{eq:rofconcthm} for a cat code of size $m$ equal to zero,
we obtain the values $p_\text{max}^\text{cat}(m)$ shown in figure \ref{fig:pmaxcatcode}.
The highest value (and therefore the best lower bound on $p_{\text{max}}$) is obtained for $m=5$,
$p_\text{max}^\text{cat}(m=5) = 19.0356\%$.
The rates $R_m(p)$ for $m=3$ and $m=5$ are shown in figure \ref{fig:catcoderates} \textit{(blue)}.
\begin{rem}
As discussed in the remark following theorem \ref{thm:concrate},
the value $p_\text{max}^\text{cat}(m=5)$ was used by Lo in \cite{Lo01} to improve the security of 6-state quantum key distribution protocol.
Since the 6-state protocol corresponds to a qubit depolarizing channel $\mathcal{D}_{\frac{3}{2}p}$, he improved the maximum tolerable bit error rate of the 6-state protocol from
$\frac{2}{3} \cdot p_\text{max}^\text{hash} = 12.6193\%$ to
$\frac{2}{3} \cdot p_\text{max}^\text{cat}(m=5) = 12.6904\%$.
\end{rem}

The concatenated cat code presented in subsection \ref{subsec:theconcatcode} is obtained by concatenating an inner cat code with an outer 'flipped' version of the cat code.
We proceed by presenting this 'flipped' cat code, whose stabilizer is obtained from the stabilizer of the 'standard' cat code described above by exchanging the $Z$ operators with $X$ operators.
\subsubsection{'Flipped' Cat Code}
The 'flipped' cat code is an $[[m,k=1]]_2$ CSS code with $k_1=m$, $k_2=m-1$ and $k=k_1-k_2=1$.
It is specified by a classical code $\mathcal{C}_2 = \vspan \{ \vec{\xi}^x_1,\dots,\vec{\xi}^x_{k_2} \}$,
where the entries of the vector $\vec{\xi}^x_i$ are given by
$(\vec{\xi}^x_i)_j = \delta_{j,1}+\delta_{j,i+1}$ for $j\in\{1,\dots,m\}$.
The corresponding stabilizer is $L = \vspan \{ (\vec{\xi}^x_1,\vec{0}),\dots,(\vec{\xi}^x_{m-1},\vec{0}) \}$.
We consider an extension to the bases
\begin{equation}
\begin{split}
            \mathbb{F}_2^n&=\{ \vec{\eta}^z_1,\dots,\vec{\eta}^z_{m-1} \,,\, \vec{\mu}^z \} \\
\text{and } \mathbb{F}_2^n&=\{ \vec{\xi}^x_1,\dots,\vec{\xi}^x_{m-1} \,,\, \vec{\mu}^x \}
\end{split}
\end{equation}
as shown in figure \ref{fig:cat-code}\subref{fig:flipped-cat-code}.
To construct the encoding $U_\text{enc}$ associated with these extensions, we set the phase factors $\theta_z(\cdot)$ and $\theta_x(\cdot)$ equal to one as it was done in subsection \ref{subsec::cssencod}.
Then, the corresponding encoded states of equation \eqref{eq:cssencodedstates} become
\begin{equation}
 \overline{ \ket{ \vec{s}^z,\vec{l}^x } }
= \overline{X}^{(\vec{s}^z,\vec{l}^x)} \overline{\ket{0\dots 0}}
= \frac{1}{\sqrt{\vert \mathcal{C}_2\vert}} \sum_{\vec{\mathfrak{v}}\in\mathcal{C}_2}
  \omega^{\vec{\mathfrak{z}}\cdot\vec{\mathfrak{v}}} \ket{\ \vec{\mathfrak{v}} + l^x\cdot\vec{\mu}^x \ },
\text{ with } \vec{\mathfrak{z}} = \sum_{i=1}^{m-1} s^z_i \vec{\eta}^z_i.
\end{equation}

\subsection{The Concatenated Cat Code}\label{subsec:theconcatcode}

\begin{figure}
\centering
\scalebox{0.7}{
\begin{pspicture}(-1.5,0.5)(16.25,-6.75)
%

\psframe[linecolor=lightgray,fillstyle=solid](0,0)(7.5,-4.5)  %
\psframe[linecolor=lightgray,fillstyle=solid,fillcolor=mygray](8.5,0)(16,-4.5) %

\psframe[linecolor=lightgray,linestyle=dotted](-0.25,0.25)(7.75,-5.75) %
\pspolygon[linecolor=lightgray,linestyle=dashed](-0.5,0.5)(8,0.5)(8,-5.75)(16.25,-5.75)(16.25,-6.75)(-0.5,-6.75) %
\psframe[linecolor=lightgray,fillstyle=solid](0,0)(2.5,-1.5)
\rput[c](1.25,-0.75){
$\begin{array}{cccc}
 Z & Z & \id & \id \\
 Z & \id & Z & \id \\
 Z & \id & \id & Z
 \end{array}$}
\psframe[linecolor=lightgray,fillstyle=solid](2.5,-1.5)(5,-3)
\rput[c](3.75,-2.25){
$\begin{array}{cccc}
 Z & Z & \id & \id \\
 Z & \id & Z & \id \\
 Z & \id & \id & Z
 \end{array}$}
\psframe[linecolor=lightgray,fillstyle=solid](5,-3)(7.5,-4.5)
\rput[c](6.25,-3.75){
$\begin{array}{cccc}
 Z & Z & \id & \id \\
 Z & \id & Z & \id \\
 Z & \id & \id & Z
 \end{array}$}
\psframe[linecolor=lightgray,fillstyle=solid,fillcolor=mygray](0,-4.5)(2.5,-5.5)
\rput[c](1.25,-5){
$\begin{array}{cccc}
 X & X & X & X \\
 X & X & X & X
 \end{array}$}
\psframe[linecolor=lightgray,fillstyle=solid,fillcolor=mygray](2.5,-4.5)(5,-5.5)
\rput[c](3.75,-5){
$\begin{array}{cccc}
 X & X & X & X \\
 \id & \id & \id &\id
 \end{array}$}
\psframe[linecolor=lightgray,fillstyle=solid,fillcolor=mygray](5,-4.5)(7.5,-5.5)
\rput[c](6.25,-5){
$\begin{array}{cccc}
 \id & \id & \id &\id \\
 X & X & X & X
 \end{array}$}

\psframe[linecolor=lightgray,fillstyle=solid](0,-6)(2.5,-6.5)
\rput[c](1.25,-6.25){
$\begin{array}{cccc}
  Z & \id & \id & \id
 \end{array}$}
\psframe[linecolor=lightgray,fillstyle=solid](2.5,-6)(5,-6.5)
\rput[c](3.75,-6.25){
$\begin{array}{cccc}
 Z & \id & \id & \id
 \end{array}$}
\psframe[linecolor=lightgray,fillstyle=solid](5,-6)(7.5,-6.5)
\rput[c](6.25,-6.25){
$\begin{array}{cccc}
 Z & \id & \id & \id
 \end{array}$}

\psframe[linecolor=lightgray,fillstyle=solid,fillcolor=mygray](8.5,0)(11,-1.5)
\rput[c](9.75,-0.75){
$\begin{array}{cccc}
 \id & X & \id & \id \\
 \id & \id & X & \id \\
  \id & \id & \id & X
 \end{array}$}
\psframe[linecolor=lightgray,fillstyle=solid,fillcolor=mygray](11,-1.5)(13.5,-3)
\rput[c](12.25,-2.25){
$\begin{array}{cccc}
 \id & X & \id & \id \\
 \id & \id & X & \id \\
  \id & \id & \id & X
 \end{array}$}
\psframe[linecolor=lightgray,fillstyle=solid,fillcolor=mygray](13.5,-3)(16,-4.5)
\rput[c](14.75,-3.75){
$\begin{array}{cccc}
 \id & X & \id & \id \\
 \id & \id & X & \id \\
  \id & \id & \id & X
 \end{array}$}
\psframe[linecolor=lightgray,fillstyle=solid](8.5,-4.5)(11,-5.5)
\rput[c](9.75,-5){
$\begin{array}{cccc}
 \id & \id & \id & \id \\
 \id & \id & \id & \id
 \end{array}$}
\psframe[linecolor=lightgray,fillstyle=solid](11,-4.5)(13.5,-5.5)
\rput[c](12.25,-5){
$\begin{array}{cccc}
 Z & \id & \id & \id \\
 \id & \id & \id & \id
 \end{array}$}
\psframe[linecolor=lightgray,fillstyle=solid](13.5,-4.5)(16,-5.5)
\rput[c](14.75,-5){
$\begin{array}{cccc}
 \id & \id & \id & \id \\
 Z & \id & \id & \id
 \end{array}$}

\psframe[linecolor=lightgray,fillstyle=solid,fillcolor=mygray](8.5,-6)(11,-6.5)
\rput[c](9.75,-6.25){
$\begin{array}{cccc}
 X & X & X & X
 \end{array}$}
\psframe[linecolor=lightgray,fillstyle=solid,fillcolor=mygray](11,-6)(13.5,-6.5)
\rput[c](12.25,-6.25){
$\begin{array}{cccc}
 \id & \id & \id & \id
 \end{array}$}
\psframe[linecolor=lightgray,fillstyle=solid,fillcolor=mygray](13.5,-6)(16,-6.5)
\rput[c](14.75,-6.25){
$\begin{array}{cccc}
 \id & \id & \id & \id
 \end{array}$}

\rput[r](-0.05,-0.75){$m_1-1 \left\{ \makebox(0,0.9)[b]{}\right.$}
\rput[r](-0.05,-5){$m_2-1 \left\{ \makebox(0,0.6)[b]{}\right.$}
\rput[B](1.25,0.6){$m_1$}
\rput[B](1.25,0.1){\rotatebox[origin=c]{-90}{$\left\{ \makebox(0,1.35)[b]{}\right.$}}
\end{pspicture}}%
\caption[Encoded Pauli operators corresponding to the concatenated cat code.]{\label{fig:concat}
The encoded Pauli operators corresponding to a hyperbolic basis of the concatenated cat code.
Here $m_1=4$ and $m_2=3$, so that one qubit is encoded into $n=m_1\times m_2$.
The first $m_2\times (m_1-1)$ operators on the left hand side are the $\{ \XZ(\vec{0},\vec{\xi}^z_i) \}$,
the next $m_2-1$ the $\{ \XZ(\vec{\xi}^x_i,\vec{0}) \}$ and the last one is $\XZ(\vec{0},\vec{\mu}^z)$.
Those on the right hand side are $\{ \XZ(\vec{\eta}^x_i,\vec{0}) \}$, $\{ \XZ(\vec{0},\vec{\eta}^z_i) \}$ and $\XZ(\vec{\mu}^x,\vec{0})$ accordingly.}
\end{figure}
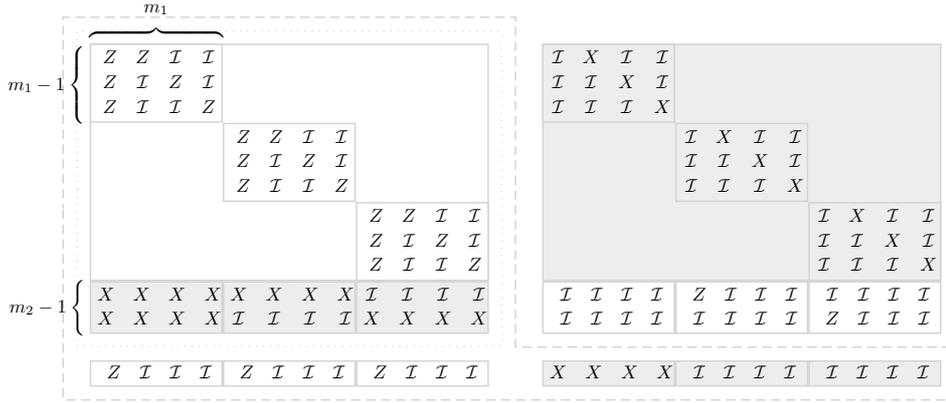

By concatenating an outer $[[m_2,1]]_q$ 'flipped' cat code with an inner $[[m_1,1]]_2$ 'standard' cat code,
we obtain the $[[m_1\times m_2,1]]_2$ code used in \cite{DiSS98,SmSm07} and shown in figure \ref{fig:concat}.
The $[[m_1\times m_2,1]]_2$ code is a CSS code with parameters $n=m_1m_2$, $k_1=m_2$, $k_2=m_2-1$ and $k=k_1-k_2=1$.
We proceed by calculating the corresponding probability distributions
$\{ P_{\!\mathcal{A}}(\vec{s}) \}$ and $\{ P_{\!\mathcal{A}}(\vec{l} \vert \vec{s}) \}$ which allow us to evaluate the achievable transmission rate given in equation \eqref{eq:rofconcthm}.

\subsubsection{Joint Probabilities of Logical Errors and Syndrome}

We are going to calculate the joint probabilities
$\{  P_{\!\mathcal{A}}(\vec{l} , \vec{s}) \}$ with $\vec{l}\in\mathbb{F}_2^2$ and $\vec{s}\in\mathbb{F}_2^{m_1m_2-1}$
defined in equation \eqref{eq:paulislxlzDie2te}
for the $[[m_1\times m_2,1]]_2$ code described above.
From $\{  P_{\!\mathcal{A}}(\vec{l} , \vec{s}) \}$ we will obtain
$\{ P_{\!\mathcal{A}}(\vec{s}) \}$ by summation over $\vec{l}$
and $\{ P_{\!\mathcal{A}}(\vec{l} \vert \vec{s}) \}$ by
$P_{\!\mathcal{A}}(\vec{l} \vert \vec{s}) = P_{\!\mathcal{A}}(\vec{l} , \vec{s})/P_{\!\mathcal{A}}(\vec{s})$.
Neither the details of these calculations nor formulas expressing the resulting probabilities have been presented in the literature \cite{DiSS98,SmSm07}.

We denote the elements of the probability distribution
$\{ P_{\!\mathcal{A}}(a^x,a^z) \}$, $(a^x,a^z)\in\mathbb{F}_2^2$, of the qubit Pauli channel $\mathcal{A}$ as $\{ p_e,p_x,p_y,p_z \}$, i.\,e.
\begin{equation}
 \mathcal{A}(\rho) = p_e \rho + p_x X\rho X^\dagger + p_y (\XZ)\rho (\XZ)^\dagger + p_z Z\rho Z^\dagger.
\end{equation}
Then, by definition,
$P_{\!\mathcal{A}}(\vec{s},\vec{n},\vec{l}) = P^n_{\!\mathcal{A}}(\vec{a})$,
where $\vec{a}\in\mathbb{F}_2^{2n}$ depends on $\vec{s},\vec{n},\vec{l}$ via the bases decomposition given in equation \eqref{eq:cssdecompoaDie2te}.
$P_{\!\mathcal{A}}(\vec{s},\vec{l})$ was defined in equation \eqref{eq:paulislxlzDie2te} as
\begin{equation}\label{eq:calcconcPA}
 P_{\!\mathcal{A}}(l^x,l^z, \vec{s}^x,\vec{s}^z) =
 \sum_{\vec{n}^x\in\mathbb{F}_q^{n-k_1}}
 \sum_{\vec{n}^z\in\mathbb{F}_q^{k_2}}
 P_{\!\mathcal{A}}(\vec{s},\vec{n},\vec{l}).
\end{equation}
Do we have to calculate this sum for all $2^{m_1m_2-1}$ distinct syndromes
$\vec{s}=(\vec{s}^x\in\mathbb{F}_2^{m_2(m_1-1)} ,\vec{s}^z\in\mathbb{F}_2^{m_2-1} )$~?
A moment's thought shows that $P_{\!\mathcal{A}}(\vec{l},\vec{s})$ actually depends only on the value of
\begin{equation}\label{eq:syndromeasab}
 ( (0,\beta_1), (\alpha_2,\beta_2), \dots, (\alpha_{m_2},\beta_{m_2}) ),
\end{equation}
where $\alpha_i = s^z_{i-1}$ and
$\beta_j$ is total number of ones in the $j$-th $(m_1-1)$-bit block in $\vec{s}^x$ (compare with figure \ref{fig:concat:add}).
\begin{figure}
\centering
\scalebox{0.9}{
\begin{pspicture}(-2.75,-2.25)(11.0,3.0)
%

\psframe[linecolor=lightgray,fillstyle=solid,fillcolor=mygray](0,1.75)(9.0,2.25)
\psframe[linecolor=lightgray,fillstyle=solid,fillcolor=mygray](0.5,1.75)(3.0,2.25)%
\psframe[linecolor=lightgray,fillstyle=solid,fillcolor=mygray](3.5,1.75)(6.0,2.25)%
\psframe[linecolor=lightgray,fillstyle=solid,fillcolor=mygray](6.5,1.75)(9.0,2.25)%
\multirput[c](1.75,2)(0.5,0){3}{1}\rput[c](2.25,2.5){%
\rotatebox[origin=c]{-90}{\rotatebox[origin=c]{90}{$\beta_1$}$\left\{\makebox(0,0.8)[b]{}\right.$}}
\multirput[c](5.25,2)(0.5,0){2}{1}\rput[c](5.50,2.5){%
\rotatebox[origin=c]{-90}{\rotatebox[origin=c]{90}{$\beta_2$}$\left\{\makebox(0,0.5)[b]{}\right.$}}
\multirput[c](7.25,2)(0.5,0){4}{1}\rput[c](8.00,2.5){%
\rotatebox[origin=c]{-90}{\rotatebox[origin=c]{90}{$\beta_3$}$\left\{\makebox(0,1.1)[b]{}\right.$}}
\rput[r](-0.25,2.0){\small $\displaystyle\sum_{i=1}^{\mathclap{m_2(m_1-1)}} s^x_i\vec{\eta}^x_i = $}
\psframe[linecolor=lightgray,fillstyle=solid,fillcolor=mygray](0,1.25)(9.0,1.75)
\psframe[linecolor=lightgray,fillstyle=solid,fillcolor=mygray](0,1.25)(3.0,1.75)
\psframe[linecolor=lightgray,fillstyle=solid,fillcolor=mygray](6,1.25)(9.0,1.75)
\multirput[c](0.25,1.5)(0.5,0){6}{1}
\multirput[c](6.25,1.5)(0.5,0){6}{1}
\rput[l](9.25,1.5){\small $\displaystyle=\sum_{j=1}^{\mathclap{m_2-1}} n^z_j\vec{\xi}^x_j$}
\psframe[linecolor=lightgray,fillstyle=solid,fillcolor=mygray](0,0.75)(9.0,1.25)
\psframe[linecolor=lightgray,fillstyle=solid,fillcolor=mygray](0,0.75)(3.0,1.25)
\multirput[c](0.25,1.0)(0.5,0){6}{$l^x$}
\rput[r](-0.25,1.0){\small $l^x\vec{\mu}^x=$}
\rput[c](-2.2,1.5){$\vec{a}^x \left\{ \makebox(0,0.9)[b]{}\right.$}

\psframe[linecolor=lightgray,fillstyle=solid](0,-0.75)(9.0,-1.25)
\psframe[linecolor=lightgray,fillstyle=solid](3,-0.75)(3.5,-1.25)
\psframe[linecolor=lightgray,fillstyle=solid](6,-0.75)(6.5,-1.25)
                    \rput[c](3.25,-0.5){%
\rotatebox[origin=c]{-90}{\rotatebox[origin=c]{90}{$\alpha_2$}$\left\{\makebox(0,0.3)[b]{}\right.$}}
\rput[c](6.25,-1){1}\rput[c](6.25,-0.5){%
\rotatebox[origin=c]{-90}{\rotatebox[origin=c]{90}{$\alpha_3$}$\left\{\makebox(0,0.3)[b]{}\right.$}}
\rput[r](-0.25,-1){\small $\displaystyle\sum_{j=1}^{\mathclap{m_2-1}} s^z_j \vec{\eta}^z_j=$}
\psframe[linecolor=lightgray,fillstyle=solid](0,-1.25)(9.0,-1.75)
\psframe[linecolor=lightgray,fillstyle=solid](0,-1.25)(3.0,-1.75)
\psframe[linecolor=lightgray,fillstyle=solid](6,-1.25)(9.0,-1.75)
\rput[c](6.25,-1.5){1}\rput[c](7.25,-1.5){1}\rput[c](8.25,-1.5){1}\rput[c](8.75,-1.5){1}
\rput[c](3.75,-1.5){1}\rput[c](4.25,-1.5){1}
\rput[c](0.25,-1.5){1}\rput[c](0.75,-1.5){1}\rput[c](1.25,-1.5){1}\rput[c](2.25,-1.5){1}
\rput[l](9.25,-1.5){\small $\displaystyle=\sum_{i=1}^{\mathclap{m_2(m_1-1)}} n^x_i \vec{\xi}^z_i$}
\psframe[linecolor=lightgray,fillstyle=solid](0,-1.75)(9.0,-2.25)
\psframe[linecolor=lightgray,fillstyle=solid](0.0,-1.75)(0.5,-2.25)%
\psframe[linecolor=lightgray,fillstyle=solid](3.0,-1.75)(3.5,-2.25)%
\psframe[linecolor=lightgray,fillstyle=solid](6.0,-1.75)(6.5,-2.25)%
\multirput[c](0.25,-2)(3,0){3}{$l^z$}
\rput[r](-0.25,-2){\small $l^z\vec{\mu}^z=$}
\rput[c](-2.2,-1.5){$\vec{a}^z \left\{ \makebox(0,0.9)[b]{}\right.$}
\end{pspicture}}%
\caption[Graphical representation of $(\vec{a}^x,\vec{a}^z)\in\mathbb{F}_2^{2n}$ for the concatenated cat code]{\label{fig:concat:add}
Graphical representation of the strings
$\vec{a}^x = \sum_is^x_i\vec{\eta}^x_i + \sum_jn^z_i\vec{\xi}^x_j + l^x\vec{\mu}^x$ and
$\vec{a}^z = \sum_in^x_i\vec{\xi}^z_i + \sum_js^z_i\vec{\eta}^z_j + l^z\vec{\mu}^z$ for the concatenated cat code with $m_1=6$ and $m_2=3$.
The structure of the $\vec{\xi}^x_j$ leads to an even number of
completely filled blocks of size $m_1$ in the middle part of $\vec{a}^x$.
Similarly, the structure of the $\vec{\xi}^z_i$ leads to an
even number of ones in each of the $m_2$ blocks of size $m_1$ in the middle part of $\vec{a}^z$.}
\end{figure}
In addition, only the frequency distribution of the $(\alpha_i,\beta_i)$ matters.

For some $\vec{s}$ which has the properties expressed by \eqref{eq:syndromeasab}, we have
\begin{align}
 P_{\!\mathcal{A}}(l^x,l^z, \vec{s}^x,\vec{s}^z) &\equiv
 P_{\!\mathcal{A}}\bigl(l^x,l^z, (0,\beta_1), (\alpha_2,\beta_2), \dots, (\alpha_{m_2},\beta_{m_2}) \bigr) \nonumber\\
&=
 \sum_{\vec{n}^x\in\mathbb{F}_q^{m_2(m_1-1)}}
 \sum_{\vec{n}^z\in\mathbb{F}_q^{m_2-1}}
 P^n_{\!\mathcal{A}}\bigl( \vec{a}^x(\vec{s}^x,\vec{n}^z,l^x), \vec{a}^z(\vec{s}^z,\vec{n}^x,l^z) \bigr).
\end{align}
The sum over $\vec{n}^z$ can be written as
\begin{equation}
 \sum_{b_1=0}^1\dots\sum_{b_{m_2}=0}^1 \frac{1+(-1)^{\sum_ib_i+l^x}}{2},
\end{equation}
which assures that the total number of completely filled blocks of size $m_1$ in $\vec{a}^x$ (compare with figure \ref{fig:concat:add}) is even for $l^x=0$ and odd for $l^x=1$.
The sum over $\vec{n}^x$ is decomposed into $m_2$ sums each of which is written using the shorthand notation
\begin{equation}
\sum_{l_i,t_i}^{(\alpha_i,\beta_i)} \equiv
\sum_{l_i=0}^{\beta_i}\sum_{t_i=0}^{m_1-\beta_i} \frac{1+(-1)^{l_i+t_i+l^z+\alpha_i}}{2}
 \binom{\beta_i}{l_i}\binom{m_1-\beta_i}{t_i}.
\end{equation}
Here, $l_i$ denotes the number of ones which are placed in a region of $\vec{a}^z$
where $\vec{a}^x$ contains ones counted by $\beta_i$,
and $t_i$ denotes the number of ones which are placed in the remaining regions of $\vec{a}^z$.
Therefore, there are
$l_i$ $Y$-errors, $\beta_i-l_i$ $X$-errors and $t_i$ $Z$-errors
if $b_i=0$,
while there are
$l_i$ $Z$-errors, $t_i$ $Y$-errors and $m_1-\beta_1-t_i$ $X$-errors
if $b_i=1$.
Altogether we obtain
\begin{multline}
P_{\!\mathcal{A}}(l^x,l^z, \vec{s}) =
 \sum_{b_1=0}^1\dots\sum_{b_{m_2}=0}^1 \frac{1+(-1)^{\sum_ib_i+l^x}}{2}
 \sum_{l_1,t_1}^{(0,\beta_1)}
 \sum_{l_2,t_2}^{(\alpha_2,\beta_2)}
 \dots
 \sum_{l_{m_2},t_{m_2}}^{(\alpha_{m_2},\beta_{m_2})} \\
 \prod_{i=1}^{m_2}
 \Bigl( p_y^{l_i}p_z^{t_i}p_x^{\beta_i-l_i}p_e^{m_1-\beta_i-t_i}\Bigr)^{1-b_i}
 \Bigl( p_z^{l_ i}p_y^{t_i}p_e^{\beta_i-l_i}p_x^{m_1-\beta_i-t_i}\Bigr)^{b_i},
\end{multline}
which can be simplified by applying the following binomial series identity,
\begin{equation}
 \sum_{k=0}^n \binom{n}{k} \frac{1+(-1)^{k+l}}{2} x^k y^{n-k} = \frac{1}{2} \bigl( (x+y)^n +(-1)^l (y-x)^n \bigr),
\end{equation}
first to each sum over $l_i$ and then to each sum over $t_i$, leading to
\begin{equation}
P_{\!\mathcal{A}}(l^x,l^z, \vec{s}) =
  \sum_{b_1=0}^1\dots\sum_{b_{m_2}=0}^1 \frac{1+(-1)^{\sum_ib_i+l^x}}{2}
  F_{b_1}(l^z,0,\beta_1) F_{b_2}(l^z,\alpha_2,\beta_2) \dots F_{b_{m_2}}(l^z,\alpha_{m_2},\beta_{m_2}),
\end{equation}
with
\begin{subequations}\label{eq:F0F1concat}
\begin{align}
 F_0(l^z,\alpha,\beta) &= \frac{1}{2}\bigl[ (p_x+p_y)^\beta(1-p_x-p_y)^{m_1-\beta}\! +
  (-1)^{l_z+\alpha}(p_x-p_y)^\beta(1-p_x-p_y-2p_z)^{m_1-\beta} \bigr] \\
 F_1(l^z,\alpha,\beta) &= \frac{1}{2}\bigl[ (1-p_x-p_y)^\beta(p_x+p_y)^{m_1-\beta}\! +
   (-1)^{l_z+\alpha}(1-p_x-p_y-2p_z)^\beta(p_x-p_y)^{m_1-\beta} \bigr].
\end{align}
\end{subequations}
In the above expressions we replaced $p_e$ by $1-p_x-p_y-p_z$.
By adding up the last remaining sums over the $b_i$,
eventually we arrive at the final result,
\begin{multline}\label{eq:Prlxlzsm1m2}
P_{\!\mathcal{A}}(l^x,l^z, \vec{s}) =\\
\frac{1}{2}\Biggl[
 \prod_{i=1}^{m_2} \bigl( F_0(l^z,\alpha_i,\beta_i)+F_1(l^z,\alpha_i,\beta_i) \bigr) +
 (-1)^{l_x }
 \prod_{i=1}^{m_2} \bigl( F_0(l^z,\alpha_i,\beta_i)-F_1(l^z,\alpha_i,\beta_i) \bigr)
 \Biggr],
\end{multline}
where $\alpha_1$ is always assumed to be zero.

The observation that only the frequency distribution of the $(\alpha_i,\beta_i)$ matters allows us to speed up the summation over all possible syndromes drastically.
To calculate the total probability of getting a certain logical error, we have to evaluate the sum over all $2^{m_1m_2-1}$ syndromes $\vec{s}$,
\begin{align}
P_{\!\mathcal{A}}(l^x,l^z)
&=
 \sum_{\mathclap{\vec{s}\in \mathbb{F}_2^{m_2m_1-1}}} P_{\!\mathcal{A}}(l^x,l^z,\vec{s}) \label{eq:sumoverallsyns}\\
&=
 \sum_{\beta_1=0}^{m_1-1} \binom{m_1-1}{\beta_1}
 \sum_{(\alpha_2,\beta_2),\dots,(\alpha_{m_2},\beta_{m_2})}
    \binom{m_1-1}{\beta_2}\hdots \binom{m_1-1}{\beta_{m_2}} \times \nonumber\\
&\qquad\qquad\qquad\qquad\qquad\qquad\qquad P_{\!\mathcal{A}}\bigl(l^x,l^z, ((0,\beta_1),(\alpha_2,\beta_2),\dots,(\alpha_{m_2},\beta_{m_2})) \bigr).
\intertext{%
Since $(\alpha_i,\beta_i)$ takes on $2m_1$ different values, this expression simplifies to}
&=
\sum_{\beta_1=0}^{m_1-1} \binom{m_1-1}{\beta_1}
 \sum_{\substack{a_1,a_2,\dots,a_{2m_1}=0\\ \text{s.\,t. } \sum_i a_i=m_2-1}}^{m_2-1}
 \frac{(m_2-1)!}{a_1! a_2! \dots a_{2m_1}!}
 \prod_{i=1}^{2m_1}\binom{m_1-1}{\beta(i)}^{a_i} \times \nonumber\\
&\qquad\qquad\qquad\qquad\qquad\qquad\qquad P_{\!\mathcal{A}}\bigl(l^x,l^z,((0,\beta_1),\{ (\alpha(j),\beta(j))^{a_j} \}_{j=1\dots 2m_1}) \bigr) \label{eq:Prlxlz} .
\end{align}
Instead of adding up $2^{m_1m_2-1}$ terms as in \eqref{eq:sumoverallsyns}, we only have to consider $m_1\cdot \binom{m_2-1+2m_1-1}{m_2-1}$ terms.

\subsubsection{Joint Probabilities for the Cat Code} %
By setting $m_2=1$ and $m_1=m$ in equation \eqref{eq:Prlxlzsm1m2},
we get the joint probabilities for the $[[m,1]]_2$ cat code of subsection \ref{subsec:thecatcode},
\begin{multline}\label{eq:probcatcode}
P_{\!\mathcal{A}}(l^x,l^z,\vec{s}^x) = \frac{1}{2} \Bigl[
(p_x+p_y)^{l^x(m-2\beta)+\beta} (1-p_x-p_y)^{(1-l^x)(m-2\beta)+\beta} + \\ (-1)^{l^z}
(p_x-p_y)^{l^x(m-2\beta)+\beta} (1-p_x-p_y-2p_z)^{(1-l^x)(m-2\beta)+\beta}  \Bigr].
\end{multline}
Here, $\beta$ denotes the number of ones in $\vec{s}=\vec{s}^x$.

\begin{rem}
If we calculate expressions like \eqref{eq:F0F1concat} or \eqref{eq:probcatcode} for the depolarizing channel $\mathcal{D}_p$, we have $p_x=p_y=p_z=p/3$ and therefore some of the products in these expressions become zero.
If such a product is exponentiated, as it is the case for the term $(p_x-p_z)^\beta$ for instance,
one has to take special care of the case $\beta=0$ in which the term is equal to one.
\end{rem}

\begin{figure}
\centering
\includegraphics[scale=0.825]{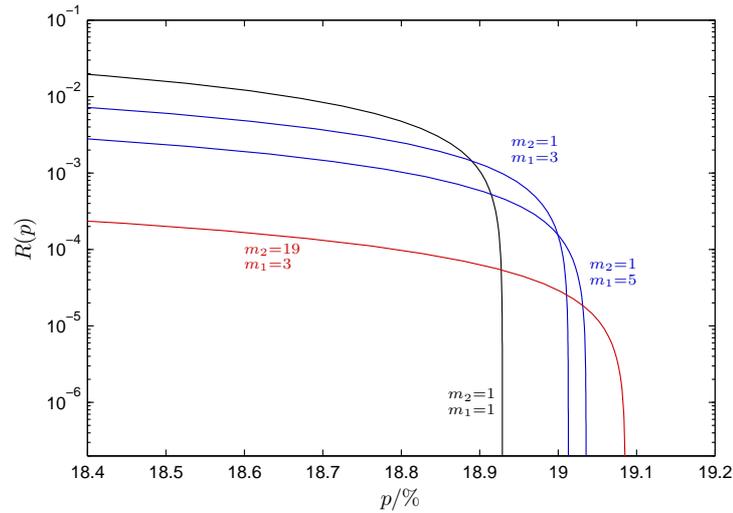}
\caption{\label{fig:catcoderates}
Achievable transmission rates for various $[[m_1\times m_2,1]]_2$ codes over the qubit depolarizing channel $\mathcal{D}_p$ plotted as function of the noise $p$:
The hashing rate (corresponding to $m_1=m_2=1$) \textit{(black)},
the cat code ($m_2=1$) with $m_1=3$ and $m_1=5$ \textit{(blue)},
the concatenated cat code with parameters $m_1=3,m_2=19$ \textit{(red)} and $m_1=5,m_2=15$ \textit{(orange)}.
}
\end{figure}

\subsubsection{Results for the Depolarizing Channel}

We use \eqref{eq:Prlxlzsm1m2} and \eqref{eq:Prlxlz} to evaluate the achievable transmission rate of equation \eqref{eq:rofconcthm} for various inner $[[m_1\times m_2,1]]_2$ concatenated cat codes concatenated with random outer codes over the qubit ($q=2$) depolarizing channel $\mathcal{D}_p$.
The hashing rate (corresponding to $m_1=m_2=1$) is compared with the rates of various concatenated cat codes in figure \ref{fig:catcoderates}.
It can be seen that the hashing rate, which equals the one-shot capacity as shown in lemma \ref{lem:oneshotdepola},
\begin{equation}
 Q^{(1)}(\mathcal{D}_p) =  \max_{\rho} I_c(\rho, \mathcal{D}_p ),
\end{equation}
is surpassed e.\,g. by the rate of the cat code ($m_2=1$) of size $m=m_1=5$ for high values of noise ($p\approx 0.19$).
Since this rate may be expressed as
\begin{equation}
  \frac{1}{5}  I_c\Bigl( \frac{1}{2}\Pi_\mathcal{C}, \mathcal{D}_p^{\otimes 5} \Bigr),
\end{equation}
where $\Pi_\mathcal{C}$ denotes the projector on one of the codespaces of the $[[5,1]]_2$ cat code (see subsection \ref{subsec:ratecoinf}),
it is clear that the limit as $n$ goes to infinity in the regularized coherent information expressing the quantum capacity of a quantum channel $\mathcal{A}$,
\begin{equation}
 Q( \mathcal{A} ) =  \lim_{n\rightarrow\infty} \frac{1}{n} \max_{\rho} I_c(\rho, \mathcal{A}^{\otimes n} ),
\end{equation}
is crucial since this example shows that in general $Q^{(n)}( \mathcal{A} ) = \max_\rho I_c( \rho, \mathcal{A}^{\otimes n} ) / n$ might be larger than $Q^{(1)}( \mathcal{A} )$.

\begin{figure}
\centering
\includegraphics[scale=0.9]{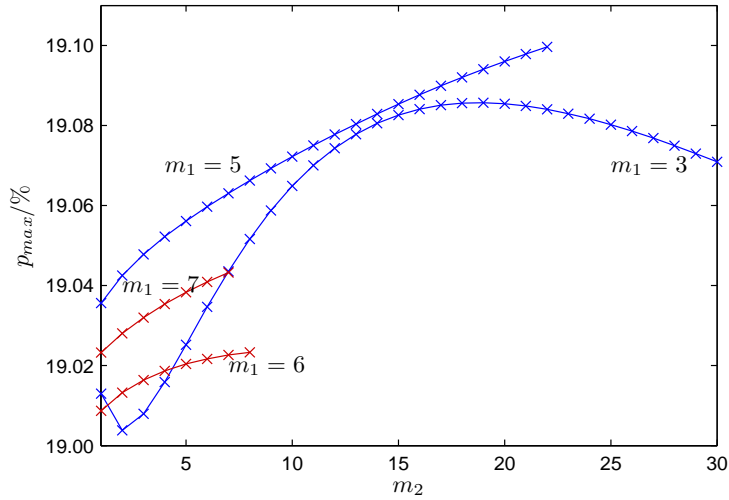}
\caption{\label{fig:pmaxconcatcode}
The maximum tolerable value of noise $p$ for the qubit depolarizing channel $\mathcal{D}_p$ as a function of the size $m_2$ of the inner $[[m_1\times m_2,1]]_2$ concatenated cat code for various values of $m_1$.
For $m_1=3$ the highest value is obtained for $m_2=19$,
$p_\text{max}^\text{conc-cat}(m_1=3,m_2=19) = 19.0857\%$.
For $m_1=5$ the highest value shown is $p_\text{max}^\text{conc-cat}(m_1=5,m_2=22) = 19.0996\%$.
}
\end{figure}

So far lower bounds on the maximum tolerable noise $p_\text{max}$ of the qubit depolarizing channel have been determined by
(i) setting the hashing rate of theorem \ref{thm:randstab} and \ref{thm:randcss} equal to zero
$\Rightarrow p_\text{max}^\text{hash} = 18.9290\%$ and
(ii) by setting the rate of \eqref{eq:rofconcthm} for a cat code of size $m$ equal to zero
$\Rightarrow p_\text{max}^\text{cat}(m=5) = 19.0356\%$.
Now we set the rate of \eqref{eq:rofconcthm} for various $[[m_1\times m_2,1]]_2$ concatenated cat codes equal to zero.
The corresponding tolerable values of noise are plotted in figure \ref{fig:pmaxconcatcode} as a function of $m_2$ for various values of $m_1$.
It can be seen that for $m_1=3$ the best lower bound is obtained for $m_2=19$,
$p_\text{max}^\text{conc-cat}(m_1=3,m_2=19) = 19.0857\%$.
Due to computational limitations (calculation of the $m_1=5,m_2=22$ point took roughly a week on a Intel core 2 duo E8500 CPU),
the $m_1=5$ curve was calculated only up to $m_2=22$ leading to the best lower bound known to date of $p_\text{max}^\text{conc-cat}(m_1=5,m_2=22) = 19.0996\%$.
This beats the highest previously known lower bound of \cite{SmSm07} which was
$p_\text{max}^\text{conc-cat}(m_1=5,m_2=16) = 19.0877\%$.
While the optimal value of $m_2$ for $m_1=5$ was conjectured in \cite{SmSm07} to be $m_2\approx 25$,
according to our new data we expect it to lie slightly higher (maybe $m_2\approx 30$).

\newcommand  {\BBP}{11.0028\%} %
\newcommand {\BBPn}{12.4120\%} %
\newcommand{\BBPub}{14.6447\%} %
\newcommand{\BBPcr}{12.9379\%} %
\newcommand{\BBPcq}{0.32656}   %
\newcommand{\BBPcm}{500}       %
\newcommand  {\SSP}{12.6193\%} %
\newcommand {\SSPd}{12.6904\%} %
\newcommand {\SSPn}{14.1119\%} %
\newcommand{\SSPcr}{14.5741\%} %
\newcommand{\SSPcq}{0.31210}   %
\newcommand{\SSPcm}{250}       %
\newcommand{\SSPer}{14.5930\%} %
\newcommand{\SSPeq}{0.31650}   %
\newcommand{\SSPem}{300}       %

\newcommand{\entropy}[1]{\mbox{$H\!\left( #1\right)$}}
\newcommand{\bentropy}[1]{\mbox{$H_2\!\left( #1\right)$}}

\chapter{Quantum Cryptography}\label{chap:crypto}

Quantum key distribution (QKD) protocols try to establish a secure and random key between two distant parties usually called Alice and Bob.
While the security of corresponding classical protocols relies on the assumption that an eavesdropper has limited computational power, the security of a QKD protocol is guaranteed by the validity of quantum mechanics.
Quantum cryptography was initiated by Bennett and Brassard in 1984 who developed the first QKD protocol, which is now called BB84 protocol \cite{BB84}.
A natural extension of BB84 which makes use of four different quantum states is the 6-state protocol \cite{Bru98} which makes use of two additional quantum states.
To prove the security of a QKD protocol, one makes the worst case assumption that the quantum channel connecting the two parties is under complete control of an eavesdropper, usually named Eve.
Since non-orthogonal quantum states cannot be cloned perfectly \cite{Dieks82,WZ82}, the two users Alice and Bob are able to detect the presence of an eavesdropper by comparing some of Bob's measurement results with Alice's preparations in a step called parameter estimation.
Depending on the result, they might either abort the protocol, or, if the action of the eavesdropper seems harmless enough, proceed with an error correction and privacy amplification step to obtain a random and private key.

Using a quantum channel to create a secret key between two parties is closely related to using the channel to send quantum information, with many results found in one area applicable in the other.
For instance, by treating the steps in a quantum key distribution (QKD) protocol coherently and viewing the entire process as an entanglement distillation scheme, one can use properties of random quantum error-correcting codes to prove the security of the BB84 and 6-state protocols up to bit error rates of $p_\text{max}^\text{BB84}=\BBP$ \cite{SP00} and $p_\text{max}^\text{6-st.}=\SSP$  \cite{Lo01}, respectively.
Conversely, the formula for the quantum channel capacity can be obtained by importing the key rate resulting from a general approach to secret key generation over a known channel~\cite{DeWi0307,De0304,DeWi0306}.

One of the surprising results related to quantum capacity is the non-optimality of random codes, in contrast to the classical case.
As it was shown in chapter \ref{chap:cecc}, the classical capacity of a channel can be achieved by using randomly-constructed block codes, and the independence of one input to the channel from the next results in a so-called single-letter formula for the capacity.
While random coding can be used to create quantum error-correcting codes as well (compare with section \ref{sec:randqcodes}), these do not always achieve the capacity.
Better performance can be achieved by structured codes which exploit the ability of quantum error-correcting
codes to correct errors without precisely identifying them, a property called degeneracy (compare with  section \ref{sec:concrandom}).

By appealing to the coherent formulation of the protocol, degenerate codes should also be useful in QKD.
This was shown to be the case in the original security proof of the 6-state protocol \cite{Lo01}, as the results of \cite{DiSS98} were used to improve the error rate threshold to $p_\text{max}^\text{6-st.}=\SSPd$.
More striking threshold improvements are possible, if counterintuitive, by simply adding noise to the raw key bits before they are processed into the final key, a procedure known as local randomization \cite{KGR05,KGR05b}.
This improves the error rate thresholds for the two protocols to $p_\text{max}^\text{BB84}=\BBPn$ and $p_\text{max}^\text{6-st.}=\SSPn$, respectively. At first glance, these results make no sense in the coherent picture of QKD, since adding more noise to already noisy entangled pairs only decreases the amount of pure entanglement which can be extracted.
The entanglement/secret-key analogy does not hold perfectly, however; entangled states are sufficient, but
not necessary, for creation of secret keys.
A broader class of states, called private states, leads to secret keys when measured~\cite{H3O03}, and these should properly be the target output of the coherent version of the QKD protocol.
Indeed, the exact error thresholds are recovered in the coherent picture when the QKD protocols with local randomization are analyzed in these terms \cite{RS07}.

With a systematic understanding of how degenerate codes and local randomization boost the key rate, it becomes sensible to combine the two methods to look for even higher thresholds.
Recently it was shown in \cite{SRS06} that doing so improves the error threshold of the BB84 protocol up to at least $p_\text{max}^\text{BB84}\approx 12.92\%$ by using the same type of structured code studied in~\cite{ShSm96,DiSS98,SmSm07} and subsection \ref{subsec:thecatcode}. These specific codes consist of the concatenation of two codes, the first a simple repetition code and the second a random code.
The repetition code, sometimes called a cat code in the context of quantum information theory since the codewords are $\ket{0}^{\otimes m}$ and $\ket{1}^{\otimes m}$, induces degeneracy in the overall code since a phase flip on any of the physical qubits leads to the same logical error, and is corrected in the same way.
In particular, blocklength $m=400$ corresponds to the threshold stated above.
Since the random code portion of the protocol corresponds to information reconciliation and privacy amplification in the classical view, the local randomization and the repetition code together become a type of \emph{preprocessing} performed before these ``usual'' steps.

In this chapter we show that the same preprocessing protocol as used in \cite{SRS06} can also be used to improve the maximum tolerable bit error rate for the 6-state protocol, up to at least $p_\text{max}^\text{6-st.}=\SSPer$ for a blocksize of $m=\SSPem$.
This is already quite close to the upper bound of  $\BBPub$ \cite{FGGNP97,KGR05,MoCL06} on the tolerable error rate for the BB84 protocol, and since the error threshold grows with blocklength, the bound is presumably exceeded at larger blocklengths, indicating the higher robustness of the 6-state protocol.
We also improve the lower bounds for the BB84 protocol presented in \cite{SRS06}.
In addition we investigate iterating the preprocessing scheme in the BB84 protocol, and show an improvement both in rate and error threshold over single-round preprocessing for even modest blocklengths.
The results presented in this chapter have been obtained in collaboration with J. Renes and have been published in \cite{KR08}.

To begin, section \ref{sec:bb6sprotocols} explains the BB84 and 6-state QKD protocols and summarizes Shor and Preskill's security proof \cite{SP00} which uses the structure of CSS codes to show the equivalence between these protocols and corresponding entanglement distillation protocols.
Section \ref{sec:combiprepro} describes the preprocessing scheme in more depth and then derives secret key rate expressions for the BB84 and the 6-state protocols.
Numerical calculations for blocklengths into the hundreds are then presented for the two protocols.
We explain how representation theory is helpful for the numerical evaluation of such key rates in both cases.
Section \ref{sec:itcombiprepro} examines the advantages of iterating the preprocessing protocol to achieve higher rates and thresholds for the same amount of effort in noise addition and block coding.

\section{BB84 and 6-State Protocols}\label{sec:bb6sprotocols}

The BB84 \cite{BB84} and the 6-state \cite{Bru98} protocol are QKD protocols of the prepare and measure type.
Their goal is to establish a random and secret key between two parties --- usually called Alice and Bob --- which are connected via a quantum channel and a classical channel.
The quantum channel is fully accessible to an eavesdropper --- traditionally called Eve --- while the classical channel is assumed to be authenticated, i.\,e. Eve can only listen to the messages, but cannot interfere.
(To authenticate the classical channel, Alice and Bob need to share a small secret key in advance.
Hence, strictly speaking, QKD protocols are secret key growing protocols.)
\begin{rem}
While Alice and Bob have to use two-way classical communication for the parameter estimation step of the protocol, this chapter deals only with protocols using one-way communication during the error correction and privacy amplification steps.
The use of two-way communication during these steps allows them to obtain a secure key for even higher levels of noise \cite{GL03}, which we assume is caused by Eve.
\end{rem}

\subsection{Description of the Protocols}\label{subsec:6sbb84}

Let $s=2$ for the BB84 protocol and $s=3$ for the 6-state protocol.
If we denote the eigenstates corresponding to eigenvalues $+1$ and $-1$ of the Pauli $Z$ matrix by $\ket{0}_z$ and $\ket{1}_z$, the corresponding eigenstates of the Pauli $X$ and $Y$ matrices are given by
\begin{align}
\ket{0}_x &= (\ket{0}_z+\ket{1}_z)/\sqrt{2} & \ket{1}_x &= (\ket{0}_z-\ket{1}_z)/\sqrt{2} \\
\ket{0}_y &= (\ket{0}_z+i\ket{1}_z)/\sqrt{2} & \ket{1}_y &= (\ket{0}_z-i\ket{1}_z)/\sqrt{2}.
\end{align}
In addition, let $B(0)=z$, $B(1)=x$ and $B(2)=y$.

Alice chooses a random sequence of zeros and ones $\vec{x}=(x_1,x_2,\dots, x_N )\in \mathbb{F}_2^N$ of length $N\gtrsim 2\cdot s \cdot n$ and a random sequence $\vec{b}=(b_1,b_2,\dots, b_N)\in \mathbb{F}_s^N$.
Then she prepares the sequence of quantum states $\bigotimes_{i=1}^N \ket{ x_i }_{\!B(b_i)}$ and sends them to Bob.
Bob chooses a random sequence $\vec{b}'=(b'_1,b'_2,\dots, b'_N)\in \mathbb{F}_s^N$ and measures the $i$-th qubit in the basis $B(b'_i)$ denoting the result as $y_i\in\mathbb{F}_2$.
After Bob finished his measurements, he announces this fact and both parties compare their strings $\vec{b}$ and $\vec{b}'$.
If $b_i\neq b'_i$ they remove the $i$-th entry from their strings $\vec{x}$ and $\vec{y}$.
The resulting strings $\vec{x}_\text{sifted}$ and $\vec{y}_\text{sifted}$ form the sifted key and are of length $2\cdot n$ approximately.
If the quantum states had been transmitted unperturbed, the sifted keys of Alice and Bob coincide, $\vec{x}_\text{sifted} = \vec{y}_\text{sifted}$.
To check whether this is the case, Alice selects half of the bits to serve as check bits, submits her choice to Bob, and both parties compare this part of their sifted key.
The resulting error rate is called the bit-error rate $p$.
If the bit-error rate $p$ is zero, they can be confident that no eavesdropper was present and may use the remaining $n$ bits $\vec{x}'_\text{sifted}$ and $\vec{y}'_\text{sifted}$ as a secure and random key.

In practice there will always be a bit-error rate $p>0$ due to imperfections of the quantum channel or the presence of an eavesdropper.
Hence the task is to proof the security of the protocols up to a certain bit-error rate $p_\text{max}$.
As long as $p < p_\text{max}$, Alice and Bob should be able to perform error correction and privacy amplification to obtain a secure key $\vec{k}$ of length $k<n$ from $\vec{x}'_\text{sifted}\in\mathbb{F}_2^n$ and from $\vec{y}'_\text{sifted}\in\mathbb{F}_2^n$.
The first simple proof of security was given by Shor and Preskill \cite{SP00}:
By treating the steps in a QKD protocol coherently and viewing the entire process as an entanglement distillation scheme, one can use properties of random quantum error-correcting codes to prove the security of the BB84 and 6-state protocols up to bit-error rates of $p_\text{max}^\text{BB84}=\BBP$ \cite{SP00} and $p_\text{max}^\text{6-st.}=\SSP$  \cite{Lo01}, respectively.

\subsection{Shor and Preskill's Security Proof}\label{subsec:spproof}

The security proof of Shor and Preskill is based on the observation of Deutsch et\,al. \cite{QPA96} and Lo and Chau \cite{LC99} that entanglement distillation protocols provide a way to establish a secret key between the two parties Alice and Bob.
If, as a result of an entanglement distillation protocol, Alice and Bob share (near) perfect states $\ket{\Phi^+}_{AB}=(\ket{00}_{AB}+\ket{11}_{AB})/\sqrt{2}$, a bipartite measurement of $\ket{\Phi^+}_{AB}$ in the z-basis results in a shared secret bit\footnote{A maximally entangled state like $\ket{\Phi^+}_{AB}$ is not necessary to provide a secret bit; so-called private states are necessary and sufficient \cite{H3O03}.}.
Shor and Preskill \cite{SP00} (see also \cite{GP01} for a more elaborate version of the proof) realized that an entanglement distillation protocol making use of CSS codes is equivalent to the BB84 protocol.
Their proof was adapted to the 6-state protocol by Lo \cite{Lo01}.
In the following we describe the corresponding entanglement distillation protocol and its reduction to a prepare and measure scheme.
For BB84, let $T=\frac{1}{\sqrt{2}}\bigl(\begin{smallmatrix} 1&1\\1& -1\end{smallmatrix}\bigr)$ denote the Hadamard matrix mapping the z-basis onto the x-basis and vice versa.
For the 6-state protocol, let
\begin{equation}
 T = \exp\Bigl( -\frac{i}{2} (X+Y+Z)/\sqrt{3}\cdot \frac{2\pi}{3} \Bigr)\cdot e^{i\pi/4} =
 \frac{1}{\sqrt{2}}\begin{pmatrix} 1 & -i\\1 & i \end{pmatrix}
\end{equation}
denote the rotation of angle $2\pi/3$ around the axis $(1,1,1)/\sqrt{3}$, mapping the z-axis to the x-axis, the x-axis to the y-axis, and the y-axis to the z-axis.

\subsubsection{Entanglement Distillation Protocol}

Alice prepares $N=2n$ maximally entangled pairs $\ket{\Phi^+}_{AB}$ and chooses a random string $\vec{b}=(b_1,\dots,b_N)\in\mathbb{F}_s^N$.
After applying the operation $T_B^{b_i}$ onto Bob's part of the $i$-th pair, she sends him his half of the states.
Bob acknowledges the reception of his qubits.
Alice picks out $n$ pairs which have to serve as check pairs and tells Bob the string $\vec{b}$ together with her choice of the check pairs.
Bob applies the operation $T_B^{-b_i}$ onto his $i$-th qubit.
Both parties measure the check pairs in the z-basis, share their results and obtain the bit-error rate $p$.
Since there is no way for Eve to know the check pairs in advance, the bit-error rate of the check bits should be a pretty good estimate for the bit error rate of the remaining $n$ pairs.

Let us assume now that Eve's attack can be described by a memoryless Pauli channel $\mathcal{E}^{\otimes N}$ where $\mathcal{E}$ is characterized by the probability distribution $\{ q_I,q_x,q_y,q_z \}$.
Of course Eve might apply any completely positive map, but, as it was pointed out in \cite{LC99}, the entanglement distillation protocol which will be used to generate $k<n$ (near) perfect pairs from the remaining $n$, commutes with a measurement of each pair in the Bell basis.
Hence the most general attack of Eve can be described by a general Pauli channel which corresponds to the twirled version of Eve's attack (compare with theorem \ref{thm:twirledcpmap}).
Furthermore, it can be shown that if the entanglement distillation protocol is capable of correcting an uncorrelated Pauli attack, it is also capable of correcting a correlated one (if Alice and Bob apply a random permutation to their qubits; see e.\,g. \cite{GL03}).
As a result of the application of the $T_B^{b_i}$ with $b_i\in\mathbb{F}_s$, parameter estimation assures us that the effective Pauli channel
\begin{equation}
 \mathcal{E}_\text{eff}(\rho) = \frac{1}{s} \sum_{j=0}^{s-1} T^{-j}\mathcal{E}( T^j \rho T^{-j}) T^j
\end{equation}
is characterized by the probability distribution $ \{ p_{uv} \} \equiv \{p_{00},p_{10},p_{11},p_{01}\}$ s.\,t.
\begin{equation}\label{eq:probdistrieff}
 \{ p_{uv} \}  = \begin{cases}
\{1-2p+t,p-t,t,p-t\}, t\in [0,p], &\text{ in case of the BB84 protocol.}\\
\{ 1-\frac{3}{2}p, \frac{p}{2}, \frac{p}{2}, \frac{p}{2}\}, &\text{ in case of the 6-state protocol.}
\end{cases}
\end{equation}

We are now going to describe the entanglement distillation protocol which is capable of distilling $k<n$ (near) perfect $\ket{\Phi^+}_{AB}$ pairs from the remaining state
$\mathcal{I}_A\otimes \mathcal{E}_{\text{eff},B}^{\otimes n}\bigl( (\ket{\Phi^+}\bra{\Phi^+})^{\otimes n} \bigr)$ as long as the bit error rate $p$ is not too high.
Let us fix a CSS code encoding $k=k_1-k_2$ qubits into $n$.
As explained in section \ref{sec:csscodes}, together with an encoding $U_\text{enc}$ such a code is specified by the two lists of vectors
\begin{align*}
&\{\vec{\xi}^z_1,\dots,\vec{\xi}^z_{n-k_1} \,,\, \vec{\eta}^z_1,\dots,\vec{\eta}^z_{k_2} \,,\,
        \vec{\mu}^z_1,\dots,\vec{\mu}^z_k\} \text{ and } \\
&\{\vec{\eta}^x_1,\dots,\vec{\eta}^x_{n-k_1} \,,\, \vec{\xi}^x_1,\dots,\vec{\xi}^x_{k_2} \,,\,
         \vec{\mu}^x_1,\dots,\vec{\mu}^x_k\},
\end{align*}
both spanning $\mathbb{F}_2^n$ and satisfying \eqref{eq:css:nineconditions}, where
$\mathcal{C}_1^\perp=\vspan\{ \vec{\xi}^z_1,\dots,\vec{\xi}^z_{n-k_1} \}$ and
$\mathcal{C}_2=\vspan\{ \vec{\xi}^x_1,\dots,\vec{\xi}^x_{k_2} \}$ are classical linear codes satisfying $\mathcal{C}_2\subseteq \mathcal{C}_1$.
Note that because of lemma \ref{lem:UoUgleich1o1},
\begin{equation}
 \ket{\Phi^+}^{\otimes n}_{AB} =
 U_{\text{enc},A}^\ast \otimes U_{\text{enc},B} \ket{\Phi^+}^{\otimes n}_{AB}  =
 \frac{1}{\sqrt{2^{n-k_1}}} \sum_{\vec{x}\in\mathbb{F}_2^{n-k_1}}
 \frac{1}{\sqrt{2^{k_2}}} \sum_{\vec{z}\in\mathbb{F}_2^{k_2}}
 \frac{1}{\sqrt{2^k}} \sum_{\vec{c}\in\mathbb{F}_2^k}
 \overline{\ket{\vec{x},\vec{z},\vec{c}}}^\ast_A \overline{\ket{\vec{x},\vec{z},\vec{c}}}_B.
\end{equation}
Alice measures her stabilizers $\{ \overline{Z}^\ast_i \}_{i=1,\dots,n-k}$, sends her resulting syndrome $\vec{s}_A = (\vec{x},\vec{z})$ to Bob, who, by measuring his stabilizers $\{ \overline{Z}_i \}_{i=1,\dots,n-k}$, obtains the syndrome $\vec{s}_B=\vec{s}_A+\vec{s}$ and calculates the relative syndrome $\vec{s}$.
Depending on $\vec{s}$, Bob performs error correction.
Eventually, Alice and Bob both measure $\{ \overline{Z}^\ast_i \}_{i=n-k+1,\dots,n}$ and $\{ \overline{Z}_i \}_{i=n-k+1,\dots,n}$, respectively, to obtain the $k$ bit key.
(Alternatively they might also decode, obtain $\ket{\Phi^+}_{AB}^{\otimes k}$, and measure in the z-basis to obtain the key.)

\subsubsection{Protocol based on Quantum Error Correction}

Since Alice might perform her measurements immediately after the preparation of $\ket{\Phi^+}_{AB}^{\otimes N}$, the following procedure is equivalent:
She chooses the syndrome $(\vec{x},\vec{z})$, the key $\vec{c}$, and the values of the check bits at random.
Then she prepares the $n$-qubit state $\overline{\ket{\vec{x},\vec{z},\vec{c}}}$ and inserts the $n$ check states prepared as $\ket{0}_z$ or $\ket{1}_z$ in random positions.
After choosing a random string $\vec{b}=(b_1,\dots,b_N)\in\mathbb{F}_s^N$,
she applies the operation $T^{b_i}$ onto the $i$-th qubit, and sends her $N=2n$ qubits to Bob.
Bob acknowledges the reception of the qubits.
Alice tells Bob the string $\vec{b}$ together with the positions of the check qubits.
Bob applies the operation $T^{-b_i}$ onto his $i$-th qubit.
He measures the check qubits in the z-basis, they share their check bit data, and, as a result, obtain the bit-error rate $p$.
At this point Bob is left with the state
$\mathcal{E}_{\text{eff}}^{\otimes n}\bigl( \overline{\ket{\vec{x},\vec{z},\vec{c}}} \overline{\bra{\vec{x},\vec{z},\vec{c}}} \bigr)$.
Alice tells him the syndrome $\vec{s}_A = (\vec{x},\vec{z})$, and Bob knows that the key is encoded in the codespace $\mathcal{C}\bigl(L(\mathcal{C}_1,\mathcal{C}_2),\vec{s}_A\bigr)$ of the CSS code.
He applies the appropriate recovery operation $\mathcal{R}_{\vec{s}_A}$ by measuring the stabilizers
$\{ \overline{Z}_i \}_{i=1,\dots,n-k}$ followed by error correction.
Eventually, Bob measures $\{ \overline{Z}_i \}_{i=n-k+1,\dots,n}$ to obtain the $k$ bit key.

The rate $k/n$ of the key they can generate this way depends only on the form of the memoryless Pauli channel $\mathcal{E}_\text{eff}$ which in turn depends only on the bit error rate $p$.
Hence, lower bounds on the rates are given by theorem \ref{thm:randcss} which states that, as long as
\begin{equation}\label{eq:spproofrate}
 \frac{k}{n} < 1 - H_{4[\log_2]} ( \{p_{uv}\} ),
\end{equation}
and for large enough $n$, there exists
a pair of codes $\mathcal{C}_2 \subset \mathcal{C}_1$ such that
for any codespace $\mathcal{C}_{(L(\mathcal{C}_1,\mathcal{C}_2),\vec{s})}$ of the corresponding CSS code with stabilizer $L(\mathcal{C}_1,\mathcal{C}_2)$,
there exists a recovery operation with minimum fidelity larger than $1-\varepsilon$ for any $\varepsilon>0$.
To obtain higher rates, they might also use concatenated CSS codes as it was done by Lo \cite{Lo01} (see the second remark following theorem \ref{thm:concrate}).
\begin{rem}
In the case of the BB84 protocol the set $\{p_{uv}\}$ is not completely known and we have to assume the worst case, i.\,e. we have to minimize the key rates over the unknown parameter $t\in[0,p]$.
\end{rem}

\subsubsection{BB84 and 6-state Protocol}
Finally we are going to show that the protocol based on quantum error correction is equivalent to the BB84 and the 6-state protocol, respectively.
The crucial observation is that the recovery operation for CSS codes decomposes into bit and phase error correction.
Since Bob obtains the key by measuring the operators $\{ \overline{Z}_i \}_{i=n-k+1,\dots,n}$, where $\overline{Z}_{n-k+j} = \XZ(\vec{0},\vec{\mu}^z_j)$ for $j=1\dots k$, he does not need to perform phase error correction.
Hence, he only needs to know the absolute bit syndrome $\vec{x}$ and the relative bit syndrome obtained by measuring the $\overline{Z}_j = \XZ(\vec{0},\vec{\xi}^z_j)$, $j=1\dots n-k_1$.
To obtain his measurement results, he might simply measure all qubits in the $z$-Basis, obtain a string $\vec{y}\in\mathbb{F}_2^n$ and reconstruct them via $\vec{\mu}^z_j\cdot \vec{y}$, $j=1\dots k$, and $\vec{\xi}^z_j\cdot \vec{y}$, $j=1\dots n-k_1$, respectively.
Alice, who in turn does not need to send the phase error syndrome $\vec{z}$, prepares on average the state
\begin{align}
 \frac{1}{2^{k_2}} \sum_{\vec{z}\in\mathbb{F}_2^{k_2}}
\overline{\ket{\vec{x},\vec{z},\vec{c}}}\overline{\bra{\vec{x},\vec{z},\vec{c}}}
 &=
\frac{1}{\vert \mathcal{C}_2\vert} \sum_{\vec{\mathfrak{v}}_1,\vec{\mathfrak{v}}_2\in\mathcal{C}_2}
 \frac{1}{2^{k_2}} \sum_{\vec{z}\in\mathbb{F}_2^{k_2}} (-1)^{\vec{\mathfrak{z}}\cdot(\vec{\mathfrak{v}}_1-\vec{\mathfrak{v}}_2)}
\ket{\vec{\mathfrak{v}}_1+\vec{\mathfrak{c}}+\vec{\mathfrak{x}}}\bra{\vec{\mathfrak{v}}_1+\vec{\mathfrak{c}}+\vec{\mathfrak{x}}} \nonumber\\
&=
\frac{1}{\vert \mathcal{C}_2\vert} \sum_{\mathfrak{v}\in\mathcal{C}_2}
\ket{\vec{\mathfrak{v}}+\vec{\mathfrak{c}}+\vec{\mathfrak{x}}}\bra{\vec{\mathfrak{v}}+\vec{\mathfrak{c}}+\vec{\mathfrak{x}}},
\end{align}
where $\vec{\mathfrak{x}}$, $\vec{\mathfrak{z}}$ and $\vec{\mathfrak{c}}$ had been defined in \eqref{eq:css:encodedstatesdefi} as
\begin{align}
\vec{\mathfrak{x}} &= \sum_{i=1}^{n-k_1} x_i \vec{\eta}^x_i,  &
\vec{\mathfrak{z}} &= \sum_{i=1}^{k_2} z_i \vec{\eta}^z_i,  &
\text{ and }
\vec{\mathfrak{c}} &= \sum_{i=1}^k c_i \vec{\mu}^x_i.
\end{align}
Note that $\vec{\mathfrak{v}}+\vec{\mathfrak{c}} \in \mathcal{C}_1$ and
$\vec{\mathfrak{v}}+\vec{\mathfrak{c}}+\vec{\mathfrak{x}} \in \mathbb{F}_2^n$
so that Alice just prepares a sequence of $n$ random states taken from the set $\{ \ket{0}_z, \ket{1}_z \}$.

In summary, we have the following secure protocol:
Alice and Bob implement the corresponding QKD protocol as described in subsection \ref{subsec:6sbb84}.
As a result they end up with Alice having the $n$ bits $\vec{x}'_\text{sifted}$, Bob having the $n$ bits $\vec{y}'_\text{sifted}$, and both knowing the bit error rate $p$.
They decide on a CSS code encoding $k$ qubits into $n$ which is able to correct the memoryless Pauli channel  $\mathcal{E}_\text{eff}^{\otimes n}$ characterized by the probability distribution of equation \eqref{eq:probdistrieff}.
Alice interprets $\vec{x}'_\text{sifted}$ as $(\vec{\mathfrak{v}}+\vec{\mathfrak{c}})+\vec{\mathfrak{x}}$
with random $(\vec{\mathfrak{v}}+\vec{\mathfrak{c}})\in\mathcal{C}_1$ and random syndrome $\vec{\mathfrak{x}}$, and tells Bob the syndrome.
Bob's data $\vec{y}'_\text{sifted}$ can be written as the sum of Alice's string plus an error, $\vec{y}'_\text{sifted}=\vec{x}'_\text{sifted}+\vec{e}$. Bob subtracts the syndrome, obtains $(\vec{\mathfrak{v}}+\vec{\mathfrak{c}})+\vec{e}$, and performs bit error correction with the classical code $\mathcal{C}_1$ to obtain $(\vec{\mathfrak{v}}+\vec{\mathfrak{c}})$.
To obtain the key, he extracts the coset of $\mathcal{C}_2$ in $\mathcal{C}_1$, $\vec{\mu}^z_i\cdot(\vec{\mathfrak{v}}+\vec{\mathfrak{c}}) = \vec{\mu}^z_i \cdot \vec{\mathfrak{c}} = c_i$.
The last step can be viewed as privacy amplification: The correct $k_1$ bits included in $(\vec{\mathfrak{v}}+\vec{\mathfrak{c}})$ are shrunk into $k=k_1-k_2$ private bits.

\section{Combined Preprocessing}\label{sec:combiprepro}

\begin{figure}
 \centering
 \includegraphics[scale=0.8]{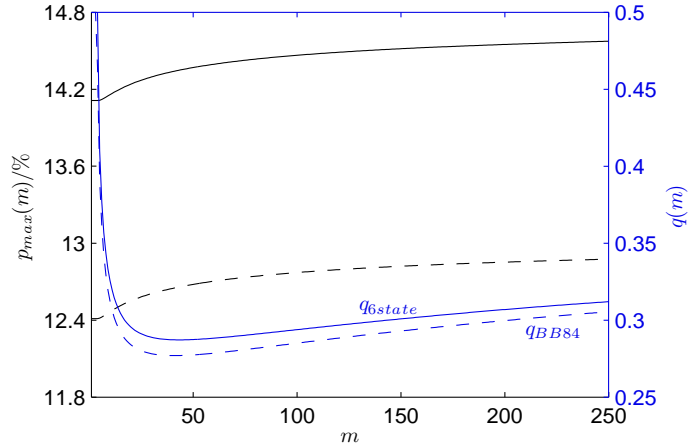}
 \caption{
Maximum tolerable bit error rate $p_\text{max}$ \textit{(left y-axis, black)} and the corresponding rate $q$ of the added noise for which it is achieved \textit{(right y-axis, blue)} versus block length $m$.
Dashed lines correspond to the BB84 protocol, solid lines to the 6-state protocol.
 \label{fig:PmaxOverM}}
\end{figure}

The preprocessing protocol proposed in \cite{SRS06} combines local randomization with the use of a degenerated quantum code.
It begins after Bob has received the quantum signals from Alice and they have sifted their raw keys to throw out mismatches between the preparation and measurement basis.
Alice then flips each of her sifted key bits $(x_1,\dots,x_n)$ with probability $q$, resulting in new bits $(\tilde{x}_1,\dots,\tilde{x}_n)$.
These are partitioned into blocks of size $m$, and for each block she computes the syndrome $(\tilde{x}_1\oplus\tilde{x}_2,\tilde{x}_1\oplus\tilde{x}_3,\dots,\tilde{x}_1\oplus\tilde{x}_m)$ and sends this information to Bob.
He computes the \emph{relative} syndrome of their blocks by adding his corresponding syndrome to Alice's, modulo two.
Alice's message is public knowledge, but the first bit of each block is still secret, so it is kept as a potential key bit.
The protocol then proceeds with the usual error correction and privacy amplification steps to transform these kept bits into a secret key, now aided by the relative syndrome of each block and knowledge of the probability $q$ of local randomization.
Without local randomization, it turns out that $m=5$ is the optimal blocklength for improving the error threshold in the 6-state protocol --- longer blocklengths have worse thresholds (compare with figure \ref{fig:pmaxcatcode} of section \ref{sec:concexamples}).
However, the results in \cite{SRS06} indicate that with the addition of noise, the highest tolerable bit error rate of BB84 grows with the blocksize $m$, and we find a similar result in the 6-state case
(see figure \ref{fig:PmaxOverM}).

\subsection{Security Proof}

We determine the secure key rates of the BB84 and 6-state one-way key distillation protocols involving the preprocessing protocol described above using the security proof of Renner \cite[corollary 6.5.2]{phdrenner}.
This proof states that the secure key rate of such a protocol is given by
\begin{equation}\label{eq:rennerrate}
 r = \frac{1}{m} \min_{\sigma_{AB} \in \Gamma} \bigl( S(X|E) - S(X|Y) \bigr)
\end{equation}
where the minimum ranges over the set of states $\Gamma$ of all density operators on the $2\times 2$ dimensional Hilbert space $\mathcal{H}_A \otimes \mathcal{H}_B$ such that the measurement performed during the parameter estimation phase of the protocol leads to a certain bit error rate $p$.
The conditional von Neumann entropies in \eqref{eq:rennerrate} are calculated for the states
\begin{equation}\label{eq:sgima_xye}
 \sigma_{XY\overline{E}} = \mathcal{E}_{XY\overline{E}\leftarrow A^mB^mE^m} (\sigma_{ABE}^{\otimes m} )
\end{equation}
which describe the processing of each block, including local randomization and syndrome calculation, 
and eventual measurement of the output qubits of the repetition code. That is, the preprocessing is treated
quantum-mechanically or coherently, but the usual processing classically.
Here $X$ denotes Alice's key outcome when measuring the output bits and $Y$ Bob's key and syndrome outcomes. 

For the BB84 protocol the set $\Gamma$ contains the states
\begin{equation}\label{eq:sigma_ab}
 \sigma_{AB} = \sum_{u,v} p_{uv} \opl{\XZ}{B}(u,v) \ketbra{\Phi^+}{\Phi^+} \XZ_B^\dagger(u,v),
\end{equation}
where $\ket{\Phi^+}_{AB}=\frac{1}{\sqrt{2}}\sum_k \ket{kk}_{AB}$ and
$\{ p_{uv} \} \equiv \{p_{00},p_{10},p_{11},p_{01}\} = \{1-2p+t,p-t,t,p-t\}$, $t\in [0,p]$.
In the 6-state protocol, meanwhile, parameter estimation assures us that $\Gamma$ contains only the single state $\sigma_{AB}$ with $\{ p_{uv} \} = \{ 1-\frac{3}{2}p, \frac{p}{2}, \frac{p}{2}, \frac{p}{2}\}$.

Using Renner's proof allows us to include the preprocessing but still only minimize over the quantum states $\sigma$ corresponding to individual signals.
The crucial simplification is that the quantum state of the block can be taken to be the product $\sigma^{\otimes m}$ without loss of generality.\
Other proof techniques would require minimization over all possible (potentially-entangled) block states, or an additional step in the parameter estimation procedure to ensure that the state does have this power form.

\subsection{Computation of the Secure Key Rate}\label{subsec:combi:compuseckeyrate}

To compute the secure key rates we make use of the fact that the difference of entropies in \eqref{eq:rennerrate} can also be written as difference of corresponding quantum mutual informations, i.\,e.~$S(X|E) - S(X|Y) = I(X:Y)-I(X:E)$.
In order to calculate these quantities, we need to determine the states $\sigma_{XY\overline{E}}$ defined in \eqref{eq:sgima_xye} for both protocols, i.\,e. for $\sigma_{AB}$ being a member of the two different sets $\Gamma$ defined in the paragraph including equation \eqref{eq:sigma_ab}.
We are going to perform the rate calculation for a general $\sigma_{AB}$ and specialize in the two different cases in the succeeding subsections.

An $m$-fold tensor product of a purification of a general Bell diagonal $\sigma_{AB}$ is given by
\begin{equation}\label{eq:sigma_abe}
\ket{\sigma}_{ABE} \equiv \ket{\sigma}^{\otimes m}_{ABE_1E_2} =
\sum_{\vec{u},\vec{v}} \sqrt{ p_{\vec{u},\vec{v}} }  \,\XZ_B(\vec{u},\vec{v})
\ket{\Phi^+}^{\otimes m}_{AB} \ket{\vec{u}}_{E_1}\ket{\vec{v}}_{E_2},
\end{equation}
where $p_{\vec{u},\vec{v}} = \prod_{i=1}^m p_{u_i,v_i}$.
We now need to calculate the state resulting from noisy preprocessing followed by a blockwise stabilizer code measurement in which the stabilizers contain Pauli $\id$ and $Z$ operators only.

\subsubsection{Local Randomization}
The first step, local randomization, can be described in a coherent way by adding a classical register $\mathbf{A'}$ (such systems will be denoted with boldface type) in the state
$\bigl( (1-q)\ketbra{0}{0}+q\ketbra{1}{1} \bigr)^{\otimes m}$ and then applying controlled not gates from the individual register states to the bits $A$.
This leads to
\begin{equation}\label{eq:sigmap}
\ket{\sigma'}_{ABE} =
 \sum_{\vec{u},\vec{v},\vec{f}} \sqrt{ p_{\vec{u},\vec{v}} q_{\vec{f}} }
 \,\XZ_B(\vec{u}+\vec{f}, \vec{v})
 \ket{\Phi^+}^{\otimes m}_{AB} \ket{\vec{f}}_{\mathbf{A'}} \ket{\vec{u}}_{E_1} Z^{\vec{f}}_{E_2}\ket{\vec{v}}_{E_2},
\end{equation}
where $\vec{f}\in\mathbb{F}_2^m$ and $q_{\vec{f}}=q^f(1-q)^{m-f}$ for $f=|\vec{f}|$,
the number of 1s in $\vec{f}$, a notation we shall use throughout.
Here we have used the fact that $X_A\ket{\Phi^+}_{AB} = X_B\ket{\Phi^+}_{AB}$ (compare with lemma \ref{lem:OTo1gleich1oO}) to simplify the expression;
this move is responsible for the $Z^{\vec{f}}$ operation applied to $E_2$.

\subsubsection{Syndrome Measurement}
In the second step, Alice and Bob both measure the $m-1$ (generators of the) stabilizers of a $\id/Z$-only stabilizer code which encodes one logical qubit into $m$ physical qubits.
Using a public (authenticated) channel, Alice sends her syndrome to Bob who calculates the relative syndrome $\vec{s}$ by adding Alice's string to his measurement outcome modulo two.
Afterwards both decode their encoded state.
Such a stabilizer code is a CSS code constructed from classical linear codes $\mathcal{C}_2\subset \mathcal{C}_1$, where $\mathcal{C}_2=\{\vec{0}\}$ contains only the zero codeword and $\mathcal{C}_1=\{\vec{0},\vec{\mu}^x_1\}$ is spanned by a single codeword $\vec{\mu}^x_1$ (compare with section \ref{sec:csscodes}).
Together with an encoding
$U_\text{enc} \ket{ \vec{e},c } = \overline{ \ket{ \vec{e},c } }$,
where
\begin{equation}\label{eq:encodedstabstate}
 \overline{\ket{\vec{e},c}} = \bigl\vert c\cdot\vec{\mu}^x_1 + \sum_{j=1}^{m-1} e_j\cdot \vec{\eta}^x_j \bigr\rangle,
\end{equation}
our CSS code is completely specified by defining two bases
$\{ \vec{\xi}^z_1,\dots,\vec{\xi}^z_{m-1}, \vec{\mu}^z_1 \}$ and
$\{ \vec{\eta}^x_1,\dots,\vec{\eta}^x_{m-1}, \vec{\mu}^x_1 \}$ both spanning $\mathbb{F}_2^m$
and satisfying condition \eqref{eq:css:nineconditions} (see section \ref{subsec::cssencod}).
In this case the stabilizers are given by
$\overline{Z}_i = \XZ(\vec{0},\vec{\xi}^z_i)$, $i=1\dots m-1$, and a measurement of these stabilizers on the encoded state \eqref{eq:encodedstabstate} will give the syndrome $\vec{e}$.
Measurement of the logical $Z$ operator $\overline{Z}_m = \XZ(\vec{0},\vec{\mu}^z_1)$ gives the value of the encoded bit $c$.
Applying one of the $\overline{X}_i = \XZ(\vec{\eta}^x_i,\vec{0})$, $i=1\dots m-1$, operators on a encoded state results in a flip of the $i$-th bit of the syndrome, while applying the logical $X$ operator $\overline{X}_m = \XZ(\vec{\mu}^x_1,\vec{0})$ flips the encoded bit, $c\mapsto c\oplus 1$.
Both the set of all $\overline{Z}_i$ and the set of all $\overline{X}_j$ are complete sets of commuting observables.
Note that because of lemma \ref{lem:UoUgleich1o1},
\begin{equation}
 \ket{\Phi^+}^{\otimes m}_{AB} =
 U_{\text{enc},A}^\ast \otimes U_{\text{enc},B} \ket{\Phi^+}^{\otimes m}_{AB}
  = \frac{1}{\sqrt{2^{m-1}}} \sum_{\vec{e}\in\mathbb{F}_2^{m-1}} \frac{1}{\sqrt{2}} \sum_{c\in\mathbb{F}_2}
 \overline{\ket{\vec{e},c}}^\ast_A \overline{\ket{\vec{e},c}}_B.
\end{equation}
In other words, the maximally-entangled state of $m$ physical qubits is the equal superposition
of a logical maximally-entangled state in all the possible encodings.
Also note that lemma \ref{lem:xzdecompocss} tells us that any $m$ fold Pauli operator can be decomposed as
\begin{equation}
\XZ(\vec{u}',\vec{v})  =  \overline{X}_m^{l^x}\overline{Z}_m^{l^z} \,
\prod_{i=1}^{m-1} \, \overline{X}_i^{s^x_i} \overline{Z}_i^{n^x_i},
\end{equation}
where $s^x_i = \vec{\xi}^z_i \cdot \vec{u}'$, $n^x_i = \vec{\eta}^x_i  \cdot \vec{v}$, and
$l^x =\vec{\mu}^z_1 \cdot \vec{u}'$ and $l^z =\vec{\mu}^x_1 \cdot \vec{v}$ are the logical bit and phase flip errors resulting when this Pauli operator is applied to an encoded state like \eqref{eq:encodedstabstate}.
Using these two facts we find that, after Bob's calculation of the relative syndrome $\vec{s}$, the tripartite state can be expressed as (up to a local unitary acting only on Eve's systems)
\begin{equation}\label{eq:sigmapp}
\ket{\sigma''}_{ABE}=
  \sum_{\vec{u},\vec{v},\vec{f}} \sqrt{ p_{\vec{u},\vec{v}} q_{\vec{f}} } \,
 \XZ_B\bigl( \vec{\mu}^z_1  \cdot (\vec{u}+\vec{f}) , \vec{\mu}^x_1 \cdot \vec{v} \bigr)
 \ket{\Phi^+}_{AB} \ket{\vec{f}}_{\mathbf{A'}}
 \ket{\vec{u}}_{E_1} Z^{\vec{f}}_{E_2}\ket{\vec{v}}_{E_2}
 \ket{ \vec{s} }_{\mathbf{B'}},
\end{equation}
where
$\vec{s}=( \vec{\xi}^z_1\cdot(\vec{u}+\vec{f}),\dots,\vec{\xi}^z_{m-1}\cdot(\vec{u}+\vec{f}) )$.
While the registers $A$ and $B$ in equation \eqref{eq:sigmap} have been $m$-qubit registers, here they contain only a single qubit each.
Alice missing $(m-1)$-qubits have been traced out since they contained only classical information about her absolute syndrome (accessible to all parties).
The rest of Bob's $m$-qubit register now contains classical information about the relative syndrome $\vec{s}$ and is labeled $\mathbf{B'}$.

\subsubsection{Key Bit Measurement}
Finally, Alice and Bob both measure their key bit.
Alice forgets about which bits she flipped by tracing out the $\mathbf{A'}$ register.
The correlations between Alice, Bob, and Eve are described by the following semiclassical state:
\begin{multline}\label{eq:sigmaxye}
 \sigma_{XY\overline{E}} =
 \frac{1}{2} \sum_x [x]_A \otimes
 \sum_{\vec{u},\vec{f}} \sum_{\vec{v}_1,\vec{v}_2}
 \sqrt{ p_{\vec{u},\vec{v}_1} p_{\vec{u},\vec{v}_2} }  q_{\vec{f}} \,
 [x+\vec{\mu}^z_1 \cdot(\vec{u}+\vec{f})]_B    \\
\otimes\, [ \vec{s} ]_{B'} \otimes  [\vec{u}]_{E_1} \otimes
 (Z^{\vec{\mu}^x_1})^x_{E_2} Z^{\vec{f}}_{E_2}
     \ketbra{\vec{v}_1}{\vec{v}_2}
 Z^{\vec{f}}_{E_2} (Z^{\vec{\mu}^x_1})^x_{E_2},
\end{multline}
where $[x]_B=\ket{x}\bra{x}_B$, etc.
Note that the state is diagonal in $E_1$ since the quantities $\vec{\xi}^z_i\cdot(\vec{u}+\vec{f})$, $i=1,\dots,m-1$, and $\vec{\mu}^z_1\cdot(\vec{u}+\vec{f})$ are all classical:
The former are already classical in \eqref{eq:sigmapp}, the latter became classical after the key bit measurements by Alice and Bob.
The $\{ \vec{\xi}^z_1,\dots,\vec{\xi}^z_{m-1},\vec{\mu}^z_1 \}$ span $\mathbb{F}_2^m$ thereby completely fixing the string $\vec{u}+\vec{f}$.

\subsubsection{The Mutual Information between Alice and Bob and Alice and Eve}
To calculate the quantum mutual information between Alice and Bob we trace out Eve and obtain
\begin{align}
 \sigma_{XY} &=
 \frac{1}{2} \sum_x [x]_A \otimes
 \sum_{\vec{u},\vec{f}} p_{\vec{u}} q_{\vec{f}} \,
 [x+\vec{\mu}^z_1 \cdot(\vec{u}+\vec{f})]_B \otimes
 \bigl[  \bigl(\vec{\xi}^z_1 \cdot(\vec{u}+\vec{f}),\dots\bigr) \bigr]_{B'} \nonumber\\
 &=
 \frac{1}{2} \sum_x [x]_A \otimes
 \sum_{\vec{u}} \tilde{p}_{\vec{u}}  \,
 [x+\vec{\mu}^z_1 \cdot\vec{u}]_B \otimes
 [  (\vec{\xi}^z_1\cdot\vec{u},\vec{\xi}^z_2\cdot\vec{u},\dots) ]_{B'} \nonumber\\
 &=\frac{1}{2} \sum_x [x]_A \otimes \sum_{l^x, \vec{s}} \tilde{P}(l^x,\vec{s})
   [x+l^x]_B \otimes [ \vec{s} ]_{B'},
\end{align}
where $\tilde{p}_{\vec{u}}$ is defined as $\tilde{p}_{\vec{u}}=\tilde{p}^u(1-\tilde{p})^{m-u}$ with $\tilde{p}=p(1-q)+(1-p)q$.
In the last step we used $\vec{u} = l^x \vec{\mu}^x_1 +\sum_{i=1}^{m-1} s_i \vec{\eta}^x_i$ to write the sum over $\vec{u}$ as a sum over $l^x$ and $\vec{s}$, where $l^x$ is the logical $X$ error, i.e.~$X$ error on the first qubit in the block; i.\,e. we have $\tilde{P}(l^x,\vec{s}) = \tilde{p}_{\vec{u}(l^x,\vec{s})}$.
This immediately yields
\begin{equation}\label{eq:iab}
 I(X:Y) = 1 - \sum_{\vec{s}\in\mathbb{F}_2^{m-1}} \tilde{P}(\vec{s}) H_2(  \tilde{P}(l^x \vert \vec{s}) ),
\end{equation}
using the binary entropy $H_2(x) = -x\log_2 x -(1-x)\log_2(1-x)$.
Note that $I(X:Y)$ does not depend on the particular values $\{p_{uv}\}$ in $\sigma_{AB}$ (see \eqref{eq:sigma_ab}),
but only depends on the bit error rate $p=p_{10}+p_{11}$.
The form of the mutual information indicates the advantage provided by the syndrome.
If Alice did not send any information, Bob's state would be averaged over the possible syndromes, and
the mutual information would involve the entropy of the average of the $\tilde{P}(l^x\vert \vec{s})$ rather
than the average of the entropies. By concavity of entropy, the latter rate is larger.

To calculate the quantum mutual information between Alice and Eve, we trace out Bob's systems and obtain
\begin{align}
 \sigma_{X\overline{E}} &=  \frac{1}{2} \sum_x [x]_A \otimes \rho_{E_1E_2}^{(x)}, \label{eq:sigma_xe1}\\
 \rho_{E_1E_2}^{(x)} &=
 \sum_{\vec{u}} p_{\vec{u}} \, [\vec{u}]_{E_1} \otimes \rho_{E_2}^{(x),\vec{u}} , && \text{and} \label{eq:14}\\
 \rho_{E_2}^{(x),\vec{u}} &=
 (Z^{\vec{\mu}^x_1})^x
 \sum_{\vec{f}} q_{\vec{f}}
   Z^{\vec{f}}  \ketbra{ \Psi_{\vert\vec{u}} }{ \Psi_{\vert\vec{u}} }  Z^{\vec{f}}
 (Z^{\vec{\mu}^x_1})^x, \label{eq:sigma_xe3}
\end{align}
with
\begin{equation}
\ket{ \Psi_{\vert\vec{u}} } = \sum_{\vec{v}}\sqrt{ p_{\vec{v}\vert \vec{u}} } \ket{\vec{v}}.
\end{equation}
It follows that the quantum mutual information between Alice and Eve is given by
\begin{equation}\label{eq:iae:general}
 I(X:E) = \sum_{\vec{u}\in\mathbb{F}_2^m} p_{\vec{u}} \Bigl[ S\Bigl(\frac{1}{2}\rho_{E_2}^{(0),\vec{u}} + \frac{1}{2} \rho_{E_2}^{(1),\vec{u}}\Bigr) - S\Bigl(\rho_{E_2}^{(0),\vec{u}}\Bigr) \Bigr].
\end{equation}

\begin{figure}
\centering
\scalebox{0.75}{
\begin{pspicture}(-1.5,0.5)(7.25,-2.75)
\psframe[linecolor=lightgray,linestyle=dotted](-0.25,0.25)(2.75,-1.75) %
\pspolygon[linecolor=lightgray,linestyle=dashed](-0.5,0.5)(3,0.5)(3,-1.75)(6.25,-1.75)(6.25,-2.75)(-0.5,-2.75) %
\psframe[linecolor=lightgray,fillstyle=solid](0,0)(2.5,-1.5)
\rput[c](1.25,-0.75){
$\begin{array}{cccc}
 Z & Z & \id & \id \\
 Z & \id & Z & \id \\
 Z & \id & \id & Z
 \end{array}$}
\psframe[linecolor=lightgray,fillstyle=solid](0,-2)(2.5,-2.5)
\rput[c](1.25,-2.25){
$\begin{array}{cccc}
  Z & \id & \id & \id
 \end{array}$}
\psframe[linecolor=lightgray,fillstyle=solid,fillcolor=mygray](3.5,0)(6,-1.5)
\rput[c](4.75,-0.75){
$\begin{array}{cccc}
 \id & X & \id & \id \\
 \id & \id & X & \id \\
  \id & \id & \id & X
 \end{array}$}
\psframe[linecolor=lightgray,fillstyle=solid,fillcolor=mygray](3.5,-2)(6,-2.5)
\rput[c](4.75,-2.25){
$\begin{array}{cccc}
 X & X & X & X
 \end{array}$}
\rput[r](-0.05,-0.75){$m-1 \left\{ \makebox(0,0.9)[b]{}\right.$}
\rput[B](1.25,0.6){$m$}
\rput[B](1.25,0.1){\rotatebox[origin=c]{-90}{$\left\{ \makebox(0,1.35)[b]{}\right.$}}
\end{pspicture}}%
\caption['standard' cat code]{\label{fig:cat}
Cat code encoding one qubit into $m=4$.
The operators on the left hand side are the
$\{ \XZ(\vec{0},\vec{\xi}^z_i) \}$ ($i=1\dots m-1$ from \textit{top} to \textit{bottom})
and $\XZ(\vec{0},\vec{\mu}^z_1)$, those on the right hand side are
$\{ \XZ(\vec{\eta}^x_i,\vec{0}) \}$ and $\XZ(\vec{\mu}^x_1,\vec{0})$.
The (generators of the) stabilizers are within the dotted line, the (generators of the) normalizers within the dashed one.}
\end{figure}
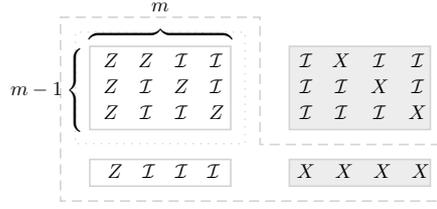

We now restrict ourselves to the cat code presented in subsection \ref{subsec:thecatcode}, which is given by
$(\vec{\xi}^z_i)_j=\delta_{1j}+\delta_{i+1,j}$ for $i=1\dots m-1$, $(\vec{\mu}^z_1)_j=\delta_{1j}$ and
$(\vec{\eta}^x_i)_j=\delta_{i+1,j}$ for $i=1\dots m-1$, $(\vec{\mu}^x_1)_j=1$ (see figure \ref{fig:cat} which is the same as figure \ref{fig:cat-code}a).
This code leads to the correct coherent description of the syndrome calculation of the combined preprocessing scheme.
The name comes from the fact that
$\alpha \overline{\ket{\vec{0},0}} + \beta \overline{\ket{\vec{0},1}} =
 \alpha \ket{00\dots0} + \beta \ket{11\dots1}$, a Schr\"odinger cat state when $\alpha=\beta=\frac{1}{\sqrt{2}}$.
For the cat code we obtain the probability distribution $\tilde{P}(l^x,\vec{s})$ in the mutual information between Alice and Bob by summing equation \eqref{eq:probcatcode} over $l^z \in \{0,1\}$,
\begin{align}
P(l^x,\vec{s}) &=
(p_{10}+p_{11})^{l^x(m-2s)+s} (1-p_{10}-p_{11})^{(1-l^x)(m-2s)+s}  \nonumber \\
&=(p^s(1-p)^{m-s})^{1-l^x} (p^{m-s}(1-p)^s)^{l^x},
\end{align}
and by replacing $p$ with $\tilde{p}$.

We proceed with the computation of the mutual information between Alice and Eve
for the BB84 and the 6-state protocol separately in the following two subsections.
Before we step into these calculations, let us examine the special case $q=0$ which can be treated without specifying the protocols:
In expression \eqref{eq:iab} for the mutual information between Alice and Bob we simply have to replace $\tilde{p}$ with $p$.
To calculate the mutual information between Alice and Eve given by \eqref{eq:iae:general}, we note that $\rho_{E_2}^{(x),\vec{u}}$ is now a pure state.
Using the fact that
$\bra{\Psi_{\vert\vec{u}}} Z^{\vec{\mu}^x_1} \ket{\Psi_{\vert\vec{u}}} = 1-2P(l^z=0\vert\vec{u})$, we find that
\begin{equation}
 I_{q=0}(X:E) = \sum_{\vec{s}\in\mathbb{F}_2^{m-1},l^x\in\mathbb{F}_2} P(\vec{s},l^x) H_2\bigl( \{P(l^z\vert l^x,\vec{s})\} \bigr).
\end{equation}
Hence, we have the following theorem which already emerged as a result of Shor and Preskill's security proof in subsection \ref{subsec:spproof}.
\begin{thm}
The secure key rate of the BB84 protocol [6-state protocol] involving only the syndrome calculation part of the combined preprocessing scheme is given by
\begin{equation}\label{eq:seckeyrateq=0}
 r_{q=0}(m,p) = \frac{1}{m} \min_{\sigma_{AB} \in \Gamma} \Bigl(
 1 - \sum_{\vec{s}\in\mathbb{F}_2^{m-1}} P(\vec{s}) \, H_{4[\log_2]}\bigl( \{P(l^x,l^z\vert \vec{s})\} \bigr)
 \Bigr),
\end{equation}
where $P(l^x,l^z\vert \vec{s}) = P(l^x,l^z,\vec{s}) / P(\vec{s})$ is the conditional error probability for the cat code, the joint probability $P(l^x,l^z,\vec{s})$ of which is given by equation \eqref{eq:probcatcode},
and the set $\Gamma$ contains the Bell diagonal states characterized by the probability distribution
$\{p_{uv}\}=\{1-2p+t,p-t,t,p-t\}_{t\in [0,p]}$
[$\{p_{uv}\}=\{ 1-\frac{3}{2}p, \frac{p}{2}, \frac{p}{2}, \frac{p}{2}\}$].
\end{thm}

\begin{rem}[i]
Note that (apart from the minimization) the secure rate of the above theorem is exactly the rate at which we can send quantum information reliably over a Pauli channel characterized by the probability distribution $\{p_{uv}\}$ when using a concatenation of a random outer CSS code with an inner cat code (see theorem \ref{thm:concrate} and the following remarks).
Therefore, as we already mentioned in subsection \ref{subsec:thecatcode},
results on the maximum tolerable noise of the qubit depolarizing channel characterized by $\{ 1-p, \frac{p}{3}, \frac{p}{3}, \frac{p}{3}\}$ can be applied to the 6-state protocol if the factor $2/3$ is taken into account \cite{Lo01}.
In particular it was shown in subsection \ref{subsec:thecatcode} that the highest robustness is obtained for $m=5$ leading to maximal tolerable bit error rate of $p_\text{max}^\text{6-st.}(m=5,q=0)=\SSPd$.
As it will be shown later, the minimum for the BB84 protocol is achieved for independent errors, $\{(1-p)^2,p(1-p),p^2,p(1-p)\}$, and it turns out that the optimal block length is $m=7$ leading to $p_\text{max}^\text{BB84}(m=7,q=0)=11.2107\%$.
\end{rem}

\begin{rem}[ii]
If we use no preprocessing at all, we obtain the secure key rates from \eqref{eq:seckeyrateq=0} by setting $m=1$,
\begin{equation}
 r_{q=0}(m=1,p) = \min_{\sigma_{AB} \in \Gamma} \Bigl(1 - H_{4[\log_2]}\bigl( \{ p_{uv} \} \bigr) \Bigr).
\end{equation}
If we leave aside the minimization, this is exactly the rate at which we can send quantum information reliably over a Pauli channel characterized by the probability distribution $\{ p_{uv} \}$ when using a random CSS code (see theorem \ref{thm:randcss} and \eqref{eq:spproofrate}).
For the BB84 protocol, the minimum is achieved for independent errors %
and we obtain the rate \cite{SP00}
\begin{equation}\label{eq:ratebb84:m=1:q=0}
 r_\text{SP}(p) = 1 - 2 H_2(p).
\end{equation}
Secure key generation becomes impossible for bit error rates higher than $p_\text{max}^\text{BB84}(m=1,q=0)=\BBP$.
For the 6-state protocol, the minimization is obsolete. We obtain the rate \cite{Lo01}
\begin{equation}\label{eq:rate6state:m=1:q=0}
 r_\text{Lo}(p) = 1 - H_2( 3p/2 ) + \frac{3p}{2}\log_2 3,
\end{equation}
and secure key generation becomes impossible for bit error rates higher than $p_\text{max}^\text{6-st.}(m=1,q=0)=\SSP$.
\end{rem}

\subsubsection{BB84}\label{sec:bb84}
To calculate the secure key rate of the combined preprocessing scheme for the BB84 protocol, we must find the minimum over all $\sigma_{AB}$ of the difference between the quantum mutual information between Alice and Bob and Alice and Eve.
Since $I(X:Y)$ does not depend on the particular structure of $\{p_{uv}\} = \{1-2p+t,p-t,t,p-t\} $, $t\in[0,p]$, in $\sigma_{AB}$, but only depends on the bit error rate $p=p_{10}+p_{11}$,
this corresponds to finding the maximum of $I(X:E)$.
Let us assume for a moment that this maximum is achieved for independent bit and phase errors, i.\,e.
we consider the state $\sigma_{AB}$ with $\{ p_{uv} \} = \{1-2p+t,p-t,t,p-t\}$ and $t=p^2$.
In this case $\ket{ \Psi_{\vert\vec{u}} }$ does not depend on $\vec{u}$, and we get
\begin{equation}
\rho_{E_2}^{(x),\vec{u}} =
 (Z^{\vec{\mu}^x_1})^x
 \rho_{pq}^{\otimes m}
 (Z^{\vec{\mu}^x_1})^x
\end{equation}
with $\rho_{pq} = (1-q)\ketbra{\varphi_+}{\varphi_+} + q \ketbra{\varphi_-}{\varphi_-}$ and
$\ket{\varphi_\pm} = \sqrt{1-p}\ket{0}\pm \sqrt{p}\ket{1}$.
Part $E_1$ and $E_2$ of the state $\rho^{(x)}_{E_1E_2}$ in \eqref{eq:14} are now completely decoupled.
As it was shown in \cite{SRS06}, the fact that $E_1$ is classical allows the corresponding state describing dependent errors to be reconstructed from this state:
After tracing out the $E_1$ part, we add an ancilla $[0]_{E_3}$, apply the isometry
$\sum_{\vec{u},\vec{v}}\sqrt{p_{\vec{u}\vert\vec{v}}} \ket{\vec{u}}_{E_3}\bra{0} \otimes [\vec{v}]_{E_2}$ and eventually dephase the ancilla.
Since quantum mutual information never increases under local operations, the maximum of $I(X:E)$ is indeed achieved for independent errors and \eqref{eq:iae:general} becomes
\begin{equation}\label{eq:iaebb84}
 I(X:E) = S\Bigl( \frac{1}{2} \rho_{pq}^{\otimes m} + \frac{1}{2} (Z\rho_{pq}Z)^{\otimes m} \Bigr)
 -m S\bigl( \rho_{pq} \bigr).
\end{equation}
Subtraction of \eqref{eq:iaebb84} from \eqref{eq:iab} gives the secure key rate:
\begin{thm}\label{thm:seckeybb84}
The secure key rate of the BB84 protocol involving the combined preprocessing scheme is given by
\begin{multline}\label{eq:ratebb84}
 r(m,p) = \max_q \frac{1}{m} \Bigl[ 1- \sum_{s=0}^{m-1} \binom{m-1}{s} \tilde{P}(s) \bentropy{\tilde{P}(l^x\vert s)} \\
- S\Bigl( \frac{1}{2} \rho_{pq}^{\otimes m} + \frac{1}{2} (Z\rho_{pq}Z)^{\otimes m} \Bigr) +m \bentropy{\frac{1}{2}(1+\sqrt{1-16p(1-p)q(1-q)})}  \Bigr].
\end{multline}
\end{thm}
\begin{rem}
Without the use of the cat code (i.\,e. if we take $m=1$) the rate reduces to \cite{KGR05,KGR05b}
\begin{equation}\label{eq:ratebb84:m=1}
 r(p) = \max_q \Bigl[1-H_2(\tilde{p})-H_2(p) + H_2\Bigl( \frac{1}{2}(1+\sqrt{1-16p(1-p)q(1-q)})\Bigr) \Bigr].
\end{equation}
\end{rem}

\begin{figure}
 \centering
 \includegraphics[scale=0.75]{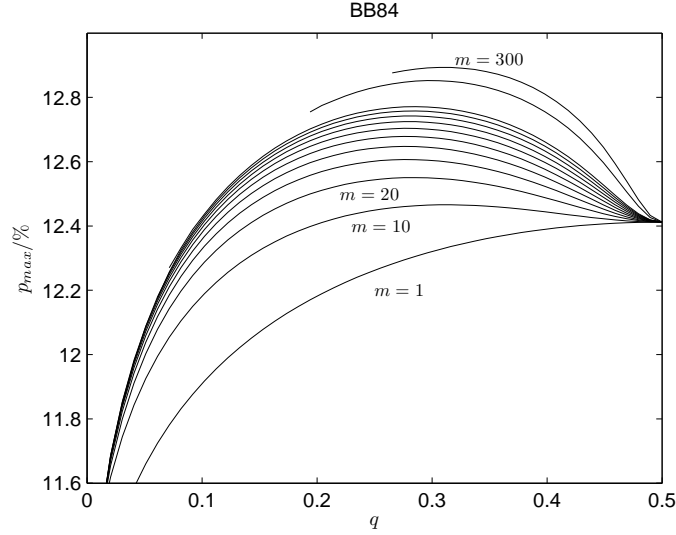}
 \caption{Highest tolerable bit error rate $p_\text{max}^\text{BB84}$ of the BB84 protocol as a function of the added noise $q$ for different block lengths $m\in\{1,10,20,\dots,90,100,200,300\}$.\label{fig:maxpoverqbb}}
\end{figure}

Omitting the maximization over $q$, the above formula \eqref{eq:ratebb84} gives the key rate $r_{m,q}(p)$ for some fixed values of $m$ and $q$ as a function of the bit error rate $p$.
By setting $r_{m,q}(p)$ equal to zero, we find $p_\text{max}^\text{BB84}(m,q)$, the maximum tolerable bit error rate for given $m$ and $q$.
For very high levels of added noise, i.\,e. for $q=\frac{1}{2}-\epsilon$, we find that for all values of $m$, the key rate becomes zero at the bit error rate $p_\text{max}^\text{BB84}(m,q=\frac{1}{2}-\epsilon ) = \BBPn$, but by adding less noise at higher values of $m$, secret keys can be generated for even larger bit error rates (compare with figure \ref{fig:maxpoverqbb}).
\begin{figure}
 \centering
 \includegraphics[scale=0.75]{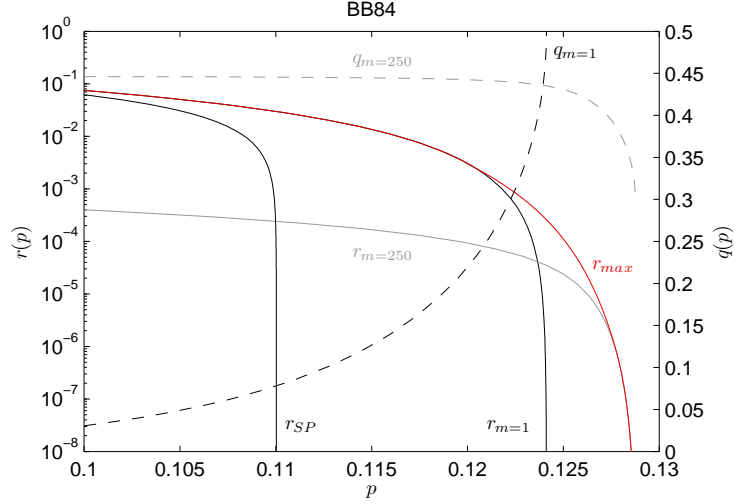}
 \caption{Secure key rate $r$ of BB84 for various types of preprocessing versus bit error rate $p$. No preprocessing corresponds to $r_{SP}$, noisy preprocessing to $r_{m=1}$, and the maximum over all block lengths $m\leq 250$ to $r_\text{max}$, shown in red.
For the rates achieved by the blocklengths $m=1$ and $m=250$, the corresponding rate of the added noise is shown on the right y axis.\label{fig:ratebb}}
\end{figure}
Figure \ref{fig:ratebb} shows plots of the key rates given by \eqref{eq:ratebb84:m=1:q=0} and \eqref{eq:ratebb84:m=1} (black) and the maximum over the key rates given by \eqref{eq:ratebb84} (red) for values of $m$ up to $250$.
The increase of the maximal tolerable bit error rate with the block length $m$ is illustrated in figure \ref{fig:PmaxOverM}.
The highest value of $m$ for which we maximized the tolerable bit error rate as function of the added noise $q$ was $m=\BBPcm$ leading to $p_\text{max}^\text{BB84}(m=500,q=\BBPcq)=\BBPcr$.

By far the most difficult part in the numerical evaluation of \eqref{eq:ratebb84} is computing the von Neumann entropy, as it contains a sum of two $m$-fold tensor products of different one qubit density operators.
Such an expression can be more efficiently calculated by taking into account its block diagonal structure which follows from permutation invariance, as detailed in the next subsection.

\subsubsection{6-State}
Since the set $\Gamma$ only contains the single state $\{ p_{uv} \} = \{ 1-\frac{3}{2}p, \frac{p}{2}, \frac{p}{2}, \frac{p}{2}\}$, minimization over $\sigma_{AB}$ is unnecessary and the secure key rate is directly given by the difference of the quantum mutual informations between Alice and Bob \eqref{eq:iab} and Alice and Eve.
Despite the simplicity of $\Gamma$, this calculation is more difficult than BB84 due to the correlation between bit and phase errors.
The corresponding conditional probabilities are given by
$p_{v=1\vert u=0} = \frac{p}{2(1-p)} = p'$, $p_{v=0\vert u=0} = 1-p'$
and $p_{v\vert u=1} = \frac{1}{2}$.
Therefore, denoting the number of ones in $\vec{u}$ as $u$ and by reordering the qubits in such a way that the first $u$ qubits are the ones with $u_i=1$,
we get
\begin{equation}
\ket{\Psi_{\vert\vec{u}}} = \sum_{\vec{v}}\sqrt{ p_{\vec{v}\vert \vec{u}} } \ket{\vec{v}}
          = \ket{+}^{\otimes u} \otimes \ket{\varphi_+'}^{\otimes m-u} = \ket{\Psi_{\vert u}}
\end{equation}
with $\ket{\pm}=\frac{1}{\sqrt{2}} ( \ket{0}\pm\ket{1} )$ and
$\ket{\varphi'_\pm} = \sqrt{p'}\ket{0} \pm \sqrt{1-p'}\ket{1}$, leading to
\begin{align}
 \rho_{E_2}^{(x),u} &=
 (Z^{\vec{\mu}^x_1})^x
 \sum_{\vec{f}} q_{\vec{f}}
   Z^{\vec{f}} [+]^{\otimes u} \otimes [\varphi'_+]^{\otimes m-u}  Z^{\vec{f}}
 (Z^{\vec{\mu}^x_1})^x \nonumber\\
 &= (Z^{\vec{\mu}^x_1})^x \sigma^{\otimes u} \otimes \gamma^{\otimes m-u} (Z^{\vec{\mu}^x_1})^x
\end{align}
with $\sigma=(1-q)[+]+q[-]$ and $\gamma=(1-q)[\varphi'_+]+q[\varphi'_-]$.
Reordering the state in 
this manner does not change the entropy, and so will not alter the rate.
Using these results the quantum mutual information between Alice and Eve \eqref{eq:iae:general} can be expressed as
\begin{multline}\label{eq:iae6state}
 I(X:E) = \sum_{u=0}^m \binom{m}{u} p^u(1-p)^{m-u} \Bigl[
 S\Bigl( \frac{1}{2}\sigma^{\otimes u}\otimes\gamma^{\otimes m-u} +
         \frac{1}{2}(Z\sigma Z)^{\otimes u}\otimes(Z\gamma Z)^{\otimes m-u} \Bigr) \\
 -u \bentropy{q} - (m-u) \bentropy{\frac{1}{2}(1+\sqrt{1-16p'(1-p')q(1-q)}) }  \Bigr].
\end{multline}
Since $\sigma$ and $Z\sigma Z$ are diagonal in the same basis we are able to write the von Neumann entropy as
\begin{equation}
\sum_{k=0}^u \binom{u}{k} S\Bigl( \frac{q^k(1-q)}{2}^{u-k}\gamma^{\otimes m-u} +
     \frac{(1-q)^kq^{u-k}}{2}(Z\gamma Z)^{\otimes m-u}  \Bigr)
\end{equation}
which is of the same form as the von Neumann entropy in \eqref{eq:iaebb84}.
Therefore the same methods for evaluation can be applied; see the next subsection.

\begin{thm}\label{thm:seckey6state}
The secure key rate of the 6-state protocol involving the combined preprocessing scheme is given by subtracting \eqref{eq:iae6state} from \eqref{eq:iab},
\begin{equation}\label{eq:rate6state}
 r(m,p) = \max_q \frac{1}{m} \Bigl[ 1- \sum_{s=0}^{m-1} \binom{m-1}{s} \tilde{P}(s)
  H_2\bigl( \tilde{P}(l^x\vert s) \bigr) - I(X:E)  \Bigr].
\end{equation}
\end{thm}
\begin{rem}
For $m=1$ the rate \eqref{eq:rate6state} reduces to \cite{KGR05,KGR05b},
\begin{equation}\label{eq:rate6state:m=1}
 r(p) = \max_q  \Bigl[ 1- \bentropy{\tilde{p}}
- \sum_u p_u \left( \bentropy{p_{v\vert u}} - \bentropy{\frac{1}{2}(1+\sqrt{1-16p_{1\vert u}(1-p_{1\vert u})q(1-q)}) } \right) \Bigr],
\end{equation}
\end{rem}

\begin{figure}
 \centering
 \includegraphics[scale=0.75]{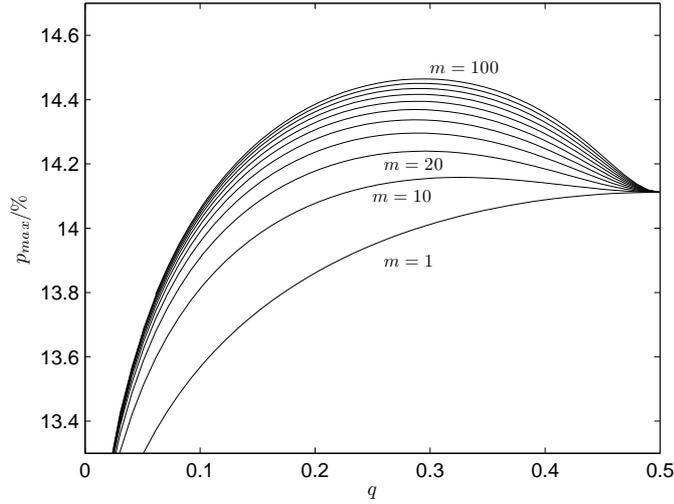}
 \caption{Highest tolerable bit error rate $p_\text{max}^\text{6-st.}$ of the 6-state protocol as a function of the added noise $q$ for different block lengths $m\in\{1,10,20,\dots,90,100\}$.\label{fig:maxpoverq6s}}
\end{figure}

\begin{figure}
 \centering
 \includegraphics[scale=0.75]{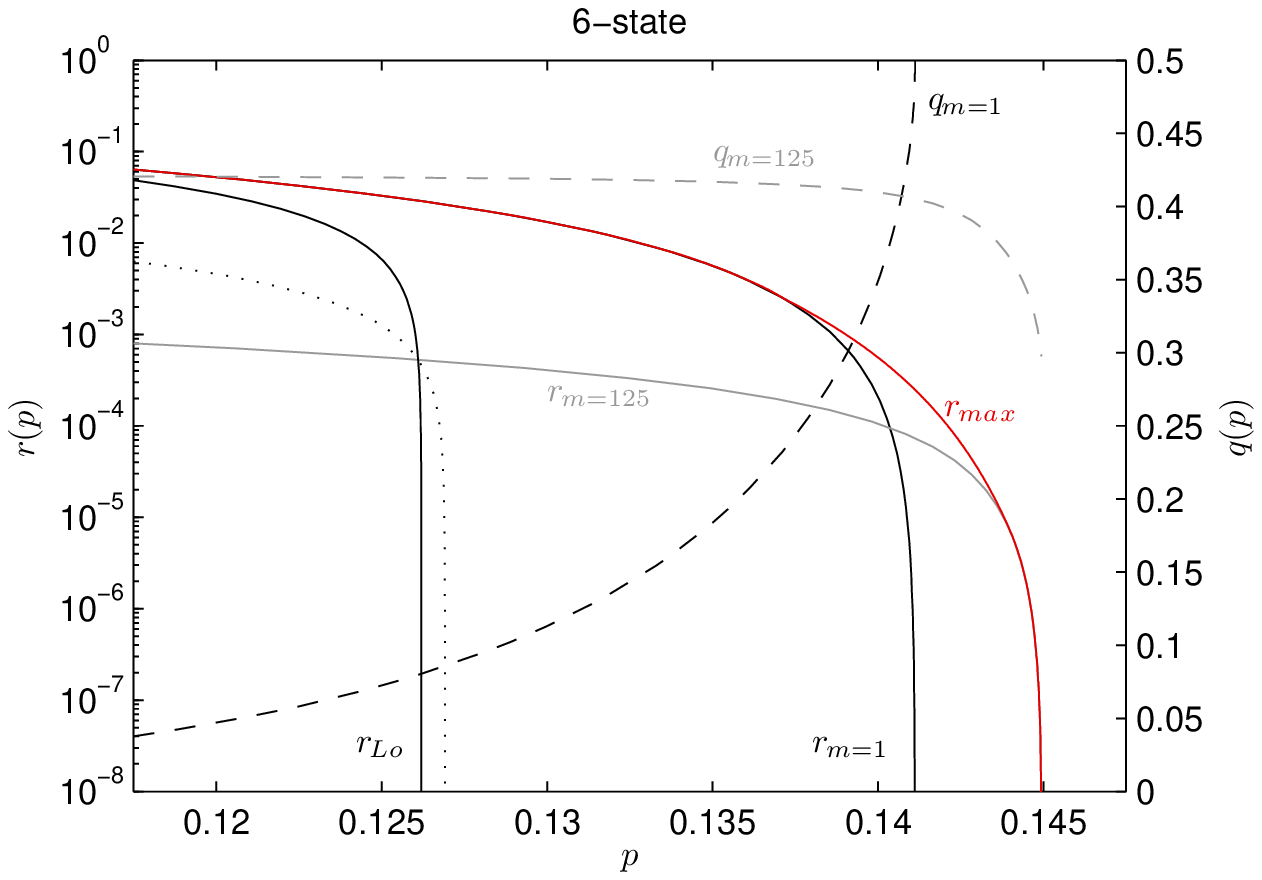}
 \caption{Secure key rate $r$ of the 6-state protocol for various types of preprocessing versus bit error rate $p$. No preprocessing corresponds to $r_{\rm Lo}$, noisy preprocessing to $r_{m=1}$, and the maximum achievable rate over all blocklengths $m\leq 125$, to $r_\text{max}$, shown in red.
For the rates achieved by the blocklengths $m=1$ and $m=125$, the corresponding rate of the added noise is shown on the right y-axis.
The dotted rate with $p_\text{max}^\text{6-st.} = \SSPd$ is due to Lo, corresponding to use of a repetition code of blocklength $m=5$ and no noisy preprocessing.
\label{fig:rate6s}}
\end{figure}

As it is the case for the BB84 protocol, the key rate becomes zero for all values of $m$ for $q\rightarrow \frac{1}{2}$ (this time at bit error rate $p_\text{max}^\text{6-st.}(m,q=\frac{1}{2}-\epsilon)=\SSPn$), but again adding less noise at higher values of $m$ gives rise to secret keys for even higher bit error rates (compare with figure \ref{fig:maxpoverq6s}).
In figure \ref{fig:rate6s} we show the key rates in these special cases as well as the general case for optimal noise and blocklengths up to $m=125$.
Included are $q=0,m=1$ (black),
$q=0,m=5$ (dotted),
and $m=1$ for the optimal $q$ (black).
The maximum over the key rates given by \eqref{eq:rate6state} for values of $m$ up to $125$ is shown in red, along with the specific case of $m=125$.
The increase of the maximal tolerable bit error rate with the block length $m$ is illustrated in figure \ref{fig:PmaxOverM}.
The highest value of $m$ for which we maximized the tolerable bit error rate as function of the added noise $q$ was $m=\SSPcm$ leading to $p_\text{max}^\text{6-st.}(m=\SSPcm,q=\SSPcq) = \SSPcr$.
Since the computation for larger blocksizes becomes rather slow, we extrapolated the value for the optimum noise leading to $q\approx \SSPeq$ for $m=\SSPem$.
By calculating the highest tolerable bit error for this value of noise we get the best lower bound $p_\text{max}^\text{6-st.}(m=\SSPem,q=\SSPeq)=\SSPer$.
It seems likely that for large blocklength ($m\approx 500$) the threshold of the 6-state protocol exceeds the lowest known upper bound on the threshold for the BB84 protocol ($\BBPub$).


\subsection{Evaluation of the Key Rates}\label{subsec:evaluation1}

To evaluate the secure key rates of the BB84 and 6-state protocols given in theorems
\ref{thm:seckeybb84} and \ref{thm:seckey6state} for a certain set of parameters $m$, $p$ and $q$,
in both cases a von Neumann entropy of the form
\begin{equation}\label{eq:vnform}
 S\bigl( \alpha \cdot \rho^{\otimes n} + \beta \cdot (Z\rho Z)^{\otimes n} \bigr)
\end{equation}
with $\alpha,\beta\in\mathbb{R}$, $n\in \{1,2,\dots,m\}$, and
\begin{align}
 \rho = \begin{cases}
  \rho_{pq} = (1-q)[\varphi_+] + q [\varphi_-] & \text{ for the BB84 protocol} \\
  \gamma = (1-q)[\varphi'_+] + q [\varphi'_-] & \text{ for the 6-state protocol}
        \end{cases}
\end{align}
with
$\ket{\varphi_{\pm}}=\sqrt{p}\ket{0}\pm\sqrt{1-p}\ket{1}$, $\ket{\varphi_{\pm}'}=\sqrt{p'}\ket{0}\pm\sqrt{1-p'}\ket{1}$ and $p'=p/(2(1-p))$, has to be computed.
In the following we restrict ourselves to the BB84 protocol, the corresponding results for the 6-state protocol are obtained simply by replacing $p$ with $p'$.
In the Bloch sphere representation, the density matrices $\rho$ and $\sigma \equiv Z\rho Z$ are represented by non-normalized vectors
\begin{equation}
 \vec{r}_\pm = \bigl( \pm2\sqrt{p(1-p)}(1-2q), 0, 1-2p\bigr),
\end{equation}
with $r = \vert \vec{r}_\pm \vert = \sqrt{ 1-16p(1-p)q(1-q)}$, such that
\begin{align}
 \rho &= \frac{1}{2}\bigl( \id + \vec{r}_+ \cdot \vec{s} \bigr), \\
 \sigma &= \frac{1}{2}\bigl( \id + \vec{r}_- \cdot \vec{s} \bigr),
\end{align}
where $\vec{s} = (X,Y,Z)$ denotes a vector containing the Pauli spin matrices.
The vector $\vec{r}_+$ is obtained from $\vec{r}_-$ by rotating $\vec{r}_-$ around the y-axis by the angle $\theta$,
\begin{equation}\label{eq:defitheta}
 \frac{\vec{r}_+}{r} \cdot \frac{\vec{r}_-}{r} = \frac{1-8p(1-p)(1-2q(1-q))}{r^2} = \cos\theta.
\end{equation}
If we diagonalize $\rho$ and $\sigma$, we obtain
\begin{align}
 \rho &= U_\rho \varrho U^\dagger_\rho, &  \varrho&=\mathsf{diag}\{ \rho_1,\rho_2 \}, \\
 \sigma &= U_\sigma \varsigma U^\dagger_\sigma, & \varsigma&=\mathsf{diag}\{ \sigma_1,\sigma_2 \},
\end{align}
and the eigenvalues $\{ \rho_1 , \rho_2 \}$ and $\{ \sigma_1 , \sigma_2 \}$ of $\rho$ and $\sigma$ are both given by  $\{ (1+r)/2, (1-r)/2 \}$.

To speed up the computation of von Neumann entropies of expressions like
$\alpha \cdot \rho^{\otimes n} + \beta \cdot (Z\rho Z)^{\otimes n}$, we make use of their permutation invariance.
As it is discussed in subsection \ref{subsec:schurbasis} of appendix \ref{chap:schur}, operators like $\rho^{\otimes n}$ become block-diagonal when expressed in the Schur basis.
In other words, the reducible representation $D(\rho) = \rho^{\otimes n}$ decomposes into a direct sum of inequivalent irreducible representations $D^{(\nu)}(\rho)$ labeled by a Young diagram $\nu$, where the irrep $D^{(\nu)}(\rho)$ occurs $h_\nu(\textsf{S}_n)$ times and is of dimension~$h_\nu(\textsf{GL}_2)$:
\begin{equation}
  D(\rho) \equiv \rho^{\otimes n} = \bigoplus_\nu D^{(\nu)}(\rho) \otimes \id_{h_\nu(\textsf{S}_n)}.
\end{equation}
In the qubit case,
the summation over the Young diagrams $\nu$ becomes a summation over the index $j$ which ranges from $0\dots \frac{n}{2}$ for even $n$ and $\frac{1}{2}\dots\frac{n}{2}$ for odd $n$.
The dimension of the irreps $D^{(j)}(\rho)$ is given by $h_j(\textsf{GL}_2)=2j+1$ and they are spanned by basis states labeled by a 'Weyl tableau' $k=-j,\dots,+j$. Their degeneracy is given by
\begin{equation}
h_j(\textsf{S}_n) = \binom{n}{n/2-j} \frac{2j+1}{n/2+j+1}.
\end{equation}
Diagonal density operators like $\varrho$ and $\varsigma$ can easily be expressed in the $j$-th representation, since they are diagonal in all these representations, too.
The action of $\varrho^{\otimes n}$ on basis states of the Schur basis becomes simply a multiplication by powers of the two eigenvalues because of the symmetry properties of these basis states:
Each basis state of the Schur basis labeled by a certain Young diagram $j$ and Weyl tableaux $k$ consists of a superposition of computational basis states which are permutations of
$\ket{01}^{\otimes(m/2-j)}\ket{0}^{\otimes(j-k)}\ket{1}^{\otimes(j+k)}$
independently of the Young tableaux (specifying degeneracy).
Hence we obtain
\begin{equation}
D^{(j)}(\varrho) = \mathsf{diag} \{
  \rho_1^{j-k}  \rho_2^{j+k}  (\rho_1 \rho_2)^{m/2-j}    \}_{k=-j\dots j} ,
\end{equation}
and an analogous expression for $D^{(j)}(\varsigma)$.
To obtain the desired non-diagonal block matrices $D^{(j)}(\rho)$ [$D^{(j)}(\sigma)$],
we have to apply the unitary $U_\rho\in\textsf{SU}_2$ [$U_\sigma\in\textsf{SU}_2$] in the irrep $j$ onto $D^{(j)}(\varrho)$ [$D^{(j)}(\varsigma)$],
\begin{equation}
 D^{(j)}(\rho) = D^{(j)}(U_\rho) \cdot D^{(j)}(\varrho) \cdot D^{\dagger(j)}(U_\rho).
\end{equation}
Since the $\textsf{SU}_2 \subset \textsf{GL}_2$ is locally equivalent to $\textsf{SO}_3$ (see e.\,g. \cite{Tung85}), the matrices $D^{(j)}( U_\rho )$ and $D^{(j)}( U_\sigma )$ are Wigner rotation matrices.
In our case these Wigner matrices describe a rotation of $\pm\theta/2$ around the y-axis (where $\theta$ is defined by eq. \eqref{eq:defitheta}) and are given simply by matrix exponentiation,
\begin{align}
 D^{(j)}( U_\rho ) &= \exp\bigl( -i  J_y \cdot \theta/2 \bigr) &
 D^{(j)}( U_\sigma ) &= \exp\bigl( +i  J_y \cdot \theta/2 \bigr),
\end{align}
where $J_y = (J_+-J_-)/(2i)$ and $J_\pm$ denotes the usual angular momentum ladder operators,
\begin{equation}
J_\pm \ket{j,k} = \sqrt{j(j+1)- k(k\pm 1)}\ket{j,k\pm 1}.
\end{equation}
This way,
\begin{equation}
S\bigl( \alpha \cdot \rho^{\otimes n} + \beta \cdot \sigma^{\otimes n} \bigr) =
 \sum_{j=0,1/2}^{n/2} \, h_j(\textsf{S}_n) \cdot S\bigl( \alpha\cdot D^{(j)}(\rho) + \beta\cdot D^{(j)}(\sigma) \bigr),
\end{equation}
and it becomes feasible to calculate such expressions for values of $n$ up to several hundreds.
(Since we are only interested in the eigenvalues of $\alpha D^{(j)}(\rho) + \beta D^{(j)}(\sigma)$, in practice we might apply only a unitary which rotates by $2\times\theta/2$ to $D^{(j)}(\varrho)$ and leave $D^{(j)}(\varsigma)$ in the diagonal form.)

\section{Iterated Preprocessing}\label{sec:itcombiprepro}

By combining local randomization with the cat code of size $m$, Alice and Bob gain an advantage over Eve and intuitively it seems this advantage might be even bigger by performing the procedure twice.
In this section we discuss such a twofold iterated protocol where Alice adds noise at a rate $q$ to $m_2$ blocks of size $m_1$ each,
and then after measuring the syndromes of these blocks, adds further noise at another rate $Q$ to the $m_2$ 'key' bits of these blocks.
Then the syndrome of these $m_2$ bits is measured and the remainder of the protocol proceeds as usual.
We restrict ourselves to the BB84 protocol for simplicity.
Using essentially the same argument as in section \ref{sec:bb84},
we find that we only need to consider independent bit and phase errors described by the state
$\sigma_{AB}$ with $\{ p_{uv} \} = \{1-2p+t,p-t,t,p-t\}$ and $t=p^2$:
(i) $I(X:Y)$ depends only on the bit error rate $p=p_{10}+p_{11}$,
(ii) therefore we have to find the maximum of $I(X:E)$,
(iii) which is achieved for independent errors.
The proof of (iii) works as in section \ref{sec:bb84}, since, as we will see, $E_1$ of $\sigma_{X\overline{E}}$ is again classical.

\subsection{Rate Calculation}
We start with an $m_2\times m_1$-fold tensor product of a purification of $\sigma_{AB}$,
$\ket{ \sigma }_{ABE}^{\otimes m_2}$, where $\ket{ \sigma }_{ABE}$ is the $m_1$-fold tensor product which was defined in equation \eqref{eq:sigma_abe}, (we now denote $m$ as $m_1$).

\subsubsection{First Iteration}
The first step of the iterated preprocessing protocol is to apply the combined preprocessing protocol of the preceeding section to each of the $m_2$ blocks of size $m_1$.
As explained in subsection \ref{subsec:combi:compuseckeyrate}, after this step, the $i$-th block of size $m_1$ is given by \eqref{eq:sigmapp},
\begin{equation}%
\ket{\sigma''}_{ABE}^{(i)} =
  \sum_{\vec{u}_i,\vec{v}_i,\vec{f}_i} \sqrt{ p_{\vec{u}_i,\vec{v}_i} q_{\vec{f}_i} } \,
 \XZ_B\bigl( \vec{\mu}^z_1\cdot(\vec{u}_i+\vec{f}_i) , \vec{\mu}^x_1 \cdot \vec{v}_i \bigr)
 \ket{\Phi^+}_{AB} \ket{\vec{f}_i}_{\mathbf{A'}}
 \ket{\vec{u}_i}_{E_1} Z^{\vec{f}_i}_{E_2}\ket{\vec{v}_i}_{E_2}
 \ket{ \vec{s}_i }_{\mathbf{B'}},
\end{equation}
where $\vec{s}_i=( \vec{\xi}^z_1\cdot(\vec{u}_i+\vec{f}_i),\dots,\vec{\xi}^z_{m_1-1}\cdot(\vec{u}_i+\vec{f}_i) )$ and we added the index $i\in\{1,\dots, m_2\}$.

\subsubsection{Second Iteration}
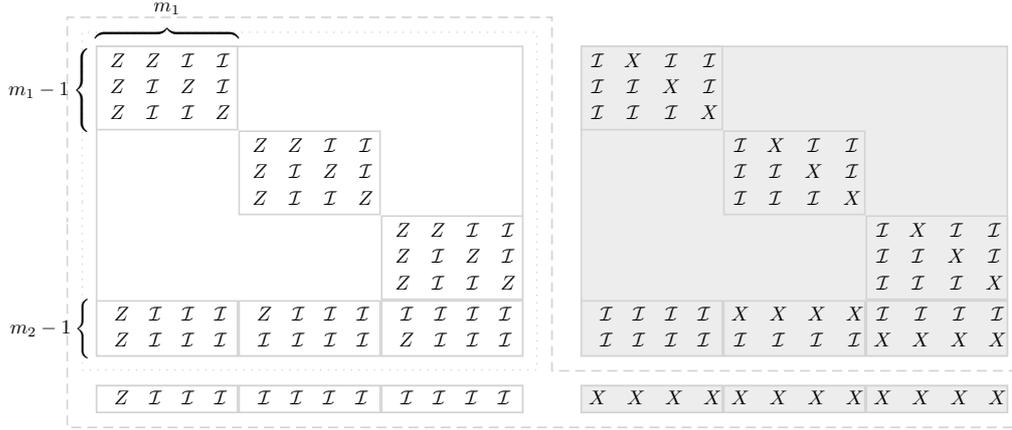
\begin{figure}
\centering
\scalebox{0.75}{
\begin{pspicture}(-1.5,0.5)(16.25,-6.75)
\psframe[linecolor=lightgray,fillstyle=solid](0,0)(7.5,-5.5)  
\psframe[linecolor=lightgray,fillstyle=solid](0,-6)(7.5,-6.5) 
\psframe[linecolor=lightgray,fillstyle=solid,fillcolor=mygray](8.5,0)(16,-5.5) 
\psframe[linecolor=lightgray,fillstyle=solid,fillcolor=mygray](8.5,-6)(16,-6.5)

\psframe[linecolor=lightgray,linestyle=dotted](-0.25,0.25)(7.75,-5.75) 
\pspolygon[linecolor=lightgray,linestyle=dashed](-0.5,0.5)(8,0.5)(8,-5.75)(16.25,-5.75)(16.25,-6.75)(-0.5,-6.75) 
\psframe[linecolor=lightgray,fillstyle=solid](0,0)(2.5,-1.5)
\rput[c](1.25,-0.75){
$\begin{array}{cccc}
 Z & Z & \id & \id \\
 Z & \id & Z & \id \\
 Z & \id & \id & Z
 \end{array}$}
\psframe[linecolor=lightgray,fillstyle=solid](2.5,-1.5)(5,-3)
\rput[c](3.75,-2.25){
$\begin{array}{cccc}
 Z & Z & \id & \id \\
 Z & \id & Z & \id \\
 Z & \id & \id & Z
 \end{array}$}
\psframe[linecolor=lightgray,fillstyle=solid](5,-3)(7.5,-4.5)
\rput[c](6.25,-3.75){
$\begin{array}{cccc}
 Z & Z & \id & \id \\
 Z & \id & Z & \id \\
 Z & \id & \id & Z
 \end{array}$}
\psframe[linecolor=lightgray,fillstyle=solid](0,-4.5)(2.5,-5.5)
\rput[c](1.25,-5){
$\begin{array}{cccc}
 Z & \id & \id & \id \\
 Z & \id & \id & \id
 \end{array}$}
\psframe[linecolor=lightgray,fillstyle=solid](2.5,-4.5)(5,-5.5)
\rput[c](3.75,-5){
$\begin{array}{cccc}
 Z & \id & \id & \id \\
 \id & \id & \id &\id
 \end{array}$}
\psframe[linecolor=lightgray,fillstyle=solid](5,-4.5)(7.5,-5.5)
\rput[c](6.25,-5){
$\begin{array}{cccc}
 \id & \id & \id &\id \\
 Z & \id & \id & \id
 \end{array}$}

\psframe[linecolor=lightgray,fillstyle=solid](0,-6)(2.5,-6.5)
\rput[c](1.25,-6.25){
$\begin{array}{cccc}
  Z & \id & \id & \id
 \end{array}$}
\psframe[linecolor=lightgray,fillstyle=solid](2.5,-6)(5,-6.5)
\rput[c](3.75,-6.25){
$\begin{array}{cccc}
 \id & \id & \id & \id
 \end{array}$}
\psframe[linecolor=lightgray,fillstyle=solid](5,-6)(7.5,-6.5)
\rput[c](6.25,-6.25){
$\begin{array}{cccc}
 \id & \id & \id & \id
 \end{array}$}

\psframe[linecolor=lightgray,fillstyle=solid,fillcolor=mygray](8.5,0)(11,-1.5)
\rput[c](9.75,-0.75){
$\begin{array}{cccc}
 \id & X & \id & \id \\
 \id & \id & X & \id \\
  \id & \id & \id & X
 \end{array}$}
\psframe[linecolor=lightgray,fillstyle=solid,fillcolor=mygray](11,-1.5)(13.5,-3)
\rput[c](12.25,-2.25){
$\begin{array}{cccc}
 \id & X & \id & \id \\
 \id & \id & X & \id \\
  \id & \id & \id & X
 \end{array}$}
\psframe[linecolor=lightgray,fillstyle=solid,fillcolor=mygray](13.5,-3)(16,-4.5)
\rput[c](14.75,-3.75){
$\begin{array}{cccc}
 \id & X & \id & \id \\
 \id & \id & X & \id \\
  \id & \id & \id & X
 \end{array}$}
\psframe[linecolor=lightgray,fillstyle=solid,fillcolor=mygray](8.5,-4.5)(11,-5.5)
\rput[c](9.75,-5){
$\begin{array}{cccc}
 \id & \id & \id & \id \\
 \id & \id & \id & \id
 \end{array}$}
\psframe[linecolor=lightgray,fillstyle=solid,fillcolor=mygray](11,-4.5)(13.5,-5.5)
\rput[c](12.25,-5){
$\begin{array}{cccc}
 X & X & X & X \\
 \id & \id & \id & \id
 \end{array}$}
\psframe[linecolor=lightgray,fillstyle=solid,fillcolor=mygray](13.5,-4.5)(16,-5.5)
\rput[c](14.75,-5){
$\begin{array}{cccc}
 \id & \id & \id & \id \\
 X & X & X & X
 \end{array}$}

\psframe[linecolor=lightgray,fillstyle=solid,fillcolor=mygray](8.5,-6)(11,-6.5)
\rput[c](9.75,-6.25){
$\begin{array}{cccc}
 X & X & X & X
 \end{array}$}
\psframe[linecolor=lightgray,fillstyle=solid,fillcolor=mygray](11,-6)(13.5,-6.5)
\rput[c](12.25,-6.25){
$\begin{array}{cccc}
 X & X & X & X
 \end{array}$}
\psframe[linecolor=lightgray,fillstyle=solid,fillcolor=mygray](13.5,-6)(16,-6.5)
\rput[c](14.75,-6.25){
$\begin{array}{cccc}
 X & X & X & X
 \end{array}$}

\rput[r](-0.05,-0.75){$m_1-1 \left\{ \makebox(0,0.9)[b]{}\right.$}
\rput[r](-0.05,-5){$m_2-1 \left\{ \makebox(0,0.6)[b]{}\right.$}
\rput[B](1.25,0.6){$m_1$}
\rput[B](1.25,0.1){\rotatebox[origin=c]{-90}{$\left\{ \makebox(0,1.35)[b]{}\right.$}}
\end{pspicture}}
\caption{\label{fig:concat:prepro}
The concatenated code for the iterated preprocessing of size $m_1=4$ and $m_2=3$ encoding one qubit into $n=m_1\times m_2$.
The operators on the left hand side are the $\{ \overline{Z}_i \}_{i=1\dots n}$ with
$\overline{Z}_i = \XZ(\vec{0},\vec{\xi}^z_i)$ for $1\leq i\leq n-1$ and
$\overline{Z}_n = \XZ(\vec{0},\vec{\mu}^z  )$,
those on the right hand side the $\{ \overline{X}_i \}_{i=1\dots n}$ with
$\overline{X}_i = \XZ(\vec{\eta}^x_i,\vec{0})$ for $1\leq i\leq n-1$ and
$\overline{X}_n = \XZ(\vec{\mu }^x  ,\vec{0})$.
The (generators of the) stabilizers are within the dotted line, the (generators of the) normalizers within the dashed one.}
\end{figure}

After adding additional noise at rate $Q$ to the key bit of each of the $m_2$ blocks the state is described as
\begin{multline}
\ket{\sigma'''}_{ABE}^{(i)} =
 \sum_{\vec{f}_i,\vec{u}_i,\vec{v}_i}
 \sum_{F_i}
 \sqrt{ p_{\vec{u}_i,\vec{v}_i} q_{\vec{f}_i}\, Q_{F_i} } \,
 \XZ_B\bigl( \vec{\mu}^z_1\cdot(\vec{u}_i+\vec{f}_i) + F_i , \vec{\mu}^x_1\cdot\vec{v}_i \bigr)
 \ket{\Phi^+}_{AB} \\
 \otimes  \ket{\vec{u}_i}_{E_1} 
 (Z^{\vec{\mu}^x_1})^{F_i} Z^{\vec{f}_i} \ket{\vec{v}_i}_{E_2} 
 \ket{ \vec{s}_i\, }_{\mathbf{B'}} \ket{\vec{f}_i\,}_{\mathbf{A'}} \ket{F_i}_{\mathbf{A''}}
\end{multline}
with classical registers $\mathbf{B'}$, $\mathbf{A'}$ and $\mathbf{A''}$.
Now we define the abbreviations
$\vec{U} = \bigl( \vec{\mu}^z_1\cdot(\vec{u}_1+\vec{f}_1), \dots,
                  \vec{\mu}^z_1\cdot(\vec{u}_{m_2}+\vec{f}_{m_2}) \bigr)$ and
$\vec{V} = \bigl( \vec{\mu}^x_1\cdot\vec{v}_1, \dots, \vec{\mu}^x_1\cdot\vec{v}_{m_2} \bigr)$.
Again Alice and Bob both measure their stabilizers (this time the cat code is of length $m_2$),
and Alice sends her result to Bob, who calculates the relative syndrome
$\vec{S} = ( \vec{\xi}_1\cdot(\vec{U}+\vec{F}), \dots, \vec{\xi}_{m_2-1}\cdot(\vec{U}+\vec{F}) )$.
Both then measure their key bit.
The tripartite semiclassical state describing the correlations is now given by
\begin{multline}\label{eq:sigmaxye:it}
\sigma_{XY\overline{E}} = \frac{1}{2} 
 \sum_{\vec{F}} Q_{\vec{F}}
 \sum_{\vec{f}_1,\dots,\vec{f}_{m_2}} q_{\vec{f}_1}\dots q_{\vec{f}_{m_2}}
 \sum_{\vec{u}_1,\dots,\vec{u}_{m_2}}
 p_{\vec{u}_1}\dots p_{\vec{u}_{m_2}} \\
 \times\, \sum_x [x]_A \otimes [x+L^x]_B \otimes [\vec{s}_1,\dots,\vec{s}_{m_2},\vec{S}]_{B'} \otimes   [\vec{u}_1,\dots,\vec{u}_{m_2}]_{E_1} \\
\otimes \,
 ( Z^{\otimes m_1m_2} )^x
  \bigotimes_{i=1}^{m_2}
  \bigl(
   (Z^{\otimes m_1})^{F_i} Z^{\vec{f}_i} \ketbra{\Psi}{\Psi} Z^{\vec{f}_i} (Z^{\otimes m_1})^{F_i}
  \bigr)
 ( Z^{\otimes m_1m_2} )^x,
\end{multline}
where $\ket{ \Psi } = \sum_{\vec{v}} \sqrt{ p_{\vec{v}} } \ket{\vec{v}}$
and $\vec{s}_i=(\vec{\xi}_1\cdot(\vec{u}_i+\vec{f}_i),\dots)$,
 $\vec{S}=(\vec{\xi}_1\cdot(\vec{U}+\vec{F}),\dots)$, and
$L^x=\vec{\mu}^z_1\cdot(\vec{U}+\vec{F})$.
Note that, as it was the case for \eqref{eq:sigmaxye}, the state is classical in $E_1$ since the quantities $\{ \vec{s}_1,\dots,\vec{s}_{m_2},\vec{S},L^x\}$ are all classical;
$\{\vec{S},L^x\}$ fixes $\vec{U}+\vec{F}$, and,
since $(\vec{s}_i)_j = \vec{\xi}_j\cdot(\vec{u}_i+\vec{f}_i) = \vec{\xi}_j\cdot(\vec{u}_i+\vec{f}_i+F_i\cdot\vec{1})$, together with $\{ \vec{s}_1,\dots,\vec{s}_{m_2} \}$ the string
 $(\vec{u}_1+\vec{f}_1+F_1\cdot\vec{1},\dots,\vec{u}_{m_2}+\vec{f}_{m_2}+F_{m_2}\cdot\vec{1})$ is fixed (compare with figure \ref{fig:concat:prepro} which shows the stabilizers of the corresponding concatenated cat code).

\subsubsection{The Quantum Mutual Informations}
To calculate the quantum mutual information between Alice and Bob we trace out Eve's systems and obtain
\begin{equation}
\sigma_{XY} = \frac{1}{2} 
 \sum_{\vec{F}} Q_{\vec{F}}
 \sum_{\vec{u}_1 \dots \vec{u}_{m_2}}
 \tilde{p}_{\vec{u}_1}\dots \tilde{p}_{\vec{u}_{m_2}}\sum_x\, [x]_A \otimes
 [x+L^x]_B \otimes [\vec{s}_1 \dots \vec{s}_{m_2},\vec{S}]_{B'},
\end{equation}
using $\tilde{p}=p(1-q)+(1-p)q$.
Since Alice's additional noise $\vec{f}$ is now combined with Eve's noise
$\vec{u}$, $\vec{f}$ no longer appears in the the syndromes $\vec{s}_i$ and $\vec{S}$:
$\vec{s}_i=(\vec{\xi}_1\cdot\vec{u}_i, \dots) = (\vec{\xi}_1\cdot(\vec{u}_i+F_i\cdot\vec{1}), \dots)$, $\vec{S}=(\vec{\xi}_1\cdot(\vec{U}'+\vec{F}),\dots)$.
Additionally, $L^x$ is now $L^x=\vec{\mu}^z_1\cdot(\vec{U}'+\vec{F}),$
with $\vec{U}' = ( \vec{\mu}^z_1\cdot\vec{u}_1, \dots, \vec{\mu}^z_1\cdot\vec{u}_{m_2} )$.
Hence,
\begin{equation}
\sigma_{XY} = \frac{1}{2} \sum_x\, [x]_A \otimes
  \sum_{\vec{u}_1 \dots \vec{u}_{m_2}}
 \tilde{P}'(\vec{u}_1,\dots,\vec{u}_{m_2})
 [x+L^x]_B \otimes [\vec{s}_1 \dots \vec{s}_{m_2},\vec{S}]_B,
\end{equation}
with
\begin{equation}
\tilde{P}'(\vec{u}_1,\dots,\vec{u}_{m_2}) =
\prod_{i=1}^{m_2} \bigl[ (1-Q)\tilde{p}_{\vec{u}_i} + Q\tilde{p}_{\vec{u}_i+\vec{1}} \bigr]
\end{equation}
and $\vec{s}_i=(\vec{\xi}_1\cdot\vec{u}_i, \dots)$, $\vec{S}=(\vec{\xi}_1\cdot\vec{U}', \dots)$ and  $L^x=\vec{\mu}^z_1\cdot\vec{U}'$, or,
\begin{equation}
\sigma_{XY} = \frac{1}{2} \sum_x\, [x]_A \otimes
 \, \sum_{\mathclap{\vec{s}_1 \dots \vec{s}_{m_2},\vec{S},L^x}} \,
 \tilde{P}'( \vec{s}_1\dots\vec{s}_{m_2}, \vec{S}, L^x )
 [x+L^x]_B \otimes [\vec{s}_1 \dots \vec{s}_{m_2},\vec{S}]_B,
\end{equation}
where the probability distribution
$ \tilde{P}'( \vec{s}_1\dots\vec{s}_{m_2}, \vec{S}, L^x )$
only depends on the number of ones in each of the syndromes $\vec{s}_i$ and $\vec{S}$
(we assume that the zeros and ones in $\vec{S}$ are ordered such that the syndromes $\vec{s}_i$, $i\in\{1,\dots,m_2-S\}$, correspond to $S_i=0$):
\begin{multline}\label{eq:itprob}
 \tilde{P}'( L^x=0, s_1\dots s_{m_2}, S) =
 \prod_{i=1}^{m_2-S}[ (1-\tilde{p})^{m_1-s_i} \tilde{p}^{s_i} (1-Q)
                     +(1-\tilde{p})^{s_i} \tilde{p}^{m_1-s_i} Q ] \times \\
 \prod_{i=m_2-S+1}^{m_2}[ (1-\tilde{p})^{s_i} \tilde{p}^{m_1-s_i} (1-Q)
                         +(1-\tilde{p})^{m_1-s_i} \tilde{p}^{s_i} Q ]  .
\end{multline}
The mutual information can therefore be written as
\begin{equation}\label{eq:iab:iter}
 I(X:Y) = 1 -
 \sum_{\vec{s}_1 \dots \vec{s}_{m_2},\vec{S}}
 \tilde{P}'(\vec{s}_1 \dots \vec{s}_{m_2},\vec{S})
 H_2\bigl( \tilde{P}'(L^x\vert \vec{s}_1 \dots \vec{s}_{m_2},\vec{S}) \bigr).
\end{equation}
In addition we see by the means of \eqref{eq:itprob} that for a given value of $S$ only the frequency distribution of the
$s_i$, $i\in\{1,\dots,m_2-S\}$, and the $s_j$, $j\in\{m_2-S+1,\dots,m_2\}$, matters.
This fact can be used to speed up the calculation of the sum over the syndromes in \eqref{eq:iab:iter},
\begin{multline}
 I(X:Y) = 1 -
 \sum_{S=0}^{m_2-1}
 \sum_{\substack{c_0,\dots,c_{m_1-1}=0 \\ \text{s.\,t. } \sum_i c_i=m_2-S}}^S
 \prod_{j=0}^{m_2-1}\binom{m_1-1}{j}^{c_j}
 \sum_{\substack{a_0,\dots,a_{m_1-1}=0 \\ \text{s.\,t. } \sum_i a_i=S}}^S
 \prod_{j=0}^{m_2-1}\binom{m_1-1}{j}^{a_j} \\
 \tilde{P}'( s_1\dots s_{m_2} , S )
 H_2\bigl( \tilde{P}'(L^x\vert s_1\dots s_{m_2},S ) \bigr),
\end{multline}
where $(s_1,\dots,s_{m_2-S})$ contains $c_0\times 0,\dots, c_{m_1-1}\times m_1-1$, and $(s_{m_2-S+1},\dots,s_{m_2})$ contains $a_0\times 0,\dots, a_{m_1-1}\times m_1-1$.

Tracing out Bob's systems from \eqref{eq:sigmaxye:it}, and writing the resulting state as in \eqref{eq:sigma_xe1}-\eqref{eq:sigma_xe3}, we obtain
\begin{equation}
 \rho_{E_2}^{(x),\vec{u}_1\dots\vec{u}_{m_2}} = ( Z^{\otimes m_1m_2} )^x
  \bigl[
    (1-Q) \rho_{pq}^{\otimes m_1} + Q (Z\rho_{pq}Z)^{\otimes m_1}
  \bigr]^{\otimes m_2}
 ( Z^{\otimes m_1m_2} )^x,
\end{equation}
which does not depend on the strings $\vec{u}_1\dots\vec{u}_{m_2}$
and the mutual information between Alice and Eve \eqref{eq:iae:general} can be seen to be
\begin{multline}\label{IAE:BB84:it}
 I(X:E) =  S\Bigl(
  \frac{1}{2}\bigl[ (1-Q)\rho_{pq}^{\otimes m_1} + Q(Z\rho_{pq}Z)^{\otimes m_1} \bigr]^{\otimes m_2}
+ \frac{1}{2}\bigl[ Q\rho_{pq}^{\otimes m_1} + (1-Q)(Z\rho_{pq}Z)^{\otimes m_1} \bigr]^{\otimes m_2}
  \Bigr) \\
 -m_2 S\bigl( (1-Q)\rho_{pq}^{\otimes m_1} + Q(Z\rho_{pq}Z)^{\otimes m_1} \bigr)
\end{multline}
Once more the secure key rate is given by the difference of these mutual informations.
\begin{thm}
The secure key rate of the BB84 protocol involving the iterated preprocessing protocol of size $m_2\times m_1$ is given by
\begin{equation}\label{eq:ratebb84:it}
 r(m_1,m_2,p)= \max_{q,Q} \frac{1}{m_1m_2} \bigl(  I(X:Y)-I(X:E) \bigr),
\end{equation}
where the mutual informations are defined in \eqref{eq:iab:iter} and \eqref{IAE:BB84:it}.
\end{thm}

Again the hardest part in the numerical evaluation of \eqref{eq:ratebb84:it} comes from the von Neumann entropies.
One contains a sum of two $m_2$-fold tensor products of different density operators, but this time these density operators are $m_1$-qubit density operators.
For more details on the evaluation of \eqref{eq:ratebb84:it} see the next subsection.

\begin{figure}
 \centering
 \includegraphics[scale=0.75]{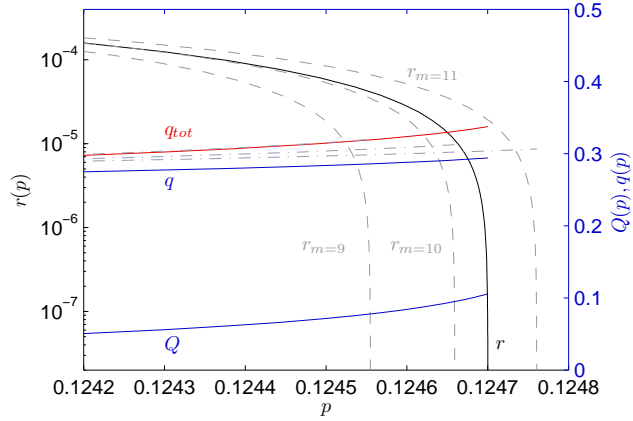}
 \caption{
Secure key rate $r$ of BB84 with iterated preprocessing of size $m_1\times m_2=3\times 3$ versus bit error rate $p$.
The \textit{right} y-axis shows the corresponding values of added noise in the first ($q$) and second iteration ($Q$) as well as values of the total amount of added noise ($q_{tot}=q(1-Q)+(1-q)Q$, red).
For comparison, the rates of the non-iterated protocol are shown for blocksizes $m\in\{9,10,11\}$ (dashed lines). The corresponding values of added noise for these cases are also shown (dash-dot lines).
\label{fig:bb84:it}}
\end{figure}

We compare the resulting key rate of the $m_1\times m_2=3\times 3$ iterated code with the key rates of the non-iterated codes of blocksizes $m\in\{9,10,11\}$ in figure \ref{fig:bb84:it}.
The entire rate curve of the $3\times 3$ code shifts to higher values than the single round $m=9$ code,
while the total amount of noise $q_\text{tot}=q(1-Q)+(1-q)Q$ added to the sifted key bits is essentially the same as in the case of one round, showing that the improvement comes from making better use of the same amount of noise.

\subsection{Rate Evaluation}
As it was the case for the non-iterated preprocessing protocol, to evaluate the mutual information between Alice and Eve (given by \eqref{eq:ratebb84:it}) von Neumann entropies of the form
\begin{equation}\label{eq:vnform2}
 S\bigl( \alpha \cdot \rho^{\otimes n} + \beta \cdot (Z\rho Z)^{\otimes n} \bigr)
\end{equation}
have to be evaluated.
This time, in addition to the case where $\rho$ is a qubit density operator,
there is also the case where $\rho$ is a qudit density operator of dimension $2^{m_1}$.
Such an expression can also be calculated more efficiently by taking into account its permutation invariance (see subsection \ref{subsec:evaluation1}),
\begin{equation}\label{eq:entroblockdiag}
S\bigl( \alpha \cdot \rho^{\otimes n} + \beta \cdot \sigma^{\otimes n} \bigr) =
 \sum_\nu \, h_\nu(\textsf{S}_n) \cdot S\bigl( \alpha\cdot D^{(\nu)}(\rho) + \beta\cdot D^{(\nu)}(\sigma) \bigr),
\end{equation}
(with $\sigma\equiv Z\rho Z$) but now we cannot determine the irreducible representations $D^{(\nu)}(\rho)$ by matrix multiplication from their diagonal counterparts $D^{(\nu)}(\varrho)$, because there is no simple way to determine the representation matrices of $\textsf{SU}_{2^{m_1}}$ for $m_1>1$.
Therefore, we explicitly calculate the Schur basis $\{ \ket{ W^{(\nu)}_k Y^{(\nu)}_m } \}$ (see section \ref{sec:schurtransform}) of $n$ qudits of dimension $2^{m_1}$ with the help of the eigenfunction method \cite{CPW02}, and obtain
\begin{equation}\label{eq:schur}
\ket{ W^{(\nu)}_k Y^{(\nu)}_m } =
  \sum_{i_1\dots i_n}
  [U_\text{Sch}]^{W^{(\nu)}_k Y^{(\nu)}_m}_{i_1\dots i_n} \ket{i_1, \dots , i_n}.
\end{equation}
Then we determine the matrix elements of both the $D^{(\nu)}(\rho)$ and the $D^{(\nu)}(\sigma)$ blocks by using the Schur basis states \eqref{eq:schur},
\begin{multline}
D^{(\nu)}_{kk'}(\rho) =
 \bra{ W^{(\nu)}_k Y^{(\nu)}_m }
 \rho^{\otimes n}
 \ket{ W^{(\nu)}_{k'} Y^{(\nu)}_m }= \\
\sum_{j_1\dots j_n} \sum_{i_1\dots i_n}
 [U^\ast_\text{Sch}]^{W^{(\nu)}_k Y^{(\nu)}_m}_{j_1\dots j_n}
 [U_\text{Sch}]^{W^{(\nu)}_{k'} Y^{(\nu)}_m}_{i_1\dots i_n}
 \bra{j_1, \dots , j_n} \rho^{\otimes n} \ket{i_1, \dots , i_n},
\end{multline}
for some arbitrary Young tableau $Y^{(\nu)}_m$ which specifies the degeneracy of the irreps $\nu$ of $\textsf{GL}_{2^{m_1}}$.

For example, to calculate the key rate of the $m_1\times m_2=3\times 3$ case presented in the last subsection, we calculated the Schur basis of
\begin{multline}
 \mathcal{H}_8^{\otimes 3} =
 \vspan \Bigl\{ \ket{ W_{k_j}^{([3])} } \Bigr\}_{j=1\dots 120}
 \otimes
 \ket{ Y_{m_1}^{([3])} }
\bigoplus  \\
 \vspan \Bigl\{ \ket{ W_{k_j}^{([2,1])} } \Bigr\}_{j=1\dots 168}
 \otimes
 \vspan \Bigl\{ \ket{ Y_{m_i}^{([2,1])} }  \Bigr\}_{i=1\dots 2}
\bigoplus  \\
 \vspan \Bigl\{ \ket{ W_{k_j}^{([1,1,1])} } \Bigr\}_{j=1\dots 56}
 \otimes
 \ket{ Y_{m_1}^{([1,1,1])} },
\end{multline}
and the calculation of the eigenvalues of a $512\times 512$ dimensional matrix in \eqref{eq:vnform2} reduces to a calculation of the eigenvalues of three matrices of dimension $120\times 120$, $168\times 168$ and $56\times 56$ in \eqref{eq:entroblockdiag}.

It may be possible to further streamline the calculation by taking into account the fact that the qudit inputs to the second round are block-diagonal themselves.
Hence more sophisticated representation-theoretic methods, in particular a Clebsch-Gordon decomposition of the states input to the second preprocessing round, should make the analysis of more rounds and larger blocksizes tractable.

\addpart{Appendix}
\appendix
\chapter{Tables of Difference Schemes and Orthogonal Arrays}\label{chap:oatables}

In this chapter of the appendix we list difference schemes based on $\mathbb{F}_2^2$ and orthogonal arrays with four levels.
Orthogonal arrays $OA(n_c,n,2,4)$ with four levels and strength two can be used to build decoupling schemes of length $n_c$ for any Hamiltonian $H_0$ describing a network of up to $n$ qubits with arbitrary qubit-qubit couplings.
If the couplings involve only terms of the form $J_x^{ij} X_i\otimes X_j + J_y^{ij} Y_i\otimes Y_j + J_z^{ij} Z_i\otimes Z_j$, decoupling schemes of smaller length can be obtained from difference schemes $D(n_c,n,4)$ based on $\mathbb{F}_2^2$.
The decoupling schemes $\{g_j\}_{j=0}^{n_c-1}$ are obtained by setting
$g_j = u_{m_{1,j+1}} \otimes u_{m_{2,j+1}} \otimes \dots \otimes u_{m_{n,j+1}}$, where $m_{ij}$ denotes the matrix elements of the corresponding orthogonal array or difference scheme,
and the set $\{ u_i \}_{i=0}^3$ denotes the set of Pauli operators $\mathcal{P}_2=\{\mathcal{I},X,Y,Z\}$.
(Alternatively, in the case of an orthogonal array, any nice error basis may be chosen to form the set $\{ u_i \}_{i=0}^3$).

\section{Difference Schemes}\label{app:dstables}
For an overview over construction methods and lower bounds on the maximal number $c\in \{2,3,\dots,4\lambda\}$ for which a difference scheme $D(4\lambda,c,4)$, $\lambda\in \mathbb{N}$, exists, we refer to \cite[chapter 6]{OABook}.
We list difference schemes $D(4\lambda, 4\lambda, 4)$ for $\lambda=\{1,2,3,4\}$ in tables \ref{ds:4} -- \ref{ds:16}.
The entries $\{0,1,2,3\}$ are to be understood as elements in $\mathbb{F}_2^2$, $0=(0,0), 1=(1,0), 2=(1,1), 3=(0,1)$.
Note that all schemes are symmetric with respect to their matrix indices, i.\,e. $m_{ij}=m_{ji}$.
The difference schemes in tables \ref{ds:4} and \ref{ds:8} are the same as those presented in \cite{SM01}, the schemes in tables \ref{ds:12} and \ref{ds:16} have been obtained by the author via a computer search.
It was conjectured in \cite{SM01} that schemes $D(4\lambda, 4\lambda, 4)$ may exist for all $\lambda\in\mathbb{N}$.
For $\lambda=5$ at the current time only a lower bound of $c\geq 10$ is known.

\begin{table}
\begin{minipage}{.24\textwidth}\centering
\begin{tabular}{cccc}
 0 & 0 & 0 & 0 \\
 0 & 1 & 2 & 3 \\
 0 & 2 & 3 & 1 \\
 0 & 3 & 1 & 2
\end{tabular}%
\caption[Difference scheme $D(4,4,4)$]{$D(4,4,4)$}\label{ds:4}
\end{minipage}%
\begin{minipage}{.31\textwidth}\centering
\small{
\begin{tabular}{cccccccc}
 0 & 0 & 0 & 0 & 0 & 0 & 0 & 0 \\
 0 & 0 & 1 & 1 & 2 & 2 & 3 & 3 \\
 0 & 1 & 2 & 3 & 0 & 1 & 2 & 3 \\
 0 & 1 & 3 & 2 & 2 & 3 & 1 & 0 \\
 0 & 2 & 0 & 2 & 3 & 1 & 3 & 1 \\
 0 & 2 & 1 & 3 & 1 & 3 & 0 & 2 \\
 0 & 3 & 2 & 1 & 3 & 0 & 1 & 2 \\
 0 & 3 & 3 & 0 & 1 & 2 & 2 & 1
\end{tabular}}%
\caption[Difference scheme $D(8,8,4)$]{$D(8,8,4)$}\label{ds:8}
\end{minipage}%
\begin{minipage}{.45\textwidth}\centering
\small{
\begin{tabular}{cccccccccccc}
 0 & 0 & 0 & 0 & 0 & 0 & 0 & 0 & 0 & 0 & 0 & 0 \\
 0 & 0 & 0 & 1 & 1 & 1 & 2 & 2 & 2 & 3 & 3 & 3 \\
 0 & 0 & 0 & 2 & 2 & 2 & 3 & 3 & 3 & 1 & 1 & 1 \\
 0 & 1 & 2 & 1 & 2 & 3 & 0 & 1 & 3 & 0 & 2 & 3 \\
 0 & 1 & 2 & 2 & 3 & 1 & 1 & 3 & 0 & 3 & 0 & 2 \\
 0 & 1 & 2 & 3 & 1 & 2 & 3 & 0 & 1 & 2 & 3 & 0 \\
 0 & 2 & 3 & 0 & 1 & 3 & 2 & 3 & 1 & 0 & 1 & 2 \\
 0 & 2 & 3 & 1 & 3 & 0 & 3 & 1 & 2 & 2 & 0 & 1 \\
 0 & 2 & 3 & 3 & 0 & 1 & 1 & 2 & 3 & 1 & 2 & 0 \\
 0 & 3 & 1 & 0 & 3 & 2 & 0 & 2 & 1 & 3 & 2 & 1 \\
 0 & 3 & 1 & 2 & 0 & 3 & 1 & 0 & 2 & 2 & 1 & 3 \\
 0 & 3 & 1 & 3 & 2 & 0 & 2 & 1 & 0 & 1 & 3 & 2
\end{tabular}}%
\caption[Difference scheme $D(12,12,4)$]{$D(12,12,4)$}\label{ds:12}
\end{minipage}
\end{table}

\begin{table}\centering\small
\begin{tabular}{
c@{\ \ }c@{\ \ }c@{\ \ }c@{\ \ \ }
c@{\ \ }c@{\ \ }c@{\ \ }c@{\ \ \ }
c@{\ \ }c@{\ \ }c@{\ \ }c@{\ \ \ }
c@{\ \ }c@{\ \ }c@{\ \ }c}
 0 & 0 & 0 & 0 & 0 & 0 & 0 & 0 & 0 & 0 & 0 & 0 & 0 & 0 & 0 & 0 \\
 0 & 0 & 0 & 0 & 1 & 1 & 1 & 1 & 2 & 2 & 2 & 2 & 3 & 3 & 3 & 3 \\
 0 & 0 & 0 & 0 & 2 & 2 & 2 & 2 & 3 & 3 & 3 & 3 & 1 & 1 & 1 & 1 \\
 0 & 0 & 0 & 0 & 3 & 3 & 3 & 3 & 1 & 1 & 1 & 1 & 2 & 2 & 2 & 2 \\
 0 & 1 & 2 & 3 & 0 & 1 & 2 & 3 & 0 & 1 & 2 & 3 & 0 & 1 & 2 & 3 \\
 0 & 1 & 2 & 3 & 1 & 0 & 3 & 2 & 2 & 3 & 0 & 1 & 3 & 2 & 1 & 0 \\
 0 & 1 & 2 & 3 & 2 & 3 & 0 & 1 & 3 & 2 & 1 & 0 & 1 & 0 & 3 & 2 \\
 0 & 1 & 2 & 3 & 3 & 2 & 1 & 0 & 1 & 0 & 3 & 2 & 2 & 3 & 0 & 1 \\
 0 & 2 & 3 & 1 & 0 & 2 & 3 & 1 & 0 & 2 & 3 & 1 & 0 & 2 & 3 & 1 \\
 0 & 2 & 3 & 1 & 1 & 3 & 2 & 0 & 2 & 0 & 1 & 3 & 3 & 1 & 0 & 2 \\
 0 & 2 & 3 & 1 & 2 & 0 & 1 & 3 & 3 & 1 & 0 & 2 & 1 & 3 & 2 & 0 \\
 0 & 2 & 3 & 1 & 3 & 1 & 0 & 2 & 1 & 3 & 2 & 0 & 2 & 0 & 1 & 3 \\
 0 & 3 & 1 & 2 & 0 & 3 & 1 & 2 & 0 & 3 & 1 & 2 & 0 & 3 & 1 & 2 \\
 0 & 3 & 1 & 2 & 1 & 2 & 0 & 3 & 2 & 1 & 3 & 0 & 3 & 0 & 2 & 1 \\
 0 & 3 & 1 & 2 & 2 & 1 & 3 & 0 & 3 & 0 & 2 & 1 & 1 & 2 & 0 & 3 \\
 0 & 3 & 1 & 2 & 3 & 0 & 2 & 1 & 1 & 2 & 0 & 3 & 2 & 1 & 3 & 0 \\
\end{tabular}
\caption[Difference scheme $D(16,16,4)$]{$D(16,16,4)$}\label{ds:16}
\end{table}

\section{Orthogonal Arrays}\label{app:oatables}
We list orthogonal arrays $OA(16,5,2,4)$, $OA(32,9,2,4)$, and $OA(48,13,2,4)$ in tables \ref{oa:16} -- \ref{oa:48}.
The arrays are constructed using the difference schemes listed in tables \ref{ds:4} -- \ref{ds:12} in connection with the construction method described in \cite[corollary 6.20]{OABook}, which, for a given difference scheme $D(n_c,n,4)$, leads to an $OA(4n_c,n+1,2,4)$.
As a result, the upper left $(n-1)\times (n-1)$ corner of any of the listed orthogonal arrays of the form $OA(n_c,n,2,4)$ is identical to the corresponding difference scheme.

\begin{table}\centering\small
\begin{tabular}{
c@{\ \ }c@{\ \ }c@{\ \ }c@{\ \ \ }
c@{\ \ }c@{\ \ }c@{\ \ }c@{\ \ \ }
c@{\ \ }c@{\ \ }c@{\ \ }c@{\ \ \ }
c@{\ \ }c@{\ \ }c@{\ \ }c}
0&0&0&0&1&1&1&1&2&2&2&2&3&3&3&3\\
0&1&2&3&1&0&3&2&2&3&0&1&3&2&1&0\\
0&2&3&1&1&3&2&0&2&0&1&3&3&1&0&2\\
0&3&1&2&1&2&0&3&2&1&3&0&3&0&2&1\\
0&1&2&3&0&1&2&3&0&1&2&3&0&1&2&3
\end{tabular}
\caption{$OA(16,5,2,4)$}\label{oa:16}
\end{table}

\begin{table}\centering\small
\begin{tabular}{
c@{\ \ }c@{\ \ }c@{\ \ }c@{\ \ }c@{\ \ }c@{\ \ }c@{\ \ }c@{\ \ \ }
c@{\ \ }c@{\ \ }c@{\ \ }c@{\ \ }c@{\ \ }c@{\ \ }c@{\ \ }c@{\ \ \ }
c@{\ \ }c@{\ \ }c@{\ \ }c@{\ \ }c@{\ \ }c@{\ \ }c@{\ \ }c@{\ \ \ }
c@{\ \ }c@{\ \ }c@{\ \ }c@{\ \ }c@{\ \ }c@{\ \ }c@{\ \ }c}
0&0&0&0&0&0&0&0&1&1&1&1&1&1&1&1&2&2&2&2&2&2&2&2&3&3&3&3&3&3&3&3\\
0&0&1&1&2&2&3&3&1&1&0&0&3&3&2&2&2&2&3&3&0&0&1&1&3&3&2&2&1&1&0&0\\
0&1&2&3&0&1&2&3&1&0&3&2&1&0&3&2&2&3&0&1&2&3&0&1&3&2&1&0&3&2&1&0\\
0&1&3&2&2&3&1&0&1&0&2&3&3&2&0&1&2&3&1&0&0&1&3&2&3&2&0&1&1&0&2&3\\
0&2&0&2&3&1&3&1&1&3&1&3&2&0&2&0&2&0&2&0&1&3&1&3&3&1&3&1&0&2&0&2\\
0&2&1&3&1&3&0&2&1&3&0&2&0&2&1&3&2&0&3&1&3&1&2&0&3&1&2&0&2&0&3&1\\
0&3&2&1&3&0&1&2&1&2&3&0&2&1&0&3&2&1&0&3&1&2&3&0&3&0&1&2&0&3&2&1\\
0&3&3&0&1&2&2&1&1&2&2&1&0&3&3&0&2&1&1&2&3&0&0&3&3&0&0&3&2&1&1&2\\
0&0&1&1&2&2&3&3&0&0&1&1&2&2&3&3&0&0&1&1&2&2&3&3&0&0&1&1&2&2&3&3
\end{tabular}
\caption{$OA(32,9,2,4)$}\label{oa:32}
\end{table}

\begin{table}\centering\small
\begin{tabular}{
c@{\ }c@{\ }c@{\ }c@{\ }c@{\ }c@{\ }c@{\ }c@{\ }c@{\ }c@{\ }c@{\ }c@{\ \ }
c@{\ }c@{\ }c@{\ }c@{\ }c@{\ }c@{\ }c@{\ }c@{\ }c@{\ }c@{\ }c@{\ }c@{\ \ }
c@{\ }c@{\ }c@{\ }c@{\ }c@{\ }c@{\ }c@{\ }c@{\ }c@{\ }c@{\ }c@{\ }c@{\ \ }
c@{\ }c@{\ }c@{\ }c@{\ }c@{\ }c@{\ }c@{\ }c@{\ }c@{\ }c@{\ }c@{\ }c}
0&0&0&0&0&0&0&0&0&0&0&0&1&1&1&1&1&1&1&1&1&1&1&1&2&2&2&2&2&2&2&2&2&2&2&2&3&3&3&3&3&3&3&3&3&3&3&3\\
0&0&0&1&1&1&2&2&2&3&3&3&1&1&1&0&0&0&3&3&3&2&2&2&2&2&2&3&3&3&0&0&0&1&1&1&3&3&3&2&2&2&1&1&1&0&0&0\\
0&0&0&2&2&2&3&3&3&1&1&1&1&1&1&3&3&3&2&2&2&0&0&0&2&2&2&0&0&0&1&1&1&3&3&3&3&3&3&1&1&1&0&0&0&2&2&2\\
0&1&2&1&2&3&0&1&3&0&2&3&1&0&3&0&3&2&1&0&2&1&3&2&2&3&0&3&0&1&2&3&1&2&0&1&3&2&1&2&1&0&3&2&0&3&1&0\\
0&1&2&2&3&1&1&3&0&3&0&2&1&0&3&3&2&0&0&2&1&2&1&3&2&3&0&0&1&3&3&1&2&1&2&0&3&2&1&1&0&2&2&0&3&0&3&1\\
0&1&2&3&1&2&3&0&1&2&3&0&1&0&3&2&0&3&2&1&0&3&2&1&2&3&0&1&3&0&1&2&3&0&1&2&3&2&1&0&2&1&0&3&2&1&0&3\\
0&2&3&0&1&3&2&3&1&0&1&2&1&3&2&1&0&2&3&2&0&1&0&3&2&0&1&2&3&1&0&1&3&2&3&0&3&1&0&3&2&0&1&0&2&3&2&1\\
0&2&3&1&3&0&3&1&2&2&0&1&1&3&2&0&2&1&2&0&3&3&1&0&2&0&1&3&1&2&1&3&0&0&2&3&3&1&0&2&0&3&0&2&1&1&3&2\\
0&2&3&3&0&1&1&2&3&1&2&0&1&3&2&2&1&0&0&3&2&0&3&1&2&0&1&1&2&3&3&0&1&3&0&2&3&1&0&0&3&2&2&1&0&2&1&3\\
0&3&1&0&3&2&0&2&1&3&2&1&1&2&0&1&2&3&1&3&0&2&3&0&2&1&3&2&1&0&2&0&3&1&0&3&3&0&2&3&0&1&3&1&2&0&1&2\\
0&3&1&2&0&3&1&0&2&2&1&3&1&2&0&3&1&2&0&1&3&3&0&2&2&1&3&0&2&1&3&2&0&0&3&1&3&0&2&1&3&0&2&3&1&1&2&0\\
0&3&1&3&2&0&2&1&0&1&3&2&1&2&0&2&3&1&3&0&1&0&2&3&2&1&3&1&0&2&0&3&2&3&1&0&3&0&2&0&1&3&1&2&3&2&0&1\\
0&0&0&1&1&1&2&2&2&3&3&3&0&0&0&1&1&1&2&2&2&3&3&3&0&0&0&1&1&1&2&2&2&3&3&3&0&0&0&1&1&1&2&2&2&3&3&3
\end{tabular}
\caption{$OA(48,13,2,4)$}\label{oa:48}
\end{table}

\chapter{Quantum Algorithms for Quantum Maps}\label{chap:qmaps}

This chapter presents quantum algorithms implementing quantum maps like the quantum sawtooth map \cite{shep129} and the quantum tent map \cite{shep144}.
These algorithms have been used in this thesis to study the error suppressing properties of the \textsf{PAREC} method in section \ref{sec:parec} and the embedded recoupling scheme in chapter \ref{chap:decrec} by means of numerical simulations.
A more elaborated discussion of such algorithms can be found in the author's diploma thesis \cite[chapter 2]{DiplKern}.
Furthermore, we define a discrete Husimi function which can be understood as the coherent state representation of a quantum state, and which can be used to illustrate quantum states.

\section{Quantum Gates}\label{sec:qgates}

Before we are going to derive a decomposition of a quantum map into a sequence of elementary one- and two-qubit gates, we have to define these gates.
Each of the one- and two-qubit gates will be represented in the standard computational basis
$\{ \ket{0}, \ket{1} \}$ and $\{ \ket{00},\ket{01},\ket{10},\ket{11} \}$, respectively.
Let us start with the one-qubit gates.

\subsection{One-Qubit Gates}

\subsubsection{Phase Gate}
The phase gate $\textsf{P}_{t}(\varphi)$ applies a phase $\varphi$ if the $t$-th qubit is in the state $\ket{1}$.
\begin{equation*} %
\text{\includegraphics[trim=0 9 0 0]{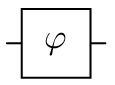}} \Leftrightarrow
\begin{pmatrix} 1&0\\ 0&e^{i\varphi} \end{pmatrix}
\end{equation*}
\subsubsection{Hadamard Gate}
The Hadamard gate $\textsf{H}_t$ generates a superposition of $\ket{0}$ and $\ket{1}$.
\begin{equation*}
\text{\includegraphics[trim=0 9 0 0]{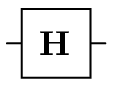}}  \Leftrightarrow
\frac{1}{\sqrt{2}}\begin{pmatrix} 1&1\\ 1&-1 \end{pmatrix}
\end{equation*}

\subsection{Two-Qubit Gates}

\subsubsection{The Controlled-Not Gate}
The controlled-not gate $\textsf{CNOT}_{c\,t}$ flips the state of the target qubit $t$ if the control qubit $c$ is in the state~$\ket{1}$.
\begin{equation*}
\text{\includegraphics[trim=0 19 0 0]{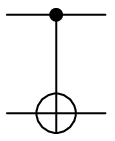}} \Leftrightarrow
\begin{pmatrix} 1&0&0&0\\ 0&1&0&0\\ 0&0&0&1\\ 0&0&1&0\end{pmatrix}
\end{equation*}

\subsubsection{The Controlled-Phase Gate}
The controlled-phase gate $\textsf{CP}_{c_1c_2}$ applies a phase $\varphi$ if the control qubits $c_1$ and $c_2$ are both in the state~$\ket{1}$.
\begin{equation*}
\text{\includegraphics[trim=0 21.5 0 0]{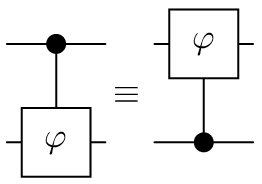}} \Leftrightarrow\
\begin{pmatrix} 1&0&0&0\\ 0&1&0&0\\ 0&0&1&0\\ 0&0&0&e^{i\varphi} \end{pmatrix}
\end{equation*}
A three-qubit controlled phase gate $\textsf{CCP}_{c_1c_2c_3}$ might be defined in a similar fashion.

\subsubsection{The Swap Gate}
The swap gate $\textsf{SWAP}_{t_1t_2}$ exchanges the state of the target qubits $t_1$ and $t_2$.
\begin{equation*}
\text{\includegraphics[trim=0 16 0 0]{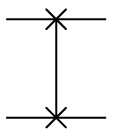}} \Leftrightarrow\
\begin{pmatrix} 1&0&0&0\\ 0&0&1&0\\ 0&1&0&0\\ 0&0&0&1 \end{pmatrix}
\end{equation*}

\section{Gate Decompositions for Quantum Maps}\label{sec:gatesqmaps}

Let us consider a quantum computer consisting of $n$ qubits.
The Hilbert space $\mathcal{H} = \mathcal{H}_2^{\otimes n}$ spanned by the computational basis $\{ \ket{ i_0,i_1,\dots, i_{n-1} } \}$, with $i_j\in\{0,1\}$ for $j=0,1,\dots,n-1$, is of dimension $d=2^n$.
A short hand notation of the basis states is given by $\ket{i} = \ket{ i_0,i_1,\dots, i_{n-1} }$ with $i = \sum_{j=0}^{n-1} i_j \cdot 2^j$.
We are going to construct a decomposition of a quantum map
\begin{equation}\label{eq:qmap}
U = \exp\Bigl( -\frac{i}{2} m^2 T \Bigr) \exp\Bigl( -i k V( q ) \Bigr),
\end{equation}
characterized by the parameters $T=2\pi/d$ and $k\in\mathbb{R}$,
into a sequence of elementary one- and two-qubit gates defined in the preceding section.
Here, $m$ denotes the momentum operator whose eigenstates form the computational basis,
$ m \ket{ i } = i\ket{ i }$,
and $q$ denotes the position operator which is related to the momentum operator via the quantum Fourier transform (QFT):
\begin{equation}
 q = U_\text{QFT}^{-1} \cdot \frac{2\pi}{d} m \cdot U_\text{QFT}.
\end{equation}
As a consequence, the quantum map can be written as the product of four unitaries
\begin{equation}
U = \exp\Bigl( -\frac{i}{2} m^2 T \Bigr) \cdot U_\text{QFT}^{-1} \cdot \exp\Bigl( -i k V\bigl( \frac{2\pi}{d} m \bigr) \Bigr) \cdot U_\text{QFT}.
\end{equation}
Each of these unitaries,
the QFT $U_\text{QFT}$,
the kick operator $\exp\bigl( -i k V( 2\pi m/d) \bigr)$,
the inverse QFT
and the free evolution operator $\exp\bigl( -\frac{i}{2} m^2 T \bigr)$,
can be decomposed into a sequence of elementary one- and two-qubit gates.
We present gate decompositions for the kick operator employing the sawtooth-potential
\begin{equation}\label{eq:Vsaw}
 V_\text{saw}(q) = -\frac{1}{2} (q-\pi)^2
\end{equation}
and the tent-potential
\begin{equation}\label{eq:Vtent}
 V_\text{tent}(q) = \begin{cases}
 -\frac{1}{2}q(q-\pi) &, 0 \leq q < \pi \\
 \frac{1}{2}(q-\pi)(q-2\pi) &, \pi \leq q < 2\pi
\end{cases}.
\end{equation}
The classical map corresponding to the quantum map \eqref{eq:qmap} is given by
\begin{equation}\label{eq:clmap}
\begin{split}
 p' &= p - K V'(q) \pmod{2\pi} \\ %
 q' &= q + p' \pmod{2\pi}
\end{split}
\end{equation}
and depends only on the single parameter $K=kT$.

\subsection{The Quantum Fourier Transform}\label{subsec:AppQFT}

\begin{figure}
\centering
\includegraphics[width=.75\textwidth, trim=0 0 0 0]{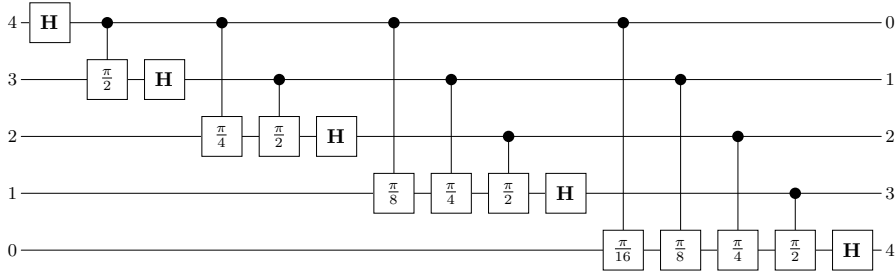}
\caption[Quantum Fourier transform]{Quantum circuit of the quantum Fourier transform for $n=5$ qubits.}
\label{fig:qft5}
\end{figure}

If we let the quantum Fourier transform (QFT) reverse the order of the qubits, i.\,e. if
\begin{equation}
U_\text{QFT} \ket{m_0} \otimes \ket{m_1} \otimes \dots  \otimes \ket{m_{n-1}} =
\frac{1}{\sqrt{d}} \sum_{x=0}^{d-1} \exp\Bigl(i\frac{2\pi}{d} mx \Bigr) \ket{x_{n-1}}\otimes \dots \otimes \ket{x_1} \otimes \ket{x_0},
\end{equation}
a decomposition of $U_\text{QFT}$ into $n(n+1)/2$ quantum gates (Hadamard gates and controlled-phase gates) is given by \cite{EkJo96}
\begin{equation}\label{eq:UQFTgates}
U_\text{QFT} =
\prod_{j=n-1}^0  \left(
 \biggl( \prod_{i=n-1}^{i>j} \textsf{CP}_{ji}\Bigl(\frac{\pi}{2^{i-j}}\Bigr) \biggr) \textsf{H}_j
\right),
\end{equation}
where the product over $j$ is non-commutative and has to be applied starting with $j=n-1$.
A corresponding quantum circuit for $n=5$ qubits is depicted in figure \ref{fig:qft5}.
The inverse operation $U_\text{QFT}^{-1}$ is obtained from \eqref{eq:UQFTgates} by multiplying each phase by the factor minus one.

\subsection{The Free Evolution Operator}

The free evolution operator $\exp\bigl( -\frac{i}{2} m^2 T \bigr)$ is implemented by a series of controlled- and uncontrolled-phase gates.
Using the binary representation $m=\sum_{j=0}^{n-1} m_j \cdot 2^j$, we obtain
\begin{equation}
\begin{split}
\exp\Bigl( -\frac{i}{2} m^2 T \Bigr)&= \exp\Bigl( -iT \sum_{v,w=0}^{n-1} m_v m_w 2^{v+w-1}\Bigr)\\
&=\prod_{v=0}^{n_q-1} \exp\left( -iT m_v 2^{2v-1}\right)\prod_{v<w}^{n_q-1} \exp\left( -iT m_v m_w 2^{v+w}\right),
\end{split}
\end{equation}
which translates to a series of $n(n+1)/2$ phase gates as follows:
\begin{equation}
\prod_{v=0}^{n-1} \textsf{P}_v\bigl( -T 2^{2v-1} \bigr) \prod_{v<w}^{n-1} \textsf{CP}_{vw}\bigl( -T 2^{v+w} \bigr).
\end{equation}

\subsection{The Kick Operator}

The gate decomposition of the kick operator $\exp\bigl( -i k V( 2\pi m/d) \bigr)$ depends on the detailed form of the potential $V$.
We start with the sawtooth potential given by \eqref{eq:Vsaw} and proceed with the tent potential given by \eqref{eq:Vtent}.

\subsubsection{Sawtooth Map}

The kick operator of the sawtooth map is given by
\begin{equation}
\exp\Bigl( i \frac{k}{2} \bigl( \frac{2\pi}{d}m-\pi \bigr)^2 \Bigr) =
\exp\Bigl( i \frac{2k\pi^2}{d^2} m^2  \Bigr)
\exp\Bigl( -i \frac{2k\pi^2}{d} m  \Bigr)
\exp\Bigl( i \frac{k\pi^2}{2}  \Bigr).
\end{equation}
Omitting the global phase, this translates into the sequence
\begin{equation}
\prod_{v=0}^{n-1} \textsf{P}_v\Bigl(- \frac{2k\pi^2}{d} 2^v %
+ \frac{2k\pi^2}{d^2} 2^{2v} \Bigr)
\prod_{v<w}^{n-1} \textsf{CP}_{vw}\Bigl( \frac{4k\pi^2}{d^2} 2^{v+w} \Bigr)
\end{equation}
consisting of $n(n+1)/2$ phase gates.

Hence, in total, the quantum algorithm implementing the quantum sawtooth map consists of $n_g = 4\times n(n+1)/2 = 2n(n+1)$ elementary quantum gates.
The algorithm presented in this section is an improved version of the algorithm proposed by Benenti et.~al. in \cite{shep129} and \cite{shep134} which makes use of a four-phase two-qubit gate which applies an individual phase to each of the states $\{ \ket{00},\ket{01},\ket{10},\ket{11}\}$ and consists of the larger number of $3n^2+n$ quantum gates in total.
\begin{figure}
\centering
\includegraphics[width=\textwidth]{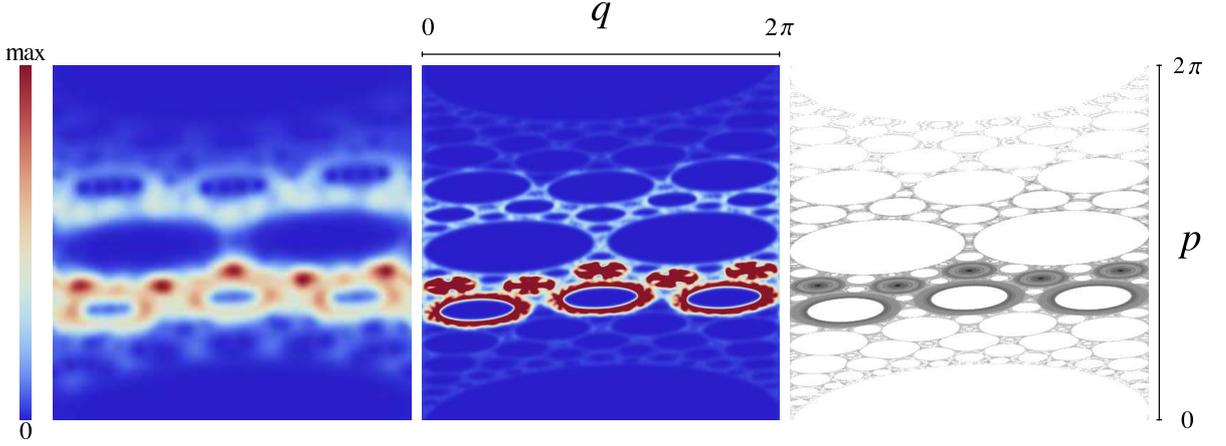}
\caption[Quantum sawtooth map]{Husimi function of the quantum sawtooth map with parameters $K=kT=-0.1$ and $T=2\pi/2^n$:
for $n=8$ qubits (\textit{left}) and $n=12$ qubits (\textit{middle}).
Classical trajectories (\textit{right}).}
\label{fig:qsaw}
\end{figure}
As an example, let us apply the sawtooth map $U$ with parameters $K=kT=-0.1$ and $T=2\pi/2^n$ on the initial state
$\ket{\Psi} = \ket{\ [0.38\cdot 2^n]\ }$.
Figure \ref{fig:qsaw} shows the average of the Husimi function of the state $U^t\ket{\Psi}$ taken over $950\leq t \leq 1000$.
The calculation was performed for $n=8$ (\textit{left part}) and $n=12$ qubits (\textit{middle part}).
The color gradient encodes the function values ranging from $0$ (\textit{blue}) up to the maximal value (\textit{red}).
For comparison, there are also $1000$ classical trajectories depicted (\textit{right part}) starting in the range $(0\leq q<2\pi , p\approx 0.38\cdot 2\pi)$ and resulting from $2000$ iterations of the classical map \eqref{eq:clmap}.

\subsubsection{Tent Map}

Setting $\bar{q}(q) = q $ if $0\leq q < \pi$ and $\bar{q}(q) = q -\pi$ if $\pi \leq q < 2\pi$,
the tent-potential becomes
\begin{equation}
 V_\text{tent}\bigl(\bar{q}(q)\bigr) = \begin{cases}
 -\frac{1}{2}\bar{q}(\bar{q}-\pi) &, 0 \leq q < \pi \\
 +\frac{1}{2}\bar{q}(\bar{q}-\pi) &, \pi \leq q < 2\pi
\end{cases}.
\end{equation}
In order to implement the kick operator $\exp\bigl( -i k V_\text{tent}( 2\pi m/d ) \bigr)$, we start by applying the operator
$\exp\bigl( i k \frac{1}{2}\bar{q}(\bar{q}-\pi) \bigr) =
\exp\bigl(i \frac{2k\pi^2}{d^2} \overline{m}^2 \bigr)
\exp\bigl(-i \frac{k\pi^2}{d} \overline{m} \bigr)$, where $\overline{m} = \sum_{j=0}^{n-2} m_j \cdot 2^j$ does not depend on the most significant qubit in position $n-1$.
This operator translates into the following sequence of phase gates:
\begin{equation}\label{eq:VtentA}
\prod_{v=0}^{n-2} \textsf{P}_v\Bigl( -\frac{k\pi^2}{d} 2^v + \frac{2k\pi^2}{d^2} 2^{2v} \Bigr)
\prod_{v<w}^{n-2} \textsf{CP}_{vw}\Bigl( \frac{4k\pi^2}{d^2} 2^{v+w} \Bigr)
\end{equation}
Since states with $q\geq \pi$ should have been multiplied with
$\exp\bigl( -i k \frac{1}{2}\bar{q}(\bar{q}-\pi) \bigr)$ instead,
the next step is to apply the operator $\exp\bigl( -i k \bar{q}(\bar{q}-\pi) \bigr)$ onto all such states.
This can be done by using the same gate sequence as in \eqref{eq:VtentA}, if each phase is multiplied by the factor $-2$, and each gate is additionally controlled by the most significant qubit in position $n-1$:
\begin{equation}
\prod_{v=0}^{n-2} \textsf{CP}_{n-1,v}\Bigl( \frac{2k\pi^2}{d} 2^v - \frac{4k\pi^2}{d^2} 2^{2v} \Bigr)
\prod_{v<w}^{n-2} \textsf{CCP}_{n-1,v,w}\Bigl( - \frac{8k\pi^2}{d^2} 2^{v+w} \Bigr).
\end{equation}
The three-qubit gate $\textsf{CCP}_{c_1 c_2 c_3}$ can be implemented by the following five qubit sequence:
\begin{equation}
\textsf{CCP}_{c_1 c_2 c_3} =
\textsf{CP}_{c_2c_1}\Bigl(\frac{\varphi}{2}\Bigr)
\textsf{CP}_{c_2c_3}\Bigl(\frac{\varphi}{2}\Bigr)
\textsf{CNOT}_{c_1c_3}
\textsf{CP}_{c_2c_3}\Bigl(-\frac{\varphi}{2}\Bigr)
\textsf{CNOT}_{c_1c_3}.
\end{equation}
As a consequence, the kick operator of the tent map is decomposed into $3n^2-7n+4$ elementary one- and two-qubit quantum gates.

In total, the quantum algorithm implementing the quantum tent map consists of $n_g = 3\times n(n+1)/2 +3n^2-7n+4 = \frac{9}{2}n^2-\frac{11}{2}n + 4$ quantum gates.
It was originally proposed by Frahm et.~al. in \cite{shep144}.
\begin{figure}
\centering
\includegraphics[width=\textwidth]{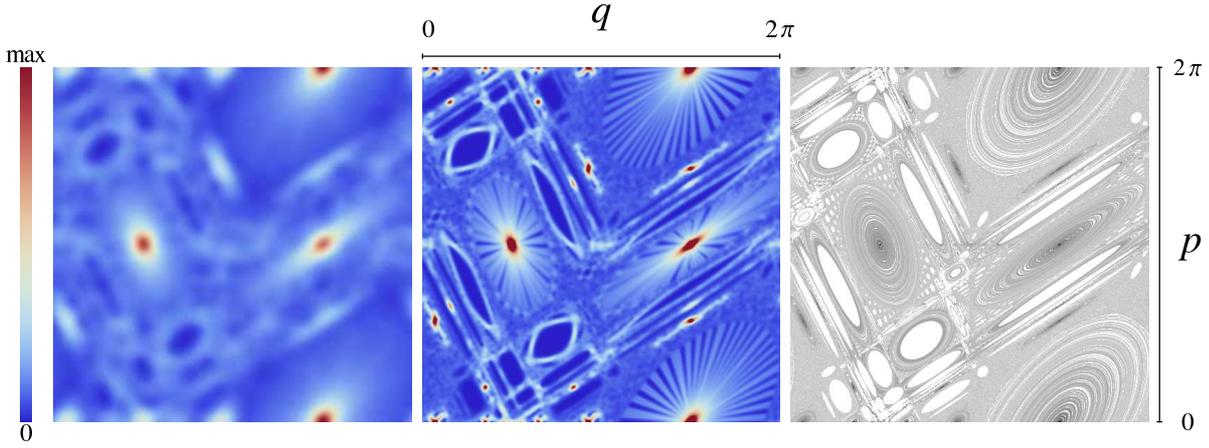}
\caption[Quantum tent map]{Husimi function of the quantum tent map with parameters $K=kT=4/3$ and $T=2\pi/2^n$:
for $n=8$ qubits (\textit{left}) and $n=12$ qubits (\textit{middle}).
Classical trajectories (\textit{right}).}
\label{fig:qtent}
\end{figure}
As an example, let us apply the tent map $U$ with parameters $K=kT=4/3$ and $T=2\pi/2^n$ on the initial state
$\ket{\Psi} = (\ket{\ 0\ }+\ket{\ 2^{n-1}\ })/\sqrt{2}$.
Figure \ref{fig:qtent} shows the average of the Husimi function of the state $U^t\ket{\Psi}$ taken over $950\leq t \leq 1000$.
The calculation was performed for $n=8$ (\textit{left part}) and $n=12$ qubits (\textit{middle part}).
For comparison, there are $1000$ classical trajectories shown (\textit{right part}) starting in the range
$(0\leq q<2\pi , p \in \{0,\pi\})$ and resulting from $2000$ iterations of the classical map~\eqref{eq:clmap}.\pagebreak

\section{Coherent States and the Husimi Function}\label{sec:csandhf}

Let us consider a quantum register consisting of $n$ qubits described by a Hilbert space $\mathcal{H}=\mathcal{H}_2^{\otimes n}$ of dimension $d=2^n$.
A coherent state in position $(0\leq q<d, 0\leq p<d)$ is defined as
\begin{equation}
\ket{\Phi (q,p)}  = \Bigl( \frac{2}{d} \Bigr)^\frac{1}{4} \sum_{j=0}^{d-1} \exp\Bigl( -i \frac{2\pi}{d} jq-\frac{\pi}{d} D^2(j,p) \Bigl) \ket{\ j\ }.
\end{equation}
Here, $D(j,p)$ denotes the difference $j-p$ mapped to the range $-d/2 \leq D < d/2$:
\begin{equation}
D(j,p) = \Bigl( j-p+\frac{d}{2} \pmod{d} \Bigr)-\frac{d}{2}.
\end{equation}
The state $\ket{\Phi(q,p)}$ is normalized in the limit of large $d$.
A quantum algorithm which prepares a coherent state in good approximation can be found in \cite{miqpaz3}.

\begin{figure}\hfill
\begin{minipage}{.4\textwidth}\centering
\includegraphics[width=.9\textwidth, trim=0 0 0 0]{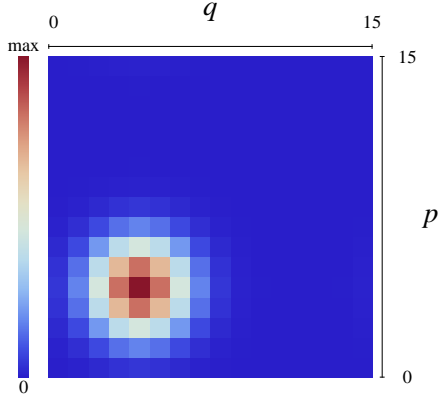}
\end{minipage}%
\hfill
\begin{minipage}{.4\textwidth}
\caption[Husimi function]{Husimi function $H(q,p)$ of the coherent $n=4$ qubit state $\ket{ \Phi(4,4) }$.\label{fig:hustest}}
\end{minipage}\hspace*{\fill}
\end{figure}

The Husimi function $H(q,p)$ of a quantum state $\ket{\Psi} = \sum_{j=0}^{d-1} \Psi_j \ket{j}$ is defined as the absolute square of the inner product between $\ket{\Psi}$ and a coherent state $\ket{\Phi (q,p)}$:
\begin{equation}\label{eq::husimi}
\begin{split}
H(q,p)&=\frac{1}{d} \bigl|\braket{\Phi(q,p)}{\Psi}\bigr|^2 \\
      &=\Bigl( \frac{2}{d^3} \Bigl)^\frac{1}{2} \biggl| \sum_{j=0}^{d-1} \exp\Bigl( i \frac{2\pi}{d} j q-\frac{\pi}{d} D^2(j,p) \Bigr)  \Psi_j \biggr|^2
\end{split}
\end{equation}
According to the above formula, a calculation of all $d^2$ values of $H(q,p)$ takes $\mathcal{O}(d^3)$ steps.
As it was recognized in \cite{shep144}, this calculation can be accelerated substantially by noting that a Fourier transformation is involved in expression \eqref{eq::husimi}:
Let us rewrite the equation as
\begin{equation}
\begin{split}
H(q,p) &= \biggl\vert \sum_{q'=0}^{d-1}  \bigl\langle q\bigr\vert
\frac{1}{\sqrt{d}} \sum_{j=0}^{d-1}  \exp\Bigl( \frac{2\pi i}{d} j q' \Bigr)
\Bigl[  \Psi_j \Bigl(\frac{2}{d}\Bigr)^\frac{1}{4} \exp\Bigl( -\frac{\pi}{d} D^2(j,p) \Bigr)  \Bigr]
\bigl\vert q'\bigr\rangle \biggr\vert^2 \\
& = \biggl\vert  \bigl\langle q\bigr\vert \sum_{q'=0}^{d-1} \tilde{\Psi}_{q'} \bigl\vert q'\bigr\rangle \biggr\vert^2,
\end{split}
\end{equation}
where the vector with entries $\tilde{\Psi}_{q'}$ denotes the Fourier transform of
\begin{equation}
\ket{\Psi'} = %
\sum_{j=0}^{d-1} \Bigl[  \Psi_j \Bigl(\frac{2}{d}\Bigr)^\frac{1}{4} \exp\Bigl( -\frac{\pi}{d} D^2(j,p) \Bigr)  \Bigr] \ket{j}.
\end{equation}
By calculating the fast Fourier transformation for the $d$ vectors $\ket{\Psi'}$ associated with $p\in\{0,1,\dots,d-1\}$, all values of $H(q,p)$ can be obtained in only $\mathcal{O}(d^2 \log_2 d)$ steps.
In the limit of large $d$, the function values of the Husimi function add up to one:
\begin{equation}
 \sum_{p=0}^{d-1} \sum_{q=0}^{d-1} H(p,q) = 1.
\end{equation}
As an example, the Husimi function of the coherent $n=4$ qubit state $\ket{ \Phi(d/4,d/4) }$ is depicted in figure~\ref{fig:hustest}.

\chapter{Technical Results}\label{chap:techres}

This chapter of the appendix contains various technical results which are referred to in part II of this thesis.
The first section proves some counting lemmas for linear codes, the second section proves the existence of good self-orthogonal codes, and the third section proves some lemmas concerning a Bell state.

\section{Linear Codes}\label{sec:app:lincodes}
This section provides two corollaries  which are needed for the proof the random coding arguments in subsections \ref{subsec:randomlincodes} and  \ref{subsec:randcss}.

Let us denote the set containing all $[n,k]_q$ codes by
\begin{equation}
 A_{n,k,q} = \{ \mathcal{C}\subseteq\mathbb{F}_q^n \sthat \mathcal{C}\text{ is an } [n,k]_q \text{-code} \},
\end{equation}
and let us denote the subset of codes in $A_{n,k,q}$ which contain a certain nonzero codeword $\vec{x}\in\mathbb{F}_q^n$ by
\begin{equation}
A_{n,k,q} (\vec{x}) = \{ \mathcal{C}\in A_{n,k,q} \sthat \vec{x} \in \mathcal{C} \}.
\end{equation}

\begin{lem}\label{lem:numberlincodes}
The total number of $[n,k]_q$ codes is given by
\begin{equation}
\vert A_{n,k,q} \vert =
\frac{ \prod_{i=0}^{k-1} (q^n-q^i) }{ \prod_{i=0}^{k-1} (q^k-q^i) } \qquad (1\leq k \leq n)
\end{equation}
and $\vert A_{n,0,q} \vert = 1$.
The number of $[n,k]_q$ codes which contain a certain nonzero vector $\vec{x}$ is given by
\begin{equation}
\vert A_{n,k,q}(\vec{x}) \vert =
\frac{ \prod_{i=1}^{k-1} (q^n-q^i) }{ \prod_{i=1}^{k-1} (q^k-q^i) }  \qquad (1 < k \leq n),
\end{equation}
and $\vert A_{n,1,q}(\vec{x}) \vert = 1$, $\vert A_{n,0,q}(\vec{x}) \vert = 0$ independently of $\vec{x} \neq \vec{0}$.
\end{lem}
\begin{proof}
To determine the total number of $[n,k]_q$ codes, we have to count all possibilities to choose $k$ linearly independent vectors from $\mathbb{F}_q^n$.
There are $q^n-1$ candidates for the first vector, there remain $q^n-q$ for the second, $q^n-q^2$ for the third, and so on.
We therefore get in total $N=(q^n-q^0)(q^n-q^1)\dots (q^n-q^{k-1})$ possibilities.
Since many of these selections of $k$ vectors span the same codespace, we have to divide $N$ by the number of ways a set of $k$ generating vectors can be found for a $k$-dimensional subspace.
This number is $(q^k-q^0)(q^k-q^1)\dots (q^k-q^{k-1})$.
The number of linear codes containing a particular nonzero $\vec{x}$ can be found in a similar fashion, but now as first independent vector we choose $\vec{x}$ itself.
\end{proof}

\begin{cor}\label{cor:numberlincodes}
One obtains from the above lemma that for any nonzero $\vec{x}\in\mathbb{F}_q^n$
\begin{equation}
 \frac{ \vert A_{n,k,q} (\vec{x}) \vert }{ \vert A_{n,k,q} \vert } = \frac{ q^k-1 }{q^n-1} \leq \frac{1}{q^{n-k}} .
\end{equation}
\end{cor}

\noindent
The following lemmas are slight generalizations of lemma \ref{lem:numberlincodes}.
\begin{lem}\label{lem:numberlincodescontaink}
Let $\mathcal{K}$ be an $[n,\kappa]_q$ code and let
\begin{equation}
A_{n,k,q} ( \mathcal{K} ) = \{ \mathcal{C}\in A_{n,k,q} \sthat  \mathcal{K} \subseteq  \mathcal{C} \}
\end{equation}
be the set of all $[n,k]_q$ codes which contain $\mathcal{K}$.
Then,
\begin{equation}
\vert A_{n,k,q}( \mathcal{K} ) \vert =
\frac{ \prod_{i=\kappa}^{k-1} (q^n-q^i) }{ \prod_{i=\kappa}^{k-1} (q^k-q^i) }  \qquad (\kappa < k \leq n),
\end{equation}
and $\vert A_{n,\kappa,q}( \mathcal{K} )\vert =1$, $\vert A_{n,k,q}( \mathcal{K} )\vert =0$ for $(k<\kappa)$.
\end{lem}

\begin{lem}\label{lem:numberlincodescontainkx}
Let $\mathcal{K}$ be an $[n,\kappa]_q$ code and let
\begin{equation}
A_{n,k,q} ( \mathcal{K},\vec{x} ) = \{ \mathcal{C}\in A_{n,k,q} \sthat \mathcal{K} \subseteq  \mathcal{C} \text{ and } \vec{x}\in \mathcal{C} \}
\end{equation}
be the set of all $[n,k]_q$ codes which contain $\mathcal{K}$ and a certain nonzero vector $\vec{x}\in\mathbb{F}_q^n$.
Then,
\begin{equation}
\vert A_{n,k,q}( \mathcal{K},\vec{x} ) \vert =
\begin{cases}
 \vert A_{n,k,q}( \mathcal{K} ) \vert & \text{ if } \vec{x}\in\mathcal{K} \\
\frac{ \prod_{i=\kappa+1}^{k-1} (q^n-q^i) }{ \prod_{i=\kappa+1}^{k-1} (q^k-q^i) } & \text{ if }
 \vec{x}\notin\mathcal{K} \text{ and } \kappa+1 < k \leq n \\
 1 & \text{ if } \vec{x}\notin\mathcal{K} \text{ and } \kappa+1 = k  \\
 0 & \text{ else }
\end{cases}.
\end{equation}
\end{lem}

\begin{cor}\label{cor:numberc2ifc1}
Let $\mathcal{K}$ be an $[n,n-k_1]_q$ code and let $\bigl\langle \,\cdot\, \bigr\rangle_{\mathcal{K}\in A_{n,n-k_1,q}}$ denote the average over all such codes. Then,
\begin{equation}
\Bigl\langle
\frac{\vert A_{n,n-k_2,q}( \mathcal{K},\vec{x} ) \vert}{\vert A_{n,n-k_2,q}( \mathcal{K} ) \vert}
\Bigr\rangle_{  \mathcal{K}\in A_{n,n-k_1,q}  } = \frac{q^{n-k_2}-1}{q^n-q} \leq \frac{1}{q^{k_2}}.
\end{equation}
\end{cor}
\begin{proof}

\begin{align*}
\Bigl\langle
\frac{\vert A_{n,n-k_2,q}( \mathcal{K},\vec{x} ) \vert}{\vert A_{n,n-k_2,q}( \mathcal{K} ) \vert}
\Bigr\rangle_{  \mathcal{K}\in A_{n,n-k_1,q}  } &=
\frac{1}{\vert A_{n,n-k_1,q} \vert}
\sum_{ \mathcal{K}\in A_{n,n-k_1,q} }
\frac{\vert A_{n,n-k_2,q}( \mathcal{K},\vec{x} ) \vert}{\vert A_{n,n-k_2,q}( \mathcal{K} ) \vert} %
\intertext{We use lemma \ref{lem:numberlincodescontaink} and \ref{lem:numberlincodescontainkx} and obtain}
&=
\frac{1}{\vert A_{n,n-k_1,q} \vert}
\sum_{ \mathcal{K}\in A_{n,n-k_1,q} }
\begin{cases}
 1 &, \text{if } \vec{x}\in \mathcal{K}\\
\frac{  q^{n-k_2}-q^{n-k_1}  }{  q^n-q^{n-k_1}  } &,\text{else }
\end{cases}\\
&=
1
\cdot \frac{ \vert A_{n,n-k_1,q} (\vec{x}) \vert }{ \vert A_{n,n-k_1,q} \vert }
+
\frac{  q^{n-k_2}-q^{n-k_1}  }{  q^n-q^{n-k_1}  }
\cdot \Bigl( 1- \frac{ \vert A_{n,n-k_1,q} (\vec{x}) \vert }{ \vert A_{n,n-k_1,q} \vert } \Bigr).
\end{align*}
Corollary \ref{cor:numberlincodes} tells us that $ \vert A_{n,n-k_1,q} (\vec{x}) \vert / \vert A_{n,n-k_1,q} \vert = (q^{n-k_1}-1)/(q^n-1)$ which leads to the desired result.
\end{proof}

\section{Self-Orthogonal Codes}\label{sec:app:goodselfortho}

In this section it is shown that good self-orthogonal codes do exist.
A self-orthogonal $q$-ary linear $[n,\textsf{k}]_q$ code $\mathcal{C}$ over the field $\mathbb{F}_q^n$ is a code which is contained in its dual $[n,n-\textsf{k}]_q$ code $\mathcal{C}^\perp$.
A code $\mathcal{C}$ is called self-dual provided that $\mathcal{C} = \mathcal{C}^\perp$ (in this case $n$ has to be even and $\textsf{k}=n/2$).
Self-orthogonal $[n,\textsf{k}]_q$ codes can be used to construct quantum CSS-codes encoding $k=n-2\textsf{k}$ qudits into $n$.
If $\mathcal{C}^\perp$ has minimum distance $d$, $\mathcal{C}$ has to be at least of the same minimum distance.
Hence the quantum CSS-code will be of distance $d$.

In the following subsections,
a Gilbert-Varshamov lower bound is established, which guarantees the existence of self-orthogonal $[n,\textsf{k},d]_q$ codes such that the dual $[n,n-\textsf{k},d]_q$ code has minimum distance $d$ and rate
\begin{equation}
 \frac{n-\textsf{k}}{n}  \geq 1 - H_{q[\log_q]}\Bigl( 1-\frac{d}{n}, \frac{d/n}{q-1},\dots, \frac{d/n}{q-1} \Bigr).
\end{equation}
For the binary case ($q=2$) this result was found by Calderbank and Shor \cite{CS96}.
The corresponding proof is given in the first section.
The nonbinary case ($q\geq 3$) has to be treated separately.
It is proven in the second section using results presented in \cite{Ha03}.

\subsection{The Binary Case}

\begin{lem}[\cite{CS96}]
For even $n\geq 2$ and $0 < \textsf{k} \leq n/2$, %
let
\begin{equation}
 A(n,\textsf{k}) = \bigl\{ \mathcal{C} \subseteq F_2^n \ \vert\ \mathcal{C} \text{ is a } [n,\textsf{k}]_2\text{-code}, \{ \vec{0},\vec{1} \} \subseteq \mathcal{C} \subseteq \mathcal{C}^\perp \bigr\}
\end{equation}
be the set of all self-orthogonal $[n,\textsf{k}]_2$ codes which include the $[n,1]_2$ subcode $\{ \vec{0},\vec{1} \}$, and let
\begin{equation}
 A_{\vec{x}} = \bigl\{ \mathcal{C}\in A(n,\textsf{k}) \ \vert\ \vec{x}\in\mathcal{C}^\perp \bigr\}
\end{equation}
be the subset of $A(n,\textsf{k})$ including only those codes whose dual codes include $\vec{x}\in\mathbb{F}_2^n$.
Then, there exists a constant $T_0$ satisfying $\vert A_{\vec{x}} \vert = T_0$
for any $\vec{x}\in\mathbb{F}_2^n$ with $\vec{x}\neq \vec{0}$, $\vec{x}\neq \vec{1}$ and $\vec{x}\cdot\vec{x}=0\pmod{2}$.
\end{lem}

\begin{rem}
For a proof we refer to \cite{CS96}.
Note that for all $\vec{x}\in\mathcal{C}^\perp$, $\wt(\vec{x})=0\pmod{2}$ which follows from $\vec{1} \cdot \vec{x}=0\pmod{2}$.
(By $\vec{1}$ we denote the vector $(1,1,\dots,1)\in\mathbb{F}_2^n$ and analogously $\vec{0}=(0,0,\dots,0)\in\mathbb{F}_2^n$.)
\end{rem}

\begin{thm}
Consider the set of codes $\Phi = \{ \mathcal{C}^\perp \ \vert\ \mathcal{C}\in A(n,\textsf{k}) \}$.
Then, as long as
\begin{equation}\label{eq:cssgvbin_condi}
 \sum_{s=1}^{2s\leq d-1} \binom{n}{2s} < \frac{2^{n-1}-2}{2^{n-\textsf{k}}-2},
\end{equation}
there exist codes of minimum distance $d$ in $\Phi$.
\end{thm}
\begin{proof}
Counting all vectors $\vec{x}$ (except $\vec{x}=\vec{0}$ and $\vec{x}=\vec{1}$) in $\Phi$ in two different ways, we get (by noting that $\vert\Phi\vert = \vert A(n,\textsf{k}) \vert$)
\begin{equation}
 \vert  A(n_,\textsf{k}) \vert \cdot (2^{n-\textsf{k}}-2) = (2^{n-1}-2) \cdot T_0.
\end{equation}
There are $\sum_{s=1}^{2s\leq d-1} \binom{n}{2s}$ nonzero vectors of even weight less than $d$.
These vectors are distributed over $\sum_{s=1}^{2s\leq d-1} \binom{n}{2s} \cdot T_0$ codes at most.
As long as this number of codes is smaller than $\vert A(n,\textsf{k}) \vert$ (the total number of codes in $\Phi$), there have to be codes in $\Phi$ which are at least of minimum distance $d$.
\end{proof}

\begin{cor}
Consider the set codes $\Phi = \{ \mathcal{C}^\perp \ \vert\ \mathcal{C}\in A(n,\textsf{k}) \}$.
Then, as long as
\begin{equation}\label{eq:corbincase}
 \frac{n-\textsf{k}}{n}  < 1 - H_2( d/n ),
\end{equation}
there exist codes of minimum distance $d$ in $\Phi$.
\end{cor}
\begin{proof}
The tail inequality gives an upper bound for the left hand side of \eqref{eq:cssgvbin_condi}:
\begin{equation*}
 \sum_{s=1}^{2s\leq d-1} \binom{n}{2s} < \sum_{j=0}^{d-1} \binom{n}{j} \leq 2^{n H_2( (d-1)/n )}
 < 2^{n H_2( d/n )}.
\end{equation*}
A lower bound for the right hand side of \eqref{eq:cssgvbin_condi} is given by $2^{n-1} / 2^{n-\textsf{k}}$.
Hence, as long as  $nH_2(d/n)+n-\textsf{k} < n-1$ condition \eqref{eq:cssgvbin_condi} will be satisfied, too.
For large $n$ this leads to condition \eqref{eq:corbincase}.
\end{proof}

\subsection{The Higher Dimensional Case} %

\begin{lem}[\cite{Ha03}]
For $q\geq 3$ let
\begin{equation}
 A(n,\textsf{k}) = \bigl\{ \mathcal{C} \subseteq F_q^n \ \vert\ \mathcal{C} \text{ is a } [n,\textsf{k}]_q\text{-code}, \mathcal{C} \subseteq \mathcal{C}^\perp \bigr\}
\end{equation}
be the set of all self-orthogonal $[n,\textsf{k}]_q$ codes, and let
\begin{equation}
 A_{\vec{x}} = \bigl\{ \mathcal{C}\in A(n,\textsf{k}) \ \vert\ \vec{x}\in\mathcal{C}^\perp \bigr\}
\end{equation}
be the subset of $A(n,\textsf{k})$ including only those codes whose dual code includes $\vec{x}\in\mathbb{F}_q^n$.
Then, for any $u\in\mathbb{F}_q$, there exists a constant $T_u$ satisfying $\vert A_{\vec{x}} \vert = T_u$
for any nonzero $\vec{x}\in\mathbb{F}_q^n$ with $\vec{x}\cdot\vec{x}=u\pmod{q}$.
\end{lem}

\begin{rem}
For a proof of the above lemma we refer to \cite[Lemma 1]{Ha03}.
The following theorem is proven using results from \cite[Corollary 1]{Ha03}.
\end{rem}

\begin{thm}
Consider the set of codes $\Phi = \{ \mathcal{C}^\perp \ \vert\ \mathcal{C}\in A(n,\textsf{k}) \}$.
Then, as long as
\begin{equation}\label{eq:cssgv_condi}
 \sum_{j=1}^{d-1} \binom{n}{j} (q-1)^j < \frac{q^{n-q+1}-1}{q^{n-\textsf{k}}-1},
\end{equation}
there exist codes of minimum distance $d$ in $\Phi$.
\end{thm}
\begin{proof}
Let $S_u=\{ \vec{x}\in\mathbb{F}_q^n \vert \vec{x}\cdot\vec{x}=u\pmod{q}, \vec{x}\neq\vec{0} \}$ for $u\in\mathbb{F}_q$.
It follows that $\vert S_u\vert \geq q^{n-q+1}-1$ since the first $n-q+1$ digits of any $\vec{x}\in S_u$ can be set in arbitrary manner (except to $(0,\dots,0)$).
Counting pairs $(\vec{x},\mathcal{C})$ such that $\vec{x}\cdot\vec{x}=u\pmod{q}$, $\vec{x}\neq\vec{0}$, and $\vec{x} \in \mathcal{C}^\perp \in \Phi$, we find that (noting that $\vert\Phi\vert = \vert A(n,\textsf{k}) \vert$)
\begin{equation}
 \vert S_u\vert \cdot T_u \leq \vert  A(n,\textsf{k}) \vert \cdot (q^{n-\textsf{k}}-1)
\end{equation}
and we get (using the upper bound on $\vert S_u\vert$)
\begin{equation}\label{eq:cssgv_adt}
 \frac{ q^{n-q+1}-1 }{ q^{n-\textsf{k}}-1 } \leq \frac{\vert  A(n,\textsf{k}) \vert}{ T_u }.
\end{equation}
There are $\sum_{j=1}^{d-1} \binom{n}{j}(q-1)^j$ nonzero vectors of weight less than $d$.
These vectors are distributed over $\sum_{j=1}^{d-1} \binom{n}{j}(q-1)^j \cdot \max_u\{T_u\}$ codes at most.
As long as this number is smaller than $\vert A(n,\textsf{k}) \vert$, there have to be codes in $\Phi$ which are at least of minimum distance $d$.
Because of \eqref{eq:cssgv_adt}, equation \eqref{eq:cssgv_condi} is a sufficient condition.
\end{proof}

\begin{cor}
Consider the set of codes $\Phi = \{ \mathcal{C}^\perp \ \vert\ \mathcal{C}\in A(n,\textsf{k}) \}$.
Then, for large enough $n$, as long as
\begin{equation}
 \frac{n-\textsf{k}}{n}  < 1 - H_{q[\log_q]}\Bigl( 1-\frac{d}{n}, \frac{d/n}{q-1},\dots, \frac{d/n}{q-1} \Bigr)
\end{equation}
there exist codes of minimum distance $d$ in $\Phi$.
\end{cor}
\begin{proof}
By using the Chernoff bound \ref{lem:chernovbound} it was shown in the proof of corollary \ref{cor:gvasymp}
that an upper bound for the left hand side of \eqref{eq:cssgv_condi} is given by
\begin{equation}
  \sum_{j=0}^{d-1} \binom{n}{j} (q-1)^j <
 \exp_q\biggl(n H_{q[\log_q]}\Bigl( 1-\frac{d}{n}, \frac{d}{n(q-1)},\dots,\frac{d}{n(q-1)}\Bigr) \biggr).
\end{equation}
A lower bound for the right hand side of \eqref{eq:cssgv_condi} is given by
\begin{equation}
 \frac{q^{n-q+1}}{q^{n-\textsf{k}}} < \frac{q^{n-q+1}-1}{q^{n-\textsf{k}}-1} .
\end{equation}
Therefore, as long as
\begin{equation}
 \frac{n-\textsf{k}}{n}  < 1 - H_{q[\log_q]}\Bigl( 1-\frac{d}{n}, \frac{d/n}{q-1},\dots, \frac{d/n}{q-1} \Bigr) -\frac{q-1}{n},
\end{equation}
condition \eqref{eq:cssgv_condi} will be satisfied, too.
For large $n$ we can neglect the $q-1$ term.
\end{proof}

\section{Bell State Lemmas}

We are going to prove two simple lemmas concerning the Bell state
$\ket{\Phi}_{AB} = q^{-\frac{1}{2}} \sum_{j=0}^{q-1} \ket{j\tilde{j}}_{AB}$ that are relevant in chapter \ref{chap:crypto}.
Here, $\ketl{i\tilde{j}}{AB} = \ketl{i}{A} \otimes \ketl{\tilde{j}}{B}$,
where $\{ \ketl{i}{A} \}_{i=0,\dots,q-1}$ and $\{ \ketl{\tilde{j}}{B} \}_{j=0,\dots,q-1}$ denote orthonormal bases of the $q$-dimensional Hilbert spaces $\mathcal{H}_A$ and $\mathcal{H}_B$, respectively.

\begin{lem}[]\label{lem:OTo1gleich1oO}
Let $\ketl{\Phi}{AB} = \frac{1}{\sqrt{q}}\sum_{j=0}^{q-1} \ketl{j\tilde{j}}{AB}$.
Then,
\begin{equation}
\optl{O}{A} \otimes \opl{\id}{B} \ketl{\Phi}{AB} = \opl{\id}{A} \otimes \opl{O}{B} \ketl{\Phi}{AB},
\end{equation}
if the transposition is with respect to the $\{ \ketl{j}{A} \}$ basis and $\opl{O}{B}$ has the same matrix elements with respect to the $\{ \ketl{\tilde{j}}{B} \}$ basis as $\opl{O}{A}$ with respect to the $\{ \ketl{j}{A} \}$ basis, i.\,e.
$\opl{O}{A} = \sum_{ij} O_{ij} \ketbral{i}{j}{A}$ and
$\opl{O}{B} = \sum_{ij} O_{ij} \ketbral{\tilde{i}}{\tilde{j}}{B}$.
\end{lem}
\begin{proof}
We obtain
\begin{equation*}
\begin{split}
 \optl{O}{A} \otimes \opl{\id}{B} \ketl{\Phi}{AB}
&=\frac{1}{\sqrt{d}} \sum_{ij} O_{ij} \ketbral{j}{i}{A} \sum_k \ketl{k}{A}\ketl{\tilde{k}}{B} \\
&=\frac{1}{\sqrt{d}} \sum_{ij} O_{ij} \ketl{j}{A}\ketl{\tilde{i}}{B} \\
&=\frac{1}{\sqrt{d}} \sum_{ij} O_{ij} \ketbral{\tilde{i}}{\tilde{j}}{B} \sum_k \ketl{k}{A}\ketl{\tilde{k}}{B} \\
&=\opl{\id}{A} \otimes \opl{O}{B} \ketl{\Phi}{AB}. \qedhere
\end{split}
\end{equation*}
\end{proof}

\begin{lem}[]\label{lem:UoUgleich1o1}
Let $\ketl{\Phi}{AB} = \frac{1}{\sqrt{q}}\sum_{j=0}^{q-1} \ketl{j\tilde{j}}{AB}$.
Then, for any unitary $U$,
\begin{equation}
\opsl{U}{A} \otimes \opl{U}{B} \ketl{\Phi}{AB} = \ketl{\Phi}{AB}
\end{equation}
if the conjugation is with respect to the $\{ \ketl{j}{A} \}$ basis and $\opl{U}{B}$ has the same matrix elements with respect to the $\{ \ketl{\tilde{j}}{B} \}$ basis as $\opl{U}{A}$ with respect to the $\{ \ketl{j}{A} \}$ basis, i.\,e.
$\opl{U}{A} = \sum_{ij} U_{ij} \ketbral{i}{j}{A}$ and
$\opl{U}{B} = \sum_{ij} U_{ij} \ketbral{\tilde{i}}{\tilde{j}}{B}$.
\end{lem}
\begin{proof}
We obtain
\begin{equation*}
\begin{split}
\opsl{U}{A} \otimes \opl{U}{B} \ketl{\Phi}{AB}
&=\frac{1}{\sqrt{d}} \sum_{ijmnk} U^\ast_{ij} \ketbral{i}{j}{A} U_{mn} \ketbral{\tilde{m}}{\tilde{n}}{B} \ketl{k}{A}\ketl{\tilde{k}}{B} \\
&=\frac{1}{\sqrt{d}} \sum_{ijmn} U^\ast_{ij} \ketl{i}{A} U_{mn} \ketl{\tilde{m}}{B} 
\braketl{\tilde{n}}{\tilde{j}}{B} \\
&=\frac{1}{\sqrt{d}} \sum_{ijm} U^\ast_{ij} \ketl{i}{A} U_{mj} \ketl{\tilde{m}}{B} \\
&=\frac{1}{\sqrt{d}} \sum_{im} \delta_{im} \ketl{i}{A} \ketl{\tilde{m}}{B} = \ketl{\Phi}{AB}. \qedhere
\end{split}
\end{equation*}
\end{proof}

\chapter{Schur Transform and Eigenfunction Method}\label{chap:schur}

The Schur transform is a unitary transformation relating the standard computational basis of $n$ qudits of dimension $q$ to a basis associated with the representation theory of the symmetric and general linear groups.
This chapter explains how the eigenfunction method \cite{CPW02} can be used to obtain a computer program which calculates the Schur transform for given values of $n$ and $q$.
As explained in section \ref{sec:efm}, the eigenfunction method decomposes a given group representation into its irreducible parts.
It is shown in section \ref{sec:schurtransform} how the Schur transform can be obtained with the help of the eigenfunction method applied to the natural representation of the symmetric group $\textsf{S}_n$.
In addition we present some examples and discuss how the Schur transform allows for efficient communication in the absence of a shared reference frame.

\section{The Eigenfunction Method}\label{sec:efm}

This section summarizes the eigenfunction method (EFM) of Chen, Ping and Wang \cite{CPW02}.
Let $R(G)$ be a $d$-dimensional representation of a finite group $G$ on an inner product space $\mathcal{V}$ over the field~$\mathbb{C}$.
The EFM can be used to decompose $\mathcal{V}$ into a direct sum of irreducible subspaces and to construct a basis for each of these subspaces which corresponds to a given canonical subgroup chain.
To achieve this decomposition of $\mathcal{V}$, a complete set of commuting observables (CSCO) $\mathfrak{C}$ is constructed, whose eigenvectors (eigenfunctions) are the desired basis vectors.
They can be identified by their eigenvalue list.
All the results presented in this section are taken from \cite{CPW02}.
While we tried to supply the proofs for the fundamental results, we sometimes give the remark 'it can be shown'. These missing proofs can be found in \cite{CPW02}.

We start with a description of the EFM for general finite groups in subsection \ref{subsec:fingr} and specialize in the symmetric group in subsection \ref{subsec:symmgr}.

\subsection{General Finite Groups}\label{subsec:fingr}

We begin with the construction of the CSCO $\mathfrak{C}$ decomposing the representation space of the regular representation of a finite group $G$.
Let $\mathcal{V}$ be an inner product space of dimension $d = n_G$ over the field $\mathbb{C}$, where $n_G = \vert G\vert$ denotes the order of the finite group $G$, and fix an orthonormal basis $\{ \ket{i} \}$ ($i=0,\dots,d-1$).
The elements of the regular representation $R(G)$ of $G$ have the property
that $\ket{i} = R_i \ket{0}$ for all $R_i\equiv R(i)$ with $i\in G$, with $R_0$ denoting the identity.
In other words,
\begin{equation}
 \bra{i} R_k \ket{j} \equiv D_{ij}(k) = \begin{cases}
 1 & \text{ if } R_k R_j = R_i\\
 0 & \text{ else }
\end{cases}.
\end{equation}
The state $\ket{0}$ is said to possess no symmetry with respect to $G$.
A state $\ket{0}'$ which remains invariant under $G$ is called totally symmetric with respect to $G$ (it would generate a one-dimensional representation). States showing an intermediate behavior are said to possess partial symmetry.

Subsequently, we show how the construction of the CSCO $\mathfrak{C}$ has to be adjusted when dealing with non-regular representations $R(G)$.
In this case the state $\ket{0}$ is invariant under a set of elements $G_\text{in}$ forming a non-trivial subgroup of $G$, i.\,e. $R_a \ket{0} = \ket{0}$ for all $a\in G_\text{in}$, and is said to possess at least partial symmetry with respect to $G$.
Naturally, the dimension $d$ of a non-regular rep space spanned by the linearly independent $\ket{ i }=R_i\ket{0}$, $R_i\in R(G)$, is smaller than $n_G$.

\subsubsection{Reduction of the Regular Representation}

Let us define a class operator $C_i$ for each of the $n_\zeta$ conjugacy classes of $G$ as
the sum over all operators in the corresponding class,
\begin{equation}
 C_i = \sum_{j=1}^{n_i} R( a_j^{(i)} ), \quad i=1\dots n_\zeta,
\end{equation}
where $a_j^{(i)}$ denotes the $j$-th element of the $i$-th class and $n_i$ denotes the total number of elements in the $i$-th class.
The class operators commute with all elements in $R(G)$, $[C_i, R_a]=0$ for all $a\in G$, and therefore with one another, $[C_i, C_j] = 0$ for $i,j=1\dots n_\zeta$.
We assume that the $\{ C_i \}_{i=1}^{n_\zeta}$ are self-adjoint (they are if the classes are ambivalent), otherwise an equivalent set of $n_\zeta$ self-adjoint operators $\{ C_i' \}_{i=1}^{n_\zeta}$ can be obtained by taking suitable linear combinations of the non-ambivalent $C_i$.
The class space is defined as the $n_\zeta$-dimensional subspace of the regular rep space $\mathcal{V}$ spanned by the orthogonal set of states
\begin{equation}
 \Bigl\{  \ket{ C_i } = \sum_{j=1}^{n_i} R( a_j^{(i)} ) \ket{0} \Bigr\}_{i=1}^{n_\zeta}
\end{equation}
with $\braket{ C_j }{ C_i } = n_i \delta_{ij}$.
It can be shown that the class space forms a so-called natural representation space of the class operators,
and that the set of $n_\zeta$ class operators $( C_1,\dots, C_{n_\zeta} )$ is a CSCO of the natural rep, reducing the natural rep to a sum of $n_\zeta$ one-dimensional irreps via the eigenvector equation
\begin{equation}
 ( C_1,\dots, C_{n_\zeta} ) \ket{ Q^{(\nu)} } = (\lambda_1^{(\nu)},\dots, \lambda_{n_\zeta}^{(\nu)}) \ket{ Q^{(\nu)} } \equiv \lambda^{(\nu)} \ket{ Q^{(\nu)} },
\end{equation}
with $\ket{ Q^{(\nu)} } = \sum_{j=1}^{n_\zeta} q_j^{(\nu)} \ket{ C_j }$ and $q_j^{(\nu)}\in\mathbb{C}$.
In general $( C_1,\dots, C_{n_\zeta} )$ is over-complete.
If a subset $C=(C_{i_1},\dots, C_{i_l})$ of the class operators $( C_1,\dots, C_{n_\zeta} )$ is a CSCO of the class space, then $C$ is called CSCO of the first kind (CSCO-I) of $G$ (different CSCO's are equivalent in the sense that they lead to the same eigenvectors $\ket{ Q^{(\nu)} }$).
It can be shown that in any representation space $\mathcal{V}$ the eigenvalues $\lambda^{(\nu)}$ of $C$ do not go beyond the $n_\zeta$ values determined in class space, and that in the regular representation space there are $n_\zeta$ and only $n_\zeta$ distinct eigenvalues $\lambda^{(\nu)}$.
By theorem \ref{thm:commuteseigenspace}, the eigenspaces of $C$ in a rep space $\mathcal{V}$ are representation spaces and the regular rep space $\mathcal{V}$ is reduced to a direct sum of $n_\zeta$ mutually orthogonal subspaces,
\begin{equation}
 \mathcal{V} = \bigoplus_{\nu=1}^{n_\zeta} \mathcal{V}_\nu,
\end{equation}
where $C \mathcal{V}_\nu = \lambda^{(\nu)} \mathcal{V}_\nu$ (in a non-regular representation space $\mathcal{V}$ one or more of the $\mathcal{V}_\nu$ might be trivial subspaces containing only the zero vector).
Using the fact that the representative of $C$ on $\mathcal{V}_\nu$ must be equal to the identity times the eigenvalue $\lambda^{(\nu)}$, it can be seen that representation spaces belonging to different eigenvalues are inequivalent.
A representation space $\mathcal{V}_\nu$ may still be reducible,
\begin{equation}\label{eq:LnudecompoLnk}
 \mathcal{V}_\nu = \mathcal{V}_{\nu,1}\oplus \dots \oplus \mathcal{V}_{\nu,\tau_\nu},
\end{equation}
where irreps $\mathcal{V}_{\nu,k}$ with the same label $\nu$ are equivalent.
The results presented so far lead to the well known result that a finite group with $n_\zeta$ classes has $n_\zeta$ and only $n_\zeta$ inequivalent irreps.
The irreps can be labeled uniquely by the eigenvalue list $\lambda^{(\nu)}$ of a CSCO-I $C$ (we use the symbol $\nu$ as label).
If a vector $\ket{\psi^{(\nu)}}$ belongs to the eigenspace $\mathcal{V}_\nu$ of a CSCO-I $C$ of $G$, the vector is said to belong to the irrep $\nu$ of~$G$.
\begin{thm}\label{thm:nsCSCO1}
A necessary and sufficient condition for a vector $\ket{\psi^{(\nu)}}$ to belong to the irrep $\nu$ of $G$ is that
\begin{equation}\label{eq:snCSCO1}
 C \ket{\psi^{(\nu)}} = \lambda^{(\nu)} \ket{\psi^{(\nu)}}.
\end{equation}
\end{thm}
\begin{proof}
The sufficiency is trivial. We prove that the condition is a necessary one.
Suppose $ \ket{\psi^{(\nu)}} $ is a vector in an irreducible subspace $\mathcal{V}_\nu$ of $G$.
It follows that $\mathcal{V}_\nu$ is an invariant subspace of $C$ and by Schur's lemma we obtain that $\mathcal{V}_\nu$ is necessarily an eigenspace of $C$.
\end{proof}
\noindent
This theorem is the corner stone of the EFM.
It allows the problem of finding the irreps of $G$ to be converted into the problem of finding the eigenspaces of a CSCO-I $C$ of $G$
(i.\,e. we have to diagonalize the operator $C$\footnote{$C$ is a set of commuting operators, but by taking a suitable linear combination of these operators, it suffices to diagonalize only one single operator.} in the reducible basis $\ket{0},\dots,\ket{d-1}$ spanning $\mathcal{V}$).

Let us now consider a canonical subgroup chain $G \supset G(s_1) \supset G(s_2) \dots $, or by using the
the abbreviation $G(s) = G(s_1) \supset G(s_2) \dots $, $G \supset G(s)$.
Analogous to theorem \ref{thm:nsCSCO1} we obtain the following theorem.
\begin{thm}\label{thm:cscoII}
A necessary and sufficient condition for a vector $\ket{\psi^{(\nu)}_{\lambda(s_1),\lambda(s_2),\dots}}$ in a rep space $\mathcal{V}$ to belong to the irreps $\nu, \lambda(s_1),\lambda(s_2),\dots $ of a subgroup chain $G\supset G(s)$ is that the vector satisfies the following eigenequations,
\begin{equation}
\begin{pmatrix}
 C \\
C(s_1) \\
C(s_2) \\
\vdots
\end{pmatrix}
\ket{\psi^{(\nu)}_{\lambda(s_1),\lambda(s_2),\dots}}
=
\begin{pmatrix}
\nu \\
\lambda(s_1)\\
\lambda(s_1)\\
\vdots
\end{pmatrix}
\ket{\psi^{(\nu)}_{\lambda(s_1),\lambda(s_2),\dots}},
\end{equation}
where $C$ is a CSCO-I of $G$ and $C(s_i)$ is a CSCO-I of $G(s_i)$.
\end{thm}
\begin{rem}
Using the abbreviations $C(s)=( C(s_1),C(s_2),\dots)$ and $m=(\lambda(s_1),\lambda(s_2),\dots)$, the eigenequation of the above theorem becomes
\begin{equation}
\begin{pmatrix}
 C \\
C(s)
\end{pmatrix}
\ket{\psi^{(\nu)}_m}
=
\begin{pmatrix}
\nu \\
m
\end{pmatrix}
\ket{\psi^{(\nu)}_m}.
\end{equation}
If the subgroup chain is canonical, the set $(C,C(s))$ is called CSCO-II of $G$.
\end{rem}
\noindent
Suppose the eigenspace $\mathcal{V}_\nu$ is an irreducible rep space of $G$.
Than the degeneracy of the eigenvalue $\lambda^{(\nu)}$ in \eqref{eq:snCSCO1} is equal to the dimension $h_\nu$ of the irrep and is totally lifted by the eigenequations of the $C(s)$ (i.\,e. the degeneracy of the eigenvalues $\{ (\nu,m_i) \}_{i=1}^{h_\nu}$ is one).
If $\mathcal{V}_\nu$ is a reducible rep space of $G$, the degeneracy of $\lambda^{(\nu)}$ is given by $\tau_\nu \times h_\nu$ and for each value $(\nu,m_i)$ there are $\tau_\nu$ linearly independent eigenvectors $\ket{\psi^{(\nu)\tau}_m}$, $\tau=1\dots\tau_\nu$, $\tau_\nu\in\{2,3,4,\dots\}$.

We now introduce the intrinsic group $\bar{G}$ of $G$ which is used to complete the set CSCO-II to a complete set of commuting observables (CSCO-III) $\mathfrak{C}$ on the representation space $\mathcal{V}$.
\begin{defi}
For each operator $g$ in $G$, we define a super-operator $\bar{g}$ acting on the elements of the group algebra $\mathcal{A} = \mathbb{C}G$ (any element $a$ in $\mathcal{A}$ can be written as $a=\sum_{g\in G} a_g g$ with $a_g\in\mathbb{C}$) by
\begin{equation}
 \bar{g} a = a g \quad \text{ for all } a \in \mathcal{A}.
\end{equation}
The group formed by all $\bar{g}$ is called the intrinsic group $\bar{G}$ of $G$.
\end{defi}
\noindent
We proceed by proving two important lemmas concerning the intrinsic group.
\begin{lem}\label{lem:barGcommuteWithG}
The operators in $\bar{G}$ commute with those in $G$.
\end{lem}
\begin{proof}
We have $s \bar{r} t = s t r = \bar{r} st$ for all $t\in \mathcal{A}$ and therefore $[\bar{r},s]=0$ for all $s\in G$ and $\bar{r}\in\bar{G}$.
\end{proof}
\begin{lem}
The group $\bar{G}$ is anti-isomorphic to $G$.
\end{lem}
\begin{proof}
Suppose the multiplication relation in $G$ is $rs = u$ for $r,s,u\in G$.
Then $\bar{s}\bar{r} t = \bar{s} tr = trs = tu = \bar{u} t$ for all $t\in \mathcal{A}$ and we have $\bar{s}\bar{r}=\bar{u}$.
\end{proof}
\noindent
If we consider the action of the elements of the intrinsic group on the representation space $\mathcal{V}$ of a representation $R(G)$ with basis $\{ \ket{i} = R_i \ket{0} \}$, we have to define a state, say $\ket{0}$, as the intrinsic state, i.\,e. the elements of $\bar{G}$ act on the basis states as
\begin{equation}\label{eq:actionIgroup}
 \bar{R}_b \ket{a} = \bar{R}_b R_a \ket{0} = R_a R_b \ket{0}.
\end{equation}
Note that if the intrinsic state is invariant under a symmetry group $G_\text{in} \subset G$,
we have
$\bar{R}_b \ket{0} = \bar{R}_b T \ket{0} = T R_b \ket{0}
=  T \ket{b}$ for all $T\in G_\text{in}$ and on the other hand
$\bar{R}_b \ket{0} = R_b \ket{0} = \ket{b}$ which is a contradiction.
Therefore, the following only holds for the regular representation $R(G)$ for which $G_\text{in}$ contains only the identity.
The anti-isomorphism between $G$ and $\bar{G}$ assures that the conclusions about $G$ apply to $\bar{G}$ as well:
\begin{enumerate}
\item
If $C = (C_{i_1},\dots, C_{i_l})$ is a CSCO-I of $G$, then
$\bar{C} = (\bar{C}_{i_1},\dots, \bar{C}_{i_l})$ is a CSCO-I of $\bar{G}$ with
\begin{equation}
 \bar{C}_i = \sum_{j=1}^{n_i} \bar{R}( a^{(i)}_j ).
\end{equation}
Note that the CSCO-I of $G$ and $\bar{G}$ are equal, since
\begin{equation}\label{eq:CG-barCG}
 \bar{C}_i R_k = \bigl( \sum_{j=1}^{n_i} \bar{R}( a^{(i)}_j ) \bigr) R_k
 = R_k \bigl( \sum_{j=1}^{n_i} R( a^{(i)}_j ) \bigr) = R_k C_i = C_i R_k,
\end{equation}
where the last identity holds because $[C_i,R]=0$ for all $R_k\in R(G)$.
\item
If $G$ has a canonical subgroup chain $G\supset G(s)$, $G(s)=G(s_1)\supset G(s_2)\supset\dots $,
with CSCO-II $\bigl(C, C(s) = ( C(s_1),C(s_2),\dots) \bigr)$,
$\bar{G}$ has a canonical subgroup chain $\bar{G}\supset \bar{G}(s)$, $\bar{G}(s)=\bar{G}(s_1)\supset \bar{G}(s_2)\supset\dots $,
with CSCO-II $\bigl(\bar{C}, \bar{C}(s) = ( \bar{C}(s_1), \bar{C}(s_2), \dots) \bigr)$.
\end{enumerate}
Because of lemma \ref{lem:barGcommuteWithG} $[C(s),\bar{C}(s)]=0$, which allows the $\bar{C}(s)$ to be added to a CSCO-II of $G$ and the following theorem to be proved.
\begin{thm}\label{thm:cscoIII}
The set $\mathfrak{C} = ( C, C(s),\bar{C}(s) )$ defined on the regular rep space $\mathcal{V}$ of a group $G$ with canonical subgroup chain $G(s)=G(s_1)\supset G(s_2)\supset\dots$ is a CSCO on $\mathcal{V}$ (called CSCO-III).
The corresponding eigenequation is given by
\begin{equation}
 \begin{pmatrix}
 C \\
 C(s) \\
\bar{C}(s)
\end{pmatrix}
\ket{\psi^{(\nu)k}_m}
=
\begin{pmatrix}
\nu \\
m \\
k
\end{pmatrix}
\ket{\psi^{(\nu)k}_m},
\end{equation}
with $k = ( \bar{\lambda}(s_1),\bar{\lambda}(s_2),\dots)$.
\end{thm}
\noindent
Because $( C, \bar{C}(s) )$ commutes with all the elements in $R(G)$, the eigenspaces
$\mathcal{V}_{\nu,k} = \vspan\bigl\{ \ket{\psi^{(\nu)k}_{m_i}} \bigr\}$, $i=1\dots h_\nu$ of $( C, \bar{C}(s) )$ are necessarily representation spaces of $R(G)$ and the degeneracy of $m_i$ is necessarily independent of $i$.
Since in addition $\mathfrak{C} = ( C, C(s),\bar{C}(s) )$ is a CSCO of $\mathcal{V}$, $( C(s),\bar{C}(s) )$ is necessarily a CSCO in each eigenspace $\mathcal{V}_\nu$, $\nu=1\dots n_\zeta$, and the degeneracy of $m_i$, $i=1\dots h_\nu$, in $\mathcal{V}_\nu$ is completely lifted by the eigenvalue $k$ of $\bar{C}(s)$.
It can be shown that the representatives of the operators $C(s_i)$ and $\bar{C}(s_i)$ in $\mathcal{V}_\nu$ are similar matrices.
Therefore, the characteristic equations of $C(s)$ and $\bar{C}(s)$ in $\mathcal{V}_\nu$ are identical and it follows that the eigenvalue $k$ takes on the values $k_i = m_i$ for $i=1\dots h_\nu$.
Equation \eqref{eq:LnudecompoLnk} in the regular rep case becomes
\begin{equation}
 \mathcal{V}_\nu = \bigoplus_{i=1}^{\tau_\nu=h_\nu} \mathcal{V}_{\nu,k_i},
\end{equation}
and we have $d= n_G = \sum_{\nu=1}^{n_\zeta} h_\nu^2$.

Since the normalized vectors $\ket{\psi^{(\nu)k}_m}$ are obtained by solving an eigenequation, they are determined only up to a phase factor.
Let the eigenvectors be expressed as
\begin{equation}
\ket{\psi^{(\nu)k}_m} = \sum_{i=0}^{n_G-1} u_{\nu mk,i} \ket{i}, \quad u_{\nu mk,i}\in\mathbb{C},
\end{equation}
or, in the basis of the $\ket{i}$, as column vector $\vec{u}_{\nu mk}$.
The standard phase choice is the convention to choose the $u_{\nu mm,0}$ to be real and positive for all $m$ (in fact it can be shown that in this case $u_{\nu mm,0} = \sqrt{h_\nu/n_G}$).
Starting with the first eigenvalue of $k$ (denoted now simply as $k=1$), the phases of the vectors $\vec{u}_{\nu m 1}$ for $m=2\dots h_\nu$ can be chosen arbitrarily.
This choice fixes the representation matrices (we demand them to be identical in equivalent representations),
which are now given by
\begin{equation}
 D_{ij}^{(\nu)} (a) = \bra{\psi^{(\nu)1}_i} R_a \ket{\psi^{(\nu)1}_j},
\end{equation}
for all $a\in G$.
When using the standard phase choice, the representation matrices can be shown to be directly related with the vectors $\ket{\psi^{(\nu)j}_i}$ via
\begin{equation}
D_{ij}^{(\nu)} (a) = \sqrt{\frac{n_G}{h_\nu}} u^\star_{\nu ij,a}.
\end{equation}
This expression allows the phases of the remaining vectors $\vec{u}_{\nu m k}$ with $k>1$ to be fixed by demanding that
$u_{\nu mk,a}$ is equal to $\sqrt{h_\nu/n_G} \cdot D_{mk}^{(\nu)} (a)^\star$ for all $m$.

\subsubsection{Reduction of Non-Regular Reps}

The construction of the CSCO-II of $G$ is the same for regular and non-regular representations.
For non-regular reps, only the completion of the CSCO-II to the CSCO-III $\mathfrak{C}$ has to be adjusted.
As it was shown in the paragraph following equation \eqref{eq:actionIgroup},
an intrinsic state which is invariant under a non-trivial symmetry group $G_\text{in} \subset G$,
leads to a contradiction which makes the definition of intrinsic group elements meaningless.
The observation which saves the day is the following.
\begin{lem}\label{lem:classopsintrinsicg}
If a class operator $C_i(s_j)$ of a subgroup $G(s_j)$ of $G$  commutes with the symmetry group $G_\text{in} \subset G$, then the class operator $\bar{C}_i(s_j)$ of the corresponding intrinsic group $\bar{G}(s_j)$ has a well defined meaning.
\end{lem}
\begin{proof}
We repeat the calculation which led to the contradiction.
On the one hand we have
$\bar{C}_i(s_j) \ket{a} = \bar{C}_i(s_j) R_a \ket{0} = R_a C_i(s_j) \ket{0}$,
on the other hand we have
$\bar{C}_i(s_j) \ket{a} = \bar{C}_i(s_j) R_a T \ket{0} = R_a T C_i(s_j) \ket{0}$.
But if the class operator and the symmetry group commute, we continue the calculation and obtain
$ \hdots = R_a C_i(s_j) T \ket{0} =  R_a C_i(s_j) \ket{0}$ for all $T\in G_\text{in}$.
The contradiction vanishes.
\end{proof}
\noindent
If we remove all subgroups from the canonical subgroup chain $G(s) = G(s_1) \supset G(s_2) \supset \dots$ whose class operators do not commute with $G_\text{in}$, we obtain the (non-canonical) subgroup chain $G(s')=G(s_{i_1})\supset G(s_{i_2})\supset\dots$ (with $i_1<i_2$) and the lemma tells us that the CSCO-I's
$\bar{C}(s') = (\bar{C}(s_{i_1}),\bar{C}(s_{i_2}),\dots)$ of $\bar{G}(s')$ still have a definite meaning.
Theorem \ref{thm:cscoIII} is replaced by:
\begin{thm}\label{thm:cscoIIInonreg}
Let $\mathcal{V} = \vspan\bigl\{ R_a \ket{0} \sthat a\in G \bigr\}$ be the rep space of a rep $R(G)$ of a group $G$ with canonical subgroup chain $G(s)$ and symmetry group $G_\text{in}$.
Then the set $\mathfrak{C} = ( C, C(s),\bar{C}(s') )$ (with $\bar{C}(s')$ as defined above) is a CSCO on $\mathcal{V}$ (called CSCO-III).
The corresponding eigenequation is given by
\begin{equation}
 \begin{pmatrix}
 C \\
 C(s) \\
\bar{C}(s')
\end{pmatrix}
\ket{\psi^{(\nu)\kappa}_m}
=
\begin{pmatrix}
\nu \\
m \\
\kappa
\end{pmatrix}
\ket{\psi^{(\nu)\kappa}_m},
\end{equation}
with $\kappa = ( \bar{\lambda}(s_{i_1}),\bar{\lambda}(s_{i_2}),\dots)$.
\end{thm}
\noindent
To set the phases of the vectors $\ket{\psi^{(\nu)\kappa}_m}$ in such a way that the representation matrices
\begin{equation}
  D_{ij}^{(\nu)\kappa} (a) = \bra{\psi^{(\nu)\kappa}_i} R_a \ket{\psi^{(\nu)\kappa}_j}
\end{equation}
do not depend on $\kappa$ and agree with those of the regular rep,
we choose the phases of $\ket{\psi^{(\nu)\kappa}_1}$ arbitrarily for all $\kappa$ and
use the known matrix elements of the regular rep matrices to determine the phases of the $\ket{\psi^{(\nu)\kappa}_m}$ with $m>1$.

\subsection{Symmetric Groups}\label{subsec:symmgr}
The results of the preceding subsection are now specialized for the case that the group $G$ under consideration is the symmetric group~$\textsf{S}_n$.
Using standard results of the representation theory of $\textsf{S}_n$, the construction of the CSCO $\mathfrak{C}$ can be simplified.

\subsubsection{Representation Spaces}
The elements of $\textsf{S}_n$ are permutations which are denoted as
\begin{equation}
p = \begin{pmatrix} 1 & 2 & \dots & n\\      p(1) & p(2) & \dots & p(n)  \end{pmatrix}.
\end{equation}
The inverse of $p\in \textsf{S}_n$ is given by
\begin{equation}
p^{-1} =
 \begin{pmatrix} p(1) & p(2) & \dots & p(n)\\      1 & 2 & \dots & n  \end{pmatrix} =
 \begin{pmatrix} 1 & 2 & \dots & n\\      p^{-1}(1) & p^{-1}(2) & \dots & p^{-1}(n)  \end{pmatrix},
\end{equation}
where the right-hand side is obtained by permuting the columns of the matrix of the left-hand side.
Let $\mathcal{H}_q$ be the Hilbert space of a qudit of dimension $q$ and let an orthonormal basis
$\{ \ket{0},\ket{1},\dots, \ket{q-1} \}$ be fixed.
In this section we will usually label these basis states using the Greek alphabet, i.\,e.
$\ket{0}\equiv\ket{\alpha}$, $\ket{1}\equiv\ket{\beta}$, $\ket{2}\equiv\ket{\gamma}$, and so on.
A representation $D(\textsf{S}_n)$ of $\textsf{S}_n$ on the $q^n$-dimensional linear vector space $\mathcal{H}_q^{\otimes n}$ is given by defining the action of a permutation $p\in \textsf{S}_n$ on a $n$-fold tensor-product of one qudit basis states by
\begin{equation}\label{eq:defiactionSn}
 D(p) \ket{ i_1 \,,\, i_2 \,,\, \dots \,,\, i_n } = \ket{ i_{p^{-1}(1)} \,,\, i_{p^{-1}(2)} \,,\, \dots \,,\, i_{p^{-1}(n)} }.
\end{equation}
We define the configuration of a standard basis vector in $\mathcal{H}_q^{\otimes n}$ as a string of integers of length $q$ counting the number of times a certain one-qudit basis state appears in the vector,
e.\,g. for $n=5$ and $q=4$ we have
\begin{equation}\label{eq:deficonfig}
 \textsf{config}( \ket{ \alpha \, \alpha \, \delta \, \alpha \, \beta } ) = (3,1,0,1).
\end{equation}
Since the configuration of basis vectors in $\mathcal{H}_q^{\otimes n}$ remains invariant under $\textsf{S}_n$, the representation space $\mathcal{H}_q^{\otimes n}$ of $\textsf{S}_n$ decomposes into a direct sum of representation spaces each of which is characterized by a certain configuration string.
The dimension of a rep space $\mathcal{V}$ with configuration $\textsf{config}=(n_0,n_1,\dots,n_{q-1})$, $\sum_i n_i=n$, is given by the multinomial coefficient $n!/(\prod_i n_i!)$.
Altogether there are $\binom{n+q-1}{q-1}$ different configurations.

The regular representation space $\mathcal{V}$ occurs only if $q=n$ and coincides with the rep space with configuration $(1,1,\dots,1)$.
Its basis vectors $\{ \ket{p} \}$ are obtained by applying the $g=n!$ elements of $\textsf{S}_n$ to the generating state
$\ket{0} := \ket{ 0 , 1, \dots, q-1}$, i.\,e. $\ket{p} = D(p) \ket{0}$ with $p\in \textsf{S}_n$.

For a non-regular representation space $\mathcal{V}$ with configuration $(n_0,n_1,\dots,n_{q-1}) \neq (1,1,\dots,1)$, we define the generating state $\ket{0}$ by
\begin{equation}\label{eq:ketphi1}
 \ket{0} := \bigl\vert  \,
 \underset{n_0}{\underbrace{0,\dots, 0}} \ , \
 \underset{n_1}{\underbrace{1,\dots, 1}} \ ,\ \dots \ ,\
 \underset{n_{q-1}}{\underbrace{q-1,\dots, q-1} }  \, \bigr\rangle.
\end{equation}
The generating state $\ket{0}$ is obviously invariant under a non-trivial symmetry group $G_\text{in} \subset \textsf{S}_n$ and the dimension $d = n!/(\prod_i n_i!) $ of the non-regular rep space $\mathcal{V}=\vspan\{ D(p)\ket{0} \sthat p\in \textsf{S}_n \}$ is smaller than $n_G=n!$.

In the remaining part of this subsection we explain how the EFM described in the last subsection is applied to a rep space $\mathcal{V}\subset \mathcal{H}_q^{\otimes n}$ of $\textsf{S}_n$ characterized by a certain configuration string.

\subsubsection{Young Diagrams \& CSCO-I}

Each permutation can be decomposed into a product of disjoint cycles,
for example $p=\bigl( \begin{smallmatrix}1&2&3&4&5&6\\3&5&1&2&4&6\end{smallmatrix} \bigr)$
can be written as the product
$p=\bigl( \begin{smallmatrix}1&3\\3&1\end{smallmatrix} \bigr) \times
\bigl( \begin{smallmatrix}2&4&5\\5&2&4\end{smallmatrix} \bigr) \times
\bigl( \begin{smallmatrix}6\\6\end{smallmatrix} \bigr) \equiv (13)(254)(6)$.
The conjugacy classes of the symmetric group $\textsf{S}_n$ are characterized by a certain cycle structure:
Each class contains only elements of one particular cycle structure.
A cycle structure corresponds to a partition of $n$.
A partition $\nu$ of $n$ is given by a set of positive integers $\nu=[\nu_1,\nu_2,\dots, \nu_v]$ such that
$\sum_i \nu_i = n$ and $\nu_i \geq \nu_{i+1}$.
It can be depicted as a Young diagram in which the $i$-th row contains $\nu_i$ boxes.
For instance, for $n=4$ there are the partitions
$[4]$, $[3,1]$, $[2,2]$, $[2,1,1]$ and $[1,1,1,1]$ which correspond to the Young diagrams
\begin{equation*}
\yng(4) \,,\,
\yng(3,1) \,,\,
\yng(2,2) \,,\,
\yng(2,1,1) \text{ and } \yng(1,1,1,1) \, .
\end{equation*}
Since the number of inequivalent representations of a group $G$ is equal to the number of its conjugacy classes, the inequivalent reps of $\textsf{S}_n$ may be labeled by Young diagrams corresponding to the partitions of $n$.
This means that there is a one-to-one correspondence between the eigenvalues of the CSCO-I $C$ of $\textsf{S}_n$ and the Young diagrams corresponding to the partitions of $n$.

If a state $\ket{\psi^{(\nu)}}$ is in a rep space $\mathcal{V}_\nu \subset \mathcal{V}$ labeled by the Young diagram $\nu=[\nu_1,\nu_2,\dots, \nu_v]$, it is necessarily an eigenstate of the class operators of $\textsf{S}_n$.
It can be shown that the eigenvalues $\lambda_2^n$ and $\lambda_3^n$ of the 2- and 3-cycle class operators $C_2^n$ and $C_3^n$ can be expressed as functions of the Young diagram $\nu$ as follows,
\begin{subequations}
\begin{align}
 \lambda_2^n &= \frac{n}{2} + \frac{1}{2} \sum_{i=1}^v \nu_i(\nu_i-2i) \\
 \lambda_3^n &= \frac{2}{3}n -\frac{1}{2}n^2 +\frac{1}{3}\sum_{i=1}^v\nu_i\bigl[\nu_i^2-(3i-3/2)\nu_i+3i(i-1) \bigr].
\end{align}
\end{subequations}
For $n<6$ the 2-cycle class operator alone forms a CSCO-I, but for $n=6$ degeneracy occurs which has to be lifted by adding for example the 3-cycle class operator.
For $n<15$ a CSCO-I $C$ of $\textsf{S}_n$ is given by the 2- and 3-cycle class operators, $C = (C_2^n,C_3^n)$.
The eigenvalues $\lambda_2^n$ and $\lambda_3^n$ of $C_2^n$ and $C_3^n$ are listed in the form $\begin{smallmatrix} \lambda^n_2 \\ \lambda^n_3 \end{smallmatrix}$ for $n=1\dots 7$ in figure \ref{fig:treediag}.

\subsubsection{Young Tableaux \& CSCO-II}

A canonical subgroup chain $\textsf{S}_n \supset G(s)$ of $\textsf{S}_n$ is given by
$G(s) = \textsf{S}_{n-1} \supset \dots \supset \textsf{S}_3 \supset \textsf{S}_2$.
According to theorem \ref{thm:cscoII}, the CSCO-II of $\textsf{S}_n$ is given by
$\bigl( C(\textsf{S}_n), C(s)=(C(\textsf{S}_{n-1}),\dots,C(\textsf{S}_2)) \bigr)$, where the operator $C(\textsf{S}_i)$ denotes the CSCO-I of $\textsf{S}_i$.
It can be shown that this set of commuting observables is over-complete and that a simpler set is given by $\bigl( C_2^n, C_2^{n-1}, \dots, C_2^2 \bigr)$ which contains only 2-cycle class operators.
While the 2-cycle eigenvalue of, say, $C_2^i$ alone is not necessarily enough to deduce the eigenvalue $\lambda(\textsf{S}_i)$ of $C(\textsf{S}_i)$, the whole set of 2-cycle eigenvalues allows us to deduce the eigenvalues $\nu$ and $m=(\lambda(\textsf{S}_{n-1}),\dots,\lambda(\textsf{S}_2))$ of $C(\textsf{S}_n)$ and $C(s)$.
The reason behind this fact is the branching law, which states that a subduced rep $D^{(\nu)}(\textsf{S}_j) \downarrow \textsf{S}_{j-1}$ of an irrep $\nu$ of $\textsf{S}_j$ decomposes into
\begin{equation}
 D^{(\nu)}(\textsf{S}_j) \downarrow \textsf{S}_{j-1}  =  \bigoplus_{\nu'}  D^{(\nu')}(\textsf{S}_{j-1}),
\end{equation}
where the $\nu'$ are obtained from the Young diagram $\nu$ by removing a single box in all possible ways, e.\,g.
\begin{equation}
\yng(4,2,1) \rightarrow \yng(4,2) \oplus \yng(4,1,1) \oplus \yng(3,2,1).
\end{equation}
\begin{sidewaysfigure}\centering
\begin{pspicture}(0,-.62\textwidth)(\textheight,0)
\psframe[linecolor=lightgray](0,-.62\textwidth)(\textheight,0)
\psset{linecolor=gray}
\rput(.5\textheight,1.6){\scalebox{0.49}{\input{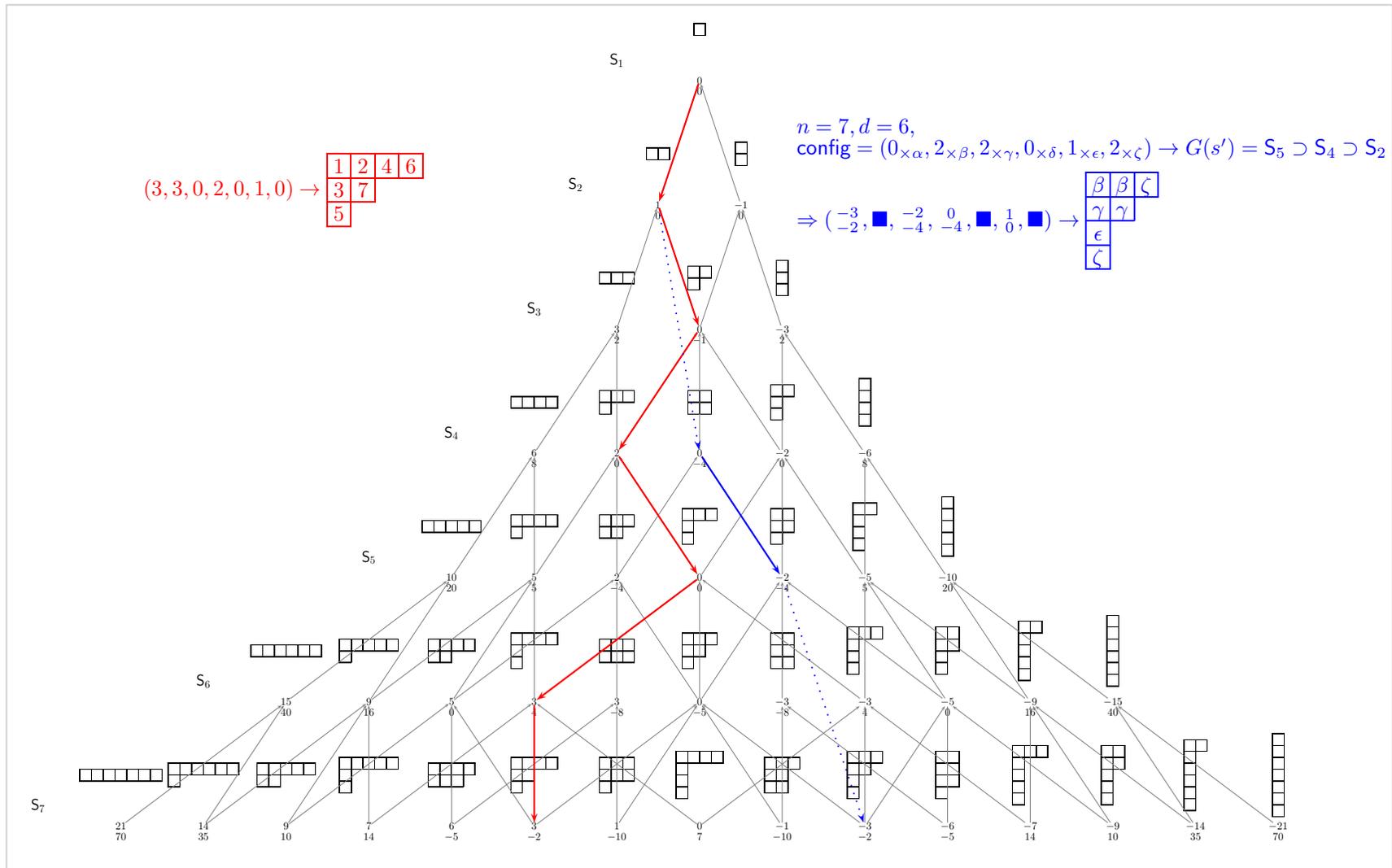}}}%
\psset{linecolor=red}
\ncline{->}{N1N1}{N2N1}
\ncline{->}{N2N1}{N3N2}
\ncline{->}{N3N2}{N4N2}
\ncline{->}{N4N2}{N5N4}
\ncline{->}{N5N4}{N6N4}
\ncline{->}{N6N4}{N7N6}
\rput(.2\textheight,-3){\red $(3,3,0,2,0,1,0) \rightarrow \young(1246,37,5)$}
\psset{linecolor=blue}
\ncline[linestyle=dotted]{->}{N2N1}{N4N3}
\ncline{->}{N4N3}{N5N5}
\ncline[linestyle=dotted]{->}{N5N5}{N7N10}
\rput[l](.57\textheight,-2){\blue $n=7,d=6,$}
\rput[l](.57\textheight,-2.35){\blue $\textsf{config} = (
0_{\times\alpha}, 2_{\times\beta}, 2_{\times\gamma}, 0_{\times\delta}, 1_{\times\epsilon}, 2_{\times\zeta}
) \rightarrow G(s')=\textsf{S}_5\supset \textsf{S}_4\supset \textsf{S}_2$}
\rput[l](.57\textheight,-3.5){\blue $\Rightarrow(
\begin{smallmatrix}-3\\-2\end{smallmatrix},
\blacksquare,
\begin{smallmatrix}-2\\-4\end{smallmatrix},
\begin{smallmatrix}0\\-4\end{smallmatrix},
\blacksquare,
\begin{smallmatrix}1\\0\end{smallmatrix},
\blacksquare
)\rightarrow \young(\beta\beta\zeta,\gamma\gamma,\epsilon,\zeta)$}
\end{pspicture}
\caption[Tree diagram]{Tree diagram showing the Young-diagrams corresponding to the partitions of $n=1$ (\textit{top row}) up to $n=7$ (\textit{bottom row}). The gray arrows indicate the branching law.
Under each Young-diagram the corresponding eigenvalues $\lambda^n_2$ of the 2-cycle class operator $C_2^n$ and $\lambda^n_3$ of the 3-cycle class operator $C_3^n$ are shown in the form $\begin{smallmatrix} \lambda^n_2 \\ \lambda^n_3 \end{smallmatrix}$.
\label{fig:treediag}}
\end{sidewaysfigure}%
In figure \ref{fig:treediag} the Young diagrams of $\textsf{S}_n$ are shown in rows from $n=1$ (\textit{top row}) to $n=7$ (\textit{bottom row}).
The branching law is indicated in the figure by gray arrows.
Under each Young diagram the eigenvalues of $\lambda_2^n$ and $\lambda_3^n$ of $C_2^n$ and $C_3^n$ are shown in the form $\begin{smallmatrix} \lambda^n_2 \\ \lambda^n_3 \end{smallmatrix}$.
As an example of how the eigenvalue list of $\bigl( C_2^n, C_2^{n-1}, \dots, C_2^2 \bigr)$ determines all Young diagrams (i.\,e. all eigenvalues $\nu$ and $m$ of $C$ and $C(s)$), let us consider the case $n=7$ with
\begin{equation}
\bigl( C_2^7, C_2^6, C_2^5, C_2^4, C_2^3, C_2^2, C_2^1 \bigr) \ket{\psi^{(\nu)}_m} =
\bigl( 3, 3, 0,2,0,1,0\bigr) \ket{\psi^{(\nu)}_m}
\end{equation}
(we added $C_2^1$ whose only eigenvalue is zero) which is shown in figure \ref{fig:treediag} in red.
Starting at the top of the tree diagram with the eigenvalue $0$ of $C_2^1$, we follow the gray arrow (in the opposite direction from top to bottom) which leads to the next eigenvalue $1$ of $C_2^2$ (the resulting path is shown in red), and so on.
By following a path provided by the branching law, any possible degeneracy of the 2-cycle eigenvalues
(in our case the degeneracy of the eigenvalue $3$ of $C_2^6$) is artificially lifted
since it can be shown that only one of them will be accessible by the preceding path.
Therefore, the eigenvalue list of the CSCO-II $\bigl( C_2^n, C_2^{n-1}, \dots, C_2^2 \bigr)$ describes a unique path connecting Young diagrams of $\textsf{S}_n,\textsf{S}_{n-1},\dots \textsf{S}_2,\textsf{S}_1$ which correspond to the eigenvalues $(\nu, m)$ of the original CSCO-II
$\bigl( C(\textsf{S}_n), C(s)=(C(\textsf{S}_{n-1}),\dots,C(\textsf{S}_2)) \bigr)$.
Instead of writing down all $n$ Young diagrams,
they are summarized in a Young tableau $Y_m^{(\nu)}$ which is obtained from the Young diagram $\nu$ corresponding to the eigenvalue of $C(\textsf{S}_n)$ by filling its boxes with the numbers $1,2,\dots n$ in such a way
that if the box with number $n$ is removed, we obtain a Young tableau which is shaped like the Young diagram corresponding to the eigenvalue $\lambda(\textsf{S}_{n-1})$ of $C(\textsf{S}_{n-1})$, and so on.
For our example we obtain
\begin{multline}
 (3,3,0,2,0,1,0) \rightarrow Y_m^{(\nu)} = \young(1246,37,5) = \\
  \left(  \nu=\yng(4,2,1), m=\yng(4,1,1),\yng(3,1,1),\yng(3,1),\yng(2,1),\yng(2),\yng(1) \right) \, .
\end{multline}
As a consequence, a Young tableaux is always filled in such a way that the successive removal of boxes corresponding to the numbers $n$, $n-1$, etc., results in valid Young diagrams:
In a Young tableau, the numbers always increase to the right and downwards.

\subsubsection{Weyl Tableaux \& CSCO-III}

Let us consider the representation space $\mathcal{V} \subset \mathcal{H}_q^{\otimes n}$ with configuration $\textsf{config}=(n_0,n_1,\dots,n_{q-1})$, $\sum_i n_i=n$.
The generating state $\ket{0}$ defined in equation \eqref{eq:ketphi1},
\begin{equation}
 \ket{0}: = \bigl\vert  \,
 \underset{n_0}{\underbrace{0,\dots, 0}} \ ,\
 \underset{n_1}{\underbrace{1,\dots, 1}} \ ,\ \dots \ ,\
 \underset{n_{q-1}}{\underbrace{q-1\dots q-1} }  \, \bigr\rangle,
\end{equation}
is invariant under the action of the subgroup $G_\text{in} \subset \textsf{S}_n$ containing $\vert G_\text{in}\vert = \prod_i n_i !$ elements.
The group $G_\text{in}$ decomposes $\textsf{S}_n$ into a disjoint set of left cosets,
$\textsf{S}_n = G_\text{in} \cup a'G_\text{in} \cup b'G_\text{in} \cup \dots$, where each coset contains $\vert G_\text{in}\vert $ elements and $\{a',b',\dots \}$ denotes a set of coset representatives.
An orthonormal basis $\{ \ket{i} \}_{i=0\dots d-1}$ of $\mathcal{V}$ is obtained by applying the
$d = \vert \textsf{S}_n\vert / \vert G_\text{in}\vert = n!/ \prod_i n_i !$ coset representatives to the generating state $\ket{0}$, i.\,e.
$\ket{i} = D(p_i) \ket{0}$, with $p_i \in \{a',b',\dots \}$.
The basis $\{ \ket{i} \}_{i=0\dots d-1}$ forms a subset of the computational basis of $\mathcal{H}_q^{\otimes n}$.
We define the string of integers $\textsf{config}'=( n_{i_1}, n_{i_2}, \dots,  n_{i_l} )$ by removing all zeros from $\textsf{config}$.
Then the structure of the symmetry group $G_\text{in} \subset \textsf{S}_n$ is given by
\begin{equation}
 G_\text{in} = \textsf{S}_{n_{i_1}} \otimes \textsf{S}_{n_{i_2}} \otimes \dots \otimes \textsf{S}_{n_{i_l}}.
\end{equation}
It is easy to see that the class operators $C_i^{n(j)}$ of the subgroups $\{ \textsf{S}_{n(j)} \}_{j=1\dots l}$ with
$n(j) = \sum_{c=1}^j n_{i_c}$ commute with all the elements in $G_\text{in}$.
Therefore, according to lemma \ref{lem:classopsintrinsicg}, the corresponding class operators $\bar{C}_i^{n(j)}$ of the intrinsic group are well defined and according to theorem \ref{thm:cscoIIInonreg}, the CSCO-II of $\mathcal{V}$ can be extended to a CSCO $\mathfrak{C}$ on $\mathcal{V}$ (called CSCO-III) by adding the set
$\bar{C}(s') = ( \bar{C}(\textsf{S}_{n(l-1)}),\dots,\bar{C}(\textsf{S}_{n(2)}),\bar{C}(\textsf{S}_{n(1)}) )$
of CSCO-I's of $\bar{G}(s') = \bar{\textsf{S}}_{n(l-1)} \supset \dots \supset \bar{\textsf{S}}_{n(2)} \supset \bar{\textsf{S}}_{n(1)}$.
Since $\bar{G}(s')$ is not a canonical subgroup chain of $\textsf{S}_n$ anymore, the set
$( \bar{C}(\textsf{S}_n), \bar{C}(s') )$ cannot be replaced by a set of 2-cycle operators as it was done for the set $( C(\textsf{S}_n), C(s) )$.

To give an example, let $n=7$, $q=6$, and let us decompose the rep space $\mathcal{V}\subset \mathcal{H}_q^{\otimes n}$ with $\textsf{config}=(0_{\times\alpha}, 2_{\times\beta}, 2_{\times\gamma}, 0_{\times\delta}, 1_{\times\epsilon}, 2_{\times\zeta})$.
For this configuration we have $G(s') = \textsf{S}_5 \supset \textsf{S}_4 \supset \textsf{S}_2$ and we consider an eigenvector $\ket{\psi^{(\nu)\kappa}}$ with eigenequation
\begin{equation}
 ( \bar{C}(\textsf{S}_n), \bar{C}(s') ) \ket{\psi^{(\nu)\kappa}} =
(\begin{smallmatrix}-3\\-2\end{smallmatrix},
\blacksquare,
\begin{smallmatrix}-2\\-4\end{smallmatrix},
\begin{smallmatrix}0\\-4\end{smallmatrix},
\blacksquare,
\begin{smallmatrix}1\\0\end{smallmatrix},
\blacksquare)
\ket{\psi^{(\nu)\kappa}} \equiv (\nu,\kappa) \ket{\psi^{(\nu)\kappa}}.
\end{equation}
(We expanded the eigenvalue list to the length $n$ by inserting black squares in the places where subgroups have been removed from $G(s)$ to obtain $G(s')$.)
Since $n<15$, $\bar{C}(\textsf{S}_{n(j)}) = (\bar{C}_2^{n(j)}, \bar{C}_3^{n(j)})$ and we wrote the corresponding eigenvalues on top of each other.
Our eigenvalue list $(\nu,\kappa)$ corresponds to a set of Young diagrams (indicated in blue in figure \ref{fig:treediag}),
\begin{equation}
(\begin{smallmatrix}-3\\-2\end{smallmatrix},
\blacksquare,
\begin{smallmatrix}-2\\-4\end{smallmatrix},
\begin{smallmatrix}0\\-4\end{smallmatrix},
\blacksquare,
\begin{smallmatrix}1\\0\end{smallmatrix},
\blacksquare)
\rightarrow W_\kappa^{(\nu)} = \young(\beta\beta\zeta,\gamma\gamma,\epsilon,\zeta) = \left( \nu=\yng(3,2,1,1), \kappa = \yng(2,2,1),\yng(2,2),\yng(2) \right)\, ,
\end{equation}
which can be summarized in a so-called Weyl tableau $W_\kappa^{(\nu)}$ as follows:
The Weyl tableau $W_\kappa^{(\nu)}$ is the Young diagram of $\textsf{S}_n=\textsf{S}_{n(l)}$ corresponding to the eigenvalue $\nu$ of $\bar{C}(\textsf{S}_n) = C(\textsf{S}_n)$ (compare with equation \eqref{eq:CG-barCG}) in which $n_i$ boxes are filled with the $i$-th letter (basis state) of the Greek alphabet,
and where the filling is done in such a way, that removing the $n_{i_l}$ boxes filled with the $i_l$-th letter results in a Weyl tableau which is shaped like the Young diagram corresponding to the eigenvalue of $\bar{C}( \textsf{S}_{n(l-1)} )$, and so on.
It follows that a Weyl tableaux is always filled in such a way that the successive removal of boxes corresponding to the $i_l$-th, $i_{l-1}$-th, etc., letter results in valid Young diagrams:
In a Weyl tableau, letters have to increase downwards and never decrease to the right.
Because of the former restriction, the maximum number of rows of a Young diagram is given by the number of letters (basis states) $q$.

\subsubsection{Final Remarks}

As a summary, the complete CSCO $\mathfrak{C}$ of $\textsf{S}_n$ on a rep space $\mathcal{V} \subset \mathcal{H}_q^{\otimes n}$ characterized by a configuration $\textsf{config}=(n_0,n_1,\dots,n_{q-1})$ is given by the set of operators
\begin{equation}
\mathfrak{C} = \bigl( C^n_2, \quad  C^{n-1}_2,\dots,C^2_2, \quad
   \bar{C}(\textsf{S}_{n(l-1)}),\dots,\bar{C}(\textsf{S}_{n(3)}),\bar{C}(\textsf{S}_{n(2)}) \bigr)
\footnote{Instead of constructing the operators $\bar{C}(\textsf{S}_{n(j)})$, in practice it is of advantage to construct the operators $C(\mathcal{S}_{n(j)})$ which correspond to the so-called state permutation group $\mathcal{S}_{n(j)}$ and which are identical to the $\bar{C}(\textsf{S}_{n(j)})$.}.
\end{equation}
(Note that $\bar{C}(\textsf{S}_{n(l)})$ and $\bar{C}(\textsf{S}_{n(1)})$ are obsolete since $\bar{C}(\textsf{S}_{n(l)}) = \bar{C}(\textsf{S}_n) = C(\textsf{S}_n)$ and $\bar{C}(\textsf{S}_{n(1)})$ always corresponds to the Young diagram of the form $\nu = [n(1)]$).
The eigenvectors $\ket{\psi^{(\nu)\kappa}_m}$ of the corresponding eigenequation
\begin{equation}\label{eq:eigeneqnukm}
\begin{split}
\mathfrak{C} \ket{\psi^{(\nu)\kappa}_m} &= (\nu,\kappa,m) \ket{\psi^{(\nu)\kappa}_m} \\
                                        &\equiv (W_\kappa^{(\nu)},Y_m^{(\nu)}) \ket{ W_\kappa^{(\nu)} Y_m^{(\nu)} }
\end{split}
\end{equation}
are labeled by
(i) a Young diagram $\nu$ labeling inequivalent rep spaces $\mathcal{V}_\nu$,
(ii) a Weyl tableau $W_\kappa^{(\nu)}$ labeling equivalent irreducible rep spaces $\mathcal{V}_{\nu,\kappa} \subset \mathcal{V}_\nu$,
(iii) a Young tableau $Y_m^{(\nu)}$ labeling the basis states of an irreducible rep space.

The orthonormal $ \ket{\psi^{(\nu)\kappa}_m} $ obtained from equation \eqref{eq:eigeneqnukm} are determined only up to a phase.
The Yamanouchi phase convention demands off-diagonal matrix elements of adjacent transpositions to be positive. It can be shown that as a result of this convention, the following rule determines the phase of a vector
\begin{equation}
 \ket{\psi^{(\nu)\kappa}_m} = \sum_{p\in \{a',b',\dots\}} u_{\nu m \kappa,p} \ket{p}.
\end{equation}
\begin{lem}
To satisfy the Yamanouchi phase convention,
the phase of a vector $\ket{\psi^{(\nu)\kappa}_m}$ has to be chosen in such a way that $u_{\nu m \kappa,\mathfrak{p}} > 0$, where $\mathfrak{p}$ is called the principal term.
The corresponding principal state $\ket{\mathfrak{p}} = \ket{ \mathfrak{p}_1,\mathfrak{p}_2, \dots, \mathfrak{p}_n} $ is constructed by setting $\mathfrak{p}_i$ equal to the Greek letter in the box of the Weyl tableau $W_\kappa^{(\nu)}$,
which is in the same position as the box in the Young tableau $Y_m^{(\nu)}$ containing the number $i$.
For example,
\begin{equation}
\ket{\psi^{(\nu)\kappa}_m} \equiv \ket{ W_\kappa^{(\nu)} Y_m^{(\nu)} } =
\left|  \young(\alpha\alpha\alpha\beta,\beta\delta,\gamma) \, \young(1245,37,6) \right\rangle
\rightarrow \ket{\mathfrak{p}} = \ket{ \alpha\alpha\beta\alpha\beta\gamma\delta  }.
\end{equation}
\end{lem}
\noindent
The basis $\{ \ket{\psi^{(\nu)\kappa}_m}  \}$ of $\mathcal{V}$ obeying the Yamanouchi phase convention is called quasi-standard basis (it is called standard basis or Young-Yamanouchi basis for the special case of $\mathcal{V}$ being the regular rep space).

We close this subsection by defining an order of the Young tableaux $Y_m^{(\nu)}$ and Weyl tableaux $W_\kappa^{(\nu)}$ corresponding to a certain Young diagram $\nu=[\nu_1,\dots,\nu_v]$, $\sum_i\nu_i=n$.
Before we start, note that the Young diagrams $\{ \nu \}$ corresponding to partitions of $n$ are ordered by sorting the $q$-digit strings given by a partition and supplemented with zeros if $v<q$ (note that $v\leq q$),
\begin{equation}\label{eq:yd_order}
( \nu_1 , \nu_2, \dots, \nu_q ) =
 \bigl( [\nu_1,\dots,\nu_v], \underset{q-v}{\underbrace{0,\dots 0,}} \bigr),
\end{equation}
in descending order.
The Young tableaux $\{ Y_{m_i}^{(\nu)} \}_{i=1\dots h_\nu}$ are enumerated by their eigenvalue list of $(C_2^n,\dots C_2^2, C_2^1)$ and are sorted in descending order.
The total number of Young tableaux (for a given Young diagram $\nu$) is equal to the dimension $h_\nu$ of the irrep labeled by $\nu$ and is given by the hook length formula \cite[page 120]{CPW02}.
To define an order of the Weyl tableaux $\{ W_{\kappa_i}^{(\nu)} \}_{i=1\dots \tau_\nu}$\footnote{Note that $\tau_\nu$ depends on the configuration of $\mathcal{V}$.}, we enumerate them by their corresponding Gel'fand symbols which are then sorted in descending order.
The Gel'fand symbol corresponding to a Weyl tableau $W_\kappa^{(\nu)}$ is defined as
the list of non-negative integers
\begin{equation}
\bigl[ \nu^{(1)}_1,\nu^{(1)}_2,\dots,\nu^{(1)}_q \,;\, \nu^{(2)}_1,\nu^{(2)}_2,\dots,\nu^{(2)}_{q-1} \,;\,
 \dots \,;\, \nu^{(q-1)}_1,\nu^{(q-1)}_2 \,;\, \nu^{(q)}_1 \bigr],
\end{equation}
where $\nu^{(1)}$ denotes the Young diagram $\nu$ (supplemented with zeros as in equation \eqref{eq:yd_order}) and
$\nu^{(j)} = [ \nu^{(j)}_1, \dots, \nu^{(j)}_{q-j+1} ]$ for $j=2\dots q$ denotes the Young diagram
which results after removing the boxes with letters $q-1, \dots, q-j+1$ from $W_\kappa^{(\nu)}$.
For instance,
\begin{equation}
\young(\beta\beta\gamma\gamma,\gamma\delta,\delta)
\leftrightarrow
 \begin{pmatrix}
  4\ 2\ 1\ 0 \\ 4\ 1\ 0 \\ 2\ 0 \\ 0
 \end{pmatrix}.
\end{equation}

\subsubsection{Example}
\begin{sidewaystable}\centering
\begin{tabular}{@{}ccc|cccccccccccc}
$\nu$ & $W_\kappa^{(\nu)}$ & $Y_m^{(\nu)}$
&\scriptsize $\ket{\alpha\gamma\delta\delta}$ &\scriptsize $\ket{\alpha\delta\gamma\delta}$
&\scriptsize $\ket{\alpha\delta\delta\gamma}$ &\scriptsize $\ket{\gamma\alpha\delta\delta}$
&\scriptsize $\ket{\gamma\delta\alpha\delta}$ &\scriptsize $\ket{\gamma\delta\delta\alpha}$
&\scriptsize $\ket{\delta\alpha\gamma\delta}$ &\scriptsize $\ket{\delta\alpha\delta\gamma}$
&\scriptsize $\ket{\delta\gamma\alpha\delta}$ &\scriptsize $\ket{\delta\gamma\delta\alpha}$
&\scriptsize $\ket{\delta\delta\alpha\gamma}$ &\scriptsize $\ket{\delta\delta\gamma\alpha}$\\
\hline\hline
%


\tiny $\yng(4)$ &\tiny $\young(\alpha\gamma\delta\delta)$ &\tiny $\young(1234)$ &\tiny $1/6\,\sqrt {3}$ &\tiny $1/6\,\sqrt {3}$ &\tiny $1/6\,\sqrt {3}$ &\tiny $1/6\,\sqrt {3}$ &\tiny $1/6\,\sqrt {3}$ &\tiny $1/6\,\sqrt {3}$ &\tiny $1/6\,\sqrt {3}$ &\tiny $1/6\,\sqrt {3}$ &\tiny $1/6\,\sqrt {3}$ &\tiny $1/6\,\sqrt {3}$ &\tiny $1/6\,\sqrt {3}$ &\tiny $1/6\,\sqrt {3}$\\[4pt]
\cline{2-15}
\hline
\tiny $\yng(3,1)$ &\tiny $\young(\alpha\gamma\delta,\delta)$ &\tiny $\young(123,4)$ &\tiny $1/6\,\sqrt {3}$ &\tiny $1/6\,\sqrt {3}$ &\tiny $-1/6\,\sqrt {3}$ &\tiny $1/6\,\sqrt {3}$ &\tiny $1/6\,\sqrt {3}$ &\tiny $-1/6\,\sqrt {3}$ &\tiny $1/6\,\sqrt {3}$ &\tiny $-1/6\,\sqrt {3}$ &\tiny $1/6\,\sqrt {3}$ &\tiny $-1/6\,\sqrt {3}$ &\tiny $-1/6\,\sqrt {3}$ &\tiny $-1/6\,\sqrt {3}$\\[4pt]
 & &\tiny $\young(124,3)$ &\tiny $1/6\,\sqrt {6}$ &\tiny $-1/12\,\sqrt {6}$ &\tiny $1/12\,\sqrt {6}$ &\tiny $1/6\,\sqrt {6}$ &\tiny $-1/12\,\sqrt {6}$ &\tiny $1/12\,\sqrt {6}$ &\tiny $-1/12\,\sqrt {6}$ &\tiny $1/12\,\sqrt {6}$ &\tiny $-1/12\,\sqrt {6}$ &\tiny $1/12\,\sqrt {6}$ &\tiny $-1/6\,\sqrt {6}$ &\tiny $-1/6\,\sqrt {6}$\\[4pt]
 & &\tiny $\young(134,2)$ &\tiny $0$ &\tiny $1/4\,\sqrt {2}$ &\tiny $1/4\,\sqrt {2}$ &\tiny $0$ &\tiny $1/4\,\sqrt {2}$ &\tiny $1/4\,\sqrt {2}$ &\tiny $-1/4\,\sqrt {2}$ &\tiny $-1/4\,\sqrt {2}$ &\tiny $-1/4\,\sqrt {2}$ &\tiny $-1/4\,\sqrt {2}$ &\tiny $0$ &\tiny $0$\\[4pt]
\cline{2-15}
 &\tiny $\young(\alpha\delta\delta,\gamma)$ &\tiny $\young(123,4)$ &\tiny $0$ &\tiny $0$ &\tiny $1/6\,\sqrt {6}$ &\tiny $0$ &\tiny $0$ &\tiny $-1/6\,\sqrt {6}$ &\tiny $0$ &\tiny $1/6\,\sqrt {6}$ &\tiny $0$ &\tiny $-1/6\,\sqrt {6}$ &\tiny $1/6\,\sqrt {6}$ &\tiny $-1/6\,\sqrt {6}$\\[4pt]
 & &\tiny $\young(124,3)$ &\tiny $0$ &\tiny $1/4\,\sqrt {3}$ &\tiny $1/12\,\sqrt {3}$ &\tiny $0$ &\tiny $-1/4\,\sqrt {3}$ &\tiny $-1/12\,\sqrt {3}$ &\tiny $1/4\,\sqrt {3}$ &\tiny $1/12\,\sqrt {3}$ &\tiny $-1/4\,\sqrt {3}$ &\tiny $-1/12\,\sqrt {3}$ &\tiny $-1/6\,\sqrt {3}$ &\tiny $1/6\,\sqrt {3}$\\[4pt]
 & &\tiny $\young(134,2)$ &\tiny $1/2$ &\tiny $1/4$ &\tiny $1/4$ &\tiny $-1/2$ &\tiny $-1/4$ &\tiny $-1/4$ &\tiny $-1/4$ &\tiny $-1/4$ &\tiny $1/4$ &\tiny $1/4$ &\tiny $0$ &\tiny $0$\\[4pt]
\cline{2-15}
\hline
\tiny $\yng(2,2)$ &\tiny $\young(\alpha\gamma,\delta\delta)$ &\tiny $\young(12,34)$ &\tiny $1/6\,\sqrt {6}$ &\tiny $-1/12\,\sqrt {6}$ &\tiny $-1/12\,\sqrt {6}$ &\tiny $1/6\,\sqrt {6}$ &\tiny $-1/12\,\sqrt {6}$ &\tiny $-1/12\,\sqrt {6}$ &\tiny $-1/12\,\sqrt {6}$ &\tiny $-1/12\,\sqrt {6}$ &\tiny $-1/12\,\sqrt {6}$ &\tiny $-1/12\,\sqrt {6}$ &\tiny $1/6\,\sqrt {6}$ &\tiny $1/6\,\sqrt {6}$\\[4pt]
 & &\tiny $\young(13,24)$ &\tiny $0$ &\tiny $1/4\,\sqrt {2}$ &\tiny $-1/4\,\sqrt {2}$ &\tiny $0$ &\tiny $1/4\,\sqrt {2}$ &\tiny $-1/4\,\sqrt {2}$ &\tiny $-1/4\,\sqrt {2}$ &\tiny $1/4\,\sqrt {2}$ &\tiny $-1/4\,\sqrt {2}$ &\tiny $1/4\,\sqrt {2}$ &\tiny $0$ &\tiny $0$\\[4pt]
\cline{2-15}
\hline
\tiny $\yng(2,1,1)$ &\tiny $\young(\alpha\delta,\gamma,\delta)$ &\tiny $\young(12,3,4)$ &\tiny $0$ &\tiny $1/4$ &\tiny $-1/4$ &\tiny $0$ &\tiny $-1/4$ &\tiny $1/4$ &\tiny $1/4$ &\tiny $-1/4$ &\tiny $-1/4$ &\tiny $1/4$ &\tiny $1/2$ &\tiny $-1/2$\\[5pt]
 & &\tiny $\young(13,2,4)$ &\tiny $1/6\,\sqrt {3}$ &\tiny $1/12\,\sqrt {3}$ &\tiny $-1/4\,\sqrt {3}$ &\tiny $-1/6\,\sqrt {3}$ &\tiny $-1/12\,\sqrt {3}$ &\tiny $1/4\,\sqrt {3}$ &\tiny $-1/12\,\sqrt {3}$ &\tiny $1/4\,\sqrt {3}$ &\tiny $1/12\,\sqrt {3}$ &\tiny $-1/4\,\sqrt {3}$ &\tiny $0$ &\tiny $0$\\[5pt]
 & &\tiny $\young(14,2,3)$ &\tiny $1/6\,\sqrt {6}$ &\tiny $-1/6\,\sqrt {6}$ &\tiny $0$ &\tiny $-1/6\,\sqrt {6}$ &\tiny $1/6\,\sqrt {6}$ &\tiny $0$ &\tiny $1/6\,\sqrt {6}$ &\tiny $0$ &\tiny $-1/6\,\sqrt {6}$ &\tiny $0$ &\tiny $0$ &\tiny $0$\\[5pt]
\cline{2-15}
\hline\hline

\end{tabular}
\caption[Quasi-standard basis]{The quasi-standard basis $\{ \ket{W_\kappa^{(\nu)} Y_m^{(\nu)}} \}$ of $\mathcal{V} \subset \mathcal{H}_4^{\otimes 4}$ with $\textsf{config}=(1,0,1,2)$.}\label{tab:quasisbasis44}
\vspace{3mm}
\begin{align*}
  D^{([3,1])}(p_{(12)}) &= \begin{pmatrix} 1 & 0 & 0\\ 0 & 1 & 0\\ 0 & 0 &-1\end{pmatrix}  &
  D^{([3,1])}(p_{(23)}) &= \begin{pmatrix} 1 & 0 & 0\\ 0 & -\frac{1}{2} & \frac{\sqrt{3}}{2}\\ 0 & \frac{\sqrt{3}}{2} & \frac{1}{2}\end{pmatrix} &
  D^{([3,1])}(p_{(34)}) &= \begin{pmatrix} -\frac{1}{3} & \frac{\sqrt{8}}{3} & 0\\ \frac{\sqrt{8}}{3} & \frac{1}{3} & 0\\ 0 & 0 &1\end{pmatrix} \\
  D^{([2,1,1])}(p_{(12)}) &= \begin{pmatrix} 1 & 0 & 0\\ 0 & -1 & 0\\ 0 & 0 &-1\end{pmatrix}  &
  D^{([2,1,1])}(p_{(23)}) &= \begin{pmatrix}-\frac{1}{2} & \frac{\sqrt{3}}{2} & 0\\ \frac{\sqrt{3}}{2} & \frac{1}{2} & 0 \\ 0 & 0 &-1\end{pmatrix} &
  D^{([2,1,1])}(p_{(34)}) &= \begin{pmatrix}  -1& 0 & 0\\ 0& -\frac{1}{3} & \frac{\sqrt{8}}{3}\\ 0 & \frac{\sqrt{8}}{3} & \frac{1}{3} \end{pmatrix}
\end{align*}
\end{sidewaystable}
To give an example, we apply the EFM to the 12-dimensional rep space $\mathcal{V} \subset \mathcal{H}_4^{\otimes 4}$ with configuration $\textsf{config}=(1,0,1,2)$.
The computational basis of $\mathcal{V}$ is given by
$\{ \ket{i} \}_{i=0\dots 11} = \{   \ket{\alpha\gamma\delta\delta},
\dots, \ket{\delta\delta\alpha\gamma},\ket{\delta\delta\gamma\alpha} \}$.
The sorted basis vectors $\ket{\psi^{(\nu)\kappa}_m}  \equiv \ket{ W_\kappa^{(\nu)} Y_m^{(\nu)} }$ of the quasi-standard basis of $\mathcal{V}$ obtained via the EFM are shown in table \ref{tab:quasisbasis44}.
As it can be seen from the table, $\mathcal{V}$ decomposes into 4 irreducible subspaces,
\begin{equation}
 \mathcal{V} = \mathcal{V}_{[4]} \oplus \mathcal{V}_{[3,1],1} \oplus \mathcal{V}_{[3,1],2} \oplus \mathcal{V}_{[2,2]} \oplus \mathcal{V}_{[2,1,1]},
\end{equation}
where the irrep $[3,1]$ is two-fold degenerated.
The dimensions of the irreps are given by $h_{[4]}=1$, $h_{[3,1]}=3$, $h_{[2,2]}=2$ and $h_{[2,1,1]}=3$.
Since $\tau_{[3,1]}=2$ and $\tau_{\nu}=1$ for the remaining $\nu$, we can easily check that $\sum_\nu \tau_\nu \times h_\nu = \dim(\mathcal{V})$.
Below the table, the representation matrices of adjacent transpositions are shown for the two 3-dimensional irreps $[3,1]$ and $[2,1,1]$.

\section{Schur Transform}\label{sec:schurtransform}

The Schur transform is a unitary transformation relating the standard computational basis of $n$ qudits of dimension $q$ to the Schur basis, a basis associated with the representation theory of the symmetric and general linear groups.
In the preceding subsection it was shown that the vector space of $n$ qudits decomposes into a direct sum of representation spaces $\mathcal{V}$ of the symmetric group, each of which is characterized by the frequency distribution of the one-qudit basis states.
In this section we show that the Schur basis is given by the collection of the quasi-standard bases of the symmetric group of all the rep spaces $\mathcal{V}$.

\subsection{The Schur Basis}\label{subsec:schurbasis}

Let $\mathcal{H}_q$ denote the Hilbert space of a qudit of dimension $q$ and let an orthonormal basis
$\{ \ket{0},\ket{1},\dots, \ket{q-1} \}$ be fixed.
Occasionally we label these $q$ basis states using letters from the Greek alphabet, i.\,e.
$\ket{0}\equiv\ket{\alpha}$, $\ket{1}\equiv\ket{\beta}$, $\ket{2}\equiv\ket{\gamma}$, and so on.
The set of invertible linear transformations on $\mathcal{H}_q$ is called the general linear group $\textsf{GL}(q,\mathbb{C}) = \textsf{GL}_q$. An element $\rho \in \textsf{GL}_q$ is defined by $q\times q$ complex numbers $\rho_{ij}$ and transforms the basis states according to
\begin{equation}
 \rho \ket{j} = \sum_{i=0}^{q-1} \rho_{ij} \ket{i}.
\end{equation}
The computational basis of the Hilbert space $\mathcal{H}_q^{\otimes n}$ of $n$ qudits of dimension $q$
is given by the set of $n$-fold product states of the one-qudit basis states,
\begin{equation}
 \mathcal{H}_q^{\otimes n} = \vspan\bigl\{ \ket{i_1,i_2,\dots,i_n} \bigr\},
\end{equation}
with $0\leq i_j<q$  for $j\in\{1,2,\dots,n\}$.
A representation $D(\textsf{GL}_q)$ of $\textsf{GL}_q$ on $\mathcal{H}_q^{\otimes n}$ is defined by
\begin{equation}
D(\rho) \ket{i_1,i_2,\dots,i_n} = \rho\otimes\rho \dots \otimes \rho \ket{i_1,i_2,\dots,i_n}
\end{equation}
for any $\rho\in\textsf{GL}_q$.
For the symmetric group $\textsf{S}_n$ a representation $D(\textsf{S}_n)$ on $\mathcal{H}_q^{\otimes n}$ was defined by equation \eqref{eq:defiactionSn}, where the action of $D(p)$ on a computational basis state was defined as
\begin{equation}
 D(p) \ket{ i_1 , i_2 , \dots , i_n } = \ket{ i_{p^{-1}(1)} , i_{p^{-1}(2)} , \dots , i_{p^{-1}(n)} }
\end{equation}
for any $p\in \textsf{S}_n$.
Hence, the $q^n$-dimensional vector space $\mathcal{H}_q^{\otimes n}$ forms a representation space for both the symmetric group $\textsf{S}_n$ and the general linear group $\textsf{GL}_q$.
An important observation is the following lemma.
\begin{lem}\label{lem:DSnDGLqCommute}
Elements of $D(\textsf{S}_n)$ and $D(\textsf{GL}_q)$ commute, i.\,e.
\begin{equation}\label{eq:DpDgCommute}
 [ D(p), D(\rho) ] = 0,
\end{equation}
for all $p\in \textsf{S}_n$ and all $\rho \in \textsf{GL}_q$.
\end{lem}

As it was discussed in subsection \ref{subsec:symmgr}, $\mathcal{H}_q^{\otimes n}$  is a direct sum of rep spaces $\mathcal{V}$ of $\textsf{S}_n$, each of which is spanned by a subset of the computational basis which is characterized by a configuration string $\textsf{config}=(n_0,n_1,\dots,n_{q-1})$ of length $q$ (with $\sum_i n_i=n$) specifying the number of one-qudit basis states (see eq. \eqref{eq:deficonfig}).
Let us now calculate the quasi-standard basis of $\textsf{S}_n$ for each of the $\binom{n+q-1}{q-1}$ different representation spaces $\mathcal{V} \subset \mathcal{H}_q^{\otimes n}$ by solving the eigenvalue equation \eqref{eq:eigeneqnukm} and applying the Yamanouchi phase convention\footnote{%
Actually we do not have to perform this calculation for all the spaces $\mathcal{V} \subset \mathcal{H}_q^{\otimes n}$.
If the non-zero elements $\textsf{config}'$ and $\widetilde{\textsf{config}}'$ of the configurations $\textsf{config}$ and $\widetilde{\textsf{config}}$ of rep spaces $\mathcal{V}$ and $\tilde{\mathcal{V}}$ are the same, the CSCO-III of $\mathcal{V}$ and $\tilde{\mathcal{V}}$ is identical and we can adopt solutions already known by relabeling the basis states and Weyl tableaux.}.
The collection of all basis states obtained this way,
\begin{equation}
 \Bigl\{  \ket{ W_{\kappa_j}^{(\nu)}  Y_{m_i}^{(\nu)}  }  \Bigr\}, \text{ with }
 \nu=\{[n],[n-1,1],\dots\}, \
 j=\{1,\dots, h_\nu(\textsf{GL}_q)\}, \
 i=\{1,\dots, h_\nu(\textsf{S}_n)\},
\end{equation}
forms the Schur basis which has the following properties:
\begin{lem}[Properties of the Schur basis]\label{lem:PropSchurBasis}\white{H}\black
\begin{enumerate}
\item The subspaces $\mathcal{V}_{\nu,\kappa}$ which are spanned by the
$\bigl\{  \ket{ W_{\kappa}^{(\nu)}   Y_{m_i}^{(\nu)} }  \bigr\}_{i=1,\dots, h_\nu(\textsf{S}_n)}$
are irreducible rep spaces of $\textsf{S}_n$. For $p\in \textsf{S}_n$ we have
\begin{equation}
 D(p) \ket{ W_{\kappa}^{(\nu)} Y_{m_i}^{(\nu)} } = \sum_{j=1}^{h_\nu(\textsf{S}_n)} D^{(\nu)}_{ji}(p)
      \ket{ W_{\kappa}^{(\nu)} Y_{m_j}^{(\nu)} }.
\end{equation}
The dimension $h_\nu(\textsf{S}_n)$ of these irreps is given by the hook length formula (see e.\,g. \cite[page 120]{CPW02}) and depends only on $\nu$.

\item The subspaces $\mathcal{V}_{\nu}^m$ which are spanned by the
$\bigl\{  \ket{ W_{\kappa_j}^{(\nu)}   Y_{m}^{(\nu)} }  \bigr\}_{j=1,\dots, h_\nu(\textsf{GL}_q)}$
are irreducible rep spaces of $\textsf{GL}_q$. For $\rho\in \textsf{GL}_q$ we have
\begin{equation}
 D(\rho) \ket{ W_{\kappa_i}^{(\nu)} Y_{m}^{(\nu)} } = \sum_{j=1}^{h_\nu(\textsf{GL}_q)} D^{(\nu)}_{ji}(\rho)
         \ket{ W_{\kappa_j}^{(\nu)} Y_{m}^{(\nu)} }.
\end{equation}
The dimension $h_\nu(\textsf{GL}_q)$ of these irreps is given by the Robinson formula (see e.\,g. \cite[page 319]{CPW02}) and depends on $\nu$ and $q$.

\end{enumerate}
\end{lem}
\begin{proof}
Part (i) was shown in detail in subsection \ref{subsec:symmgr}.
To give a (partial) prove of part (ii), we recall that an over-complete CSCO-III of $\textsf{S}_n$ on a subspace $\mathcal{V}\subset \mathcal{H}_q^{\otimes n}$ characterized by a certain configuration $\textsf{config}$ is given by the union of $( C(\textsf{S}_n), C(s) )$ and $( C(\textsf{S}_n), \bar{C}(s') )$ (see subsection \ref{subsec:symmgr}),
and that the $\ket{ W_{\kappa}^{(\nu)} Y_{m}^{(\nu)} }$ are the eigenstates of the CSCO-III with eigenvalues $Y_m^{(\nu)}$ and $W_{\kappa}^{(\nu)}$,
\begin{align*}
\bigl( C(\textsf{S}_n), C(s) \bigr) \ket{ W_{\kappa}^{(\nu)} Y_{m}^{(\nu)} } &= Y_m^{(\nu)} \ket{ W_{\kappa}^{(\nu)} Y_{m}^{(\nu)} } \\
\bigl( C(\textsf{S}_n), \bar{C}(s') \bigr) \ket{ W_{\kappa}^{(\nu)} Y_{m}^{(\nu)} } &= W_\kappa^{(\nu)} \ket{ W_{\kappa}^{(\nu)} Y_{m}^{(\nu)} }.
\end{align*}
Let us now assume that the operators $( C(\textsf{S}_n), C(s) )$ are extended to the corresponding operators on $\mathcal{H}_q^{\otimes n}$.
By lemma \ref{lem:DSnDGLqCommute}, any $D(\rho)$ with $\rho\in \textsf{GL}_q$ commutes with
$( C(\textsf{S}_n) , C(s) )$,
\begin{equation*}
\bigl( C(\textsf{S}_n), C(s) \bigr) D(\rho) \ket{ W_{\kappa}^{(\nu)} Y_{m}^{(\nu)} } =
      Y_m^{(\nu)} D(\rho) \ket{ W_{\kappa}^{(\nu)} Y_{m}^{(\nu)} },
\end{equation*}
and we conclude that
\begin{equation*}
 D(\rho) \ket{ W_{\kappa_i}^{(\nu)} Y_{m}^{(\nu)} } = \sum_{j=1}^{h_\nu(\textsf{GL}_q)} D^{(\nu)m}_{ji}(\rho)
         \ket{ W_{\kappa_j}^{(\nu)} Y_{m}^{(\nu)} }.
\end{equation*}
Eventually, it can be shown that the representations $D^{(\nu)m}(\rho)$ do not depend on $m$ and that they are irreducible.
\end{proof}

\noindent
It follows from lemma \ref{lem:PropSchurBasis} that the common representation space $\mathcal{H}_q^{\otimes n}$ decomposes into a direct sum of tensor spaces,
\begin{equation}\label{eq:hqn:decomposes}
 \mathcal{H}_q^{\otimes n} = \bigoplus_\nu
 \vspan \Bigl\{ \ket{ W_{\kappa_j}^{(\nu)} } \Bigr\}_{j=1,\dots, h_\nu(\textsf{GL}_q)} 
 \bigotimes
 \vspan \Bigl\{ \ket{ Y_{m_i}^{(\nu)} }  \Bigr\}_{i=1,\dots, h_\nu(\textsf{S}_n)} ,
\end{equation}
where the sum over the Young diagrams $\nu$ runs over all diagrams with at most $q$ rows.
Any product of operators $D(p)$ and $D(\rho)$ becomes block-diagonal,
\begin{equation}
 D(\rho)D(p)  = \bigoplus_\nu  D^{(\nu)}(\rho) \otimes D^{(\nu)}(p),
\end{equation}
for any $p\in \textsf{S}_n$ and any $\rho\in \textsf{GL}_q$.

\subsubsection{The Special Case of Qubits}

For $q=2$ things are simpler, as Young diagrams $\{ \nu=[\nu_1,\nu_2] \}$, with $\nu_1\geq \nu_2\geq 0$ and $\nu_1+\nu_2=n$, consist of at most two rows and can be labeled by an index $j$, where $2j$ is the number of columns consisting of one row only ($j=0\dots \frac{n}{2}$ if $n$ is even, $j=\frac{1}{2}\dots \frac{n}{2}$ if $n$ is odd).
\begin{center}
\scalebox{1}{
\begin{pspicture}(-3.50,0)(5.8,1.8)
\rput[c](-1.4,0.6){$j\leftrightarrow \nu=[\frac{n}{2}+j,\frac{n}{2}-j] =$}
\rput[t](1,1){\yng(2,2)}
\rput[c](1.8,0.6){$\dots$}
\rput[t](3,1){\yng(4,2)}
\rput[c](4.2,0.8){$\dots$}
\rput[t](5,1){\yng(2)}
\rput[c](1.8,1.55){$\frac{n}{2}-j$}
\rput[B](1.8,1.1){\rotatebox[origin=c]{-90}{$\left\{ \makebox(0,1.3)[b]{}\right.$}}
\rput[c](4.2,1.55){$2j$}
\rput[B](4.2,1.1){\rotatebox[origin=c]{-90}{$\left\{ \makebox(0,1.3)[b]{}\right.$}}
\end{pspicture}}%
\end{center}
The dimension of the irrep $j$ of $\textsf{S}_n$ is given by the hook length formula, which in this case yields
\begin{subequations}
\begin{equation}
h_j(\textsf{S}_n) = \binom{n}{n/2-j} \frac{2j+1}{n/2+j+1}.
\end{equation}
The Weyl tableaux $W_k^{(\nu=j)}$ are now labeled by $k=-j,\dots,j$, where $j+k$ denotes the number of $\beta$'s (ones) in the first row of the Weyl tableaux:
\begin{center}
\scalebox{1}{
\begin{pspicture}(0.0,-0.1)(6,1.8)
\rput[t](1,1){\young(\alpha\alpha,\beta\beta)}
\rput[c](1.8,0.6){$\dots$}
\rput[t](3,1){\young(\alpha\alpha\alpha\alpha,\beta\beta)}
\rput[c](4.2,0.8){$\dots$}
\rput[t](5,1){\young(\beta\beta)}
\rput[c](3.625,0.1){$j-k$}
\rput[B](3.625,0.3){\rotatebox[origin=c]{90}{$\left\{ \makebox(0,0.67)[b]{}\right.$}}
\rput[c](4.825,0.1){$j+k$}
\rput[B](4.825,0.3){\rotatebox[origin=c]{90}{$\left\{ \makebox(0,0.67)[b]{}\right.$}}
\rput[c](1.8,1.5){$\frac{n}{2}-j$}
\rput[B](1.8,1.1){\rotatebox[origin=c]{-90}{$\left\{ \makebox(0,1.3)[b]{}\right.$}}
\rput[c](4.2,1.5){$2j$}
\rput[B](4.2,1.1){\rotatebox[origin=c]{-90}{$\left\{ \makebox(0,1.3)[b]{}\right.$}}
\end{pspicture}}%
\end{center}
The total number of Weyl tableaux for a given $j$ is
\begin{equation}
 h_j(\textsf{GL}_2)=2j+1.
\end{equation}
\end{subequations}
Equation \eqref{eq:hqn:decomposes} becomes
\begin{equation}\label{eq:h2n:decomposes}
 \mathcal{H}_2^{\otimes n} = \bigoplus_{j=0,1/2}^{n/2}
 \vspan \Bigl\{ \ket{ W_k^{(j)} } \Bigr\}_{k=-j\dots +j}
 \otimes
 \vspan \Bigl\{ \ket{ Y_{m_i}^{(j)} }  \Bigr\}_{i=1\dots h_j(\textsf{S}_n)}.
\end{equation}

\subsection{Examples}%

Within the framework of this theses,
a matlab program has been developed which obtains the Schur basis for given values of $n$ and $q$ by implementing the ideas presented in subsections \ref{subsec:symmgr} and \ref{subsec:schurbasis}.
To give some examples, we present some of the Schur bases obtained by the program.

The first example is the Schur basis for $q=3$ and $n=3$. The Hilbert space of the three qudits decomposes as
\begin{multline}
\mathcal{H}_3^{\otimes 3} =
 \vspan \Bigl\{ \ket{ W_{\kappa_j}^{([3])} } \Bigr\}_{j=1\dots 10}
 \otimes
 \ket{ Y_{m_1}^{([3])} }
\bigoplus \\
 \vspan \Bigl\{ \ket{ W_{\kappa_j}^{([2,1])} } \Bigr\}_{j=1\dots 8} 
 \otimes
 \vspan \Bigl\{ \ket{ Y_{m_i}^{([2,1])} }  \Bigr\}_{i=1\dots 2}
\bigoplus
 \ket{ W_{\kappa_1}^{([1,1,1])} }
 \otimes
 \ket{ Y_{m_1}^{([1,1,1])} } ,
\end{multline}
and the Schur-basis-vectors are listed in table \ref{tab:schurq3n3}.
\begin{table}\centering
\begin{tabular}{ccc|cccccc}
$\nu$ &  $W_\kappa^{(\nu)}$ & $Y_m^{(\nu)}$ & \\
\hline\hline
 & & &\scriptsize $\ket{\alpha\alpha\alpha}$\\[-3pt]
\tiny $\yng(3)$ &\tiny $\young(\alpha\alpha\alpha)$ &\tiny $\young(123)$ &\tiny $1$\\
\cline{2-9}
 & & &\scriptsize $\ket{\alpha\alpha\beta}$ &\scriptsize $\ket{\alpha\beta\alpha}$ &\scriptsize $\ket{\beta\alpha\alpha}$\\[-3pt]
 &\tiny $\young(\alpha\alpha\beta)$ &\tiny $\young(123)$ &\tiny $1/3\,\sqrt {3}$ &\tiny $1/3\,\sqrt {3}$ &\tiny $1/3\,\sqrt {3}$\\
\cline{2-9}
 & & &\scriptsize $\ket{\alpha\beta\beta}$ &\scriptsize $\ket{\beta\alpha\beta}$ &\scriptsize $\ket{\beta\beta\alpha}$\\[-3pt]
 &\tiny $\young(\alpha\beta\beta)$ &\tiny $\young(123)$ &\tiny $1/3\,\sqrt {3}$ &\tiny $1/3\,\sqrt {3}$ &\tiny $1/3\,\sqrt {3}$\\
\cline{2-9}
 & & &\scriptsize $\ket{\beta\beta\beta}$\\[-3pt]
 &\tiny $\young(\beta\beta\beta)$ &\tiny $\young(123)$ &\tiny $1$\\
\cline{2-9}
 & & &\scriptsize $\ket{\alpha\alpha\gamma}$ &\scriptsize $\ket{\alpha\gamma\alpha}$ &\scriptsize $\ket{\gamma\alpha\alpha}$\\[-3pt]
 &\tiny $\young(\alpha\alpha\gamma)$ &\tiny $\young(123)$ &\tiny $1/3\,\sqrt {3}$ &\tiny $1/3\,\sqrt {3}$ &\tiny $1/3\,\sqrt {3}$\\
\cline{2-9}
 & & &\scriptsize $\ket{\alpha\beta\gamma}$ &\scriptsize $\ket{\alpha\gamma\beta}$ &\scriptsize $\ket{\beta\alpha\gamma}$ &\scriptsize $\ket{\beta\gamma\alpha}$ &\scriptsize $\ket{\gamma\alpha\beta}$ &\scriptsize $\ket{\gamma\beta\alpha}$\\[-3pt]
 &\tiny $\young(\alpha\beta\gamma)$ &\tiny $\young(123)$ &\tiny $1/6\,\sqrt {6}$ &\tiny $1/6\,\sqrt {6}$ &\tiny $1/6\,\sqrt {6}$ &\tiny $1/6\,\sqrt {6}$ &\tiny $1/6\,\sqrt {6}$ &\tiny $1/6\,\sqrt {6}$\\[1pt]
\cline{2-9}
 & & &\scriptsize $\ket{\beta\beta\gamma}$ &\scriptsize $\ket{\beta\gamma\beta}$ &\scriptsize $\ket{\gamma\beta\beta}$\\[-3pt]
 &\tiny $\young(\beta\beta\gamma)$ &\tiny $\young(123)$ &\tiny $1/3\,\sqrt {3}$ &\tiny $1/3\,\sqrt {3}$ &\tiny $1/3\,\sqrt {3}$\\
\cline{2-9}
 & & &\scriptsize $\ket{\alpha\gamma\gamma}$ &\scriptsize $\ket{\gamma\alpha\gamma}$ &\scriptsize $\ket{\gamma\gamma\alpha}$\\[-3pt]
 &\tiny $\young(\alpha\gamma\gamma)$ &\tiny $\young(123)$ &\tiny $1/3\,\sqrt {3}$ &\tiny $1/3\,\sqrt {3}$ &\tiny $1/3\,\sqrt {3}$\\
\cline{2-9}
 & & &\scriptsize $\ket{\beta\gamma\gamma}$ &\scriptsize $\ket{\gamma\beta\gamma}$ &\scriptsize $\ket{\gamma\gamma\beta}$\\[-3pt]
 &\tiny $\young(\beta\gamma\gamma)$ &\tiny $\young(123)$ &\tiny $1/3\,\sqrt {3}$ &\tiny $1/3\,\sqrt {3}$ &\tiny $1/3\,\sqrt {3}$\\
\cline{2-9}
 & & &\scriptsize $\ket{\gamma\gamma\gamma}$\\[-3pt]
 &\tiny $\young(\gamma\gamma\gamma)$ &\tiny $\young(123)$ &\tiny $1$\\
\cline{2-9}
\hline
 & & &\scriptsize $\ket{\alpha\alpha\beta}$ &\scriptsize $\ket{\alpha\beta\alpha}$ &\scriptsize $\ket{\beta\alpha\alpha}$\\[-2pt]
\tiny $\yng(2,1)$ &\tiny $\young(\alpha\alpha,\beta)$ &\tiny $\young(12,3)$ &\tiny $1/3\,\sqrt {6}$ &\tiny $-1/6\,\sqrt {6}$ &\tiny $-1/6\,\sqrt {6}$\\
 & &\tiny $\young(13,2)$ &\tiny $0$ &\tiny $1/2\,\sqrt {2}$ &\tiny $-1/2\,\sqrt {2}$\\[2pt]
\cline{2-9}
 & & &\scriptsize $\ket{\alpha\beta\beta}$ &\scriptsize $\ket{\beta\alpha\beta}$ &\scriptsize $\ket{\beta\beta\alpha}$\\[-2pt]
 &\tiny $\young(\alpha\beta,\beta)$ &\tiny $\young(12,3)$ &\tiny $1/6\,\sqrt {6}$ &\tiny $1/6\,\sqrt {6}$ &\tiny $-1/3\,\sqrt {6}$\\
 & &\tiny $\young(13,2)$ &\tiny $1/2\,\sqrt {2}$ &\tiny $-1/2\,\sqrt {2}$ &\tiny $0$\\[2pt]
\cline{2-9}
 & & &\scriptsize $\ket{\alpha\alpha\gamma}$ &\scriptsize $\ket{\alpha\gamma\alpha}$ &\scriptsize $\ket{\gamma\alpha\alpha}$\\[-2pt]
 &\tiny $\young(\alpha\alpha,\gamma)$ &\tiny $\young(12,3)$ &\tiny $1/3\,\sqrt {6}$ &\tiny $-1/6\,\sqrt {6}$ &\tiny $-1/6\,\sqrt {6}$\\
 & &\tiny $\young(13,2)$ &\tiny $0$ &\tiny $1/2\,\sqrt {2}$ &\tiny $-1/2\,\sqrt {2}$\\[2pt]
\cline{2-9}
 & & &\scriptsize $\ket{\alpha\beta\gamma}$ &\scriptsize $\ket{\alpha\gamma\beta}$ &\scriptsize $\ket{\beta\alpha\gamma}$ &\scriptsize $\ket{\beta\gamma\alpha}$ &\scriptsize $\ket{\gamma\alpha\beta}$ &\scriptsize $\ket{\gamma\beta\alpha}$\\[-2pt]
 &\tiny $\young(\alpha\beta,\gamma)$ &\tiny $\young(12,3)$ &\tiny $1/3\,\sqrt {3}$ &\tiny $-1/6\,\sqrt {3}$ &\tiny $1/3\,\sqrt {3}$ &\tiny $-1/6\,\sqrt {3}$ &\tiny $-1/6\,\sqrt {3}$ &\tiny $-1/6\,\sqrt {3}$\\
 & &\tiny $\young(13,2)$ &\tiny $0$ &\tiny $1/2$ &\tiny $0$ &\tiny $1/2$ &\tiny $-1/2$ &\tiny $-1/2$\\[2pt]
\cline{2-9}
 & & &\scriptsize $\ket{\beta\beta\gamma}$ &\scriptsize $\ket{\beta\gamma\beta}$ &\scriptsize $\ket{\gamma\beta\beta}$\\[-2pt]
 &\tiny $\young(\beta\beta,\gamma)$ &\tiny $\young(12,3)$ &\tiny $1/3\,\sqrt {6}$ &\tiny $-1/6\,\sqrt {6}$ &\tiny $-1/6\,\sqrt {6}$\\
 & &\tiny $\young(13,2)$ &\tiny $0$ &\tiny $1/2\,\sqrt {2}$ &\tiny $-1/2\,\sqrt {2}$\\[2pt]
\cline{2-9}
 & & &\scriptsize $\ket{\alpha\beta\gamma}$ &\scriptsize $\ket{\alpha\gamma\beta}$ &\scriptsize $\ket{\beta\alpha\gamma}$ &\scriptsize $\ket{\beta\gamma\alpha}$ &\scriptsize $\ket{\gamma\alpha\beta}$ &\scriptsize $\ket{\gamma\beta\alpha}$\\[-2pt]
 &\tiny $\young(\alpha\gamma,\beta)$ &\tiny $\young(12,3)$ &\tiny $0$ &\tiny $1/2$ &\tiny $0$ &\tiny $-1/2$ &\tiny $1/2$ &\tiny $-1/2$\\
 & &\tiny $\young(13,2)$ &\tiny $1/3\,\sqrt {3}$ &\tiny $1/6\,\sqrt {3}$ &\tiny $-1/3\,\sqrt {3}$ &\tiny $-1/6\,\sqrt {3}$ &\tiny $-1/6\,\sqrt {3}$ &\tiny $1/6\,\sqrt {3}$\\[2pt]
\cline{2-9}
 & & &\scriptsize $\ket{\alpha\gamma\gamma}$ &\scriptsize $\ket{\gamma\alpha\gamma}$ &\scriptsize $\ket{\gamma\gamma\alpha}$\\[-2pt]
 &\tiny $\young(\alpha\gamma,\gamma)$ &\tiny $\young(12,3)$ &\tiny $1/6\,\sqrt {6}$ &\tiny $1/6\,\sqrt {6}$ &\tiny $-1/3\,\sqrt {6}$\\
 & &\tiny $\young(13,2)$ &\tiny $1/2\,\sqrt {2}$ &\tiny $-1/2\,\sqrt {2}$ &\tiny $0$\\[2pt]
\cline{2-9}
 & & &\scriptsize $\ket{\beta\gamma\gamma}$ &\scriptsize $\ket{\gamma\beta\gamma}$ &\scriptsize $\ket{\gamma\gamma\beta}$\\[-2pt]
 &\tiny $\young(\beta\gamma,\gamma)$ &\tiny $\young(12,3)$ &\tiny $1/6\,\sqrt {6}$ &\tiny $1/6\,\sqrt {6}$ &\tiny $-1/3\,\sqrt {6}$\\
 & &\tiny $\young(13,2)$ &\tiny $1/2\,\sqrt {2}$ &\tiny $-1/2\,\sqrt {2}$ &\tiny $0$\\[2pt]
\cline{2-9}
\hline
 & & &\scriptsize $\ket{\alpha\beta\gamma}$ &\scriptsize $\ket{\alpha\gamma\beta}$ &\scriptsize $\ket{\beta\alpha\gamma}$ &\scriptsize $\ket{\beta\gamma\alpha}$ &\scriptsize $\ket{\gamma\alpha\beta}$ &\scriptsize $\ket{\gamma\beta\alpha}$\\[-2pt]
\tiny $\yng(1,1,1)$ &\tiny $\young(\alpha,\beta,\gamma)$ &\tiny $\young(1,2,3)$ &\tiny $1/6\,\sqrt {6}$ &\tiny $-1/6\,\sqrt {6}$ &\tiny $-1/6\,\sqrt {6}$ &\tiny $1/6\,\sqrt {6}$ &\tiny $1/6\,\sqrt {6}$ &\tiny $-1/6\,\sqrt {6}$\\[5pt]
\cline{2-9}
\hline\hline

\end{tabular}%
\caption[Schur basis of $\mathcal{H}_3^{\otimes 3}$]{Schur basis $\{ \ket{W_\kappa^{(\nu)} Y_m^{(\nu)}} \}$ of $\mathcal{H}_3^{\otimes 3}$.}\label{tab:schurq3n3}
\end{table}

As a second example, we consider $q=2$ and $n=1,2,3,4,5$.
The resulting Schur-basis vectors are listed in table \ref{tab:schurq2-4} for $n=2,3,4$ and in table \ref{tab:schurq2n5} for $n=5$.
The Schur basis of $\mathcal{H}_2^{\otimes 1}$ coincides with the computational basis, i.\,e. we have
$\bigl\vert \young(\alpha)\,\young(1)\, \bigr\rangle = \ket{\alpha}$ and
$\bigl\vert \young(\beta)\,\young(1)\,  \bigr\rangle = \ket{\beta}$.
The Schur basis of $\mathcal{H}_2^{\otimes 2}$ is given by
\begin{equation}
\begin{split}
\Bigl\vert \young(\alpha\alpha)\,\young(12)\, \Bigr\rangle &= \ket{\alpha\alpha},\\
\Bigl\vert \young(\alpha\beta)\,\young(12)\, \Bigr\rangle &= (\ket{\alpha\beta}+\ket{\beta\alpha})/\sqrt{2},\\
\Bigl\vert \young(\beta\beta)\,\young(12)\, \Bigr\rangle &= \ket{\beta\beta}, \text{ and }\\
\Bigl\vert \, \young(\alpha,\beta) \ \young(1,2)\, \Bigr\rangle &= (\ket{\alpha\beta}-\ket{\beta\alpha})/\sqrt{2}.
\end{split}
\end{equation}
The Hilbert space of the $n=5$ qubits decomposes as in equation \eqref{eq:h2n:decomposes} with
$h_{1/2}(\textsf{S}_5)=5$, $h_{3/2}(\textsf{S}_5)=4$ and $h_{5/2}(\textsf{S}_5)=1$.

\begin{table}
\begin{minipage}{.5\textwidth}\centering
\begin{tabular}{@{}ccc|c@{}c@{}c@{}c@{}c@{}c@{}c@{}c@{}c@{}c@{}}
$j/\nu$ &  $k/W_\kappa^{(\nu)}$ & $Y_m^{(\nu)}$ & \\
\hline\hline
\scriptsize $1$ &\scriptsize $1$ & &\scriptsize $\ket{\alpha\alpha}$\\[-2pt]
\tiny $\yng(2)$ &\tiny $\young(\alpha\alpha)$ &\tiny $\young(12)$ &\tiny $1$\\[4pt]
\cline{2-5}
 &\scriptsize $0$ & &\scriptsize $\ket{\alpha\beta}$ &\scriptsize $\ket{\beta\alpha}$\\[-2pt]
 &\tiny $\young(\alpha\beta)$ &\tiny $\young(12)$ &\tiny $1/2\,\sqrt {2}$ &\tiny $1/2\,\sqrt {2}$\\[4pt]
\cline{2-5}
 &\scriptsize $-1$ & &\scriptsize $\ket{\beta\beta}$\\[-2pt]
 &\tiny $\young(\beta\beta)$ &\tiny $\young(12)$ &\tiny $1$\\[4pt]
\cline{2-5}
\hline
\scriptsize $0$ &\scriptsize $0$ & &\scriptsize $\ket{\alpha\beta}$ &\scriptsize $\ket{\beta\alpha}$\\[-2pt]
\tiny $\yng(1,1)$ &\tiny $\young(\alpha,\beta)$ &\tiny $\young(1,2)$ &\tiny $1/2\,\sqrt {2}$ &\tiny $-1/2\,\sqrt {2}$\\[4pt]
\cline{2-5}
\hline

\end{tabular}
\end{minipage}%
\begin{minipage}{.5\textwidth}\centering
\begin{tabular}{@{}ccc|c@{}c@{}c@{}c@{}c@{}c@{}c@{}c@{}c@{}c@{}}
$j/\nu$ &  $k/W_\kappa^{(\nu)}$ & $Y_m^{(\nu)}$ & \\
\hline\hline
\scriptsize $3/2$ &\scriptsize $3/2$ & &\scriptsize $\ket{\alpha\alpha\alpha}$\\[-2pt]
\tiny $\yng(3)$ &\tiny $\young(\alpha\alpha\alpha)$ &\tiny $\young(123)$ &\tiny $1$\\[4pt]
\cline{2-6}
 &\scriptsize $1/2$ & &\scriptsize $\ket{\alpha\alpha\beta}$ &\scriptsize $\ket{\alpha\beta\alpha}$ &\scriptsize $\ket{\beta\alpha\alpha}$\\[-2pt]
 &\tiny $\young(\alpha\alpha\beta)$ &\tiny $\young(123)$ &\tiny $1/3\,\sqrt {3}$ &\tiny $1/3\,\sqrt {3}$ &\tiny $1/3\,\sqrt {3}$\\[4pt]
\cline{2-6}
 &\scriptsize $-1/2$ & &\scriptsize $\ket{\alpha\beta\beta}$ &\scriptsize $\ket{\beta\alpha\beta}$ &\scriptsize $\ket{\beta\beta\alpha}$\\[-2pt]
 &\tiny $\young(\alpha\beta\beta)$ &\tiny $\young(123)$ &\tiny $1/3\,\sqrt {3}$ &\tiny $1/3\,\sqrt {3}$ &\tiny $1/3\,\sqrt {3}$\\[4pt]
\cline{2-6}
 &\scriptsize $-3/2$ & &\scriptsize $\ket{\beta\beta\beta}$\\[-2pt]
 &\tiny $\young(\beta\beta\beta)$ &\tiny $\young(123)$ &\tiny $1$\\[4pt]
\cline{2-6}
\hline
\scriptsize $1/2$ &\scriptsize $1/2$ & &\scriptsize $\ket{\alpha\alpha\beta}$ &\scriptsize $\ket{\alpha\beta\alpha}$ &\scriptsize $\ket{\beta\alpha\alpha}$\\[-2pt]
\tiny $\yng(2,1)$ &\tiny $\young(\alpha\alpha,\beta)$ &\tiny $\young(12,3)$ &\tiny $1/3\,\sqrt {6}$ &\tiny $-1/6\,\sqrt {6}$ &\tiny $-1/6\,\sqrt {6}$\\[4pt]
 & &\tiny $\young(13,2)$ &\tiny $0$ &\tiny $1/2\,\sqrt {2}$ &\tiny $-1/2\,\sqrt {2}$\\[4pt]
\cline{2-6}
 &\scriptsize $-1/2$ & &\scriptsize $\ket{\alpha\beta\beta}$ &\scriptsize $\ket{\beta\alpha\beta}$ &\scriptsize $\ket{\beta\beta\alpha}$\\[-2pt]
 &\tiny $\young(\alpha\beta,\beta)$ &\tiny $\young(12,3)$ &\tiny $1/6\,\sqrt {6}$ &\tiny $1/6\,\sqrt {6}$ &\tiny $-1/3\,\sqrt {6}$\\[4pt]
 & &\tiny $\young(13,2)$ &\tiny $1/2\,\sqrt {2}$ &\tiny $-1/2\,\sqrt {2}$ &\tiny $0$\\[4pt]
\cline{2-6}
\hline

\end{tabular}
\end{minipage}
\begin{minipage}{\textwidth}\centering
\begin{tabular}{@{}ccc|c@{}c@{}c@{}c@{}c@{}c@{}c@{}c@{}c@{}c@{}}
$j/\nu$ &  $k/W_\kappa^{(\nu)}$ & $Y_m^{(\nu)}$ & \\
\hline\hline
\scriptsize $2$ &\scriptsize $2$ & &\scriptsize $\ket{\alpha\alpha\alpha\alpha}$\\[-2pt]
\tiny $\yng(4)$ &\tiny $\young(\alpha\alpha\alpha\alpha)$ &\tiny $\young(1234)$ &\tiny $1$\\[2pt]
\cline{2-9}
 &\scriptsize $1$  & &\scriptsize $\ket{\alpha\alpha\alpha\beta}$ &\scriptsize $\ket{\alpha\alpha\beta\alpha}$ &\scriptsize $\ket{\alpha\beta\alpha\alpha}$ &\scriptsize $\ket{\beta\alpha\alpha\alpha}$\\[-2pt]
 &\tiny $\young(\alpha\alpha\alpha\beta)$ &\tiny $\young(1234)$ &\tiny $1/2$ &\tiny $1/2$ &\tiny $1/2$ &\tiny $1/2$\\[2pt]
\cline{2-9}
 &\scriptsize $0$ & &\scriptsize $\ket{\alpha\alpha\beta\beta}$ &\scriptsize $\ket{\alpha\beta\alpha\beta}$ &\scriptsize $\ket{\alpha\beta\beta\alpha}$ &\scriptsize $\ket{\beta\alpha\alpha\beta}$ &\scriptsize $\ket{\beta\alpha\beta\alpha}$ &\scriptsize $\ket{\beta\beta\alpha\alpha}$\\[-2pt]
 &\tiny $\young(\alpha\alpha\beta\beta)$ &\tiny $\young(1234)$ &\tiny $1/6\,\sqrt {6}$ &\tiny $1/6\,\sqrt {6}$ &\tiny $1/6\,\sqrt {6}$ &\tiny $1/6\,\sqrt {6}$ &\tiny $1/6\,\sqrt {6}$ &\tiny $1/6\,\sqrt {6}$\\[2pt]
\cline{2-9}
 &\scriptsize $-1$ & &\scriptsize $\ket{\alpha\beta\beta\beta}$ &\scriptsize $\ket{\beta\alpha\beta\beta}$ &\scriptsize $\ket{\beta\beta\alpha\beta}$ &\scriptsize $\ket{\beta\beta\beta\alpha}$\\[-2pt]
 &\tiny $\young(\alpha\beta\beta\beta)$ &\tiny $\young(1234)$ &\tiny $1/2$ &\tiny $1/2$ &\tiny $1/2$ &\tiny $1/2$\\[2pt]
\cline{2-9}
 &\scriptsize $-2$ & &\scriptsize $\ket{\beta\beta\beta\beta}$\\[-2pt]
 &\tiny $\young(\beta\beta\beta\beta)$ &\tiny $\young(1234)$ &\tiny $1$\\[2pt]
\cline{2-9}
\hline
\scriptsize $1$ &\scriptsize $1$ & &\scriptsize $\ket{\alpha\alpha\alpha\beta}$ &\scriptsize $\ket{\alpha\alpha\beta\alpha}$ &\scriptsize $\ket{\alpha\beta\alpha\alpha}$ &\scriptsize $\ket{\beta\alpha\alpha\alpha}$\\[-2pt]
\tiny $\yng(3,1)$ &\tiny $\young(\alpha\alpha\alpha,\beta)$ &\tiny $\young(123,4)$ &\tiny $1/2\,\sqrt {3}$ &\tiny $-1/6\,\sqrt {3}$ &\tiny $-1/6\,\sqrt {3}$ &\tiny $-1/6\,\sqrt {3}$\\[4pt]
 & &\tiny $\young(124,3)$ &\tiny $0$ &\tiny $1/3\,\sqrt {6}$ &\tiny $-1/6\,\sqrt {6}$ &\tiny $-1/6\,\sqrt {6}$\\[4pt]
 & &\tiny $\young(134,2)$ &\tiny $0$ &\tiny $0$ &\tiny $1/2\,\sqrt {2}$ &\tiny $-1/2\,\sqrt {2}$\\[4pt]
\cline{2-9}
 &\scriptsize $0$ & &\scriptsize $\ket{\alpha\alpha\beta\beta}$ &\scriptsize $\ket{\alpha\beta\alpha\beta}$ &\scriptsize $\ket{\alpha\beta\beta\alpha}$ &\scriptsize $\ket{\beta\alpha\alpha\beta}$ &\scriptsize $\ket{\beta\alpha\beta\alpha}$ &\scriptsize $\ket{\beta\beta\alpha\alpha}$\\[-2pt]
 &\tiny $\young(\alpha\alpha\beta,\beta)$ &\tiny $\young(123,4)$ &\tiny $1/6\,\sqrt {6}$ &\tiny $1/6\,\sqrt {6}$ &\tiny $-1/6\,\sqrt {6}$ &\tiny $1/6\,\sqrt {6}$ &\tiny $-1/6\,\sqrt {6}$ &\tiny $-1/6\,\sqrt {6}$\\[4pt]
 & &\tiny $\young(124,3)$ &\tiny $1/3\,\sqrt {3}$ &\tiny $-1/6\,\sqrt {3}$ &\tiny $1/6\,\sqrt {3}$ &\tiny $-1/6\,\sqrt {3}$ &\tiny $1/6\,\sqrt {3}$ &\tiny $-1/3\,\sqrt {3}$\\[4pt]
 & &\tiny $\young(134,2)$ &\tiny $0$ &\tiny $1/2$ &\tiny $1/2$ &\tiny $-1/2$ &\tiny $-1/2$ &\tiny $0$\\[4pt]
\cline{2-9}
 &\scriptsize $-1$ & &\scriptsize $\ket{\alpha\beta\beta\beta}$ &\scriptsize $\ket{\beta\alpha\beta\beta}$ &\scriptsize $\ket{\beta\beta\alpha\beta}$ &\scriptsize $\ket{\beta\beta\beta\alpha}$\\[-2pt]
 &\tiny $\young(\alpha\beta\beta,\beta)$ &\tiny $\young(123,4)$ &\tiny $1/6\,\sqrt {3}$ &\tiny $1/6\,\sqrt {3}$ &\tiny $1/6\,\sqrt {3}$ &\tiny $-1/2\,\sqrt {3}$\\[4pt]
 & &\tiny $\young(124,3)$ &\tiny $1/6\,\sqrt {6}$ &\tiny $1/6\,\sqrt {6}$ &\tiny $-1/3\,\sqrt {6}$ &\tiny $0$\\[4pt]
 & &\tiny $\young(134,2)$ &\tiny $1/2\,\sqrt {2}$ &\tiny $-1/2\,\sqrt {2}$ &\tiny $0$ &\tiny $0$\\[4pt]
\cline{2-9}
\hline
\scriptsize $0$ &\scriptsize $0$ & &\scriptsize $\ket{\alpha\alpha\beta\beta}$ &\scriptsize $\ket{\alpha\beta\alpha\beta}$ &\scriptsize $\ket{\alpha\beta\beta\alpha}$ &\scriptsize $\ket{\beta\alpha\alpha\beta}$ &\scriptsize $\ket{\beta\alpha\beta\alpha}$ &\scriptsize $\ket{\beta\beta\alpha\alpha}$\\[-2pt]
\tiny $\yng(2,2)$ &\tiny $\young(\alpha\alpha,\beta\beta)$ &\tiny $\young(12,34)$ &\tiny $1/3\,\sqrt {3}$ &\tiny $-1/6\,\sqrt {3}$ &\tiny $-1/6\,\sqrt {3}$ &\tiny $-1/6\,\sqrt {3}$ &\tiny $-1/6\,\sqrt {3}$ &\tiny $1/3\,\sqrt {3}$\\[4pt]
 & &\tiny $\young(13,24)$ &\tiny $0$ &\tiny $1/2$ &\tiny $-1/2$ &\tiny $-1/2$ &\tiny $1/2$ &\tiny $0$\\[4pt]
\cline{2-9}
\hline

\end{tabular}
\end{minipage}
\caption[Schur bases of $\mathcal{H}_2^n$, $n=1,2,3,4$]{Schur bases $\{ \ket{W_\kappa^{(\nu)} Y_m^{(\nu)}} \}$ of $\mathcal{H}_2^{\otimes 2}$ (\textit{upper left corner}), $\mathcal{H}_2^{\otimes 3}$ (\textit{upper right corner}), and $\mathcal{H}_2^{\otimes 4}$ (\textit{bottom}).}\label{tab:schurq2-4}
\end{table}

\begin{table}\centering
\begin{tabular}{@{}ccc|c@{}c@{}c@{}c@{}c@{}c@{}c@{}c@{}c@{}c@{}}
$j/\nu$ &  $k/W_\kappa^{(\nu)}$ & $Y_m^{(\nu)}$ & \\
\hline\hline
\scriptsize $5/2$ & \scriptsize $-5/2$ & &\scriptsize $\ket{\alpha\alpha\alpha\alpha\alpha}$\\[-2pt]
\tiny $\yng(5)$ &\tiny $\young(\alpha\alpha\alpha\alpha\alpha)$ &\tiny $\young(12345)$ &\tiny $1$\\[1pt]
\cline{2-13}
 &\scriptsize $-3/2$ & &\scriptsize $\ket{\alpha\alpha\alpha\alpha\beta}$ &\scriptsize $\ket{\alpha\alpha\alpha\beta\alpha}$ &\scriptsize $\ket{\alpha\alpha\beta\alpha\alpha}$ &\scriptsize $\ket{\alpha\beta\alpha\alpha\alpha}$ &\scriptsize $\ket{\beta\alpha\alpha\alpha\alpha}$\\[-2pt]
 &\tiny $\young(\alpha\alpha\alpha\alpha\beta)$ &\tiny $\young(12345)$ &\tiny $1/5\,\sqrt {5}$ &\tiny $1/5\,\sqrt {5}$ &\tiny $1/5\,\sqrt {5}$ &\tiny $1/5\,\sqrt {5}$ &\tiny $1/5\,\sqrt {5}$\\[1pt]
\cline{2-13}
 &\scriptsize $-1/2$ & &\scriptsize $\ket{\alpha\alpha\alpha\beta\beta}$ &\scriptsize $\ket{\alpha\alpha\beta\alpha\beta}$ &\scriptsize $\ket{\alpha\alpha\beta\beta\alpha}$ &\scriptsize $\ket{\alpha\beta\alpha\alpha\beta}$ &\scriptsize $\ket{\alpha\beta\alpha\beta\alpha}$ &\scriptsize $\ket{\alpha\beta\beta\alpha\alpha}$ &\scriptsize $\ket{\beta\alpha\alpha\alpha\beta}$ &\scriptsize $\ket{\beta\alpha\alpha\beta\alpha}$ &\scriptsize $\ket{\beta\alpha\beta\alpha\alpha}$ &\scriptsize $\ket{\beta\beta\alpha\alpha\alpha}$\\[-2pt]
 &\tiny $\young(\alpha\alpha\alpha\beta\beta)$ &\tiny $\young(12345)$ &\tiny $1/10\,\sqrt {10}$ &\tiny $1/10\,\sqrt {10}$ &\tiny $1/10\,\sqrt {10}$ &\tiny $1/10\,\sqrt {10}$ &\tiny $1/10\,\sqrt {10}$ &\tiny $1/10\,\sqrt {10}$ &\tiny $1/10\,\sqrt {10}$ &\tiny $1/10\,\sqrt {10}$ &\tiny $1/10\,\sqrt {10}$ &\tiny $1/10\,\sqrt {10}$\\[1pt]
\cline{2-13}
 &\scriptsize $1/2$ & &\scriptsize $\ket{\alpha\alpha\beta\beta\beta}$ &\scriptsize $\ket{\alpha\beta\alpha\beta\beta}$ &\scriptsize $\ket{\alpha\beta\beta\alpha\beta}$ &\scriptsize $\ket{\alpha\beta\beta\beta\alpha}$ &\scriptsize $\ket{\beta\alpha\alpha\beta\beta}$ &\scriptsize $\ket{\beta\alpha\beta\alpha\beta}$ &\scriptsize $\ket{\beta\alpha\beta\beta\alpha}$ &\scriptsize $\ket{\beta\beta\alpha\alpha\beta}$ &\scriptsize $\ket{\beta\beta\alpha\beta\alpha}$ &\scriptsize $\ket{\beta\beta\beta\alpha\alpha}$\\[-2pt]
 &\tiny $\young(\alpha\alpha\beta\beta\beta)$ &\tiny $\young(12345)$ &\tiny $1/10\,\sqrt {10}$ &\tiny $1/10\,\sqrt {10}$ &\tiny $1/10\,\sqrt {10}$ &\tiny $1/10\,\sqrt {10}$ &\tiny $1/10\,\sqrt {10}$ &\tiny $1/10\,\sqrt {10}$ &\tiny $1/10\,\sqrt {10}$ &\tiny $1/10\,\sqrt {10}$ &\tiny $1/10\,\sqrt {10}$ &\tiny $1/10\,\sqrt {10}$\\[1pt]
\cline{2-13}
 &\scriptsize $3/2$ & &\scriptsize $\ket{\alpha\beta\beta\beta\beta}$ &\scriptsize $\ket{\beta\alpha\beta\beta\beta}$ &\scriptsize $\ket{\beta\beta\alpha\beta\beta}$ &\scriptsize $\ket{\beta\beta\beta\alpha\beta}$ &\scriptsize $\ket{\beta\beta\beta\beta\alpha}$\\[-2pt]
 &\tiny $\young(\alpha\beta\beta\beta\beta)$ &\tiny $\young(12345)$ &\tiny $1/5\,\sqrt {5}$ &\tiny $1/5\,\sqrt {5}$ &\tiny $1/5\,\sqrt {5}$ &\tiny $1/5\,\sqrt {5}$ &\tiny $1/5\,\sqrt {5}$\\[1pt]
\cline{2-13}
 &\scriptsize $5/2$ & &\scriptsize $\ket{\beta\beta\beta\beta\beta}$\\[-2pt]
 &\tiny $\young(\beta\beta\beta\beta\beta)$ &\tiny $\young(12345)$ &\tiny $1$\\[1pt]
\cline{2-13}
\hline
\scriptsize $3/2$ &\scriptsize $-3/2$ & &\scriptsize $\ket{\alpha\alpha\alpha\alpha\beta}$ &\scriptsize $\ket{\alpha\alpha\alpha\beta\alpha}$ &\scriptsize $\ket{\alpha\alpha\beta\alpha\alpha}$ &\scriptsize $\ket{\alpha\beta\alpha\alpha\alpha}$ &\scriptsize $\ket{\beta\alpha\alpha\alpha\alpha}$\\[-2pt]
\tiny $\yng(4,1)$ &\tiny $\young(\alpha\alpha\alpha\alpha,\beta)$ &\tiny $\young(1234,5)$ &\tiny $2/5\,\sqrt {5}$ &\tiny $-1/10\,\sqrt {5}$ &\tiny $-1/10\,\sqrt {5}$ &\tiny $-1/10\,\sqrt {5}$ &\tiny $-1/10\,\sqrt {5}$\\[1pt]
 & &\tiny $\young(1235,4)$ &\tiny $0$ &\tiny $1/2\,\sqrt {3}$ &\tiny $-1/6\,\sqrt {3}$ &\tiny $-1/6\,\sqrt {3}$ &\tiny $-1/6\,\sqrt {3}$\\[1pt]
 & &\tiny $\young(1245,3)$ &\tiny $0$ &\tiny $0$ &\tiny $1/3\,\sqrt {6}$ &\tiny $-1/6\,\sqrt {6}$ &\tiny $-1/6\,\sqrt {6}$\\[1pt]
 & &\tiny $\young(1345,2)$ &\tiny $0$ &\tiny $0$ &\tiny $0$ &\tiny $1/2\,\sqrt {2}$ &\tiny $-1/2\,\sqrt {2}$\\[1pt]
\cline{2-13}
 &\scriptsize $-1/2$ & &\scriptsize $\ket{\alpha\alpha\alpha\beta\beta}$ &\scriptsize $\ket{\alpha\alpha\beta\alpha\beta}$ &\scriptsize $\ket{\alpha\alpha\beta\beta\alpha}$ &\scriptsize $\ket{\alpha\beta\alpha\alpha\beta}$ &\scriptsize $\ket{\alpha\beta\alpha\beta\alpha}$ &\scriptsize $\ket{\alpha\beta\beta\alpha\alpha}$ &\scriptsize $\ket{\beta\alpha\alpha\alpha\beta}$ &\scriptsize $\ket{\beta\alpha\alpha\beta\alpha}$ &\scriptsize $\ket{\beta\alpha\beta\alpha\alpha}$ &\scriptsize $\ket{\beta\beta\alpha\alpha\alpha}$\\[-2pt]
 &\tiny $\young(\alpha\alpha\alpha\beta,\beta)$ &\tiny $\young(1234,5)$ &\tiny $1/10\,\sqrt {15}$ &\tiny $1/10\,\sqrt {15}$ &\tiny $-1/15\,\sqrt {15}$ &\tiny $1/10\,\sqrt {15}$ &\tiny $-1/15\,\sqrt {15}$ &\tiny $-1/15\,\sqrt {15}$ &\tiny $1/10\,\sqrt {15}$ &\tiny $-1/15\,\sqrt {15}$ &\tiny $-1/15\,\sqrt {15}$ &\tiny $-1/15\,\sqrt {15}$\\[1pt]
 & &\tiny $\young(1235,4)$ &\tiny $1/2$ &\tiny $-1/6$ &\tiny $1/3$ &\tiny $-1/6$ &\tiny $1/3$ &\tiny $-1/3$ &\tiny $-1/6$ &\tiny $1/3$ &\tiny $-1/3$ &\tiny $-1/3$\\[1pt]
 & &\tiny $\young(1245,3)$ &\tiny $0$ &\tiny $1/3\,\sqrt {2}$ &\tiny $1/3\,\sqrt {2}$ &\tiny $-1/6\,\sqrt {2}$ &\tiny $-1/6\,\sqrt {2}$ &\tiny $1/6\,\sqrt {2}$ &\tiny $-1/6\,\sqrt {2}$ &\tiny $-1/6\,\sqrt {2}$ &\tiny $1/6\,\sqrt {2}$ &\tiny $-1/3\,\sqrt {2}$\\[1pt]
 & &\tiny $\young(1345,2)$ &\tiny $0$ &\tiny $0$ &\tiny $0$ &\tiny $1/6\,\sqrt {6}$ &\tiny $1/6\,\sqrt {6}$ &\tiny $1/6\,\sqrt {6}$ &\tiny $-1/6\,\sqrt {6}$ &\tiny $-1/6\,\sqrt {6}$ &\tiny $-1/6\,\sqrt {6}$ &\tiny $0$\\[1pt]
\cline{2-13}
 &\scriptsize $1/2$ & &\scriptsize $\ket{\alpha\alpha\beta\beta\beta}$ &\scriptsize $\ket{\alpha\beta\alpha\beta\beta}$ &\scriptsize $\ket{\alpha\beta\beta\alpha\beta}$ &\scriptsize $\ket{\alpha\beta\beta\beta\alpha}$ &\scriptsize $\ket{\beta\alpha\alpha\beta\beta}$ &\scriptsize $\ket{\beta\alpha\beta\alpha\beta}$ &\scriptsize $\ket{\beta\alpha\beta\beta\alpha}$ &\scriptsize $\ket{\beta\beta\alpha\alpha\beta}$ &\scriptsize $\ket{\beta\beta\alpha\beta\alpha}$ &\scriptsize $\ket{\beta\beta\beta\alpha\alpha}$\\[-2pt]
 &\tiny $\young(\alpha\alpha\beta\beta,\beta)$ &\tiny $\young(1234,5)$ &\tiny $1/15\,\sqrt {15}$ &\tiny $1/15\,\sqrt {15}$ &\tiny $1/15\,\sqrt {15}$ &\tiny $-1/10\,\sqrt {15}$ &\tiny $1/15\,\sqrt {15}$ &\tiny $1/15\,\sqrt {15}$ &\tiny $-1/10\,\sqrt {15}$ &\tiny $1/15\,\sqrt {15}$ &\tiny $-1/10\,\sqrt {15}$ &\tiny $-1/10\,\sqrt {15}$\\[1pt]
 & &\tiny $\young(1235,4)$ &\tiny $1/3$ &\tiny $1/3$ &\tiny $-1/3$ &\tiny $1/6$ &\tiny $1/3$ &\tiny $-1/3$ &\tiny $1/6$ &\tiny $-1/3$ &\tiny $1/6$ &\tiny $-1/2$\\[1pt]
 & &\tiny $\young(1245,3)$ &\tiny $1/3\,\sqrt {2}$ &\tiny $-1/6\,\sqrt {2}$ &\tiny $1/6\,\sqrt {2}$ &\tiny $1/6\,\sqrt {2}$ &\tiny $-1/6\,\sqrt {2}$ &\tiny $1/6\,\sqrt {2}$ &\tiny $1/6\,\sqrt {2}$ &\tiny $-1/3\,\sqrt {2}$ &\tiny $-1/3\,\sqrt {2}$ &\tiny $0$\\[1pt]
 & &\tiny $\young(1345,2)$ &\tiny $0$ &\tiny $1/6\,\sqrt {6}$ &\tiny $1/6\,\sqrt {6}$ &\tiny $1/6\,\sqrt {6}$ &\tiny $-1/6\,\sqrt {6}$ &\tiny $-1/6\,\sqrt {6}$ &\tiny $-1/6\,\sqrt {6}$ &\tiny $0$ &\tiny $0$ &\tiny $0$\\[1pt]
\cline{2-13}
 &\scriptsize $3/2$ & &\scriptsize $\ket{\alpha\beta\beta\beta\beta}$ &\scriptsize $\ket{\beta\alpha\beta\beta\beta}$ &\scriptsize $\ket{\beta\beta\alpha\beta\beta}$ &\scriptsize $\ket{\beta\beta\beta\alpha\beta}$ &\scriptsize $\ket{\beta\beta\beta\beta\alpha}$\\[-2pt]
 &\tiny $\young(\alpha\beta\beta\beta,\beta)$ &\tiny $\young(1234,5)$ &\tiny $1/10\,\sqrt {5}$ &\tiny $1/10\,\sqrt {5}$ &\tiny $1/10\,\sqrt {5}$ &\tiny $1/10\,\sqrt {5}$ &\tiny $-2/5\,\sqrt {5}$\\[1pt]
 & &\tiny $\young(1235,4)$ &\tiny $1/6\,\sqrt {3}$ &\tiny $1/6\,\sqrt {3}$ &\tiny $1/6\,\sqrt {3}$ &\tiny $-1/2\,\sqrt {3}$ &\tiny $0$\\[1pt]
 & &\tiny $\young(1245,3)$ &\tiny $1/6\,\sqrt {6}$ &\tiny $1/6\,\sqrt {6}$ &\tiny $-1/3\,\sqrt {6}$ &\tiny $0$ &\tiny $0$\\[1pt]
 & &\tiny $\young(1345,2)$ &\tiny $1/2\,\sqrt {2}$ &\tiny $-1/2\,\sqrt {2}$ &\tiny $0$ &\tiny $0$ &\tiny $0$\\[1pt]
\cline{2-13}
\hline
\scriptsize $1/2$  &\scriptsize $-1/2$ & &\scriptsize $\ket{\alpha\alpha\alpha\beta\beta}$ &\scriptsize $\ket{\alpha\alpha\beta\alpha\beta}$ &\scriptsize $\ket{\alpha\alpha\beta\beta\alpha}$ &\scriptsize $\ket{\alpha\beta\alpha\alpha\beta}$ &\scriptsize $\ket{\alpha\beta\alpha\beta\alpha}$ &\scriptsize $\ket{\alpha\beta\beta\alpha\alpha}$ &\scriptsize $\ket{\beta\alpha\alpha\alpha\beta}$ &\scriptsize $\ket{\beta\alpha\alpha\beta\alpha}$ &\scriptsize $\ket{\beta\alpha\beta\alpha\alpha}$ &\scriptsize $\ket{\beta\beta\alpha\alpha\alpha}$\\[-2pt]
\tiny $\yng(3,2)$ &\tiny $\young(\alpha\alpha\alpha,\beta\beta)$ &\tiny $\young(123,45)$ &\tiny $1/2\,\sqrt {2}$ &\tiny $-1/6\,\sqrt {2}$ &\tiny $-1/6\,\sqrt {2}$ &\tiny $-1/6\,\sqrt {2}$ &\tiny $-1/6\,\sqrt {2}$ &\tiny $1/6\,\sqrt {2}$ &\tiny $-1/6\,\sqrt {2}$ &\tiny $-1/6\,\sqrt {2}$ &\tiny $1/6\,\sqrt {2}$ &\tiny $1/6\,\sqrt {2}$\\[1pt]
 & &\tiny $\young(124,35)$ &\tiny $0$ &\tiny $2/3$ &\tiny $-1/3$ &\tiny $-1/3$ &\tiny $1/6$ &\tiny $-1/6$ &\tiny $-1/3$ &\tiny $1/6$ &\tiny $-1/6$ &\tiny $1/3$\\[1pt]
 & &\tiny $\young(134,25)$ &\tiny $0$ &\tiny $0$ &\tiny $0$ &\tiny $1/3\,\sqrt {3}$ &\tiny $-1/6\,\sqrt {3}$ &\tiny $-1/6\,\sqrt {3}$ &\tiny $-1/3\,\sqrt {3}$ &\tiny $1/6\,\sqrt {3}$ &\tiny $1/6\,\sqrt {3}$ &\tiny $0$\\[1pt]
 & &\tiny $\young(125,34)$ &\tiny $0$ &\tiny $0$ &\tiny $1/3\,\sqrt {3}$ &\tiny $0$ &\tiny $-1/6\,\sqrt {3}$ &\tiny $-1/6\,\sqrt {3}$ &\tiny $0$ &\tiny $-1/6\,\sqrt {3}$ &\tiny $-1/6\,\sqrt {3}$ &\tiny $1/3\,\sqrt {3}$\\[1pt]
 & &\tiny $\young(135,24)$ &\tiny $0$ &\tiny $0$ &\tiny $0$ &\tiny $0$ &\tiny $1/2$ &\tiny $-1/2$ &\tiny $0$ &\tiny $-1/2$ &\tiny $1/2$ &\tiny $0$\\[1pt]
\cline{2-13}
 &\scriptsize $1/2$ & &\scriptsize $\ket{\alpha\alpha\beta\beta\beta}$ &\scriptsize $\ket{\alpha\beta\alpha\beta\beta}$ &\scriptsize $\ket{\alpha\beta\beta\alpha\beta}$ &\scriptsize $\ket{\alpha\beta\beta\beta\alpha}$ &\scriptsize $\ket{\beta\alpha\alpha\beta\beta}$ &\scriptsize $\ket{\beta\alpha\beta\alpha\beta}$ &\scriptsize $\ket{\beta\alpha\beta\beta\alpha}$ &\scriptsize $\ket{\beta\beta\alpha\alpha\beta}$ &\scriptsize $\ket{\beta\beta\alpha\beta\alpha}$ &\scriptsize $\ket{\beta\beta\beta\alpha\alpha}$\\[-2pt]
 &\tiny $\young(\alpha\alpha\beta,\beta\beta)$ &\tiny $\young(123,45)$ &\tiny $1/6\,\sqrt {2}$ &\tiny $1/6\,\sqrt {2}$ &\tiny $-1/6\,\sqrt {2}$ &\tiny $-1/6\,\sqrt {2}$ &\tiny $1/6\,\sqrt {2}$ &\tiny $-1/6\,\sqrt {2}$ &\tiny $-1/6\,\sqrt {2}$ &\tiny $-1/6\,\sqrt {2}$ &\tiny $-1/6\,\sqrt {2}$ &\tiny $1/2\,\sqrt {2}$\\[1pt]
 & &\tiny $\young(124,35)$ &\tiny $1/3$ &\tiny $-1/6$ &\tiny $1/6$ &\tiny $-1/3$ &\tiny $-1/6$ &\tiny $1/6$ &\tiny $-1/3$ &\tiny $-1/3$ &\tiny $2/3$ &\tiny $0$\\[1pt]
 & &\tiny $\young(134,25)$ &\tiny $0$ &\tiny $1/6\,\sqrt {3}$ &\tiny $1/6\,\sqrt {3}$ &\tiny $-1/3\,\sqrt {3}$ &\tiny $-1/6\,\sqrt {3}$ &\tiny $-1/6\,\sqrt {3}$ &\tiny $1/3\,\sqrt {3}$ &\tiny $0$ &\tiny $0$ &\tiny $0$\\[1pt]
 & &\tiny $\young(125,34)$ &\tiny $1/3\,\sqrt {3}$ &\tiny $-1/6\,\sqrt {3}$ &\tiny $-1/6\,\sqrt {3}$ &\tiny $0$ &\tiny $-1/6\,\sqrt {3}$ &\tiny $-1/6\,\sqrt {3}$ &\tiny $0$ &\tiny $1/3\,\sqrt {3}$ &\tiny $0$ &\tiny $0$\\[1pt]
 & &\tiny $\young(135,24)$ &\tiny $0$ &\tiny $1/2$ &\tiny $-1/2$ &\tiny $0$ &\tiny $-1/2$ &\tiny $1/2$ &\tiny $0$ &\tiny $0$ &\tiny $0$ &\tiny $0$\\[1pt]
\cline{2-13}
\hline\hline

\end{tabular}%
\caption[Schur basis of $\mathcal{H}_2^5$]{Schur basis $\{ \ket{W_\kappa^{(\nu)} Y_m^{(\nu)}} \}$ of $\mathcal{H}_2^{\otimes 5}$.}\label{tab:schurq2n5}
\end{table}

\subsection{Application: Communication without a Shared Reference Frame}
An important application for the Schur transform in the context of quantum information theory is classical and quantum communication without a shared reference frame \cite{BRS03,BRS07}.
Let us restrict ourselves to the case where two parties, say Alice and Bob, are connected via an ideal quantum channel transmitting qubits.
If they don't share a common reference frame, the action of the quantum channel is to apply a random change of the computational basis spanning the Hilbert space $\mathcal{H}_2$ of the qubits. When Alice sends $n$ qubits in the state $\rho\in\mathcal{S}(\mathcal{H}_2^{\otimes n})$, Bob receives the state
\begin{equation}
 \mathcal{M}_n (\rho) = \int U^{\otimes n } \, \rho \, U^{\dagger\otimes n} \, dU.
\end{equation}
Using the Schur basis, the Hilbert space of $n$ qubits decomposes as in equation \eqref{eq:h2n:decomposes} and Schur's lemma assures that the action of $\mathcal{M}_n$ can be written as
\begin{equation}
\mathcal{M}_n  =
\sum_{j=0,1/2}^{n/2}
 \bigl(
 \mathcal{D}^{(j)}_{2j+1}
 \otimes
 \mathcal{I}^{(j)}_{h_j(\textsf{S}_n)}
 \bigr)\cdot \Pi_j,
\end{equation}
where $\Pi_j$ denotes the projection on the Young diagram $j$ and $\mathcal{D}^{(j)}_{2j+1}$ denotes the complete depolarizing channel on the $2j+1$ dimensional tensor space of the irreps of $\textsf{SU}_2$.

To transmit classical information to Bob, Alice chooses a normalized state
\begin{equation}
\ket{\Gamma^{(j)}_{m_i}} = \sum_{k=-j}^{+j} \alpha_k \ket{ W^{(j)}_k } \ket{ Y^{(j)}_{m_i} }, \qquad \alpha_k\in\mathbb{C},
\end{equation}
for each Young diagram $j=0,1/2\, \dots \, n$ and Young tableau $Y^{(j)}_{m_i}$, $i=1\dots h_j(\textsf{S}_n)$ (for example $\alpha_k=\delta_{k,+j}$).
Altogether there are
\begin{equation}
 c_n = \sum_{j=0,1/2}^{n/2} h_j(\textsf{S}_n) = \begin{cases}
 \binom{n}{n/2} & \text{if $n$ is even} \\
 \binom{n}{n/2+1/2} & \text{if $n$ is odd}
                                       \end{cases}
\end{equation}
such states. Bob can identify these states by a measuring the Young diagram and Young tableau since
\begin{equation}
\mathcal{M}_n \bigl( \ket{\Gamma^{(j)}_{m_i}}\bra{\Gamma^{(j)}_{m_i}} \bigr) = \frac{1}{2j+1}\mathcal{I}\otimes  \ket{ Y^{(j)}_{m_i} }\bra{ Y^{(j)}_{m_i} }.
\end{equation}
Asymptotically, the rate at which Alice is able to send classical information to Bob tends to one,
\begin{equation}
 \lim_{n\rightarrow\infty} \frac{\log_2(c_n)}{n} \approx 1-\frac{1}{2n}\log_2(n).
\end{equation}
As an example, consider the Schur basis of five qubits in table \ref{tab:schurq2n5}. Here, $c_5=1+4+5=10$ and Alice is able to send classical information at a rate $\approx 0.66$.

To transmit quantum information to Bob, Alice encodes the information into the subsystem spanned by the $\{ \ket{ Y^{(j)}_{m_i} } \}_{i=1\dots h_j(\textsf{S}_n)}$ with the largest dimension $h_j(\textsf{S}_n)$, i.\,e. she prepares a state
\begin{equation}
 \sigma\otimes\rho = \sum_{k,k'=-j}^{+j} \sigma_{kk'} \ket{W^{(j)}_k}\bra{W^{(j)}_{k'}} \otimes \sum_{i,i'=1}^{h_j(\textsf{S}_n)} \rho_{ii'} \ket{Y^{(j)}_{m_i}}\bra{Y^{(j)}_{m_{i'}}}
\end{equation}
with arbitrary $\sigma_{kk'}$.
Bob receives the state
\begin{equation}
 \mathcal{M}_n (\sigma\otimes\rho)=
 \frac{1}{2j+1}\sum_{k=-j}^{+j} \ket{W^{(j)}_k}\bra{W^{(j)}_k} \otimes \sum_{i,i'=1}^{h_j(\textsf{S}_n)} \rho_{ii'} \ket{Y^{(j)}_{m_i}}\bra{Y^{(j)}_{m_{i'}}} = \frac{1}{2j+1} \mathcal{I} \otimes \rho.
\end{equation}
For large $n$, $h_j(\textsf{S}_n)$ becomes maximal for $j_\text{max}=\sqrt{n}/2$.
Again the rate at which Alice is able to send quantum information to Bob asymptotically tends to one,
\begin{equation}
 \lim_{n\rightarrow\infty} \frac{\log_2\bigl( h_{j_\text{max}}(\textsf{S}_n) \bigr)}{n} \approx 1-\frac{1}{2n}\log_2(n).
\end{equation}
For our example of five qubits, the largest dimension is $h_{1/2}(\textsf{S}_5)=5$ and Alice is able to send qubits at a rate $\approx 0.46$.

\backmatter

\bibliography{/home/okern/Documents/references}

\selectlanguage{german}
\chapter{Danksagung}

Die vorliegende Arbeit wurde in der Arbeitsgruppe von Herrn Prof. Gernot Alber angefertigt,
dem ich an dieser Stelle daf\"ur danken m\"ochte, mir die Gelegenheit gegeben zu haben, in seiner Arbeitsgruppe mitzuarbeiten.

Des Weiteren gilt mein Dank Herrn Prof. Dima L. Shepelyansky f\"ur die produktive Zusammenarbeit im Rahmen des EU Projekts EDIQIP, und f\"ur die Gelegenheit neben seiner Arbeitsgruppe \glqq Quantware\grqq{} in Toulouse auch die folgenden Veranstaltungen zu besuchen:
Im Rahmen der International School of Physics \glqq Enrico Fermi\grqq{} in Varenna das Programm \glqq Quantum Computers, Algorithms and Chaos\grqq{} vom 5. bis 15. Juli 2005, und das Trimester \glqq Quantum information, computation, and complexity\grqq{} am Institut Henri Poincar\'e in Paris vom 4. Januar bis 7. April 2006.

Bedanken m\"ochte ich mich auch bei Herrn Prof. Igor Jex, dessen Arbeitsgruppe in Prag ich mehrfach besuchen konnte, und insbesondere bei seinen Studenten Stanislav Vym\u{e}tal und Pavel Ba\u{z}ant f\"ur interessante Diskussionen.

Herrn Prof. Thomas H. Seligman gilt mein Dank f\"ur die Einladung zur Konferenz \glqq Decoherence: Measures, models and semi-classics\grqq{} in Cuernavaca, Mexiko, vom 9. bis 22. September 2007.

Mein besonderer Dank gilt nat\"urlich allen Mitgliedern meiner Arbeitsgruppe f\"ur die nette Zusammenarbeit und die zahlreichen Diskussionen.
F\"ur das Korrekturlesen samt hilfreichen Kommentaren seien (in alphabetischer Reihenfolge) Kedar Ranade, Joseph Renes und Ulrich Seyfarth nochmal gesondert erw\"ahnt.
Ebenfalls besonderer Dank gilt Herrn Prof. J\"urgen Berges f\"ur die \"Ubernahme des Korreferats.

\selectlanguage{american}
\chapter{Curriculum Vitae}
\begin{cv}{} %
  \begin{cvlist}{Personal Data}
  \item Oliver Kern

  \item  Email: oliver.kern@physik.tu-darmstadt.de

  \item Born: March 6th, 1978 in Mainz (Germany)\\
   German citizen
  \end{cvlist}
  \begin{cvlist}{Education}
  \item[06/1984--07/1988]Friedrich Fr\"obel Schule, Primary School, (Grundschule des Kreises Offenbach)
  \item[08/1988--07/1994]Hermann Hesse Schule, Secondary School, (Gesamtschule des Kreises Offenbach)
  \item[08/1994--06/1997]Claus von Stauffenberg Schule, Secondary School, (Gymnasiale Oberstufenschule des Kreises Offenbach),\\
  Higher Education Entrance Qualification\\
  (Main Subjects: Mathematics and Physics)
  \end{cvlist}
  \begin{cvlist}{Civilian Service}
  \item[09/1997--09/1998]German Red Cross Blood Donation Service
  \end{cvlist}
  \begin{cvlist}{Higher Education}
  \item[10/1998--09/2000] Pre-Diploma in Physics, Technical University Darmstadt
  \item[10/2000--09/2004] Diploma in Physics, Technical University Darmstadt, Institute of Applied Physics\\
  (Main focus: Quantum Information Theory)
  \end{cvlist}
  \begin{cvlist}{Ph.\,D. Studies}
  \item[09/2004] Beginning of Ph.\,D. Studies at the Technical University Darmstadt
   under supervision of Prof. Dr. G. Alber.
  \item[01/2006--04/2006] Marie Curie fellowship within the program `quantum information, computation, and complexity' which took place at the Institut Henri Poincar\'e in Paris.
  \end{cvlist}
\renewcommand*{\cvbibname}{List of Publications\\}

\end{cv}

\selectlanguage{german}
\addchap*{Erkl\"arung}
\vspace{\baselineskip}

Hiermit erkl\"are ich an Eides Statt, da\ss{} ich
die vorliegende Dissertation selbst\"andig, nur unter Verwendung der angegebenen Quellen und Hilfsmittel verfa\ss{}t habe.
Ich habe bisher keinen Versuch unternommen, an einer
anderen Hochschule das Promotionsverfahren einzuleiten.

\vspace{2\baselineskip}
\noindent
Darmstadt, den \einreichdatelang \hfill Oliver Kern

\end{document}